\documentclass[a4paper,twoside,11pt,useAMS,usergraphicx]{Classes/CUEDthesisPSnPDF1}

\usepackage{ulem}
\usepackage{bm}

\usepackage{amssymb,amsthm}
\usepackage{StyleFiles/watermark}
\usepackage{lscape}
\usepackage{graphicx}
\usepackage{pdfpages}
\usepackage{wasysym}
\usepackage{contour}
\usepackage{xspace}	
\usepackage{endnotes}
\usepackage[sort]{natbib}
\usepackage{enumitem}
\usepackage{multirow}
\usepackage{pifont}

\usepackage{hyperref}

\newcounter{daggerfootnote}

\usepackage{multicol}
\usepackage{lipsum}
\usepackage{float}
\usepackage{adjustbox}

\newcommand\blfootnote[1]{%
  \begingroup
  \renewcommand\thefootnote{}\footnote{#1}%
  \addtocounter{footnote}{-1}%
  \endgroup
}
\bibpunct{(}{)}{;}{a}{,}{,}
\usepackage{subfig}
\usepackage{setspace}
\usepackage{booktabs, caption, makecell}

\usepackage{latexsym}
\usepackage{threeparttable}
\usepackage{longtable}

\def\msun{{M}_{\odot}}

\defcitealias{sle76}{SLE76}
\defcitealias{kc18}{KSC18b}
\defcitealias{kc18jaa}{KSC18a}

\definecolor{dark-gray}{gray}{0.1}
\begin{document}

\renewcommand\baselinestretch{1.2}
\baselineskip=18pt plus1pt

\setcounter{secnumdepth}{3}
\setcounter{tocdepth}{3}

\setcounter{secnumdepth}{4}
\setcounter{tocdepth}{4}
\frontmatter 

\cleardoublepage
\thispagestyle{empty}
\begin{titlepage}
\centering

\begin{doublespacing}

{\huge \bf Multiwavelength Observations of Gamma Ray Bursts}\\

\vspace*{1.4cm}
\includegraphics[width = 40mm]{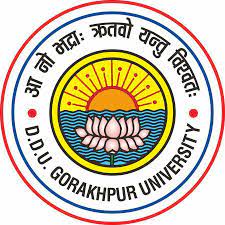} \\
\vspace{0.30cm}
{\large \bf THESIS} \\
{\large \sc Submitted For the Degree of } \\
\vspace{0.30cm}
{\LARGE \bf DOCTOR OF PHILOSOPHY}\\
\vspace{0.15cm}
{\large \sc in } \\
\vspace{0.25cm}
{\LARGE \bf PHYSICS} \\
\vspace{0.30cm}
{\large \sc by} \\
\vspace{0.5cm}
{\large \bf RAHUL GUPTA} \\
{\large \bf DEPARTMENT OF PHYSICS}\\
{\large \bf  DEEN DAYAL UPADHYAYA GORAKHPUR UNIVERSITY}\\
{\large \bf  GORAKHPUR-273009 (U.P.) INDIA}\\
\vspace{0.60cm}

{\large \sc Research Centre}\\
\vspace{0.1cm}
{\large \bf ARYABHATTA RESEARCH INSTITUTE OF OBSERVATIONAL SCIENCES (ARIES)} \\
{\large \bf MANORA PEAK, NAINITAL-263001 INDIA}\\
\end{doublespacing}

\vspace{1cm}
{\large \bf SEPTEMBER 2023}\\

\end{titlepage}

\cleardoublepage
\thispagestyle{empty}

\vspace*{8cm}
\begin{center}
\centering   
\hspace*{0.0cm}{\fontsize{80}{80} \selectfont \it  Dedicated to,}

\vspace*{1.0cm}

\hspace*{4.0cm}{\fontsize{80}{80} \selectfont \it  My Family ......}\\

\end{center}

\pagenumbering{roman}


\begin{center}
\LARGE \bf {Acknowledgements} \\
\bigskip
\normalsize
\end{center}

\noindent 

I belong to a small village called ``Balua" in District Chandauli, Uttar Pradesh, India. This village is located near the bank of river Ganga and connects District Chandauli to Varanasi (also known as Banaras, ``The spiritual capital of India"). From the very beginning of my high school days, I was highly motivated toward science in general and astronomy in particular. My family always supported me in fulfilling my goals. I joined as a Ph.D. student on 16-07-2018 at Aryabhatta Research Institute of Observational Sciences (ARIES), Nainital, India, and started working under the supervision of Dr. Shashi Bhushan Pandey. The early days were challenging for me; however, I worked hard day and night and achieved the major goals of my Ph.D. synopsis. I express my deepest thanks to my Ph.D. supervisor {\bf Dr. Shashi Bhushan Pandey} (Scientist-F, ARIES Nainital), and Co-supervisor, {\bf Prof. Sugriva Nath Tiwari} (Former Head of the Department of Physics, Dean of the Faculty of Science and Dean of the Faculty of Engineering \& Technology, Deen Dayal Upadhyay Gorakhpur University, Gorakhpur), for their continuous support and guidance in this exciting journey. Their valuable suggestion and guidelines helped me a lot to complete this thesis in a beautiful manner. I am deeply grateful to them for allowing me to explore my intellectual curiosity in my work and providing me with a critical foundation. I will always be thankful to my supervisor Dr. S. B. Pandey, for encouraging me to work hard and providing an excellent opportunity to explore and collaborate with leading experts in the GRB field. I am also very thankful to Dr. S. B. Pandey's wife (ma'am) and their children for giving me a family-like environment at ARIES. I am very grateful to my co-supervisor, Prof. Sugriva Nath Tiwari; he helped me a lot during my Ph.D. at the Department of Physics, DDUGU, Gorakhpur. 

I am also very thankful to {\bf Prof. Alberto J. Castro-Tirado, Prof. A. R. Rao, Prof. Dipankar Bhattacharya, Dr. S. R. Oates, Prof. Soebur Razzaque, Dr. Tanmoy Chattopadhyay, Prof. Judith Racusin, Prof. Bin-Bin Zhang, Dr. MD Caballero-García, Dr. Youdong Hu, Dr. Vidushi Sharma, Dr. Rupal Basak, Dr. Jagdish Joshi, Dr. Vikash Chand, Prof. Varun Bhalerao, Dr. Shabnam Iyyani, Prof. Gudlaugur Johannesson, Prof. Nicola Omodei, Prof. Peter Veres, Prof. V. Lipunov, Prof. David Buckley, Prof. Alexei Pozanenko, Prof. Martino Marisaldi, Prof. Vladimir V. Sokolov, Dr. Azamat F. Valeev, Dr. Pavel Minaev, Dr. Brajesh Kumar, Miss. Soumya Gupta, Mr. Harsh Kumar, and Miss. Divita Saraogi} for their kind support and help at various stages during my Ph.D. I have greatly benefited from my association with them. I am incredibly grateful to Prof. Alberto J. Castro-Tirado and Dr. S. R. Oates; they were very supportive throughout this exciting journey. Further, I would like to thank the whole {\bf AstroSat CZTI, Fermi-LAT, ASIM-ISS, GTC, and DOT teams} for giving me an excellent platform to work under the guidance of well-known experts in the field. 

I thank all the faculty members ({\bf Prof. Dipankar Banerjee, Dr. Brijesh Kumar, Dr. Alok Chandra Gupta, Dr. Jeewan C. Pandey, Dr. Ramakant S. Yadav, Dr. Saurabh, Dr. Kuntal Misra, Dr. Neelam Panwar, Dr. Indranil Chattopadhyay, Dr. Snehlata, Dr. Santosh Joshi, Dr. Yogesh Chandra Joshi, Dr. Manish K. Naja, Dr. Narendra Singh, Prof. Shantanu Rastogi (Registrar, DDUGU, Gorakhpur), Prof. Ravi Shankar Singh, Prof. Umesh Yadava, Dr. Apara Tripathi, Dr. Prabhunath Prasad, Dr. Dipendra Sharma, and others}), post-doctoral fellows, research scholars ({\bf Mr. Prayagraj Singh, Dr. Rahul Kumar Anand, Mr. Vishnu Patel, Mr. Atul Kumar Singh, Mr. Vaibhav Kumar, Mr. Teekendra Kumar Sahu, Mr. Bakhtawar H. Abdullah, Mr. Sumit Kumar, and others}), and staff members {\bf (Amar Singh Meena, Abhishek Sharma, Himanshu Vidyarthi, Ram Dayal Bhatt, Hemant Kumar, Arjun Singh, and others)} of ARIES Nainital and the Department of Physics of DDU Gorakhpur University Gorakhpur profusely for their kind help and support throughout my study period. In addition, I acknowledge the financial support of the {\bf BRICS grant and the AstroSat archival Data utilization program funded by the Department of Science and Technology (DST), New Delhi, India, and the Indian Space Research Organisation (ISRO), Bengaluru, India, respectively.} I am also thankful to {\bf Science and Engineering Research Board (SERB)} for providing financial support to attend the 10th international Fermi symposium under International Travel Support (ITS) Scheme.

I thank my group members ({\bf Amar Aryan, Amit Kumar, and Amit Kumar Ror}). They helped me a lot in this journey. I am very thankful to my dear friend {\bf Amar}, with whom I enjoyed my stay at Nainital. We used to discuss academic as well as personal matters positively. We have a lot of good memories from ARIES Nainital. I also thank my batch-mates ({\bf Amar, Raj, Mahendar, Nikita, and Akanksha}), seniors ({\bf Mukesh bhaiya, Anjasha di, Ashwani bhaiya, Vineet bhaiya, Pankaj bhaiya, Ankur, Arpan, Krishan, Alex, Prajjwal, Aditya, Vinit, Jaydeep, Vibhore, and others)}, juniors ({\bf Amit Kumar Ror, Devanand, Mrinmoy, Bhavya, Kiran, Shubham, Tushar, Nitin, Gurpreet, Naveen, Aayushi, Shivangi, Dibya, and others}) and canteen members at ARIES and Devasthal campus, who have made this journey even more memorable. 

I want to express my gratitude to my family for their unwavering support throughout my Ph.D. years and during all of the major decisions in my life. My respected grandmother ({\bf Sharada devi}), my father ({\bf Surendra Kumar Gupta}), mother ({\bf Kanchan Gupta}), brothers ({\bf Rajan bhaiya, Vishal, Deepu bhaiya, Rinku bhaiya, and others}), my sister-in-law ({\bf Soni Gupta}), sisters ({\bf Pooja di, Baby di, and Amrita}), uncle ({\bf Prem Shankar Gupta}), maternal uncle ({\bf Shashi Kant Gupta, Amit Gupta and others}) and other relatives have been strong support throughout my life. This journey was also made possible by my supportive family despite several difficulties and medical emergencies in my family. I want to thank my sweet love {\bf Madhu} and my friends {\bf Manish, Manvendra, Raj Kumar, Nauratan, Pankaj, Shashank, Nagendra, Nitish Ji, Ashwani, Poonam, Pooja, Amrit Lal, Abhishek, Amalesh, Ajay, Shyam, Bhartendu, Jayashree, Rashmita, Neda, Lalit, Soumyarup, Subhadip, Subhranil, Suchit, Rakesh (Jimmy), Happy, Rambahadur, Amrit, Amit, Sharad, Piyush, Rajat, Prabhakar, Prince, Akhilesh, Ram Ashish, Sheshanath, Subham, Debasish, and many others} for all their care, help and support. 

I am very much grateful to my learned teacher {\bf Prof. T. V. Ramakrishnan} (FRS), for his valuable inspiration for pursuing Physics in higher studies during my B.Sc. days at Banaras Hindu University (BHU). I express my sincere gratitude to {\bf Dr. Kaushal Kumar} and all the faculty members of the Department of Applied Physics, Indian Institute of Technology (Indian School of Mines), Dhanbad, for their valuable inspiration during my master's days. I am also grateful to {\bf Prof. Nandita Srivastava} and all the members ({\bf Dr. Alok Ranjan Tiwari, Dr. Upendra Kumar Singh Kushwaha, Dr. Sajal Kumar Dhara, Dr. Rahul Yadav, and others}) of the Udaipur Solar Observatory, Physical Research Laboratory, for inspiring me towards research.  

As a researcher, it has been an emotional adventure that includes highs and lows, and keeping my sanity during the journey was quite difficult. In every situation, I always relied on myself and was confident that I would be able to solve the problem. Thank everyone who helped me along the way - without your support, this thesis would have been impossible. I express my deepest gratitude to my dear friends who helped me complete this thesis. Finally, I would like to thank God for everything.






\begin{center}
{\LARGE \bf ABSTRACT} \\
\bigskip
\end{center}

\noindent 

Gamma-ray bursts (GRBs) are fascinating sources studied in modern astronomy. They are extremely luminous electromagnetic ($\rm L_{\gamma, iso}$ $\sim$ $10^{48}$ $-$ $10 ^{54}$ erg/s) explosions in the Universe observed from cosmological distances. These unique characteristics provide a marvellous chance to study the evolution of massive stars and probe the rarely explored early Universe. In addition, the central source's compactness and the high bulk Lorentz factor in GRB's ultra-relativistic jets make them efficient laboratories for studying high-energy astrophysics. GRBs are the only astrophysical sources observed in two distinct signals: gravitational and electromagnetic waves. GRBs are believed to be produced from a ``fireball" moving at a relativistic speed, launched by a fast-rotating black hole or magnetar. GRBs emit radiation in two phases: the initial gamma/hard X-rays prompt emission, the duration of which ranges from a few seconds to hours, followed by the multi-wavelength and long-lived afterglow phase. Based on the observed time frame of GRB prompt emission, astronomers have generally categorized GRBs into two groups: long ($>$ 2 s) and short ($<$ 2 s) bursts. Short GRBs are typically produced when two compact objects merge, while long GRBs could result from a collapsing massive star. Despite the discovery of GRBs in the late 1960s, their origin is still a great mystery. There are several open questions related to GRBs, such as: What are the possible progenitors? What powers the GRBs jets/central engine? How to classify them? The short bursts originating from the collapse or long bursts presenting features characteristic of compact binary mergers have posed a new challenge to our understanding of possible progenitors and the origin of these events. What is the jet composition? Is it a baryon-dominated or Poynting-flux-dominated outflow? What is the underlying emission process that gives rise to observed radiation? Where and how does the energy dissipation occur in the outflow? Is it via internal shocks or magnetic reconnections? How to solve the radiative efficiency problem? What are the possible causes of Dark GRBs and orphan afterglows? How to investigate the local environment of GRBs? etc. In this thesis, we explored some of these open problems using multi-wavelength observations. 

To examine the jet composition and radiation physics of prompt emission, we have used a unique time-resolved spectro-polarimetric technique using the prompt observations of some of the bright GRBs discovered using {\it Fermi} and {\it AstroSat} Cadmium Zinc Telluride Imager (CZTI) instruments. We conducted a time-resolved spectral analysis of a very luminous burst (GRB 210619B) detected using {\it Fermi} Gamma-ray Burst Monitor (GBM) and Atmosphere-Space Interactions Monitor (ASIM) instruments. This GRB has a very bright and hard pulse followed by a fainter and softer pulse. Our spectral analysis shows that the first pulse has very hard values of low-energy photon indices followed by softer values during the weaker episodes. This indicates a possible thermal to non-thermal transition of the jet composition within the burst. For another one of the most energetic bursts (GRB 190530A), we carried out a detailed time-resolved spectro-polarimetric analysis using simultaneous {\it Fermi} and {\it AstroSat} CZTI observations. During the first two pulses, the values of low-energy photon indices were in agreement with poynting flux-dominated synchrotron scenarios. Our polarization analysis was also consistent with the same scenarios, i.e., synchrotron emission in an ordered orientation of magnetic field vectors. Our study suggests that spectro-polarimetry of the prompt emission of GRBs can solve the emission mechanisms and jet composition of GRBs.       

The internal shock model of GRBs is inefficient in converting the jet's kinetic energy into gamma-ray radiation, known as a low-efficiency problem. This problem can be resolved using very early optical afterglow follow-up observations and modelling of rarely observed reverse shock emissions in earlier phases. For GRB 140102A, we detected very early optical afterglow, even before the X-Ray Telescope (XRT) and Ultra-violet Optical Telescope (UVOT), using the BOOTES robotic telescope. Our broadband afterglow modelling shows that early optical emission is dominated by a reverse shock moving toward ejecta. Late-time emission is dominated by forward shock emission driving in the opposite direction of ejecta. Our modelling constrains a lower electron equipartition parameter of the reverse shock component, resulting in a very high value of radiative efficiency. Therefore, early observations and modelling of such cases may help to lighten the low-efficiency problem. Furthermore, we compared the physical properties of GRB 140102A with a complete sample of thin shell reverse shock-dominated GRBs and noted that such GRBs might have a wide range of magnetization.  

We studied the dark GRBs and orphan afterglow properties using late-time afterglow data. We examined the characteristics of two dark bursts (GRB 150309A and GRB 210205A). The detailed analysis of these GRBs reveals that GRB 150309A is one of the most dust-extinguished GRBs to date, and local dust in the host galaxy might be the potential reason for its optical darkness. In the case of GRB 210205A, we estimated a lower value of dust extinction in the local host using joint optical to X-ray SED analysis; therefore, either intrinsic faintness or high redshift might be the possible origin of its optical darkness. We examined an orphan afterglow's characteristics (AT20221any) and compared its brightness with other known cases of orphan afterglows. Our detailed multi-wavelength modelling of AT20221any suggests that it was observed on-axis. Still, no gamma-ray detecting satellite could detect the prompt emission either because the source was not in their field of view or due to limited sensitivity for fainter GRB detection.  

Further, we studied a sample of host galaxies of five peculiar GRBs observed using 3.6\,m Devasthal Optical Telescope (DOT) of ARIES to explore the environment of GRBs. We compared the physical parameters of these galaxies with well-studied host galaxies of long and short GRBs. We noted that most of the bursts in our sample have a massive galaxy with a high star formation rate (SFR), and only one burst (GRB 030329) belongs to a rare low-mass host galaxy with a low SFR. Our study demonstrated the capabilities of 3.6\,m DOT for faint sources observations such as host galaxies of GRBs.


The community is actively developing several larger optical telescopes, such as the Extremely Large Telescope (ELT) and Thirty Meter Telescope (TMT), alongside currently operational 10-meter class telescopes like 10.4m GTC. Simultaneously, on the high-energy front, the community is working towards the next gamma-ray missions (e.g., COSI, POLAR2, DAKSHA). The synchronous observations of GRBs and related transients utilizing these upcoming major ground and space-based facilities will play a pivotal role in unravelling the intricate details of these energetic and cosmic sources. Our study on the prompt emission of GRBs, as well as the comprehensive analysis of their early-to-late time afterglow observations, modelling, and host galaxy investigations, provides valuable insights for future observations of similar sources using upcoming larger telescopes. Overall, the research presented in this thesis contributes significantly to the field of GRB and paves the way for further progress in our understanding of this extraordinary cosmic phenomenon under a larger perspective of time-domain astronomy. Looking ahead, the combined observations of GRBs across various messengers hold immense potential for unravelling the mysteries of GRBs. These multi-messenger observations of GRBs and related transients will shed light on the fundamental processes occurring in the Universe, opening new avenues of exploration and discovery.

\label{NOTATIONS AND ABBREVIATIONS}

\begin{center}
\Large{\bf Abbreviations and Acronyms}
\end{center}

\begin{tabbing}
cccccccccccccccccccccccccccccccc\=ccccccccccccccccccccccc\=cccccccccccccccccccccccccccc
cccccccccccccccccccccccccccc \kill

\textbf{ARIES} \> \textbf{A}ryabhatta \textbf{R}esearch \textbf{I}nstitute of observational Scienc\textbf{ES} \> \\    
\textbf{BAT} \> \textbf{B}urst \textbf{A}lert \textbf{T}elescope \> \\    
\textbf{BATSE} \> \textbf{B}urst \textbf{a}nd \textbf{T}ransient \textbf{S}ource \textbf{E}xperiment \> \\    
\textbf{BH} \>\textbf{B}lack \textbf{H}ole \> \\
\textbf{CCD}(s) \> \textbf{C}harge \textbf{C}ouple \textbf{D}evice(s) \> \\    
\textbf{CGRO} \> \textbf{C}ompton \textbf{G}amma-\textbf{R}ay \textbf{O}bservatory \> \\    
\textbf{Dec} \> \textbf{Dec}lination \> \\    
\textbf{DAOPHOT} \> \textbf{D}ominion \textbf{A}strophysical \textbf{O}bservatory \textbf{Phot}ometry \> \\    
\textbf{DFOT} \> \textbf{D}evasthal \textbf{F}ast \textbf{O}ptical \textbf{T}elescope \> \\    
\textbf{EM} \> \textbf{E}lectro\textbf{m}agnetic \> \\    
\textbf{Eq} \>\textbf{Eq}uation\> \\
\textbf{Fig} \>\textbf{Fig}ure\> \\
\textbf{FITS} \> \textbf{F}lexible \textbf{I}mage \textbf{T}ransport \textbf{S}ystem \> \\    
\textbf{FoV} \> \textbf{F}ield \textbf{o}f \textbf{V}iew \> \\    
\textbf{FWHM} \> \textbf{F}ull \textbf{W}idth at \textbf{H}alf \textbf{M}axima \> \\    
\textbf{GRBs} \>\textbf{G}amma-\textbf{R}ay \textbf{B}urst\textbf{s}\> \\
\textbf{GCN} \> \textbf{G}RB \textbf{C}oordinates \textbf{N}etwork \> \\    
\textbf{GeV} \> \textbf{G}iga-\textbf{e}lectron \textbf{V}olt \> \\    
\textbf{HCT} \> \textbf{H}imalayan \textbf{C}handra \textbf{T}elescope \> \\    
\textbf{HEASARC} \> \textbf{H}igh \textbf{E}nergy \textbf{A}strophysics \textbf{S}cience \textbf{A}rchive \textbf{R}esearch \textbf{C}enter \> \\  
\textbf{HESS} \> \textbf{H}igh \textbf{E}nergy \textbf{S}tereoscopic \textbf{S}ystem \> \\    
\textbf{HR} \> \textbf{H}ardness \textbf{R}atio \> \\    
\textbf{HST} \> \textbf{H}ubble \textbf{S}pace \textbf{T}elescope \> \\    
\textbf{HETE}-II \> \textbf{H}igh \textbf{E}nergy \textbf{T}ransient \textbf{E}xplorer-II \> \\    
\textbf{INTEGRAL} \> \textbf{Inte}rnational \textbf{G}amma-\textbf{r}ay \textbf{A}strophysics \textbf{L}aboratory \> \\    
\textbf{IPN} \> \textbf{I}nter \textbf{P}lanetary \textbf{N}etwork \> \\    
\textbf{IR} \> \textbf{I}nfra\textbf{r}ed \> \\    
\textbf{IRAF} \> \textbf{I}mage \textbf{R}eduction \textbf{A}nalysis \textbf{F}acilities \> \\    
\textbf{ISM} \> \textbf{I}nter \textbf{S}tellar \textbf{M}edium \> \\    
\textbf{ISRO} \>\textbf{I}ndian \textbf{S}pace \textbf{R}esearch \textbf{O}rganisation \> \\
\textbf{JD} \> \textbf{J}ulian \textbf{d}ate \> \\    
\textbf{kpc} \> \textbf{K}ilo-\textbf{p}arse\textbf{c} \> \\    
\textbf{MAGIC} \>  \textbf{M}ajor \textbf{A}tmospheric \textbf{G}amma-ray \textbf{I}maging \\ \hspace{5.35cm} \textbf{C}herenkov Telescope \> \\    
\textbf{MIDAS} \> \textbf{M}unich \textbf{I}mage and \textbf{D}ata  \textbf{A}nalysis \textbf{S}ystem \> \\    
\textbf{NASA} \> \textbf{N}ational \textbf{A}eronautics and \textbf{S}pace \textbf{A}dministration \> \\
\textbf{NIR} \> \textbf{N}ear-\textbf{i}nfra\textbf{r}ed \> \\    
\textbf{NS} \>\textbf{N}eutron \textbf{S}tar \> \\
\textbf{PSF}(s) \> \textbf{P}oint \textbf{S}pread \textbf{F}unction(s) \> \\    
\textbf{RA} \> \textbf{R}ight \textbf{A}scension \> \\
\textbf{SED}(s) \> \textbf{S}pectral \textbf{E}nergy \textbf{D}istribution(s) \> \\    
\textbf{SN} \> \textbf{S}uper\textbf{n}ova \> \\    
\textbf{SNe}\> \textbf{S}uper\textbf{n}ova\textbf{e} \> \\    
\textbf{SSC} \> \textbf{S}ynchrotron \textbf{S}elf \textbf{C}ompton \> \\    
\textbf{ST} \> \textbf{S}ampurnanand \textbf{T}elescope \> \\    
\textbf{TeV} \> \textbf{T}era-\textbf{e}lectron \textbf{V}olt \> \\    
\textbf{USNO} \> \textbf{U}nited \textbf{S}tates \textbf{N}aval \textbf{O}bservatory \> \\    
\textbf{UT} \> \textbf{U}niversal \textbf{T}ime \> \\    
\textbf{UV} \> \textbf{U}ltra \textbf{V}iolet \> \\    
\textbf{UVOT} \> \textbf{U}ltra \textbf{V}iolet \textbf{O}ptical \textbf{T}elescope\> \\     
\textbf{VHE} \> \textbf{V}ery \textbf{H}igh \textbf{E}nergy \> \\    
\textbf{WD} \>\textbf{W}hite \textbf{D}warf \> \\
\textbf{XRT} \> \textbf{X}- \textbf{R}ay \textbf{T}elescope \> \\    

\end{tabbing}

\cleardoublepage
\thispagestyle{empty}

\vspace*{5cm}


\hspace*{1.0cm}



%

\hspace*{0.7cm} {{{\Large ``Education is the most powerful weapon which you can use to change the world."}\\}

\smallskip
\vspace{0.05cm}
\hfill{----- Dr. A.P.J. Abdul Kalam}\\

\hspace*{8.0cm}


%


\let\cleardoublepage=\clearpage

\tableofcontents
\listoffigures
\addcontentsline{toc}{chapter}{\sc List~of~figures} 
\listoftables
\addcontentsline{toc}{chapter}{\sc List~of~tables} 

\mainmatter

\renewcommand{\url}[1]{}
\newcommand{\AstroSat}{{\it AstroSat}\xspace}
\newcommand{\fermi}{{\it Fermi}\xspace}
\newcommand{\kw}{{\it Konus}-Wind\xspace}
\newcommand{\kwT}{{T$_{\rm kw,0}$}\xspace}
\newcommand{\swiftT}{{T$_{\rm 0}$}\xspace}
\newcommand{\fermiT}{{T$_{\rm 0}$}\xspace}
\newcommand{\keV}{{\rm keV}\xspace}
\newcommand{\erg}{{\rm ~erg}\xspace}
\newcommand{\s}{{\rm ~s}\xspace}
\newcommand{\swift}{{\it Swift}\xspace}
\newcommand{\tninty}{{$T_{\rm 90}$}\xspace}
\newcommand{\mvts}{{$t_{\rm mvts}$}\xspace}
\newcommand{\lmin}{{$\Gamma_{\rm min}$}\xspace}
\newcommand{\Ep}{$E_{\rm p}$\xspace}
\newcommand{\sw}[1]{\texttt{#1}}

\chapter{\sc GRBs: The brightest electromagnetic cosmic stellar transients}
\label{Ch1}
\ifpdf
    \graphicspath{{Chapter1/Chapter1Figs/PNG/}{Chapter1/Chapter1Figs/PDF/}{Chapter1/Chapter1Figs/}}
\else
    \graphicspath{{Chapter1/Chapter1Figs/EPS/}{Chapter1/Chapter1Figs/}}
\fi

\ifpdf
    \graphicspath{{Chapter1/Chapter1Figs/JPG/}{Chapter1/Chapter1Figs/PDF/}{Chapter1/Chapter1Figs/}}
\else
    \graphicspath{{Chapter1/Chapter1Figs/EPS/}{Chapter1/Chapter1Figs/}}
\fi

Gamma-ray bursts (GRBs) stand out as the most powerful stellar-scale sources studied in modern astronomy. The name of these sources consists of two words: ``Gamma-ray" and ``bursts," and scientists have defined GRBs based on the meaning of these two words. So first, what are gamma rays? Gamma rays are a form of light with the shortest wavelength and highest energy of the electromagnetic spectrum. They are invisible to the human eye and have so much energy that they could harm people on Earth. Due to the absorption of gamma rays by the Earth's atmosphere, we use satellites to study the gamma-ray emission from astrophysical sources. On the other hand, $\it bursts$ means sudden and violent explosions. GRBs are defined as sudden ($\sim$ 10$^{-3}$ to 10$^3$ seconds), violent, and cosmic explosions of gamma rays \citep{1973ApJ...182L..85K, Strong1974, Paczynski:1986ApJ}. GRBs' name is given in the following patterns ``GRB YYMMDD". Suppose a GRB was detected on January 23, 1999; It is named GRB 990123A. If more than one burst is detected on the same day, we change the A for B, C, etc. GRBs emit radiation in two consecutive phases: the initial gamma/hard X-rays ($\sim$ 10 \keV) prompt emission which exhibits durations ranging from a few seconds to several hours, followed by the multi-wavelength and long-lived afterglow phase \citep{1995ARA&A..33..415F, 2000ARA&A..38..379V, 2013FrPhy...8..661G, 2015PhR...561....1K}. 

GRBs are extremely luminous ($\rm L_{\gamma, iso}$ $\sim$ $10^{48}$ $-$ $10 ^{54}$ erg/s) explosions in the Universe observed at cosmological distances \citep{1999PhR...314..575P}. How such a large amount of gamma-ray energies are produced? The observed high energy radiation from these sources and rapid variability in their prompt emission light curves suggest that GRBs may originate from stellar catastrophic events. GRBs are believed to be produced from a ``fireball" moving at a relativistic speed (jetted emission), launched by a fast-rotating black hole (BH) or magnetar \citep{1999PhR...314..575P, 2004RvMP...76.1143P, 2006RPPh...69.2259M}. Although how the energy is distributed within the jet is not known, it is assumed isotropic, but this might not be the case \citep{granot07}. These unique characteristics offer a remarkable opportunity to investigate the evolution of massive stars and probe the rarely explored early Universe. In addition, the compactness of the central source and the ultra-relativistic jets (the high bulk Lorentz factor) of GRBs make them efficient cosmic laboratories for studying high-energy astrophysics in extreme conditions. Also, GRBs are the only astrophysical sources observed in two distinct signals: gravitational and electromagnetic waves. Moreover, GRBs are also thought sources of neutrinos and ultra-high cosmic rays \citep{1997PhRvL..78.2292W, 2008ApJ...677..432W}. Although, no direct evidence of the association between GRBs with neutrinos and ultra-high cosmic rays has been confirmed.

\section{Historical remarks: from Discovery to Multi-Messenger Era (Observational advancement)}
\label{GRB_History} 

The history of GRBs starts during the Cold War in the early 1960s. The United States and the Soviet Union were carrying out larger and larger nuclear tests. Eventually, they realized this was not sustainable, so they signed the limited nuclear test ban treaty in 1963. The critical part of the treaty was that it prohibits carrying out any nuclear weapon test in the atmosphere beyond a considered limit, including outer space. However, despite the treaty, there was a minimal trust level between the U.S. and the Soviet Union, particularly regarding nuclear tests in outer space. Therefore, the Department of Defence of the U.S. started a space program known as Vela satellites \citep{1993AcAau..29..723L}. This military satellite was designed to detect nuclear test signatures in space. They launched these satellites in pairs (beginning in 1963) to triangulate the direction of nuclear tests. In 1967, they suddenly detected large bursts of gamma rays, and many similar events were discovered by Vela satellites in their lifetime.
Utilizing the triangulation capabilities, they noted that these explosions were not due to nuclear tests from outer space or the moon. This information was declassified, and the results were published in 1973 regarding the observations of GRBs originating from cosmic sources. Therefore it was concluding the accidental and serendipitous discovery of GRBs by U.S. military satellites. Furthermore, they noted that these bursts lasted between a tenth of a second to about 30 seconds \citep{1973ApJ...182L..85K, Strong1974}. During this time, the American solar mission named {\it Reuven Ramaty High Energy Solar Spectroscopic Imager} (RHESSI, \citealt{1973ApJ...185L...1C}) and the Soviet space mission called Konus onboard {\it Venera} \citep{1974PZETF..19R.126M, 1981Ap&SS..80....3M} detected several GRBs.

After the discovery of GRBs in 1973, the first two decades (from 1973-1993) are considered a dark era in GRB research until the advent of the {\it Compton Gamma-ray Observatory} launch. As a result of the limited localization capability ($\sim$ few degrees) of GRB-detecting satellites, no counterpart of GRBs could be discovered at longer wavelengths. In this dark era, there was no significant observational development. Still, there were around 118 different developed theories by the early 1990s \citep{1975NYASA.262..164R, 1994ComAp..17..189N}, and at that time, there were less than 100 GRBs known, i.e., more than one theory per event. Most of these theories had some common themes, and the reason is the two essential characteristics of GRBs (short durations and extreme energy) that were known by that time. Some theories suggest new types of exploding stars, giant flares from distant stars, matter/anti-matter annihilation, neutron star-neutron star (NS-NS) collisions, etc \citep{1975NYASA.262..164R, 1981Ap&SS..75..109P, Paczynski:1986ApJ}. There were many theories because of one fundamental problem: the uncertainty in the distance scale of gamma-ray bursts. Gamma-ray satellites were very efficient at detecting these bursts of gamma rays but not precisely localizing them. 

Whether GRBs come from our Milky Way galaxy or distant galaxies was questioned during the dark era. If GRBs come from our Milky Way galaxy, they should line up with the plane of the disk. Therefore, by plotting the distribution of GRBs over the sky, we can get an indication of the distance scale of GRBs without measuring their true distance through spectroscopy. In 1991, NASA launched the Burst and Transient Source Experiment (BATSE) instrument as part of the {\it Compton gamma-ray Observatory} (CGRO) mission \citep{1993A&AS...97....5G}). One of the main objectives of this mission was to discover a significant number of GRBs and localize them \citep{fish94}. BATSE could localize the brighter GRBs with error circles of $\sim$ 0.2 degrees and fainter GRBs with error circles of $\sim$ 18 degrees \citep{1999ApJS..122..503B}. After several years of observations, they built a map with the localization of 2704 GRBs (see Figure \ref{BATSE_distribution}). The all-sky GRBs distribution (isotropic) obtained using BATSE suggests that GRBs might have a cosmological origin \citep{1992Natur.355..143M, 1996ApJ...459...40B}, although it could not solve the mystery related to their distance scale. Using the duration distribution obtained from the BATSE observations, two classes (short and long) of GRBs were found with a partition line at around two seconds \citep{1993ApJ...413L.101K}. Previous observations taken in the pre-BATSE era also hinted at two classes \citep{1981Ap&SS..80....3M, 1984Natur.308..434N}.

\begin{figure}[ht!]
\centering
\includegraphics[scale=0.45]{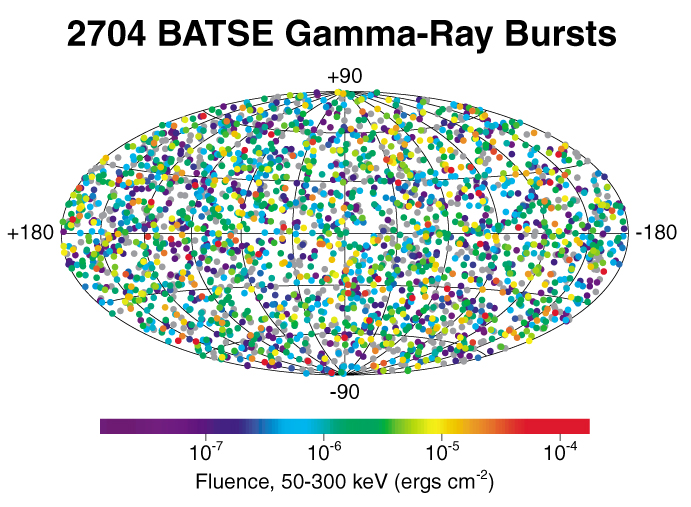}
\caption{The map of the location of around 2700 GRBs discovered by the Burst and Transient Source Experiment. Figure credit: \href{https://heasarc.gsfc.nasa.gov/docs/cgro/batse/}{CGRO Science Support Center}.}
\label{BATSE_distribution}
\end{figure}

Although redshift measurements of GRBs could not be performed at that time, one thing was recognized by many theoretical astronomers. GRBs could act like fire (a lot of energy released in a very short period), which would be followed up by an afterglow emission like glowing embers that would be extinguished after a few hours or days. By studying these afterglows, we have access to longer time scales for GRB studies that we otherwise would not. 
Because they produce most of their radiation at longer wavelengths, we can pinpoint the positions of GRBs much more accurately than using only gamma rays to decipher their origin. The idea of capturing the afterglow came to fruition in 1996 when Dutch and Italian scientists launched a satellite named $BeppoSAX$ \citep{1995SPIE.2517..169P, jager97}. $BeppoSAX$ had both Wide Field Instruments and Narrow Field Instruments, providing the unique opportunity to search large error circles to find the X-ray afterglows of GRBs. $BeppoSAX$ had its first success in February 1997; it discovered a GRB and soon after was re-pointed to focus its X-ray telescope on looking for the afterglow. $BeppoSAX$ mission discovered the first X-ray counterpart of a burst (GRB~970228, \citealt{1997Natur.387..783C}). $BeppoSAX$, thus vindicating the idea that by studying the afterglows, we can determine the origin of GRBs. Subsequently, ground-based follow-up observations for the same burst successfully detected an optical counterpart, which enabled a redshift to be measured at redshift $z$ = 0.695 \citep{1997Natur.386..686V}. These observations confirmed the cosmological origin of GRBs and helped to find the host galaxies/counterparts at different wavelengths \citep{1997Natur.389..261F, 1997Natur.387..878M}. The prolonged duration of afterglows, compared to the prompt emission of GRBs, offers a multi-band platform for studying these energetic bursts in greater detail. Another GRB mission, named High Energy Transient Explorer ($HETE-2$), was launched on 09$^{th}$ October 2000 \citep{2003AIPC..662....3R}. The main objective of this mission was to discover new GRBs and perform follow-up observations to search for afterglows of GRBs \citep{2004NewAR..48..423L}. These two satellites ($BeppoSAX$ and $HETE-2$) localized more than a few hundred bursts and helped to discover their X-ray/optical counterparts and redshift measurements. These missions also helped to solve the mystery of long GRBs. In the error circle of $BeppoSAX$ detected burst (GRB 980425), ground-based optical follow-up observations revealed the detection of type Ic broad line supernova (SN 1998bw; \citealt{1999ApJ...516..788W}) in a nearby galaxy \citep{1998Natur.395..670G}. After the discovery of GRB 980425/SN 1998bw, the late-time optical follow-up observations revealed a supernova bump in the optical light curves of many other nearby GRBs \citep{1999Natur.401..453B, 2000ApJ...536..185G}. After a few years, in 2003, in the error box of $HETE-2$ detected GRB 030329, a robust association of GRB-Supernova (GRB 030329/SN 2003dh) was confirmed using optical spectroscopy \citep{2003ApJ...591L..17S, 2003Natur.423..847H} at the redshift of 0.167. These SNe (SN 1998bw and SN 2003dh) were the first two spectroscopically confirmed GRBs-SNe connection cases. Between these two cases, there were two other nearby GRBs (GRB 011121 connected with SN 2001ke and GRB 021211 connected with SN 2002lt) for which SN bumps were detected based on only photometrically measurements. The discovery of the GRB-Supernova association indicates that long-duration GRBs are formed in active regions of star-forming galaxies and produced from the death of massive stars \citep{2006Natur.441..463F, 2017AdAst2017E...5C}. During $BeppoSAX/HETE-2$ era, the broadband follow-up observations and analysis of afterglows of GRBs were found to be compatible with the theoretical foretelling based on the standard afterglow model \citep{Rees:1992, 1993ApJ...405..278M, 1997ApJ...476..232M, 1998ApJ...497L..17S, Sari:1999ApJ, 1999ApJ...517L.109S, 1999MNRAS.306L..39M}.   

The modern era of GRB research began following the deployment of the Neil Gehrels \swift observatory \citep{2004ApJ...611.1005G} on November 20, 2004 (see more details in chapter 2) with three payloads: the hard X-ray instrument named Burst Alert Telescope (BAT; \citealt{2005SSRv..120..143B}), the soft X-ray instrument named X-ray telescope \citep[XRT;][]{2005SSRv..120..165B}, and Ultra-Violet/Optical instrument named Ultra-Violet/Optical Telescope \citep[UVOT;][]{2005SSRv..120...95R}. \swift's unique slewing capability helped to discover X-ray and optical counterparts for $\sim$ 90 \% and $\sim$ 50 \% bursts (discovered by \swift), respectively \citep{2005ApJ...634..501B, Gehrels2009ARA} using XRT and UVOT. Additionally, arcsec localization information of afterglows obtained using XRT and UVOT is sent to other space-based and ground-based observatories using the Gamma-ray Coordinates Network (GCN\footnote{http://gcn.gsfc.nasa.gov/gcn/}) for quick search and follow-up of the afterglow at different frequencies. The \swift mission was the first to detect the X-ray counterpart of short GRBs \citep{2005Natur.437..845F}. The precise localization of short GRBs further helped to explore the environment of these bursts, and scientists noted that they exhibit notable distinctions compared to those observed in long bursts \citep{2005Natur.437..851G,  2006ApJ...638..354B, 2005Natur.438..994B}. These findings provided further support for the hypothesis that short bursts may originate from a distinct source or mechanism compared to long bursts. In addition, \swift mission identified the canonical nature of X-ray afterglows \citep{2006ApJ...642..354Z}.

In this \swift era, \fermi \citep{2022arXiv221012875T} was launched by NASA in 2008 (see more details in chapter 2). \fermi discovered delayed (longer than prompt emission) GeV emission ($>$ 100 MeV) that has an external shock source \citep{2009MNRAS.400L..75K, 2010MNRAS.409..226K, 2010MNRAS.403..926G}. The composition of the jet (whether it is baryon-dominated, Poynting-flux-dominated, or a hybrid of the two) and physical mechanisms (photospheric, synchrotron, or hybrid) of GRBs are still open questions (see section \ref{Observed_characterises} for more information). The broad spectral coverage (8 \keV -300 GeV) of \fermi helped to explore the jet composition and physical mechanisms of GRBs \citep{Abdo:2009, 2009Sci...323.1688A}. \fermi revealed that three possible spectral components (thermal photospheric, non-thermal \sw{Band}, and sub-GeV cutoff; see section \ref{Observed_characterises} for more details) are present in the GRBs' spectrum \citep{2011ApJ...730..141Z}. \fermi also played a crucial role in the simultaneous discovery of a short GRB (GRB 170817A) and gravitational wave (GW 170817) in a nearby galaxy. This result supports the idea that short GRBs may originate as a consequence of the merger between two compact sources \citep{Abbott_2017, 2017ApJ...848L..14G}. This ground-breaking discovery of a GRB with simultaneous GW emission started the era of ``multi-messenger astronomy". Along with \swift and \fermi, many other successful GRB missions have been launched, for example, the $AGILE$ mission launched by the Italian Space Agency in 2007 \citep{2009AA...502..995T}, and \AstroSat mission \citep{singh:2014} launched by ISRO in 2015 (see more details in chapter 2), etc. In particular, the \AstroSat mission has enhanced our understanding of prompt emission by using polarization measurements for many GRBs \citep{2019ApJ...884..123C, 2022MNRAS.511.1694G, 2022ApJ...936...12C}.

\section{Observed characteristics of GRBs}
\label{Observed_characterises}

GRBs are observed across the electromagnetic spectrum by various space and ground-based telescopes. A list of some of these telescopes is shown in Figure \ref{GRBs_instruments}. This section outlines the distinctive features of both the prompt emission and afterglows observed in GRBs in the context of the present thesis work (see chapters \ref{ch:3}, \ref{ch:5}, \ref{ch:61}, and \ref{ch:6}, for details).

\begin{figure}[ht!]
\centering
\includegraphics[scale=1.4]{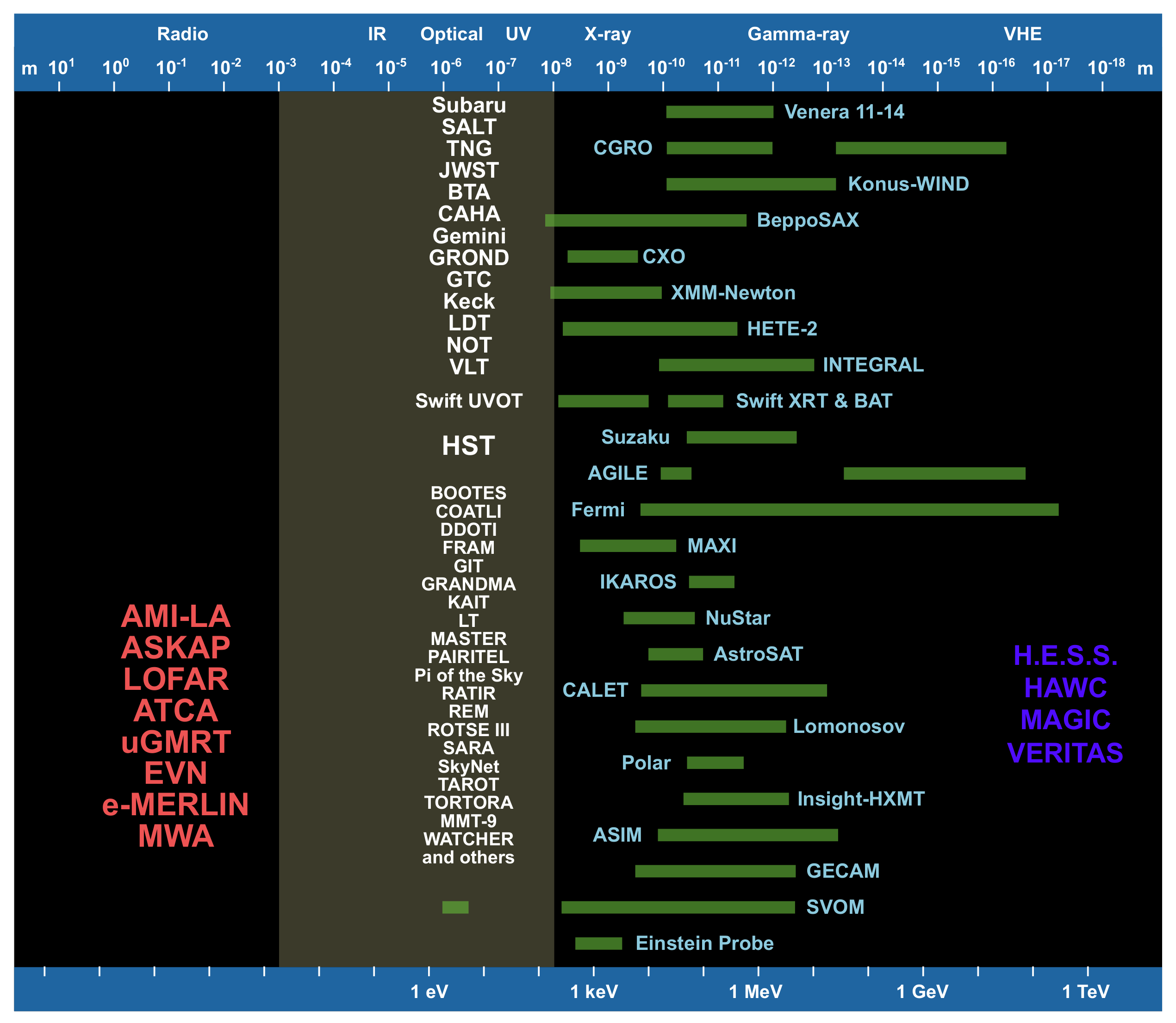}
\caption{Key GRBs observing space and ground-based instruments/telescopes. Figure credit: \cite{2022Univ....8..373T}.}
\label{GRBs_instruments}
\end{figure}

\subsection{Prompt emission}

The prompt emission (short-lived, sub-MeV emissions) of GRBs is primarily detected by different gamma-ray space telescopes, for example, BATSE, Burst Alert Telescope, and the Gamma-ray Burst Monitor (GBM), and Large Area Telescope (LAT). However, it remains a mystery how these bursts produce their prompt emission \citep{1995ARA&A..33..415F, 1999PhR...314..575P, 2014IJMPD..2330002Z, 2015AdAst2015E..22P, 2015PhR...561....1K}. After 50 years of their discovery, this problem has yet to be solved. The prompt emission in GRBs is expected to originate in a relativistic jet via energy dissipation either in internal shocks when fast-moving shells collide with slower shells or due to catastrophic reconfiguration of magnetic fields \citep{2015MNRAS.450..183S}. The comprehension of temporal and spectral properties related to prompt emission has experienced significant advancements following the deployment of dedicated space-based observatories such as \swift and \fermi.

\subsubsection{Temporal properties}

The light curves of the prompt emission of GRBs are noteworthy irregular in nature. A gallery of the prompt emission light curves observed using the BATSE mission is shown in Figure \ref{Prompt_LC_sample}. Some of them are very smooth; some are extremely chaotic; some have fast-rising exponential-decay (FRED) pattern \citep{1996ApJ...459..393N, nor05}, some have single emission episodes, some have multiple emission episodes, some have precursor emission \citep{1995ApJ...452..145K}, some last hundreds of seconds, and some other last for a few milliseconds. To date, a few thousand GRBs have been discovered. However, no two bursts have similar prompt emission light curves. The observed temporal features indicate erratic central engine activity \citep{2014ApJ...789..145H}.  

\begin{figure}[ht!]
\centering
\includegraphics[scale=0.07]{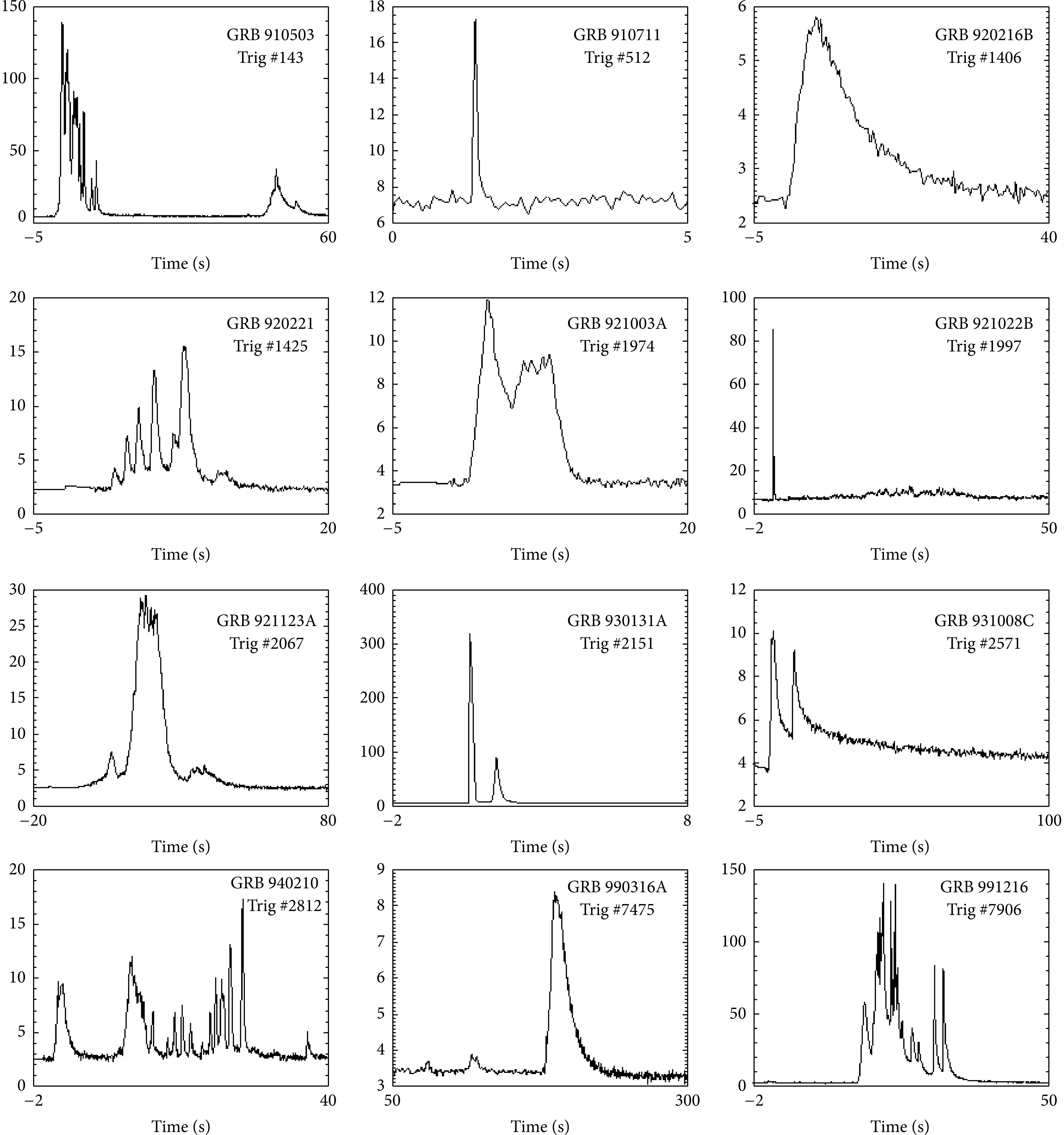}
\caption{A gallery of prompt emission light curves (the detection of gamma-ray photons as a function of time) of twelve different GRBs. Figure credit: Daniel Perley, who utilizes the \href{http://gammaray.msfc.nasa.gov/batse/grb/catalog/}{BATSE data archive }.}
\label{Prompt_LC_sample}
\end{figure}

{\bf Classification :} GRBs have two categories, and they are differentiated based on the duration distribution (\tninty\footnote{the temporal window during which from 5 \% to 95 \% of the total gamma-ray energy fluence is seen}). Some GRBs have a duration shorter than two seconds, and another group has a duration exceeding two seconds \citep{1993ApJ...413L.101K}. As astronomers are very inventive in names, they defined them as short ($<$2 sec) and long ($>$2 sec) GRBs. Long GRBs are more commonly detected than short GRBs. A distribution of duration for a larger sample of BATSE detected short (peak $\sim$ 0.2-0.3 sec) and long (peak $\sim$ 20-30 sec) GRBs is shown in Figure \ref{T90_distribution}. Furthermore, they also have different hardness values; short GRBs are, on average, harder (more photons are detected in the hard energy range than soft energy range) than long-duration GRBs \citep{2003A&A...401..129B}. The two categories are known as ``short/hard" and ``long/soft" GRBs based on hardness-duration properties.

\begin{figure}[ht!]
\centering
\includegraphics[scale=0.5]{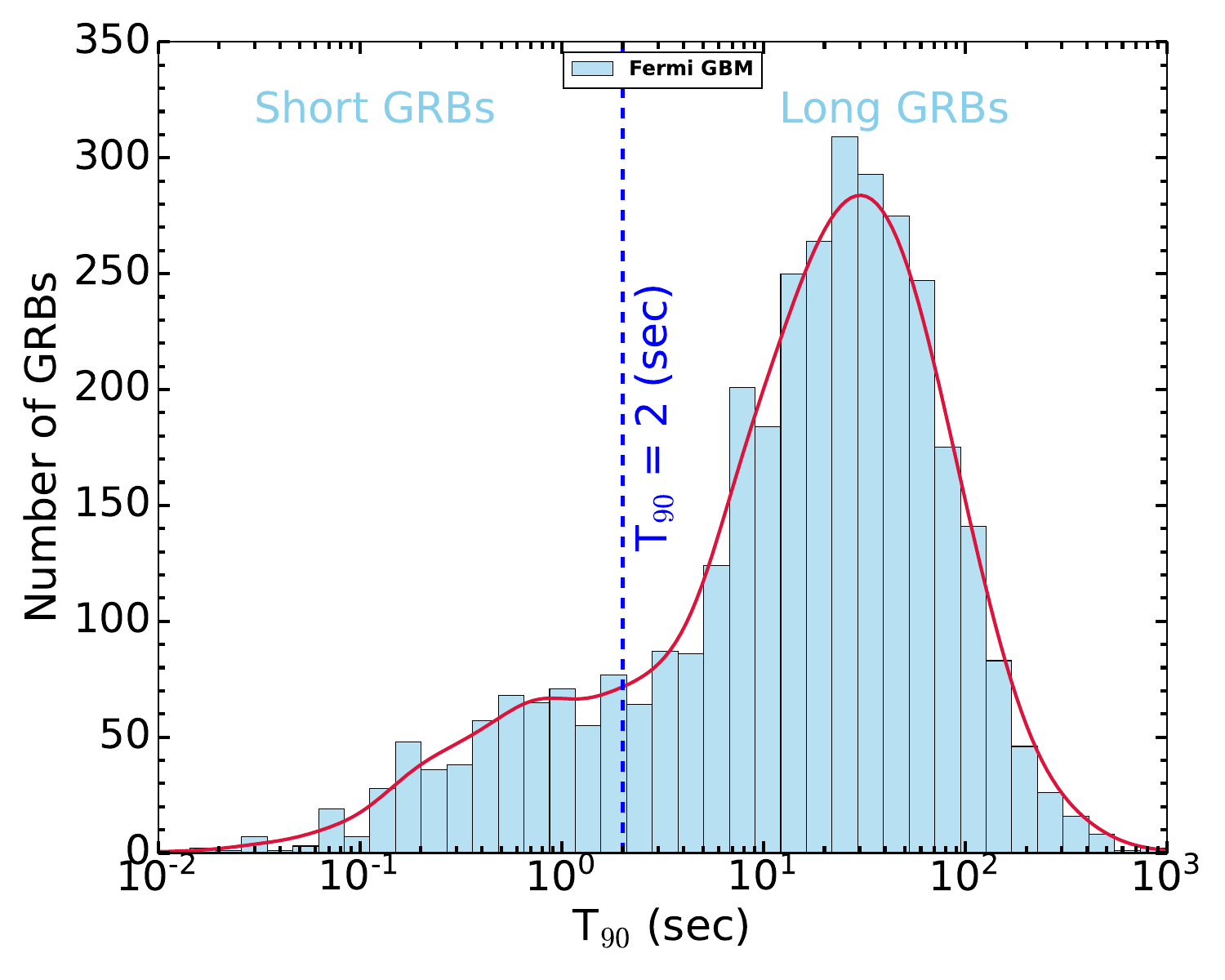}
\caption{Two categories of GRBs based on the bimodal duration distribution using \fermi GBM data. The crimson curve shows the kernel density estimation of the GRB \tninty distribution histogram.}
\label{T90_distribution}
\end{figure}

However, this traditional classification has some limitations; for example, the bimodal distribution of GRBs has significant overlap close to the boundary between long and short GRBs, duration measurements dependent on the sensitivity and energy coverage of the instruments, etc \citep{1993ApJ...413L.101K, 2008ApJS..175..179S, 2011ApJS..195....2S, 2012ApJS..199...18P}. Some authors also proposed a third intermediate class of GRBs \citep{1998ApJ...508..314M, 2006A&A...447...23H, 2008A&A...489L...1H, 2010AstBu..65..326M, 2016Ap&SS.361..155H, 2003ApJ...582..320H, 2010ApJ...725.1955V, 2011A&A...525A.109D}. Considering these limitations, \cite{2020MNRAS.492.1919M} gave a new way to classify GRBs (with known redshift) using the rest frame peak energy and isotropic gamma-ray energy relation (Amati relation). They noted that a new parameter, Energy-Hardness-Duration (EHD), is the most reliable parameter for classifying GRBs with/without redshift measurement. The distribution of EHD also follows a bimodal distribution, although the overlapping regions between short and long bursts have significantly reduced compared to the typical \tninty bimodal distribution of GRBs. \cite{2006Natur.444.1010Z, 2007ApJ...655L..25Z} classified GRBs into type I (originating from mergers) and type II (originating from collapsars) based on their physical progenitors. Recently, some other authors classified GRBs in type I and type II based on machine learning algorithms \citep{2020ApJ...896L..20J, 2022arXiv221116451L, 2023arXiv230100820S}.

\subsubsection{Spectral properties}

The prompt emission usually generates an X-ray/gamma-ray non-thermal spectrum. Generally, the prompt emission spectral shape is generally described using an empirical function known as the \sw{Band} function (two power laws connecting smoothly). The \sw{Band} function consists of three spectral parameters: the peak of the spectrum is known as peak energy, the power law index of low-energy photons and high-energy photons \citep{Band:1993}. A typical view of the \sw{GRB} function is displayed in Figure \ref{Band}. The average values of the peak energy, the power law index for low-energy photons, and the power law index for high-energy photons are found $\sim$ 200 -300 \keV, $\sim$ -1, and $\sim$ -2, respectively \citep{Preece:2000ApJS, Goldstein:2013, 2011A&A...530A..21N}. 

\begin{figure}[ht!]
\centering
\includegraphics[scale=0.8]{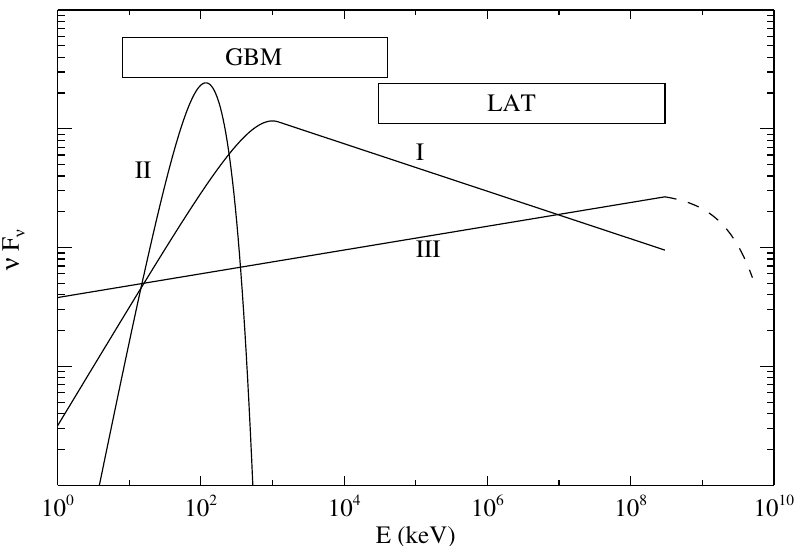}
\caption{The typical \sw{Band} spectrum of GRB prompt emission (I); \sw{Blackbody} component (II); and high energy component (III). Figure credit: \cite{2011ApJ...730..141Z}.}
\label{Band}
\end{figure}

In addition to the \sw{Band} function, we also see evidence of a \sw{Blackbody} or photospheric component (see Figure \ref{Band}) in the observed spectra of some GRBs \citep{2005ApJ...625L..95R, 2009ApJ...702.1211R, Abdo:2009, 2011MNRAS.416.2078P, 2010ApJ...716.1178A}. Furthermore, we also see evidence of an additional high energy component (see Figure \ref{Band}) in the observed spectra of some GRBs \citep{2003Natur.424..749G, 2010ApJ...716.1178A, 2011ApJ...729..114A}. The spectrum evolves rapidly as the burst evolves, indicating possible emission mechanisms. As time proceeds, typically, the spectrum becomes softer (peak energy moves towards lower energy), defined as hard-to-soft change \citep{1986ApJ...301..213N}, but in the case of multi-pulsed GRBs, the peak energy can follow the intensity tracking (the peak energy becomes harder or softer as the intensity of the burst increases or decreases within the burst) behaviour \citep{1983Natur.306..451G}. Theoretical simulations suggest that apparent intensity tracking behaviour is also possible due to the superposition of pulses having hard-to-soft behaviour \citep{2012ApJ...756..112L}.

{\bf Emission mechanisms :} The radiation mechanism behind the prompt emission remains a mystery \citep{2004ApJ...613..460B, 2011CRPhy..12..206Z, 2022Galax..10...38B}. Synchrotron emission (the radiation emitted when relativistic electrons are being accelerated in magnetic fields \citep{Rybicki:1979}) is generally believed to be responsible for the prompt emission spectral shape \citep{2014NatPh..10..351U, 2019A&A...628A..59O, Tavani:1996, 2020NatAs...4..210Z}. Synchrotron emission can produce power law features observed in the prompt spectrum. Since shocks are abundant in the outflow, they can accelerate relativistic electrons and generate a magnetic field via the Weible instability \citep{2015NatPh..11..173H}. The magnetic field may also originate from the central source. Therefore, the prompt emission may be attributed to synchrotron radiation from a cooling population of particles \citep{2014ApJ...784...17B, 2020NatAs...4..174B, 2018ApJ...853...43X, 2020A&A...636A..82G}. The low energy spectral slope ($\alpha$) of the \sw{Band} model serves as a diagnostic to identify the potential radiation physics of GRBs. In the case of fast cooling (the relativistic electrons emit all the energy right away as soon as they are accelerated), synchrotron emission \citep{2000ApJ...534L.163G}, the theoretically predicted value of $\alpha$ is -3/2. However, suppose that we plot the distribution of $\alpha$ for many GRBs observed with different telescopes, e.g., by {\it CGRO}/BATSE and \fermi/GBM. In that case, we find that a significant fraction of GRBs is inconsistent with the expectations of synchrotron emission \citep{Preece:1998}. Therefore, some other mechanisms may be playing a role in producing some or all of the emissions. For example, the physical models of photospheric emission are also directly found to fit the data \citep{2017IJMPD..2630018P, 2017SSRv..207...87B, Zeynep:2020arXiv, 2012ApJ...755L...6F}. Thermal photospheric spectra do not need to be exactly \sw{Blackbody}; if dissipation occurs just below the photosphere, then such a process may broaden the spectrum with respect to the typical \sw{Blackbody} spectrum \citep{Beloborodov2017, Ahlgren:2019ApJ, 2005ApJ...628..847R, 2011MNRAS.415.3693R}. 

In the last few years, there have been some groundbreaking developments using broadband spectroscopy of prompt emission to explore radiation physics. \cite{2017ApJ...846..137O, 2018A&A...616A.138O} conducted a joint spectral analysis involving a sample of 34 bright GRBs for which \swift BAT and XRT instruments jointly observed the prompt gamma-ray emission. The joint spectral analysis revealed a notable energy break at lower frequencies and the typical peak energy break. Exceptionally, the values of $\alpha_{1}$ (photon index below the low energy break) and $\alpha_{2}$ (photon index above the low energy break) were in agreement with the predictions of synchrotron theory. A similar spectral behaviour was observed for bright \fermi long GRBs by \cite{2018A&A...613A..16R, 2019A&A...625A..60R}, although the spectral shape is not observed for bright \fermi short GRBs. We also observed similar spectral behaviour for one of the brightest long-duration GRBs detected by the \fermi observatory (GRB 190530A; \citealt{2022MNRAS.511.1694G}). Moreover, \cite{2019A&A...628A..59O} extended the energy range down to the optical band and performed the spectral analysis using a synchrotron physical model and concluded that the synchrotron spectral shape is a good fit to the data from Gamma-ray to optical bands. The previous discussion suggests that the simultaneous observations of prompt emission from the optical to GeV energy band may play a crucial role in understanding the emission mechanisms. However, such simultaneous observations are challenging due to the extremely short (we don't have time to turn our optical/X-ray instruments to the location before the prompt emission ends) and variable nature of the prompt emission.

\begin{figure}[ht!]
\centering
\includegraphics[scale=1.2]{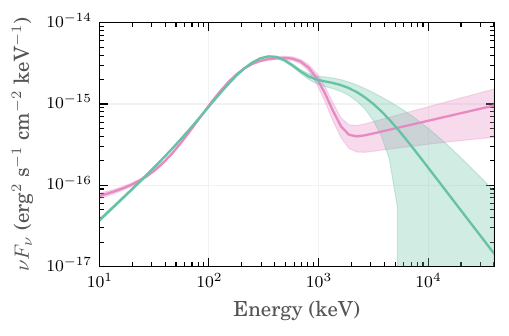}
\includegraphics[scale=1.2]{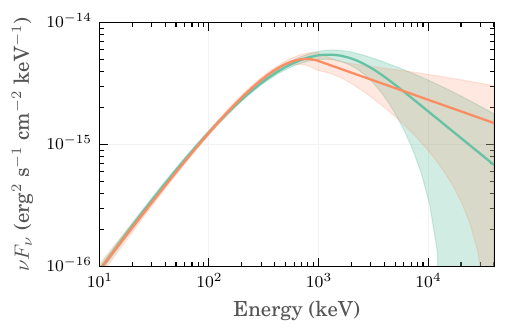}
\caption{{Top panel:} The best-fitted energy spectra plot in the model space for GRB 110920A  using the two blackbodies + power law (shown with pink) and synchrotron physical model + blackbody (shown with green). {Bottom panel:} The best-fitted energy spectra plot in the model space for GRB 081110A using the \sw{Band} function (shown in orange) and synchrotron physical model (shown in green). In both plots, the corresponding shaded regions show the associated uncertainty (68\% confidence level) of spectral shape. Figures are obtained from \cite{2016MNRAS.456.2157I}.}
\label{degeneracy}
\end{figure}

Presently, in the spectral analysis of GRBs, one of the main issues is the degeneracy between various spectral models. Sometimes the same data can be fitted with different spectral models, which provide equally good statistics \citep{2016MNRAS.456.2157I}. To show this issue in the left of Figure \ref{degeneracy}, we have demonstrated a GRB (GRB 110920A) for which the same spectrum has been modelled using two different models with different shapes. Similarly, in the bottom plot of Figure \ref{degeneracy}, we can see that the \sw{Band} function, i.e., shown in orange, has a very narrow spectral width, which means that it is inconsistent with the synchrotron. However, the observed spectrum (GRB 081110A) can also be accurately modelled using a synchrotron physical model through spectral fitting. Therefore, there is a dire need for more constraining observables such as, e.g., the polarization \citep{2022JApA...43...37I, Toma:2013arXiv, 2020MNRAS.491.3343G}. The above discussion suggests that the limited temporal and spectral information available from onboard GRB detectors is insufficient to resolve this issue unambiguously, and we need more observing constraints.

\subsubsection{Polarimetric properties}

Typically, investigations into the radiation physics of GRBs primarily rely on analyzing the temporal/spectral information of the prompt emission. However, they need to be more observable to fully solve the problem of the radiation mechanisms of prompt emission. Other than temporal and spectral information, prompt emission polarization measurements constitute a unique tool for probing prompt emission \citep{2021Galax...9...82G, McConnell:2017NewAR}. 

\begin{figure}[ht!]
\centering
\includegraphics[scale=0.48]{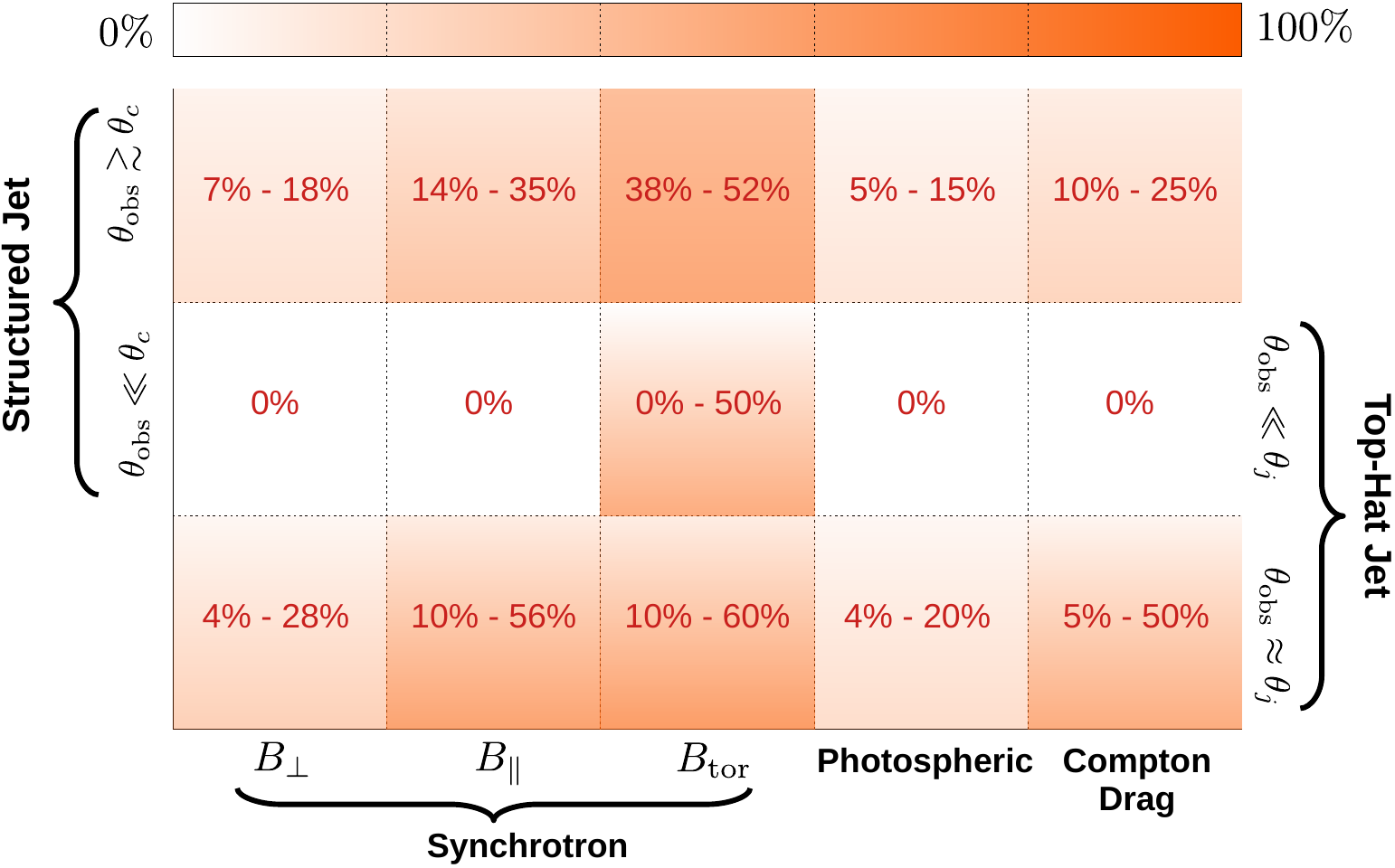}
\caption{Polarization fraction expected for different radiation models (Compton drag model, Photospheric model, Synchrotron ordered (toroidal magnetic field) model ($B_{\rm tor}$), Synchrotron ordered (magnetic field oriented towards the local velocity vector) model ($B_{\rm \parallel}$), and Synchrotron random model ($B_{\rm \perp}$) ) and jet structures (Top-Hat jet, Structured jet) along with jet viewing angle geometry. Figure credit: \cite{2021Galax...9...82G}.}
\label{PF}
\end{figure} 

The measurement of polarization (direction of the electric field) in X-ray astronomy is highly challenging for transient events like the prompt emission of GRBs because of the deficiency of incident photons \citep{2021JApA...42..106C, 2019NatAs...3..258Z}. So, studying the polarization of the prompt emission is difficult and has yet to be done in detail \citep{2021Galax...9...82G}. As of now, polarization measurements have been attempted only for a limited number of bursts using the Reuven Ramaty High Energy Solar Spectroscopic Imager ($RHESSI$, \citealt{2003Natur.423..415C, 2004MNRAS.350.1288R}), the BATSE Albedo Polarimetry System (BAPS, \citealt{2005A&A...439..245W}), the INTErnational Gamma-Ray Astrophysics Laboratory ($INTEGRAL$, \citealt{2007ApJS..169...75K, 2007A&A...466..895M}), the GAmma-ray burst Polarimeter (GAP, \citealt{2011ApJ...743L..30Y, 2011PASJ...63..625Y, 2012ApJ...758L...1Y}), the Cadmium Zinc Telluride Imager onboard \AstroSat (see chapter 2 for more detail) and POLAR \citep{2020A&A...644A.124K, 2019A&A...627A.105B}, and most of the obtained results have limited statistical significance \citep{2021JApA...42..106C}. Before \AstroSat, there have been only around 20 GRBs (including 14 GRBs using POLAR) for which polarization measurements have been reported. This low number is essential because there have not been many GRB polarimeters in the sky, and GRB polarimetry or polarimetry in X-rays, in general, is challenging to conduct \citep{2022arXiv220502072B}.

Polarization measurements can provide a robust discriminator between different possible radiation models. This is because different prompt emission radiation models predict distinct polarization fractions depending on the jet geometry. Different degrees of polarization are expected from synchrotron radiation originating from magnetic field structures ordered on different scales and depending upon the viewing geometry of the jet. A meager polarization fraction is predicted for inverse Compton and photospheric emission except when the jet is viewed off-axis \citep{2009ApJ...698.1042T}. Figure \ref{PF} shows the polarization fraction expected for different radiation models (Compton drag model, Synchrotron ordered (magnetic field vectors are aligned) model, Synchrotron random (magnetic field vectors are not aligned) model, and Photospheric model) and jet structures (Top-Hat jet, Structured jet) along with jet viewing angle geometry. So, when $\theta_{\rm obs}$ (viewing angle) is equal to $\theta_{\rm j}$ (jet opening angle) or less, it means that we are viewing within the jet cone, and when it is greater than $\theta_{\rm j}$, it means we are viewing outside the jet cone.
As shown in Figure \ref{PF}, intrinsic polarization fractions of up to $\sim$ 60\% could be produced in synchrotron radiation generated by shock-accelerated relativistic electrons in an ordered magnetic field. Polarization as high as $\sim$ 50\% may result from bulk inverse Compton scattering viewed at a suitable angle. Photospheric dissipation, on the other hand, is thought to produce up to $\sim$ 20\% polarization via unscattered synchrotron photons \citep{2021Galax...9...82G, McConnell:2017NewAR}. Therefore, if we can measure the time-resolved polarization for many bursts, we can probe something concrete about the emission mechanisms of GRBs. Generally, some asymmetry in the emitting region or viewing geometry results in linearly polarised emission. Hence, simultaneously with spectroscopy, polarization measurement can break the degeneracy between various spectral models. In addition, polarization variability is an essential aspect that affects the underlying emission mechanism and jet geometry. The measurement of polarization and its variability thus provides valuable information to distinguish between different models of GRBs and the radiation mechanisms at work. Further, the variation of polarization with time is a crucial tool for understanding the temporal dynamics of the jet.

\subsection{Broadband afterglow emission}

\begin{figure}[ht!]
\centering
\includegraphics[scale=0.22]{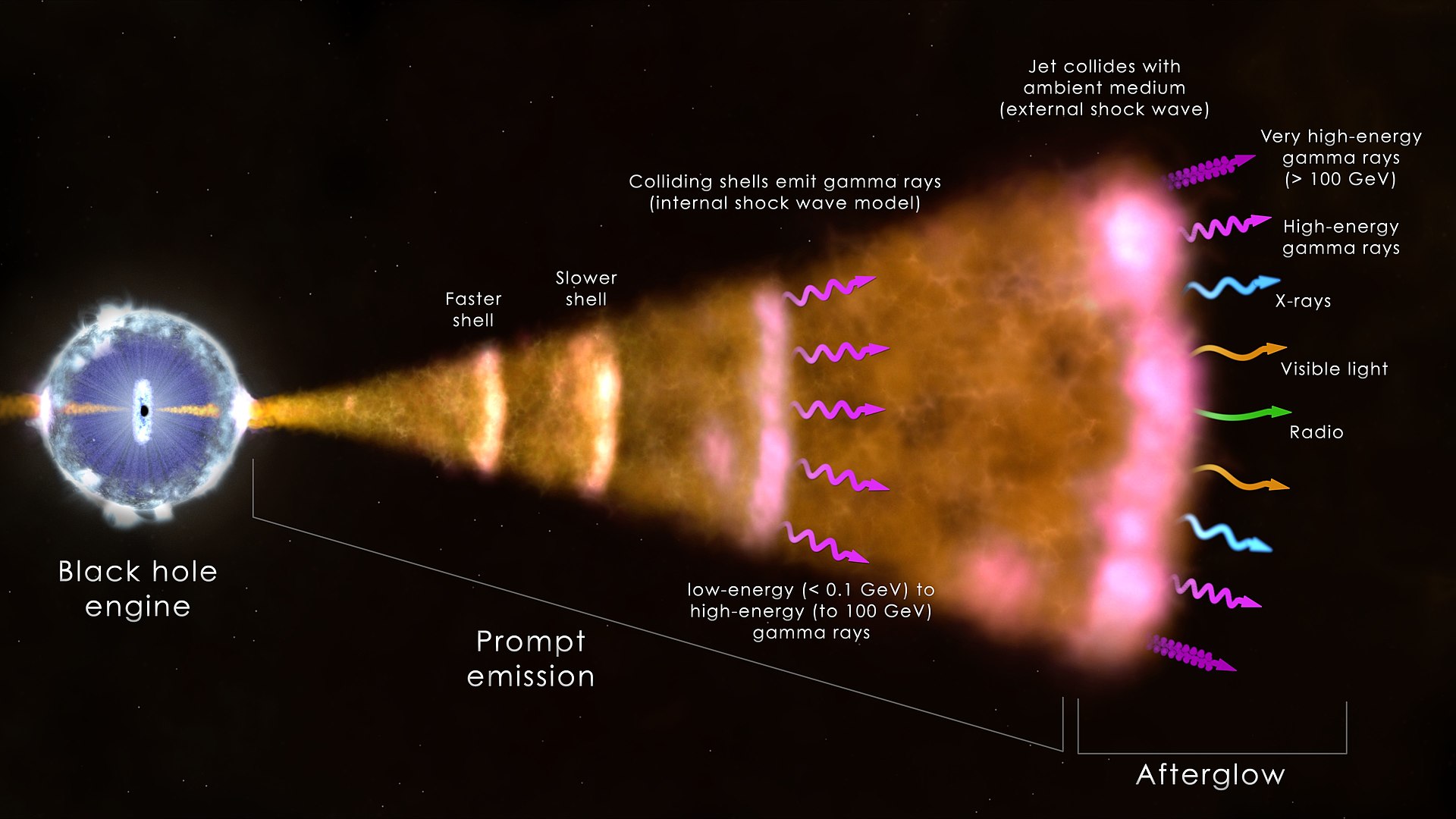}
\caption{A schematic view of the fireball model for GRBs. Figure credit: \href{https://earthsky.org/upl/2019/11/gamma-ray-burst-mechanism-lg.jpg}{NASA/Goddard Space Flight Center}.}
\label{Fireball_model}
\end{figure}

The fading afterglows (phase post prompt emission) were theoretically expected before their discovery \citep{1993ApJ...418L...5P, Katz:1994, 1997ApJ...476..232M}. The  afterglow emission is explained in terms of external shock between the jetted relativistic outflow and ambient medium (see Figure \ref{Fireball_model} and section \ref{Theories of Afterglows} for more details). The evidence of highly beamed jetted outflows is confirmed using the jet break feature in the afterglows light curves. GRBs typically have a jet opening angle ($\theta_{j}$) of a few degrees \citep{rho97, 1999ApJ...525..737R, pan99}. Considering that we are observing a GRB with viewing angle $\theta_{obs}$, and at some point, the beaming cone becomes wide enough so that we start witnessing the edge of the jet. In such a case, the afterglow light curve should show a steepening, referred to as a jet break \citep{1999ApJ...525..737R}. A schematic diagram illustrating the jet break in the afterglow light curves is shown in the top panel of Figure \ref{Jet_Break}. It has been noted that the lateral spreading of the jet could also explain the jet break due to the deceleration of outflow \citep{Sari99:jet, granot07, 2012MNRAS.421..570G}. The jet break is a geometric effect and should not affect the emission spectrum \citep{1999ApJ...525..737R}. Since it is a geometric effect, it should involve all the different wave bands (X-ray, optical, radio), so the jet break steeping is achromatic \citep{Sari99:jet, wang18:jet}. An example of a jet break observed for GRB 030226 is displayed in the bottom panel of Figure \ref{Jet_Break} \citep{2004A&A...417..919P}. The detection of a jet break provides valuable constraints on both the jet opening angle and the true energy \citep{2009ApJ...698...43R, 2004A&A...417..919P}. 

\begin{figure}[ht!]
\centering
\includegraphics[scale=0.5]{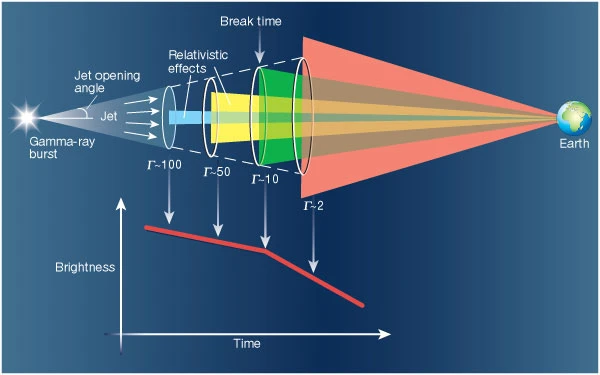}
\includegraphics[scale=0.5]{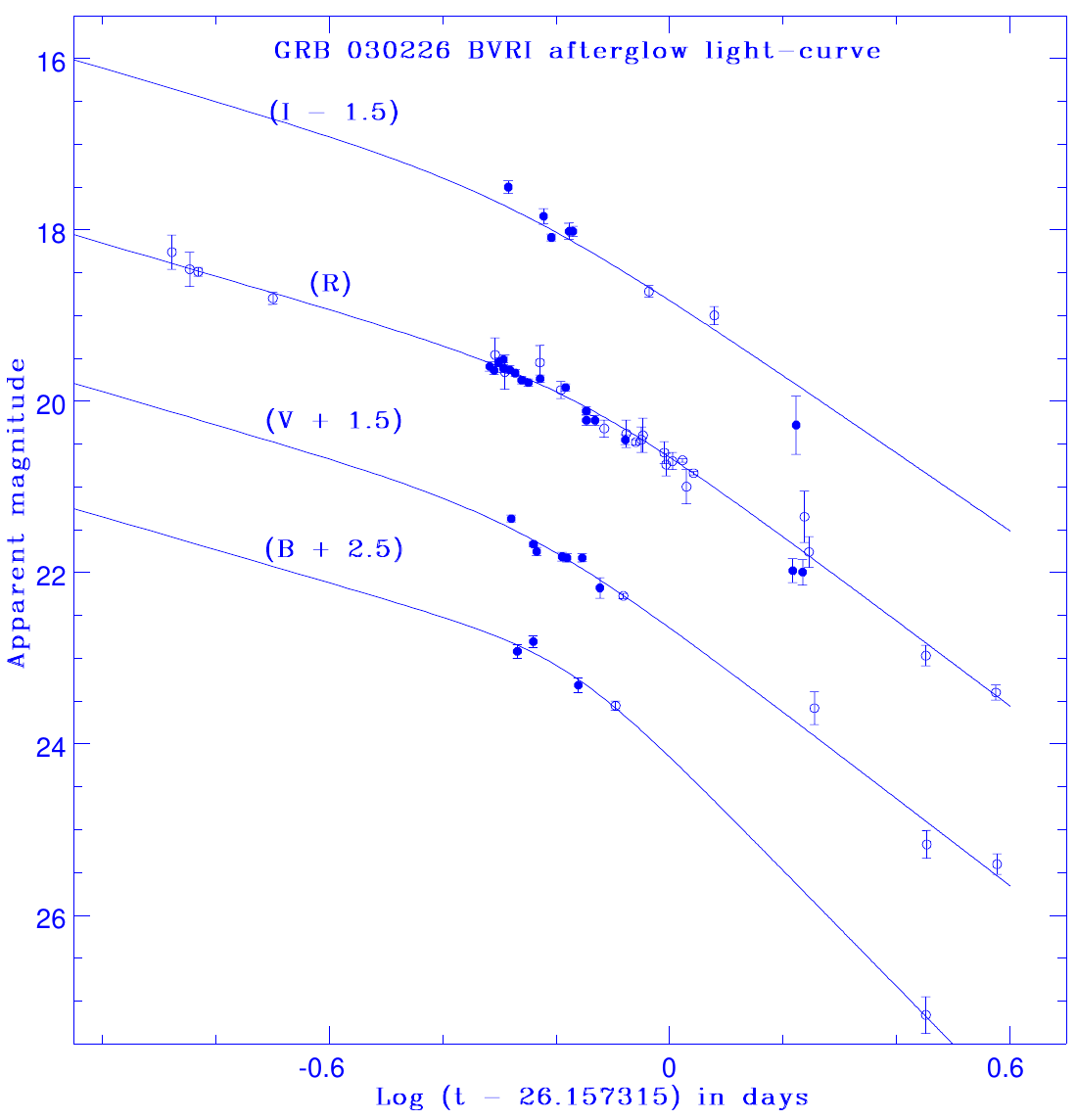}
\caption{{Top panel:} A schematic diagram elucidating the observed jet break phenomenon in GRB afterglow light curves. Figure credit: \cite{2001Natur.414..853W}. {Bottom panel:} An example of Jet break in the light curve of GRB 030226. Figure credit: \cite{2004A&A...417..919P}.}
\label{Jet_Break}
\end{figure}

As per the prediction of the external shock model, the GRB afterglows give emission from GeV energy range to radio wavelength \citep{1997ApJ...476..232M, 1998ApJ...497L..17S, 2001548787S, 2001ApJ...559..110Z}. Recently, the afterglows of a few bursts have been observed at exceptionally high energies, ranging from approximately 100 GeV to 100 TeV, using ground-based Cerenkov telescopes \citep{2019Natur.575..455M, 2019Natur.575..464A, 2021Sci...372.1081H, 2023ApJ...942...34R, 2022Galax..10...66M}. The duration of afterglows is much longer than prompt emission, and we can detect afterglow radiation for days, weeks, or even months. The properties of the jet, including its energy, structure, bulk Lorentz factor, the density of the external medium, and the physics of ultra-relativistic shocks, play a significant role in determining the characteristics of the afterglow emission. These factors also influence the efficiency of electron acceleration within the shocks. Using the broadband afterglow modelling, we can constrain these physical parameters \citep{2004RvMP...76.1143P, 2015PhR...561....1K, Meszaros:2019MmSAI}. The observed flux of the afterglows depends on the temporal and spectral indices. The afterglow can be described by the expression F$_{\nu,t}$ $\propto$ $\nu^{\beta}$~$t^{\alpha}$, where $\alpha$ represents the temporal index and $\beta$ represents the spectral index of the afterglow.

\subsubsection{Temporal properties} 

\begin{figure}[ht!]
\centering
\includegraphics[scale=0.23]{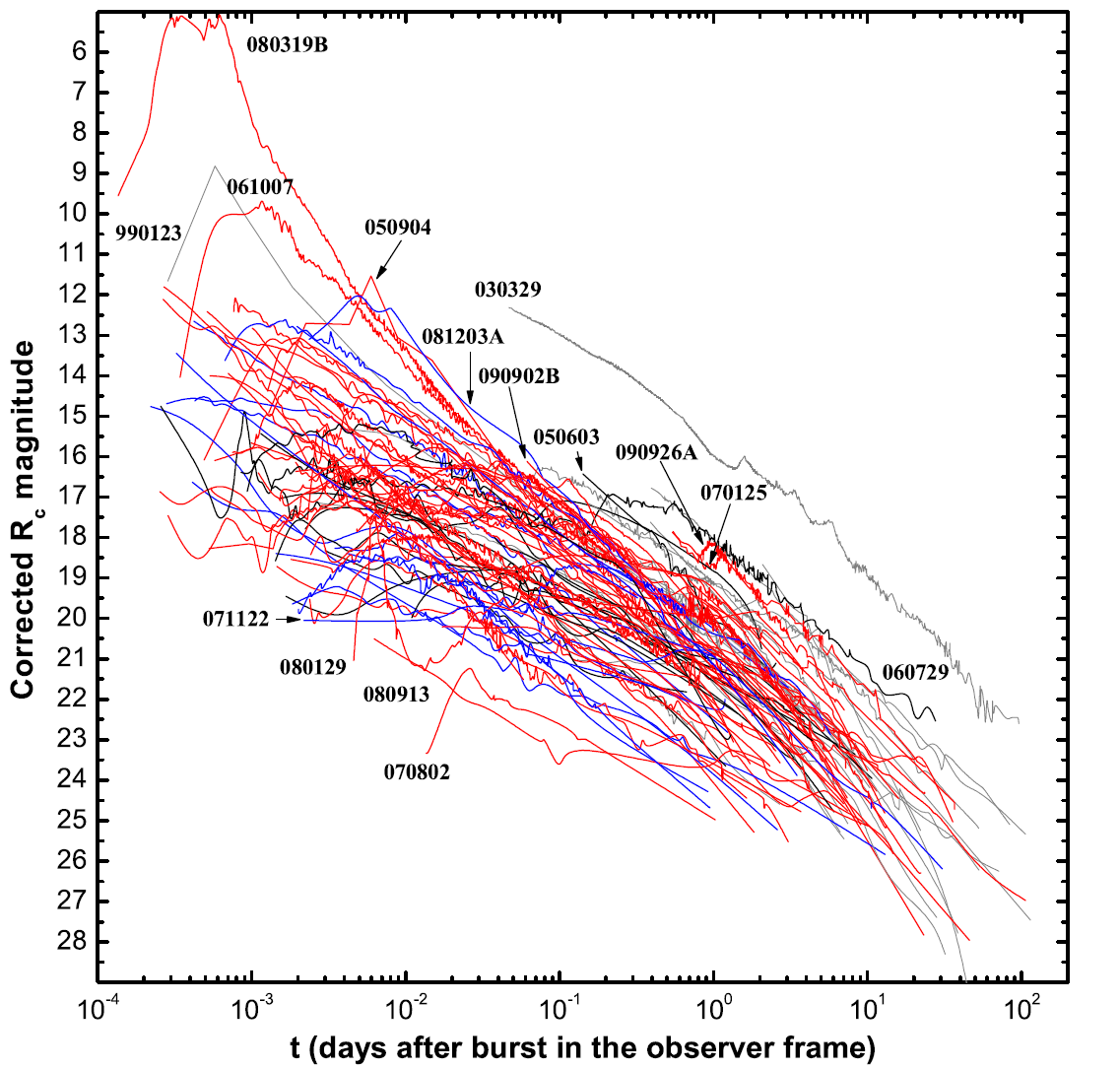}
\includegraphics[scale=.25]{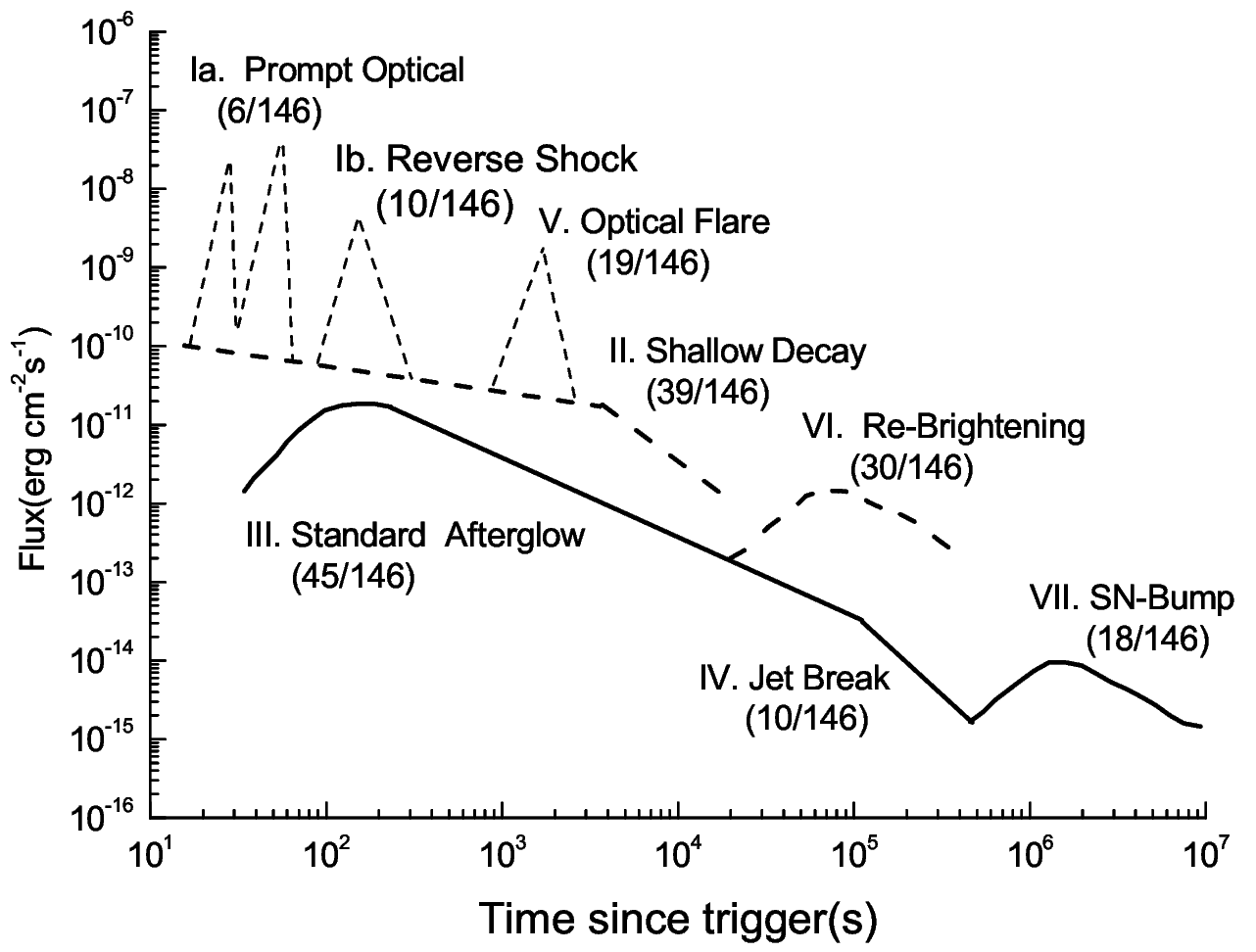}
\caption{{Top panel:} Optical afterglow light curves gallery for a large number of GRBs detected in the \swift era. Figure credit: \cite{2010ApJ...720.1513K}. {Bottom panel:} A synthetic optical afterglow light curve of GRBs. Figure credit: \cite{2012ApJ...758...27L}.}
\label{optical_LC}
\end{figure} 

The light curves of afterglows are typically well characterized by a power-law or broken power-law function, so the flux decays smoothly without the strong variability characterizing the prompt emission. I will be focusing on the optical afterglows first. The late-time (a few hours post-detection) light curves of optical afterglows usually follow a temporal decay index of $\sim$ -1 \citep{2022ApJ...940..169D}. If the optical counterparts of GRBs are bright enough (for initially faint afterglows, even much fainter optical emission post-jet break is difficult to detect using current generation optical telescopes), the later-time optical follow-up observations usually unveil a steep decay phase (temporal decay index of $\sim$ -2) expected due to jet break in the external forward shock model \citep{1999ApJ...523L.121H}. Late-time observations of GRB afterglows may also reveal re-brightening, and supernovae bump (typically with long GRBs) features \citep{2011A&A...531A..39N, 2006ARA&A..44..507W} in the optical afterglow light curves. On the other hand, the early optical follow-up observations unveil more features. The early bump is expected in the afterglow light curves due to the onset of afterglow \citep{Molinari:2007, lia13}. Furthermore, the external forward shock is accompanied by the reverse shock, which moves towards the ejecta. If the observing conditions are suitable, we sometimes see the signature of the reverse shock (see chapter \ref{ch:5} for more details) in the early optical afterglow light curves \citep{1997ApJ...476..232M, Sari:1999ApJ}. In such cases, the afterglow light curves follow a steep rise first and then steep temporal decay index \citep{2003Natur.422..284F, 2000ApJ...542..819K, 2000ApJ...545..807K, 2003ApJ...597..455K, 2014ApJ...785...84J}. The deceleration time marks the occurrence of the peak in the reverse shock \citep{2003ApJ...595..950Z, 2004MNRAS.353..647N, 2015AdAst2015E..13G}. The early optical follow-up observations also unveil the plateau phase in the optical afterglow light curve \citep{gru07, 2007A&A...470..105M, 2011MNRAS.414.3537P, 2017A&A...600A..98D}. The optical light curves might also contain flares in the optical bands \citep{2013ApJ...774....2S, 2014PhDT.......392S, 2017ApJ...844...79Y}. An example of optical afterglow light curves for several bursts detected during the \swift era is illustrated in the top panel of Figure \ref{optical_LC} \citep{2010ApJ...720.1513K, 2011ApJ...734...96K}. The bottom panel of Figure \ref{optical_LC} exhibits a synthetic optical afterglow light curve. The different possible optical emission components are shown in this figure \citep{2012ApJ...758...27L}.
The afterglow light curves for an observer on axis or whose line of sight intersects the jet aperture are very different compared to an observer who is viewing off-axis \citep{2002ApJ...570L..61G, 2022MNRAS.515..555B}.

The $BeppoSAX$ instrument conducted late-time follow-up observations of X-ray afterglows for several bursts. However, the immediate slewing capabilities of the \swift satellite enable the X-ray telescope to capture the evolution of X-ray afterglow light curves routinely from early to late epochs. A gallery of X-ray afterglow light curves for a large number of BAT, GBM, and LAT-detected GRBs is illustrated in the top panel of Figure \ref{XRT_LC}. The X-ray afterglow light curves comprise several of one or more (\citealt{2009MNRAS.397.1177E} found that only 41 \% of X-ray afterglows have canonical behaviour) characteristic features (see bottom panel of Figure \ref{XRT_LC}). The first component is the initial steep decay phase with a typical decay index of $\sim$ -3 \citep{2005Natur.436..985T}. This component can be ascribed to the late phase, or tail, of the prompt emission \citep{2005ApJ...635L.133B}. The discovery of a rapid decline phase in the initial XRT light curves helps to understand the internal and external origin of prompt emission and afterglow emission of GRBs \citep{2006ApJ...642..354Z}. The steep decay phase is generally followed by a shallow segment with a typical decay index of $\sim$ 0 (plateau) to -0.7 in the XRT light curve. These plateaus can be attributed to the energy injection from the central engine, mainly if a magnetar-like central engine gives a continuous energy injection that can produce a plateau phase \citep{1998PhRvL..81.4301D, 2001ApJ...552L..35Z}. After the plateau phase, slow decay (normal) X-ray emission is expected within the framework of external forward shock due to jet interaction with the surrounding material \citep{2006ApJ...642..354Z, pan06}. The spectral analyses of shallow and normal decay phases indicate that the break between the segments is due to geometrical effects. No spectral evolution has been observed during these phases \citep{2006ApJ...638..920V}. The normal decay phase at some point is followed by a steeper decay segment with a temporal decay index of $\sim$ -2 or steeper due to the jet break effect (see details below). In addition to the above phases, we frequently see flares in the XRT light curves \citep{bur05b, 2010MNRAS.406.2149M}. These flares are believed to have an internal origin and signatures of energy injection but are episodic \citep{2006ApJ...642..354Z}. Detailed properties of X-ray counterpart of GRBs are studied in \cite{obr06, willingale07, mar13, 2007A&A...469..379E, 2009MNRAS.397.1177E, liang07, liang08, 2009ApJ...707..328L}. 

\begin{figure}[ht!]
\centering
\includegraphics[height=9cm, width=9.5cm]{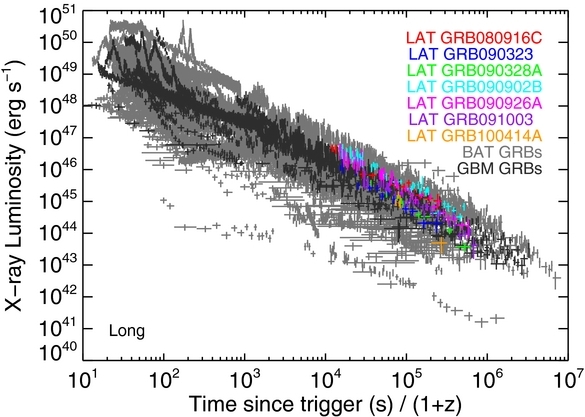}
\includegraphics[scale=0.33]{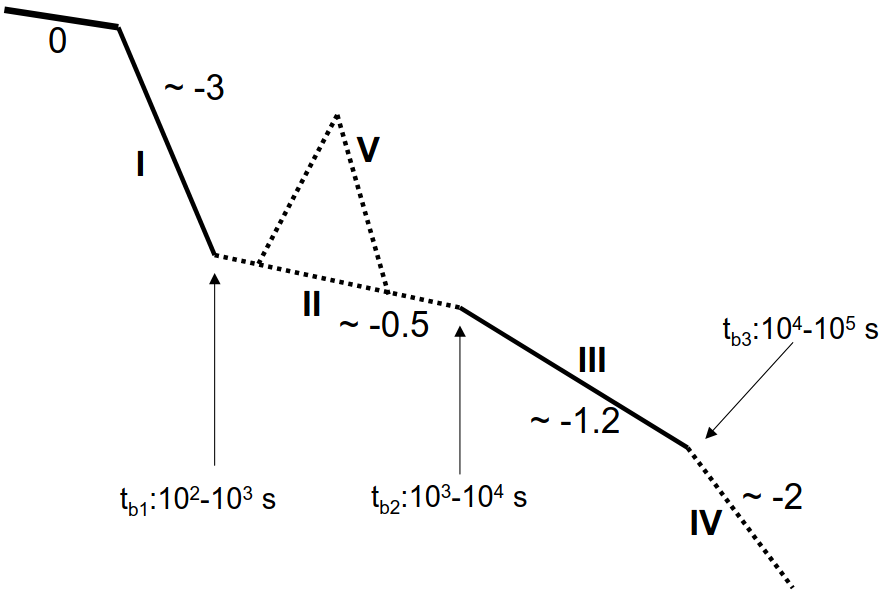}
\caption{{Top panel:} X-ray afterglow light curves gallery for a large number of BAT, GBM, and LAT detected GRBs. Figure credit: \cite{Racusin:2011ApJ}.  
{Bottom panel:} \swift XRT discovered the canonical X-ray afterglows. Figure credit: \cite{2006ApJ...642..354Z}.}
\label{XRT_LC}
\end{figure}

\subsubsection{Spectral properties}
\label{Spectral properties afterglows}

The afterglow is understood as the result of the interaction between the expanding ultra-relativistic jet and the surrounding external medium \citep{Paczynski:1986ApJ}. The interaction creates a forward shock moving towards the external medium, accelerating the particles in the external environment. This results in non-thermal accelerated electrons and a strong magnetic field, so we expect synchrotron radiation \citep{Katz:1994, Cohen:1997, 1993ApJ...418L..59M, 2007MNRAS.379..331P, Wang_2015, 2021MNRAS.505.1718J}. The energy and cooling time scale of photons emitted due to synchrotron radiation depends on the strength of the magnetic field and the bulk Lorentz factor of electrons. The afterglow radiation (from radio to GeV) is mainly explained by synchrotron emission from relativistic blast waves \citep{1999ApJ...513..679G, 2000ApJ...534L.163G, 2004RvMP...76.1143P}.

The synchrotron spectrum obtained using a single electron is given by a power-law function ($F_\nu$ $\propto$ $\nu^{-1/3}$). However, to calculate the synchrotron spectrum obtained using all the relativistic electrons, we need to integrate the flux over the distribution of Lorentz factor ($\gamma_{e}$), considering a power-law distribution of energy of these relativistic electrons N($\gamma_{e}$) $\propto$ $\gamma_{e}^{-p}$, where the power-law index ($p$) represents the energy distribution of relativistic electrons, as described by \cite{1998ApJ...497L..17S}. The afterglow spectrum typically consists of different power-law segments at different regimes of the electromagnetic spectrum. The afterglow spectrum is illustrated by three critical frequencies (cooling frequency $\nu_{c}$, frequency corresponding to maximum energy $\nu_{m}$, and self-absorption frequency $\nu_{a}$). The temporal evolution of these critical frequencies shapes the afterglow light curve \citep{Sari:1996, Sari:1997, Sari:1999ApJ, 1998ApJ...497L..17S, 1999PhR...314..575P}. Two spectral regimes are possible for the synchrotron emission based on the cooling time scale of accelerated electrons \citep{1998ApJ...497L..17S}. If the electrons are cooling rapidly, it means as soon as they are accelerated, they emit all the energy right away, and they are in a fast cooling spectral regime ($\nu_m$ $>$ $\nu_c$). If the cooling time is much longer than the dynamical time, electrons are in a slow cooling regime ($\nu_c$ $>$ $\nu_m$). The top and bottom panels of Figure \ref{Afterglow_spectrum} illustrate the temporal evolution of the critical frequencies and the corresponding expected spectral indices for the fast and slow cooling synchrotron scenarios, respectively. Generally, most GRB afterglows follow the slow cooling spectral regime and $p ~ > 2$ in their emission \citep{2013NewAR..57..141G, 2019ApJ...883..134T}. At least by the time we start observing, the synchrotron self-absorption ($\nu_a$) frequencies generally do not affect the high-frequency emission such as the X-ray/optical but will affect the radio afterglows \citep{1999ApJ...527..236G}. The dependency of different frequencies in different spectral regimes is also shown in Figure \ref{Afterglow_spectrum}.

\begin{figure}[ht!]
\centering
\includegraphics[scale=0.45]{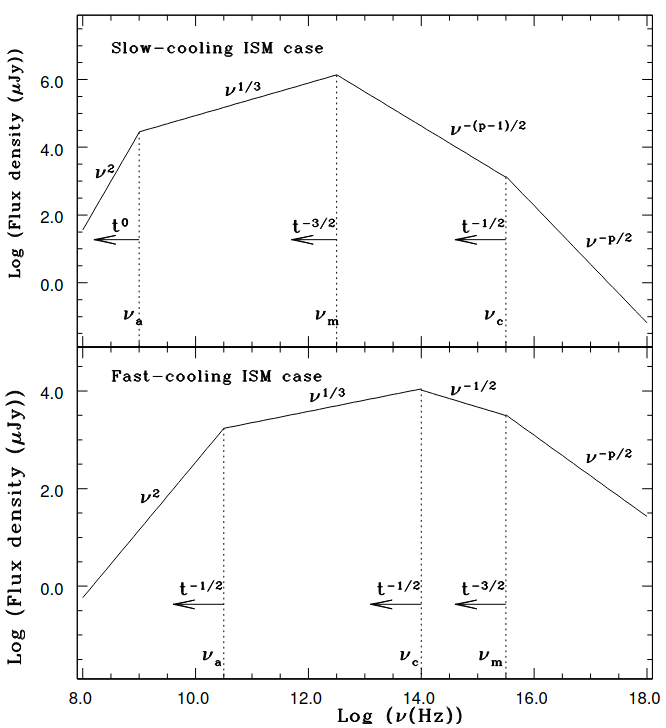}
\caption{The temporal evolution of critical frequencies and expected spectral indices for fast and slow cooling synchrotron scenarios. Figure credit: \cite{1998ApJ...497L..17S}.}
\label{Afterglow_spectrum}
\end{figure} 

Further, the relativistic electrons are up-scattered, and synchrotron photons produce an inverse Compton or synchrotron self-Compton component at higher energies in the GeV-TeV energy range. The presence of a synchrotron self-Compton (GeV-TeV) component has been observed in several bright GRBs detected at extremely high energies \citep{2019Natur.575..455M, 2019Natur.575..464A}.

\section{Dark GRBs and Orphan Afterglows:} 

There is a subclass of GRBs afterglows, known as `dark bursts' for which no optical counterpart is detected based on deep follow-up in $<$ 24 hours after the trigger. Dark bursts comprise a substantial fraction of the overall GRB population \citep{2004ApJ...617L..21J, 2011A&A...526A..30G}. From 1997 to 2021, about 66\,\% (1526/2311) of well-localized bursts have been observed with an X-ray afterglow; however, till yet, only 38\,\% (874/2311) bursts have been observed with an optical afterglow\footnote{https://www.mpe.mpg.de/~jcg/grbgen.html}. In some cases, optical afterglows could not be detected due to the lack of early observations \citep{2003A&A...408L..21P}. But in other cases, deep and early observations reveal that true dark GRBs exist and have a typical fraction of around 25-40\,\% (dependent on the telescope used) of all the bursts \citep{2016ApJ...817....7P}. 

Several methods exist in the literature to classify a GRB as being dark. At the early stage of their studies, GRBs without an optical afterglow were classified as dark GRBs \citep{1998ApJ...493L..27G}. In later studies \citep{2003ApJ...592.1018D, 2004ApJ...617L..21J}, an analogy between the X-ray and optical/near-IR afterglow properties was used for their classification. \cite{2003ApJ...592.1018D} used the flux ratio of optical and X-ray afterglows, and \cite{2004ApJ...617L..21J} used the spectral index obtained using optical to X-ray spectral energy distribution to classify these bursts ($\beta_{\rm OX} < 0.5$). \cite{2009ApJ...699.1087V} suggested a more general method and considered dark GRBs those with $\beta_{\rm OX} < \beta_{\rm X}-0.5$ in the framework of the fireball model. 

There may be many potential reasons for the optical darkness of GRB afterglows. 1) Dark GRB afterglows could be intrinsically faint \citep{Fynbo01}, e.g., if the relativistic fireball is decelerated in a low-density surrounding environment \citep{1998ApJ...497L..17S}. 2) The high redshift origin \citep{Taylor98} of GRB afterglows, in such case the Ly-$\alpha$ forest will absorb optical radiation \citep{Lamb00}. The interpretation of the optical darkness due to a high redshift origin is less common, and it is only expected for $\sim$ 10-20\,\% of dark bursts \citep{2016ApJ...817....7P}. 3) Obscuration scenario, this could be because of a high column density of gas or dust in their host galaxies at larger distances so that the optical afterglow is very reddened \citep{Fynbo01}. The last scenario is expected to be the most favoured origin for the optical darkness of GRB afterglows, as most of the host galaxies of dark bursts are detected in the optical.

A few afterglows are discovered independently without prior detection of prompt emission; such afterglows are known as orphan afterglows. However, the discovery of orphan afterglow is more challenging than the normal afterglows of GRBs due to the unavailability of localization (there is no prior X-ray or gamma-ray emission/location, which helps to point the optical telescopes for optical afterglows observations). Moreover, \cite{2002ApJ...564..209D} suggested that orphan afterglows are typically faint to detect because they are in the post-jet break phase. In the current era of survey telescopes equipped with wide field of view capabilities, a few orphan afterglows have been discovered using the ZTF survey in the optical band. The community is also developing many other survey telescopes, including Vera C. Rubin Observatory. One of the primary scientific objectives of such survey telescopes is the discovery of orphan afterglows. More discoveries of orphan afterglows can help to constrain the true rate of GRBs \citep{rho97, 2015A&A...578A..71G}. Further details about orphan afterglows are presented in chapter \ref{ch:61}.

\section{Theories of Afterglows: The Fireball model}
\label{Theories of Afterglows}

Various models have tried to explain the structure and progenitor of GRBs, but they have yet to successfully implement a robust classification scheme. The most famous relativistic ``fireball model" was given even before the first distance measurement of GRBs \citep{Goodman:1986ApJ, Paczynski:1986ApJ, Rees:1992}. The relativistic ``fireball model" uses two different shock mechanisms. The internal shock mechanism is responsible for generating the prompt emission of GRBs, while the external shock mechanism is responsible for producing GRB afterglows (see \citealt{1995ARA&A..33..415F, 2004RvMP...76.1143P} for review). An artist's view of the fireball model for GRBs (considering a BH or a magnetar as a possible central engine) is shown in Figure \ref{Fireball_model}. In accordance with the fireball model, the formation of a central BH surrounded by an accretion disk can be attributed to either the core collapse of a massive star or the merging of two compact objects. The accretion of matter triggers the formation of two highly collimated relativistic jets, in which the matter is accelerated to very high velocities \citep{1993ApJ...405..278M, 1999PhR...314..575P}. Although It is difficult to estimate the bulk Lorentz factor, the estimated values of the bulk Lorentz factor are always around 100 to 1000 (in the most extreme cases) using different techniques \citep{2018A&A...609A.112G}, making GRBs one of the most energetic cosmic sources. This indicates that the jet is ultra-relativistic in nature. Initially, the relativistic fireball is optically thick in nature and at some distance from the central engine ($\sim$ 10$^{13}$-10$^{15}$ cm \citealt{1998MNRAS.296..275D, Rees:1994}), the dissipation of the energy occurs (through collisionless shocks \citealt{Sari:1997, kob97} or magnetic reconnection \citealt{Beniamini:2017, 2016ApJ...816L..20G}) within the jet (fraction of jet energy converted into internal energy of the particles). The accelerated electrons are expected to produce prompt emission \citep{Rees:1994}. There are two dissipation mechanisms that may be involved in producing the prompt emission. One is internal shocks \citep{kob97, 2011ApJ...739..103A}; in this case, the energy of the jet is carried mainly by the particles. In this case, we expect to extract energy by collisions between different shells of the jets that move at slightly different velocities. At some point, they collide with each other and produce the dissipation of energy. These internal shocks produce the prompt emission of GRBs. The other possibility, if the jet is magnetically dominated, is the dissipation of energy through the reconnection of magnetic field lines \citep{Beniamini:2017, Uzdensky:2011SSRv.160.45U, Zhang:2011ApJ}. In both cases, we expect accelerated particles, so a non-thermal population of electrons in a strong magnetic field; therefore, we expect synchrotron radiation. However, many open issues remain in interpreting the prompt emission of GRBs; for example, the exact dissipation mechanism needs to be identified, at which distance from the central engine radiation is produced, and the radiative mechanisms responsible for the prompt emission are not well-known \citep{2011CRPhy..12..206Z}.   

Later, the relativistically moving blast wave inevitably collides with the external medium (external shocks), and the jet starts to decelerate and eventually becomes non-relativistic at late times \citep{1997MNRAS.288L..51W, 1997ApJ...476..232M, Katz:1994, 2007MNRAS.379..331P, Wang_2015}. The geometry of the fireball does not affect the transition from a relativistic (Blandford-Mckee solution, \citealt{1976PhFl...19.1130B}) to a non-relativistic (Sedov-Taylor solution, originally proposed by \citealt{Sedov} and \citealt{Von}, and further developed by \citealt{1950RSPSA.201..159T}) regime. In such condition, the radiation at lower frequencies (X-ray to radio, sometimes at the GeV band) is produced due to shock deceleration. It gives rise to the afterglow emission of GRBs (see S. B. Pandey, 2005, Ph.D. thesis\footnote{http://dspace.aries.res.in/xmlui/handle/123456789/20} and references therein). The afterglow phenomenon is elucidated by the relativistic forward shock scenario, which occurs as the relativistic outflow encounters and propagates into the ambient medium and creates the long-lasting broadband counterpart of GRBs. The interaction between ultra-relativistic fireball and ambient medium also produces a reverse shock. The reverse shock that propagates into the ejecta may produce an optical flash for a short time, together with a radio flare \citep{1999Natur.398..400A, 2015AdAst2015E..13G}. The analytical description of the relativistic fireball model is studied in detail by \cite{1998ApJ...497L..17S, sari97, sar00, 2001ApJ...559..110Z,1999ApJ...513..679G}.

\section{Central engines and possible progenitors of GRBs}

The energy released in the GRB explosion is very high, indicating that an unknown central engine presumably powers the jets of GRBs \citep{woo00}. A couple of possibilities have been suggested (see \citealt{2018ApJS..236...26L, 2020MNRAS.496.2910P} and the citations provided within these sources). The first one is BH accretion \citep{Woosley99}; in this case, we consider that the compact object produced is a BH, and there is some amount of residual matter (which would have been spinning very fast) that stays around the BH. The spinning matter accretes into a spinning BH, and the interaction between the accreting matter and the BH can extract the spin energy of the BH \citep{1999ApJ...518..356P}. A popular mechanism is the Blandford–Znajek mechanism, in which magnetic field lines present in the accretion disk thread the BH \citep{1977MNRAS.179..433B, 1999ApJ...512..100L}. The electromagnetic process can extract the spin energy of the BH and direct it into a low-mass jet, accelerating it to relativistic energies \citep{1977MNRAS.179..433B}. Another possibility could be fast-spinning, strongly magnetized neutron stars produced in the events \citep{2017ApJ835, 2020ApJ...896...42Z}. In such a case, a magnetar can dump its rotational energy in pointing flux, which could accelerate a small amount of residual matter to very high energies, producing the relativistic jet. BH accretion and magnetar are both possibilities for powering the central engine which is still a matter of strong debate. However, some authors used the diagnostic that the upper limit of the magnetar energy budget available to form a GRBs jet is $\sim$ 10$^{52}$ erg to constrain the central engine \citep{2012A&A...539A...3B, 2021ApJ...908L...2S, 2018ApJS..236...26L}.

Two possible progenitors exist for the two classes (long and short) of GRBs. An artist's view of possible progenitors is displayed in Figure \ref{progenitors}. Although the duration-based classification is not clean, recently, some hybrid cases have been discovered. That challenges our understanding and suggests that a multi-wavelength criterion is needed to find the true class/ possible progenitor of GRBs (see section \ref{hybrid}).

\begin{figure}[ht!]
\centering
\includegraphics[scale=0.35]{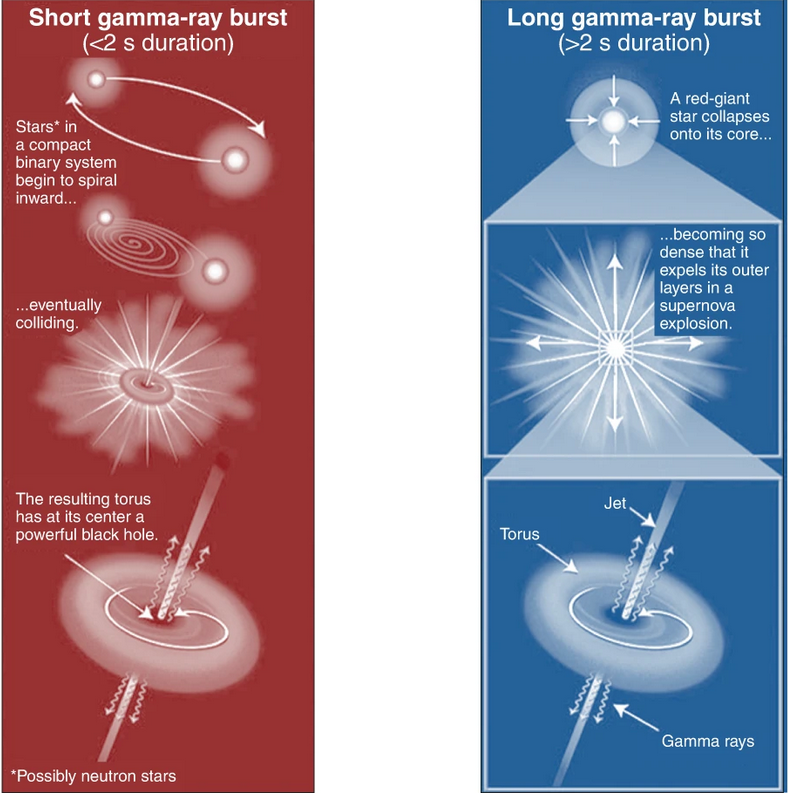}
\caption{A artist's view of possible progenitors of GRBs \citep{1993ApJ...405..273W, 1998Natur.395..663K, 2006ARA&A..44..507W, 1998ApJ...507L..59L}. Figure credit: \cite{2021NatAs...5..877A}.}
\label{progenitors}
\end{figure}

\subsection{Collapsar model}

The nearby long bursts are generally found to be connected with supernovae (broad-line Type Ic supernovae), and this provides a piece of indirect evidence that long bursts are originated by the collapse of massive stars \citep{1998Natur.395..663K, 2006ARA&A..44..507W, mac01}. For a handful of cases, the GRB-supernovae association has been confirmed using optical spectroscopy \citep{2003ApJ...591L..17S, 2003Natur.423..847H, 2006Natur.442.1011P, 2021A&A...646A..50H, 2011ApJ...735L..24S, 2012A&A...547A..82M, 2013ApJ...776...98X, 2011MNRAS.411.2792S}. The bursts should be nearby for the detection of associated supernovae (otherwise, they will be very faint to be observed using current-generation optical telescopes), and the optical afterglow should be relatively faint. Such features are typically consistent with low-luminosity GRBs \citep{2006Natur.442.1008C, 2006Natur.442.1014S}. Therefore, the majority of bursts connected to supernovae are low-luminosity GRBs, although a few high-luminosity GRBs are also found to be connected with supernovae \citep{2013ApJ...776...98X, 2017AdAst2017E...5C}.  
Long GRBs are considered to be described using the collapsar (collapse of massive stars) model, first given by \cite{1993ApJ...405..273W}. According to this model, a massive star ($>$ 30 $\msun$) burns all of its fuel, and it is so massive that it can no longer withstand the force of gravity through internal pressure. In such a scenario, the massive star collapses on itself, producing a BH at the center. The residual matter around the BH forms the accretion disk. The accretion process (accretion of matter on the BH) powers the relativistic jets of material that penetrates out of the surrounding material of the collapsing star and produces a burst of gamma rays \citep{1999ApJ...524..262M, mac01}. The collapsar origin of long GRBs is also confirmed using their location in star-forming galaxies \citep{2006Natur.441..463F}.

\subsection{Merger model}

The short-duration GRBs origin is considered to be described using the binary merger model (it could be the merger events involving either two neutron stars (NS) or the merger between a NS and a BH, \citealt{2005Natur.438..994B, 1998ApJ...507L..59L, 2005astro.ph.10256K, 2013ApJ...775...18B, 2010MNRAS.406.2650M}) and sometimes accompanied by kilonova emission \citep{2013Natur.500..547T, 2017Natur.551...75S, 2017Natur.551...64A}. A kilonova is typically fainter than a supernova and is expected to be produced due to r-process nuclei synthesis of neutron-rich material \citep{1999ApJ...525L.121F, 2013PhRvD..87b4001H}. The supernova emission associated with short-duration GRBs is ruled out based on the deep searches using larger telescopes \citep{2011ApJ...734...96K, 2014ARA&A..52...43B}. The non-detection of associated supernova emission is in agreement with the compact merger origin. According to the merger model, two compact objects, such as NS (extremely dense sources), spiral toward each other, and gravity pulls them closer and closer. Finally, these compact objects merge and release gamma rays and gravity waves from them at nearly the same (the GRB emission occurs after the merger, whilst the gravity waves occur before the merger and also shortly after the merger) time \citep{1998ApJ...507L..59L, 2017ApJ...848L..13A, 2017ApJ...850L...1L}. The merger model for short-duration GRBs was recently established through gravitational wave detection (GW 170817) in August 2017 and associated with the short-duration GRB (GRB 170817A) and a kilonovae \citep{ 2017Natur.551...75S, Coulter:2017Sci.358.1556C, 2017Natur.551...64A}.

\subsection{Recent hybrid cases}
\label{hybrid}

On August 26, 2020, \fermi detected a nearby short GRB (GRB 200826A) lasting about 1 second. Less than a day after the GRB, astronomers identified a fading visible light source using the Zwicky Transient Facility at Palomar Observatory. After a few weeks, when the afterglow had decayed, ground-based observatories discovered the brightening of the unexpected supernova. This suggests that GRB 200826A was produced due to a massive collapsing star, not a merger. The results also indicate that a few short GRBs are likely misclassified as mergers when, instead, they were created from the jets that nearly failed to drill through collapsing stars \citep{2021NatAs...5..911Z, 2021NatAs...5..917A}. On December 11, 2021, \fermi and \swift detected a nearby long GRB (GRB 211211A) lasting about a minute. However, there were many prompt and afterglow properties similar to short bursts. Later optical/NIR follow-up observations showed evidence for a kilonova (see Figure \ref{kilonova_GRB211211A}), indicating that the origin of these GRBs can be attributed to the merger of neutron stars \citep{2022Natur.612..223R, 2022Natur.612..228T, 2022Natur.612..232Y, 2022Natur.612..236M, 2023NatAs...7...67G}. These discoveries indicate hybrid cases where we could see supernovae emission from short bursts and kilonovae emission from long bursts. The above discoveries also suggest that the traditional classification scheme of GRBs has several issues. A multiwavelength analysis is required to confirm the physical class of the bursts and when supporting a hybrid model (one can not rely 100 \% on the classification done only with the gamma-ray emission) for their possible progenitors.

\begin{figure}[ht!]
\centering
\includegraphics[scale=0.21]{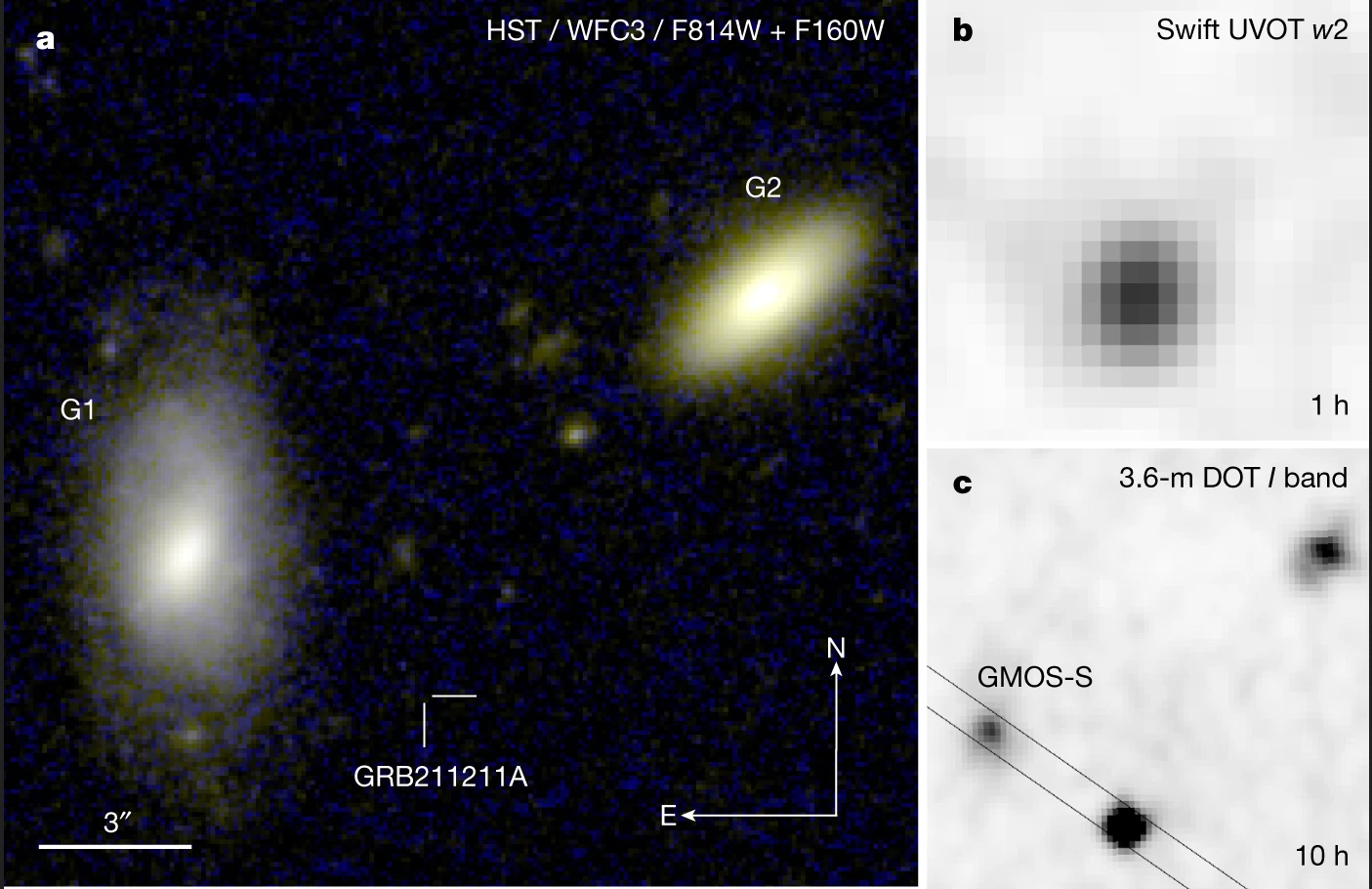}
\includegraphics[height=8cm, width=10.7cm]{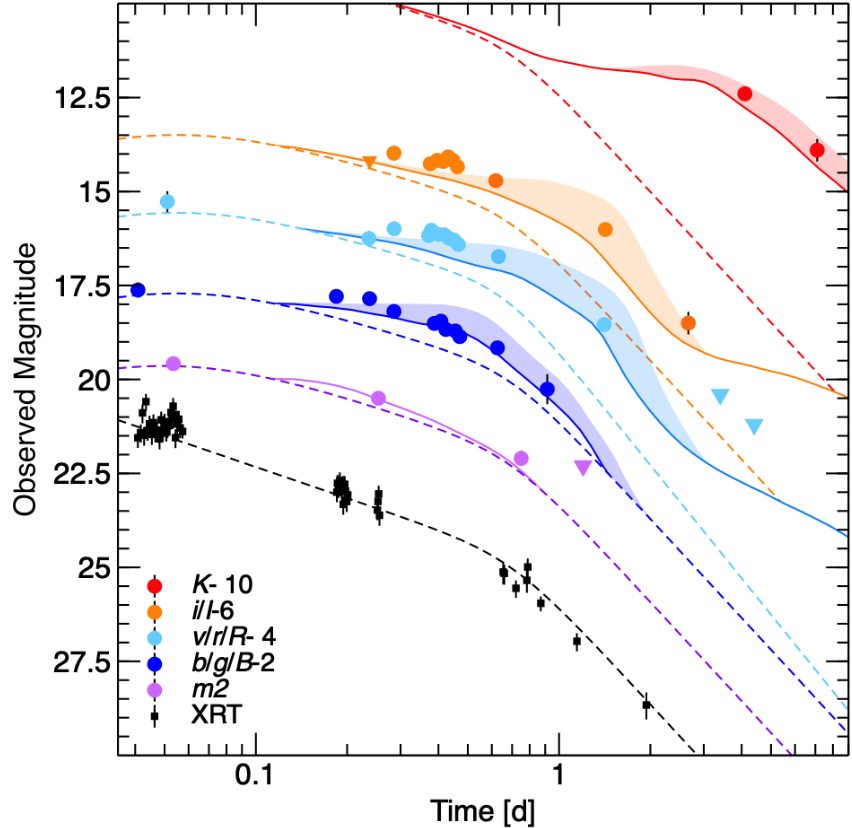}
\caption{Top panel: The observed field of long GRB 211211A accompanied by kilonova emission, taken using HST, 3.6\,m DOT, Gemini, and \swift UVOT telescopes. Bottom panel: Temporal evolution and modelling of kilonova component of GRB 211211A. Figure credit: \citep{2022Natur.612..228T}.}
\label{kilonova_GRB211211A}
\end{figure}

\section{Host galaxies and environments of GRBs}
\label{host_galaxies_environments}

The study of observed characteristics of underlying host galaxies provides information on the environment, possible progenitors, morphologies, and the distance scale or redshift measurement \citep{2006AIPC..836..540S, wai07}. The distance scale measurements are useful, specifically in the case of short GRBs, as the optical afterglows of short GRBs (median redshift $\sim$ 0.5) are very faint to perform spectroscopy measurements \citep{2006ApJ...638..354B, 2009ApJ...690..231B, 2022MNRAS.515.4890O}. GRB host galaxies offer a unique opportunity to investigate the relation between ionized and neutral gas within distant star-forming galaxies \citep{2001A&A...370..398C, 2006AIPC..836..540S}. 

Long bursts typically take place in faint and star-forming host galaxies \citep{1998ApJ...507L..25B}. They are generally found close to the bright UV regions of the host galaxy and relatively low offset from the centre region of the underlying host galaxies \citep{2002AJ....123.1111B}. They usually occur in irregular, small, low-metalicity galaxies \citep{fyn03}. Long GRBs are associated with massive stars (short-lived) found within these galaxies \citep{2003A&A...400..127G, 2017MNRAS.467.1795L, 2013ApJ...778..128P}. Currently, the galaxies hosting long-duration GRBs have been found and studied at the redshifts as high as 9.4 \citep{2011ApJ...736....7C}. A collection of host galaxies of long GRBs is shown in Figure \ref{host_galaxies}. Whereas short GRBs have very diverse environments, they can occur in early-type as well as late-type galaxies with ongoing star formation. But short GRBs are not found coincident with the star formation \citep{2005Natur.437..851G}. They are even located outside the light of their galaxy (due to natal kicks) by a few kiloparsecs, which is consistent with the predictions of binary population synthesis for neutron star mergers. They have relatively larger offsets (i.e., the distance of afterglow position from the centre region of the host galaxies, with a median offset $\sim$ five kpc) with respect to long GRBs \citep{fon10}. Therefore, studying host galaxies is essential because it tells us how binary neutron stars merge and provides insight into the amount of ejected mass \citep{2009ApJ...690..231B}. More details about the host galaxies of GRBs and their environment are given in Chapter 6. 

\begin{figure}
\centering
\includegraphics[scale=0.7]{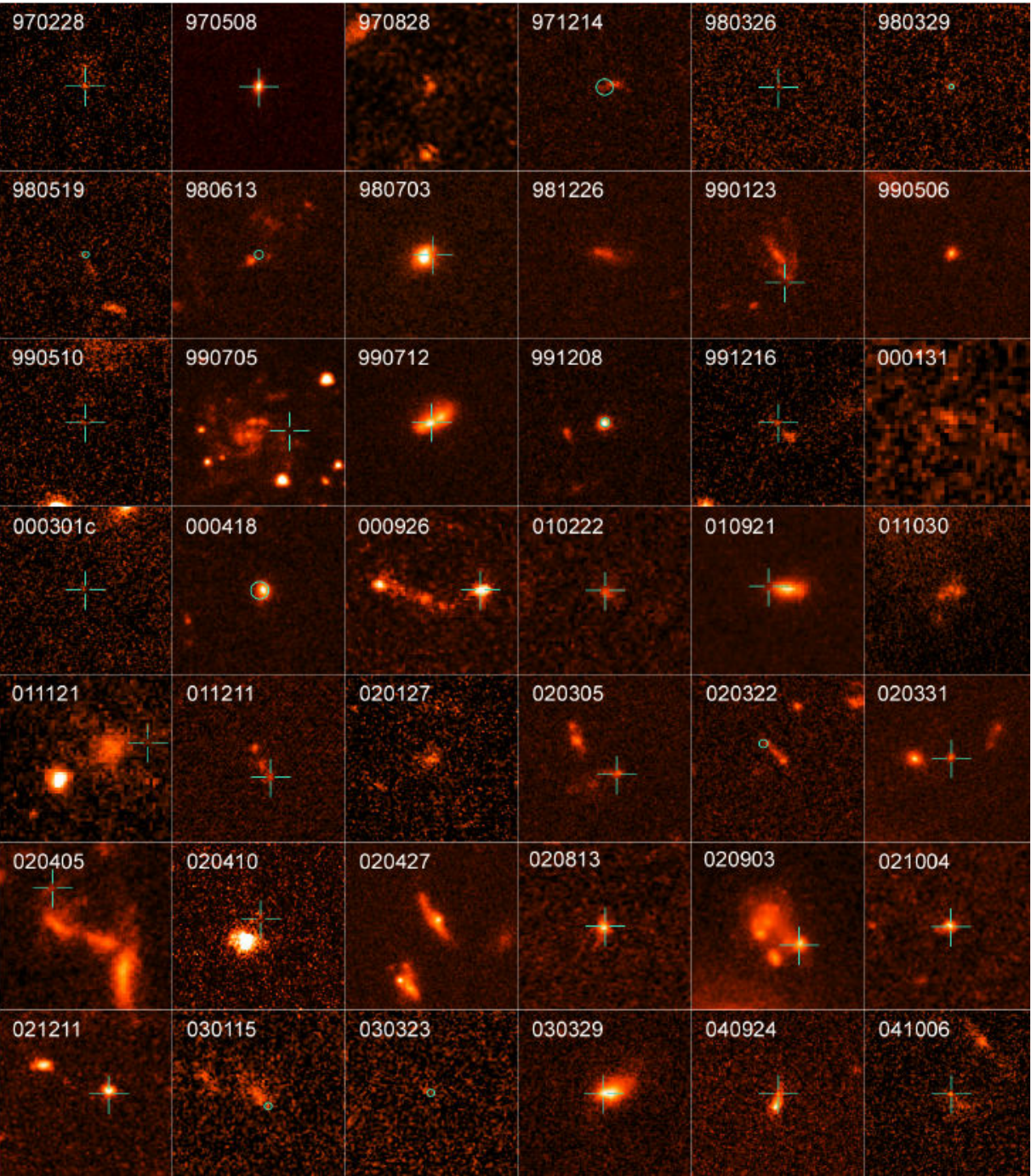}
\caption{A collection of host galaxies of long GRBs. The green marker shows the position of GRBs.  Figure credit: \cite{2006Natur.441..463F}.}
\label{host_galaxies}
\end{figure}

In the case of Dark bursts, the observations of the underlying host galaxy can determine the distance of the source of the burst \citep{2001ApJ...562..654D, 2009AJ....138.1690P}. Investigating host galaxies of dark GRBs may help us to determine the nature of the darkness of GRBs \citep{2001ApJ...562..654D, 2003A&A...400..127G, 2009AJ....138.1690P, 2022JApA...43...82G}. Usually, the galaxies hosting dark GRBs are faint and have a small angular size. Comparing the properties of host galaxies of dark and bright GRBs indicates that dark GRBs predominantly manifest within red galaxies \citep{2007A&A...475..101C, 2009AJ....138.1690P}. Moreover, the average host extinction for dark GRBs indicates more obscured and dusty galaxies \citep{2009AJ....138.1690P, 2012A&A...537A..15S, sav03, 2010MNRAS.401.2773S}. In contrast, non-dark GRBs prefer less dusty galaxies. Dark GRBs are present in galaxies with a less homogeneous medium distribution where individual line-of-sight absorption to intrinsic host absorption may differ. Dark bursts are often found in galaxies with higher star formation rates than typical galaxies of bright long GRBs \citep{2011A&A...534A.108K, 2015ApJ...806..250H}.

\section{Thesis Organization} 

The primary aim of this thesis is to study different aspects of GRBs from prompt emission to the afterglow phase, including host galaxies properties of GRBs. For this purpose, multiwavelength data encompassing a broad range of frequencies, including high-energy gamma rays to low-frequency optical bands, were employed in this study. These data were acquired from diverse space-based missions (such as \swift, \fermi, \AstroSat-CZTI, ASIM) as well as ground-based telescopes, including 3.6\,m Devasthal Optical Telescope (DOT) of Aryabhatta Research Institute of Observational Sciences (ARIES) Nainital. Using multiwavelength investigation of prompt emission and afterglow of GRBs, we examined the nature of possible progenitors, emission mechanisms, and nature of the outflows, which are not unambiguously established for GRBs. These results were then compared with extensively studied samples of GRBs for further analysis and validation. We investigated the observed properties of prompt emission and afterglow light curves, as well as the spectral energy distributions (SEDs) of GRBs. We analyzed these characteristics in relation to the geometry of ejection from the central engine of the burst. This thesis also presents the prompt emission spectro-polarimetric analyses done using joint \fermi and \AstroSat/CZTI observations. Based on the above objectives, we aimed to understand the radiation mechanisms of GRBs, their possible progenitors, and burst environments. 
The present thesis consists of seven chapters. The outline for different chapters is presented below:

{\bf Chapter 1:} In this chapter, we have introduced various GRB features from their first discovery to their present understanding and set a basis for the present thesis. We have introduced the research problem of prompt emissions, such as the emission mechanism and jet compositions, and our methods to resolve such issues. We have also introduced afterglows and host galaxies characterized in the context of our study, presented in later chapters. 

{\bf Chapter 2:} This chapter presents the details of various space and ground-based telescopes used to fulfil the goals of the present thesis. We have also given the methods of data analysis techniques for these missions in this chapter. 

{\bf Chapter 3:} This chapter is based on the results from our detailed analysis of prompt emission observations of bright GRBs detected by \fermi, \swift, \AstroSat, and ASIM missions. We have used spectral analysis and polarization measurements to understand the physical mechanisms of the prompt emission. Our results emphasize that the spectro-polarimetric approach offers unique opportunities to resolve the open issues related to the prompt emission of GRBs.

{\bf Chapter 4:} This chapter is based on the results of our detailed analysis of the very early optical afterglow data of GRB 140102A. We have used the robotic telescope named BOOTES (Burst Observer and Optical Transient Exploring System) for the very early follow-up observations of this GRB. We carried out a detailed modelling of multiwavelength data, and our modelling of GRB 140102A reveals that very early optical emission is explained using reverse shock. 

{\bf Chapter 5:} This chapter is based on the results of our detailed analysis of late-time broadband afterglows data of dark (GRB 150309A, GRB 210205A) and orphan GRBs (AT2021any). We have used a meter to the 10-meter class telescopes for the follow-up observations of these bursts. Then, we carried out detailed modelling of multiwavelength data of these peculiar sources. GRB 150309A and GRB 210205A belong to the dark GRBs class, and AT2021any is the case of the orphan burst. We also explored the possible physical causes of their dark/orphan nature.

{\bf Chapter 6:} In this chapter, the properties of the host galaxy for a selected sample of observed bursts were investigated using 3.6\,m DOT and 10.4\,m GTC optical telescopes to understand the environments of GRBs. We carried out our detailed modelling of photometric data of these host galaxies and the obtained results were compared with a well-studied sample of GRB host galaxies. Our results emphasize that 3.6\,m DOT has a unique potential for faint source observations.

{\bf Chapter 7:} In this chapter, we highlight the final summary of this thesis and give the conclusions to our work. This chapter also provides information on the future prospects of the present thesis work.               
\chapter{\sc Multiwavelength observations and data analysis}\label{Ch2}
\ifpdf
    \graphicspath{{Chapter2/Chapter2Figs/PNG/}{Chapter2/Chapter2Figs/PDF/}{Chapter2/Chapter2Figs/}}
\else
    \graphicspath{{Chapter2/Chapter2Figs/EPS/}{Chapter2/Chapter2Figs/}}
\fi

\ifpdf
    \graphicspath{{Chapter2/Chapter2Figs/JPG/}{Chapter2/Chapter2Figs/PDF/}{Chapter2/Chapter2Figs/}}
\else
    \graphicspath{{Chapter2/Chapter2Figs/EPS/}{Chapter2/Chapter2Figs/}}
\fi

To fulfil the goals of this thesis, prompt emission and afterglow data of GRBs have been taken from the following space and ground-based astronomical facilities:

\begin{itemize}
    \item {NASA's \fermi space observatory:} \fermi has broadened the spectral coverage for detailed spectral analysis of GRBs.
    \item {ESA's ASIM observatory:} ASIM has an excellent temporal resolution for precise temporal analysis of GRBs.
    \item {India's \AstroSat observatory:} \AstroSat has unique polarization measurement capabilities of prompt emission.
    \item {NASA's \swift observatory:} \swift has very fast slewing capabilities to observe both prompt and afterglow emissions nearly simultaneously.
    \item {Optical observatories:} Observations using ground-based optical observatories provide information about the nature of afterglow and their environment, including their distance measurement. 
\end{itemize}

The present chapter describes these space and ground-based astronomical observatories and their data analysis technique.

\section{\fermi space observatory}

NASA's \fermi space observatory was floated into the sky on 11 June 2008 to observe the gamma-ray sky from the lower earth orbit. The name of this high-energy satellite (earlier referred to as Gamma-ray Large Area Space Telescope) was changed in August 2008 in honour of Prof. Enrico Fermi, a pioneer high-energy physicist. \fermi is composed of two observing instruments onboard: Large Area Telescope (LAT) and Gamma-ray Burst Monitor (GBM). \fermi is used to perform all-sky surveys in the gamma-ray energy range. In the last 14 years of scientific observations, \fermi has discovered around 8000 GRBs across the gamma-ray sky.   

\subsection{Fermi LAT instrument}

The \fermi LAT instrument operates in the energy range of 20\,MeV to 300\,GeV, providing observations of a wide range of high-energy phenomena. With its wide field of view, it captures approximately 20\% of the sky simultaneously and completes a full sky coverage every three hours \citep{2009ApJ...697.1071A}. It has an astrometric resolution of only a few arc minutes. LAT is made of thin metal sheets, and when the high energy gamma rays hit the metal sheets, the photons are converted into an electron-positron pair (pair production). These are charged particles that travel through the instrument, which is then filled with layers of silicon micro-strip detectors. As the electrically charged electron and positron pair pass through these detectors, they create tiny electric currents whenever they hit. Using the data from many layers of the instrument, we can then build up the path of these particles that have travelled, and the endpoint for these particles is the detector that tells us the particle's total energy. These paths then tell us where gamma-ray photons must come from in the sky. A schematic view of \fermi mission along with both the high energy detectors (GBM and LAT) is shown in Figure \ref{FERMI_mission}. \fermi LAT, on average, detects around 15 bursts per year.

\begin{figure}
\centering
\includegraphics[scale=0.37]{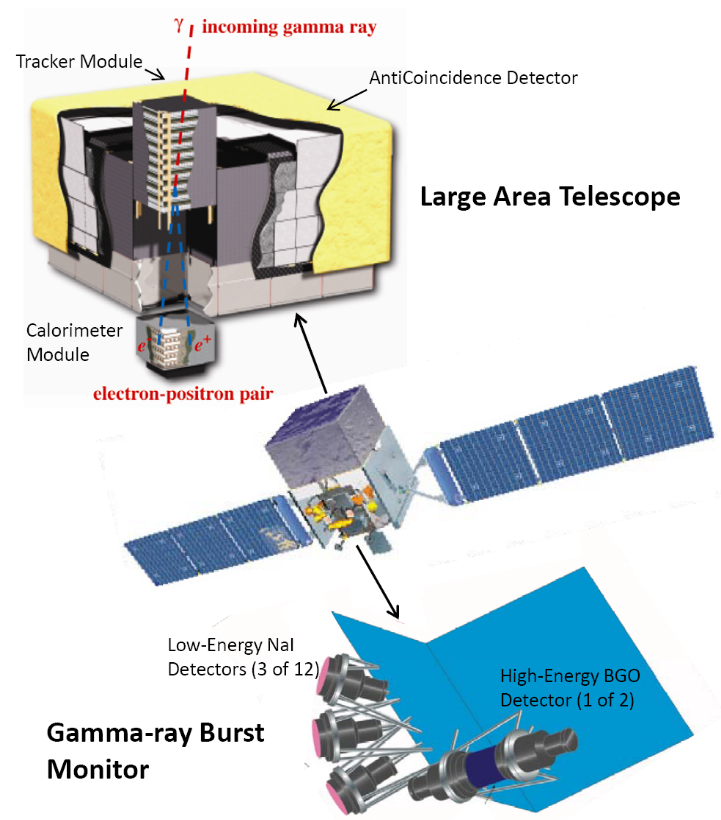}
\caption{A schematic view of \fermi mission along with both the high energy detectors (GBM and LAT). Figure credit: \cite{2010RPPh...73g4901M}.}
\label{FERMI_mission}
\end{figure}

\subsection{Fermi GBM instrument}

GBM is mainly optimized for finding the bright flashes of GRBs. The GBM contains 14 detectors and is expected to pick up about 200 GRBs annually. Of these 14 detectors, it is composed twelve low energy detectors (made up of sodium iodide (NaI)) and two high energy detectors (made up of bismuth germanate (BGO)). NaI detectors observe the Universe in the energy range of 8 \keV to 1\,MeV, while the BGO detectors cover the range of 200 \keV to 40\,MeV. These 14 detectors are positioned in various orientations to provide full sky coverage \citep{2009ApJ...702..791M}. The GBM uses a few simple scintillation processes to collect the data. When gamma rays enter these detectors, they interact with crystals in the instrument and produce scintillation. The more energetic the gamma-ray is, the more light the crystals produce. By seeing which crystals light up, the GBM can tell from which direction the GRBs are coming. This process is known as localization. Although, due to a focusing issue with the gamma-rays, GRB localization of \fermi mission is not very good. GBM observations of the brightest explosion in the Universe allow scientists to understand these unique sources better. 

\fermi GBM science data are available in three formats: CTIME, CSPEC, and Time-tagged events (TTE). CTIME and CSPEC are pre-binned science data files with temporal resolutions of 0.256\,sec and 1.024\,sec, respectively. On the other hand, TTE files are unbinned files with a very high temporal resolution of 64\,ms. Additionally, TTE and CSPEC data files have a good spectral resolution, with spectral binning in all 128 energy channels. Therefore, we mostly used TTE mode data for GBM data analysis.

{\bf Data Analysis:}

We have employed standard tools for the analysis of temporal and spectral properties utilizing the \fermi data. For example, we have used \sw{RMFIT} and Fermi GBM Data Tools \citep{GbmDataTools} to extract the timing information of \fermi GBM observations. Additionally, we have employed \sw{gtburst} tool for the analysis of \fermi LAT data and extraction of spectra using \fermi GBM data. We have utilized X-ray Spectral Fitting Package \citep[\sw{XSPEC};][]{1996ASPC..101...17A} as well as an advanced Bayesian-based tool called as Multi-Mission Maximum Likelihood framework \sw{3ML\footnote{https://threeml.readthedocs.io/en/latest/}} for the spectral fitting of time-averaged and time-resolved spectra of GRBs \citep{{2015arXiv150708343V}}. A more detailed method for analyzing \fermi data is given in chapter \ref{ch:3} of the thesis.

\section{ASIM observatory onboard ISS}

\begin{figure}[ht!]
\centering
\includegraphics[scale=0.3]{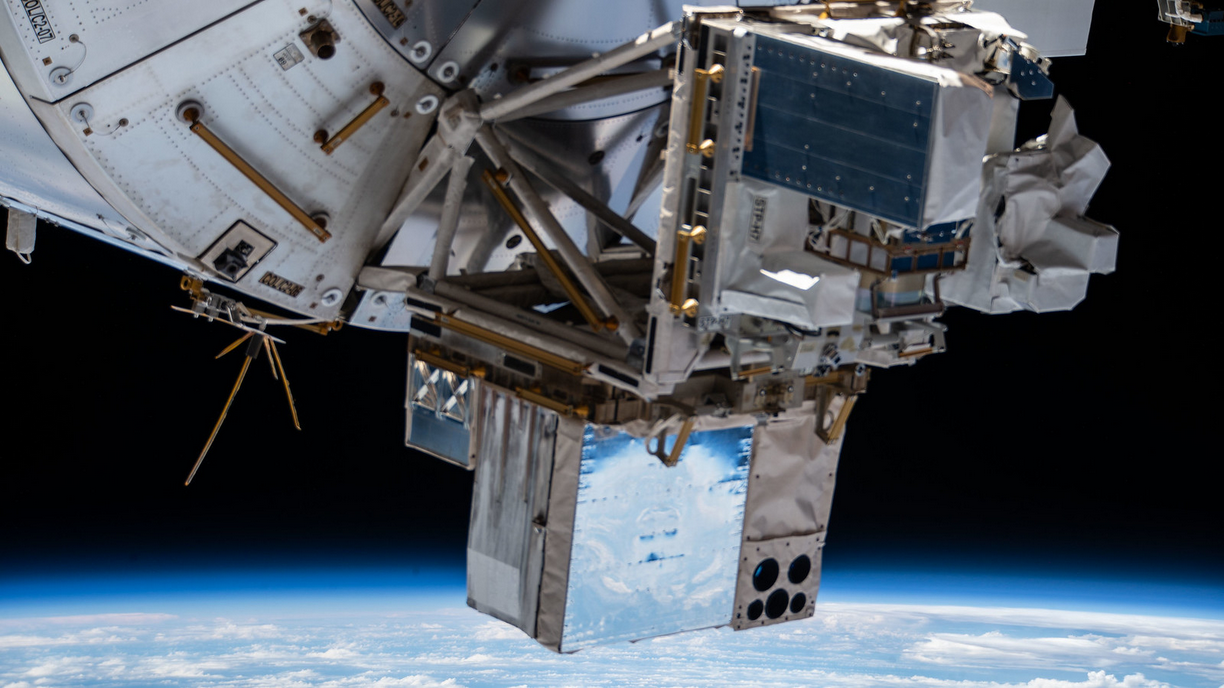}
\caption{ASIM deployed on the ({\it ISS}). Figure credit: \href{https://asim.dk/images.php}{ASIM team}.}
\label{ASIM_mission}
\end{figure}

European Space Agency's Atmosphere-Space Interactions Monitor (ASIM) is developed as a climate observatory on the International Space Station ({\it ISS}). ASIM was launched and installed in April 2018 \citep{2019arXiv190612178N}. The main objectives of the ASIM are to study the luminous transient events, terrestrial gamma-ray flashes, and upper atmospheric lightning. In addition to these objectives, ASIM can also discover GRBs and similar energetic sources \citep{2021Natur.600..621C}. This mission comprises two scientific instruments: the Modular X and Gamma-ray Instrument or MXGS for short \citep{2019SSRv..215...23O} and the Modular Multispectral Imaging Assembly or MMIA for short \citep{2019SSRv..215...28C}. The first instrument, i.e. MXGS, is used explicitly for gamma-ray detection, and MMIA is specifically used for optical imaging. The MXGS instrument is composed of the Low Energy Detector or LED for short, and the High-Energy Detector or HED for short. The LED is made up of Cadmium-Zinc Telluride (CZT) to observe between 50-400 \keV, and HED is made up of Bismuth Germanium Oxide crystals to observe between 300\,keV and $>$30\,MeV energy range, respectively. Figure \ref{ASIM_mission} illustrates an image of the ASIM deployed on ({\it ISS}). The data analysis method of ASIM data is given in chapter \ref{ch:3} of the thesis.

\section{\AstroSat observatory}

\AstroSat is the inaugural space-based observatory dedicated to scientific research from India, launched on 28$^{th}$ September 2015. \AstroSat observes the Universe from the Ultra-Violet/optical band to the hard X-rays regime with its five instruments (see Figure \ref{AstroSat}). The Cadmium Zinc Telluride Imager (CZTI) is the crucial instrument for the present thesis work; it is a hard X-ray coded aperture mask telescope consisting of an array of CZT detectors and Cesium Iodide (CsI) based scintillators as Veto detectors. It is mainly an X-ray imaging and spectroscopic instrument with GRB detection capabilities utilizing its wide field of view at energies above 100 \keV. It also has a unique ability for polarization measurement in the 100-380 \keV energy range and thus gives a unique chance to perform spectro-polarimetric studies of GRBs. Currently, it is the only space-based observatory actively measuring GRB polarisation in hard X-rays. A schematic diagram of the CZTI, along with different components of the individual quadrant, is shown in Figure \ref{CZTI}.

\begin{figure}[ht!]
\centering
\includegraphics[scale=1.5]{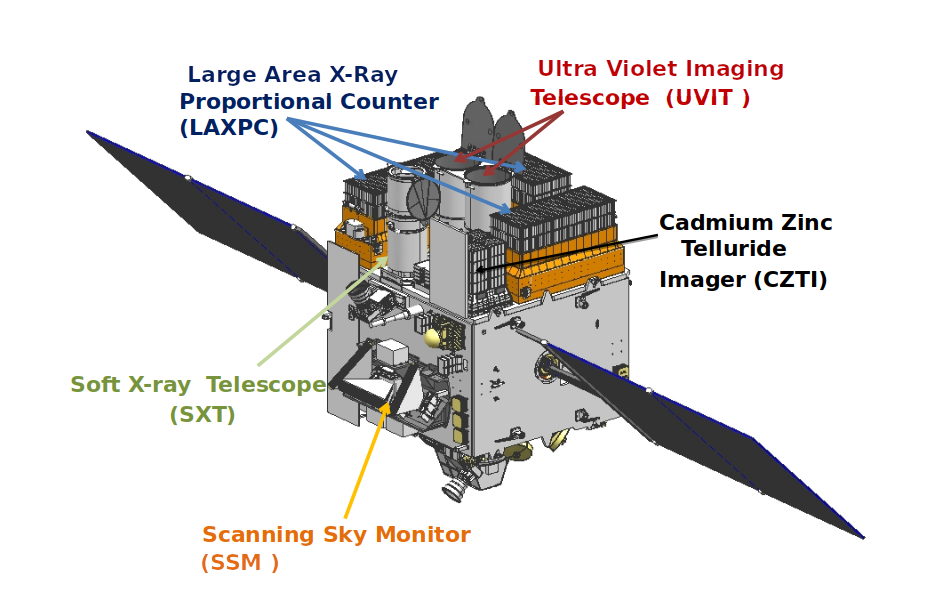}
\caption{India's \AstroSat mission. The five instruments covering the Ultra-Violet/optical band to the hard X-rays regime of the electromagnetic spectrum are also shown. Figure credit: \href{https://astrobrowse.issdc.gov.in/astro\_archive/archive/Home.jsp}{ISRO Science Data Archive}.}
\label{AstroSat}
\end{figure}

\begin{figure}[ht!]
\centering
\includegraphics[scale=.6]{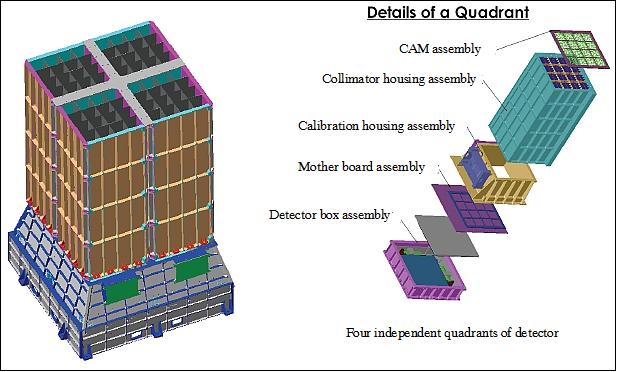}
\caption{A schematic diagram of the CZTI. The different components of a quadrant are also shown. Figure credit: AstroSat CZTI team.}
\label{CZTI}
\end{figure}

\subsection{Polarimetric data analysis of CZTI data}

\AstroSat CZTI is primarily a hard X-ray imaging/spectroscopy detector with a wide field of view. Additionally, it has shown the polarization measurement capabilities for on-axis sources during the instrument's ground calibration. Recently, \cite{2022JATIS...8c8005V} experimentally verified the off-axis polarization measurement capabilities of bright sources like GRBs using CZTI. Above 100 \keV, CZTI has a significant probability of Compton scattering and utilizing the pixilated nature of CZT detectors, it works as Compton Polarimeter. Considering its unique hard X-ray polarization measurement capabilities, the CZTI team reported the polarization measurements of persistent (Crab pulsar and nebula) as well as transient (GRBs) X-ray sources \citep{2016ApJ...833...86R, 2018NatAs...2...50V, 2019ApJ...884..123C}. Nevertheless, in the case of energetic transient sources like GRBs, the simultaneous availability of pre and post-burst backgrounds with a higher signal-to-noise ratio makes them potential X-ray sources for polarization measurements despite moderate brightness. A detailed technique of polarization analysis of GRBs using CZTI data is given in \cite{2022ApJ...936...12C}. In this section, we provide brief steps for the polarization analysis of CZTI data.

\begin{itemize}
    \item {Selection of Compton counts:} For the polarization analysis using the CZTI, we initially selected the double events detected within 20 $\mu$s temporal window. Further, we applied Compton criteria (ratio of energies received on neighbouring pixels) to filter these double events due to chance coincidence.  
    
    \item {Creating background-subtracted azimuthal angle distribution:} We implemented the selection of Compton events both on GRB emission region as well as pre and post-burst regions, i.e., the background regions. While selecting the pre and post-background regions, we removed the spacecraft's South Atlantic Anomaly crossing. Then, we subtracted the raw azimuthal angle distribution of the GRB emission region from the background distribution of the azimuthal angle and created the final background-subtracted azimuthal angle distribution of GRB. 
    
    \item {Correction for geometric effects:} The background-subtracted azimuthal angle distribution of GRB is affected due to the systematic errors coming from geometric effects and off-axis detection of GRBs. Therefore, we simulated the unpolarised azimuthal angle distribution utilizing the Geant4 toolkit and the \AstroSat mass model, considering the same distribution of photons observed from GRB spectra at the same orientation of the \AstroSat spacecraft. Then, we normalized GRB's observed background-subtracted azimuthal angle distribution using the simulated unpolarised azimuthal angle distribution.
    
    \item {Calculation of modulation amplitude and polarization angle:} We utilized a sinusoidal function to fit the observed background-subtracted and geometry-corrected azimuthal angle distribution of the GRB. This allowed us to determine the modulation factor ($\mu$) and polarization angle within the \AstroSat CZTI plane. We utilized the Markov chain Monte Carlo (MCMC) for the sinusoidal function fitting.
    
    \item {Calculation of PF:} To determine the polarization fraction, we need to normalize the modulation factor ($\mu$) with the simulated modulation amplitude for 100 \% polarized radiation $\mu_{100}$, obtained utilizing the Geant4 toolkit \AstroSat mass model simulation for the same direction and the same observed spectral parameters.
Finally, we calculated the PF by normalizing the $\mu$ with $\mu_{100}$ for those bursts with the Bayes factor greater than 2. If the Bayes factor is found to be below 2, we establish a constraint on the polarization fraction by setting a 2$\sigma$ upper limit (see \cite{2022ApJ...936...12C} for more details). 
\end{itemize}

\section{\swift observatory}

NASA's \swift space observatory was launched on 20$^{th}$ November 2004. It is a multiwavelength mission with three scientific instruments covering gamma-ray/hard X-ray to optical, electromagnetic spectrum wavelengths: 1. Burst Alert Telescope (BAT; \citealt{2005SSRv..120..143B}); 2. X-ray telescope \citep[XRT;][]{2005SSRv..120..165B}, 3. Ultra-Violet and Optical Telescope \citep[UVOT;][]{2005SSRv..120...95R}. The main objective of this mission is to study the GRB prompt emission and afterglow observations \citep{2004ApJ...611.1005G}, but \swift has evolved into the kind of a transient response observatory for the whole astronomical community in a lot of different scientific areas. The name of this mission was given in honour of Prof. Neil Gehrels, principal investigator (PI) of the mission and a pioneer of gamma-ray astronomy. \swift is discovering around 100 GRBs per year with arc minutes localization. A schematic view of Neil Gehrels \swift observatory is displayed in Figure \ref{SWIFT_mission}.

\begin{figure}[ht!]
\centering
\includegraphics[scale=0.55]{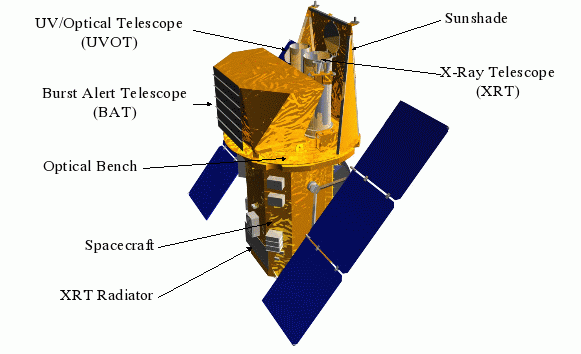}
\caption{A schematic illustration of Neil Gehrels \swift observatory along with three detectors (BAT, XRT and UVOT). Figure credit: \href{https://www.swift.ac.uk/about/instruments.php}{UK Swift Science Data Centre}.}
\label{SWIFT_mission}
\end{figure}

\subsection{\swift BAT instrument}

The BAT instrument on board the \swift mission has a wide field of view, covering approximately 2 steradians of the sky. It operates in an energy range of 15-350\,\keV. When there is a transient or a gamma-ray burst within the field, BAT detects it and determines its position within a couple of arc minutes. Then, the position is sent to the spacecraft, which autonomously slews ($\sim$ 20-70 s) and points the X-ray telescope (XRT) and the optical telescope (UVOT) towards the corresponding position to look for the afterglow of the burst. Rapid localization using \swift inform other telescopes in space and on the ground where to observe. BAT uses a coded mask aperture technique to detect GRBs/transients. A coded mask aperture is a simple method to detect hard X-ray/gamma rays because this light can not be focused as in the case of optical/ conventional telescopes. Consider a mask or imaging plane populated with absorptive materials, small tiles, for example, lead. If an incoming X-ray photon hits one of these tiles, it will be absorbed by the tiles (very dense and thick material). But this mask also has gaps, where X-ray photons can penetrate directly through an illuminated detector. If we place these tiles in some pattern, depending on the source location in the sky, it will generate a very different shadow pattern on the detector plane. Therefore, using this technique, we can infer by the shadow pattern generated on the detector plane where the source is located in the sky.         

\subsection{\swift XRT instrument}

The XRT instrument, part of the \swift mission, operates within the energy range of 0.3-10 \keV and possesses a narrow field of view. Soon after the discovery by the BAT instrument with arc minutes localization, XRT determines the position of the burst to a couple of arc seconds. In addition to better positional information, XRT provides flux measurement, temporal information, and a spectrum to study GRBs and their afterglow in soft X-ray energy bands. The telescope used is a Walter 1 telescope, which has an effective area of 120\,cm$^{2}$ at an energy of 1.5\,\keV. The XRT works on the principle of grazing incident, also known as Walter one type optics. Instead of reflecting at 90 degrees, X-ray mirrors are reflected at very shallow angles. For soft X-rays, in particular, these angles are of the order of one to two degrees, and by coming at these very small angles, X-ray photons are not going to be absorbed by the material; instead, they can be reflected similarly to optical light. If we have two sets of particularly shaped mirrors (a parabolic segment and a hyperbolic segment), the incoming plane wave will be reflected down to focus. In this way, we can get a proper focusing image similar to the optical wavelengths.    

XRT has four different readout modes of operations to cover the rapidly evolving nature of the X-ray afterglows of GRBs:

\begin{itemize}
    \item Imaging (IM) mode: In this mode, XRT only collects the image of the GRB position. 
    
    \item Photo-Diode mode (PD): This mode was developed for a high temporal resolution of 0.14\,ms and full energy resolution. This mode is inactive as the \swift telescope was shocked by a micro meteorite on May 2005. 
    
    \item Windowed Timing (WT) mode: XRT obtains a high temporal resolution of 1.8\,ms with one-dimensional imaging in this mode.
    
    \item Photon Counting (PC) mode: In PC mode, XRT obtains the source's complete spectral and spacial (two-dimensional imaging) information.
    
\end{itemize}

\subsection{\swift UVOT instrument}

The optical counterparts of GRBs usually decay rapidly as t~$^{-1}$ to t~$^{-2}$; therefore, quick follow-up observations of such sources are needed. The Ultra-Violet and Optical Telescope instrument of \swift mission is primarily developed for the quick (utilizing its slewing capability) UV and optical observations of GRBs and their afterglows. The UVOT provides a sub-arcsec localization of the afterglow. Rapid follow-up observations and sub-arcsec localization also help to determine the redshift of these energetic objects. The UVOT consists of several UV (uvw1, uvw2, uvm2) and broadband optical filter (white, v, b, u) systems, covering 170-650\,nm of the electromagnetic spectrum. The UVOT instrument offers a field of view measuring 17' X 17'.

UVOT observations are taken in Image (no time-tagged photons)/Event (time-tagged photons) mode or a combination of both modes (i.e., Image \& Event mode). 

{\bf Data Analysis:}

The \swift prompt emission and afterglow data of GRBs is obtained from the Swift Archive Download Portal\footnote{https://www.swift.ac.uk/swift\_portal/}. 
To analyze the data from \swift BAT, XRT, and UVOT, we utilized various algorithms/packages within the HEASOFT tool developed by the \swift team. We have used the latest calibration files from HEASARC Calibration Database\footnote{https://heasarc.gsfc.NASA.gov/docs/heasarc/caldb/caldb\_intro.html} for each of these \swift instruments. We have used the \sw{caldbinfo} task to confirm if the calibration files for the individual instruments are correctly installed. A brief method for the data analysis of individual instruments is given below. A more detailed technique of \swift data analysis is given in chapters \ref{ch:3} and \ref{ch:5} of the thesis. 

\begin{itemize}
    \item {{\bf \swift BAT:} Initially, we used the \sw{fkeyprint} task on the BAT event file to confirm if Gain and offset corrections are implemented on raw event files. We further identified the hot pixels and applied the mask-weighting using the \sw{bathotpix} and \sw{batmaskwtevt} tasks, respectively. After these corrections, we created the light curve and spectrum from BAT observations using the standard \sw{batbinevt} task. Further, we carried out BAT spectral analysis using \sw{3ML} or \sw{XSPEC} tools.}
    
    \item {{\bf \swift XRT:} The XRT team provides an online repository for the standard XRT data products (count rate/flux light curves and time-averaged spectrum) of GRBs X-ray afterglows\footnote{https://www.swift.ac.uk/xrt\_products/index.php}. They also provide an online repository\footnote{https://www.swift.ac.uk/xrt\_spectra/} to get customized data products (time-resolved spectra). This thesis used these repositories to obtain the XRT light curves and spectra of the GRBs studied. Further, we carried out XRT spectral analysis using \sw{XSPEC} package within the HEASOFT software.}
    
    \item {{\bf \swift UVOT:} We initially determine the position of UV/optical afterglows using \sw{uvotdetect} task on UVOT images. Further, we select the source (within 3 arcsecs) and background (within 20 arcsecs) regions to determine the magnitude of the afterglows. We obtained the standard UVOT data products using the \sw{uvotproduct} task. If the afterglow is very faint, we initially summed the UVOT images using \sw{uvotimsum} for individual filters, then determine the magnitude and flux using the \sw{uvotsource} task. We also used the \sw{uvotmaghist} task to create the UVOT light curves in different filters.}
\end{itemize}

\section{Ground-based optical follow-up observations}
\label{GBtelescopes}

Soon after the detection of prompt emission and its X-ray counterpart, an arc-sec localization of the GRB field is received using the General Coordinates Network (GCN) circulars\footnote{https://gcn.nasa.gov/circulars}. Ground-based robotic optical telescopes start searching the optical afterglow using follow-up observations and provide sub-arcsecond localization. Sometimes, if the optical afterglow is bright enough, \swift UVOT quickly provides the sub-arcsecond localization of the optical afterglow. After receiving the arc second or sub-arc second localization of the optical afterglows, we also used our collaborative facilities to conduct follow-up observations of the optical counterparts.  


In this thesis, we have carried out optical observations of GRBs optical afterglows using the following ground-based telescopes: 1) 3.6\,m Devasthal Optical Telescope (DOT), Devasthal Observatory, Nainital, India (see section \ref{3.6m DOT} for detailed information); 2) 10.4\,m Gran Telescopio Canarias (GTC) telescope (OSIRIS and CIRCE), Canary Island, Spain; 3) 1.3\,m Devasthal Fast Optical Telescope (DFOT), Nainital, India; 4) 3.5\,m Centro Astronómico Hispano-Alemán (CAHA) telescope, Calar Alto Astronomical Observatory, Spain; 5) 0.6\,m Burst Observer and Optical Transient Exploring Optical System (BOOTES-4)/MET robotic telescope, Lijiang, China; 6) 0.6\,m BOOTES-2/TELMA robotic telescope, IHSM La Mayora, Malaga, Spain; 7) MASTER Global Robotic Net telescope, Russia; 8) 1.5\,m Observatorio de Sierra Nevada (OSN) telescope, Sierra Nevada Observatory, Granada, Spain; 9) 0.8\,m Ritchy-Chrétien (RC80) robotic telescope, Konkoly Observatory, Hungary; 10) 0.7\,m GROWTH-India Telescope (GIT), Indian Astronomical Observatory (IAO), India; 11) 2\,m Himalayan Chandra Telescope (HCT), Indian Astronomical Observatory (IAO), India; 12) 2.2\,m Centro Astronómico Hispano-Alemán (CAHA) telescope, Calar Alto Astronomical Observatory, Spain; 13) 1\,m Special Astrophysical Observatory (SAO) telescope, Russian Academy of Sciences, Russia. The key characterises of these telescopes are given in Table \ref{telescope}.

\begin{table*} 
\scriptsize
\begin{center} 
\caption{The important features of the optical/near-infrared telescopes employed for conducting the follow-up observations of optical counterparts in this thesis.}
\begin{tabular}{@{}lcccccc@{}} 
\hline
Sr. No. &Telescope & Instrument & Filter & Readout noise & Gain & Location\\ \hline
1 & 3.6\,m DOT & 4K $\times$ 4K IMAGER & UBVRI & 10.0& 5.0& India\\
2 & 10.4\,m GTC  & OSIRIS  & griz & 4.5 & 0.95& Spain \\
3 & 10.4\,m GTC  & CIRCE  & $JHK_{\rm S}$ & 50 & 5.4 & Spain \\
4 & 1.3\,m DFOT & 2K $\times$ 2K Andor & VRI&7.5 &2.0 & India\\
5 & 3.5\,m CAHA & 4K $\times$ 4K CCD &R H & --& --& Spain\\
6 & 0.6\,m BOOTES 4 & 1K $\times$ 1K Andor iXon & CgriZY &-- &-- & China \\
7 & 0.6\,m BOOTES 2 &  1K $\times$ 1K Andor iXon & CiZY&-- &-- & Spain\\
8 & MASTER & 4K $\times$ 4K CCD & unfiltered & --& --& Russia \\
9 & 1.5\,m OSN & 2K $\times$ 2K Andor ikon & BVRI &-- & --& Spain\\
10 & RC80  & 2K $\times$ 2K FLI PL230 CCD & r & --& --& Hungary \\
11 & 0.7\,m GIT & 4K $\times$ 4K Andor iKon & r & 12.0 & 1.04 & India \\
12 & 2\,m HCT & 2K $\times$ 4K HFOSC & R & 5.75 & 0.28 & India\\
13 & 2.2\,m CAHA & 2K $\times$ 2K CAFOS  & R & 8 & 1.7 & Spain \\
14& 1\,m SAO & 2K $\times$ 2K MMPP & R & 9.0 & 1.15 & Russian\\
\hline
\label{telescope} 
\end{tabular} 
\end{center} 
\end{table*}

\subsection{Importance of the 3.6m DOT for afterglow follow-up observations}
\label{3.6m DOT}

\begin{figure}[ht!]
\centering
\includegraphics[scale=0.70]{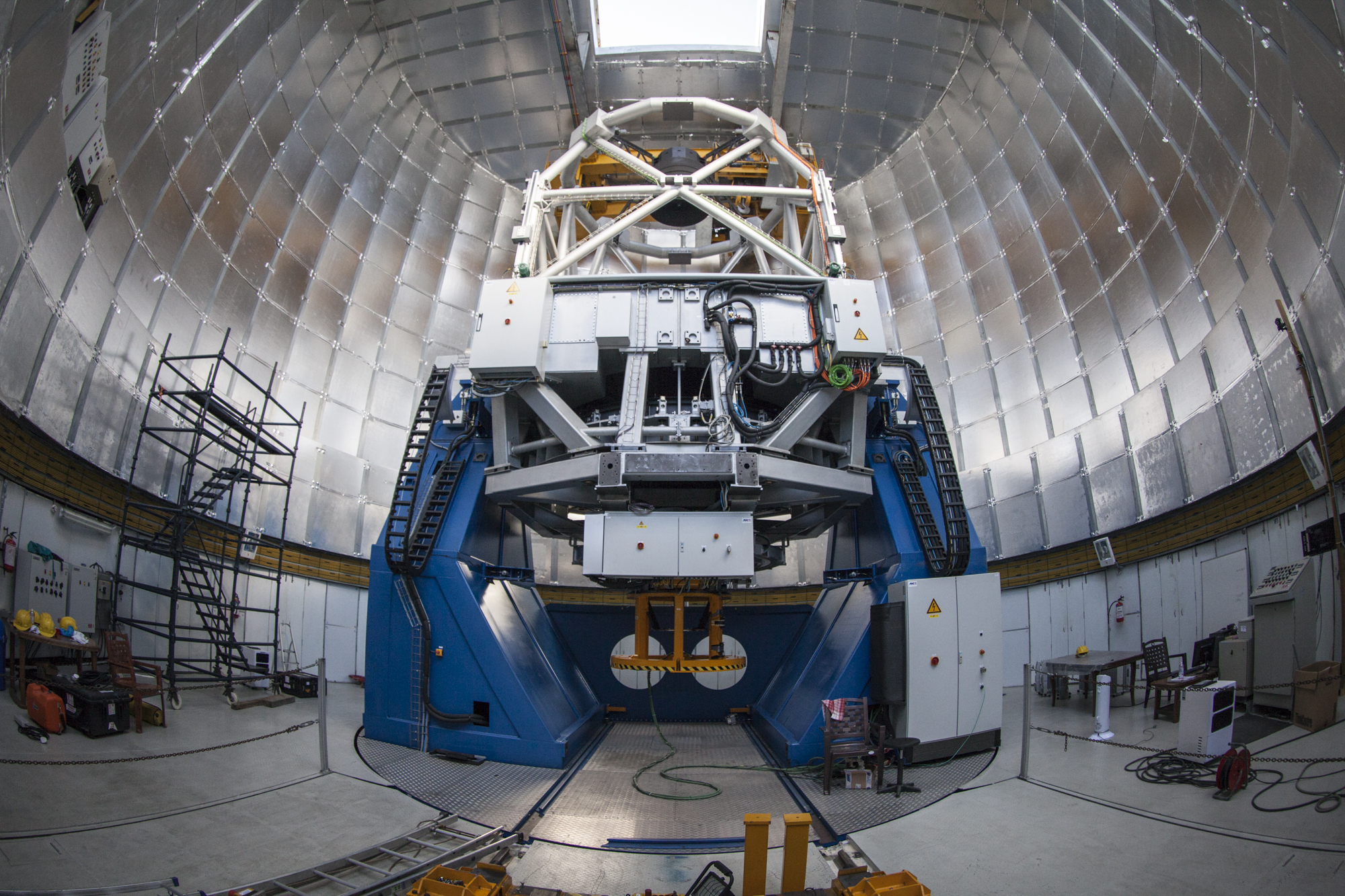}
\caption{3.6\,m Devasthal Optical Telescope installed at Devasthal observatory of Aryabhatta Research Institute of Observational Sciences (ARIES), India. Figure credit: \href{https://en.wikipedia.org/wiki/3.6m\_Devasthal\_Optical\_Telescope}{Ajay Talwar}.}
\label{DOT_ARIES}
\end{figure}

India's most giant 3.6-meter DOT is installed at the Devasthal observatory of Aryabhatta Research Institute of Observational Sciences (ARIES) in the district of Nainital, Uttarakhand, India (see Figure \ref{DOT_ARIES}). The observatory has a longitudinal advantage (Located within the central region of the 180-degree wide belt between the Canary Islands (20°W) and Eastern Australia (160°E)) for time-critical observations like afterglows of GRBs and other transients. Utilizing India's longitudinal advantages, ARIES plays a crucial role in the study and follow-up observations of the optical afterglows and has a rich history spanning over two decades of such observations. The back-end instruments of the 3.6\,m DOT offer valuable spectral and imaging capabilities in the visible and near-infrared ranges. These capabilities play a crucial role in conducting deep observations of afterglows and other rapidly fading extragalactic transient events. Four first-generation back-end instruments are currently available for broadband photometric and spectroscopy observations. (1) 4K $\times$ 4K CCD Imager for deeper photometric observations, (2) ARIES Devasthal Faint Object Spectrograph and  Camera (ADFOSC), (3) TIFR near-infrared imaging camera (TIRCAM2), (4) a TIFR-ARIES near-infrared spectrometer (TANSPEC). The telescope's big aperture size and a wide field of view of 4K $\times$ 4K CCD Imager and ADFOSC play a pivotal role in deep and faint observations of GRBs afterglows and host galaxies at optical frequencies. During the last four cycles of DOT, we observed more than two dozen afterglow/host galaxies of GRBs. Deep optical/near-IR photometric observations of afterglows are of utmost importance for various reasons. They allow us to identify any associated supernova bumps observed in nearby long bursts, shed light on the nature of afterglows and their darkness, study the occurrence of jet breaks, and investigate the environment surrounding these events. Additionally, spectroscopic observations play a crucial role in accurately determining the redshift measurements of GRBs. The broadband modelling of afterglows using multi-band photometric data taken with 3.6\,m DOT and data at other wavelengths such as X-ray from \swift XRT and radio from  Giant Metrewave Radio Telescope (GMRT) is crucial to probe the nature of potential progenitors, analyzing the geometry of the outflows, studying the emission mechanisms involved, and exploring the environments in which these bursts occur. In addition, prompt emission detections using \AstroSat (the first Indian multi-wavelength satellite), \fermi, and \swift could provide clues to the open questions on GRBs.

\subsubsection{The Filter}  
It isn't easy to measure the bolometric magnitude of astronomical sources. To measure the bolometric magnitude, we need a detector that can measure the flux at all wavelengths. Certainly, it is very hard to build such a telescope that can measure X-ray, optical, and radio emissions together. Therefore, we generally measure the flux over a finite bandpass using optical filter systems. Optical filters allow to collect the light of a particular wavelength and block the light at other wavelengths. Many photometric systems exist based on well-defined band passes and their well-characterized sensitivity. Photometric systems are divided into three different classes based on the bandwidths of the filters:
\begin{itemize}
    
\item Narrow band filters (passband $<$ 10\,nm)
\item Intermediate band filters (passband between 10\,nm to 30\,nm)
\item Broadband filters (passband $>$ 30\,nm) 
\end{itemize}

In this thesis, we have mainly used broad band filters: Johnson-Cousin $UBVRI$ and Sloan Digital Sky Survey (SDSS) $ugriz$ bandpasses systems to observe the optical afterglows/host galaxies of GRBs.

\subsubsection{4K X 4K CCD Imager}

In this thesis, we have primarily used optical observations of GRBs afterglows and their underlying host galaxies using 4K $\times$ 4K CCD Imager (see Figure \ref{4K_4K}), in-house developed and installed as the first light observing instrument at the primary port of Devasthal optical telescope. The 4K $\times$ 4K CCD Imager is designed for the photometric observations of faint astronomical sources in the optical wavelengths from around 4000 angstroms to 9000 angstroms. The 4K $\times$ 4K CCD Imager has two sets of filter wheels; One set comprises broadband Bessel U, B, V, R, I filters, while the second set comprises broadband SDSS u, g, r, i, z filters. This instrument has a CCD with 4096 $\times$ 4096 pixels$^{2}$ and with an FoV of 6.5 x 6.5 arcmin$^2$. The CCD has different gain, readout, and binning modes, providing unique capabilities for a range of observations of astronomical sources. The throughput simulations of the instrument suggest that the CCD imager can detect faint objects up to around 25.2 mag, 24.6 mag, and 24.0 mag in g (exposure time of 3600 sec), r (exposure time of  4320 sec), B (exposure time of  1200 sec) filters, respectively. More detailed information about the 4K $\times$ 4K CCD Imager and its filter systems are available in \cite{2017arXiv171105422P, 2022JApA...43...27K}.

\begin{figure}[ht!]
\centering
\includegraphics[scale=0.70]{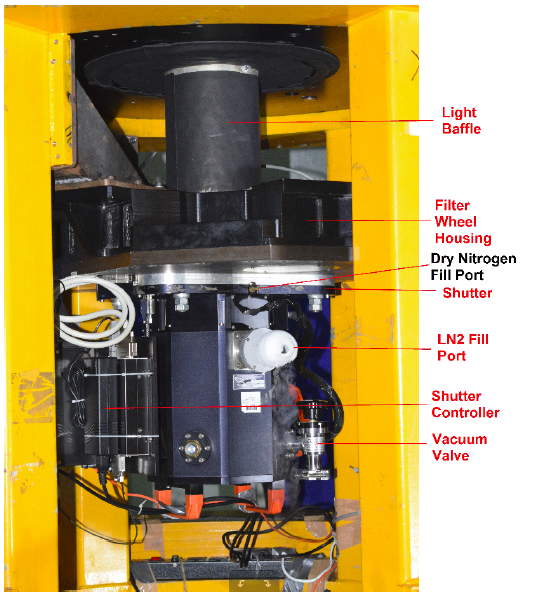}
\caption{The 4K $\times$ 4K CCD Imager, which is installed at the primary port of DOT. Figure credit: \cite{2022JApA...43...27K}.}
\label{4K_4K}
\end{figure}

\subsection{Data reduction:}

We have used different telescopes/imaging instruments to obtain many GRBs' well-sampled optical afterglow light curves. A detail of all these telescopes is given in section \ref{GBtelescopes}. The photometric images obtained using charge-coupled devices (CCDs)\footnote{They are widely used detectors for optical astronomy due to good quantum efficiency and linearity.} are in FITS format. They need to be analyzed to calculate the brightness of the optical afterglows. For the analysis of optical images, we have used the following image data analysis packages: Image Reduction and Analysis Facility (\sw{IRAF})\footnote{IRAF is distributed by the National Optical Astronomy Observatories which is operated by the Association of Universities for research in Astronomy, Inc., under cooperative agreement with the National Science Foundation.}, and Dominion Astrophysical Observatory Photometry (\sw{DAOPHOT}, \citealt{1987PASP...99..191S}). In this section, we provide detailed information on the techniques used for the analysis of photometric data of GRB afterglows studied in this thesis. Image quality taken using CCDs is usually affected by the electronic noises of CCD, optical system, and atmospheric turbulence. The optical analysis of the images helps to minimize these noises and extract the true brightness/information of the afterglows. There are three steps involved in the optical analysis of images obtained using CCD mounted at the telescopes: Pre-processing, Processing, and Post-processing.

\subsubsection{Pre-processing}

The observed raw images (the portion of sky recorded on the detector) obtained using telescopes are affected by various kinds of noises either due to CCD electronics (readout noise), thermal electrons (thermal noise), cosmic rays, pixel-to-pixel sensitivity variation, etc. The observed raw frames need to be corrected for these noises before calculating the magnitude of the sources. The process of cleaning raw images is known as the pre-processing step of optical photometric data analysis. This step incorporates a correction for bias frames, correction for flat fielding of raw images, and removal of cosmic rays. We have used standard packages included in \sw{IRAF} for these corrections. Details of these corrections are given below:

{\bf Bias correction:} 

Bias frames are zero-second exposure images to read out the electrons due to the base voltage existing in the potential well of CCDs. These bias frames are obtained while the shutter of the CCD is closed, and we just read out the contribution from the CCD chip. We have obtained multiple bias frames to reduce the noise from minor base voltage variations throughout the night. Then we combined (median) all the bias frames to create a master bias frame to reduce the noise due to readout and spurious pixels in individual frames. We have used the \sw{ZEROCOMBINE} task of the \sw{IRAF} package to create the master bias. We have used the master bias frame to subtract it from the science frames to calculate the precise brightness of the source. We have used the \sw{CCDPROC} task of the \sw{IRAF} package to subtract the master bias from all the science and flat (see details below) frames.

{\bf Flat correction:} 

The pixels of CCDs are not uniform due to differences in their sensitivity. Therefore, we need to make all the pixels to a uniform response before calculating the brightness of the source. To correct the pixel-to-pixel variation, we have exposed the CCD to a uniform light source using the sky at the time of twilight so that each pixel receives the same flux. After taking the flat frames, we subtracted the bias frames from each image. We have used the \sw{CCDPROC} task of the \sw{IRAF} package to subtract the master bias from all the flat frames. After the bias subtraction of individual flat images, we performed the medium combination of bias-subtracted flat images to get a master flat in each filter. We have used the \sw{FLATCOMBINE} task of the \sw{IRAF} package to a medium combined flat image. Further, we normalized the master flat frame so that counts from the source do not change by the counts of flat fields. Finally, we have divided the bias-subtracted science images by the normalized master flat image using the \sw{CCDPROC} task of \sw{IRAF}.

{\bf Cosmic Rays removal:}

The next step of the pre-processing of photometric data reduction is the removal of cosmic rays from images. The cosmic ray events (high energy charged particles) create bright specks on the science images. The intensity profiles of cosmic rays (usually Dirac delta function) are very different from the typical stellar profile (usually Gaussian function). The different profiles of stars and cosmic rays help to identify the cosmic ray hits on the CCD pixels. We have utilized the \sw{COSMICRAYS} task \sw{IRAF} package to remove the cosmic rays from bias and flat-corrected science frames.

{\bf Image stacking:}

Generally, the optical afterglows of GRBs are quite faint due to their rapid decay nature. Therefore, detecting such faint objects in short exposures is very hard. To increase the signal-to-noise ratio of science frames, we need to align and combine the images for individual filters. We have used the \sw{IMALIGN} task of \sw{IRAF} for the alignment of the cleaned images. Finally, we utilized the \sw{IMCOMBINE} task for combining the aligned images.

\subsubsection{Processing}

After the raw images' pre-processing (bias, flat, and cosmic rays corrections), the next step is to perform the photometry to calculate the optical afterglows' brightness (instrumental magnitudes) in the CCD frame. We have used \sw{DAOPHOT II} software to process the optical cleaned images and determine the positions and instrumental magnitudes of the optical afterglows of GRBs.  

{\bf Object detection:}

Initially, we attached the cleaned FITS images utilizing the \sw{ATTACH} task from \sw{DAOPHOT II}. The next step is to identify the source's centroids present in the cleaned FITS files of the science frames utilizing the automatic algorithm \sw{FIND} from \sw{DAOPHOT II}. The \sw{FIND} task uses the readout noise and gain of the detector and helps to identify and locate the stellar sources and differentiate them from extended sources such as galaxies. The \sw{FIND} task performs the Gaussian profile fitting to the observed brightness values over the CCD pixels for the localization of stellar sources in the frames. Finally, the identified positions of the stellar sources in the images are written in a file.

{\bf Photometry:}

Photometry in optical astronomy is a method to determine the brightness of astronomical sources in different filter systems in flux units. Let us consider two stellar sources with apparent flux values of $f_{1}$ and $f_{2}$, and their corresponding apparent magnitude values are $m_{1}$ and $m_{2}$, respectively. The calculated fluxes and apparent magnitudes are related using the following formula:

\begin{equation}
m_{1} - m_{2} = -2.5 log (f_{1}/f_{2})
\end{equation}

Using the above relation, from the ratio of integrated fluxes of two stars, one can determine the difference in their magnitudes. If one star has a magnitude of zero, then the above equation is defined as apparent magnitude (m) and given as:

\begin{equation}
m = -2.5 log (f/f_{0}) 
\end{equation}

Where f$_{0}$ is the flux from a zeroth magnitude star (Vega). The above equation can be written as m = -2.5 log (f) + Z, where Z denotes the zero-point, and the term -2.5 log (f) is the instrumental magnitude of the star. Therefore, this equation relates the star's apparent magnitude to its instrumental magnitude and the zero-point, provided that both of these quantities were measured with the same instrument and in the same manner. There are different techniques to measure the integrated flux obtained from the celestial point or extended sources. For the measurement of brightness (instrumental magnitudes) of optical afterglows of GRBs, we have used (1) Aperture photometry and (2) Point spread function (PSF) photometry.

{\bf Aperture photometry:}

The aperture photometry of images is mainly used for brighter astronomical objects located in a less crowded field (the background changes linearly). In the case of aperture photometry, the source brightness is not affected by contaminated objects, for example, other stellar objects, sky, etc. We define an aperture (circular geometric region) centred on the source. We have selected the aperture sizes such that they cover the total area of the source, starting from the FWHM of the source, and the later aperture sizes are in multiple of FWHM of the source. We added the total photons observed within the aperture and subtracted it from those sky photons (background) observed from star free region to determine the source brightness in the CCD frame. We have used the \sw{PHOT} task of \sw{DAOPHOT II} to perform the aperture photometry of images. The output of the \sw{PHOT} task of \sw{DAOPHOT II} is written in the ".ap" file with details of instrumental magnitudes and associated errors for all the sources present in the frame. The aperture photometry does not work well for fainter and more crowded fields.

{\bf PSF photometry:}

The PSF photometry of images is mainly used for fainter astronomical objects located in a crowded field. For crowded regions, the background variation does not follow a linear pattern. In such cases, it is difficult to define an aperture with a precise aperture size to measure the total flux of the stellar object present in a crowded field. The brightest sources present in crowded regions can contribute to the intrinsic brightness of fainter sources in the frame. Therefore, we have utilized the profile fitting method to determine the brightness of fainter objects in crowded regions. The images taken using ground-based telescopes are usually affected by atmospheric seeing, camera optics, and other defects of the CCDs. The PSFs (stellar profiles imaged on two-dimensional arrays) of stellar sources within the frame can be given by Gaussian, modified Lorentzian, and Moffat mathematical functions. These individual mathematical functions or their combinations describe the PSF of stellar sources present in the images. Generally, PSF fitting (chi-squared method) with the Gaussian function better represents the PSF of the brightest stellar sources with respect to the other two functions if the seeing conditions over the night are good. We have used \sw{PICK} and \sw{PSF} tasks of \sw{DAOPHOT II} to select the appropriate stars in the crowded field and to create their PSF profiles. Finally, after calculating the PSF of around twenty bright stars, we determine the positions, instrumental magnitudes, and associated errors of all the sources present in the frame using the \sw{ALLSTAR} routine from the \sw{DAOPHOT II} software.     

\subsubsection{Post-processing}
The magnitude obtained from processing the raw images is the instrumental magnitude. We need to convert these instrumental magnitudes into standard magnitudes to get the exact information about the brightness of the optical afterglows of GRBs and study the temporal evolution of the afterglows using the data obtained from various other telescopes. For the calibration of the instrumental magnitudes, we have used the differential photometry method (see details below). 

{\bf Differential Photometry:}

The photometric data of GRBs afterglows are obtained as soon as possible after receiving the arcsec localization of the afterglows from the GCN circulars. Sometimes, if the afterglow decay is slow, we have also performed follow-up observations of the afterglows on consecutive nights. In such cases, the differential photometry technique is a unique tool to calibrate the instrumental magnitudes of optical afterglows. The differential photometry technique also helps to remove the effects due to the variations in sky condition, airmass, and other related to the optical telescopes. Before starting the differential photometry, we need to perform the astrometry of the images to identify the sources in sky coordinates. Then, we identified the local standard stars (non-variable stars) observed in the image. The calibrated magnitudes of these local standard stars present in the field are already reported by surveys such as United States Naval Observatory (USNO), Sloan Digital Sky Survey (SDSS), etc. We selected those stars as our photometric standard stars for which the variation in instrumental and catalogue magnitudes are nearly the same order. We calculated the mean of the difference between instrumental and catalogue magnitudes of selected photometric standard stars of the field. We added this offset (mean difference) to the measured instrumental magnitude of the optical afterglow and obtained the observed magnitude measurement of the source. In this way, we have calibrated all the photometry of our source in different filters.                                   

\newcommand{\thisgrb}{GRB~190530A\xspace}
\newcommand{\thisgrbB}{GRB~210619B\xspace}

\chapter{\sc Prompt emission: Physical mechanisms}\label{ch:3}
\blfootnote{The present chapter is based on the results published in: \textbf{{Gupta}, Rahul} et al., 2022, {\textit{MNRAS}, {\textbf{511}}, 2} AND {M.~D. Caballero-Garc\'{i}a}, \textbf{{Gupta}, Rahul} et al., 2023,  {\textit{MNRAS}, {\textbf{519}}, 3}.}
\ifpdf
    \graphicspath{{Chapter3/Chapter3Figs/JPG/}{Chapter3/Chapter3Figs/PDF/}{Chapter3/Chapter3Figs/}}
\else
    \graphicspath{{Chapter3/Chapter3Figs/EPS/}{Chapter3/Chapter3Figs/}}
\fi

\normalsize

The radiation mechanism of the prompt emission of GRBs is still an open question. The polarization measurement is a powerful tool for investigating the radiation mechanisms in the prompt emission. Another effective technique to examine the emission mechanisms of GRBs is to study the spectral evolution of the prompt emission. The characteristics of the evolution of \Ep and $\alpha_{\rm pt}$ have been studied by many authors \citep{1986ApJ...301..213N, 1995ApJ...439..307F, 1997ApJ...479L..39C, 2012ApJ...756..112L}. Three general patterns in the evolution of \Ep have been observed: 
(i) an `intensity-tracking' evolution, where \Ep increases/decreases as the flux increases/decreases \citep{1999ApJ...512..693R}; (ii) a `hard-to-soft' evolution, where \Ep decreases continuously \citep{1986ApJ...301..213N}; (iii) a `soft-to-hard' evolution or disordered evolution, where \Ep increases continuously or does not show any correlation with intensity \citep{1994ApJ...422..260K}. The $\alpha_{\rm pt}$ also evolves with time but does not display any typical trends \citep{1997ApJ...479L..39C}. Recently, \cite{2019ApJ...884..109L} and \cite{2021MNRAS.505.4086G} found that the \Ep and $\alpha_{\rm pt}$ have same tracking behaviour with flux (``double-tracking'') for GRB 131231A and GRB 140102A.

Spectral information from the prompt emission, together with prompt emission polarization, is a powerful tool that can provide a clear view of the long-debated mystery of the emission mechanisms of GRBs. However, we should always be aware of the challenges of polarization measurements. The first detection of prompt emission polarization was reported by {\it RHESSI} satellite for GRB 021206 \citep[highly linearly polarized;][]{2003Natur.423..415C}. This result was challenged in a subsequent study \citep{2004MNRAS.350.1288R}. Since then, prompt emission polarization measurements have only been performed in a handful of bursts using: {\it INTEGRAL} \citep{2007A&A...466..895M, 2009ApJ...695L.208G, 2013MNRAS.431.3550G, 2014MNRAS.444.2776G}, GAP onboard {\it IKAROS} \citep{2011PASJ...63..625Y, 2011ApJ...743L..30Y,2012ApJ...758L...1Y}, {\it POLAR} onboard Tiangong-2 space laboratory \citep{2019NatAs...3..258Z, 2019A&A...627A.105B, 2020A&A...644A.124K}, and CZTI onboard \AstroSat \citep{2016ApJ...833...86R, 2019ApJ...884..123C, 2018ApJ...862..154C,2019ApJ...874...70C, 2019ApJ...882L..10S, 2020MNRAS.493.5218S}. 

Recent studies on prompt emission polarization suggest the presence of time-varying (rapid changes in the polarization angle) linear polarization within the burst \citep{2009ApJ...695L.208G, 2019ApJ...882L..10S, 2017Natur.547..425T, 2020A&A...644A.124K}. It indicates that observations of time-integrated polarization could be an artefact of summing over the varying polarization signal \citep{2020A&A...644A.124K}. Therefore, a detailed time-resolved polarization is crucial to understand the radiation mechanisms of GRBs \citep{2021Galax...9...82G}. In this work, we present the time-integrated as well as time-resolved spectro-polarimetric results for the sixth most fluent GBM burst, \thisgrb\footnote{GRB 130427A, GRB 160625B, GRB 160821A, GRB 171010A, and GRB 190114C have higher fluence than \thisgrb, and more importantly all of them are well-studied bursts.}. Our spectro-polarimetric analysis is based on observations of \thisgrb performed by \fermi and \AstroSat-CZTI.

In this chapter, we present the high energy observations and analysis of \thisgrb (which includes prompt spectro-polarimetric observations using \AstroSat CZTI), and \thisgrbB (which includes prompt spectroscopic observations using \fermi and ASIM).
For \thisgrb, the very bright prompt emission along with LAT GeV photons, inspired us to investigate this burst in detail. We find a hint (detection significance $\leq$ 3 $\sigma$) of a high degree of polarization at \keV energy range and a ``double-tracking'' characteristic spectral evolution during the prompt emission phase of \thisgrb. For \thisgrbB, it was promptly detected by the Modular X- and Gamma-ray Sensor \citep{2019SSRv..215...23O} on the Atmosphere-Space Interactions Monitor ({\it ASIM}; \citealt{2019arXiv190612178N}) installed on the International Space Station ({\it ISS}) and the Gamma-ray Burst Monitor (GBM; \citealt{2009ApJ...702..791M}) onboard the \fermi mission.
This chapter is organized as follows. In section~\ref{multiwaveength observations and data reduction}, we discuss the high energy observations and data reduction. The main results and discussion are presented in section~\ref{results_190530A}. 
Finally, the summary and conclusion are given in section~\ref{conclusion_190530A}. All the uncertainties are quoted at 1\,$\sigma$ throughout this paper unless mentioned otherwise. The temporal ($\rm \alpha$) and spectral indices ($\rm \beta$) for the afterglow are given by the expression $\rm F (t,\nu)\propto t^{-\alpha}\nu^{-\beta}$. We consider the Hubble parameter $\rm H_{0}$ = 70\,km $\rm s^{-1}$ $\rm Mpc^{-1}$, density parameters $\rm \Omega_{\Lambda}= 0.73$, and $\rm \Omega_m= 0.27$ \citep{2011ApJS..192...14J}.

\section{High energy Observations}
\label{multiwaveength observations and data reduction}

\subsection{\thisgrb}

In this section, we present the high energy observations and data reduction for \thisgrb. In Figure \ref{timeline}, we provide a timeline depicting when the various space and ground-based observatories performed observations.

\begin{figure}
\centering
\includegraphics[scale=0.36]{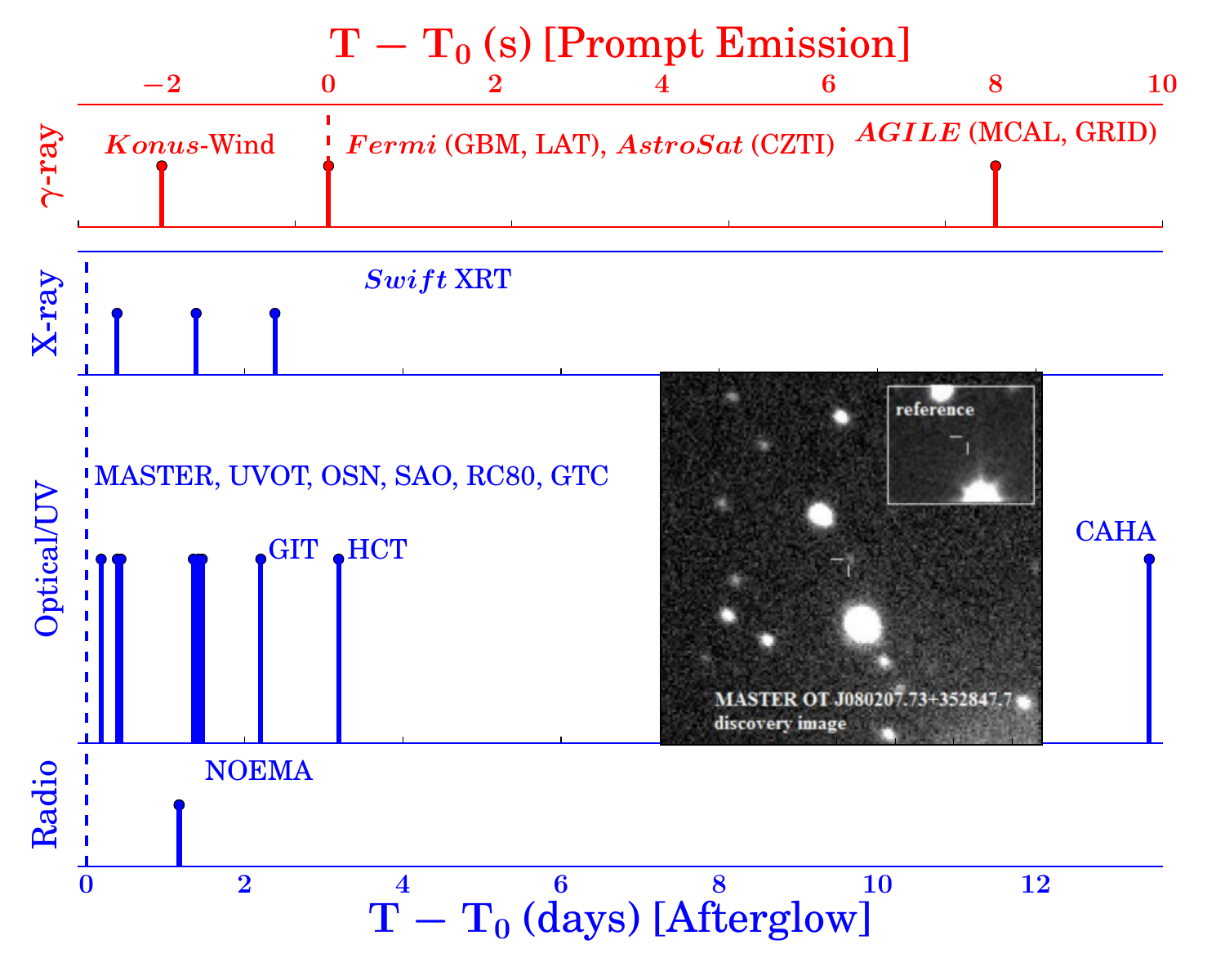}
\caption{A timeline of events for \thisgrb: The epochs of prompt (red) and afterglow (blue) observations were taken by various space-based and ground-based facilities. The sky image is the MASTER-Kislovodsk discovery image of the optical afterglow, MASTER OT J080207.73+352847.7, and the inset is the reference image (observed on 2010-12-07 01:15:41 UT with unfiltered limiting magnitude m$_{\rm lim}$=21.9 mag). The red and blue vertical dashed line indicates the \fermiT.}
\label{timeline}
\end{figure}

\begin{figure}[ht!]
\centering
\includegraphics[scale=0.29]{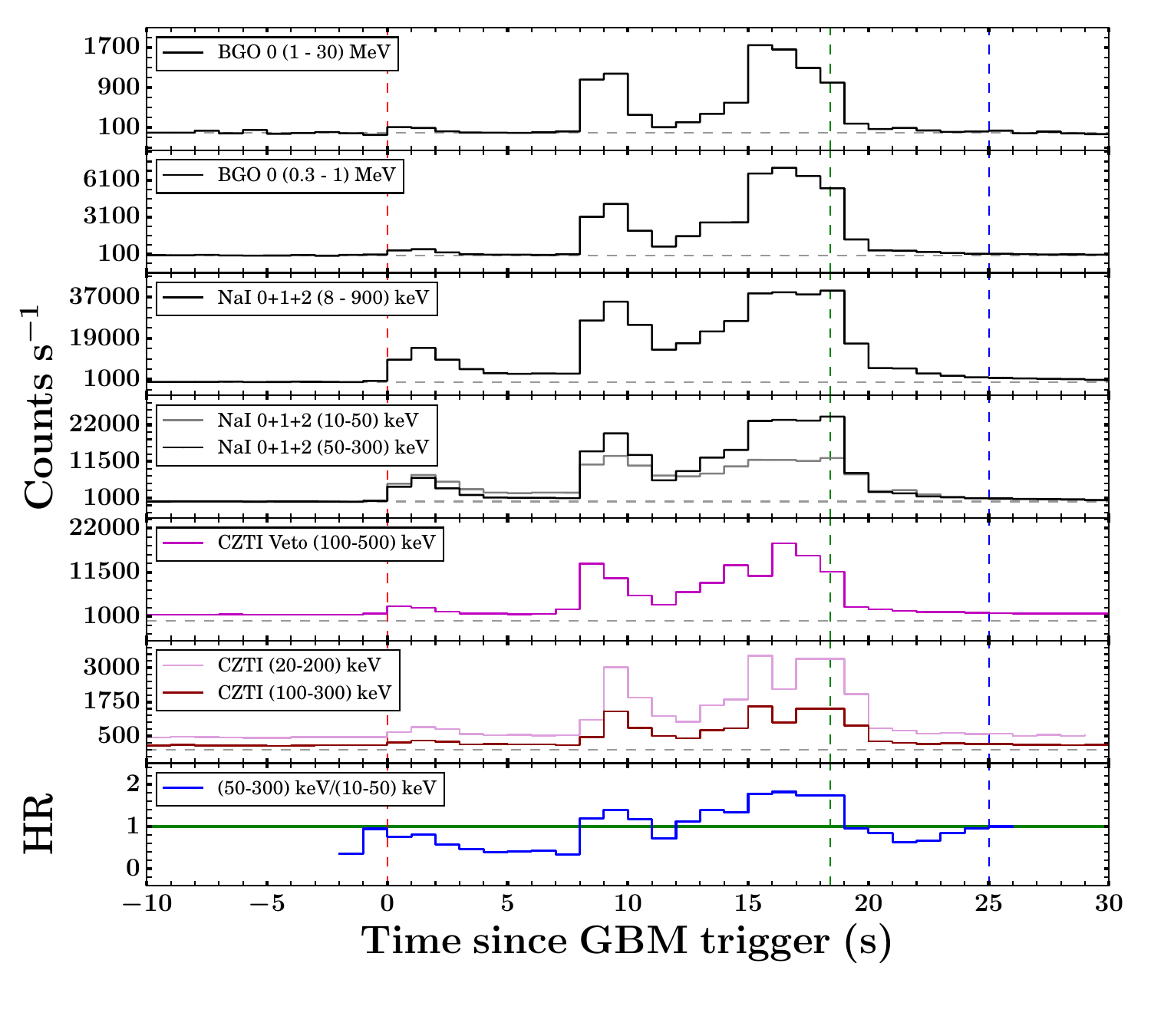}
\includegraphics[scale=0.29]{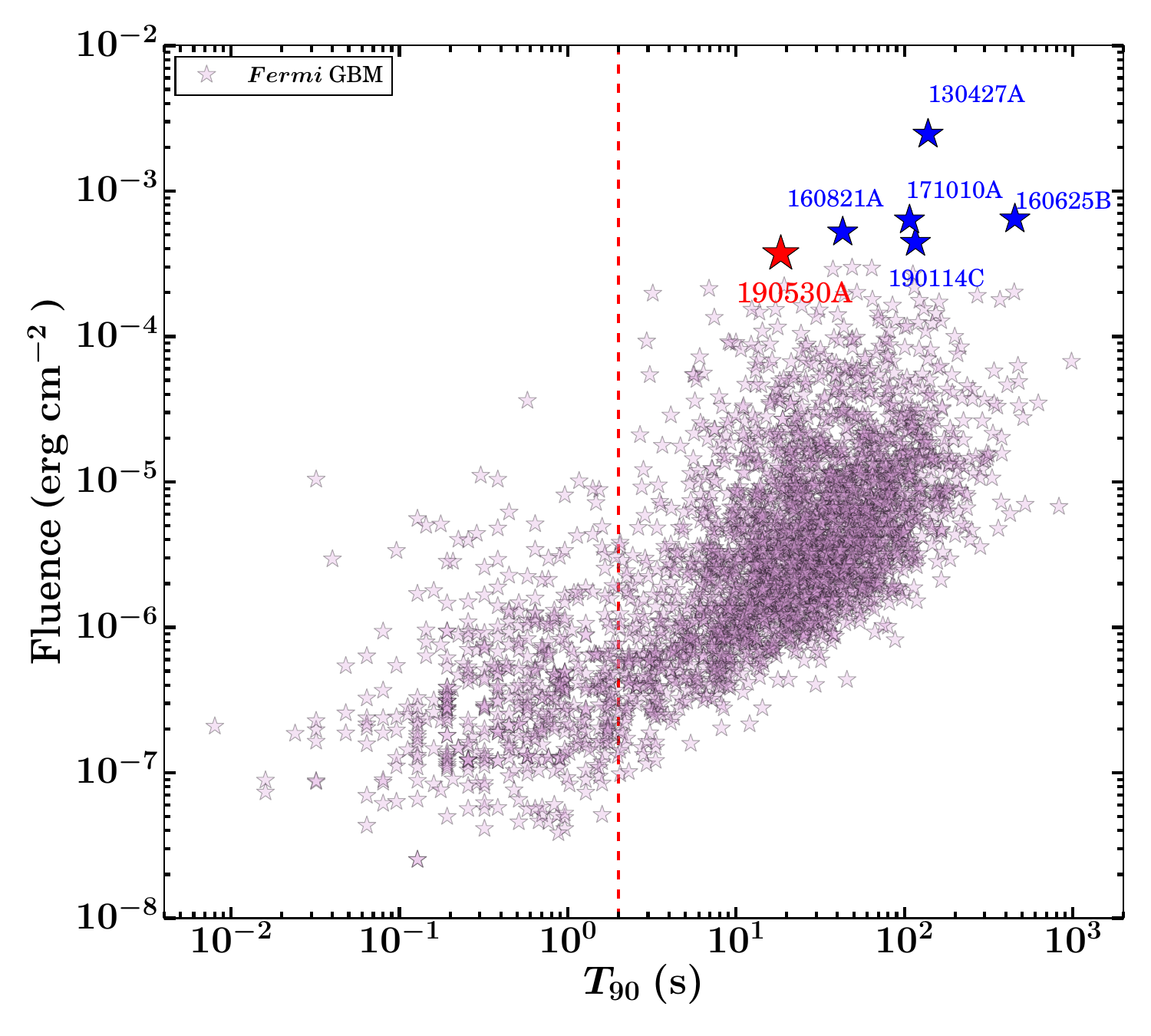}
\caption{Left: Energy-resolved prompt emission light curves of \thisgrb: The background-subtracted 1 s binned light curves of \fermi GBM and \AstroSat CZTI detectors provide in multiple energy channels (given in the first six panels). The \fermi trigger time (\fermiT) and \tninty durations for the \fermi GBM detector in the 50 - 300 \keV energy range are given by the red, and green vertical dashed lines, respectively. The start and stop times used for the time-averaged spectral analysis are provided by \fermiT and the blue vertical dashed line. The horizontal grey solid lines differentiate between the signal and background (at a count rate equal to zero). {Evolution of hardness ratio (HR) :} The bottom panel shows the evolution of HR in hard (50 - 300 $\keV$) to soft (10 - 50 \keV) energy channels of the NaI 1 detector. The horizontal green solid line corresponds to HR equal to one. {Right: Fluence distribution for GRBs:} \tninty duration as a function of energy fluence for \fermi detected GRBs in the observer frame. \thisgrb (shown with a red star) is the sixth most fluent GBM burst. The other five most fluent bursts are also highlighted with blue stars. The vertical black dashed line represents the classical boundary between long and short bursts.}
\label{promptlc_GRB190530A}
\end{figure}

\thisgrb simultaneously triggered GBM \citep{2009ApJ...702..791M} and LAT \citep{2009ApJ...697.1071A} onboard \fermi at 10:19:08 UT on May 30, 2019 (\fermiT). The best on-ground \fermi GBM position is RA, DEC = 116.9, 34.0 degrees (J2000) with an uncertainty radius of 1$^{\circ}$ \citep{2019GCN.24676....1F, 2019GCN.24679....1L}. The GBM light curve comprises of multiple bright emission peaks with a \tninty duration of 18.4 s (in 50 - 300 \keV energy channel, see Figure \ref{promptlc_GRB190530A}). For the time interval \fermiT to \fermiT + 20 s, the time-averaged \fermi GBM spectrum is best fitted with \sw{band} (GRB) function with a low energy spectral index ($\alpha_{\rm pt}$) = -1.00 $\pm$ 0.01, a high energy spectral index ($\beta_{\rm pt}$) = -3.64 $\pm$ 0.12 and a spectral peak energy (\Ep) = 900 $\pm$ 10 \keV. For this time interval, the fluence is $3.72 \pm 0.01 \times 10^{-4}$~erg~cm$^{-2}$, which is calculated in the 10 \keV ~- 10 MeV energy band \citep{2019GCN.24692....1B}. With this fluence, \thisgrb is the sixth brightest burst observed by \fermi-GBM (see Figure \ref{promptlc_GRB190530A}, other GBM data points are obtained from GRBweb page
. This brightness also implies this GRB is suitable for detailed analysis. The best \fermi-LAT on-ground position (RA, DEC = 120.76, 35.5 degrees (J2000) with an uncertainty radius of 0.12$^{\circ}$) was at 63$^{\circ}$ from the LAT boresight angle at the time of \fermiT. The \fermi LAT data show a significant increase in the event rate that is temporally correlated with the GBM \keV emission with high significance \citep{2019GCN.24679....1L}. \thisgrb also triggered \AstroSat-CZTI with a \tninty duration of 23.9 s in the CZTI energy channel \citep{2019GCN.24694....1G}. \thisgrb was also detected by several other $\gamma$-ray/hard X-ray space missions, including the Mini-CALorimeter and Gamma-Ray Imaging Detector onboard AGILE \citep{2019GCN.24678....1L, 2019GCN.24683....1V}, Insight-HXMT/HE \citep{2019GCN.24714....1Y}  and \kw \citep{2019GCN.24715....1F}. \kw obtained a total energy fluence of $5.57 \pm 0.15 \times 10^{-4}$~erg~cm$^{-2}$ in the 20 - 10000 \keV energy band; it is amongst the highest fluence event detected by \kw \citep{2019GCN.24715....1F}. The prompt emission characteristics of \thisgrb are listed in Table \ref{tab:prompt_properties_GRB190530A}.

\begin{table}[ht!]
\caption{Prompt emission characteristics of \thisgrb. \tninty: Duration from GBM observations in 50 - 300 \keV; \mvts: minimum variability time scale in 8 - 900 \keV; HR: ratio of the counts in hard (50 - 300 \keV) to the counts in soft (10 - 50  \keV) energy range; \Ep:  peak energy obtained using joint \fermi GBM and LAT observations from \fermiT to \fermiT+25 s; $F_{\rm p}$: peak flux in $\rm 10^{-6} erg ~cm^{-2}$ using GBM data in the 1 \keV -10 MeV energy range in the rest frame; $E_{\rm \gamma, iso}$: Isotropic $\gamma$-ray energy in the rest frame; $L_{\rm p, iso}$: Isotropic $\gamma$-ray peak luminosity in the rest frame; $z$: redshift of the burst obtained using GTC spectrum.}
\label{tab:prompt_properties_GRB190530A}
\begin{center}
\begin{tabular}{|c|c|c|}
\hline
\bf {Prompt Properties} & \bf {\thisgrb }& \bf {Detector} \\
\hline 
\hline
\tninty (s) & 18.43 $\pm$ 0.36 & GBM \\ \hline 
\mvts (s) &  $ \sim $ 0.50 & GBM \\ \hline 
HR  &   1.35 & GBM\\ \hline
\Ep (\keV) & $888.36_{-11.94}^{+12.71}$ & GBM+LAT \\ \hline
$F_{\rm p}$  & 135.38 & GBM \\ \hline
$E_{\rm \gamma, iso}$ ($\rm erg$) & $6.05 \times 10^{54}$ &-\\ \hline
$L_{\rm p, iso}$ ($\rm erg ~s^{-1}$) & 6.26 $ \times 10^{53}$ & -\\ \hline
Redshift $z$ & 0.9386 & GTC  \\ \hline
\end{tabular}
\end{center}
\end{table}

\subsubsection{\fermi Large Area Telescope analysis}
\label{section:LAT}

For \thisgrb, the \fermi LAT data from \fermiT to \fermiT+10ks was retrieved from the \fermi LAT data server \footnote{https://fermi.gsfc.nasa.gov/cgi-bin/ssc/LAT/LATDataQuery.cgi}
using the \sw{gtburst} \footnote{https://fermi.gsfc.nasa.gov/ssc/data/analysis/scitools/gtburst.html}
GUI software. We analyzed the \fermi LAT data using the same software. To carry out an unbinned likelihood investigation, we selected a region of interest (ROI) of $\rm 12^{\circ}$ around the enhanced \swift XRT position \citep{2019GCN.24689....1M}. We cleaned the LAT data by placing an energy cut, selecting only those photons in the energy range 100 MeV - 300 GeV. In addition, we applied an angular cut of 120$^{\circ}$ between the GRB location and zenith of the satellite to reduce the contamination of photons arriving from the Earth limb, based on the navigation plot. For the full-time intervals, we employed the \sw{P8R3\_SOURCE\_V2} response (useful for longer durations $\rm \sim 10^{3} ~s$), and for short temporal bins, we used the \sw{ P8R2\_TRANSIENT020E\_V6} response (useful for small durations $<$ 100 s). We calculated the probability of the high-energy photons being related to the source with the help of the \sw{gtsrcprob} tool. In Figure \ref{fig:LAT_LCs_GRB190530A}, we have shown the temporal distribution of the LAT photons for a total duration of 10 ks since \fermiT. LAT observed the high energy emission simultaneously with the \fermi GBM. LAT detected few photons with energy above 1 GeV and the highest-energy photon with an energy of 8.7 GeV (the emitted photon energy is 16.87 GeV in the rest frame at $z$ = 0.9386), which is observed 96 s after \fermiT \citep{2019GCN.24679....1L}. In the time interval of our analysis (\fermiT to \fermiT+10ks), we calculated the energy and photon flux in 100 MeV - 10 GeV energy range of $(5.60 \pm 1.02) ~\times 10^{-9}$  $\rm ~erg ~cm^{-2} ~s^{-1}$ and $(9.78 \pm 1.73) ~ \times 10^{-6}$  $\rm ~ph. ~cm^{-2} ~ s^{-1}$, respectively. For this temporal window, the LAT photon index ($\it \Gamma_{\rm LAT}$) is $-2.21 \pm 0.14$ with a test-statistic (TS) of detection 149. The LAT spectral index, $\beta_{\rm LAT} = \it \Gamma_{\rm LAT} + 1$, is $-1.21 \pm 0.14$. 

\begin{figure}[ht!]
\centering
\includegraphics[scale=0.39]{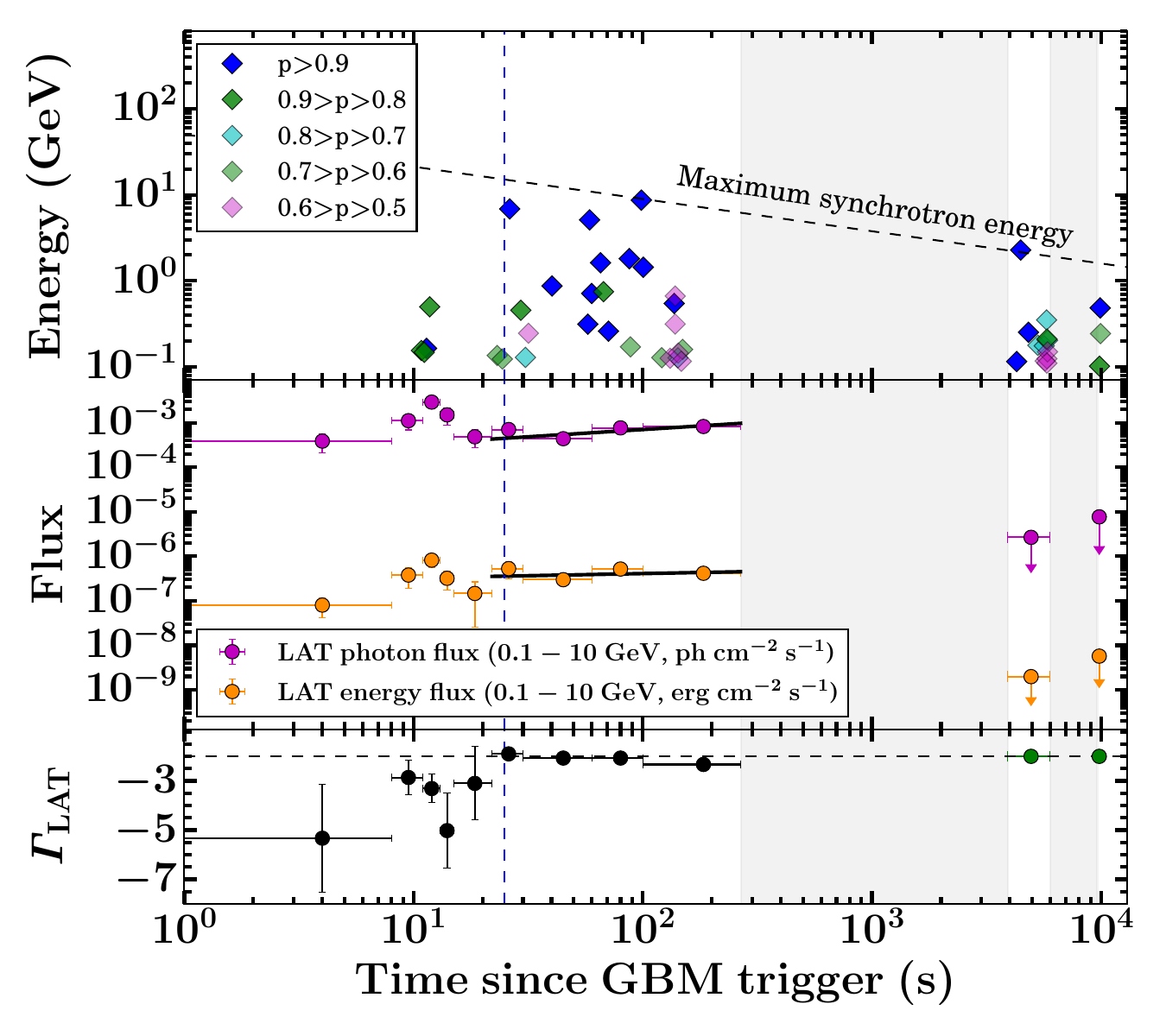}
\caption{\textit {Top panel}: Temporal distribution of \fermi-LAT photons with energies $> 100$ MeV and their association probabilities with \thisgrb. \textit{Middle panel}: Evolution of the \fermi LAT energy and photon fluxes in 0.1 - 10 GeV energy range. For the last two temporal bins, the LAT photon index was fixed to $-2$ to get an upper limit on the flux values. The black lines indicate the simple power-law fit to the extended (the photons detected after the end of the prompt emission) \fermi LAT photon and the energy flux light curves. \textit{Bottom panel}: Temporal evolution of the \fermi LAT photon indices in the 0.1 - 10 GeV range. The vertical blue dashed line represents the end epoch of the prompt emission phase (at \fermiT+25 s). Grey regions show the intervals having angle between the GRB position and the boresight of the \fermi LAT (off-axis angle) greater than 65$^{\circ}$.}
\label{fig:LAT_LCs_GRB190530A}
\end{figure}

\subsubsection{\fermi Gamma-ray Burst Monitor and Joint spectral analysis}

We obtained the time-tagged event (TTE) mode \fermi GBM data from the \fermi GBM trigger catalogue\footnote{https://heasarc.gsfc.nasa.gov/W3Browse/fermi/fermigtrig.html}
using \sw{gtburst} software. TTE data have high time precision in all 128 energy channels. We studied the temporal and spectral prompt emission properties of \thisgrb using the three brightest sodium iodide detectors (NaI 0, 1, and 2) with source observing angles, NaI 0: $\rm 39^\circ$ degree, NaI 1: $\rm 15^\circ$ degree, NaI 2: $\rm 34^\circ$ degree, respectively. We also selected the brightest bismuth germanate detector (BGO 0) as this BGO detector is closer to the direction of the burst (an observing angle of $\rm 49^\circ$ degree). The angle restrictions are to ignore the systematics coming due to uncertainty in the response at large angles.

We used \sw{RMFIT} version 4.3.2 software\footnote{{https://fermi.gsfc.nasa.gov/ssc/data/analysis/rmfit/}}
to create the energy-resolved prompt emission light curve using \fermi GBM observations.
The \fermi GBM energy-resolved (background-subtracted) light curves along with the evolution of hardness ratio (HR) are presented in Figure \ref{promptlc_GRB190530A}. The prompt emission light curve consists of three bright overlappings peaked structures, a soft and faint peak (lasting up to $\sim$ 4 s after \fermiT) followed by two merging hard peaks with a total duration of $\sim$ 18 s. The hardness ratio (HR) evolution indicates that the peaks are increasing HR (softer to harder trend), which is also evident from the very low signal for the first peak in the BGO data.

For the spectral analysis, we used the same NaI and BG0 detectors as used for the temporal analysis. We reduced the time-averaged \fermi GBM spectra (from \fermiT to \fermiT+ 25 s) using \sw{Make spectra for XSPEC} tool of \sw{gtburst} software from \emph{Fermi Science Tools}. The background (around the burst main emission) is fitted by selecting two temporal intervals, one interval before the GRB emission and another after the GRB emission. We performed the modelling of the joint GBM and LAT time-averaged spectra using the Multi-Mission Maximum Likelihood framework \citep[\sw{3ML}\footnote{{https://threeml.readthedocs.io/en/latest/}}
]{2015arXiv150708343V} software to investigate the possible emission mechanisms of \thisgrb. We began by modelling the time-averaged GBM spectrum with the \sw{Band} or \sw{GRB} function \citep{Band:1993}, and included various other models such as \sw{Black Body} in addition to the \sw{Band} function to search for thermal component in the burst; a power-law with two breaks (\sw{bkn2pow}\footnote{{https://heasarc.gsfc.nasa.gov/xanadu/xspec/manual/node140.html}}
), and cutoff-power law model (\sw{cutoffpl}) or their combinations based upon model fit, residuals of the data, and their parameters (see Table \ref{tab:TAS_GRB190530A} of the appendix). The \sw{bkn2pow} is a continual model that consists of two sharp spectral breaks (hereafter $E_{\rm break, 1}$, and $E_{\rm break, 2}$ or \Ep, respectively) and three power-laws indices (hereafter $\alpha_{1}$, $\alpha_{2}$, and $\alpha_{3}$ respectively). Where $\alpha_{1}$ is power-law index below the $E_{\rm break, 1}$, $\alpha_{2}$ is power-law index between $E_{\rm break, 1}$ and $E_{\rm break, 2}$, and $\alpha_{3}$ is power-law index above the $E_{\rm break, 2}$, respectively. The statistics Bayesian information criteria \citep[BIC;][]{Kass:1995}, and Log (likelihood) is used for optimization, testing, and to find the best-fit model of the various models used. Furthermore, we also calculated the goodness of fit value using the \sw{GoodnessOfFit}\footnote{{https://threeml.readthedocs.io/en/v2.2.4/notebooks/gof\_lrt.html}}
class of the Multi-Mission Maximum Likelihood framework. We consider the GBM spectrum over the energy range of 8 - 900 \keV (NaI detectors) and 250 - 30000 \keV (BGO detectors) for the spectral analysis. However, we ignore the 33–40 \keV energy range due to the presence of the iodine K-edge at 33.17 \keV while analyzing the NaI data. We consider 100 MeV - 100 GeV energy channels for the \fermi LAT observations.  

\begin{figure}[ht!]
\centering
\includegraphics[scale=0.45]{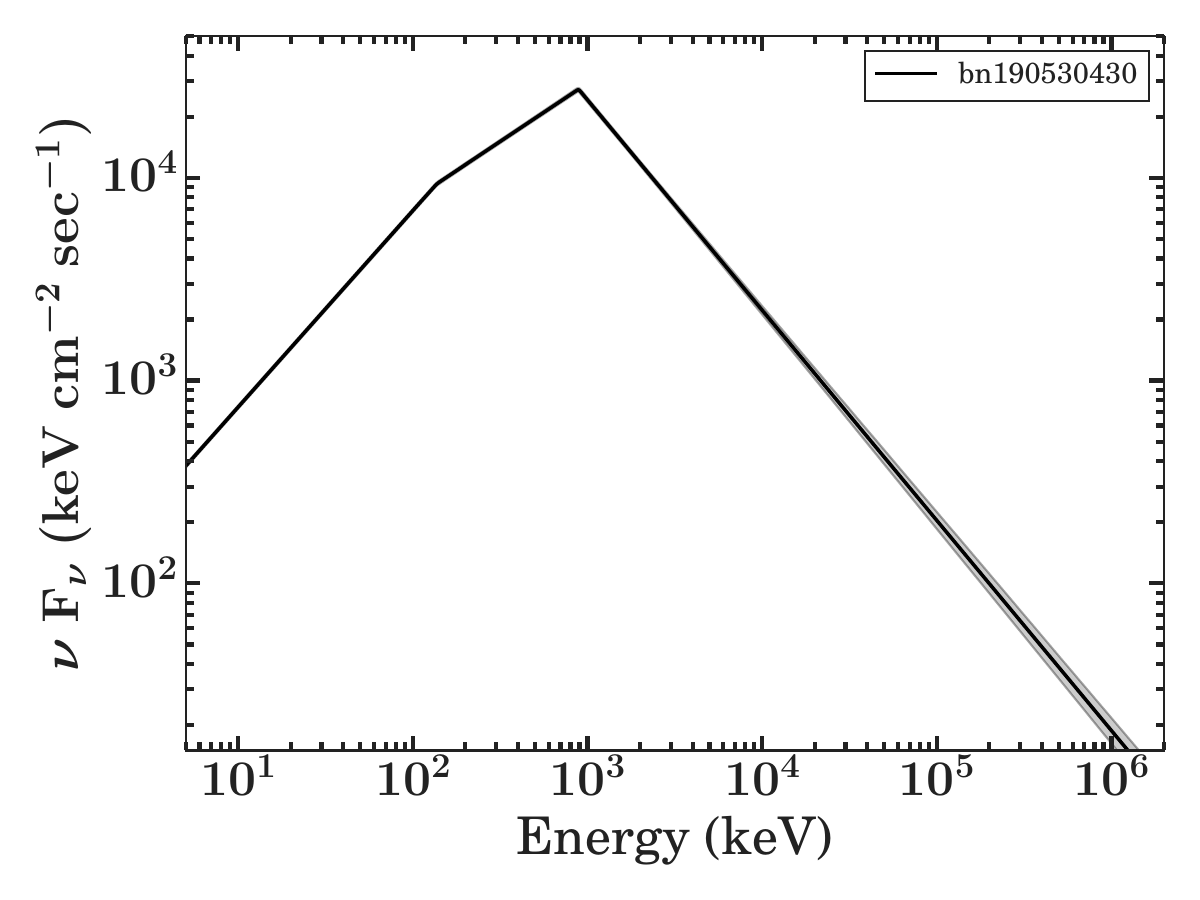}
\caption{The time-integrated best-fit energy spectrum of \thisgrb in model space modelled with a \sw{bkn2pow} model, a broken power-law model with two sharp breaks for an interval of 25 s (from \fermiT to \fermiT + 25 s) using joint spectral analysis of \fermi GBM and LAT data. The shaded grey region shows the 1 $\sigma$ uncertainty region. The legend indicates the \fermi trigger name of \thisgrb.}
\label{TAS_bkn2pow}
\end{figure} 

The best-fit spectral parameters of the joint analysis are presented in the appendix. We found that of all of the eight models used, the \sw{bkn2pow} model, a broken power-law model with two sharp breaks has the lowest BIC value. Therefore, we conclude that the time-averaged spectrum of \thisgrb is best described with \sw{bkn2pow} function with $\alpha_{1}$= 1.03$^{+0.01}_{-0.01}$, $\alpha_{2}$=1.42$^{+0.01}_{-0.01}$, $\it \beta_{\rm pt}$= 3.04$^{+0.02}_{-0.02}$, low-energy spectral break ($E_{\rm break, 1}$) = 136.65$^{+2.90}_{-2.88}$, and high-energy spectral break or peak energy ($E_{\rm break, 2}$) = 888.36$^{+12.71}_{-11.94}$. We noticed that the values of $\alpha_{1, 2}$ are consistent with the power-law indices expected for synchrotron emission. The best-fit time-averaged spectra in model space are shown in Figure \ref{TAS_bkn2pow}. Next, we perform a detailed time-resolved analysis to search the low-energy spectral break with two different (coarser and finer) bin sizes.

\subsubsection{{Time-resolved Spectroscopy of \thisgrb and Spectral Parameters Evolution}}
\label{trs_section}

The mechanisms producing the GRB prompt emission is still an open question \citep{2015AdAst2015E..22P}. The emission can be equally well described by a non-thermal synchrotron model \citep{2020NatAs...4..174B} as well as a thermal photospheric model \citep{2007MNRAS.379...73G}. Time-resolved spectral analysis of prompt emission is a propitious method to study the possible radiation mechanisms and investigate correlations between different spectral parameters. There are several methods to bin the prompt emission light curve, such as constant cadence, signal-to-noise (S/N), Bayesian blocks, and Knuth bins. Of these methods, the Bayesian blocks algorithm is the best method to identify the intrinsic intensity change in the prompt emission light curve \citep{2014MNRAS.445.2589B}.

Initially, we rebinned the total emission interval (from \fermiT to \fermiT+25 s) based on the constant cadence method with a coarse bin size of 1 s to perform the time-resolved spectral analysis. This provides a total of 25 spectra; however, the last five seconds of binned spectra do not have significant counts to be modelled. We used \sw{gtburst} to produce the 25 spectra. We modelled each spectrum with a \sw{Band} function and included various other models (\sw{Black Body}, and \sw{bkn2pow} or their combinations with \sw{Band} function) as we did for time-averaged spectral analysis if required. We find that out of twenty modelled spectra, four spectra (0-1 s, 8-9 s, 9-10 s, and 14-15 s) were best fit by \sw{bkn2pow} model, indicating the presence of a low-energy spectral break, and the rest of the temporal bins are well described with the \sw{Band} function only.\footnote{8-9 s bin is equally described with both functions.} For the best fit \sw{bkn2pow} model, the calculated mean values of the four spectra are $<\alpha_{1}>$ = 0.93 (with $\sigma$ = 0.03), $<\alpha_{2}>$ = 1.38 (with $\sigma$ = 0.03), and $E_{\rm break, 1}$ = 106.00 (with $\sigma$ = 3.14), where $\sigma$ denotes the standard deviation. When calculating mean values, we have excluded the first bin spectrum (0-1 s) as it has $E_{\rm break, 1}$ less than 20 \keV (close to the lower edge of the GBM detector). The calculated mean values of $<\alpha_{1}>$ and $<\alpha_{2}>$ are consistent with the power-law indices expected for synchrotron emission. The spectral parameters and their associated errors are listed in Tables \ref{TRS_Table_coarser} and \ref{TRS_Table_coarser_bkn2pow} of the appendix.

Furthermore, we rebinned the light curve for the detector with a maximum illumination (i.e. NaI 1) based on the Bayesian blocks algorithm integrated over the 8 - 900 \keV. This provides 53 spectra; however, some of the temporal bins do not have sufficient counts to be modelled. Therefore, we combined these intervals, resulting in a total of 41 spectra for time-resolved spectroscopy. We find that out of 41 modelled spectra, five spectra have a significant requirement for a low-energy spectral break ($\Delta$BIC$_{\rm Band/Black Body-Bkn2pow} \ge$6), three spectra are equally fitted with \sw{Band}+ \sw{Black Body} or \sw{bkn2pow} models  \break ($\Delta$BIC$_{\rm Bkn2pow-Band+Black Body} \leq$6), and six spectra are equally fitted with \sw{Band} or \sw{bkn2pow} models ($\Delta$BIC$_{\rm Bkn2pow-Band} \leq$6). The rest of the temporal bins are well described with the \sw{Band} function only. For the bins with signature of low-energy spectral break, the calculated mean values of the fourteen spectra are $<\alpha_{1}>$ = 0.84 (with $\sigma$ = 0.04), $<\alpha_{2}>$ = 1.43 (with $\sigma$ = 0.06), and $E_{\rm break, 1}$ = 79.51 (with $\sigma$ = 11.57). In this case, also, the calculated mean values are $<\alpha_{1}>$ and $<\alpha_{2}>$ are consistent with the power-law indices expected for synchrotron emission in a marginally fast cooling spectral regime. We also calculated the various spectral parameters such as \Ep, $\alpha_{\rm pt}$, and $\beta_{\rm pt}$ by modelling each spectra using \sw{3ML} software. The spectral parameters and their associated errors are listed in Tables \ref{TRS_Table_band} and \ref{TRS_Table_bkn2pow} of the appendix. Figure \ref{TRS_FIG_190530A} shows the evolution of spectral parameters such as \Ep, $\alpha_{\rm pt}$, and $\beta_{\rm pt}$ along with the light curve for the three brightest NaI detector in 8 - 900 \keV energy ranges. The value of \Ep changes throughout the burst. The \Ep evolution follows an intensity-tracking trend throughout the emission episodes. The evolution of $\alpha_{\rm pt}$ also follows an intensity-tracking trend, and it is within the synchrotron fast cooling and line of death for synchrotron slow cooling (though, in the case of the third pulse, in some of the bins, $\alpha_{\rm pt}$ becomes shallower and exceeds the line of death for synchrotron slow cooling); therefore, the emission of \thisgrb may have a synchrotron origin for the first two-pulses. 

\begin{figure}[ht!]
\centering
\includegraphics[scale=0.3]{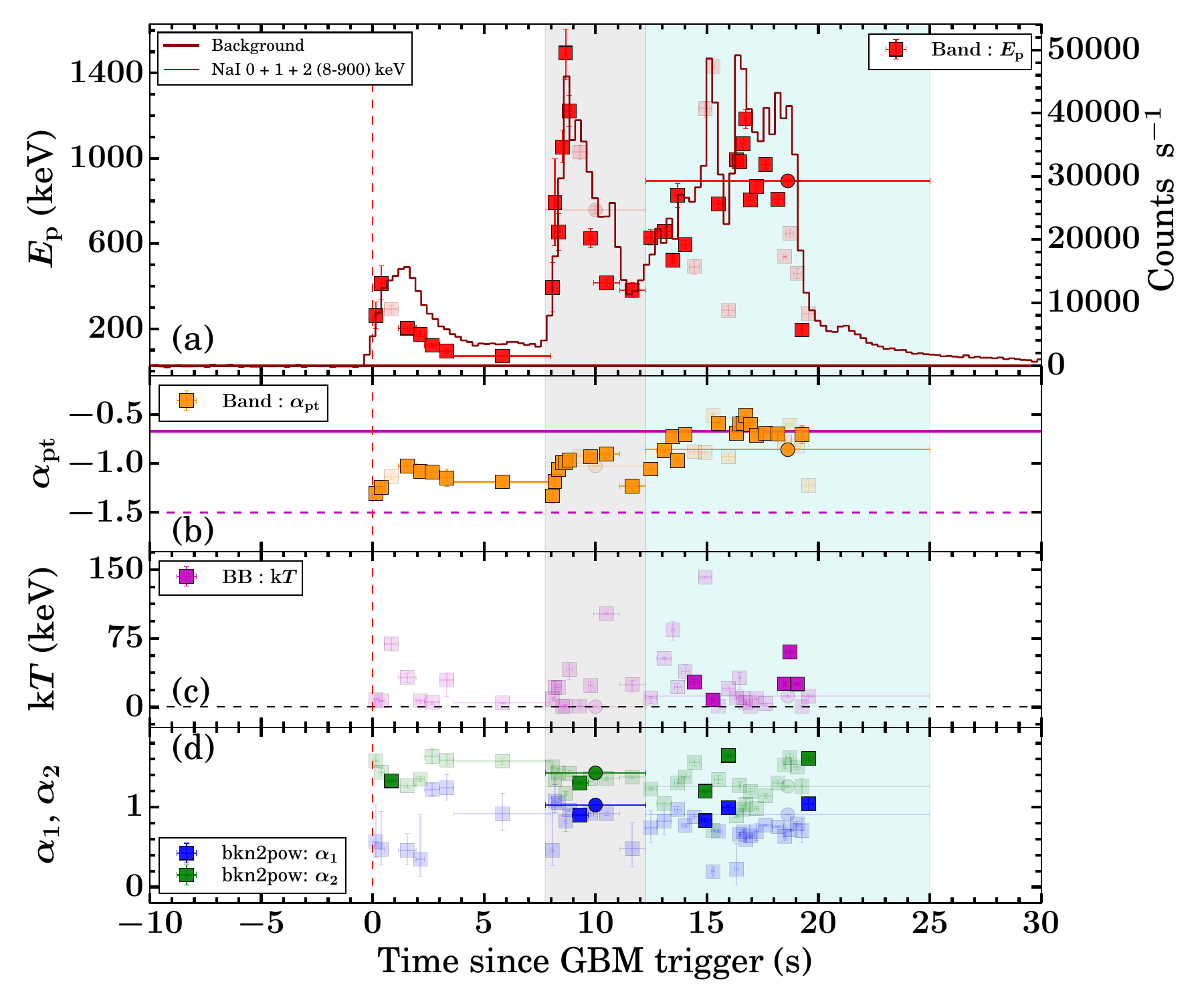}
\caption{{Evolution of the spectral parameters using \fermi GBM data.} (a) The peak energy evolves with time and follows intensity-tracking trends through the main emission episode. (b) The low energy spectral index also evolves and follows intensity-tracking behaviour. The two horizontal lines are the line of death for synchrotron fast cooling ($\alpha_{\rm pt}$ = -3/2, magenta dashed line) and the line of synchrotron slow cooling ($\alpha_{\rm pt}$ = -2/3, magenta solid line). (c) The evolution of k$T$ (\keV) is obtained from the \sw{Black Body} component. The horizontal black dashed line shows k$T$ = 0 \keV. (d) Evolution of the photon indices ($\alpha_{1}$ and $\alpha_{2}$) for the \sw{bkn2pow} model. In respective panels, if a particular model is best fitted to a particular bin, data points are highlighted with dark colour, otherwise shown with light colour. The temporal binning has been performed based on the Bayesian block algorithm. The red dashed line shows the \fermiT. The vertical grey (best fit with the \sw{bkn2pow} model) and cyan (best fit with the \sw{Band} model) shaded regions show the intervals used for time-resolved polarization measurements using CZTI data, respectively. For these bins, spectral parameters are shown with circles in respective sub-panels (a,b,c, and d).}
\label{TRS_FIG_190530A}
\end{figure}

\subsubsection{\AstroSat-Cadmium Zinc Telluride Imager and  Polarization Measurements}

\thisgrb detection had been confirmed from the ground analysis of the data of \AstroSat CZTI. The CZTI light curve observed multiple pulses of emission likewise followed by \fermi GBM prompt emission (see Figure \ref{promptlc_GRB190530A}). The substantial peak was detected at 101:19:25.5 UT, having a count rate of 2745 counts per second of the combined data of all the four quadrants of the CZTI above the background \citep{2019GCN.24694....1G}. We calculated \tninty duration 23.9 s using the cumulative count rate. We found that 1246 Compton events are associated with this burst within the time-integrated duration. In addition, the CsI anticoincidence (Veto) detector working in the energy range of 100-500 \keV also detected this burst.


During its ground calibration, the \AstroSat CZTI was shown to be a sensitive on-axis GRB polarimeter in the 100 - 350 \keV energy range \citep{2014ExA....37..555C, 2015A&A...578A..73V}. The azimuthal angle distribution of the Compton scattering events between the CZTI pixels is used to estimate the polarization. The detection of the polarization in Crab pulsar and nebula in the energy range of 100-380 \keV provided the first onboard verification of its X-ray polarimetry capability \citep{2018NatAs...2...50V}. CZTI later reported the measurement of polarization for a sample of 11 bright GRBs from the first year \AstroSat GRB polarization catalogue \citep{2019ApJ...884..123C}. The availability of simultaneous background before and after the GRB's prompt emission and the significantly higher signal-to-background contrast for GRBs compared to the persistent X-ray sources makes CZTI sensitive for polarimetry measurements even for the moderately bright GRBs. To estimate the polarization fraction (PF) and to correct for the azimuthal angle distribution for the inherent asymmetry of the CZTI pixel geometry \citep{2014ExA....37..555C}, polarization analysis with CZTI for GRBs (see below) involves a Geant4 simulation of the \AstroSat mass model. Recently, a detailed study was carried out on a large GRB sample covering the full sky based on imaging and spectroscopic analysis to validate the mass model (see \citet{2021JApA...42...93M, 2021JApA...42...82C}). The results are encouraging and boost confidence in the GRB polarization analysis. \citet{2019ApJ...884..123C} discusses the GRB polarimetry methodology in detail. Here we only give a brief description of the steps involved in polarization analysis for \thisgrb.  

\begin{enumerate}

\item The polarization analysis procedure begins by selecting the valid Compton events that are first identified as double-pixel events occurring within the 20 $\mu$s time window. The double-pixel events are further filtered against several Compton kinematics conditions like the energy of the events and distance between the hit pixels \citep{2014ExA....37..555C, 2019ApJ...884..123C}.

\item The above step is applied on both the burst region obtained from the light curve of \thisgrb (see \S~\ref{Prompt emission Polarization}) and at least 300 seconds of pre and post-burst background interval. The raw azimuthal angle distribution from the valid event list for the background region is subtracted from the GRB region.
    
\item The background-subtracted prompt emission azimuthal distribution is then normalized by an unpolarized raw azimuthal angle distribution to correct for the CZTI detector pixel geometry induced anisotropy seen in the distribution \citep{2014ExA....37..555C}. The unpolarized distribution is obtained from the \AstroSat mass model by simulating 10$^9$ unpolarized photons in Geant4 with the incident photon energy distribution the same as the GRB spectral distribution (modelled as Band function) and for the same orientation with respect to the spacecraft.
    
\item A sinusoidal function fits the corrected azimuthal angle distribution to calculate the modulation amplitude ($\mu$) and polarization angle (PA) in the CZTI plane using Markov chain Monte Carlo (MCMC) simulation.

\item To determine that the GRB is polarized, we calculate the Bayes factor for the sinusoidal and a constant model representing the polarized and unpolarized radiation \citep[see section 2.5.3 of][]{2019ApJ...884..123C}. Suppose the Bayes factor is found to be greater than 2. In that case, we estimate the polarization fraction by normalizing $\mu$ with $\mu_{100}$ (where $\mu_{100}$ is the modulation factor for 100 \% polarized photons obtained from Geant4 simulation of the \AstroSat mass model for 100 \% polarized radiation (10$^9$ photons) for the same GRB spectral distribution and orientation). For a GRB with Bayes factor $<$ 2, an upper limit of polarization is computed \citep[see][for the details of the upper limit of calculation]{2019ApJ...884..123C}. 

\end{enumerate}

\subsection{\thisgrbB}

\thisgrbB was initially detected by the Gamma-Ray Burst Monitor (GBM) and the Burst Alert Telescope (BAT; \citealt{2005SSRv..120..143B}), being on-board the \fermi and \swift satellites, respectively \citep{2021GCN.30279....1P, 2021GCN.30261....1D}. This section presents the prompt observational and data analysis details carried out by \fermi, ASIM-{\it ISS}, and \swift ~missions. 

\subsubsection{\fermi GBM observations and analysis}
\label{GBM}

The \fermi GBM detected \thisgrbB (see the light curve in Figure~\ref{promptlc_GRB210619B}) at 23:59:25.60 UT on 19 June 2021 \citep{2021GCN.30279....1P}. We used the time-tagged events (TTE) mode \fermi GBM data\footnote{https://heasarc.gsfc.nasa.gov/W3Browse/fermi/fermigbrst.html} for the temporal and spectral analysis of \thisgrbB. The TTE mode observations have a good temporal and spectral resolution. For the GBM temporal and spectral analysis, we considered the three brightest sodium iodide (NaI) and one of the brightest bismuth germanate (BGO) detectors. We have used \sw{RMFIT} version 4.3.2 software\footnote{https://fermi.gsfc.nasa.gov/ssc/data/analysis/rmfit/} to create the prompt emission light curve of \thisgrbB in different energy channels. The background-subtracted \fermi GBM light curve of \thisgrbB in different energy channels along with the hardness-ratio (HR) evolution is shown in Figure~\ref{promptlc_GRB210619B}. The shaded region in Figure~\ref{promptlc_GRB210619B} shows the time interval (\fermiT to \fermiT+67.38\,s) for the time-averaged spectral analysis.

\begin{figure}[ht!]
\centering
\includegraphics[scale=0.35]{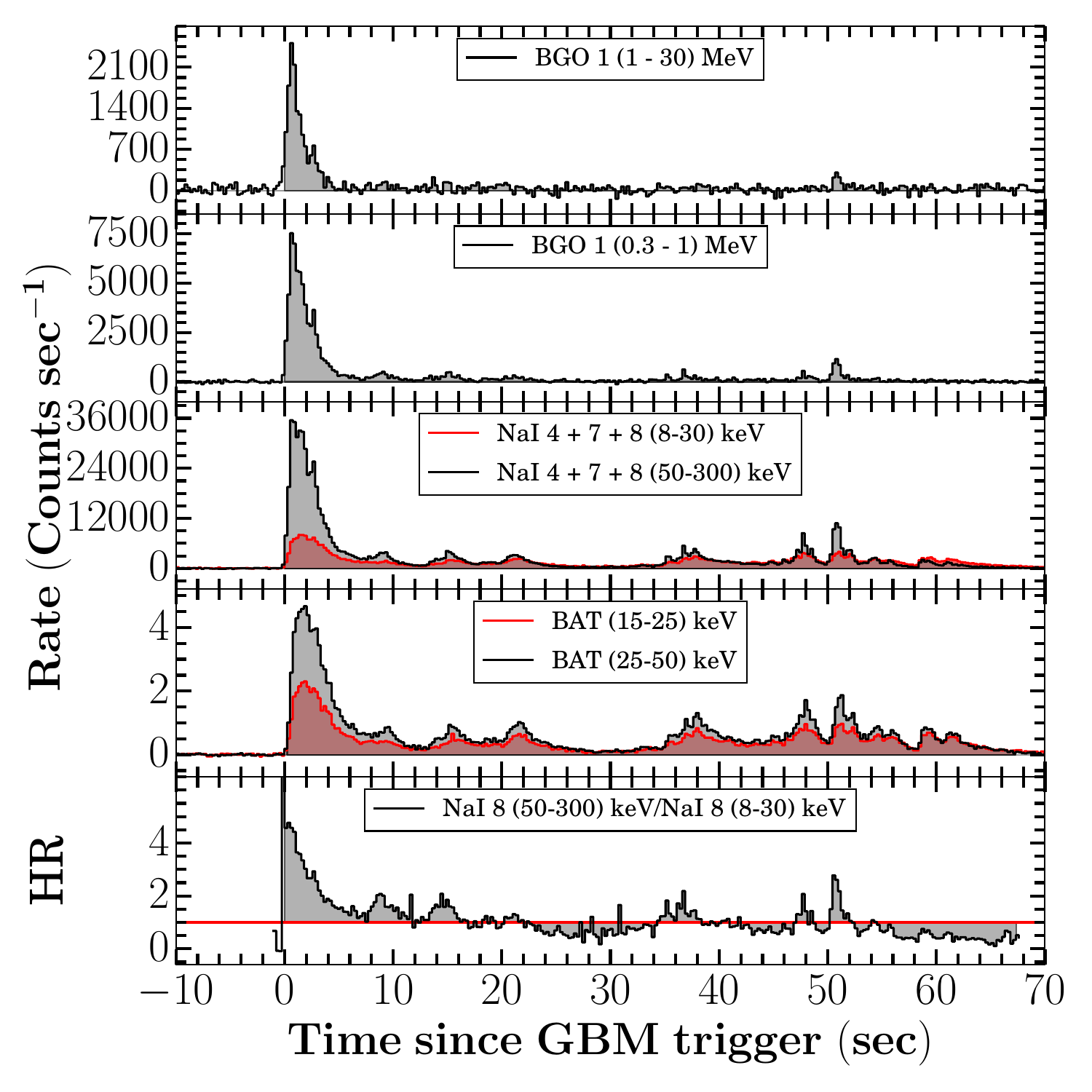}
\caption{{Prompt light curve of \thisgrbB:} The top four panels show the background-subtracted multi-channel prompt emission light curve of \thisgrbB obtained using \fermi GBM and \swift BAT observations (256\,ms bin size). The grey and red shading regions show the time interval (\fermiT to \fermiT+ 67.38\,s) of the light curve used for the time-averaged spectral analysis. The bottom-most panel shows the hardness ratio, obtained using the ratio of the count-rate light curve in the hard and soft energy range. The solid red line indicates the hardness ratio equal to one.}
\label{promptlc_GRB210619B}
\end{figure}

In addition to this, for the spectral analysis, we have used the same NaI and BGO detectors and reduced the spectrum using the \sw{gtburst}\footnote{https://fermi.gsfc.nasa.gov/ssc/data/analysis/scitools/gtburst.html} software. We selected the total emission interval (\fermiT to \fermiT+67.38\,s) for the time-averaged spectral analysis using a Bayesian block algorithm. For the spectral modelling of both time-averaged as well as time-resolved spectra, we have used the Multi-Mission Maximum Likelihood framework \citep[\sw{3ML}\footnote{https://threeml.readthedocs.io/en/latest/}]{2015arXiv150708343V} software since it is optimized for Bayesian analysis\footnote{https://threeml.readthedocs.io/en/stable/xspec\_users.html}. We selected the GBM spectrum over 8-900 \keV and 250-40000 \keV for NaI detectors and BGO detectors, respectively. In addition, we have ignored the NaI K-edge (30-40\,keV) energy range for both time-averaged and time-resolved spectral analysis. For the time-averaged spectral modelling of \thisgrbB, we have used the empirical \sw{Band} function and \sw{Cutoff power-law} models. We adopted the deviation information criterion (DIC; \citealt{Kass:1995, spiegelhalter2002bayesian}) to know the best-fit function among the \sw{Band}, and \sw{Cutoff power-law} models and selected the best model with the least DIC value ($\Delta$ DIC $<$ -10; \citealt{2019ApJ...886...20Y, 2021ApJS..254...35L}). This suggests that the \sw{Band} function is better fitting with respect to the \sw{Cutoff power-law} model. To know about the best-fit model, we utilized the following condition:

 $\Delta$DIC = $\Delta$DIC$_{\rm Band}$ - $\Delta$DIC$_{\rm CPL}$

The negative values of $\Delta DIC$ indicate that there is an improvement in the spectral fit, and if the difference of the deviation information criterion is less than - 10, in such a scenario, the existence of \sw{Band} function in the spectrum is confirmed. The time-averaged spectral parameters for both models are tabulated in Table \ref{tab:TAS_GRB210619B}.

\begin{table*}
\tiny
\caption{The time-averaged (\fermiT to \fermiT+ 67.38\,s) spectral analysis results of \thisgrbB. The best-fit model (\sw{Band}) is highlighted with boldface.}
\begin{center}
\label{tab:TAS_GRB210619B}
\begin{tabular}{cccccccc}
\hline
\textbf{Model} & \multicolumn{4}{c|}{\textbf{Spectral parameters}} & \textbf{-Log(posterior)} & \textbf{DIC} & \textbf{$\rm \bf \Delta~ {DIC}$} \\ \hline 
\sw{\bf Band} &\bf $\bf \alpha_{\rm \bf pt}$=  -0.90$\bf ^{+0.01}_{-0.01}$ & \multicolumn{2}{c|}{\bf $\bf \beta_{\rm \bf pt}$= -2.05$\bf ^{+0.02}_{-0.02}$} &\bf \Ep = 216.08$^{+4.20}_{-4.20}$ &\bf -4347.76 &\bf 8693.99 & --\\ \hline 
\sw{CPL} & $\it \alpha_{\rm pt}$=  -1.03$^{+0.01}_{-0.01}$ & \multicolumn{3}{c|}{$E_{0}$ =320.30$^{+5.74}_{-5.63}$} & -4633.92 & 9251.24 & -557.25\\ \hline 
\end{tabular}
\end{center}
\end{table*}

{\bf Time-resolved spectral analysis:} We performed the time-resolved spectroscopy of \thisgrbB using the same NaI and BGO detectors used for temporal/time-integrated analysis and the \sw{3ML} software. To select the temporal bins for the time-resolved spectral analysis, we used the Bayesian blocks binning algorithm \citep{2013arXiv1304.2818S}. \cite{2014MNRAS.445.2589B} suggested that the Bayesian blocks algorithm is crucial to find the finest temporal bins and intrinsic spectral evolution. However, it could result in a low signal-to-noise ratio due to high variability in the light curve. Therefore, we further selected only those bins which have statistical significance (S) $\geq$ 30. After implementing the Bayesian blocks and signal-to-noise ratio algorithm, we find a total of 59 spectra for the time-resolved spectroscopy of \thisgrbB (see Table \ref{TRS_Table_Bayesian_GRB210619B} of the appendix). We modelled each of the time-resolved spectra using individual empirical \sw{Band}, and \sw{Cutoff power-law} models.

\subsubsection{ASIM observations and analysis}
\label{ASIM_spectra}

The Atmosphere-Space Interactions Monitor (ASIM; \citealt{2019arXiv190612178N}) triggered on \thisgrbB on June 19 2021 \citep{2021GCN.30315....1M}. The ASIM reference time used in the rest of the analysis is T$_{\rm 0, ASIM}$ = 23:59:24.915550 UT. At trigger time, it was local daytime, so only High-Energy Detector (HED) data are available for this event. The burst direction was $135^{\circ}$ off-axis with respect to the Modular X- and Gamma-ray Sensor (MXGS; \citealt{2019SSRv..215...23O}) pointing direction. This is outside the instrument's nominal field of view, but the instrument is sensitive to the full solid angle because of its characteristics and high energy range. Photon interaction in the surrounding material and in the Columbus module is accounted for in the Monte Carlo simulation model of the instrument performed using the GEANT4 simulation toolkit \citep{Agostinelli2003}. This model was used to generate the Detector Response Matrix (DRM) of the instrument used for the spectral analysis.

Instrumental effects are carefully accounted for in HED data analysis by the implementation of the ``safety time" criterion, described in detail in \cite{Lindanger2021}. This procedure removes any count which is closer in time to the previous count in the same detector than a specific time interval, ranging from 0 and $30~\mu s$ depending on the energy of the previous count. This procedure is introduced to guarantee the best possible energy estimate, which is to some extent affected by the energy of the previous counts. This effect is particularly relevant for Terrestrial Gamma-ray Flashes (TGFs), which exhibit very high fluxes on time scales of a few hundred of microseconds. In the case of \thisgrbB, this effect is almost negligible and the ``safety time" criterion removes only about 0.7\% of the counts.

Figure~\ref{ASIM_LC_SED} shows the light curve of ASIM HED data in the (0.3-30\,MeV) energy range and with a time bining of 50\,ms. Because of the characteristics of the ASIM trigger logic, only part of the burst has been collected. A data gap is evident between 4.5 and 8\,s after ASIM reference time (T$_{\rm 0, ASIM}$). Only limited data are available prior to T$_{\rm 0, ASIM}$, therefore the background estimation for spectral analysis is based on the data in the time interval ($-420$ to 0\,ms) with respect to T$_{\rm 0, ASIM}$. To assess the spectral evolution of the burst, ASIM HED data were divided into seven-time intervals, as shown in Figure~\ref{ASIM_LC_SED}. The choice of the intervals has been made in order to include in separate intervals all the peaks evident in the light curve (intervals 2, 3, 4, and 5). The spectral fitting has been carried out using the X-ray Spectral Fitting Package \citep[\sw{XSPEC};][]{1996ASPC..101...17A} software (version 12.12). To estimate the flux, the convolution model {\tt cflux} has been used. All time intervals from 1 to 6 can be well fit with a simple power law in the energy range 0.5 to 10\,MeV. The best-fit parameters and the fluxes in the energy range 0.5 to 10\,MeV are shown in Table~\ref{tab:asim}.

\begin{figure}[ht!]
\centering
\includegraphics[angle=0,scale=0.35]{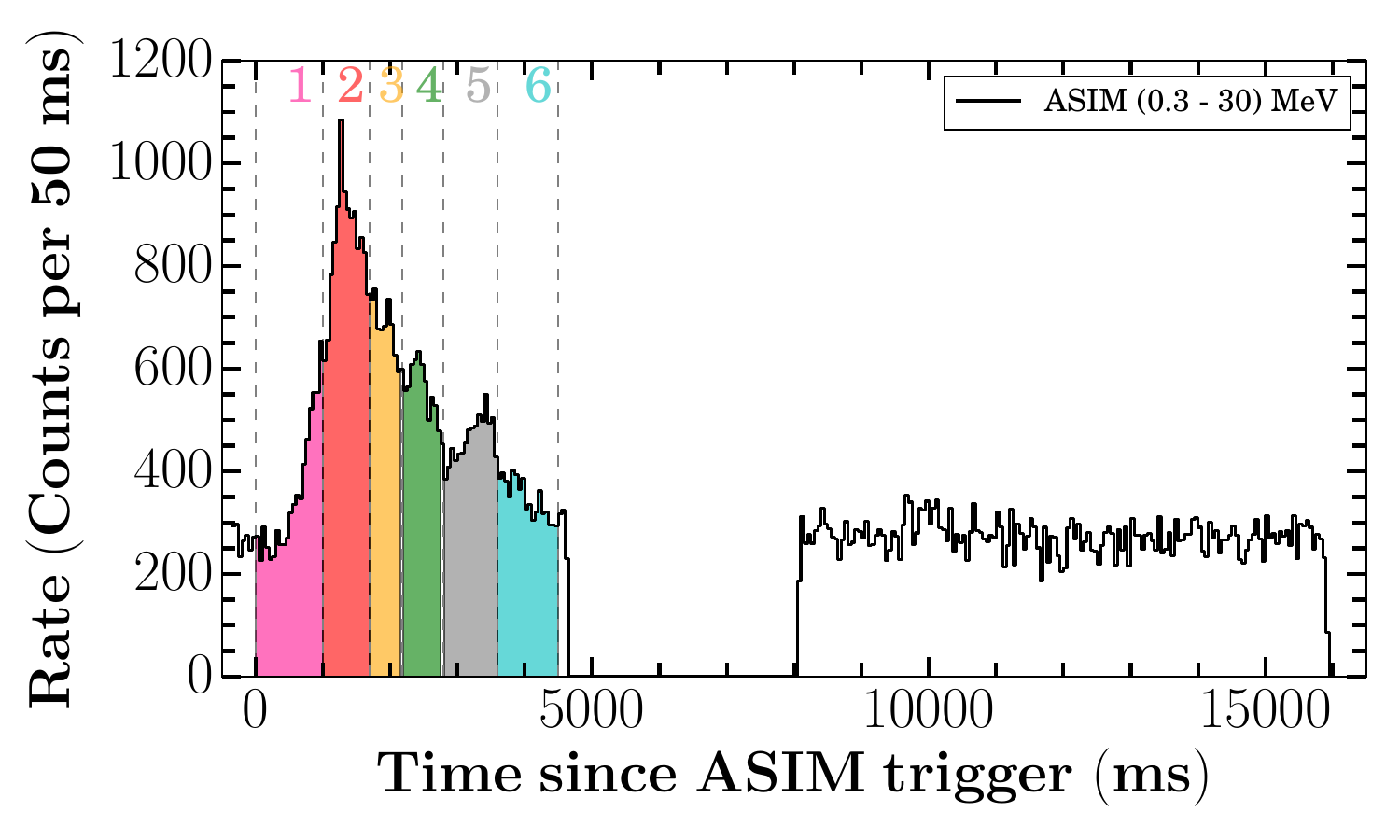}
\caption{The prompt emission (0.3 - 30)\,MeV light curve of the burst as seen by ASIM/HED. Vertical coloured shaded regions (lines) separate the time intervals used for spectral analysis, labelled by numbers on top of the panel.}
\label{ASIM_LC_SED}
\end{figure}

\subsubsection{\swift BAT observations and analysis}

The BAT detected \thisgrbB (see the mask-weighted light curve in Figure~\ref{promptlc_GRB210619B}) at 23:59:25 UT on 19 June 2021 at the following location in the sky: RA = 319.713, DEC = +33.860 deg. (J2000) with an error radius of 3\,arcmin \citep{2021GCN.30261....1D}. The \swift BAT and \fermi GBM positions were consistent for \thisgrbB.  

We downloaded the \swift BAT observations data of \thisgrbB from the \swift database portal\footnote{https://www.swift.ac.uk/swift\_portal/}. We processed the BAT data using the HEASOFT software (version 6.25). We reduced the \swift BAT data following the methodology presented in \cite{2021MNRAS.505.4086G} to obtain the energy-resolved mask-weighted light curve. The BAT light curves in the 25-50 \keV and 50-100 \keV energy ranges are shown in Figure~\ref{promptlc_GRB210619B}.

The prompt emission of \thisgrbB was also detected by the Gravitational-wave high energy Electromagnetic Counterpart All-sky Monitor ({\it GECAM}; \citealt{2021GCN.30264....1Z}), the \kw Experiment \citep{2021GCN.30276....1S}, the {\it CALET} Gamma-ray Burst Monitor (CGBM; \citealt{2021GCN.30284....1K}), the Mikhail Pavlinsky ART-XC telescope on board the {\it Spektr}-{\it RG} observatory \citep{2021GCN.30283....1L}, and the SPI-ACS/{\it INTEGRAL} \citep{2021GCN.30304....1M} satellites.

\subsubsection{Joint BAT-GBM-ASIM spectral analysis}

\begin{figure}[ht!]
\centering
\includegraphics[scale=0.32]{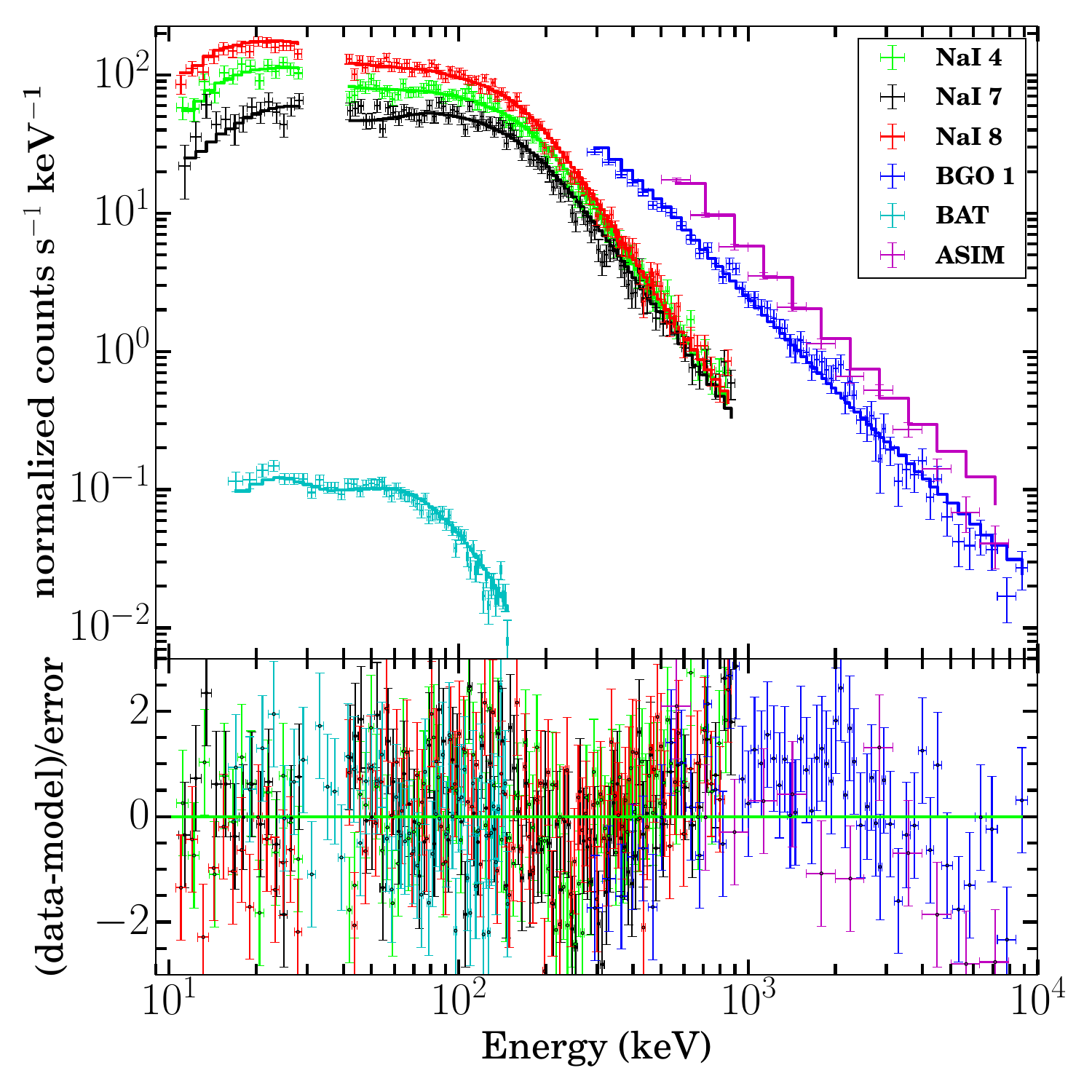}
\caption{{Top panel:} The best fit (\sw{Band}) joint \swift BAT (15-150 \keV), \fermi GBM (8 \keV-40 MeV), and ASIM (500 \keV to 10 MeV) spectrum for the temporal bin 2 (T$_{\rm 0, ASIM}$+1.00 to T$_{\rm 0, ASIM}$+ 1.70\,s) of \thisgrbB. {Bottom panel:} The corresponding residuals of the model fitting.}
\label{TRS_joint}
\end{figure}

We performed joint spectral analysis of \swift BAT, \fermi GBM, and ASIM observations during the main emission pulse of \thisgrbB, using the {\tt XSPEC} software. We used the Chi-squared statistics since the data sets are well sampled. For this purpose, we created the time-sliced (six bins) spectra using \swift BAT and \fermi GBM data for the same temporal segments of ASIM (time-resolved analysis) discussed in section \ref{ASIM_spectra} for the joint spectral analysis. We obtained the continuous spectra in
the overlapping energy regions of \swift BAT (15-150 \,keV), \fermi GBM and BGO (8 \,keV-40\,MeV), and ASIM (500 \,keV to 10\,MeV) instruments. For the spectral fitting, we ignored the NaI K-edge (30-40 \keV). Therefore, we used the data in the spectral ranges of 8-30,40-900\, KeV; 250\,keV-40\,MeV; 15-150\,KeV and 500\,-10 MeV for the \fermi/GBM, \fermi/BGO, \swift/BAT and ASIM, respectively. The spectra were rebined using the {\tt grppha} task to have at least 20 counts for each background-subtracted spectral bin. 

We initially fitted each spectrum with two different models: 1) Band, and 2) cutoff power-law models. In all the time intervals except 1 and 6 (where we got ${\chi}^{2}/{\nu}=1.45,1.57$ for ${\nu}=353,363$, respectively), we obtain bad fits ${\chi}^{2}/{\nu}{\gtrsim}2.0$
with the CPL ({\tt cutoffpl} in XSPEC) model (see Table ~\ref{Joint_spectralanalysis_Table}). On the other hand all the joint BAT-GBM-ASIM spectra are reasonably well fitted using a \sw{Band} ({\tt ngrbep}\footnote{https://fermi.gsfc.nasa.gov/science/mtgs/workshops/da2010\_nov10/swift.html} in XSPEC) 
function (${\chi}^{2}/{\nu}=1.0-1.5$, see Table ~\ref{Joint_spectralanalysis_Table}). \par

However, there are still noticeable wavy residuals towards the high-energy data points. We tried to improve them by adding a {\tt power-law} component in the previous model and did not get any significant improvement (see Table ~\ref{Joint_spectralanalysis_Table}). We also added a {\tt power-law} component
to the {\tt cutoffpl} model and whilst results improved the fits were statistically unacceptable (with the exception of Obs.~1 and 6, see comments above). We notice that
Obs.~1 and 6 are the ones showing the lowest (and similar) fluxes so it is difficult to disentangle between models in statistical terms (both {\tt Band} and {\tt cutoffpl} give similar
fits for these two observations). Therefore, we conclude that the {\tt Band} is the best fit for Obs.~1-6 presented in this paper. 

The joint \swift BAT, \fermi GBM, and ASIM count rate spectrum along with the best-fit model and residuals for one of the time bins (T$_{\rm 0, ASIM}$+1.00 to
T$_{\rm 0, ASIM}$+ 1.70\,s, i.e., the spectrum with the highest flux, Obs.~2) is shown in Figure \ref{TRS_joint}. The spectral parameters and the fluxes obtained for all six bins are given in
Table~\ref{Joint_spectralanalysis_Table}.
The fluxes are calculated in the interval 8\,keV-10\,MeV using the multiplicative model
{\tt cflux} in {\tt XSPEC}. The cross-calibration constants between the instruments were left as free parameters in the fitting process. The resulting values for GBM NaI, BGO and ASIM HED are consistent with unity within a few 10\%.

\section{Results and Discussion}
\label{results_190530A}

\subsection{\thisgrb}

Prompt emission properties of \thisgrb using \fermi and \AstroSat data are discussed and compared the observed properties using other well-studied samples of long GRBs.

\subsubsection{Prompt emission polarization}
\label{Prompt emission Polarization}

\thisgrb with a total duration of around 25 seconds (since \fermiT) registers $\sim$ 1250 Compton counts in \AstroSat CZTI. This corresponds to a high level of polarimetry sensitivity, making this GRB suitable for polarization analysis \citep[see section 2.5.3 of][]{2019ApJ...884..123C}. 

\begin{figure*}
\centering
    \includegraphics[width=0.3\linewidth]{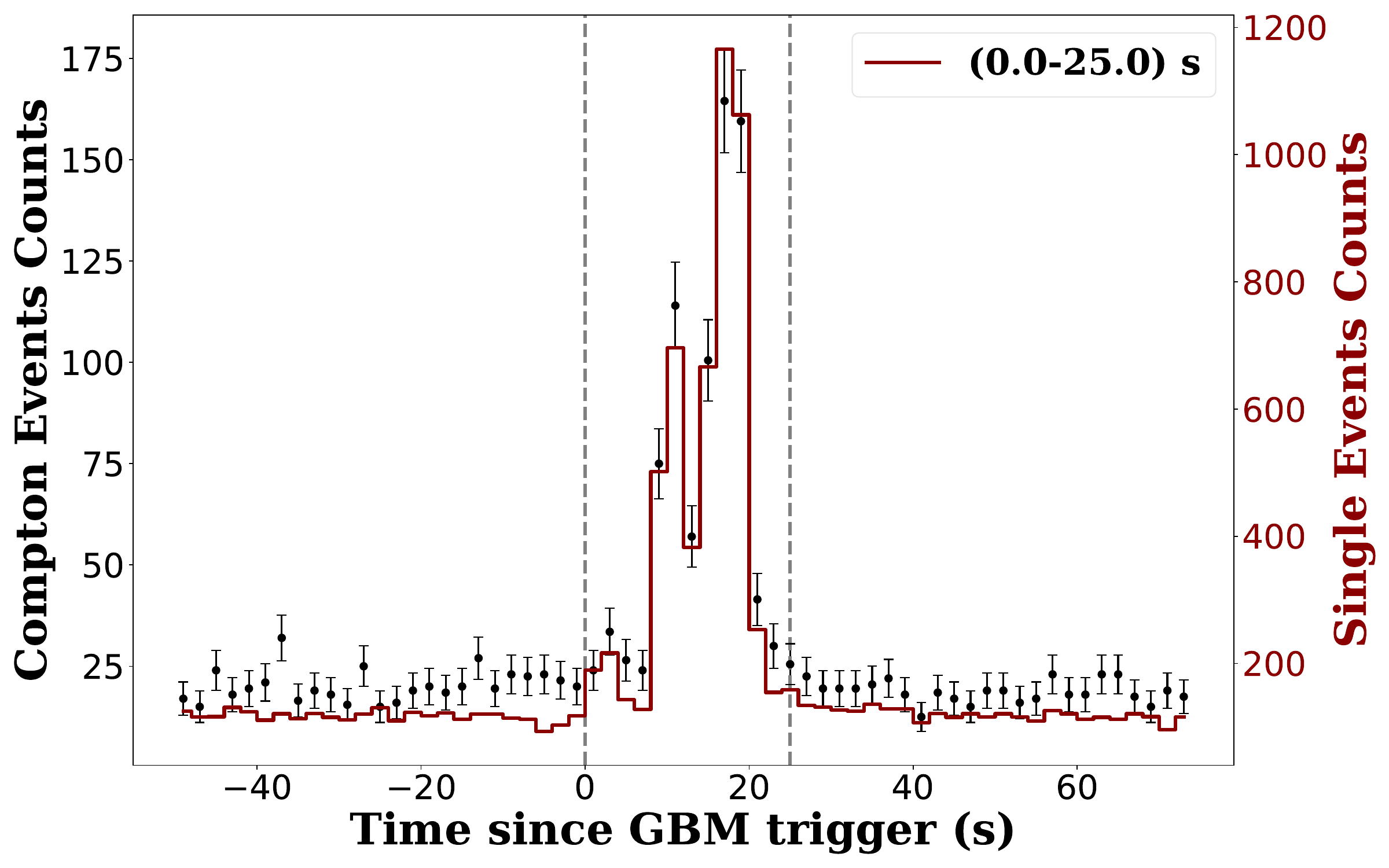}
    \includegraphics[width=0.3\linewidth]{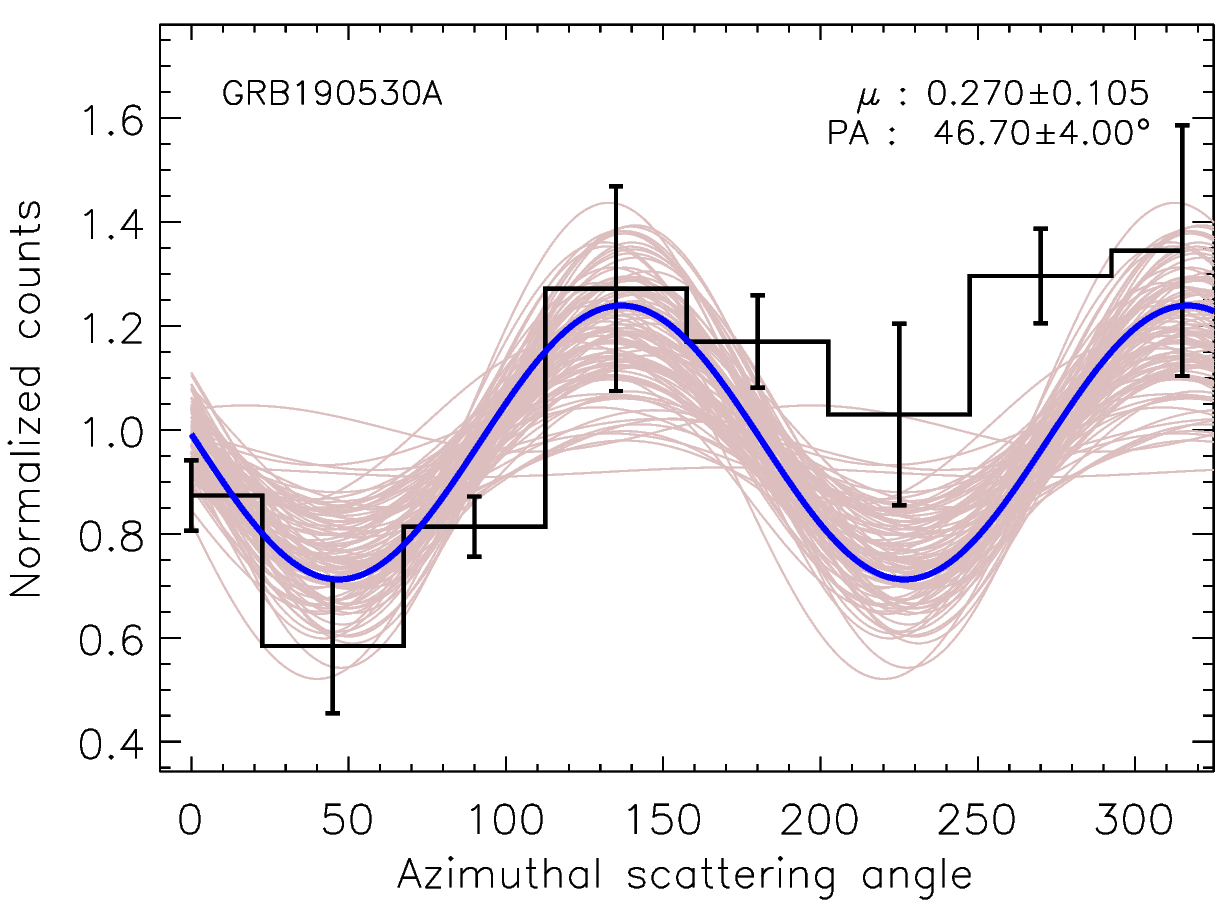}
    \includegraphics[width=0.3\linewidth]{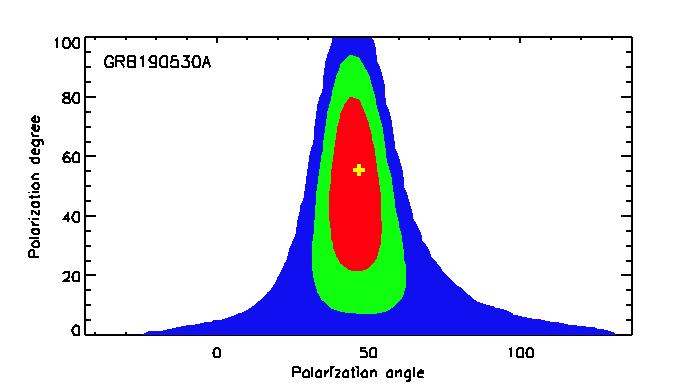}
    \includegraphics[width=0.3\linewidth]{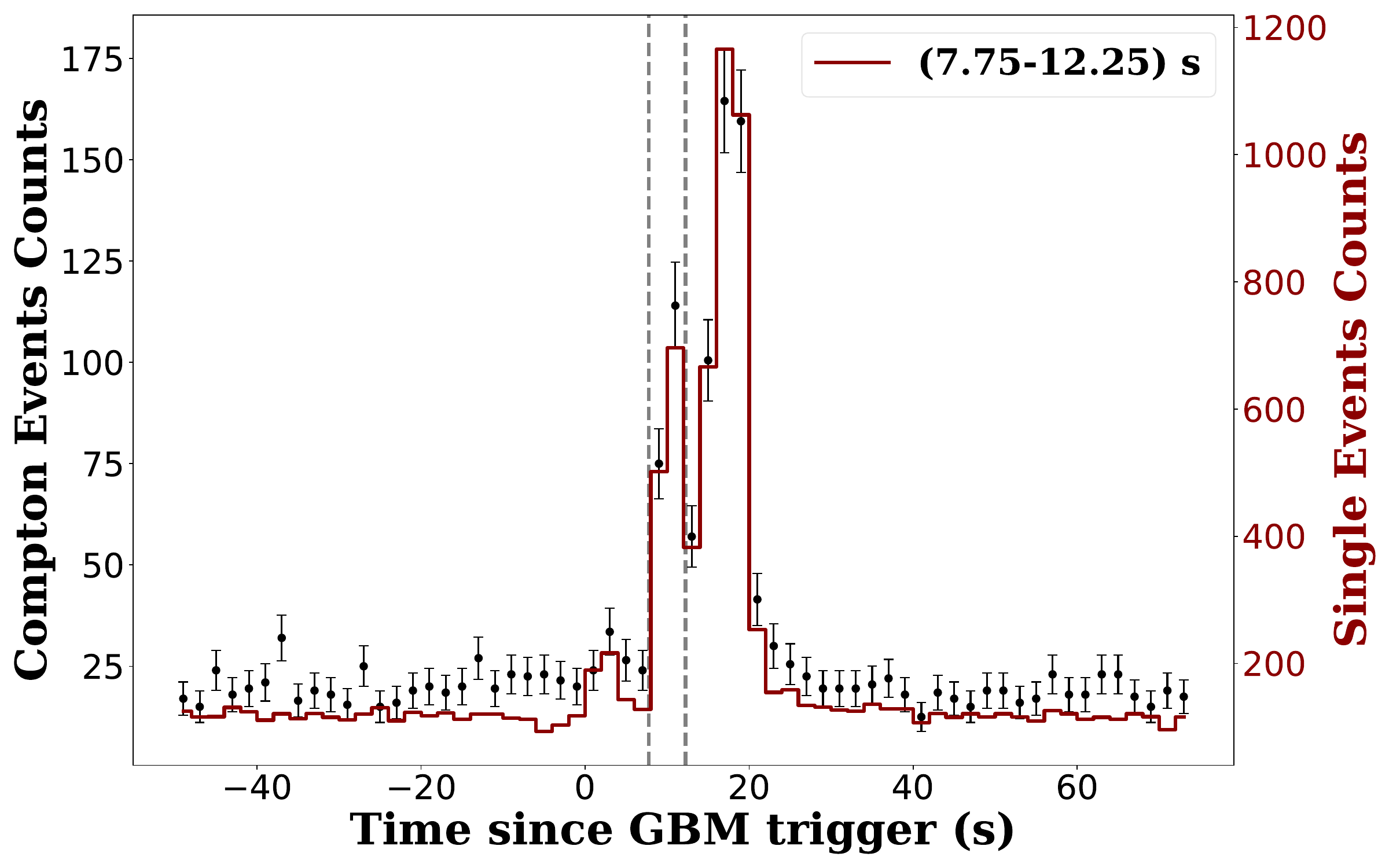}
    \includegraphics[width=0.3\linewidth]{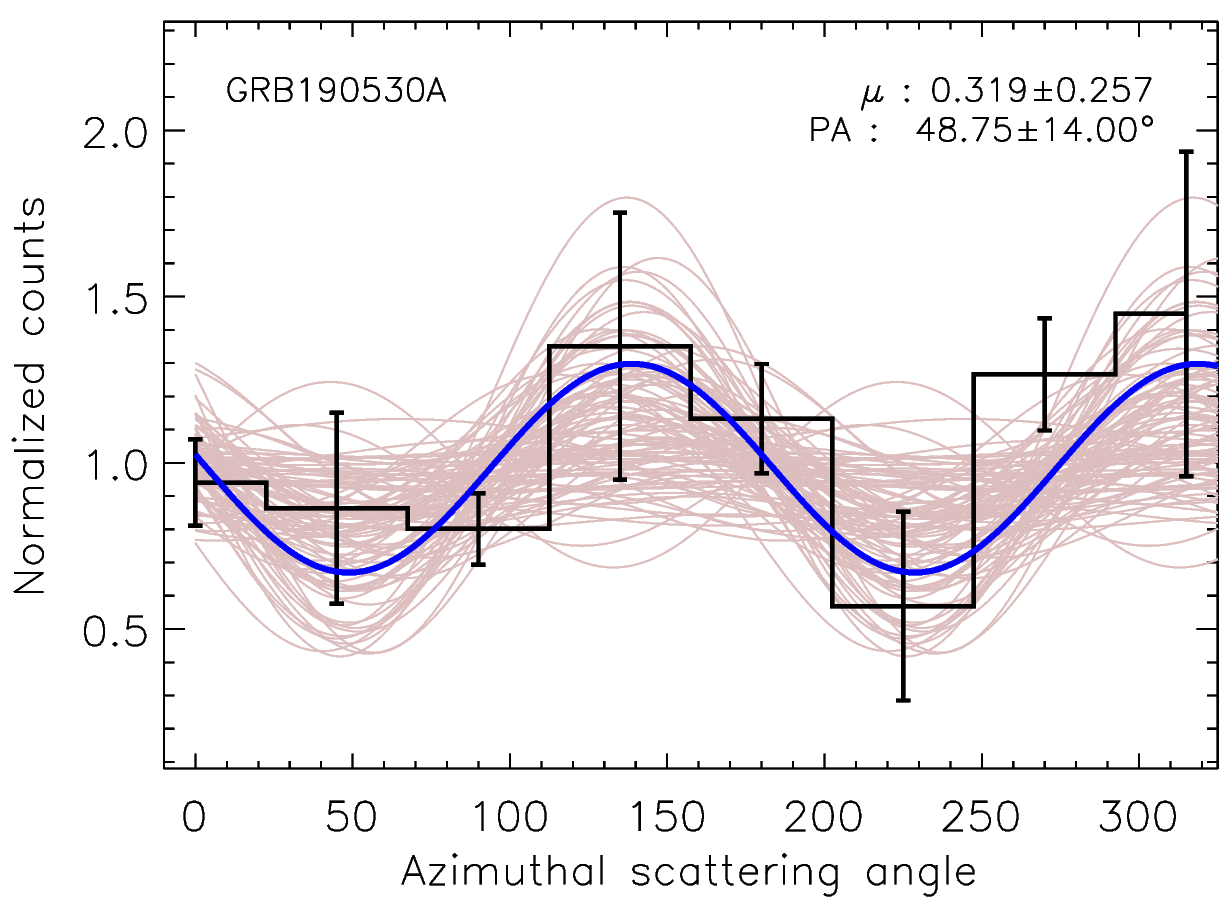}
    \includegraphics[width=0.3\linewidth]{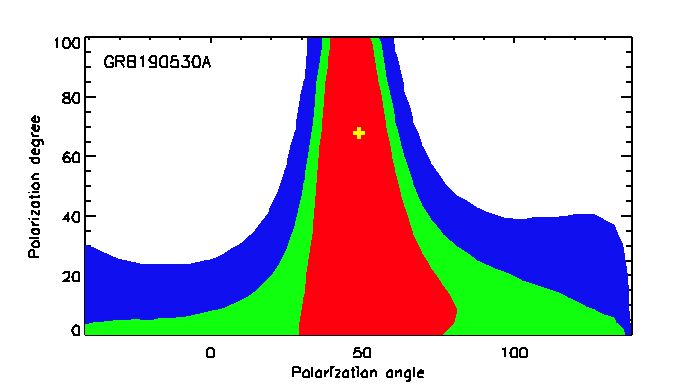}
    \includegraphics[width=0.3\linewidth]{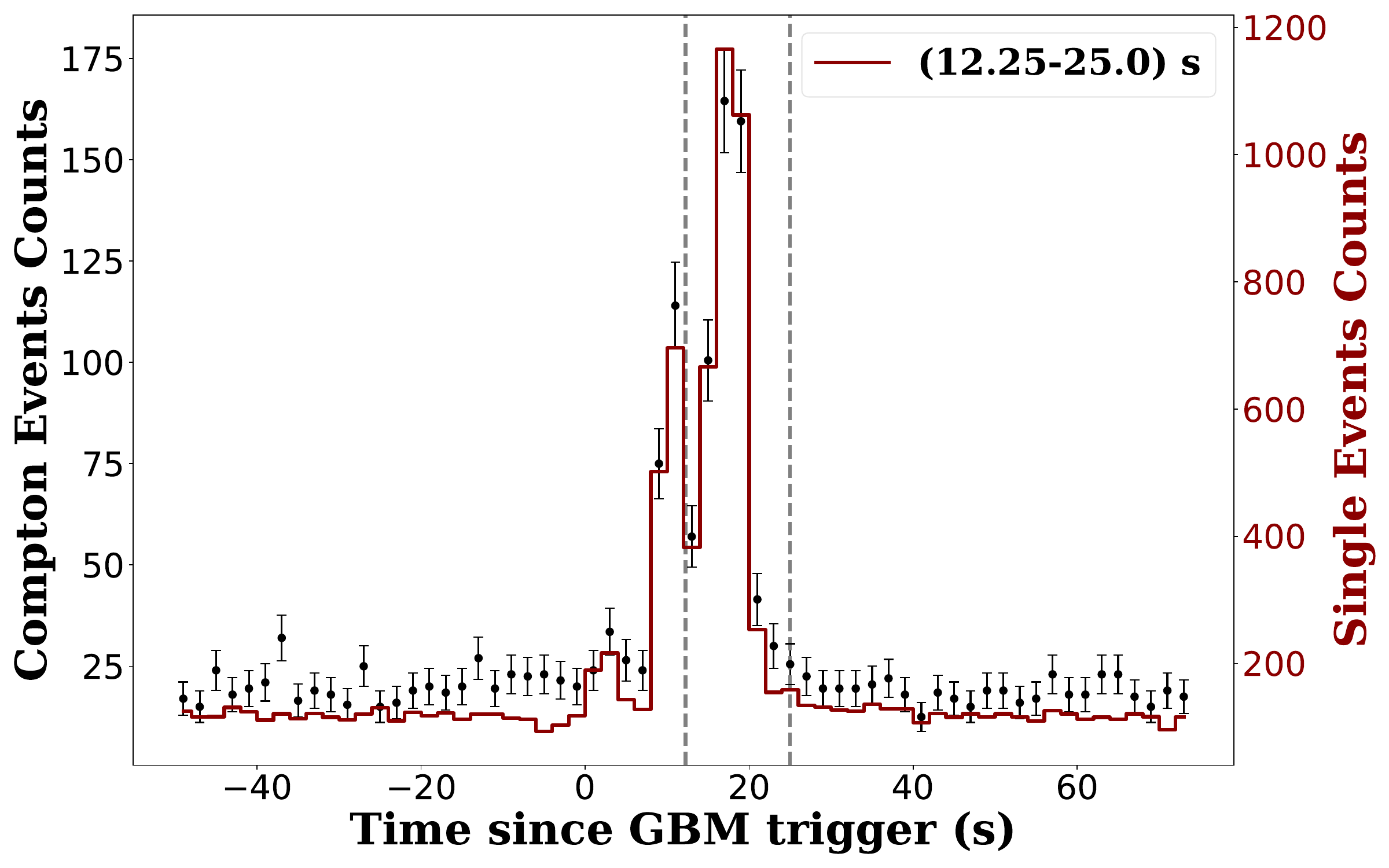}
    \includegraphics[width=0.3\linewidth]{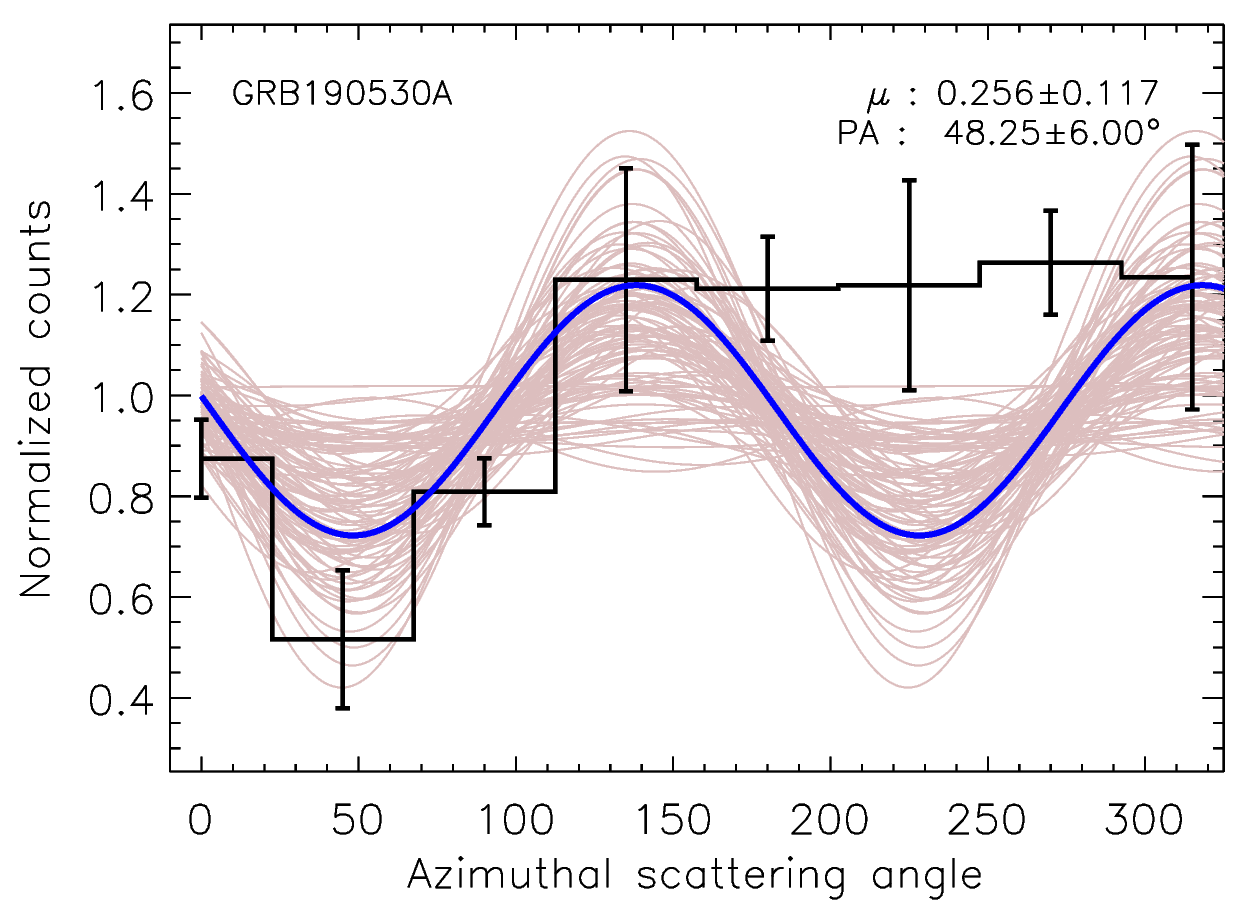}
    \includegraphics[width=0.3\linewidth]{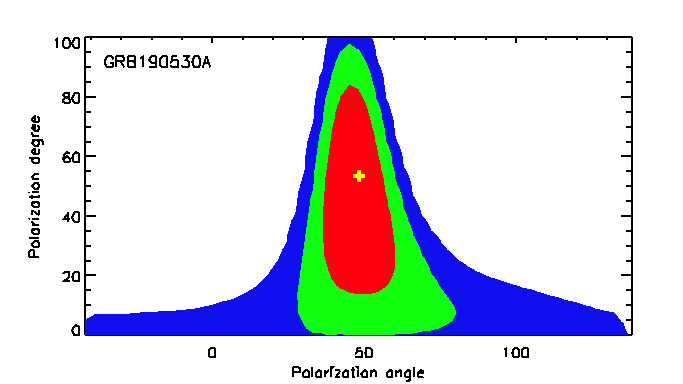}
    \includegraphics[width=0.3\linewidth]{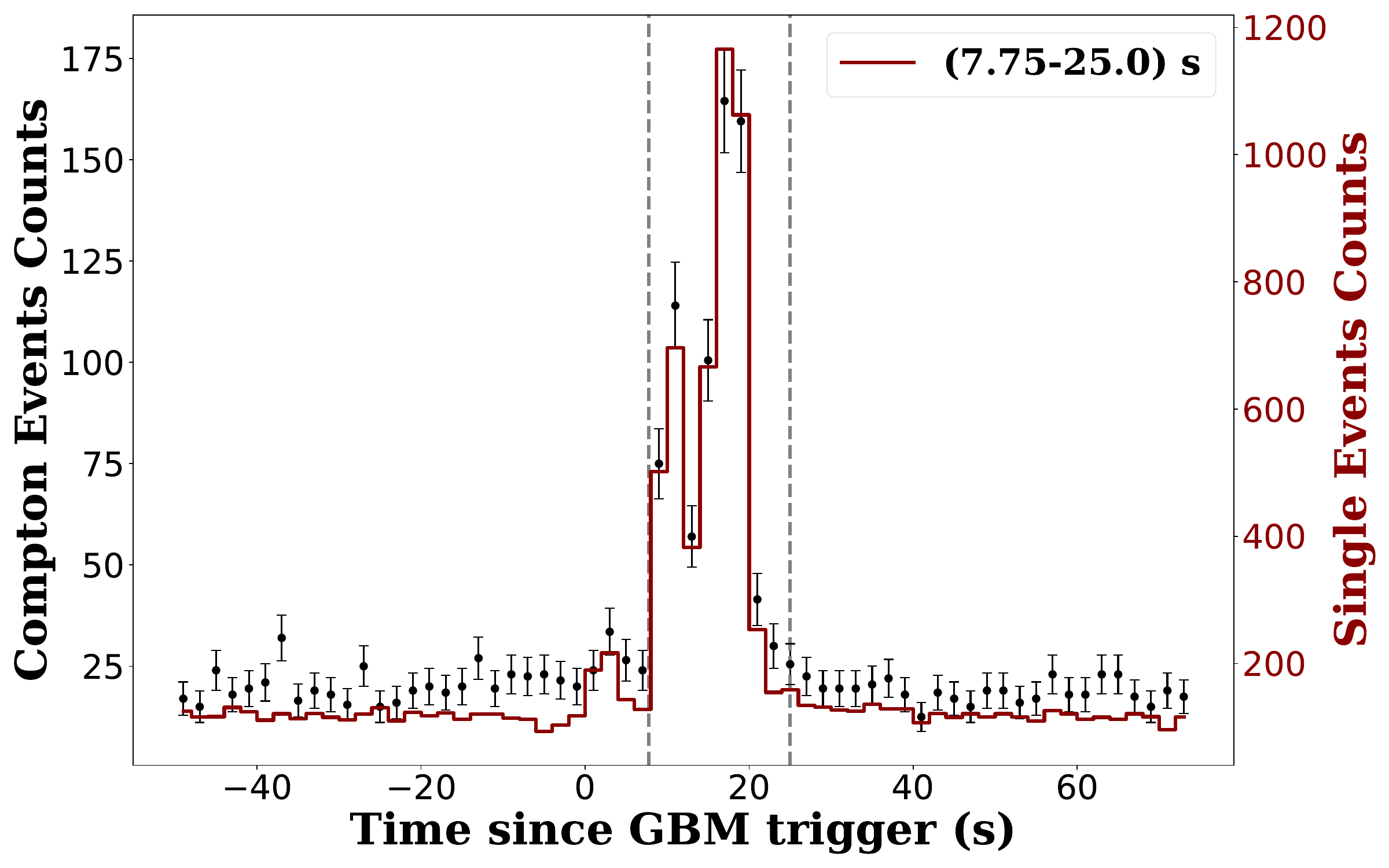}
    \includegraphics[width=0.3\linewidth]{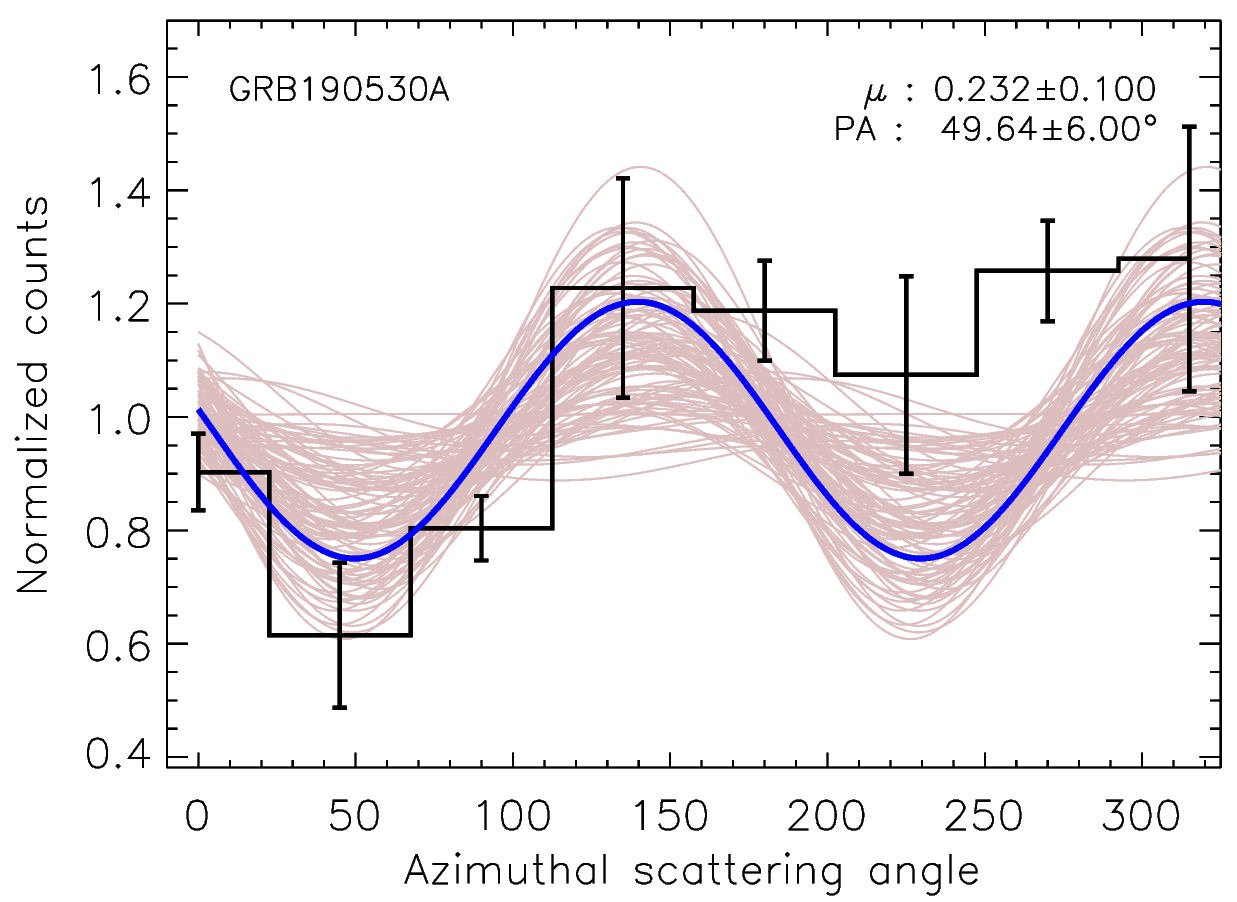}
    \includegraphics[width=0.3\linewidth]{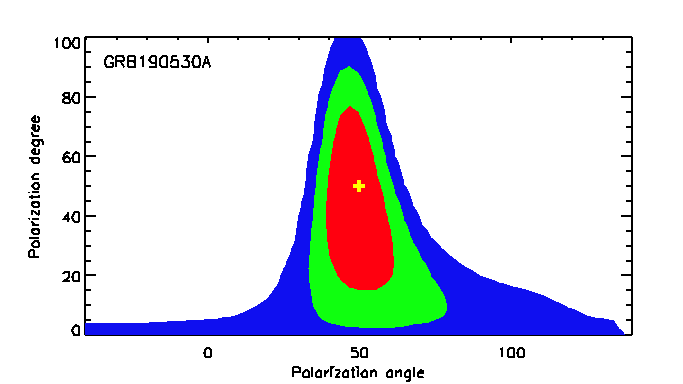}
    \includegraphics[width=0.3\linewidth]{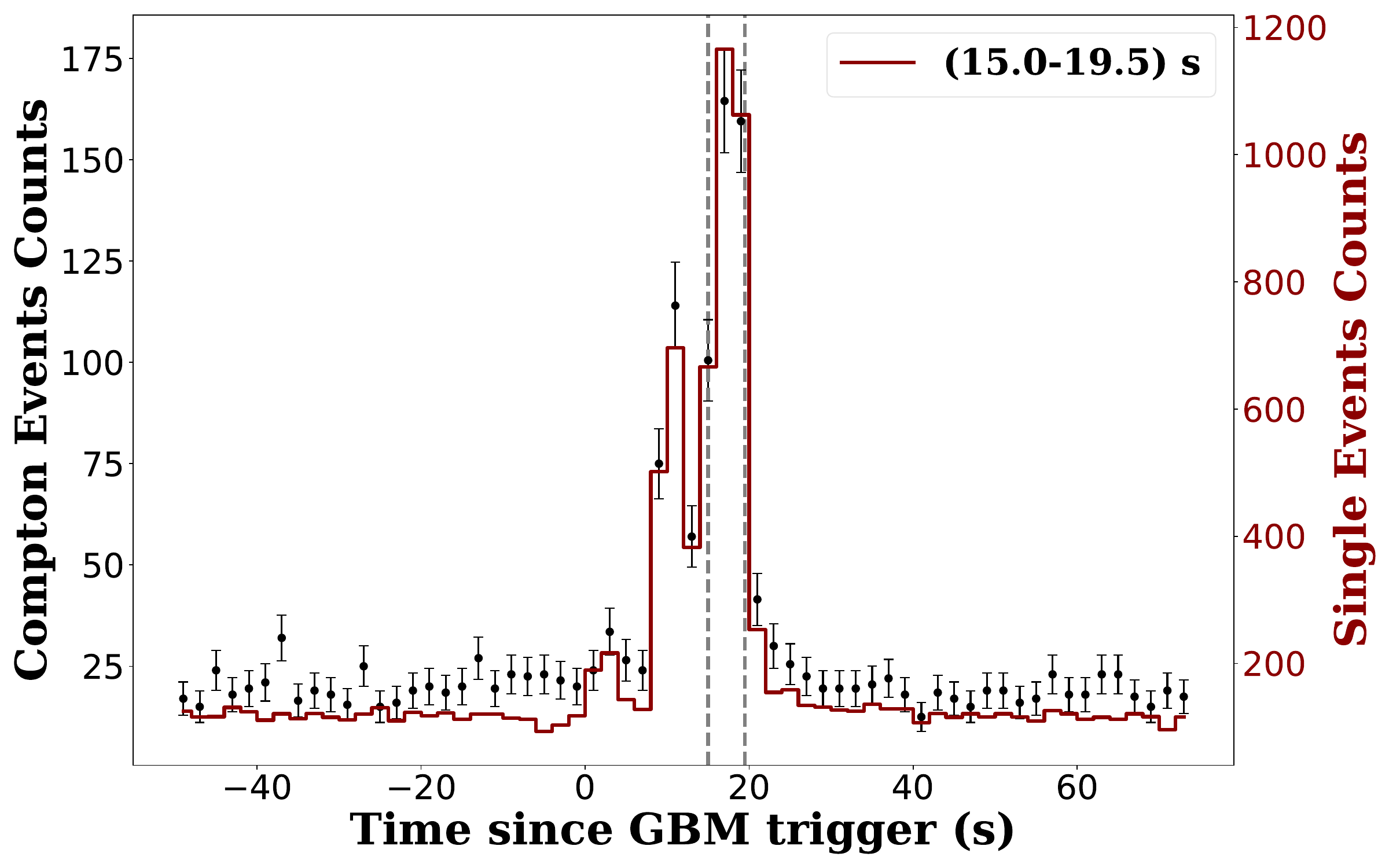}
    \includegraphics[width=0.3\linewidth]{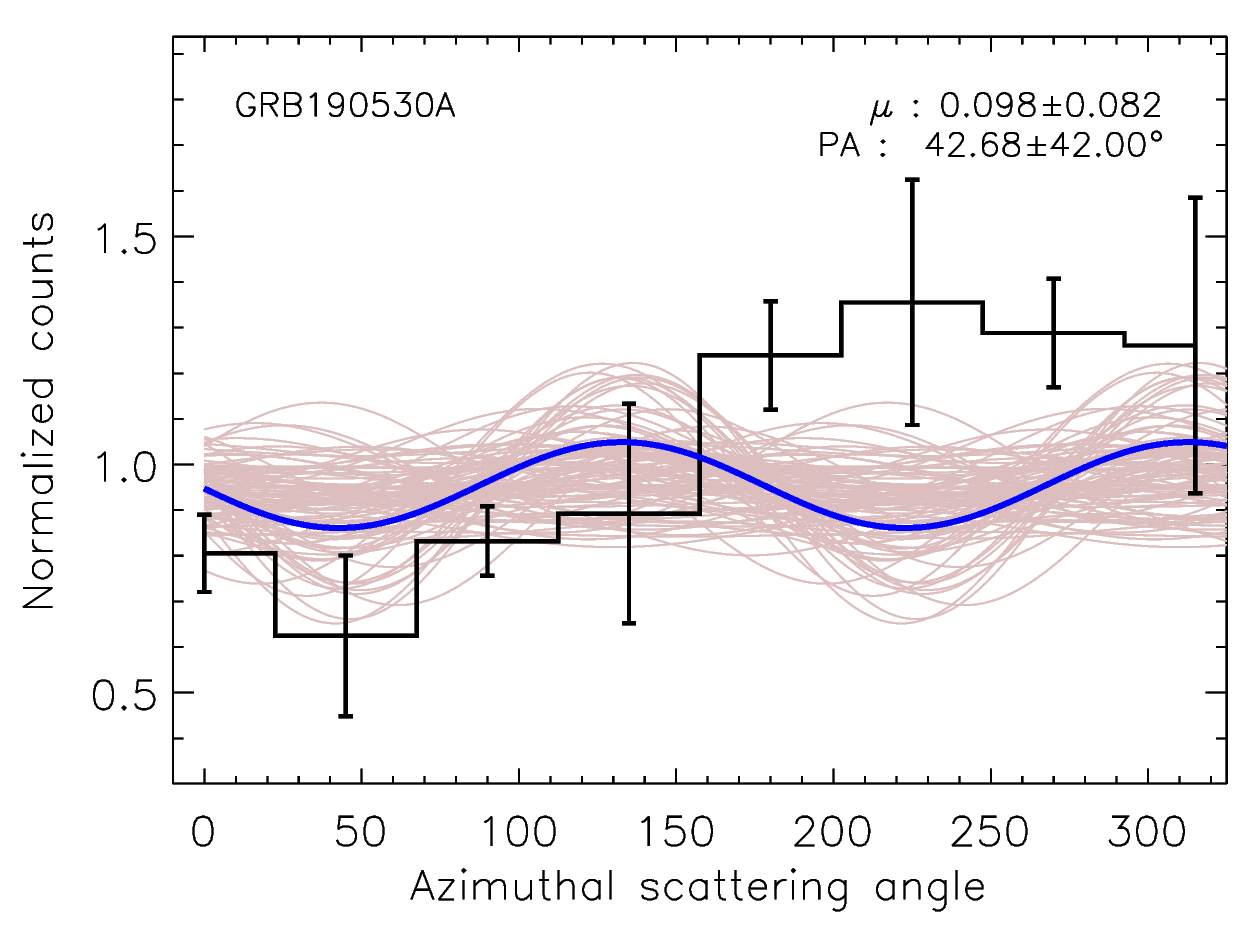}
    \includegraphics[width=0.3\linewidth]{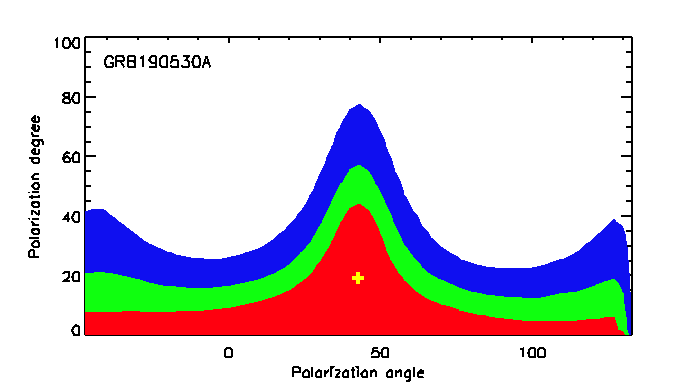}
\caption{The left panels depict the light curve of \thisgrb for the single-pixel (marked in red), and double-pixel (marked in black) counts in 100-300 \keV energy range with a temporal bin size of 2 s, obtained using \AstroSat CZTI data. The vertical grey dashed lines indicate the time intervals used for the time-integrated (in the top row) and time-resolved polarization (in the last four rows) measurements within the burst. The middle and right panels depict the contour plots of polarization fraction and angle and modulation curves for the corresponding intervals. Detailed info about the figure has been discussed in \S~\ref{Prompt emission Polarization}.}
\label{pol_190530}
\end{figure*}

\begin{table*}
\begin{tiny}
\caption{The \AstroSat CZTI polarimetry results of \thisgrb in the time-integrated and time-resolved temporal window in 100-300 \keV energy range.}
\label{polarization_table} 
\begin{center}
\begin{tabular}{cccccc}
\hline
\bf Burst interval & \bf  Compton events  &\bf Modulation amplitude &\bf PA  &\bf Bayes factor & \bf PF\\
\bf (s)& (no.)  &\bf ($\mu$) & \bf ($\bf ^\circ$) &  &  \\
\hline
0.0 $-$ 25.0&1246 &0.27 $\pm$ 0.10&46.74 $\pm$ 4.0&3.51& 55.43 \% $\pm$ 21.30\% \\\\
7.75$-$12.25 &319&0.32 $\pm$ 0.26&--&1.08& $<$64.40\% (95\%) \\\\
12.25$-$25.0 &870&0.26 $\pm$ 0.12&48.17 $\pm$ 6.0&2.11& 53.95 \% $\pm$ 24.13\%\\\\
7.75$-$25.0 &1189&0.23 $\pm$ 0.10&49.61 $\pm$ 6.0&2.52& 49.99 \% $\pm$ 21.80\%\\\\
15.0 $-$ 19.5&577 &0.09 $\pm$ 0.08&--&0.71& $<$65.29\% (95\%)  \\
\hline
\end{tabular}
\end{center}
\end{tiny}
\end{table*}

Figure \ref{pol_190530} left panels show the CZTI light curves of \thisgrb in 100-300 \keV energy range for both 1-pixel (marked in red) and 2-pixel Compton events (marked in black). This shows that the burst can be seen in Compton events. The grey dashed lines show the time intervals used for polarization analysis. The mean background registered is around 20 counts per second. The middle panels show respective azimuthal angle distributions for the burst obtained during the same time intervals. As shown in Figure \ref{pol_190530}, for \thisgrb, polarization analysis has been performed for the full burst (\fermiT to \fermiT+25 s, see top panel of Figure \ref{pol_190530}) as well as for the two brightest emission episodes (referred as 2$^{nd}$ and 3$^{rd}$ emission episodes, respectively, see panels in the second row and panels in the third row of Figure \ref{pol_190530}). The 2$^{nd}$ episode (time interval: \fermiT+7.75 to \fermiT+12.25 s) and 3$^{rd}$ emission episode (time interval: \fermiT+12.25 to \fermiT+25 s) recorded around 319 and 870 Compton events, respectively in CZTI detector. For the complete burst, we estimate a polarization fraction of 55.43 $\pm$ 21.30 \% with a Bayes factor around 3.5 (see the panel in the first row of Figure \ref{pol_190530}). We also see the polarized signature in the azimuthal angle distribution for the 2$^{nd}$ episode (see the panel in the second row of Figure \ref{pol_190530}). However, the polarization could not be constrained because of the small Compton events (Bayes factor of 1.08). We estimated the 2$\sigma$ upper limit on polarization fraction around 64 \% for this episode (see Table \ref{polarization_table}). On the other hand, the 3$^{rd}$ episode (see the panel in the third row of Figure \ref{pol_190530}) with a relatively more significant number of Compton events yields a hint of polarization in this region, having a polarization fraction around 53.95 $\pm$ 24.13\% with a Bayes factor value around 2. The panel in the fourth row light curve and azimuthal angle distribution in Figure \ref{pol_190530} is the combined analysis of the 2$^{nd}$ and 3$^{rd}$ episodes which yield a hint of polarization fraction of 49.99 $\pm$ 21.80\% with a Bayes factor around 2.5. Furthermore, we attempted to measure polarization for a temporal window (see the panel in the last row of Figure \ref{pol_190530}) where the low energy spectral index is found to be harder. However, we could only constrain the limits during this window due to a low number of Compton events. A hint of high polarization signature for both  time-integrated (with a Bayes factor of around 3.5) as well as time-resolved analysis confirms that polarization properties remain independent across the burst. This can be further verified because the polarization angles obtained for different burst intervals are within their error bar, indicating no significant change in the polarization properties with burst evolution.

\subsubsection{Empirical correlations}

{\bf Amati and Yonetoku correlations:}

The Amati correlation \citep{2006MNRAS.372..233A} is a correlation between the time-integrated peak energy in the source frame ($E_{\rm p,z}$) and isotropic equivalent $\gamma$-ray energy ($ E_{\gamma,\rm iso}$) of GRBs. The $E_{\gamma,\rm iso}$ depends on time-integrated bolometric (1-10,000 \keV in the rest frame) energy fluence. In the case of \thisgrb, we calculated the rest frame peak energy and $E_{\gamma,\rm iso}$ using the joint GBM and LAT spectral analysis (\fermiT to \fermiT + 25 s). The calculated values of these parameters are listed in Table \ref{tab:prompt_properties_GRB190530A}) and shown in Figure \ref{fig:prompt_properties_amati_yonetoku_GRB190530A} (a) along with other data points for long and short bursts published in \cite{2020MNRAS.492.1919M}. We noticed that \thisgrb lies towards the upper right edge and is consistent with the Amati correlation of long bursts. We compared the energetic of \thisgrb with a large sample of GBM detected GRBs with a measured redshift \citep{2021ApJ...908L...2S}. We noticed that \thisgrb is one of the most energetic GRBs ever detected, with only GRB 140423A and GRB 160625B reported as more energetic.

\begin{figure}[ht!]
\centering
\includegraphics[scale=0.28]{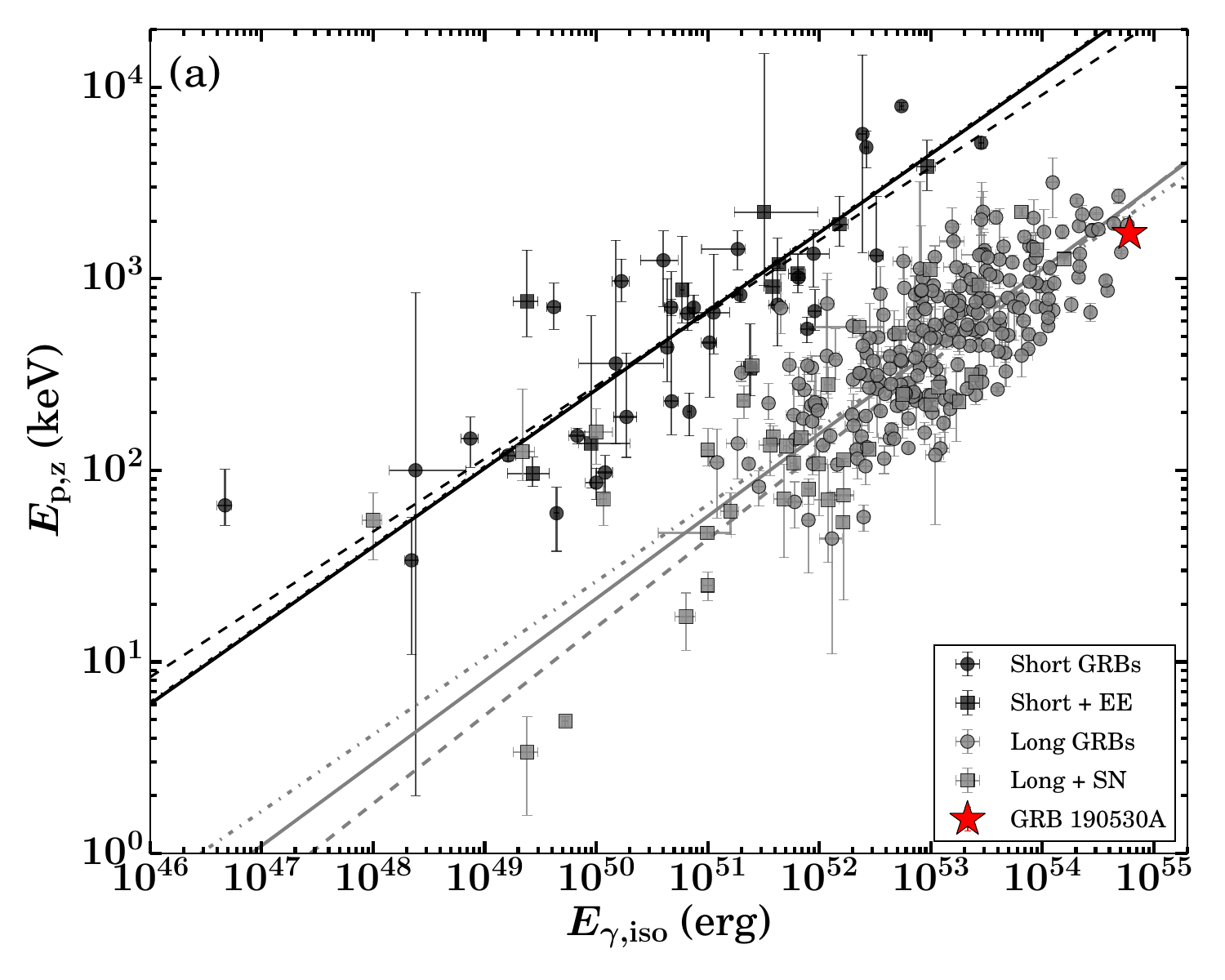}
\includegraphics[scale=0.28]{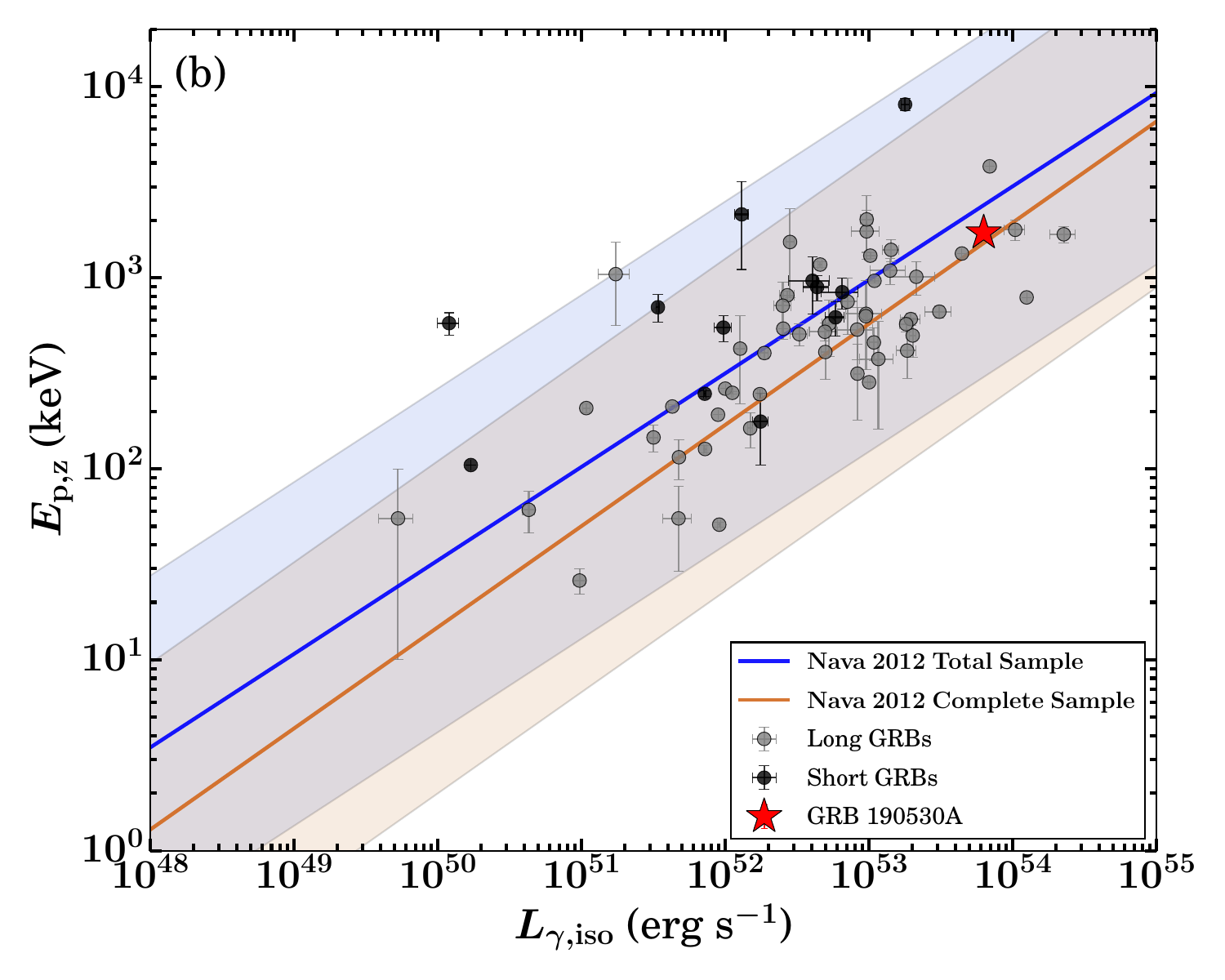}
\caption{{Prompt emission characteristics of \thisgrb (shown with a red star):}  (a) Amati correlation:  \thisgrb along with the data points for long (grey circles for typical LGRBs and grey squares for LGRBs with associated supernovae) and short GRBs (black circles for typical SGRBs and black squares for SGRBs with extended emission) published in \protect\cite{2020MNRAS.492.1919M}. Grey colour solid, dashed, and dashed-dotted lines correspond to the best-fit lines for the complete sample of LGRBs, for LGRBs with and without associated supernovae, respectively. Similarly, black colour solid, dashed, and dashed-dotted lines correspond to the best-fit lines for the complete sample of SGRBs, for SGRBs with and without extended emission, respectively. (b) Yonetoku correlation: \thisgrb along with the data points for long (grey circles) and short GRBs (black circles) published in \protect\cite{2012MNRAS.421.1256N}. The coloured solid lines indicate the best fit and the shaded region represents the 3$\sigma$ scatter of the correlations \citep{2012MNRAS.421.1256N}.}
\label{fig:prompt_properties_amati_yonetoku_GRB190530A}
\end{figure}

Furthermore, we also examined the location of \thisgrb on the Yonetoku correlation \citep{2004ApJ...609..935Y}. This is a correlation between the time-integrated peak energy in the source frame ($E_{\rm p,z}$) and isotropic peak luminosity ($L_{\gamma,\rm iso}$). To calculate the value of $L_{\gamma,\rm iso}$, we measured the peak flux in 1-10,000 \keV energy range for \thisgrb. The calculated value of $L_{\gamma,\rm iso}$ is listed in Table \ref{tab:prompt_properties_GRB190530A}). The position of \thisgrb on the Yonetoku relation is given in Figure \ref{fig:prompt_properties_amati_yonetoku_GRB190530A} (b) together with data points for other short and long GRBs, published in \cite{2012MNRAS.421.1256N}. In this plane, \thisgrb lies within the 3 $\sigma$ scatter of the total and complete samples\footnote{{https://www.mpe.mpg.de/events/GRB2012/pdfs/talks/GRB2012\_Nava.pdf}}
of GRBs studied by \cite{2012MNRAS.421.1256N}. \thisgrb is one of the bursts with the largest $L_{\gamma,\rm iso}$.

{\bf Correlation between spectral parameters:}

The prompt emission spectral parameter correlations play an important role in investigating the intrinsic behaviour of GRBs. In the case of \thisgrb, we investigated the correlation between \Ep - flux, $\alpha_{\rm pt}$-flux, and \Ep-$\alpha_{\rm pt}$ obtained using the \sw{Band} function based on time-resolved analysis of the GBM data (for each bin obtained from the Bayesian Block binning algorithm). We noticed a strong correlation between the \Ep and the flux in 8 \keV - 30 MeV energy range with a Pearson coefficient (r) and \sw{p-value} of 0.82 and 4.00 $\times$ $10^{-11}$, respectively. We also noticed a strong correlation between $\alpha_{\rm pt}$ and flux with r and \sw{p-value} of 0.76 and 9.91 $\times$ $10^{-9}$. As \Ep and $\alpha_{\rm pt}$ show a strong correlation with flux, we investigated the correlation between \Ep and $\alpha_{\rm pt}$. They also show a moderate correlation with r and \sw{p-value} of 0.52 and 4.92 $\times$ $10^{-4}$. Therefore, \thisgrb is consistent with being a ``Double tracking'' GRB (Both $\alpha_{\rm pt}$ and the \Ep follow the “intensity-tracking” trend) similar to GRB 131231A \citep{2019ApJ...884..109L} and GRB 140102A \citep{2021MNRAS.505.4086G}. The correlation results are shown in Figure \ref{spc_GRB190530A} in the appendix.

\subsubsection{Prompt emission mechanism}
\label{pem}

The different emission processes invoked to explain the prompt emission of GRBs is associated with unique polarization signatures. In the case of \thisgrb, we found a hint of high polarization fraction in both the time-integrated and time-resolved polarization measurements. We do not notice any significant variation in polarization fraction and polarization angle in our time-resolved polarization analysis, supporting the synchrotron emission model, an ordered magnetic field produced in shocks \citep{2003ApJ...597..998L} for the first two pulses. Such high polarization ($\sim$ 40-70 \%) cloud also be produced using synchrotron emission with a random magnetic field, in the case of a narrow jetted emission ($\Gamma_{0}$ $\theta_{\rm j}$ $\sim$ 1, where $\Gamma_{0}$ is the bulk Lorentz factor and $\theta_{\rm j}$ is the jet opening angle) and seen along the edge. To verify both possibilities, we calculated bulk Lorentz factor $\Gamma_{0}$ of the fireball ejecta and $\theta_{\rm j}$ (see \S~3.3.3 of \cite{2022MNRAS.511.1694G}). There are several methods to calculate $\Gamma_{0}$ using both prompt emission and afterglow properties \citep{2018A&A...609A.112G}. We calculated the value of the Lorentz factor using the prompt emission correlation between $\Gamma_{0}$-$E_{\gamma, \rm iso}$\footnote{$\Gamma_{0}$ $\approx$ 182 $\times$ $E_{\gamma, \rm iso, 52}^{0.25 \pm 0.03}$} \citep{2010ApJ...725.2209L} as $\Gamma_{0}$ decreases towards afterglow phase. The calculated value of $\Gamma_{0}$ is 902.63$^{+191.23}_{-157.80}$ using the normalization and slope of the $\Gamma_{0}$-$E_{\gamma, \rm iso}$ correlation. The calculated value of $\theta_{j}$ is 0.062 radian (3.55$^\circ$) derived from the broadband afterglow modelling (see \S~3.3.3 of \cite{2022MNRAS.511.1694G}). We obtained $\Gamma_{0}$ $\theta_{\rm j}$ equal to $\sim$ 56, which supports the synchrotron emission model with an ordered magnetic field \citep{2009ApJ...698.1042T}. We also calculated the beaming angle ($\theta_{\rm beam}$) of the emission equal to 0.001 radian (0.06$^\circ$) using the relation between Lorentz factor and $\theta_{\rm beam}$, i.e., $\theta_{\rm beam}$= 1/$\Gamma_{0}$. Thus, \thisgrb had a wider jetted emission with a narrow beaming angle.

In addition to polarization results, our time-resolved spectral analysis indicates that the low-energy spectral indices of the \sw{Band} function are consistent with the prediction of synchrotron emission for the first two pulses. Moreover, the presence of a low-energy spectral break in the time-integrated and time-resolved spectra with power-law indices consistent with the prediction of the synchrotron emission model confirms synchrotron emission as the mechanism dominating during the first two pulses of \thisgrb. However, during the third pulse, the low-energy spectral indices become harder and exceed the synchrotron death line in a few bins. During this window, we find a signature of the thermal component along with the synchrotron component in our time-resolved spectral analysis, suggesting some contribution to the emission from the photosphere.

\subsubsection{Nature of Central Engine}

Based on the properties of prompt and afterglow emission, e.g. variability in the gamma-ray light curves and X-ray flares and plateau \citep{2012A&A...539A...3B, 2020ApJ...896...42Z}, there are two types of the object thought to be powering the central engine: a stellar-mass black hole (BH), and a rapidly spinning, highly magnetized 'magnetar.' Recently, \cite{2018ApJS..236...26L} studied the X-ray light curves sample of 101 bursts with a plateau phase and measured redshift. They calculated the isotropic kinetic energies and the isotropic X-ray energies for each burst. They compared them with the maximum possible rotational energy budget of the magnetar ($10^{52}$ ergs). They found only $\sim$ 20 \% of GRBs were consistent with having a magnetar central engine. The rest of the bursts were consistent with having a BH as the central engine. More recently, \cite{2021ApJ...908L...2S} also identified GRBs with BH central engines based on the maximum rotational energy of the magnetar that powers the GRB, i.e., the upper limit of rotational energy of magnetar. They analyzed the sample of \fermi-detected GRBs with a measured redshift. They calculated the beaming corrected isotropic gamma-ray energies and compared them with the magnetars' maximum possible energy budget. In the case of \thisgrb, we could not follow the methodology discussed by \cite{2018ApJS..236...26L} due to the absence of plateau phase in the X-ray light curves; therefore, we follow the methods discussed by \cite{2021ApJ...908L...2S}. We calculated the beaming corrected energy assuming the fraction of forward shock energy into the electric field $\epsilon_{e}$ = 0.1. We performed broadband afterglow modelling to constrain the limiting value of the jet opening angle (see \S~3.3 of \cite{2022MNRAS.511.1694G}). We find beaming corrected energy for \thisgrb equal to 1.16 $\times$ $10^{52}$ ergs, and this value is well above the mean energy of the sample studied by \cite{2021ApJ...908L...2S}, see also Figure \ref{central engine}. In addition, this value is also higher than the maximum possible energy budget of the magnetar. No flares, or plateau features are present in the X-ray light curve. We proposed that BH could be the possible central engine of \thisgrb. However, the magnetar option could also be feasible as beaming corrected energy for \thisgrb is close to the upper limit of the magnetar's rotational energy, and no early X-ray observations of the X-ray afterglow of this burst are available.

\begin{figure}[ht!]
\centering
\includegraphics[scale=0.35]{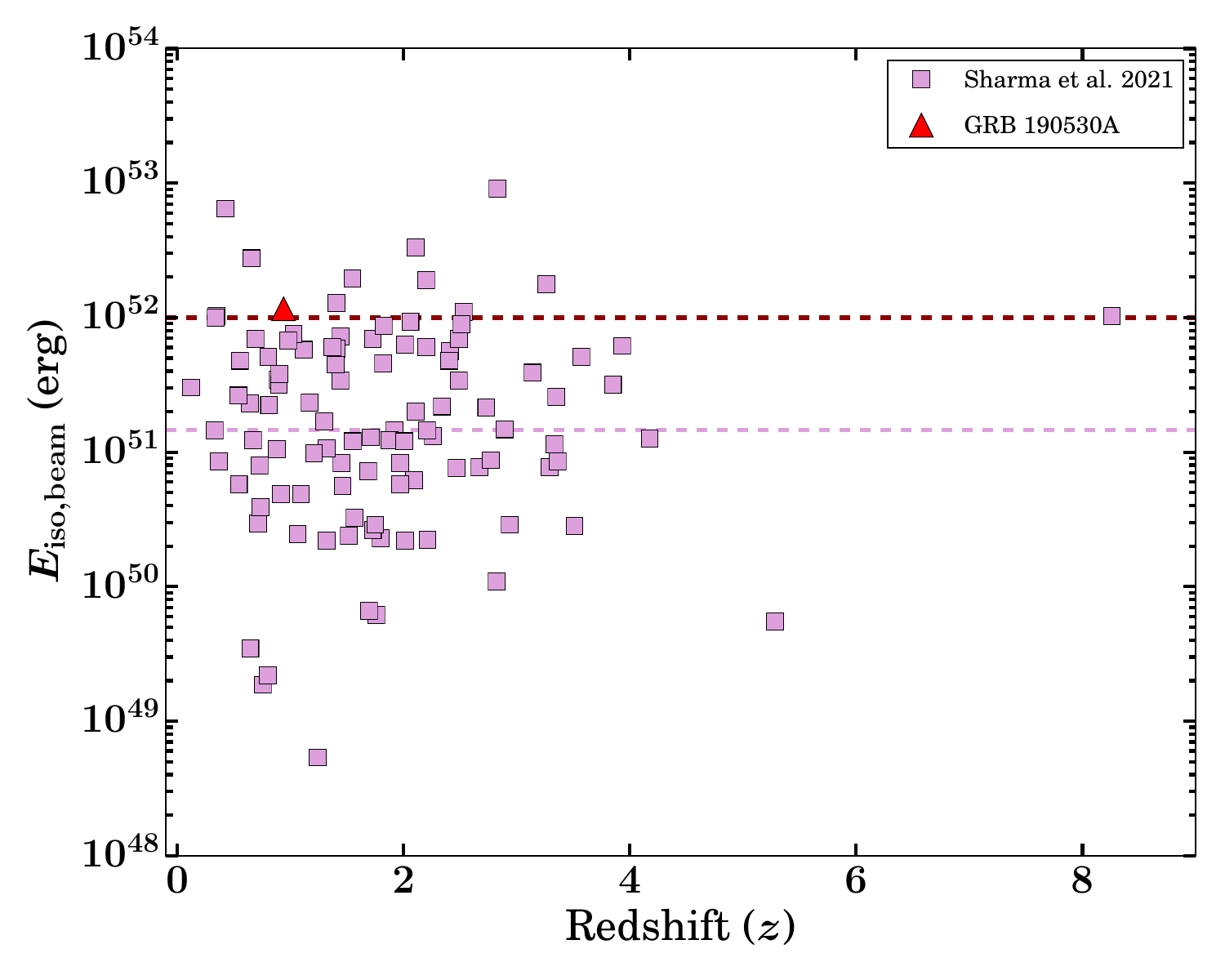}
\caption{{The central engine of \thisgrb:} Redshift distribution as a function of beaming corrected isotropic $\gamma$-ray energy for the \fermi detected bursts, data points taken from \protect \cite{2021ApJ...908L...2S}. \thisgrb is shown with a red triangle. The horizontal red and pink dashed lines indicate the maximum possible energy budget of the magnetars and the median value of beaming corrected isotropic $\gamma$-ray energy of the sample studied by \protect \cite{2021ApJ...908L...2S}, respectively.}
\label{central engine}
\end{figure}

\subsection{\thisgrbB}

The prompt emission results of \thisgrbB obtained using \fermi and ASIM data are presented in this section.

\subsubsection{Light curve and time-integrated spectrum}

\begin{figure}[ht!]
\centering
\includegraphics[scale=0.45]{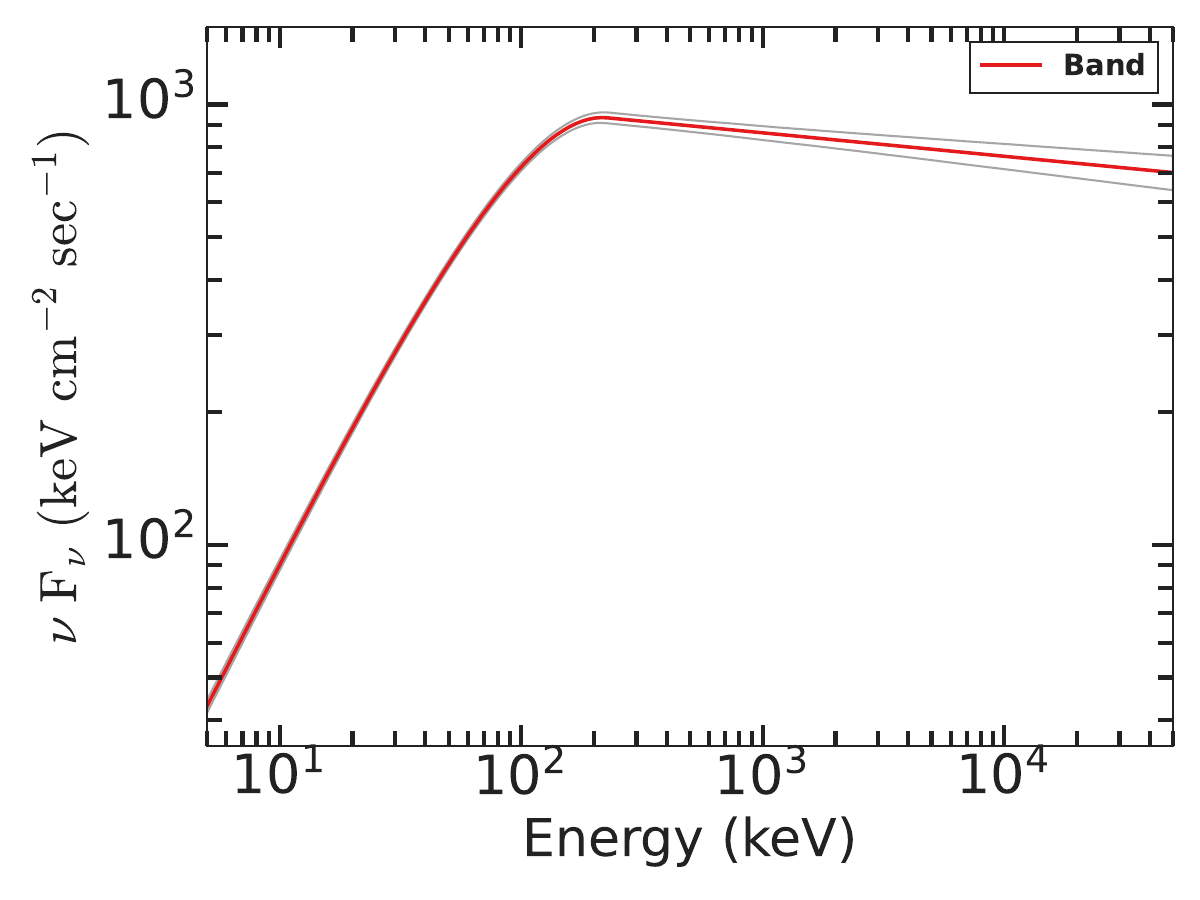}
\caption{The best fit time-averaged (\fermiT to \fermiT+ 67.38\,s) spectrum (\sw{Band}) of \thisgrbB in model space. The white-shaded regions denote the 2 $\sigma$ uncertainty regions associated with the spectral model parameters.} 
\label{TAS}
\end{figure}

The prompt \fermi GBM and \swift BAT light curve of \thisgrbB consists of a very bright main pulse followed by a long fainter emission activity up to $\sim$ 70\,s post-trigger time. On the other hand, ASIM observed only the main bright pulse, displaying a Fast-Rise Exponential Decay (FRED) type structure (see Figure~\ref{ASIM_LC_SED}). Based on the DIC condition given in section \ref{GBM}, we find that the time-averaged spectra (\fermiT to \fermiT+67.38\,s) of \thisgrbB in the energy range from 8 \keV to 40 MeV could be best fitted using the \sw{Band} model. We obtained the following best fit spectral parameters: $\alpha_{\rm pt}$=  -0.90$^{+0.01}_{-0.01}$, $\beta_{\rm pt}$= -2.05$^{+0.02}_{-0.02}$, and the spectral peak energy \Ep = 216.08$^{+4.20}_{-4.20}$ \keV. The best fit time-averaged spectrum of \thisgrbB along with the corresponding corner plot is shown in Figure~\ref{TAS} and Figure~\ref{TAS_corner_GRB210619B} of the appendix, respectively. 

\subsubsection{Empirical correlations}

{\bf Amati and Yonetoku correlations:}

\begin{figure}[ht!]
\centering
\includegraphics[scale=0.29]{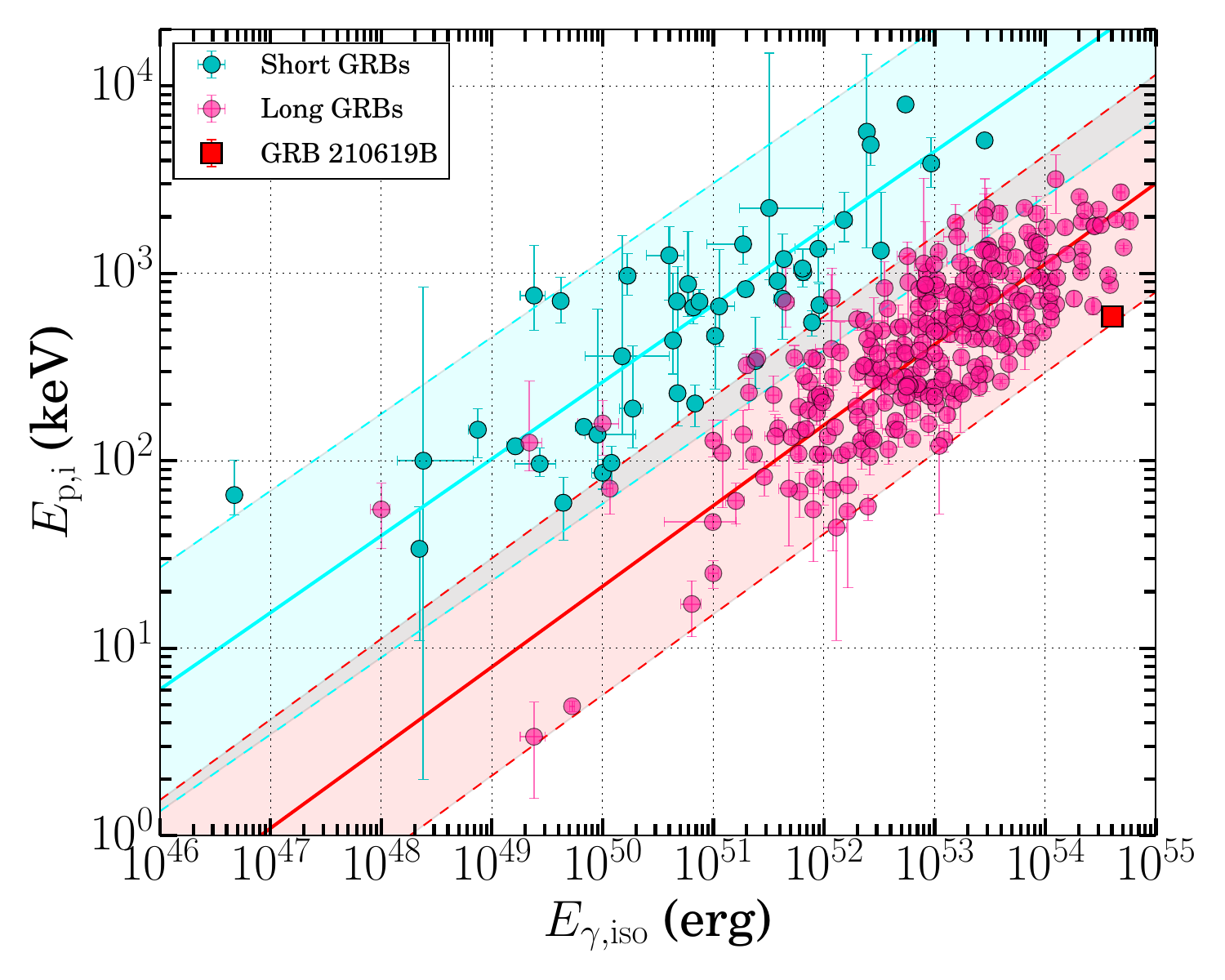}
\includegraphics[scale=0.29]{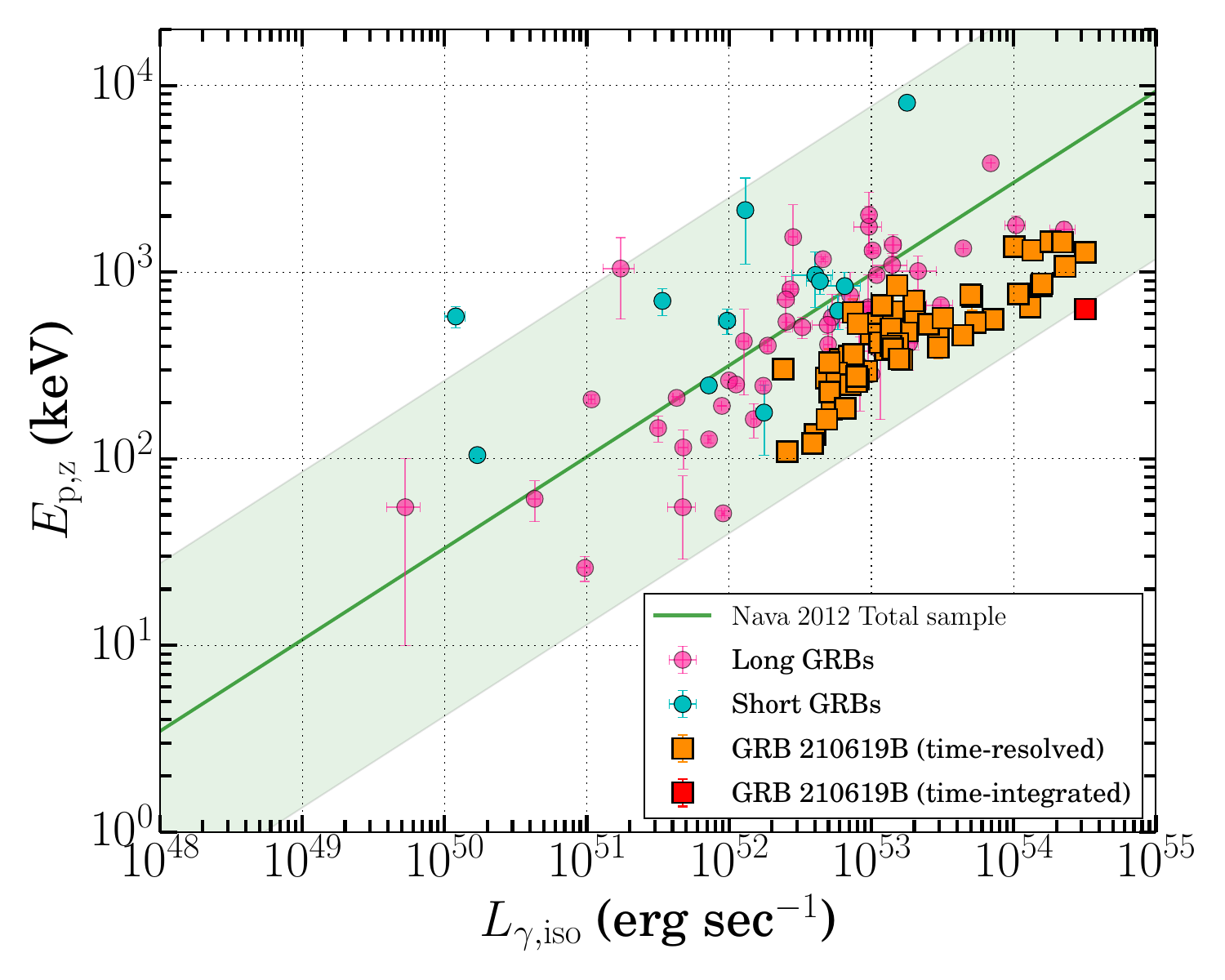}
\caption{{Left panel:} \thisgrbB in Amati correlation plane (marked with a red square). The cyan and pink circles denote the long and short GRBs taken from \protect\cite{2020MNRAS.492.1919M}. The solid cyan and pink lines show the linear fit lines for short and long GRBs, respectively. The parallel shaded regions show the 3$\sigma$ scatter of the correlations. {Right panel:} \thisgrbB in Yonetoku correlation plane (marked with a red square for the time-integrated duration and marked with orange squares for each of 59 spectra obtained using the time-resolved Bayesian bins with statistical significance $\geq$ 30). The cyan and pink circles denote the long and short GRBs taken from \protect\cite{2012MNRAS.421.1256N}. The parallel shaded regions show the 3$\sigma$ scatter of the correlations.}
\label{fig:prompt_properties_amati_yonetoku}
\end{figure}

The prompt properties of GRBs follow a few global correlations. These correlations are also used to characterize and classify them into long and short bursts \citep{2020MNRAS.492.1919M}. For example, the time-integrated spectral peak energy (in the source frame) of the burst is positively correlated with the isotropic equivalent gamma-ray energy ($E_{\rm \gamma, iso}$), known as Amati correlation \citep{2006MNRAS.372..233A}. We calculated $E_{\rm \gamma, iso}$ and restframe peak energy for \thisgrbB using the best fit time-integrated model parameters and compared the results with a larger sample of GRBs.
For the time-integrated interval, we calculated the energy flux equal to 6.27 $\rm \times ~10^{-6} erg ~cm^{-2} ~s^{-1}$ in the source frame (0.34 \keV to 3404.8 \keV) which is equivalent to $E_{\rm \gamma, iso}$= 4.05 $\rm \times ~10^{54} erg$. The $E_{\rm p, z}$-$E_{\rm \gamma, iso}$ correlation for \thisgrbB along with other data points taken from \cite{2020MNRAS.492.1919M} are shown in Figure~\ref{fig:prompt_properties_amati_yonetoku}. \thisgrbB is one of the brightest bursts ever detected by \fermi GBM and marginally satisfies the Amati correlation. In the literature, there is no consensus on the physical explanation for the Amati correlation. However, some studies indicate that the Amati correlation could be explained by the viewing angle effect within the framework of optically thin synchrotron emission \citep{2004ApJ...606L..33Y, 2004ApJ...614L..13E, 2005ApJ...629L..13L}. More recently, \cite{2021ApJ...908....9V} shows that the back-scattering dominated prompt emission model can naturally explain the Amati correlation. Furthermore, we also placed \thisgrbB in the $E_{\rm \gamma, iso}$ and redshift distribution plane. The comparison with other well-studied samples \protect\citep{2020MNRAS.492.1919M, 2021ApJ...908L...2S} indicates that \thisgrbB is one of the brightest bursts at its measured redshift.

\begin{figure}[ht!]
\centering
\includegraphics[scale=0.43]{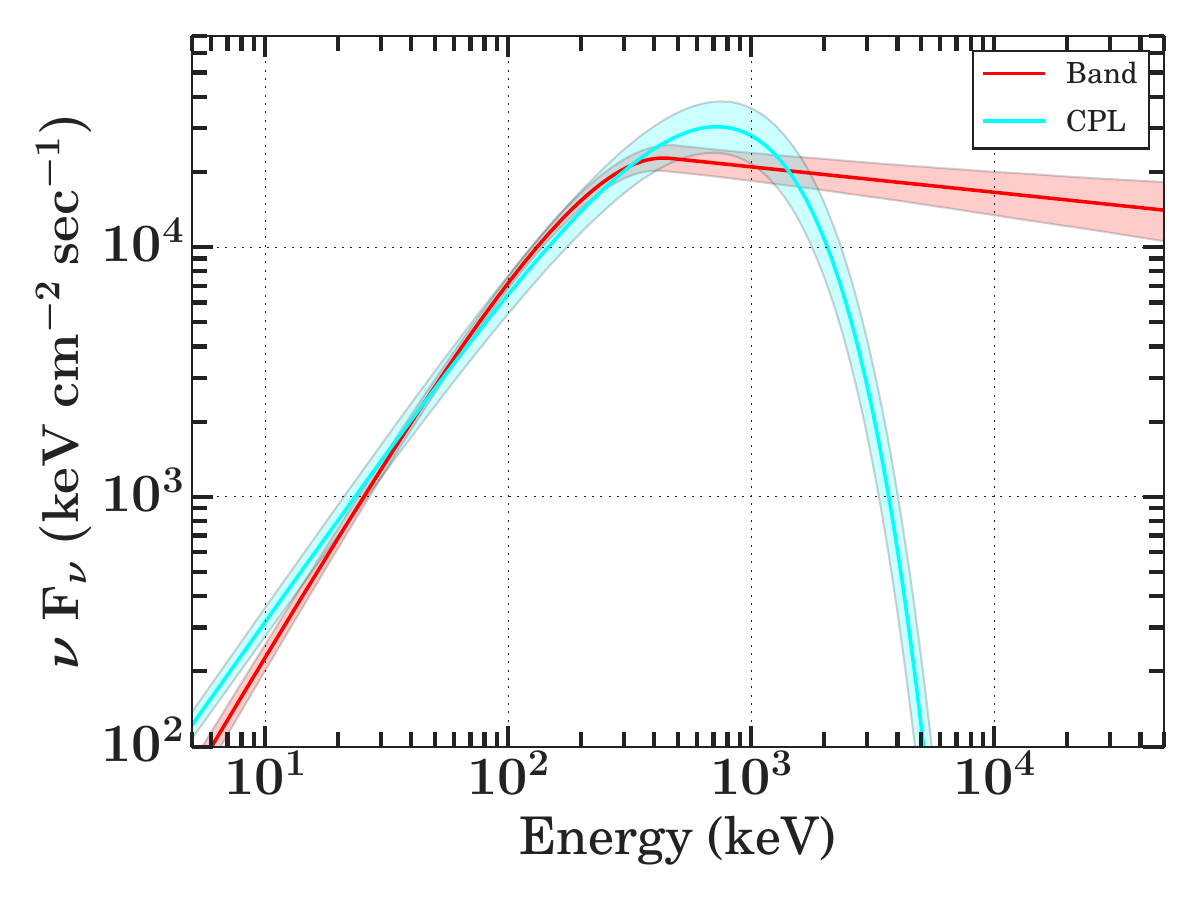}
\caption{The comparison between different empirical models (the \sw{Band} function shown in red, and the \sw{CPL} function shown in cyan) used to fit the peak spectra (time-interval from \fermiT+0.50\,s to \fermiT+ 1.01\,s, and energy range from 8 \keV to 40\,MeV) of \thisgrbB in model space. The corresponding shaded colours show 95 \% confidence levels.}
\label{Peak_spectra}
\end{figure}

Similarly, the time-integrated peak energy (in the source frame) of the burst is also correlated with the isotropic gamma-ray luminosity $L_{\rm \gamma, iso}$, known as Yonetoku correlation \citep{2010PASJ...62.1495Y}. The empirical Yonetoku correlation could be explained using the photospheric dissipation model with the consideration that the subphotospheric dissipation takes place far above the central engine \citep{2005ApJ...628..847R, 2007ApJ...666.1012T}. Recently, \cite{2019NatCo..10.1504I} proposed that the empirical Yonetoku correlation can also be interpreted as a natural consequence of viewing angle. We modelled the peak spectrum (\fermiT+0.50\,s to \fermiT+1.01\,s; see Figure \ref{Peak_spectra}) of \thisgrbB to calculate the $L_{\rm \gamma, iso}$. For the peak time-interval, we calculated the energy flux equal to 1.14 $\rm \times ~10^{-4} erg ~cm^{-2} ~s^{-1}$ in the source frame (0.34 \keV to 3404.8 \keV) and it is equivalent to $L_{\rm \gamma, iso}$= 3.20 $\rm \times ~10^{54} erg ~s^{-1}$. The Yonetoku correlation for \thisgrbB along with other data points taken from \citet{2012MNRAS.421.1256N} are shown in Figure~\ref{fig:prompt_properties_amati_yonetoku}. \thisgrbB is one of the most luminous bursts ever detected by \fermi GBM and marginally satisfied the Yonetoku correlation. In addition to time-integrated $E_{\rm p, z}$-$L_{\rm \gamma, iso}$ correlation, we also examined the time-resolved $E_{\rm p, z}$-$L_{\rm \gamma, iso}$ relation for \thisgrbB. For this purpose, we calculated the $L_{\rm \gamma, iso}$ values for each of the Bayesian bins with statistical significance (S) $\geq$ 30 used for the time-resolved spectral analysis. Figure~\ref{fig:prompt_properties_amati_yonetoku} shows the time-resolved $E_{\rm p, z}$-$L_{\rm \gamma, iso}$ correlation for \thisgrbB. We noticed that the time-resolved Yonetoku correlation for \thisgrbB is well consistent with that from the time-integrated Yonetoku correlation studied by \cite{2012MNRAS.421.1256N}. It confirmed this would indicate that the Yonetoku relation is intrinsic and not due to observational biases.

{\bf \texorpdfstring{\tninty}{}-hardness distribution:}

\begin{figure}[ht!]
\centering
\includegraphics[scale=0.28]{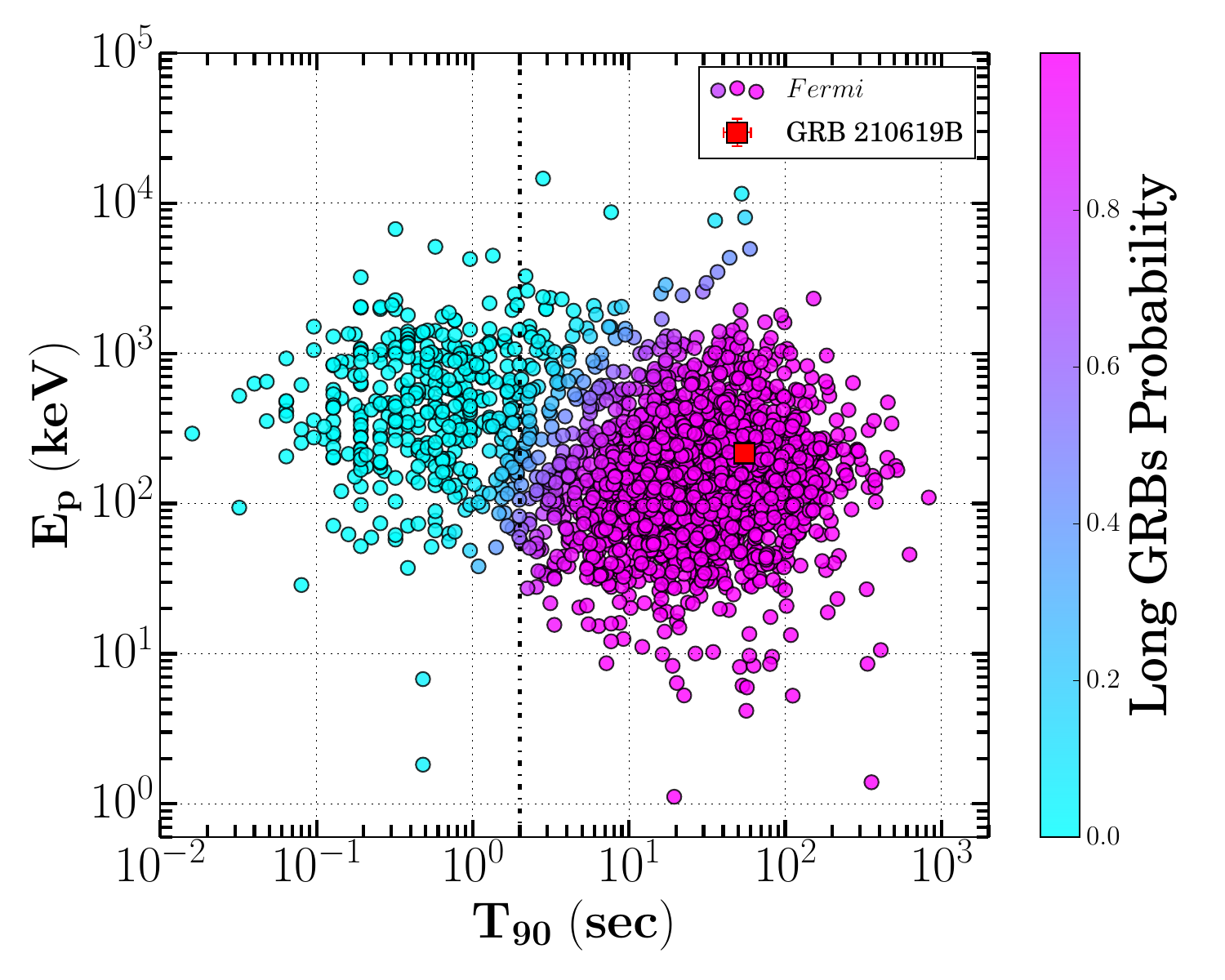}
\includegraphics[scale=0.28]{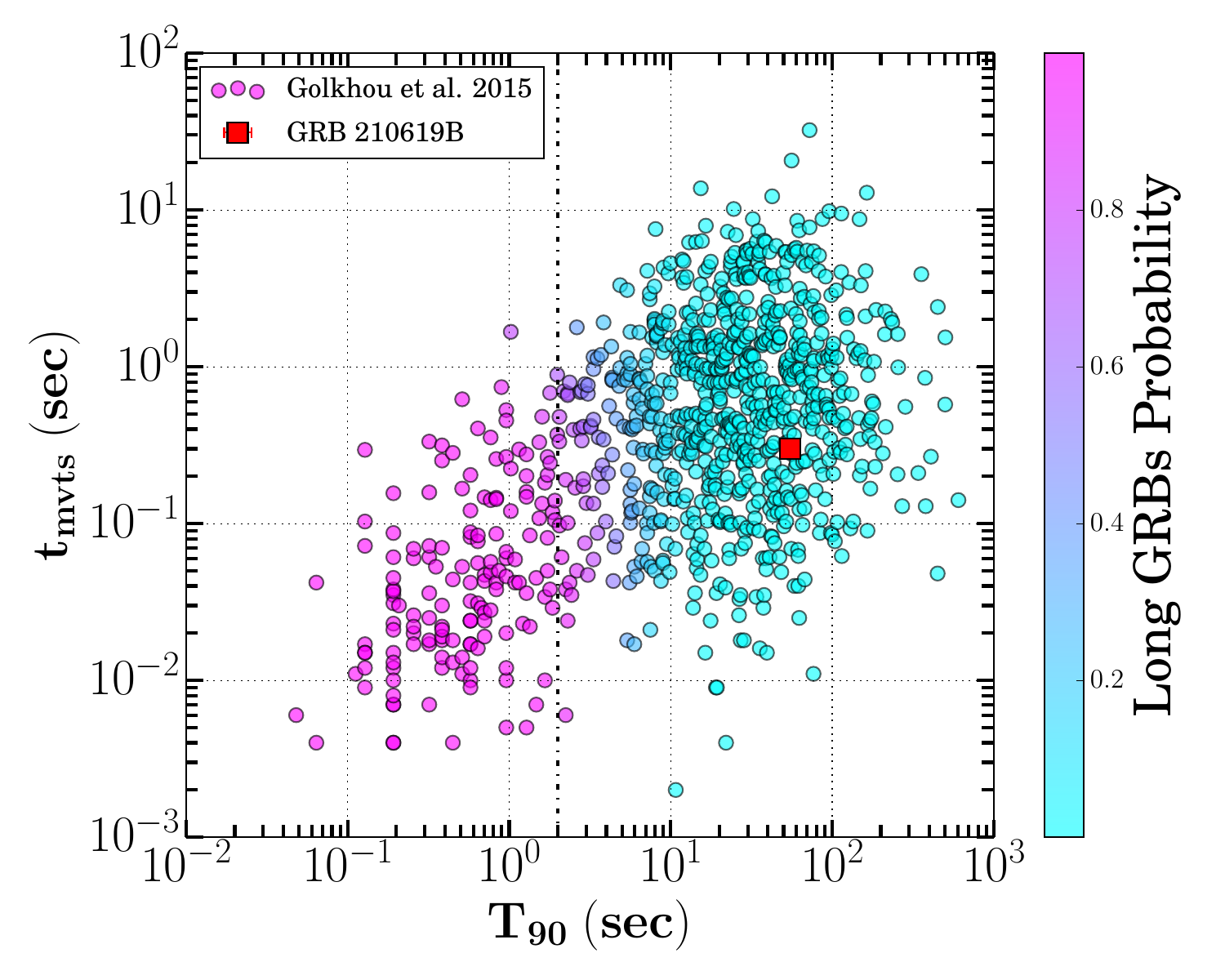}
\caption{{Left panel:} Spectral peak energy as a function of \tninty duration for \thisgrbB along with all the bursts detected by \fermi GBM.
{Right panel:} Minimum variability time scale as a function of \tninty duration for \thisgrbB along with other long and short duration GRBs taken from \protect\cite{2015ApJ...811...93G}. The right side colour bar (Y-scale) shows the probability of long GRBs in respective plots. The vertical dashed-dotted black line denotes the boundary of two GRB families.}
\label{promptproperties_GRB210619B}
\end{figure}

GRBs are mainly classified based on the prompt emission properties such as \tninty duration and hardness ratio. Long-duration GRBs have a softer spectrum in contrast to  short-duration GRBs; therefore, long and short GRBs are positioned at different places in the \tninty-hardness distribution plane of GRBs. We collected the \tninty duration and peak energy values of all the bursts detected by \fermi GBM from the GBM burst catalog\footnote{https://heasarc.gsfc.nasa.gov/W3Browse/fermi/fermigbrst.html} \citep{2014ApJS..211...12G,2016yCat..22230028B,2020ApJ...893...46V}. The time-integrated \tninty-hardness distribution for \fermi GBM detected GRBs is shown in Figure~\ref{promptproperties_GRB210619B}. The position of \thisgrbB is shown with a red square, and \thisgrbB follows the distribution of long GRBs. Furthermore, we also calculated the time-integrated hardness ratio (HR) by taking the ratio of counts in hard (50-300 \keV) and soft (10-50 \keV) energy channels. We find HR= 1.16 for \thisgrbB. Figure~\ref{promptproperties_GRB210619B} shows the \tninty-HR distribution for \thisgrbB along with other long and short duration GRBs taken from \citet{2017ApJ...848L..14G}.

{\bf \texorpdfstring{\tninty}{}-Minimum variability time}

The prompt emission light curve of GRBs is variable in nature due to being originating from internal shocks \citep{2015AdAst2015E..22P}. The minimum variability time for long (less variable) and short (more variable) GRBs follows a different distribution due to diverse compact central sources. The minimum time variability timescale, \mvts, is useful to constrain the minimum value of the bulk Lorentz factor ($\rm \Gamma_{\rm min}$) and the emission radius ($\rm R_{\rm c}$). We calculated the \mvts ($\sim$ 0.3\,s) for \thisgrbB following the continuous wavelet transforms\footnote{https://github.com/giacomov/mvts} methodology given in \cite{2018ApJ...864..163V}. The \tninty-\mvts distribution for long and short GRBs (taken from \citealt{2015ApJ...811...93G}) along with \thisgrbB are shown in Figure~\ref{promptproperties_GRB210619B} (right panel).

Furthermore, we estimated $\rm \Gamma_{\rm min}$ and $\rm R_{\rm c}$ using the following equations from \citet{2015ApJ...811...93G}:

\begin{equation}
\rm \Gamma_{\rm min} \gtrsim 110 \, \left (\frac{L_{\rm \gamma, iso}}{10^{51} \, \rm erg/s} \, \frac{1+z}{\rm t_{\rm mvts} / 0.1 \, \rm s } \right )^{1/5}
\label{gamma_min}
\end{equation}
\begin{equation}
\rm R_c \simeq 7.3 {\times} 10^{13} \, \left (\frac{L_{\rm \gamma, iso}}{10^{51} \, \rm erg/s} \right )^{2/5} \left (\frac{ t_{\rm mvts} / 0.1 \, \rm s }{1+z} \right )^{3/5} \, \rm cm.
\label{minimum_source}
\end{equation}

We calculated the minimum value of the bulk Lorentz factor and the emission radius, giving $\gtrsim$ 550 and $\simeq$ 1.87 $\times$ $10^{15}$\,cm respectively, for \thisgrbB. We also calculated the bulk Lorentz factor value ($\rm \Gamma$ = 817) for \thisgrbB using the $\Gamma_{0}$-$E_{\gamma, \rm iso}$\footnote{$\Gamma_{0}$ $\approx$ 182 $\times$ $E_{\gamma, \rm iso, 52}^{0.25 \pm 0.03}$} correlation \citep{2010ApJ...725.2209L} and noticed that the minimum value of the bulk Lorentz factor constrained using the minimum variability time scale is consistent with the value of the Lorentz factor found using the correlation. The calculated radius for \thisgrbB is much larger than the typical emission radius of the photosphere, indicating that the emission took place in an optically thin region away from the central engine \citep{2016ApJ...825...97U, 2019A&A...625A..60R, 2020NatAs...4..174B}.

{\bf Spectral lag-luminosity correlation:}
\label{lag}

\begin{figure}
\centering
\includegraphics[scale=0.3]{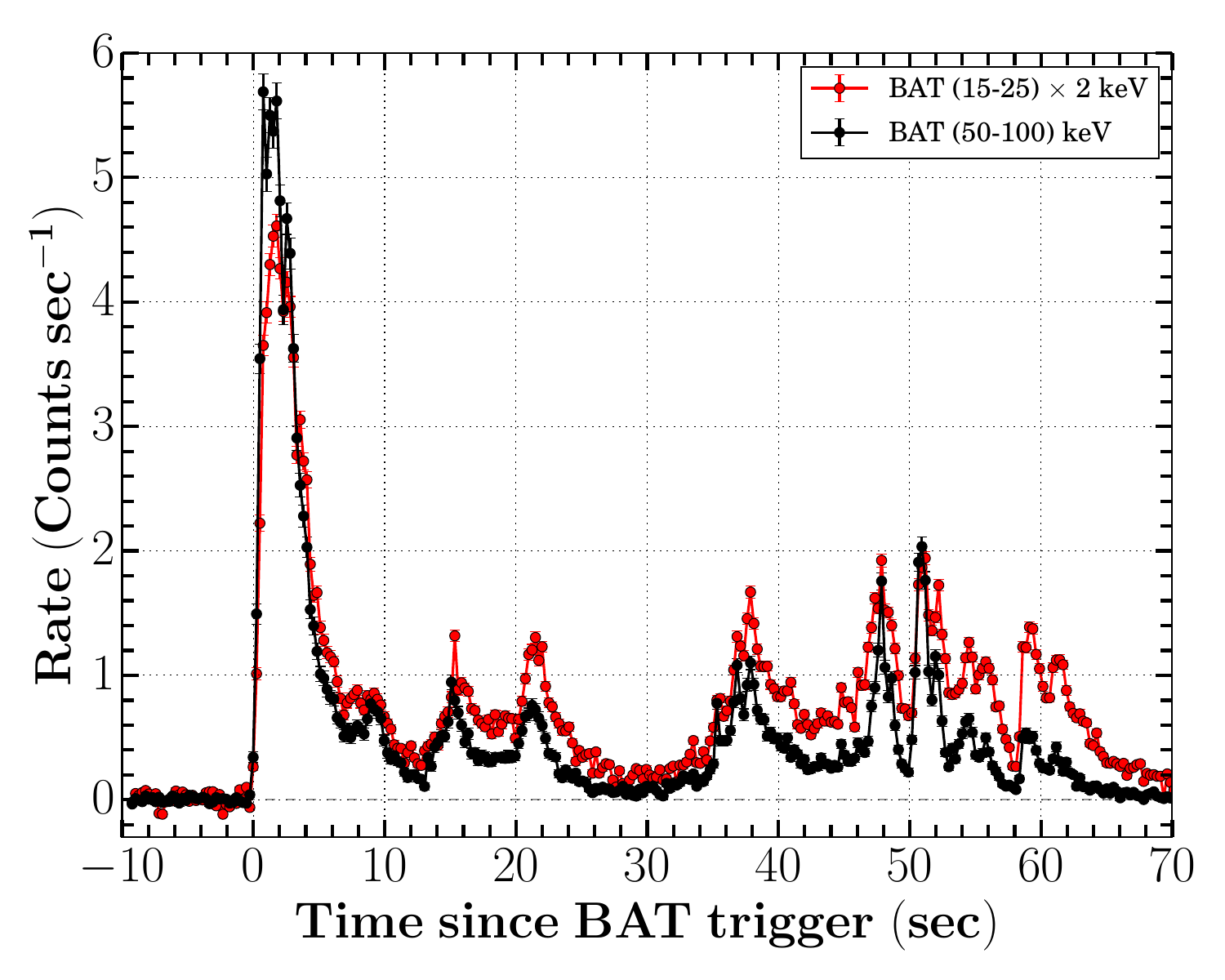}
\includegraphics[scale=0.32]{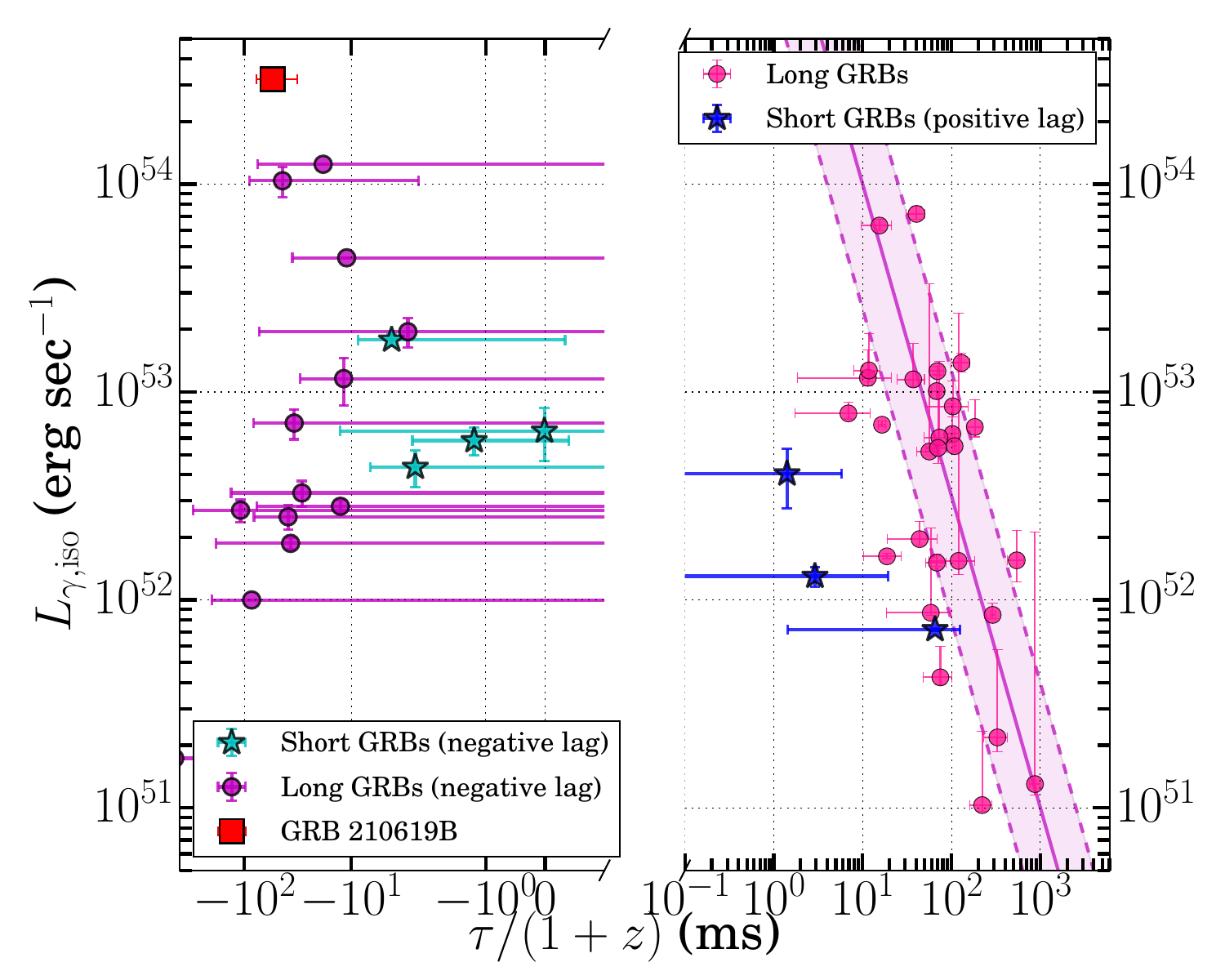}
\includegraphics[scale=0.3]{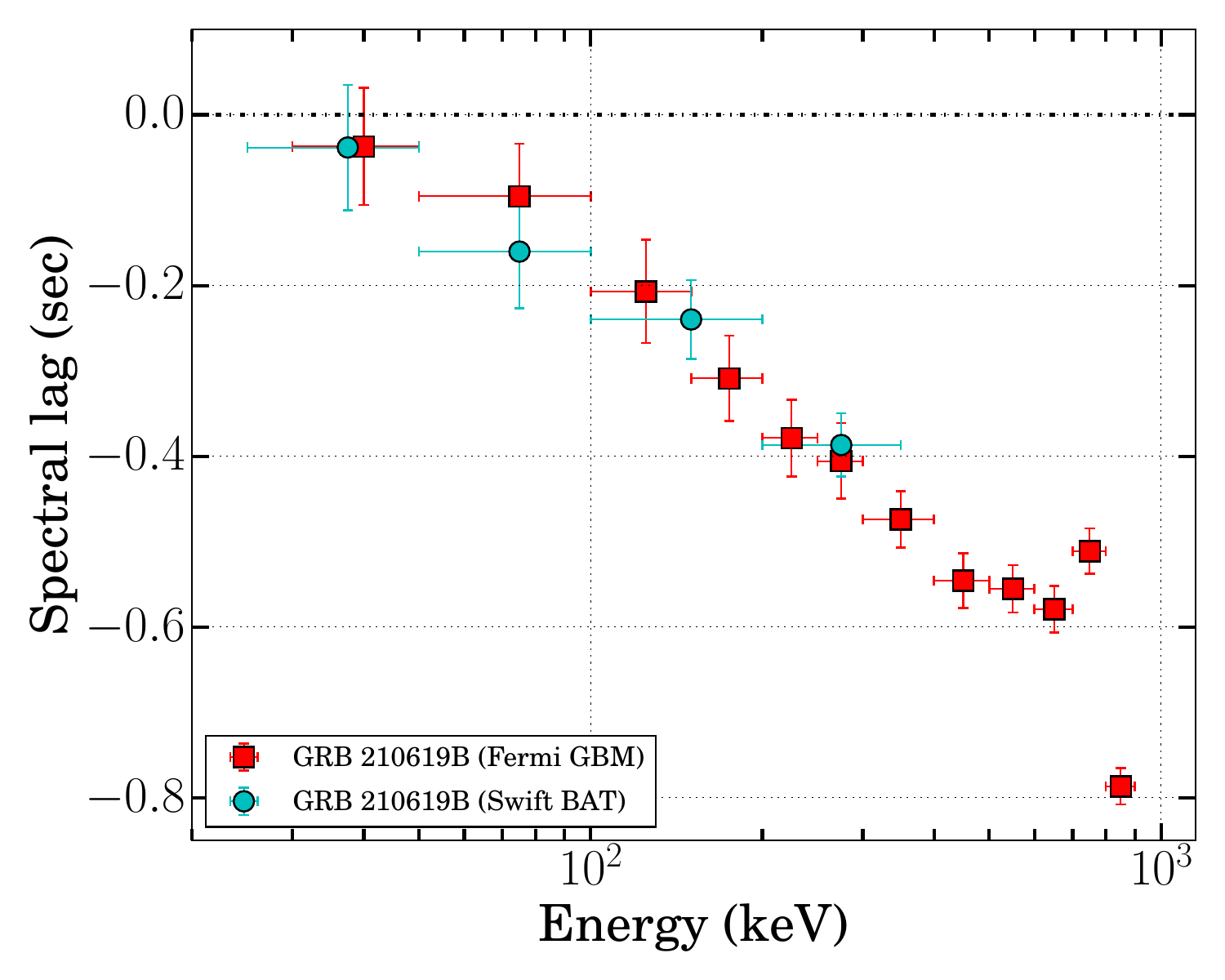}
\caption{Spectral lag of \thisgrbB: {Top panel:} \swift BAT count rate light curves in 15-25 \keV (shown with red) and 50-100 \keV (shown with black) energy range, with a bin size of 256 ms. {Middle panel:} Lag-luminosity correlation for \thisgrbB in \swift BAT 50-100 \keV and 15-25 \keV energy ranges along with other data points taken from \protect\cite{Ukwatta_2010, 2015MNRAS.446.1129B}. {Bottom panel:} Energy-resolved spectral lag analysis of \thisgrbB using \swift BAT and \fermi GBM observations.}
\label{fig:lag}
\end{figure}

The prompt emission light curves of long GRBs (short GRBs) show significant (zero) delays in two different energy ranges, and this characteristic is known as spectral lag. If the high-energy photons of GRBs come before the low-energy photons, it is defined as positive spectral lag. On the other hand, we conventionally defined the lag as negative if the low-energy photons precede the high-energy photons. The observed spectral lag is usually explained in terms of the prompt intrinsic spectral evolution (mainly temporal evolution of \Ep) or due to the curvature effect of relativistic moving shocked shells \citep{2004ApJ...614..284D, 2016ApJ...825...97U}. \cite{2000ApJ...534..248N} reported the anti-correlation between the spectral lag and isotropic peak luminosity of GRBs using a limited long bursts sample (with known redshift) observed using the {\it BATSE} mission.

In the case of \thisgrbB, we initially estimated the time-integrated (\fermiT to \fermiT+67.38\,s) spectral/timing lag for the \swift BAT light curves (see Figure~\ref{fig:lag}) in two different energy ranges (15-25 \keV and 50-100 \keV) using the cross-correlation function \citep[CCF;][]{2000ApJ...534..248N}. We followed the detailed methodology presented in \cite{2015MNRAS.446.1129B}. We have used \sw{emcee} package \citep{2013PASP..125..306F} to fit the cross-correlation function using an asymmetric Gaussian function. We obtained a negative spectral lag = -160$^{+67}_{-66}$\,ms and placed it in the anti-correlation relationship between the spectral lag and isotropic peak luminosity of GRBs (data taken from \citealt{Ukwatta_2010} in the same energy range). We noticed that \thisgrbB is an outlier of the lag-luminosity anti-correlation relationship (see middle panel of Figure~\ref{fig:lag}). Furthermore, we examined the literature and found that many long GRBs (GRB 060814, GRB 061021, GRB 070306, GRB 080721, GRB 080804, GRB 090426C, GRB 100728B, GRB 110205A, GRB 140102A, GRB 150213A, and many others) are reported with such a large value ($>$ -100\,ms, although with large error bars) for the negative spectral lag \citep{2015MNRAS.446.1129B, 2018JHEAp..18...15C, 2021MNRAS.505.4086G}. \cite{2015MNRAS.446.1129B} studied a larger sample of long and short GRBs and noticed that the bursts with high luminosity might have smaller lags. \cite{2014AstL...40..235M} proposed that the GRBs independent emission pulses exhibit hard-to-soft spectral evolution of the peak energy, suggesting a positive spectral lag \citep{2018ApJ...869..100U}. However, a strong/complex spectral evolution of the peak energy could result in superposition effects, and this superposition effect could explain the negative spectral lag observed for \thisgrbB.

In addition, we also performed the energy-resolved spectral lag analysis (\fermiT to \fermiT+67.38\,s) of \thisgrbB using \swift BAT and \fermi GBM observations. We considered the 15-25 \keV and 8-30 \keV light curves as a reference to calculate the lag using \swift BAT and \fermi GBM respectively. We noticed that the energy-resolved lag analysis also shows the negative lag, suggesting that the low-energy photons precede the high-energy photons for \thisgrbB. Figure~\ref{fig:lag} (bottom panel) shows the spectral evolution of the lags for \thisgrbB. The energy-resolved spectral lags obtained using BAT, and GBM data are tabulated in Table \ref{tab:spectral lag_GRB210619B} of the appendix.

\subsubsection{Spectral evolution}

The prompt emission is believed to consist of various spectral components, the combination of which provides the shape of the observed spectrum. To constrain these components for \thisgrbB, we performed the time-resolved spectral analysis (see section \ref{GBM}). Figure~\ref{delta_DIC} shows the difference of deviation information criterion values from \sw{Band} and \sw{Cutoff-power law} functions for each of the time-resolved bins. The comparison of differences in the DIC values for each bin suggests that most of the bins (except five) support the traditional \sw{Band} model. Figure \ref{parameter_dist} shows the distribution of spectral parameters (\Ep, $\alpha_{\rm pt}$, $\beta_{\rm pt}$, and flux) obtained using the \sw{Band} model. We have also shown the kernel density estimation (KDE) for each of the parameters. The averaged and standard deviation values of individual spectral parameter distributions are given in Table \ref{tab:mean_std}.

\begin{table*}
\begin{scriptsize}
\begin{center}
\caption{The averaged and standard deviation of individual spectral parameters distributions for \thisgrbB.}
\label{tab:mean_std}
\begin{tabular}{ccccc} \hline 
\textbf{Model} & \multicolumn{4}{c}{\textbf{Spectral parameters}}  \\ \hline
 & $\bf \alpha_{\rm \bf pt}$ & \bf \Ep /$\bf E_{\it c}$ (\keV) & $\bf \beta_{\rm \bf pt}$ & Flux ($\times 10^{-05}$\,${\rm erg}\,{\rm s}^{-1}$)\\ \hline
\sw{Band}&$-0.69 \pm 0.22$&$183.47 \pm 113.90$&$-2.43 \pm 0.28$& 2.37 $\pm$ 3.88 \\
\sw{CPL}&$-0.88 \pm 0.20$&$214.52 \pm 128.63$&-& 1.45 $\pm$ 2.34 \\\hline
\end{tabular}
\end{center}
\end{scriptsize}
\end{table*}

\begin{figure}[ht!]
\centering
\includegraphics[scale=0.32]{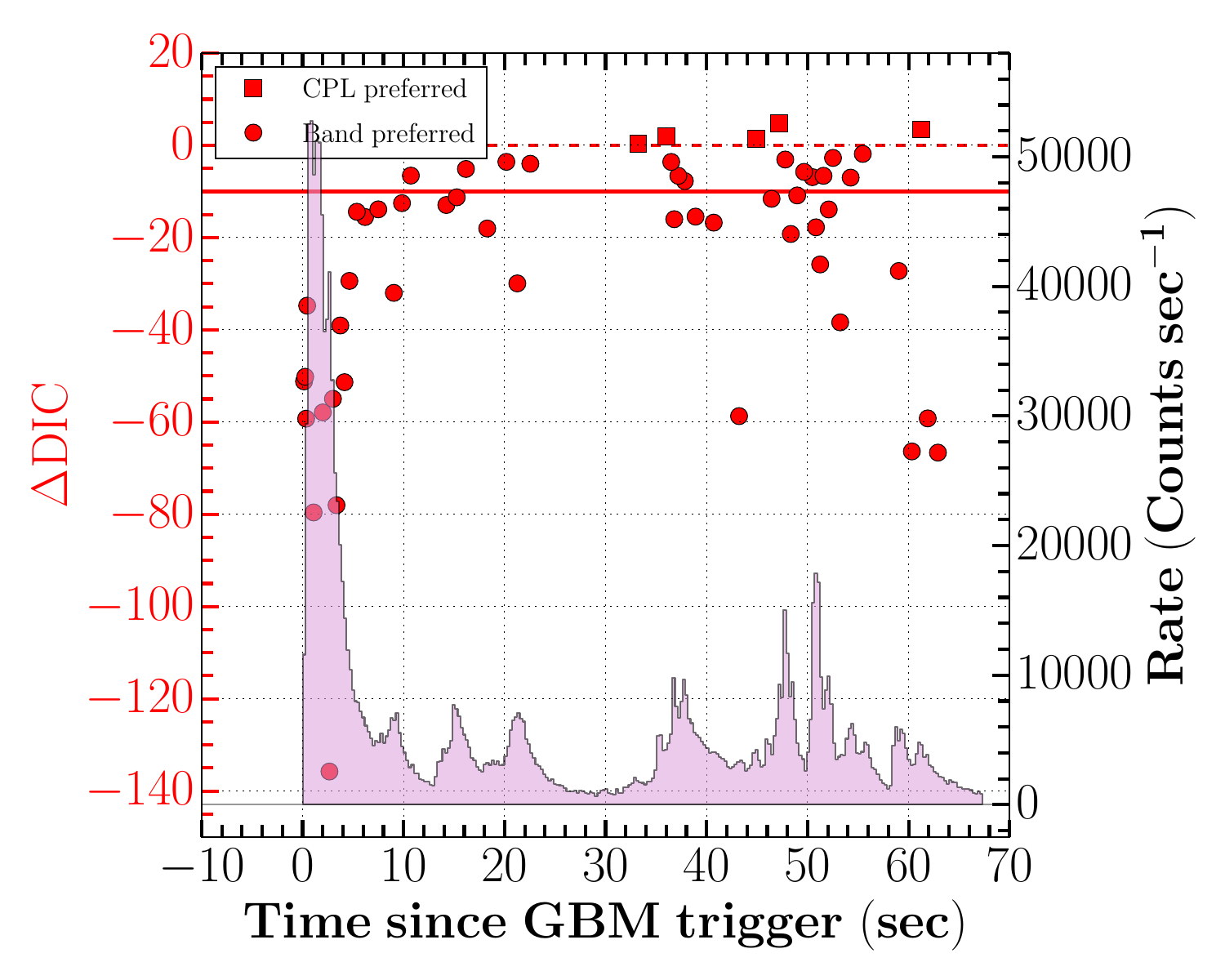}
\caption{The difference between DIC values of \sw{Band} and \sw{Cutoff-power law} models for each of the time-resolved bins obtained using Bayesian block analysis (with S $\geq$ 30). The horizontal red dashed, and solid red lines show the difference between DIC values equal zero and -10, respectively.}
\label{delta_DIC}
\end{figure}

\begin{figure*}[ht!]
    \centering
    \includegraphics[scale=0.4]{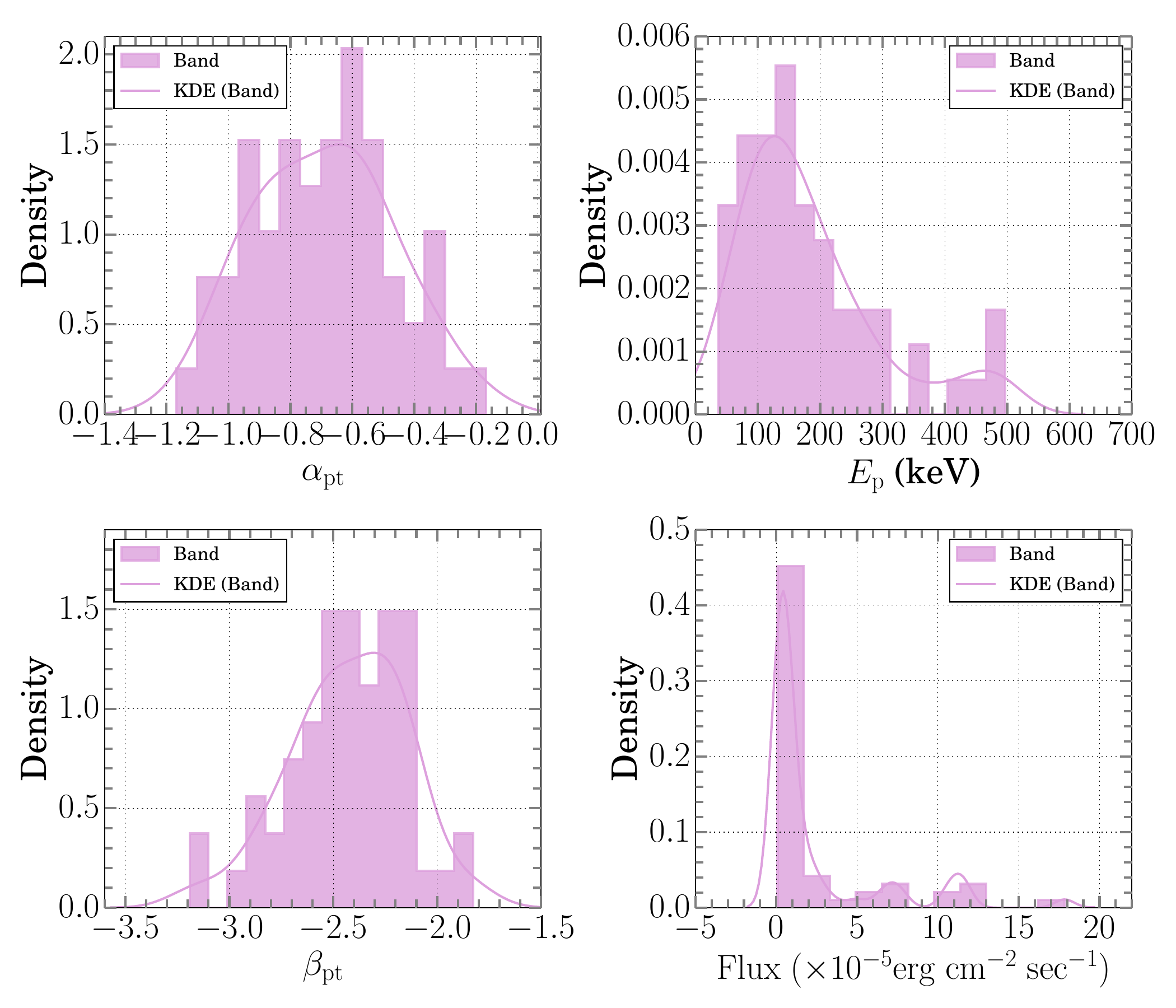}
    \caption{The spectral parameters ($\alpha_{\rm pt}$, \Ep, $\beta_{\rm pt}$, and flux) distributions for \thisgrbB obtained using the time-resolved spectral analysis of \fermi GBM data using the \sw{Band} model. The plum curves show the kernel density estimation (KDE) of the distributions of the respective parameters fitted using the \sw{Band} model.}
    \label{parameter_dist}
\end{figure*}

\begin{figure}
\centering
\includegraphics[scale=0.37]{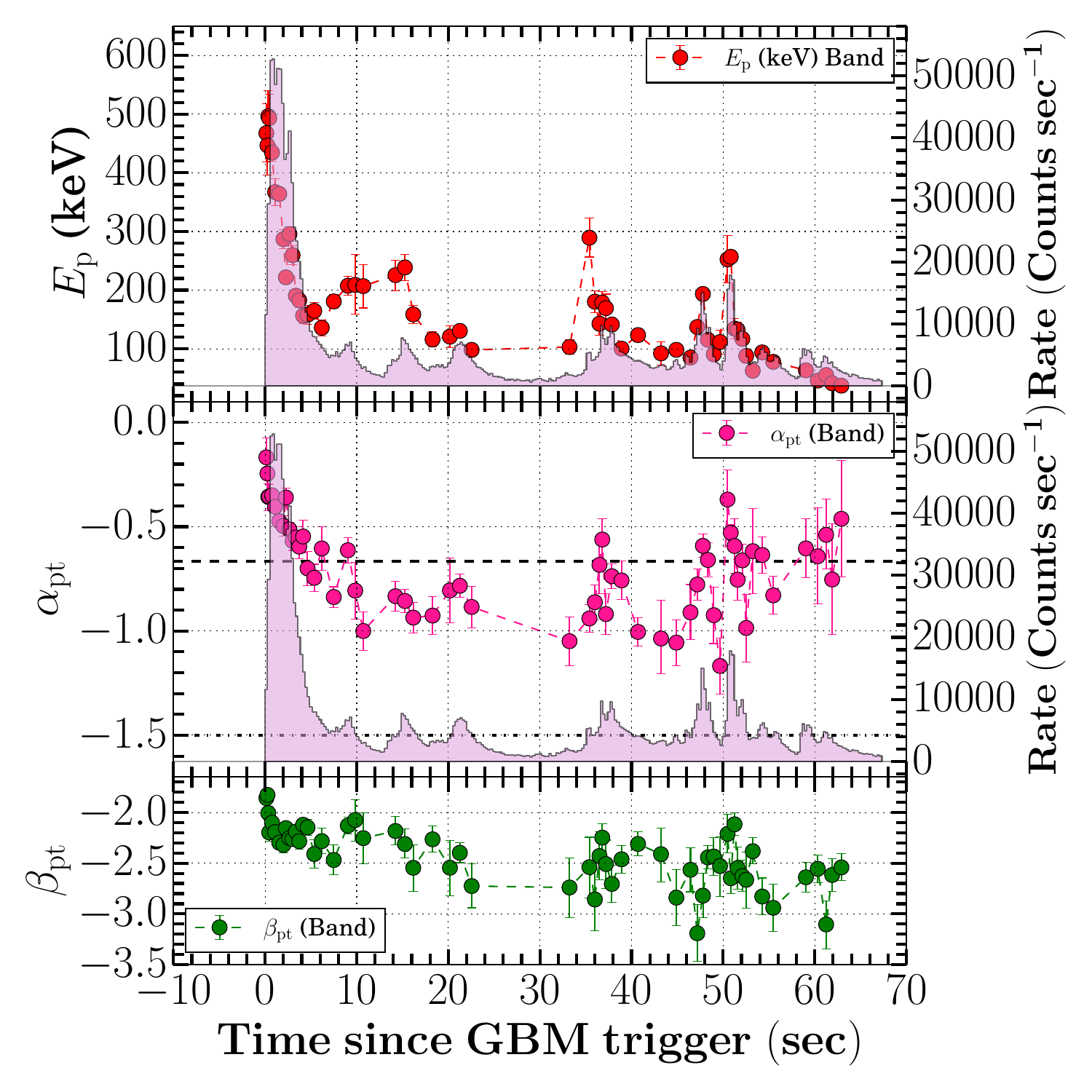}
\caption{Temporal evolution of spectral parameters obtained using time-resolved spectral modelling with the \sw{Band} model: {Top panel:} Evolution of the peak energy and it tracks the intensity of the burst. {Medium panel:}  Evolution of the low-energy spectral index. The horizontal dashed and dotted-dashed lines represent the synchrotron line of death ($\alpha_{\rm pt}$ = -2/3) and the synchrotron fast cooling line ($\alpha_{\rm pt}$ = -3/2). {Bottom panel:} Evolution of high-energy spectral index obtained using the \sw{Band} spectral model.}
\label{TRS_evolution_band}
\end{figure}

The prompt emission spectral parameters of GRBs exhibit a strong spectral evolution and are useful to understand the radiation mechanisms and jet composition of GRBs. Figure~\ref{TRS_evolution_band} shows the evolution of spectral parameters (\Ep, $\alpha_{\rm pt}$, $\beta_{\rm pt}$, and flux) for \thisgrbB obtained using the 
the \sw{Band} model. In the Figure, the evolution of spectral parameters is over-plotted on the count-rate prompt emission light curve of \thisgrbB. We noticed that the evolution of observed \Ep exhibits an intensity tracking pattern for \thisgrbB. The observed negative spectral lag (see section \ref{lag}) and intensity tracking the behaviour of \Ep for \thisgrbB is consistent with the prediction of a connection between the spectral lag and spectral evolution \citep{2018ApJ...869..100U}. Moreover, the low-energy index also exhibits an intensity-tracking trend, suggesting the double-tracking characteristics seen in a few cases of GRBs. Further, we examined the overall spectral parameters correlations. 

\begin{table*}
\caption{The correlation between the spectral parameters obtained using the time-resolved spectral analysis of \thisgrbB. We have used Pearson correlation to calculate the correlation strengths (Pearson correlation coefficient, r) and the probability of a null hypothesis (p) for individual parameters correlation.}
\label{tab:correlation}
\centering
\begin{tabular}{ccccccc} \hline 
\textbf{Model} & \multicolumn{2}{c|}{\textbf{log (Flux)-log (\Ep)}} & \multicolumn{2}{c|}{\textbf{log (Flux)-$\alpha_{\rm pt}$}} & \multicolumn{2}{c}{\textbf{log (\Ep)-$\alpha_{\rm pt}$}} \\ \hline
 & \bf r & \bf p &  \bf r & \bf p &  \bf r & \bf p  \\ \hline
\sw{Band}& 0.87 & 2.07 $\times 10^{-19}$& 0.75 & 1.10 $\times 10^{-11}$& 0.47 & 1.85 $\times 10^{-4}$ \\ \hline
\end{tabular}
\end{table*}

We studied the correlation between the following spectral parameters: (a) log (Flux)-log (\Ep), (b) log (Flux)-$\alpha_{\rm pt}$, and (c) log (\Ep)-$\alpha_{\rm pt}$. We have utilized Pearson correlation to calculate the correlation strengths (Pearson correlation coefficient, r) and the probability of a null hypothesis (p) for individual parameters correlation \footnote{The null hypothesis probability (p-value) we mention here is the probability of the correlation occurring by chance, i.e. with respect to a constant relationship. If the p-value of a statistical test is small enough we can say that the null hypothesis (i.e. constant) is false and the correlation is a good model.}. We have noticed strong correlations of (a) log (Flux)-log (\Ep) and (b) log (Flux)-$\alpha_{\rm pt}$. On the other hand, log (\Ep)-$\alpha_{\rm pt}$ shows a moderate degree of correlation (see Table \ref{tab:correlation}). Figure \ref{TRS_correlation_band} shows the correlations between the spectral parameters obtained using the time-resolved spectral modelling with the \sw{Band} model. The time-resolved spectral analysis (using ASIM data) of the main bright pulse observed with ASIM shows evidence for hard-to-soft spectral evolution (softening when brightening behaviour), which is significant at 95\% confidence level for the first two (three) time intervals (see Table \ref{tab:asim} in the appendix). This characteristic is consistent with \fermi time-resolved spectral analysis results during the main bright pulse; therefore it shows agreement between the two observations.

\begin{figure}
\centering
\includegraphics[scale=0.32]{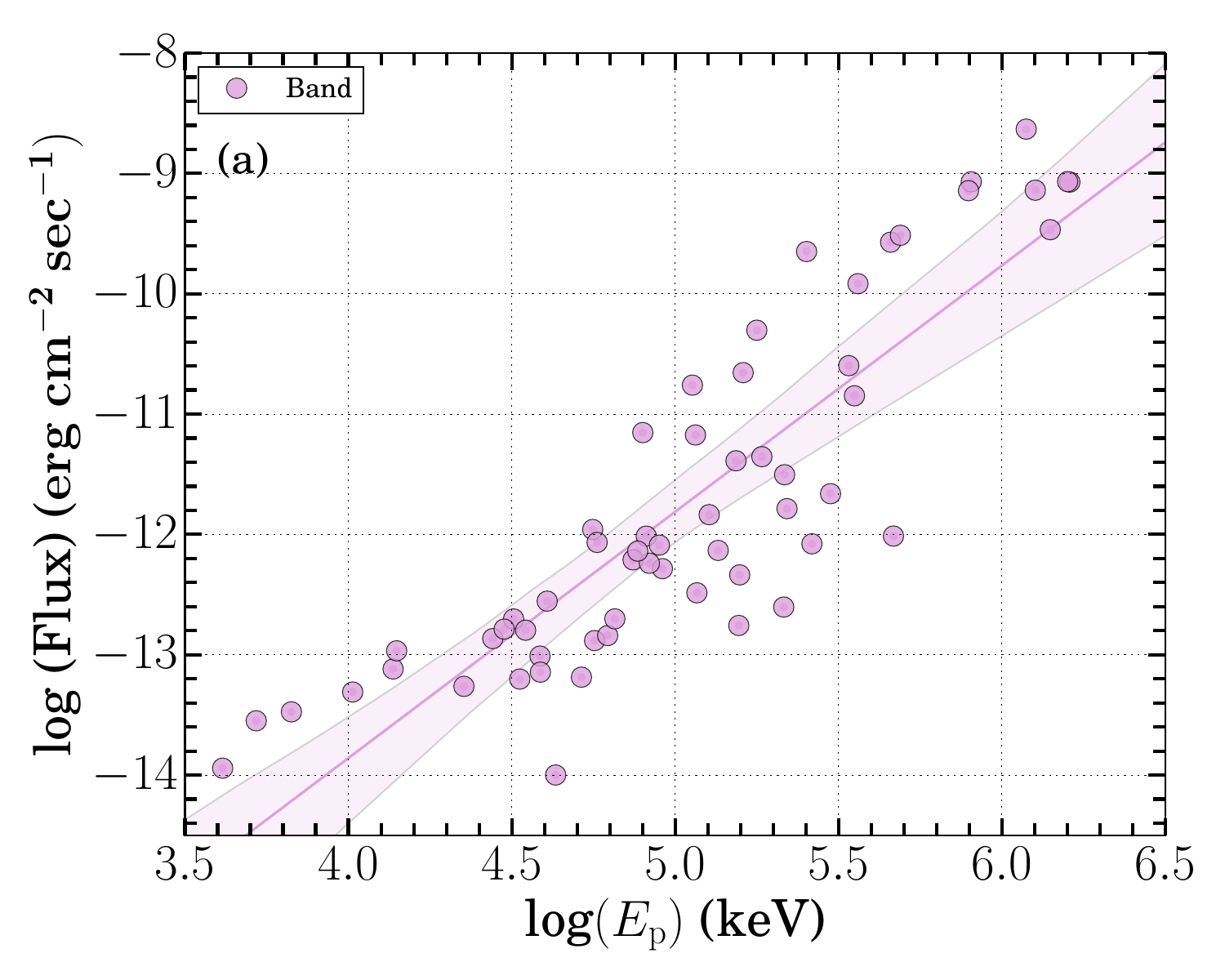}
\includegraphics[scale=0.32]{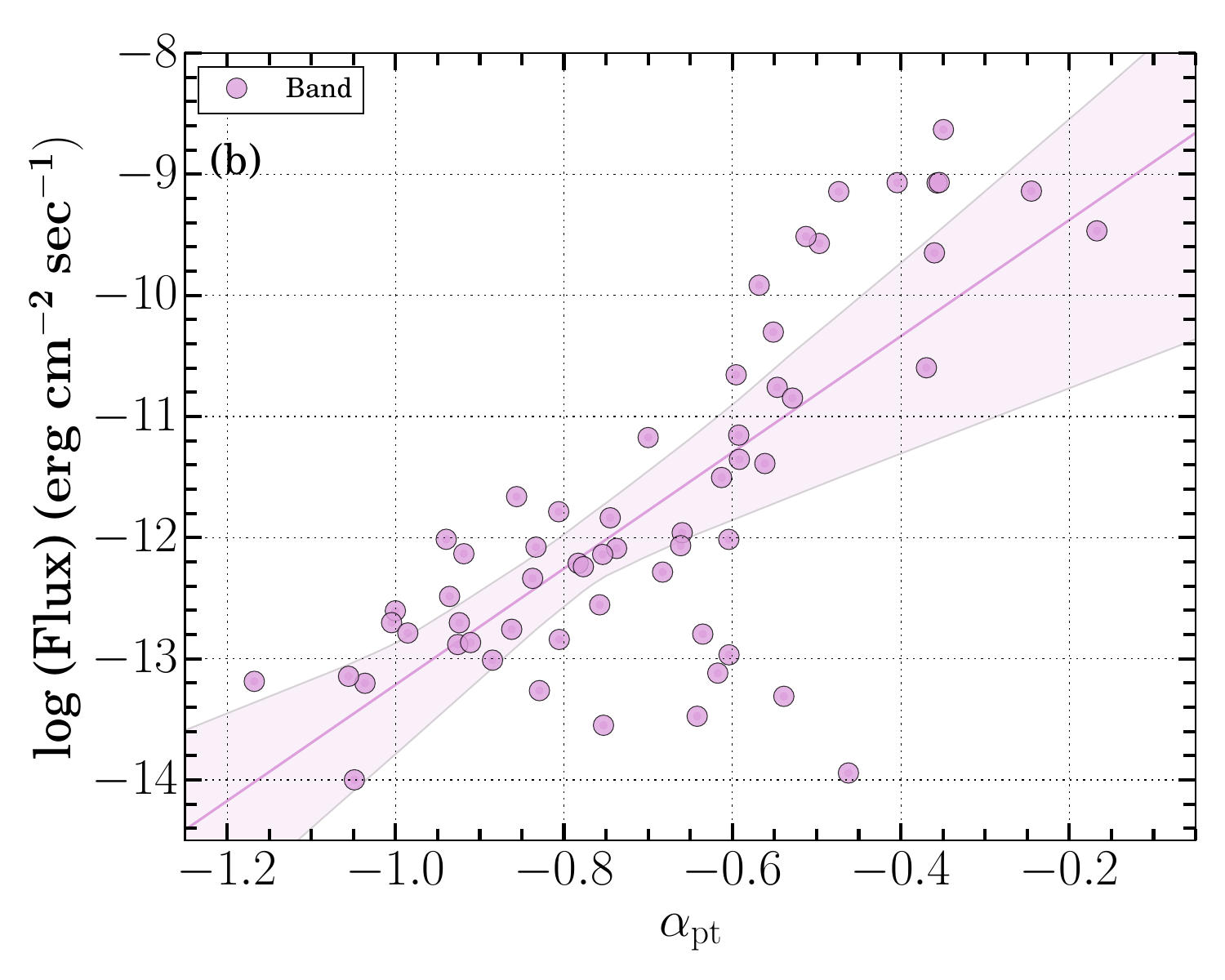}
\includegraphics[scale=0.32]{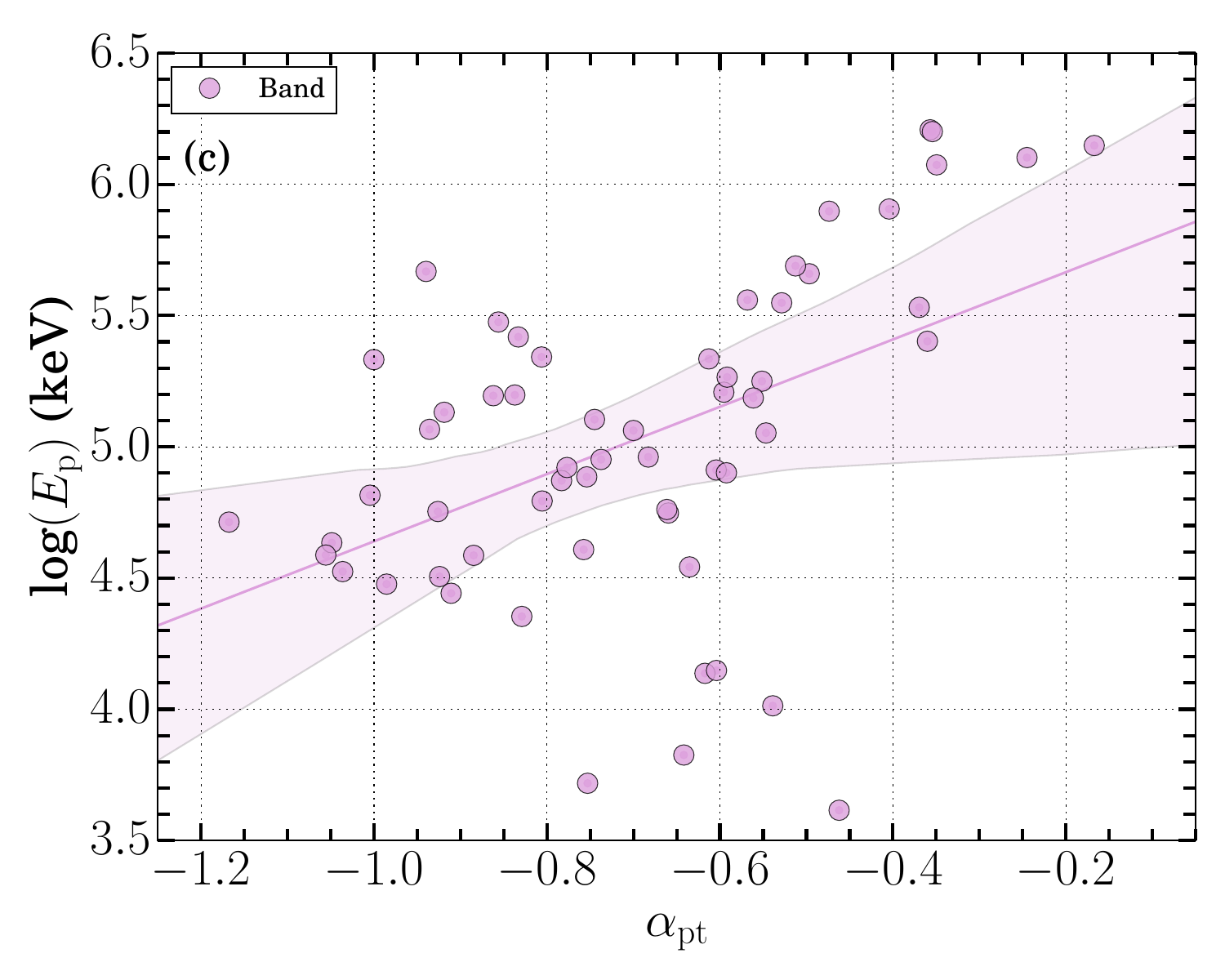}
\caption{{Correlations between the spectral parameters obtained using time-resolved spectral modelling with the \sw{Band} model:} (a) Correlation between the peak energy and flux. (b) Correlation between the low-energy spectral index and flux. (c) Correlation between the peak energy and the low-energy spectral index.}
\label{TRS_correlation_band}
\end{figure}

{\bf Comparison of spectral parameters with a larger sample:}

We collected the spectral parameters of single and multi-episodic bursts and compared them with those of \thisgrbB. Figure~\ref{TRS_comparison} (top panel) shows the distribution of time-resolved peak energy as a function of energy flux of \thisgrbB along with a large sample of GRBs. We noticed that the mean values of energy flux from \thisgrbB are significantly larger than the mean value of similar multi-pulsed GRBs; however, the mean values of the peak energy from \thisgrbB are softer in comparison with the mean values of multi-pulsed GRBs. Figure~\ref{TRS_comparison} (middle panel) shows the distribution of time-resolved low-energy photon indices as a function of energy flux of \thisgrbB along with a large sample of GRBs. The $\alpha_{\rm pt}$ values of \thisgrbB exceed the synchrotron LOD during the brighter phase. We also found that some of the single and multi-pulsed GRBs even exceed the low-energy photon index, predicted from jitter radiation ($\alpha_{\rm pt}$ = +0.5; \citealt{2000ApJ...540..704M}). Figure~\ref{TRS_comparison} (bottom panel) shows the distribution of time-resolved peak energy as a function of $\alpha_{\rm pt}$ of \thisgrbB along with a large sample of GRBs. 
\par
Furthermore, we also compared the spectral parameters of \thisgrbB with those bursts (GRB 131231A, GRB 140102A, GRB 190530A, and others) having ``Double-tracking" evolution pattern of \Ep and $\alpha_{\rm pt}$ \citep{2019ApJ...884..109L, 2021MNRAS.505.4086G, 2022MNRAS.511.1694G}. We noticed that some bursts are consistent with the non-thermal thin shell synchrotron emission model, however, some of them required a hybrid non-thermal synchrotron plus thermal photospheric emission. This suggests that the observed ``Double-tracking" evolution characteristics could be independent of the emission mechanisms of GRBs, though many more such bright bursts are needed to be analysed and physically modelled to confirm this.   

\begin{figure}[ht!]
\centering
\includegraphics[scale=0.29]{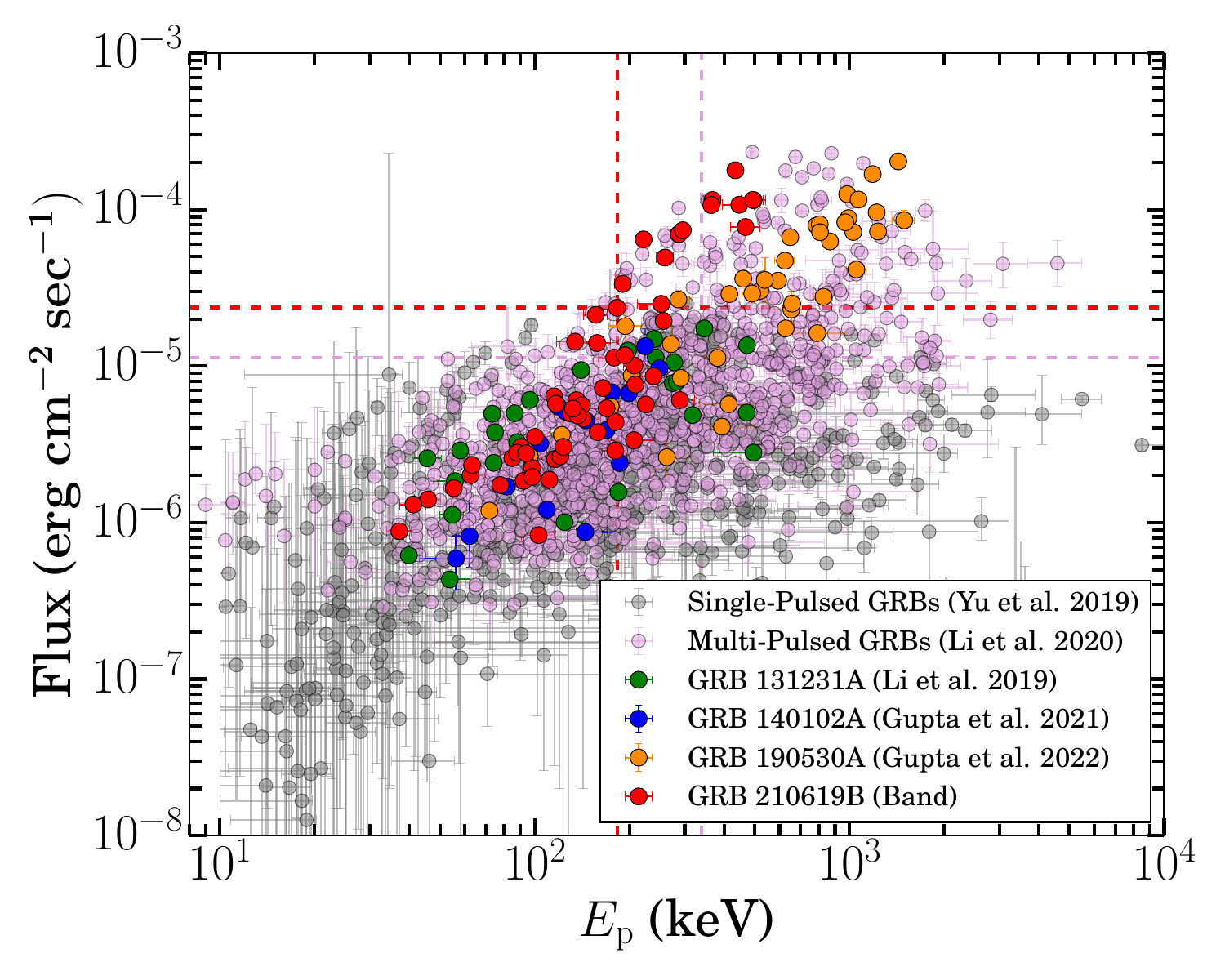}
\includegraphics[scale=0.29]{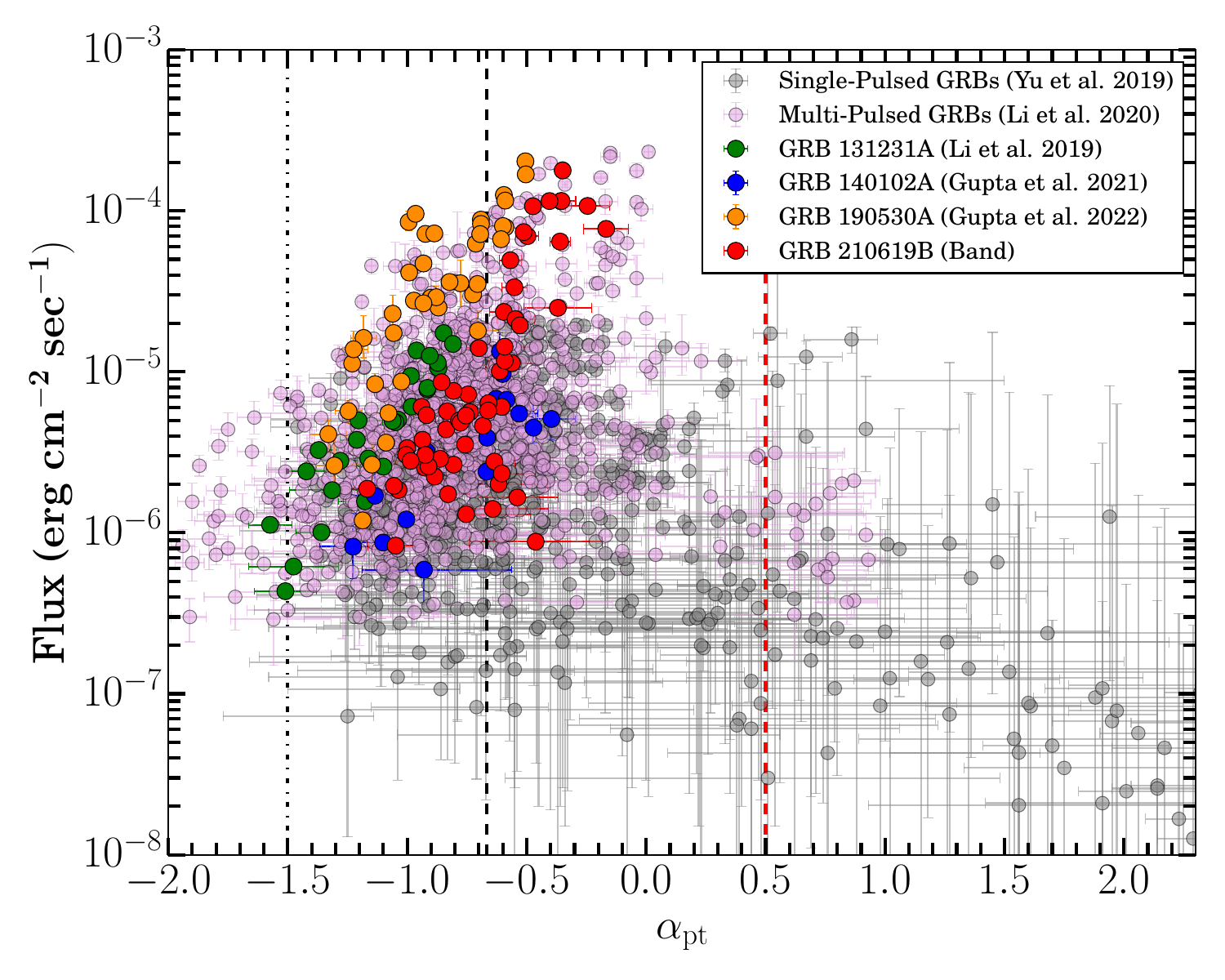}
\includegraphics[scale=0.29]{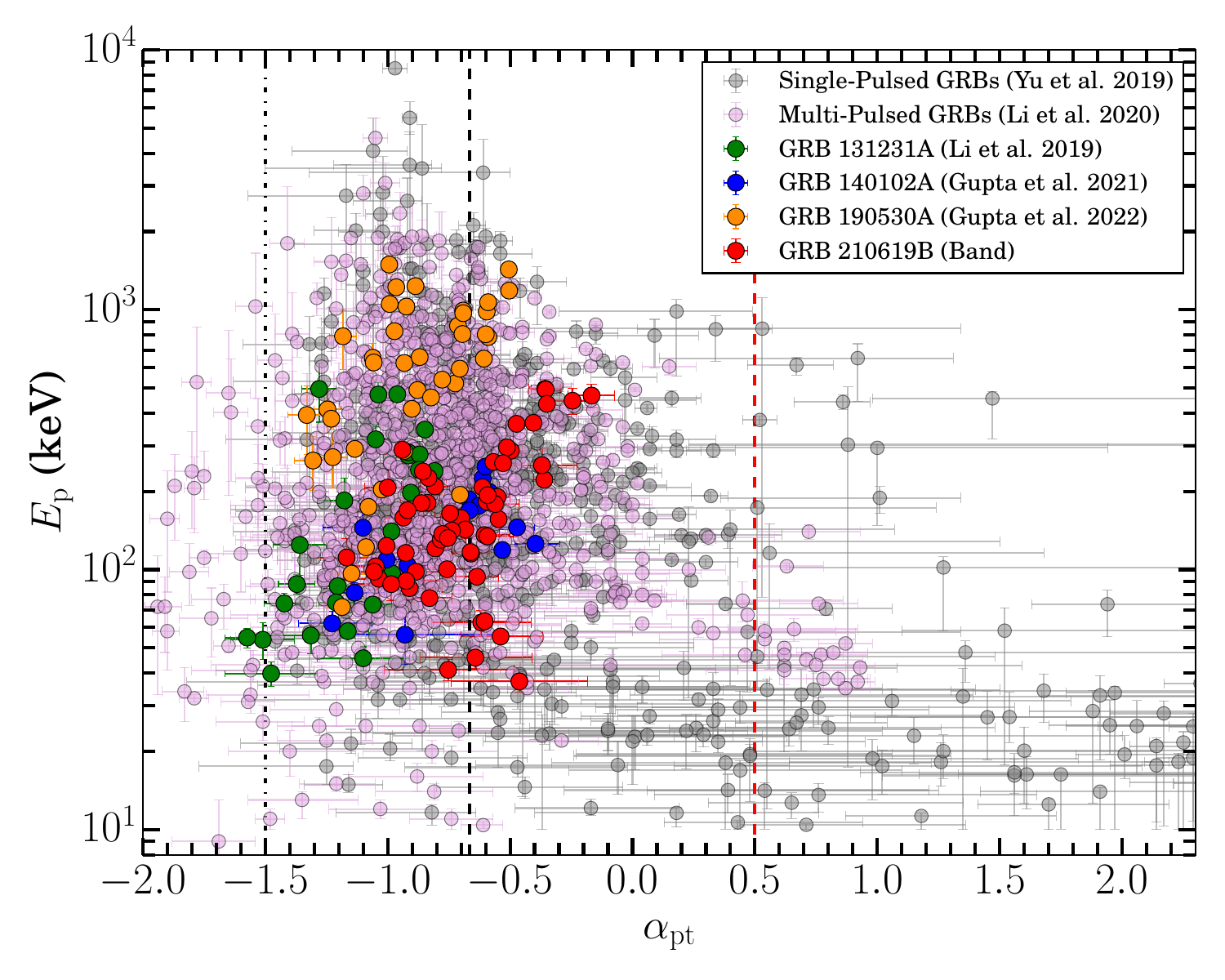}
\caption{{Comparison between the spectral parameters:} {Top panel:} Comparison between the peak energy and energy flux obtained using time-resolved spectral modelling with the 
\sw{Band} model for \thisgrbB with a well-studied sample of single, multi-pulsed, and Double tracking bursts. The vertical/horizontal pink and plum lines show the mean values of \Ep/flux of \thisgrbB and the sample of multi-pulsed GRBs, respectively. {Middle panel:} Comparison between the low-energy power-law index and energy flux. {Bottom panel:} Comparison between the low energy power-law index and peak energy. The vertical black dotted-dashed, black dashed, and red dashed lines show the synchrotron's fast cooling, slow cooling, and jitter radiation.}
\label{TRS_comparison}
\end{figure}

\subsubsection{Prompt emission mechanism and jet composition}

Regarding the observed prompt emission, the spectrum was explored to find the possible radiation mechanism of GRBs. There are mainly two possibilities, non-thermal thin shell synchrotron emission or photospheric thermal emission \citep{2015AdAst2015E..22P, 2020NatAs...4..210Z}. The non-thermal thin shell synchrotron emission suggests that the low-energy photon indices could not be harder than the -2/3 value, also known as the ``synchrotron line of death (LOD)”. However, in the case of \thisgrbB, the evolution of the low-energy photon indices exceeds the synchrotron LOD during the main bright/hard pulse, and during the softer/longer pulses, the low-energy photon indices become softer and remain consistent with this limit (see Figure~\ref{TRS_evolution_band}). The observed hard values of the low-energy photon indices (during the main pulse emission) suggest that the observed prompt emission spectrum of \thisgrbB is inconsistent with the non-thermal thin shell synchrotron emission model (both in slow and fast cooling cases). This hard $\alpha_{\rm pt}$ also suggests that our viewing angle for the burst is not significantly off the burst emission axis and that the central engine has been long-time active. The hard $\alpha_{\rm pt}$ could be explained using thermal ``photospheric emission" \citep{2015AdAst2015E..22P, 2018JApA...39...75I}. 
\par
On the other hand, the softer values of the low-energy photon indices during the longer emission phase are consistent with the non-thermal thin shell synchrotron emission model. This suggests that the radiation process responsible for \thisgrbB shows a transition between photospheric thermal emission (hard $\alpha_{\rm pt}$) and non-thermal synchrotron emission (soft $\alpha_{\rm pt}$). This further supports a transition in the jet composition of \thisgrbB, between a matter-dominated hot fireball \citep{1992Natur.357..472U, 2015AdAst2015E..22P} to Poynting flux dominated outflow \citep{1997ApJ...482L..29M, 2001MNRAS.321..177L}. Such a transition has been observed only for a few bursts \citep{2018NatAs...2...69Z, 2019ApJS..242...16L}. However, this is not the only possible scenario to explain the above observations in the spectral evolution of \thisgrbB. A few complex theoretical models have been proposed that can produce a hard $\alpha_{\rm pt}$ (as observed during the main emission phase of \thisgrbB) within the framework of synchrotron emission model, such as synchrotron emission in decaying magnetic field \citep{2006ApJ...653..454P, 2014NatPh..10..351U}, time-dependent cooling of synchrotron electrons \citep{2018MNRAS.476.1785B, 2020NatAs...4..174B}, etc.
\par
In addition, the non-thermal synchrotron emission model might have a narrow spectral width (like the Planck function) similar to the case of the photosphere model as for example GRB 081110A \citep{2018JApA...39...75I}, results in a harder $\alpha_{\rm pt}$. On the other hand, a soft $\alpha_{\rm pt}$ (as observed during the longer emission phase of \thisgrbB) can be produced within the framework of the photospheric emission model. The photosphere model does not always look like the Planck function (a narrow spectral width) but may also have a similar shape to the case of non-thermal synchrotron emission from the region above the photosphere \citep{2005ApJ...633.1018P, 2005ApJ...628..847R, 2015MNRAS.454L..31A, 2019MNRAS.485..474A}, for example, GRB 090902B \citep{2010ApJ...709L.172R, 2011MNRAS.415.3693R} and GRB 110920A \citep{2015MNRAS.450.1651I}. In some cases (for example GRB110920A, \citealt{2018JApA...39...75I}), it is shown that two different models having different shapes equally fit the same observed spectrum. These arguments suggest that many times there is a degeneracy between the spectral models. Therefore spectroscopy alone may not be sufficient to test the viability of the proposed models. The limitation of spectral fitting can be resolved using prompt emission polarization measurements \citep{2022JApA...43...37I, 2022MNRAS.511.1694G}. Nevertheless, such a work is out of the scope of this paper.

\section{Summary \& Conclusion}
\label{conclusion_190530A}

We studied the temporal, spectral, and polarization characteristics of the prompt emission of \thisgrb using \fermi and \AstroSat CZTI observations. \thisgrb (the sixth brightest burst ever observed by GBM) consists of three peaks with increasing hardness ratio. We noticed that the time-averaged spectrum (\fermiT to \fermiT+25 s) has a peculiar low-energy break in addition to the typical \Ep break. Such a low-energy break in addition to \Ep has only been seen in a few of the brightest GBM detected long bursts \citep{2019A&A...625A..60R, 2017ApJ...846..137O, 2018A&A...616A.138O, 2019A&A...628A..59O}. We performed a time-resolved analysis based on coarse (constant cadence) and fine bins (Bayesian algorithm) techniques to study the spectral evolution and search for the low-energy spectral break. Low-energy breaks were detected in some of the time-resolved bins with mean photon indices $<\alpha_{1}>$ = 0.84 (with $\sigma$ = 0.04) and $<\alpha_{2}>$ = 1.43 (with $\sigma$ = 0.06), consistent with the power-law indices expected by synchrotron emission in a marginally fast cooling spectral regime. Taking the low-energy break as due to the synchrotron cooling frequency, we constrain a limit on the co-moving magnetic field (B) following equation 8 of \cite{2018A&A...613A..16R}. We calculated B $\leq$ 9 gauss for \thisgrb. However, this value is small and not consistent with the expected value for a typical emitting region located at $\sim$ 10$^{14}$ cm \citep{2018A&A...613A..16R}.

In addition, we also found interesting spectral evolution within the \sw{Band} spectral parameters obtained using detailed time-resolved spectroscopy. The spectral evolution of \Ep tracks the intensity of the GBM light curve and exhibits a strong correlation. Usually, the $\alpha_{\rm pt}$ evolution does not have any particular trend, but for \thisgrb, we found that it also tracks the intensity of the burst; therefore, \thisgrb exhibits characteristics of a double-tracking burst. So far, this tracking behaviour has only been found in a few GRBs, i.e. in GRB 131231A \citep{2019ApJ...884..109L} and GRB 140102A \citep{2021MNRAS.505.4086G}, the low energy spectral index remains in the synchrotron limit ($\alpha_{pt} =-2/3$). Similarly, in the case of \thisgrb, $\alpha_{\rm pt}$ values are within the synchrotron limits for the first two pulses. However, during the third and the brightest pulse, $\alpha_{\rm pt}$ values become harder and exceed the synchrotron line of death in a few bins. During this temporal window, we found a signature of a thermal component along with a synchrotron one in our time-resolved spectral analysis, suggesting an additional contribution from the photosphere.

For \thisgrb, we found a hint of high polarization fraction in our time-integrated (55.43 $\pm$ 21.30 \%; 2.60 $\sigma$) as well as time-resolved (53.95 $\pm$ 24.13 \%; 2.24 $\sigma$ for the third pulse) polarization measurements in the 100-300 \keV energy range, based on our observations with \AstroSat CZTI. We investigated the origin of a high degree of polarization fraction and found that a synchrotron model with an ordered magnetic field could explain such a high polarization fraction. Our time-resolved polarization analysis does not show any substantial variation in the polarization fraction or angles. Based on our detailed spectro-polarimetric analysis, we suggest that the first two pulses of \thisgrb have a synchrotron origin, and it lies within a small subset of long GRBs with the credible signature of a high degree of prompt emission polarization \citep{2019ApJ...884..123C, 2020A&A...644A.124K}.

Apart from the prompt emission hard X-ray polarization measurements, we also constrained optical afterglow polarization using MASTER telescope data, making the burst the first case where both prompt emission and afterglow polarization measurements are constrained \citep{2022MNRAS.511.1694G}. These observations were carried out at different times: the \AstroSat CZTI data refer to the active stage of the gamma-ray burst and the MASTER optical observations to the afterglow. Relatively high polarization of the intrinsic prompt hard X-ray radiation demonstrates a high ordering of the magnetic field in the region close to the jet base. It is apparently associated with the radiation of colliding relativistic plasma flows under conditions of multiple internal shocks. The optical afterglow is formed behind the shock in the driven plasma of the progenitor stellar wind \citep{1998ApJ...497L..17S, Sari:1999ApJ, 2000ApJ...545..807K}. The absence of significant afterglow optical polarization of more than 1\% indicates that the jet's own magnetic field has decayed due to the expansion of the radiation region, and the raked-up chaotic magnetic field averaged, and the radiation ceased to be polarized for the same reason \citep{2004MNRAS.347L...1L}. Overall, \thisgrb provides a detailed insight into the prompt spectral evolution and emission polarization and challenges the traditionally used spectral model.


We also investigated the nature of the central engine of \thisgrb using the methodology discussed by \cite{2021ApJ...908L...2S}. We find beaming corrected energy for \thisgrb equal to 1.16 $\times$ $10^{52}$ ergs, larger than the mean beaming energy of a sample of GRBs studied by \cite{2021ApJ...908L...2S}. This energy is higher than the maximum possible energy budget of a magnetar, and no flares/plateau features are present in the X-ray light curve. This possibly supports a BH-based central engine for this GRB. We also constrain the radiative gamma-ray efficiency using the formula $\eta$= $E_{\rm \gamma, iso}$/($E_{\rm \gamma, iso}$+ $E_{\rm k}$)), finding $\eta$ $<$ 0.45 for \thisgrb. We conclude that the prompt emission polarization analysis, along with spectral and temporal information, has a unique capability to solve the long-debatable topic of the emission mechanisms of GRBs.

In this work, we present a detailed analysis of the prompt emission plus thermal and spectral characteristics of a long, luminous, and peculiar multi-structured burst \thisgrbB (which was also observed by the Atmosphere-Space Interactions Monitor installed on the International Space Station). The prompt light curve shows a very bright and brief pulse ($\sim$ 4\,s) followed by softer and rather long emission episodes up to $\sim$ 65\,s. A detailed time-resolved spectral analysis indicates hard low-energy photon indices, exceeding the synchrotron LOD in most of the Bayesian bins with high significance during brighter pulses, and it becomes softer (within the error bar) during the longer/softer emission phase. Harder $\alpha_{\rm pt}$ values observed during brighter emission pulses could be attributed to thermal emission arising from photospheric regions. On the other hand, the softer values of the low-energy photon indices during the longer emission phase are consistent with the non-thermal thin shell synchrotron emission model. This suggests that the radiation process responsible for \thisgrbB shows a transition between photospheric thermal emission (hard $\alpha_{\rm pt}$) and non-thermal synchrotron emission (soft $\alpha_{\rm pt}$). \par

In addition, we noticed a peculiar spectral evolution of \Ep and $\alpha_{\rm pt}$, where both of these spectral parameters exhibit the `flux tracking' pattern. We found strong/moderate positive correlations among various parameters: log (Flux)-log (\Ep), log (Flux)-$\alpha_{\rm pt}$, and log (\Ep)-$\alpha_{\rm pt}$, supporting the observed tracking pattern of the \Ep and $\alpha_{\rm pt}$. The flux tracking behaviour of \Ep could be understood in terms of cooling and expansion of the fireball. In such a scenario, during the cooling of relativistic electrons, the magnetic field will decrease, resulting in a lower intensity and \Ep values; however, during the expansion of the fireball, the magnetic field will increase, giving rise to higher intensity and \Ep values \citep{2021MNRAS.505.4086G}. Recently, \cite{2021NatCo..12.4040R} found a relation between the spectral index and the flux by examining the spectral evolution during the steep decay phase (observed with \swift XRT) of bright \swift BAT GRB pulses. They found that all the cases of their sample show a strong spectral softening, which is consistent with a shift of the peak energy. They demonstrated that such behaviour is well explained assuming an adiabatic cooling process of the emitting particles and a decay of the magnetic field. \par

Furthermore, \cite{2021A&A...656A.134G} performed numerical simulations and tracked the evolution of electrons to examine the observed evolution patterns of \Ep from GRBs. They suggested that the observed strong correlation between \Ep and flux can be explained if electrons are dominated by adiabatic cooling (considering a constant Lorentz factor). These studies support that adiabatic cooling could play a major role to understand the observed intensity tracking pattern of the peak energy. On the other hand, \cite{2019MNRAS.484.1912R} presented the possible physical interpretation of the flux tracking behaviour of $\alpha_{\rm pt}$ for the bursts having photospheric signatures (for example GRB 090719, GRB 100707A, GRB 160910A, plus many more) in their observed spectra. However, the possible physical interpretation of the flux tracking behaviour of $\alpha_{\rm pt}$ for the bursts having synchrotron signature (for example GRB 131231A, and GRB 140102A) in their observed spectra is still an open question. In the case of GRB 210619B, hard $\alpha_{\rm pt}$ values during the main pulse mean that the spectrum has a different origin than the synchrotron at the beginning (although, due to limitation of spectroscopy, it could be confirmed using additional prompt emission polarisation measurements). Therefore, assuming photospheric signatures in the early spectra of GRB 210619B, the flux tracking behaviour of $\alpha_{\rm pt}$ could be understood in terms of photospheric heating in a flow with a varying entropy \citep{2019MNRAS.484.1912R}. \par

We also observed a negative spectral lag for this burst, thus being an outlier of the typical known lag-luminosity anti-correlation of long bursts. The observed negative spectral lag could be explained in terms of a superposition of effects. Recently, some authors explained the observed spectral lag using emission mechanism models such as the photosphere \citep{2021MNRAS.505L..26J} and the optically-thin synchrotron model \citep{2016ApJ...825...97U} of GRBs. However, according to the photosphere jet model, only positive spectral lags of long GRBs can be explained. The observed negative lag for \thisgrbB might be explained using an optically-thin synchrotron model with the following assumptions: a curved spectrum, an emission radius, a decaying magnetic field (expanding jet), and the acceleration of the emitting regions \citep{2016ApJ...825...97U}. Moreover, the spectral lag is believed to be connected with the spectral evolution (correlation between \Ep and flux): (1) only the positive lags are possible for the hard-to-soft evolution of \Ep, (2) both the positive and negative lags can occur in the
case of intensity tracking pattern of \Ep. Thus, the observed negative lag and intensity tracking behaviour of \Ep for \thisgrbB are consistent with each other \citep{2018ApJ...869..100U}.  \par

More recently, \cite{2022arXiv220813821M} reported the polarization measurement of the optical afterglow of \thisgrbB in the time window \fermiT+5967 -- \fermiT +8245\,s and found changes in polarization properties such as polarization degree and angle. Such changes in the polarization properties have been reported for a few cases during the prompt emission indicating some evidence for a change in the order of magnetic field \citep{2019ApJ...882L..10S}. However, this has been the first case where a change in afterglow polarization properties has been observed. The temporal, spectral, and polarimetric observations of more such bright GRBs using space (\fermi, {\it INTEGRAL}, ASIM, {\it AstroSat})/ground-based instruments) will be useful to decipher the evolution of spectral properties, effects of hard electron energy index $p$ (giving rise to additional pieces of information to understand particle acceleration and shock physics in more detail) and changes in polarization properties during both prompt and afterglow phases.


\newcommand{\thisgrbE}{GRB~140102A\xspace}

\chapter{\sc Early Afterglow: Physical mechanisms}
\label{ch:5} 
\blfootnote{This chapter is based on the results published in: \textbf{Gupta,  Rahul} et al., 2021, {\textit{MNRAS}, {\textbf{505}}, 4086–4105}.}

\ifpdf
    \graphicspath{{Chapter4/Chapter4Figs/PNG/}{Chapter4/Chapter4Figs/PDF/}{Chapter4/Chapter4Figs/}}
\else
    \graphicspath{{Chapter4/Chapter4Figs/EPS/}{Chapter4/Chapter4Figs/}}
\fi

\normalsize

According to the standard afterglow model, the external shocks can be divided into two forms: a long-lived forward shock (FS) that propagates into the circumburst medium and produces a broadband afterglow and a short-lived reverse shock (RS) that propagates into the ejecta and produces a short-lived optical flash and a radio flare \citep{2003ApJ...595..950Z, 2004MNRAS.353..647N, 2015AdAst2015E..13G}. For most GRBs, the FS component can generally explain the observed afterglow. Investigations of the afterglow using the FS model offer detailed information about the late-time afterglow emission, jet geometry, circumburst medium, and total energy \citep{2007MNRAS.379..331P, Wang_2015, 2021MNRAS.505.1718J}. On the other hand, the short-lived RS emission is useful in probing the nature of magnetization and composition of GRB ejecta from the central engine \citep{2015AdAst2015E..13G}. RS emission has mainly two types of evolution \citep{2000ApJ...542..819K}. In the first situation, RS is relativistic enough to decelerate the shell (thick shell case, \tninty $>$ $T_{\rm dec}$  where $T_{\rm dec}$ is the deceleration time defined as $T_{\rm dec}$ =(3$E_{\rm k} (1+z)^3/32 \pi n_0 $ $\rm m_{\rm p}$ $\Gamma_0^8$ $\rm c^5)^{1/3}$, for blast-wave kinetic energy ($E_{\rm k}$), initial Lorentz factor ($\Gamma_0$), traversing into a constant density circumburst medium with density ($n_0$) \citep{2003ApJ...595..950Z}. The blast-wave radius evolves with time, and it is defined as $R_{\rm dec}(t = T_{\rm dec}) \approx 2$$\rm c$$T_{\rm dec}\Gamma_0^2/(1+z)$.  On the other hand, for the thin shell case (\tninty $<$ $T_{\rm dec}$), the RS could be sub-relativistic and too weak to decelerate the shell \citep{2015ApJ...810..160G}. It has been found that most of the GRBs with early optical RS signatures could be explained well within the thin shell case \citep{2014ApJ...785...84J, 2015ApJ...810..160G, 2020ApJ...895...94Y}.  

Fast optical follow-up of GRBs is vital for detecting and studying the relatively early and short-lived RS component. GRB 990123, the first burst with a quasi-simultaneous (with respect to the X-ray and/or gamma-ray emission) optical flash as the signature of RS, was detected using the Robotic Optical Transient Search Experiment (ROTSE)-I telescope \citep{1999Natur.398..400A, 1999MNRAS.306L..39M}. The RS emission has been observed for only a few GRBs, even after rapid follow-up observations by the \swift Ultra-Violet and Optical Telescope \citep[UVOT;][]{2005SSRv..120...95R} and other ground-based robotic telescopes network \citep{2003ApJ...595..950Z, oates09}. The lack of observed RS could be due to strongly magnetized outflows such that the RS component of the external shock is suppressed, either the RS emission component peaks at frequencies lower than optical frequencies and is thus generally missed, and/or the RS emission originating in external shock is masked by the prompt emission originating in an internal shock \citep{2005ApJ...628..315Z, 2016ApJ...825...48R, 2013ApJ...772...73K}. 

In a comprehensive sample (118 GRBs with known redshift) by \cite{2014ApJ...785...84J}, 10 bursts\footnote{GRBs 990123, 021004, 021211, 060908, 061126, 080319B, 081007, 090102, 090424, and 130427A.} showed dominant RS signatures originated from external shock with most of these having an ISM-like external medium (constant density). At late times ($>$ 10 ks), the RS emission showed magnetization parameters\footnote{It is the ratio between $\epsilon_{\rm B,r}$ (fraction of reverse shock energy into the magnetic field) and $\epsilon_{\rm B,f}$ (fraction of forward shock energy into the magnetic field), respectively.} ($R_{\rm B}$ $\sim$ 2 to $10^4$) and was fainter than average optical FS emission \citep{2014ApJ...785...84J}. \cite{2015ApJ...810..160G} identified 15 GRBs with RS signatures based on a morphological analysis of the early optical afterglow light curves of 63 GRBs and an estimated $R_{\rm B}$ $\sim$ 100. More recently, \cite{2020ApJ...895...94Y} studied the early optical afterglows of 11 GRBs\footnote{GRBs 990123, 041219A, 060607A, 061007, 081007, 081008, 090102, 110205A, 130427A, 140512A, and 161023A.} with RS emission signatures. They found that the external medium density w.r.t. blast-wave radius follows a power-law type behaviour with index ($k$) ranges between 0 to 1.5. Among other shock parameters, the densities of the external medium for thin shell RS-dominated bursts are compared in this work (see \S~\ref{sample comparison}) and shown to vary in a range of 0.1-500 $\rm cm^{-3}$. 

In this chapter, we present multiwavelength data and analysis of \thisgrbE including our early optical afterglow observations using the Burst Observer and Optical Transient Exploring System (BOOTES)-4 robotic telescope starting $\sim$ 29 sec after the \fermi GBM (also the \swift BAT at the same time) trigger (\fermiT). The very early detection of optical emission from \thisgrbE, along with GeV Large Area Telescope (LAT) detection inspired us to study this burst in detail and compare it with other similar events. This paper is organized as follows. In \S~\ref{multiwaveength observations and data analyisis}, we discuss the multiwavelength observations and data analysis. The main results and discussion are presented in \S~\ref{results}. 
Finally, the summary and conclusion are given in \S~\ref{conclusions}. All the uncertainties are quoted at 1\,$\sigma$ throughout this paper unless otherwise mentioned. 

\section{Multiwavelength Afterglow observations}
\label{multiwaveength observations and data analyisis}

\subsection{Gamma-ray observations}

\thisgrbE triggered the BAT \citep{2005SSRv..120..143B} on board the {\em Neil Gehrels Swift} observatory \citep{2004ApJ...611.1005G} on January 02, 2014 at 21:17:37 UT. The best on-ground location was found to be RA, Dec = 211.902, +1.331 degrees (J2000) with 3' uncertainty with 90 \% containment \citep{2014GCN.15653....1H}. 
The GBM \citep{2009ApJ...702..791M} on board \fermi Gamma-Ray Space Telescope triggered at 21:17:37.81 UT (\fermiT). 
The \fermi GBM light curve shows two bright overlapping peaks with a \tninty duration of $3.6\pm0.1$ sec in the 50 - 300 \keV energy range \citep{Bhat:2016ApJS..223...28N}. For a time interval from \fermiT + 0.4 to \fermiT + 4 sec, the time-averaged spectrum is best described with the \sw{Band} model \citep{Band:1993} with $\it \alpha_{\rm pt}$ = -0.71 $\pm$ 0.02, high energy spectral index $\beta_{\rm pt}$ = -2.49 $\pm$ 0.07, and \Ep = 186 $\pm$ 5 \keV.  
In this time interval, an energy fluence of 1.78 $\pm$ 0.03 $\times$ 10$^{-5}$ $\rm erg$ $\rm cm^{-2}$ is calculated in the 10-10000 \keV energy band \citep{2014GCN.15669....1Z}. This fluence value is among the top 12 per cent most bright GRBs observed by the \fermi-GBM, making this burst suitable for detailed analysis. The LAT \citep{2009ApJ...697.1071A} on board \fermi triggered at 21:17:37.64 UT and detected high energy emission from \thisgrb. 
The best \fermi-LAT on-ground location (RA, DEC = 211.88, 1.36 (J2000)) was at 47$^{\circ}$ from the LAT boresight angle at \fermiT and the highest-energy photon with an energy of 8 GeV is detected 520 sec after \fermiT \citep{2014GCN.15659....1S}.

{\bf \fermi-LAT analysis:}

\begin{figure}[ht!]
\centering
\includegraphics[scale=0.32]{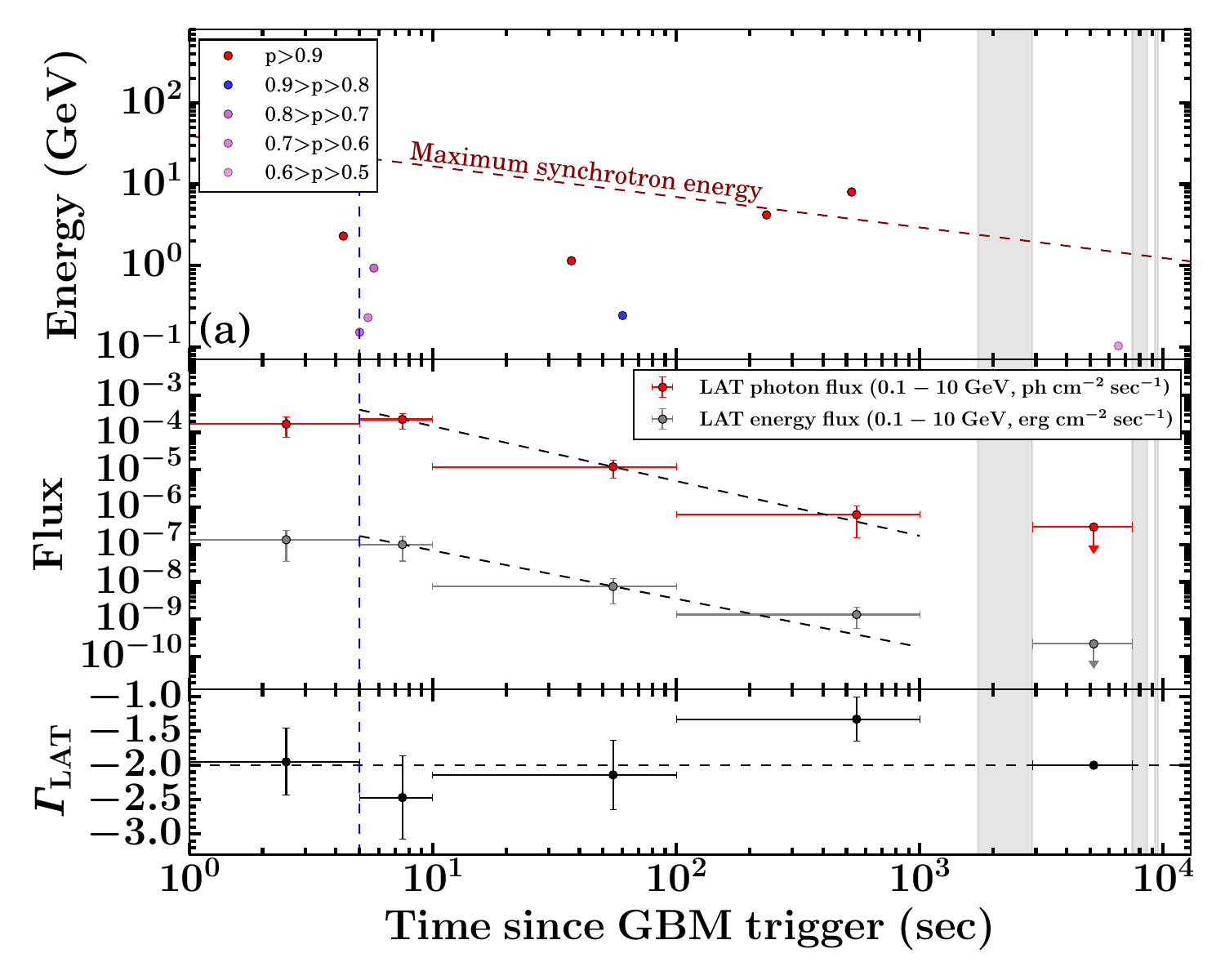}
\includegraphics[scale=0.32]{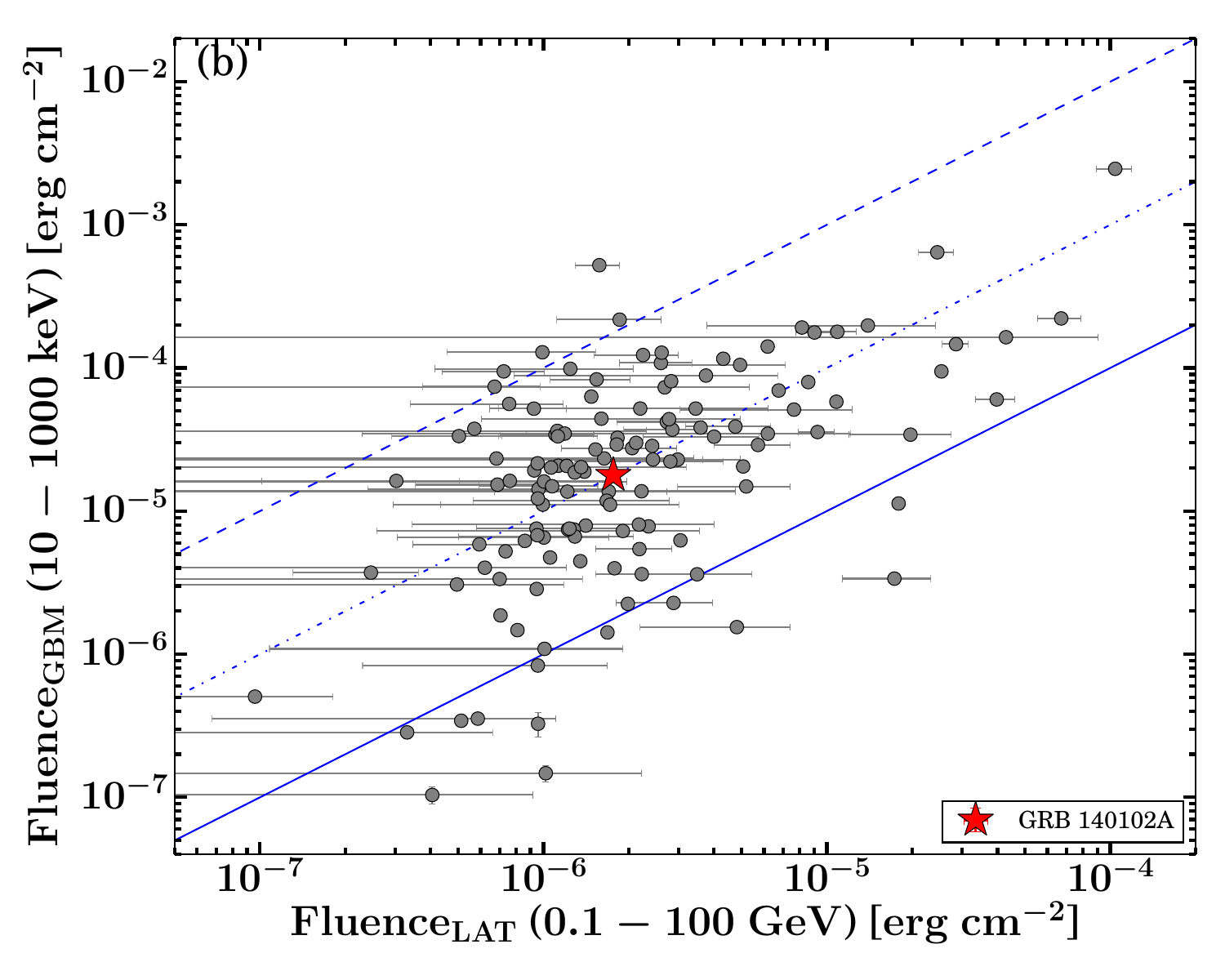}
\caption{{ (a) Extended \fermi-LAT emission:} \textit{Top panel}: Temporal distribution of high energy LAT photons with energies $> 100$ MeV. The different colours of the photons represent their associated probabilities of originating from \thisgrbE. \textit{Middle panel}: Evolution of \fermi-LAT energy (grey) and photon (red) fluxes in 100-10000 MeV range. In the last time bin, the LAT photon index was fixed to $-2$ to get an upper limit on the fluxes. The black  lines provide a power-law fit to the temporally extended photon flux and the energy flux light curves (for the photons detected 5 sec after \fermiT).
\textit{Bottom panel (b)}: Temporal evolution of \fermi-LAT photon index in the 0.1-300 GeV range. The vertical blue dashed line indicates the end time of the time-averaged spectral analysis, with the start time taken as \fermiT. The grey-shaded region represents the source off-axis angle $ > 65^\circ$. (b) Comparison of GBM and LAT fluence values for \thisgrbE along with LAT detected GRBs \citep{2019ApJ...878...52A}. The solid line represents equal fluence, and the dashed-dotted, dashed lines denote the fluence shifted by factors of 10 and 100, respectively.}
\label{fig:LAT_LCs}
\end{figure}

We obtained the \fermi-LAT \citep{2009ApJ...697.1071A} data within a temporal window extending up to 10000 sec after \fermiT from \fermi-LAT data server\footnote{https://fermi.gsfc.nasa.gov/cgi-bin/ssc/LAT/LATDataQuery.cgi} using \sw{gtburst}\footnote{https://fermi.gsfc.nasa.gov/ssc/data/analysis/scitools/gtburst.html} version 02-03-00p33 software. We performed unbinned likelihood analysis and selected a region of interest (ROI) of $\rm 12^{\circ}$ around the enhanced \swift XRT position \citep{2014GCN.15657....1G}. Further, we filtered the high energy LAT emission by cutting on photons with energies in the range of 100 MeV-300 GeV. We also applied an angular cut of $\rm 100^{\circ}$ between the source and zenith of the satellite in order to reduce the contamination of photons coming from the Earth limb. For the time-integrated duration, we used the \sw{P8R3\_SOURCE\_V2 response}, which is appropriate for long durations ($\rm \sim 10^{3} ~sec$), and for the time-resolved bins, we used \sw{ P8R2\_TRANSIENT020E\_V6} response, which is appropriate for small durations \citep{2018arXiv181011394B}. We included a point source (for GRB) at the location of the burst, considering a power-law spectrum, an isotropic component (to show the extragalactic background emission) \sw{iso\_P8R3\_SOURCE\_V2} and a Galactic diffuse component (to represent the Galactic diffuse emission) \sw{gll\_iem\_v07} \footnote{https://fermi.gsfc.nasa.gov/ssc/data/access/lat/BackgroundModels.html}. The probability of association of the photons with \thisgrbE is calculated using the \sw{gtsrcprob} tool. For the total duration of 0-10000 sec, the energy and photon flux in 100-10000 MeV energy range are $(5.96 \pm 2.37) ~\times 10^{-10}$ $\rm ~erg ~cm^{-2} ~ sec^{-1}$ and $(5.01 \pm 2.35) ~ \times 10^{-7}$ $\rm ~ph. ~cm^{-2} ~ sec^{-1}$, respectively, with a test-statistic (TS)\footnote{ It is defined as TS = -2ln$L_{\rm max,0}$/$L_{\rm max,1}$, where $L_{\rm max,0}$ and $L_{\rm max,1}$ are the maximum likelihood value for a model without and with an additional source, respectively.} of detection 46. The temporal distribution of \fermi-LAT photons with energies $> 100$ MeV along with the photon flux and the energy flux light curves is shown in Figure \ref{fig:LAT_LCs} (a). During the GBM time window, we obtained the LAT fluence value equal to $(0.18\pm0.10) ~\times$ $10^{-5}$ erg $\rm cm^{-2}$ in 0.1-100 GeV energy range. We compared this value with the GBM fluence value for \thisgrbE along with other LAT detected GRBs \citep[ the second GRB LAT catalogue (2FLGC);][]{2019ApJ...878...52A, 2011ApJ...726...22A}. \thisgrbE lies on the line for which GBM fluence is 10 times brighter than LAT fluence for all the LAT detected sample (see Figure \ref{fig:LAT_LCs} (b)). Further, we also perform (see \S~\ref{lat_TRS}) the time-resolved spectral analysis to investigate the origin of the high energy LAT photons (see Table \ref{tab:lat_sed_GRB140102A} in appendix).

\subsection{\swift XRT observations}

The X-ray Telescope \citep[XRT;][]{2005SSRv..120..165B} began observing the field of \thisgrbE on January 02, 2014 at 21:18:34.3 UT, 56.8 sec after the BAT trigger. The XRT detected a bright, uncatalogued X-ray source located RA, and DEC = 211.9190, 1.3333 degrees (J2000) with an uncertainty of 4.8$\rm "$ (radius, 90 $\%$ containment). This location is 61$\rm "$ from the BAT onboard position but within the BAT error circle \citep{2014GCN.15653....1H}. In total, there are 9.5 ks of XRT data for \thisgrbE, from 47 sec to 205 ks after the BAT trigger. The data comprise $\sim$ 2.2 ks in Windowed Timing (WT) mode (the first $\sim$ 8 sec were taken while \swift was slewing) with the remainder in Photon Counting (PC) mode. In this paper, we used X-ray data products (both light curve and spectrum) available on the \swift online repository\footnote{https://www.swift.ac.uk/} hosted by the University of Leicester \citep{2007A&A...469..379E, 2009MNRAS.397.1177E}.

We modelled the X-ray light curve using a simple power-law function and  broken power-law model (BPL). We find that the XRT light curve could be best described with a broken power-law model (see \S~\ref{xray_optical lc fitting}). We calculated $\alpha_{\rm x1}$ = $-1.09^{+0.01}_{-0.01}$, $\alpha_{\rm x2}$ = $-1.50^{+0.02}_{-0.02}$, and $t_{\rm bx}$ = $1298^{+108}_{-74}$ sec, where $\alpha_{\rm x1}$ is the temporal index before break ($t_{\rm bx}$), and $ \alpha_{\rm x2}$ corresponds to the temporal index after break. All the temporal parameters are also listed in Table \ref{lcfits_GRB140102A}.

We analyzed the \swift XRT spectra using the XSPEC package using an absorbed power-law model in 0.3-10 \keV energy channels.
The absorption includes photoelectric absorption from our Galaxy and the host galaxy of the GRB using the XSPEC components \sw{phabs} and \sw{zphabs} together with the source spectral model. We considered the Galactic hydrogen column density fixed at $\rm NH_{\rm Gal}= 3.04 \times 10^{20}{\rm cm}^{-2}$ \citep{2013MNRAS.431..394W}, and a free intrinsic hydrogen column density located at the host redshift ($\rm NH_{\rm z}$). The XRT spectra were grouped with a minimum of 1 count per bin unless mentioned otherwise. The statistics \sw{C-Stat} is used for optimization and testing the various models. The redshift of the second absorption component was fixed at 2.02 as discussed in \S~\ref{photoz}. We also search for an additional thermal (\sw{Black Body}) and other possible components in the early time WT mode spectra (63-200 sec), but we did not find any signature of thermal component based on the BIC value comparison obtained for simple absorption power-law model. All the spectral parameters for the absorbed power-law model have been listed in Table \ref{xspectra_GRB140102A}.

\begin{table}[ht!]
\scriptsize
  \caption{The best-fit models to the X-ray, combined optical light curves, and individual optical filters.}
  \begin{tabular}{|c|c|c|c|c|c|c|}
  \hline
 \bf Wavelength & \bf Model &$\bf \alpha_1$ &\bf  $\bf \alpha_2 $ or $\bf \alpha_{\rm x1}$ &\bf  $\rm \bf {t_{break}}$ or $\rm \bf t_{\rm bx}$ (sec) & $ \bf \alpha_3$ or $\bf \alpha_{\rm \bf x2}$ & $\bf \chi^2/dof$ \\
  \hline
  \hline
 X-ray             & broken power-law              & ---                   & $-1.09\pm0.01$  & $1298^{+108}_{-74}$ & $-1.50\pm0.02$           &  575/543\\
 Optical/IR         & power + broken power           & $-1.72\pm0.04$         & $-0.47\pm0.03$  & $6160^{+740}_{-240}$   & -$1.11\pm0.15$ & 109/33 \\ 
 \hline                
  BOOTES Clear      & 2x power-law                  & $-1.72\pm0.04$        & $-0.41\pm0.04$  & ---                & ---                      & 199/104 \\
 UVOT $white$        & 2x power-law                  & $-2.19^{+0.27}_{-0.41}$ & $-0.86\pm0.07$  & ---                & ---                      & 41/48   \\
  BOOTES Clear      & Broken power-law              & ---                   & $-1.49\pm0.01$  & $475\pm15$         & $ -0.70\pm0.01$          & 315/106 \\
 UVOT $white$  & Broken power-law              & ---                   & $-1.49\pm0.05$  & $294\pm58$         & $ -0.88\pm0.07$          & 43/48   \\
 \hline
  \end{tabular}
  \label{lcfits_GRB140102A}
\end{table}

Initially, we created two time-sliced spectra (before and after the break time) using the UK Swift Science Data Centre webpages\footnote{https://www.swift.ac.uk/xrt\_spectra/addspec.php?targ=00582760} and found spectral decay index $\beta_{\rm x1}$ = $0.56^{+0.02}_{-0.02}$ (before break) and  $\beta_{\rm x2}$ = $0.67^{+0.06}_{-0.06}$ (after break). Further, we created three more time-sliced spectra, using data between 63-200 sec, 200-2200 sec, and 5600- $2 \times 10^5$ sec to observe the evolution of decay slope in the X-ray band ($\beta_{\rm x}$). We notice, $\beta_{\rm x}$ evolves and increases continuously in each temporal bin.  However, the photon index obtained from the Swift Burst Analyser webpage \footnote{https://www.swift.ac.uk/burst\_analyser/00582760/} could be fitted with a constant (\sw{CONS}) model\footnote{https://www.ira.inaf.it/Computing/manuals/xanadu/xronos/node93.html}, and at very late times the photon index is different and seems to have dropped (see Figure \ref{Xray_optical_afterglow} (c)). We observed a late time ($\sim$ 4.7 $\times$ $10^{4}$ to $\sim$ 9.3 $\times$ $10^{4}$ sec) re-brightening activity in the unabsorbed X-ray flux light curve (at 10 \keV).

\begin{table}[ht!]
\centering
\large
\caption{The best-fit values of Photon indices from the X-ray afterglow spectral modelling for different temporal segments. We have frozen the host hydrogen column density, obtained from SED 3 (see \S~\ref{SED}).}
\begin{tabular}{|c|c|c|c|c}
\hline
\bf Time (sec) &$\rm \bf Photon ~index$ &  \bf Mode \\
\hline
\hline
63-65196 & $ 1.59 \pm 0.09$ &  WT+PC \\
63-200   & $ 1.51 \pm 0.03$ & WT\\
200-2200   & $ 1.61 \pm 0.02$ & WT \\
5600-$2 \times 10^5$& $ 1.63 \pm 0.09$ & PC\\
\hline
\end{tabular}
\label{xspectra_GRB140102A}
\end{table}

\subsection{UV/Optical Observations}

Ultra-violet and optical observations were carried out using \swift-UVOT, BOOTES robotic, and 1.3m DFOT telescopes as a part of the present work. Details of these observations are given below in respective sub-sections. Multi-band light curves obtained from these observations are presented in Figure \ref{optical_multiband}.

\begin{figure}[ht!]
\centering
\includegraphics[scale=0.36]{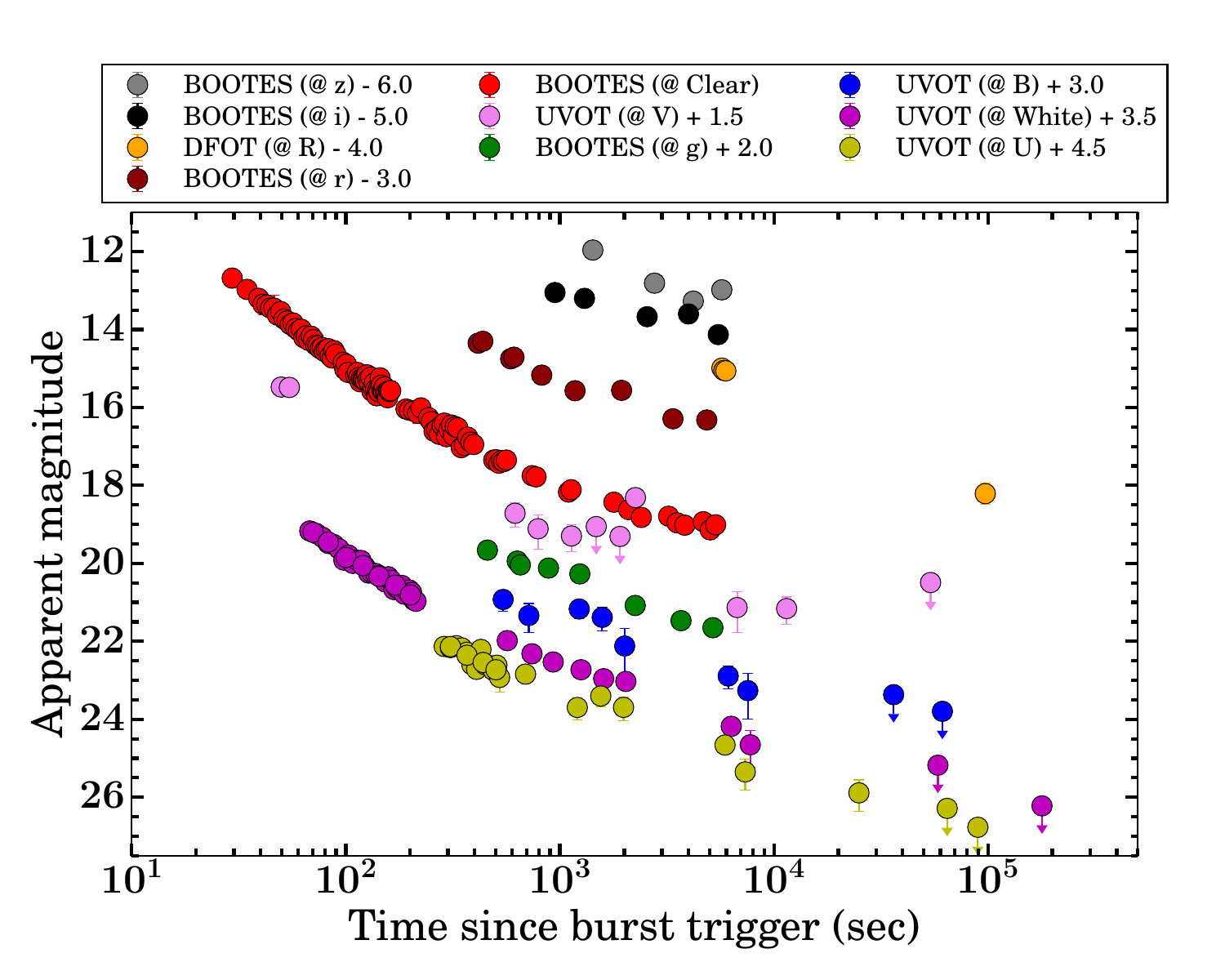}
\caption{{Multi-band optical light curve:} Multi-band light curves of the afterglow of \thisgrbE using UVOT, BOOTES, and Devasthal fast optical telescope (DFOT) data as a part of the present analysis as tabulated in appendix A. Magnitudes are not corrected for the extinctions.}
\label{optical_multiband}
\end{figure}

\subsubsection{{\swift Ultra-Violet and Optical Telescope}}

The \swift UVOT began observing the field of \thisgrbE 65 sec after the BAT trigger \citep{2014GCN.15653....1H}. The afterglow was detected in 5 UVOT filters. UVOT observations were obtained in both image and event modes. Before extracting count rates from the event lists, the astrometric corrections were refined following the methodology described in \cite{oates09}. For both the event and image mode data, the source counts were extracted using a region of a 3" radius. In order to be consistent with the \swift UVOT calibration, the count rates were then corrected to 5" using the curve of growth contained in the calibration files using standard methods. Background counts were estimated using a circular region of radius 20" from a blank area of sky situated near the source position. The count rates were obtained from the event and image lists using the \swift standard tools \sw{uvotevtlc} and \sw{uvotsource}, respectively\footnote{https://www.swift.ac.uk/analysis/uvot/}. Later, these counts were converted to magnitudes using the UVOT photometric zero points \citep{bre11}. The UVOT data for this burst is provided in Appendix A. As with data from all the following telescopes, the resulting afterglow photometry is given in AB magnitudes and not corrected for Galactic reddening of E(B-V)= 0.03 \citep{2011ApJ...737..103S}.

\subsubsection{{BOOTES Robotic Telescope Network}}
The BOOTES-4/MET robotic telescope in Lijiang, China \citep{cast12} automatically responded to the BAT trigger \citep[][]{2014GCN.15653....1H} with observations starting at 21:18:05 UT on January 02, 2014, 28 sec after the trigger. For the first 5 minutes, observations were taken with a series of 0.5-sec clear filter exposures, after which observations were taken with a systematic increase in exposure. Starting from $\sim 6$
minutes after the trigger, observations were also performed in rotation with the $g$, $r$, $i$, $Z$, and $Y$ filters. The images were dark-subtracted and flat-fielded using custom IRAF routines. The aperture photometry was extracted using the standard IRAF software, and field calibration was conducted using the SDSS DR12 catalogue \cite{ala15}. Colour transforms of \cite{hew06} were used when calibrating the $Z$ and $Y$ filters. The BOOTES data is provided in Appendix A.

\subsubsection{{1.3m DFOT Telescope}}
The 1.3m DFOT at ARIES, Nainital started observing the field of \thisgrbE $\sim$ 1.5 hours after the trigger. Several frames in V, $\rm R_c$ and $\rm I_c$ pass-bands were obtained in clear sky conditions. Images were dark-subtracted but not flat fielded since there were no available flats taken on the same night and flat fielding using those taken on a different night made the science images worse. Aperture photometry was extracted using the standard IRAF software, and field calibration
was conducted using the SDSS DR12 catalogue \cite{ala15} and the colour transforms of Robert Lupton in the SDSS online documentation\footnote{http://www.sdss.org/dr12/algoritms/sdssUBVRITransform/\#\#Lupton2005}. The optical data using the 1.3m DFOT telescope, along with other data sets, are provided in Appendix D.

\subsection{Combined Optical Light curve}
\label{combLC} 
In order to maximize the SNR of the optical light curves, we followed the methodology of \cite{oates09} to
combine the individual filter light curves into one single filter \citep{2010MNRAS.401.2773S, 2010ApJ...720.1513K, oates12, 2017ApJS..228...13R, 2019A&A...632A.100H}. First, ground-based photometry was converted from
magnitudes to count rate using an arbitrary zero-point. The light curves from the different filters were then normalized to the r filter. The normalization was determined by fitting a power law to each of the light curves in a given time range simultaneously. The power-law indices were constrained to be the same for all the filters, and the normalizations were allowed to vary between the filters. For the light curve of \thisgrbE, the power-law was fitted between 600-6000 sec since this is the epoch which maximized the number of filters, and the behaviour in each filter appeared to be similar. However, the normalization led to a slight offset between the UVOT $white$ and $V$ band data in comparison to the BOOTES-clear filter. We will discuss the consequences on model fitting in \S~\ref{optical_afterglow_modelling}. After the light curves were normalized, they were binned by taking the weighted average of the normalized count rates in time bins of $\Delta T/T = 0.2$. The combined optical light curve is shown in Figure \ref{Xray_optical_afterglow}.

\begin{figure*}
\centering
 \includegraphics[height=7.8cm,width=7.3cm,angle=0]{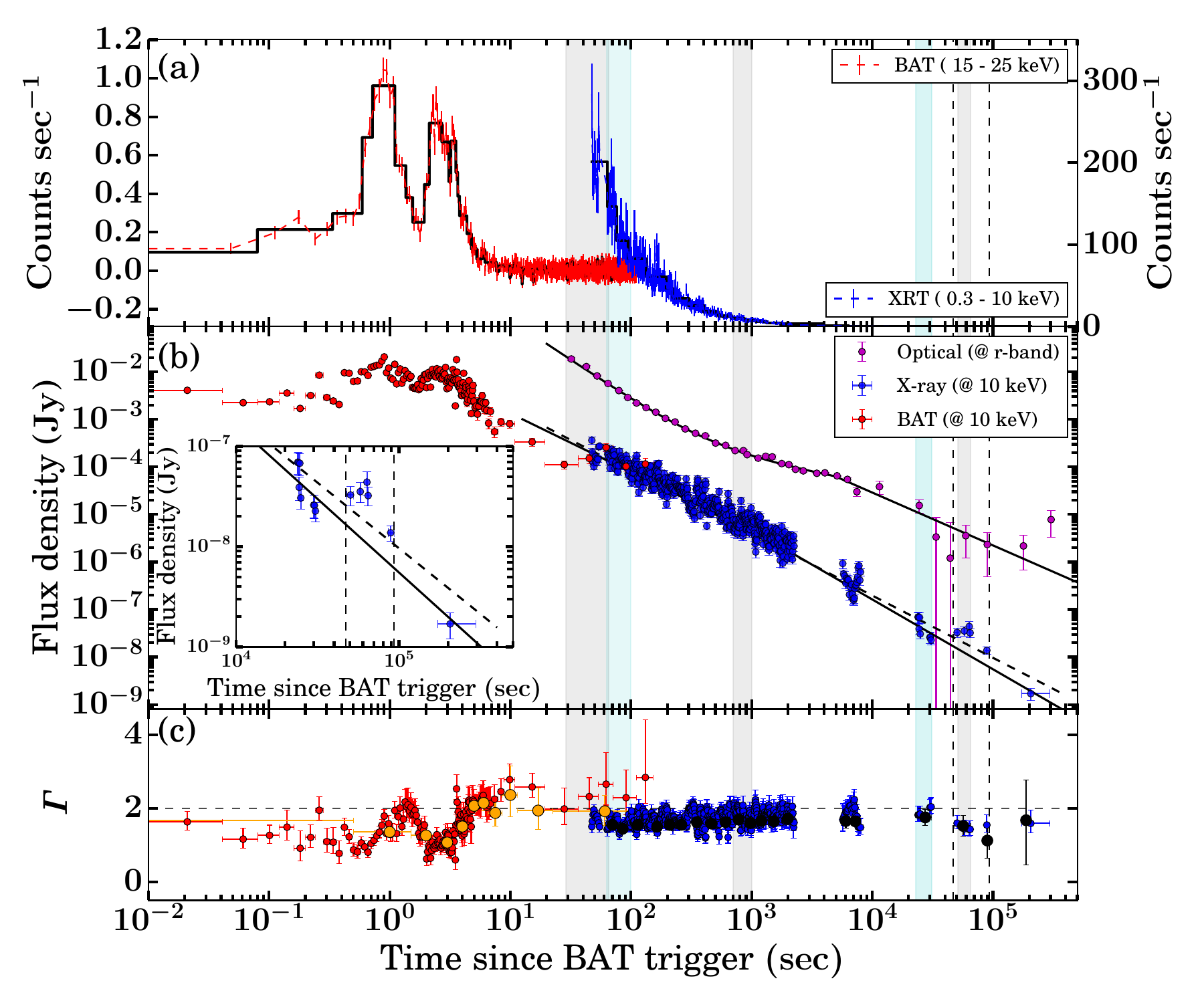}
\includegraphics[height=7.8cm,width=7.3cm,angle=0]{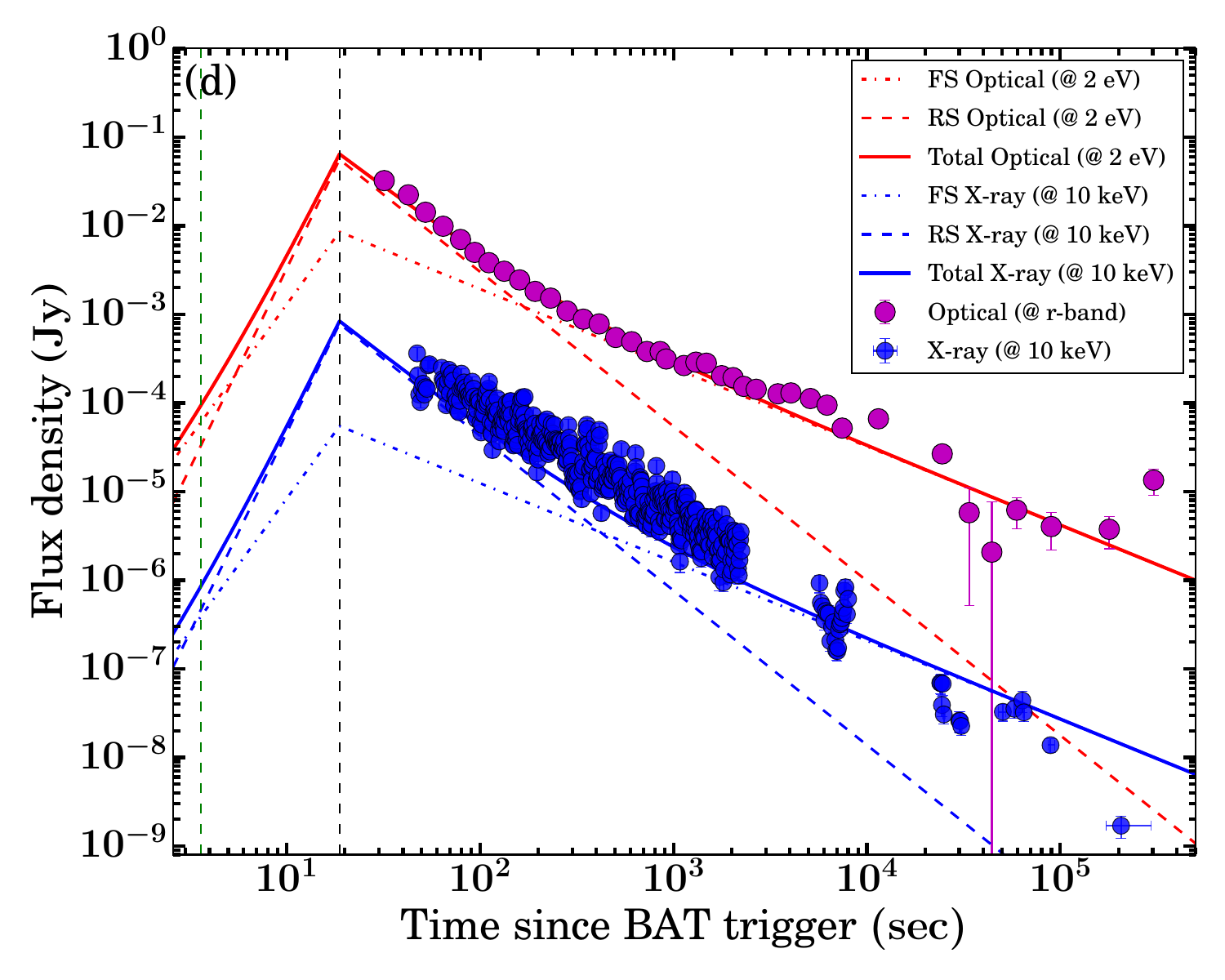}
\caption[]{{Multiwavelength light curve of \thisgrbE :} (a) Count rate BAT and XRT light curves overlaid with the Bayesian Block analysis. (b) The X-ray and optical/UV afterglow light curves of \thisgrbE overlaid with the best-fit models: a broken power-law (X-ray, solid black line) or simple power-law (X-ray,  black dashed line) or its combinations (optical, black solid line). The inset plot shows the late re-brightening activity in the X-ray light curve. (c) Evolution of photon indices within the BAT (red) and XRT (blue) window\protect\footnotemark. The orange and black circles indicate the evolution of BAT and XRT photon indices obtained using our spectral analysis of Bayesian Blocks bins. The shaded vertical grey and cyan colour bars represent the epochs used to create the spectral energy distributions (SEDs) of \thisgrbE afterglow. The vertical dashed lines indicate the late re-brightening activity in the X-ray light curve (at 10 keV). (d) A combined forward and reverse shock model is used for the interpretation of the optical emission from \thisgrbE. The peak flux is obtained at the crossing time $t_{\rm x}$ = 18.79 sec, and for $t> t_{\rm x}$ power-law behaviour of the afterglow model is followed. The optical flux is explained using the sum of reverse and forward shock components. The same set of parameters produces a lower amount of early X-ray flux in this model. The vertical green and black dashed lines indicate the end time of \tninty duration and deceleration time, respectively. The model parameters are given in Table \ref{sample_Modelling}.}
\label{Xray_optical_afterglow}
\end{figure*}

\footnotetext{https://www.swift.ac.uk/burst\_analyser/00582760/}

\subsection{Spectral Energy Distributions}
\label{SED}

SEDs were produced at five epochs (29-65 sec, 63-100 sec, 700-1000 sec, 23-31 ks, and 51-65 ks) following the prescription within \cite{depas07}, which is based on the methodology of \cite{schady07}. We used XSPEC \citep{1996ASPC..101...17A} to fit the optical and X-ray data. We adopted two models, a simple power-law model and a broken-power law model. The difference between the indices of the broken power-law model was fixed at 0.5, consistent with the expectation of a synchrotron cooling break \citep[see \S~\ref{optical_afterglow_modelling},][]{zhang04}. In each model, we include a Galactic and intrinsic absorber using the XSPEC models \sw{phabs} and \sw{zphabs}. The Galactic absorption is fixed to $\rm NH_{Gal} = 3.04 \times 10^{20} {\rm cm^{-2}}$ \citep{2013MNRAS.431..394W}. We also include Galactic and intrinsic dust components using the XSPEC model
\sw{zdust}, one at redshift $z$= 0, and the other is free to vary. The Galactic reddening was fixed at E(B-V)= 0.0295 according to the map of \cite{2011ApJ...737..103S}. For the extinction at the redshift of the burst, we test Milky Way, Large and Small Magellanic Clouds (MW, LMC and SMC) extinction laws \citep{pei92}. All five SEDs are shown in Figure \ref{SED_fig}. We extrapolated the best-fit model towards the LAT frequencies to search for a possible spectral break between X-ray and GeV. All the results of SEDs are listed in Table \ref{SED_table}. 

\begin{figure}[ht!]
\centering
\includegraphics[scale=0.28]{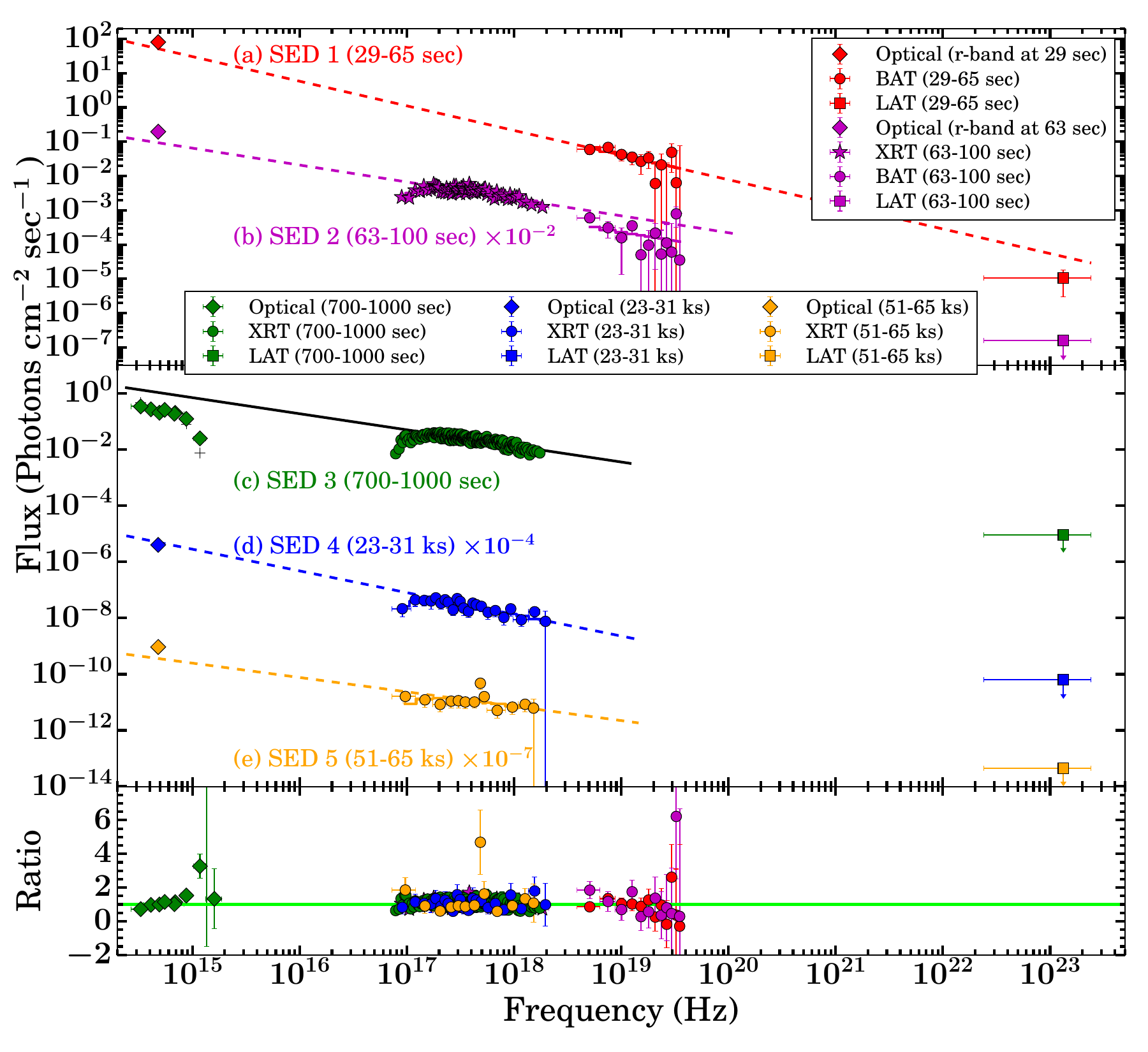}
\caption{{Spectral energy distributions:} (a) SED 1 at 29-65 sec, the black dashed lines show the extrapolations of the power-law fit to the BAT spectra to the optical and LAT frequencies. The observed LAT flux is lower than the extrapolated value, and it indicates the presence of a possible spectral break between BAT and LAT frequencies. (b) SED 2 at 63-100 sec, the black dashed lines show the extrapolations of the unabsorbed power-law model to the optical frequencies. (c) SED 3 at 700-1000 sec, used to constrain the redshift of the burst. The solid black line is the best fit by a simple power law from the joint X-ray and Optical spectral fit. (d) SED 4 23-31 ks. (e) SED 51-65 ks. The bottom panel shows the ratio of data to the model. The horizontal lime green solid line corresponds to the ratio equal to 1.}
\label{SED_fig}
\end{figure}

\begin{table}
\begin{center}
\caption{The best-fit results of optical and X-ray spectral indices at different epoch SEDs and their best describe spectral regime. $p$ is the mean value of the electron distribution indices calculated from the observed value of $\alpha_{\rm opt}$/$\alpha_{\rm x-ray}$ and $\beta_{\rm opt}$/$\beta_{\rm x-ray}$ of best describe spectral regime.}
\label{SED_table}
\begin{scriptsize}
\begin{tabular}{|c|c|c|c|c|}
\hline
\textbf{SED} & \textbf{Time interval (sec)} & \textbf{$\bf \beta_{\rm \bf X-ray/opt}$}  & \textbf{\begin{tabular}[c]{@{}c@{}} $p$\\ (Spectral regime)\end{tabular}} & \textbf{$\bf \chi_r^{2}$} \\ \hline
1 & 29-65 & 0.72$^{+0.45}_{-0.42}$ & \begin{tabular}[c]{@{}c@{}}2.33 $\pm$ 0.21 \\ ($\nu_{\rm opt}$ $<$ $\nu_{\rm x-ray}$ $<$ $\nu_{\rm c}$)  \end{tabular} & 0.88 \\ \hline
2 & 63-100 &  0.49$^{+0.05}_{-0.05}$  &  \begin{tabular}[c]{@{}c@{}}2.10 $\pm$ 0.21\\ ($\nu_{\rm opt}$ $<$ $\nu_{\rm x-ray}$ $<$ $\nu_{\rm c}$)  \end{tabular}  &  1.01\\ \hline
3 & 700-1000 & 0.57$^{+0.02}_{-0.02}$ & \begin{tabular}[c]{@{}c@{}}2.09 $\pm$ 0.29\\ ($\nu_{\rm opt}$ $<$ $\nu_{\rm x-ray}$ $<$ $\nu_{\rm c}$) \end{tabular}  & 1.05 \\ \hline
4 & 23000-31000 & 0.77$^{+0.22}_{-0.22}$ &  \begin{tabular}[c]{@{}c@{}}2.64 $\pm$ 0.21\\ ($\nu_{\rm opt}$ $<$ $\nu_{\rm x-ray}$ $<$ $\nu_{\rm c}$) \end{tabular}  &  0.78 \\ \hline
5 & 51000-65000 &  0.51$^{+0.32}_{-0.32}$ &  \begin{tabular}[c]{@{}c@{}}2.38 $\pm$ 0.40 \\ ($\nu_{\rm opt}$ $<$ $\nu_{\rm x-ray}$ $<$ $\nu_{\rm c}$) \end{tabular}  & 1.27 \\ \hline

\end{tabular}
\end{scriptsize}

\end{center}

We calculated $\rm NH_{\rm z}  (0.61^{+0.11}_{-0.10} \times 10^{22}{\rm cm}^{-2}$), host extinction ($0.21^{+0.02}_{-0.02}$ mag) and redshift ($2.02_{-0.05}^{+0.05}$) using the SED 3. $\chi_r^{2}$ notify the reduced chi-square values for SEDs at different epochs.  Uncertainty in the calculation of $p$ is obtained with a confidence level of 95 \%.
\end{table}

{{\bf Photometric Redshift:}}
\label{photoz}
A spectroscopic redshift for the GRB was not reported and could not be determined from GTC spectroscopic observations of the host galaxy (see \S~2.6 of \cite{2021MNRAS.505.4086G}). To determine a photometric redshift, we therefore, created a SED using the afterglow data (X-ray and Optical) between 700 sec and 1000 sec. During this time, there is no break in the X-ray light curve; also, no spectral evolution is observed. The evolution of the X-ray photon index ($\it \Gamma_{\rm XRT}$) measured during this time window is consistent with not changing (see Figure \ref{Xray_optical_afterglow}). Using the methodology outlined in \S~\ref{SED}, we fit the SED with the simplest model, a power-law, which we find to be statistically unacceptable with \sw{$\chi^2$} for the  LMC and SMC 282 and 316 respectively for 213 degrees of freedom. However, for the MW model, we find a better \sw{$\chi^2$} (223) for the same number of degrees of freedom. Further, we fit the SED with the broken power-law model. For the MW, LMC, and SMC, all the fits are statistically acceptable with \sw{$\chi^2$} of 169, 178, 185, respectively for 212 degrees of freedom.  We notice that for this SED (700-1000 sec), the MW model both for power-law and broken power-law fit show the smallest reduced \sw{$\chi^2$}; however, the reduced \sw{$\chi^2$} is less than 1 for broken power-law fit indicating that the model is "over-fitting". In the case of the power-law model, the reduced \sw{$\chi^2$} is close to one. It confirms that the power-law model is the best fit. The redshift values for
all three extinction laws are similar and consistent within 3 $\sigma$. Taking the redshift from the MW model, we find a photometric redshift of $2.02^{+0.05}_{-0.05}$ for \thisgrbE. We adopted this value for all our subsequent analyses.

\section{Results and Discussion}
\label{results}
Based on the multi-band observations obtained from various space and ground-based facilities of \thisgrbE, we discuss the results based on the present analysis of afterglows. 

\subsection{Multi-wavelength Afterglow behaviour using closure relations}
\label{closure relation}

In this subsection, we present the afterglow properties of \thisgrbE.

\subsubsection{Origin of the high energy LAT photons}
\label{lat_TRS}

For \thisgrbE, the extended LAT emission becomes softer and marginally brighter (consistent with statistical fluctuation) after the end of GBM \keV-MeV emission. This indicates that the LAT GeV emission originated later than the GBM \keV-MeV emission and from a different spatial region. In this section, we study the possible external origin and radiation mechanism of detected LAT emission. 

 To investigate the radiation mechanism of high energy GeV LAT photons, we calculated the maximum photon energy emitted by the synchrotron radiation mechanism in an adiabatic external forward shock during the decelerating phase in the case of ISM external medium. We used the following expression (see equation \ref{maxsy_energy}) from \cite{2010ApJ...718L..63P}:
 \begin{equation}
 \label{maxsy_energy}
{\rm h} \nu_{max}=
9{\rm~GeV~} \left(\frac{E_{\rm iso,54}}{n_0}\right)^{1/8}
\left(\frac{1+z}{2}\right)^{-5/8} \left(\frac{t}{100}\right)^{-3/8}   
\end{equation}

 Where $E_{\rm iso,54}= E_{\rm iso}/10^{54}$ in ergs, $t$ is the arrival time of the event since \fermiT in sec, and $n_0$ is the ISM density. We consider $n_0$ =  0.70 {$\rm cm^{-3}$} for the present analysis (see Table \ref{sample_Modelling}). We noticed that some of the late-time photons (even one photon with source association probability $>$ 90 $\%$) lie above the maximum synchrotron energy. It indicates the non-synchrotron origin of these photons. Photons above the maximum synchrotron energy from recent VHE detected GRBs confirm the Synchrotron self-Compton origin for these high energy photons \citep{2019Natur.575..455M, 2019Natur.575..464A, 2019Natur575448Z}. 
 
 Further, we fitted the LAT energy and photon flux light curves (see Figure \ref{fig:LAT_LCs}) obtained from the time-resolved analysis with a simple power-law decay model, which shows that emission could be continuously decreasing with time in both energy and photon fluxes during the temporal bins after the prompt phase (\fermiT + 5 sec):  $5 - 10$ sec, $10 - 100$ sec, and $100 - 1000$ sec. 
 The \fermi LAT photon flux light curve shows temporal variation as a power law with an index $-1.47 \pm 0.01$
 and the energy flux light curve shows temporal variation as a power law with an index $-1.29\pm 0.06$.  
 The LAT photon index 
($\it \Gamma_{\rm lat}$) is $-2.18\pm 0.08$ from a spectral fit obtained by fitting the first $10^{5}$ sec data. 
This gives spectral index $\beta_{\rm lat} = \it \Gamma_{\rm lat} + 1$ to be $-1.18 \pm 0.08$. The time-resolved spectra do not show strong temporal variation in the photon index in the first four temporal bins.

In the external shock model, for $\nu~>~max \{ \nu_{\rm m},~\nu_{\rm c} \}$ which is generally true for reasonable shock parameters we can derive
the power-law index of the shocked electrons by $f_\nu \propto \nu^{-p/2}$. We 
have synchrotron energy flux $\rm f_{\rm lat} \propto \nu^{-\beta_{\rm lat}} t^{-\alpha_{\rm lat}} $. We found $\alpha_{\rm lat} = -1.29 \pm 0.06$ (see the LAT light curve in Figure \ref{fig:LAT_LCs}) and $\beta_{\rm lat} = -1.18 \pm 0.08$. The value of $\beta_{\rm lat}$ gives us $p=2.36 \pm 0.16$. Thus, the power-law index for the energy flux decay can be predicted by using $f_{\rm lat}\propto t^{(2-3p)/4}$. 
The calculated value of $\alpha_{\rm lat}$ is $-1.27 \pm 0.16$, which agrees well with the observed value of $-1.29 \pm 0.06$. Hence, we can conclude that for \thisgrbE, the extended LAT high energy afterglow is formed in an external forward shock.

\subsubsection{{X-ray and Optical afterglow light curves}}
\label{xray_optical lc fitting}

The X-ray afterglow light curve, shown in Figure \ref{Xray_optical_afterglow}, declines from the beginning of observations and shows no features, such as steep and shallow decay phases or any flaring activity. A fit of a power law to the light curve gives a $\chi^2$/dof = 1190/545, suggesting a more complex model is required to fit the data. A broken power-law model is a better fit with $\chi^2$/dof = 575/543;
according to the {\it F}-test, the broken power-law model gives an improvement at the $>5\sigma$ confidence level with respect to the power-law model. We also tested to see if a further break would provide an improvement. The double broken power-law model results in $\chi^2$/dof = 565/541. The {\it F}-test suggests the additional break is not required since the improvement is only at the 2.6 $\sigma$ level. The best-fit model is given in Table \ref{lcfits_GRB140102A}.

The optical multi-band light curve is shown in Figure \ref{optical_multiband}. Observations in the UVOT $white$ and BOOTES-clear filters started before most of the other optical filters and have the best sampling and SNR during the first few thousand seconds. A power-law fit to both of these light curves results in $\chi^2/dof=1754/108$ and $\chi^2/dof=89/50$ for
BOOTES-clear and UVOT $white$, respectively. Both of which are statistically unacceptable at the $>3\sigma$ level.
Since the light curves display an early steep to shallow transition, suggesting the presence of a reverse dominated shock \citep{2015ApJ...810..160G} when fitting a more complex model such as a broken power-law, the sum of two power-law models, or their combinations. Both fit give slightly better values, with the broken power-law resulting in $\chi^2/dof=315/106$ and $\chi^2/dof=43/48$ for BOOTES-clear and UVOT $white$ respectively and for the two power-law components, we achieved $\chi^2/dof=203/106$ and $\chi^2/dof=43/48$. While both models fit the UVOT $white$, for the higher sampled BOOTES-clear filter, both fits are still unacceptable at $>3\sigma$ level, although the two power-law model results in a smaller reduced $\chi^2$. The two power-law model appears to fit the data well, it is likely that the $\chi^2$ is large for BOOTES-clear due to the significant scatter in the data about the best-fit model.

The BOOTES photometry ends at 6ks, and in UVOT, there is an observing gap between 2ks and 5ks, after which the S/N of the light curve is poor. Therefore we created a single filter light curve by combining the optical light curves from BOOTES and UVOT in order to create a light curve with better sampling and high SNR as described in \S~\ref{combLC}. A broken power-law fit to this light curve results in an unacceptable $\chi^2/dof=284/35$. We, therefore, fit the light curve with a double broken power law. This resulted in a slightly improved $\chi^2/dof=256/33$; the {\it F}-test suggests the addition of a break is not required since the improvement is $<2\sigma$. However, on the basis that such a steep decay in the early optical light curve may be due to the reverse shock, resulting in two components producing the observed optical afterglow \citep{Wang_2015}, we also tested two-component models. Firstly we tested two power laws, which resulted in $\chi^2/dof=192/35$. The fit is an improvement on the previous models but is still unacceptable. We, therefore, also tried a power-law plus broken power-law model. The fit results in a $\chi^2/dof=109/33$ with the {\it F}-test suggesting the addition of a power-law to the broken power-law model is an improvement at $3\sigma$ confidence.

We note that when constructing the combined light curve, colour evolution was observed between the BOOTES and UVOT data,
with the UVOT $white$ and $V$ band data, during the initial steep decay, systematically lower than the BOOTES clear data.
The CCD of BOOTES is sensitive to much redder photons, covering the range 3000 {\AA} to 11000 {\AA} \footnote{http://www.andor.com/scientific-cameras/ixon-emccd-camera-series/ixon-ultra-888}, while the UVOT $white$ filter covers a wavelength range of between 1700-6500 {\AA} and the UVOT $V$ filter covers a range 5000-6000 {\AA}. This would suggest that the spectrum during the steep decay is redder in comparison to the rest of the afterglow. After normalization, the data were grouped together, weighted by the errors; since the early BOOTES and UVOT data do not align well, this may affect the measurement of temporal indices. However, we see that the initial steep decay
of the combined light curve is consistent with the tightly constrained steep decay measured with BOOTES-clear
and the less well-constrained UVOT $white$. The steep decay of the combined light curve is most similar to that of the BOOTES clear filter, as the BOOTES light curve has better SNR than the UVOT $white$ filter and thus is the dominant component when creating the weighted binned light curve.

 We noticed a break in the X-ray light curve at $\sim$ 2 ks. The temporal break may be due to the change in spectral behaviour or due to energy injection from the central engine, which we noticed as a transition from a shallow decay phase and followed by a steep decay ($1.50^{+0.02}_{-0.02}$). We sliced the X-ray light curve (count rate) into small temporal bins (see Figure \ref{Xray_optical_afterglow} (a)) to study the possible origin of this break. This resulted in 20 spectra; however, five bins do not have significant counts for the spectral modelling. The evolution of photon indices is shown in Figure \ref{Xray_optical_afterglow} (c, black circles). It indicates that there is no significant change in $\beta_{\rm x}$ (as it is expected for $\nu_{\rm c}$ passing through) close to the break in the X-ray light curve. When we fitted the photon index with a constant (\sw{CONS}) model, we got a very good fit, suggesting that there is no evolution. If we consider that $\nu_{\rm c}$ is between the BAT and LAT frequencies as suggested from SED 1 (29-65 sec, see \S~\ref{closure relation}), $\nu_{\rm c}$ could not pass through the X-ray band, inconsistent with expectations (i. e. t$^{-0.5}$), indicating that X-ray break is not due to the spectral break. We also tested the other possibilities of the origin of this break, such as energy injection. Considering adiabatic cooling with energy injection from the central engine, the inferred value of $\alpha_{\rm x}$ ($1.08 \pm 0.02 $) from observed $\beta_{\rm x}$ ($0.56 \pm 0.02$) matches with the observed value of $\alpha_{\rm x-ray}$ ($1.09 \pm 0.01$) for the spectral regime $\nu_m < \nu_{\rm x-ray} < \nu_c$ and for the slow cooling case in the ISM-like medium. We estimate the value of electron distribution index $ p = 2.12 \pm 0.04$ from $p= (2\beta + 1)$.  We notice that the inferred value of $p$ is close to what we obtained from modelling. We observed an early excess X-ray flux similar to those found in many other RS-dominated bursts. It could be possible to explain early X-ray excess as an energy injection episode lasting up to 2 ks. Though we did not find an achromatic break in the X-ray and optical light curves as expected due to the end of the energy injection episode \citep{2015ApJ...814....1L}, however, we notice a break in the optical data $\sim$ 6 ks (see Table \ref{lcfits_GRB140102A} and Figure \ref{Xray_optical_afterglow} (b)). 
 
 During the late time re-brightening phase in the unabsorbed X-ray flux light curve (at 10 keV), we notice the hardening in the photon index (see Figure \ref{Xray_optical_afterglow}) and in the hardness ratio (see Figure \ref{STE}). This could not be explained by the frequency crossing the X-ray band as the photon index reverses back to the original position after this phase. This unusual emission could be originating due to patchy shells or refreshed shocks \citep{2020ApJ...898...42C, 2021A&A...646A..50H}.

{\bf Soft tail emission:}

The observed soft tail emission from GRBs is useful to constrain the transition time from prompt emission to afterglow phase \citep{2007ApJ...669.1115S}. \swift BAT observed the soft emission from \thisgrbE until $\sim$ 200 sec. To understand the origin of this emission, we binned the BAT count-rate light curve based on Bayesian Blocks and performed the spectral analysis of each interval using \sw{pegpwrlw} model (a simple power-law model with pegged normalization \footnote{https://heasarc.gsfc.nasa.gov/xanadu/xspec/manual/node207.html}). The evolution of photon indices (black circle) calculated as mentioned above in BAT soft (15-25 \keV) energy channels are shown in Figure \ref{Xray_optical_afterglow} (c).  Also, spectral indices constrained using SEDs at 29-65 sec and 63-100 sec indicate agreement between BAT and optical emissions (after correcting for the RS contribution, in the case of SED at 29-65 sec, the optical data point is already within the BAT power-law uncertainty region due to large uncertainty on the index as the presence of a low signal in BAT). Furthermore, We notice that BAT photon indices do not show a rapid fluctuation and are consistent with those estimated at XRT frequencies during the soft emission phase. We also use early X-ray observations to investigate the underlying tail emission. We created the XRT light curve in soft (0.5-2.0 \keV) and hard (2.0-10.0 \keV) energy channels and examined the evolution of HR to identify spectral evolution at early epochs (see Figure \ref{STE}). We performed the temporal and spectral analysis for this early epoch ($\sim$ 47 sec to 200 sec post burst) XRT (WT mode) data. We find a temporal decay of 0.99$^{+0.03}_{-0.03}$ and spectral index equal to 0.59$^{+0.04}_{-0.04}$ during this epoch. These values are found to disagree with the expected closure relation for early X-ray observations to be prompt tail emission as described by \cite{2006ApJ...642..354Z} ($\alpha$ = 2 + $\beta$). Based on the above, we suggest that soft tail emission observed from \thisgrbE had possibly afterglow origin.

\begin{figure}[ht!]
\centering
\includegraphics[scale=0.33]{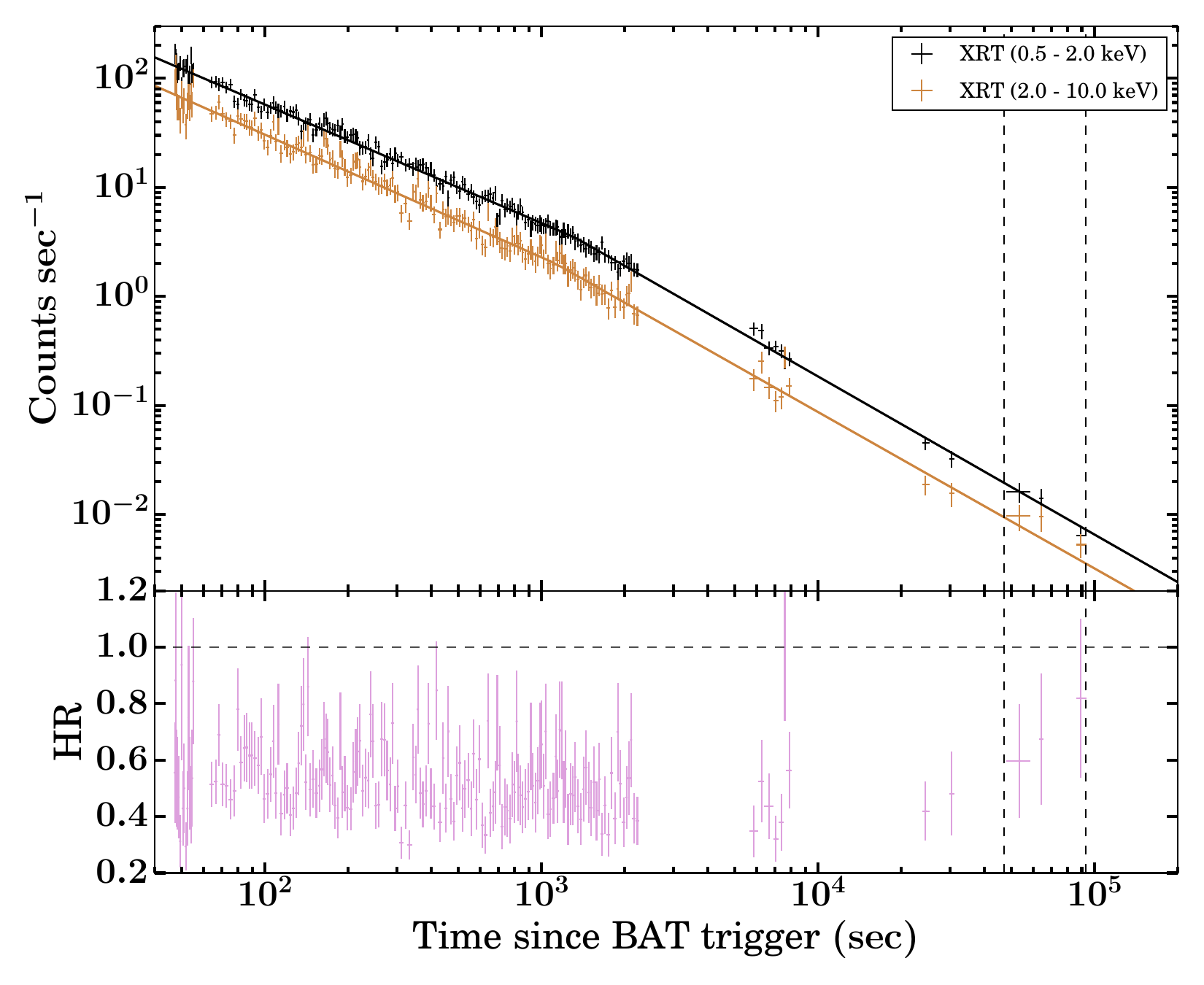}
\caption{The X-ray count rate light curves in soft (0.5-2.0 \keV) and hard (2.0-10.0 \keV) energy channels. The solid lines show the best-fit (broken power-law model) line to both the light curves. The bottom panel shows the evolution of HR in the same XRT energy bands. The vertical black dashed lines show the epoch of re-brightening activity in the X-ray light curve (at 10 \keV).}
\label{STE}
\end{figure}

{\bf Early Optical Afterglow and Reverse shock:}
The optical afterglow of \thisgrbE faded continuously after the first detection (\fermiT+ 29 sec, using BOOTES-4 robotic telescope) and showed a steep to shallow transition, not a simple power-law behaviour as is more usually observed in the light curve of optical afterglow. It is likely that the early optical afterglow light curve is produced by the RS emission, and $\sim$ \fermiT+ 100 sec presents the beginning of the dominance of the FS emission. In the case of \thisgrbE, we do not have optical observations before the peak time, but we constrain the decaying index after the peak as -1.72 $\pm$ 0.04 ($\sim$ \fermiT+100), and this decay index is consistent with the expected value for RS emission due to a thin shell expanding into the ISM-like medium. The closure relation for the RS II class i.e. $\alpha^{R}_{\rm dec, opt}$ = -(27p+7)/35 = - 1.74 considering the p = 2.00 (see Table \ref{sample_Modelling}), which is  consistent with the observed value. Furthermore, we also estimated the expected value of the rising index of the RS component using the closure relation i.e. $\alpha^{R}_{\rm ris, opt}$ = (6p-3)/2 =  4.50 considering the same value of $p$.

\subsection{Multi-wavelength Afterglow behaviour using physical modelling}
 
\subsubsection{{Optical and X-ray afterglow light curve modelling}}
\label{optical_afterglow_modelling}

The fireball synchrotron model for afterglows is currently the favoured scenario in terms of producing the observed multi-wavelength emission. In this model, the afterglow is a natural consequence of the beamed ejecta ploughing through the external medium and interacting with it, producing the observed synchrotron emission. The spectral and temporal behaviour of the afterglow emission could be described by several closure relations \citep[][see the last reference for a comprehensive list]{1998ApJ...497L..17S, 1997ApJ...476..232M, Sari:1999ApJ, dai01, 2009ApJ...698...43R}. This set of relations connects the values of $\alpha$ and $\beta$ to the power-law index of the electron energy distribution $p$, making it possible to estimate its value from the observations, which is typically found to be between 2 and 3 \citep{2002ApJ...571..779P,sta08,cur09}; the density profile of the external medium (constant or wind-like), and the relative positions of breaks in the synchrotron spectrum, primarily the synchrotron cooling frequency $\nu_{\rm c}$ and the synchrotron peak frequency $\nu_{\rm m}$. Another break frequency is called the synchrotron self-absorption frequency, though it does not influence the optical/UV or the X-rays during the timescales studied here. There are also closure relations describing temporal and spectral indices in the scenario that the afterglow is injected by an additional energy component \citep{2006ApJ...642..354Z}.

For jets having a single component, it is derived that the optical/UV and X-ray emissions are produced within the same region and therefore are explained by the same synchrotron spectrum, with the possibility that one or more of the break frequencies are between the two observing bands. This translates that the optical/UV and X-ray temporal indices, determined from an afterglow, should be described by closure relations that rely on the same assumptions about the ambient media, the electron energy index, $p$, and the energy injection parameter $q$.

The thin shell case ($T_{\rm dec} >$ \tninty) Type-II features  (also called flattening type, the peak of forward shock is below the reverse shock component) discussed by \cite{2015ApJ...810..160G} are used to interpret the optical emission. However, the forward shock emission dominates after $\sim 100$ sec. The optical emission in between $\sim 30-100$ sec follows the emission predicted by the reverse shock emission. This makes the early optical emission from this GRB a combination of RS+FS components. The RS model parameters for the early optical emission are useful to understand the magnetic energy available in the jet and the source environment \citep{2003ApJ...597..455K, 2000ApJ...542..819K}.  
 
The reverse shock crossing time is defined as $t_{x}$ = $\rm max (T_{\rm dec,}$ \tninty), which is important to estimate the peak emission for reverse shock. The break frequency evolution with time is $\nu_m^r \propto t^6$ for $ t< t_x$ and $ \nu_m^r \propto t^{-54/35}$ for $ t>t_x$, however, $\nu_m^f$ is constant  before crossing time and decreases with time after this as $t^{-3/2}$, where $r$ denotes the reverse shock and $f$ is used for forward shock \citep{2015ApJ...810..160G}. The cooling frequency $\nu_c^r \propto t^{-2}$ for $t< t_x$ and $\nu_c^r \propto t^{-54/35}$ after $t_x$. The same for forward shock is $\nu_c^f \propto t^{-2}$ for $t< t_x$ and $\nu_c^f \propto t^{-1/2}$ after crossing time. The maximum synchrotron flux is defined as $f_{\rm max}^r \propto t^{3/2}$ for $t<t_x$ and after crossing time it decreases as $t^{-34/35}$. Similarly, $f_{\rm max}^f \propto t^{3}$ for $t<t_x$ and independent of time after crossing time \cite{2015ApJ...810..160G}.  We model the afterglow emission and the parameters obtained to explain \thisgrbE are listed in Table \ref{sample_Modelling}. We have used Bayesian analysis \sw{PyMultiNest} software \citep{2014A&A...564A.125B} to estimate the afterglow modelling parameters and associated errors. A corner plot showing the analysis results is shown in Appendix A. In the right panel of Figure \ref{Xray_optical_afterglow}, we have shown the optical and X-ray light curves based on afterglow modelling. The parameters electron energy index $p$ (2.00 for RS and 2.19 for FS, respectively), micro-physical parameters $\epsilon_e$ and $\epsilon_B$ for the RS and FS are constrained using optical and X-ray data. The model can explain the optical emission and produces a slightly lower X-ray flux than observed at early times (the excess can be explained in terms of energy injection, see \S~\ref{xray_optical lc fitting}). However, in a few cases of GRBs, this kind of feature (excess X-ray emission) has been seen, and possible scenarios like (i) wavelength-dependent origin and (ii) mass loss evolution dependence are discussed by \cite{xray_access, 2020ApJ...896....4X}.

\subsubsection{Physical parameters of \thisgrbE and other thin shell cases}

\label{sample comparison}

In this section, derived parameters of \thisgrbE ($T_{\rm dec}$ and $ R_{\rm B}$) are compared with other well-studied thin shell cases of optical RS emission.
We collected the sample of confirmed optical RS cases consistent with the thin shell in the ISM medium from the literature (see Table \ref{sample_Modelling}) to the completeness of the sample. We estimated the expected decay index ($\alpha^{R}_{\rm dec, opt}$) using the closure relation i.e. $\alpha^{R}_{\rm dec, opt}$ = -(27p+7)/35 for RS component, deceleration time ($T_{\rm dec}$) and radius ($R_{\rm dec}$) of the blast-wave for each of such events. We used blast-wave kinetic energy, Lorentz factor, circumburst medium density, and redshift parameters from the literature to calculate the $T_{\rm dec}$ and $R_{\rm dec}$ of these bursts. We show the distribution of $\alpha^{R}_{\rm dec, opt}$ and $\rm R_{B}$  with ratio of deceleration time ($T_{\rm dec}$), and \tninty duration. Interestingly, we find that GRB 990123 and GRB 061126 do not follow the criteria of the thin shell, i.e., $T_{\rm dec}$ $>$ \tninty. However, these events have a large value of magnetization parameter ($R_{\rm B}$ $>>$ 1), and ($\alpha^{R}_{\rm dec, opt}$) $\sim$ -2 as expected from thin shell case of RS component in the ISM medium. This could be because of the dependency of \tninty duration on energy range and detector sensitivity. It could also be possible due to the prolonged central engine activity. In the sample of GRBs shown in Figure \ref{thin_sample}, bursts having RS emission ($T_{\rm dec} \ge$ \tninty) are considered. It is obvious from the figure that the observed $T_{\rm dec}$ values are spread over more than two magnitudes in time. We also notice that the deceleration radius of these events ranges from 1.25 $\times 10^{16}$ - 2.82 $\times 10^{17}$ cm, suggesting a diverse behaviour of ejecta surrounding possible progenitors. A larger sample of GRBs with RS detection is fruitful to understand if a thin shell case leads to dominant RS emission. If the prompt emission is bright, then any sub-dominant early optical afterglow may not be observed. In the case of \thisgrbE, the RS early optical afterglow is found with magnetization parameter $R_{\rm B} \sim$ 18, and this value lies {towards lower side} of the distribution of magnetization parameter for the sample of RS dominated bursts.

\begin{figure}[ht!]
\centering
\includegraphics[height=7.8cm,width=8.9cm]{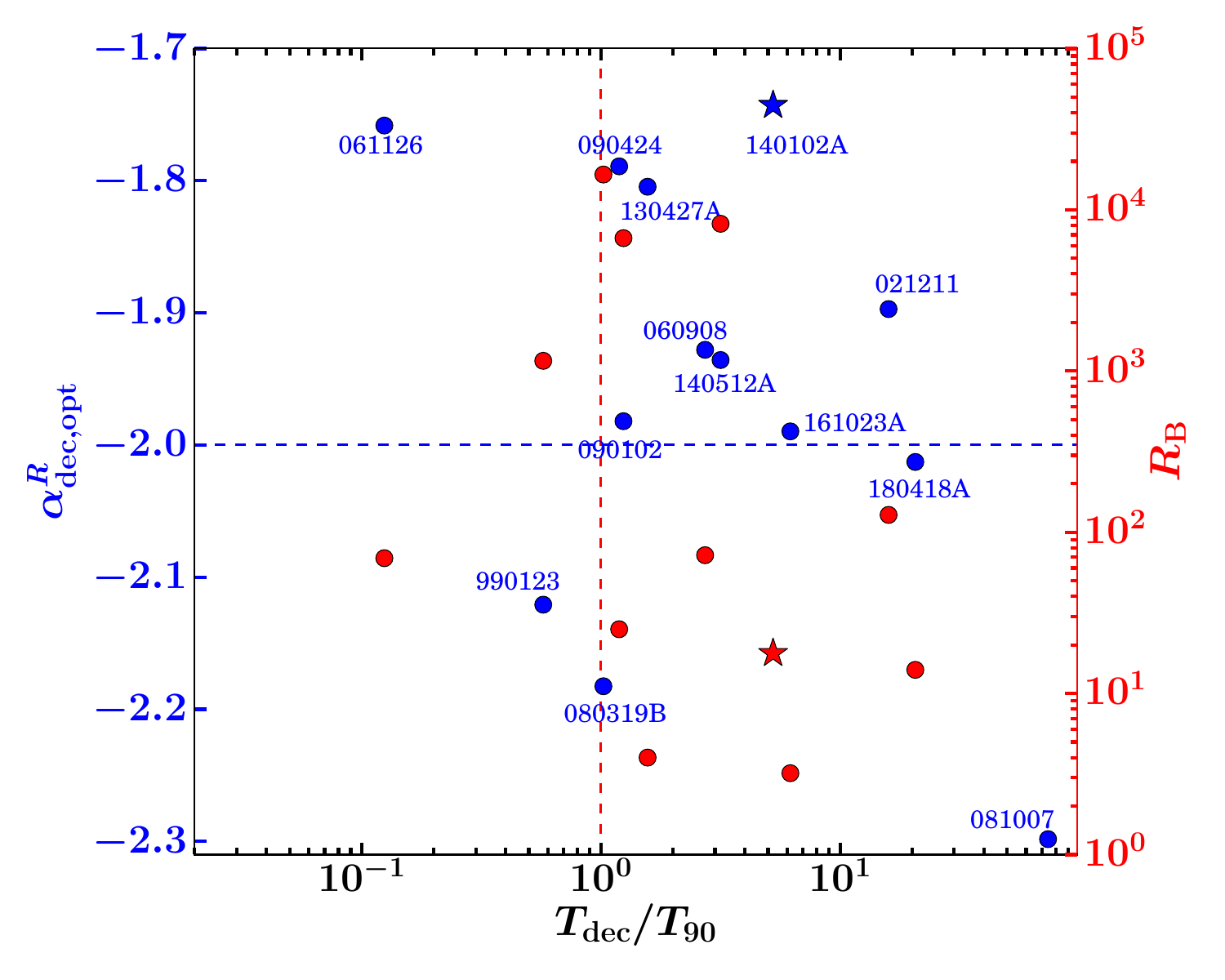}
\caption {A plot of decaying temporal indices of the RS emission ($\alpha^{\rm R}_{\rm dec, opt}$) versus the ratio of deceleration time ($T_{\rm dec}$) and \tninty duration (left side Y scale) taken from the literature (see Table \protect\ref{sample_Modelling}). The right side Y-scale corresponds to the distribution of the magnetization parameter. Blue and red stars show the position of \thisgrbE in this plane. The vertical red dashed line shows a ratio of $T_{\rm dec}$ and \tninty equal to one. The horizontal blue dashed line represents a line for $\alpha^{\rm R}_{\rm dec, opt}$ = -2.0.}
\label{thin_sample}
\end{figure}


Comparison of the temporal slopes derived using power-law fits between the optical and X-ray light curves of \thisgrbE throughout reveals an apparent mismatch. The optical emission has a steep to shallow transition, as predicted in the standard external shock model of RS and FS for a thin shell case. The optical light curve also displays a late time break. The X-ray emission exhibits a normal decay followed by a steeper decay. The optical and X-ray light curves are chromatic in behaviour.

To further investigate the chromatic nature of X-ray and optical afterglow, we performed a joint spectral analysis of the available simultaneous multiwavelength data. We created SEDs in five temporal segments, shown in Figure \ref{Xray_optical_afterglow}. The first two temporal segments are for the RS dominant phase (SED 1 and SED 2), and the last three segments are for the FS dominant phase (SED 3, SED 4, and SED 5). We describe the joint spectral analysis method in \S~\ref{SED} and present the results in Figure \ref{SED_fig}. The SED 1 is produced from the data observed with \swift BAT and extrapolated the index towards lower (optical) and higher ($\gamma$-ray) energies, which covers from 1.98 $\times 10^{-3}$ \keV to 5.5 $\times 10^{5}$ \keV. We used $\alpha_{\rm opt}-\beta_{\rm opt}$, $\alpha_{\rm x-ray}-\beta_{\rm x-ray}$ closure relations of RS and FS \citep{2013NewAR..57..141G} to constrain the $p$ value and position of the cooling-break frequency ($\nu_c$) in the slow cooling case of synchrotron spectrum for an ISM-like medium. We notice that the observed LAT flux is lower than the extrapolated value during SED 1, indicating the presence of a possible spectral break ($\nu_c$) between BAT and LAT frequencies. This is in agreement with spectral regime ($\nu~>~max \{ \nu_m,~\nu_c \}$) for the LAT photons discussed in \S~\ref{lat_TRS}.
 The optical and X-ray emission is consistent with $\nu_{\rm opt}$ $<$ $\nu_{\rm x-ray}$ $<$ $\nu_{\rm c}$ spectral regime. \cite{2018ApJ863138A} found that if the cooling break is either in between the XRT or LAT threshold energy or above this then the source can be detected in the LAT band which can be modelled using synchrotron emission. We calculated the $p$ value using the observed value of $\alpha_{\rm opt}-\beta_{\rm opt}$, $\alpha_{\rm x-ray}-\beta_{\rm x-ray}$ and find $p$ = 2.33 $\pm$ 0.21, this is consistent with that calculated from afterglow modelling (see \S~\ref{optical_afterglow_modelling}). The SED 2 is produced from the data observed with BOOTES optical telescope, \swift BAT, and \swift XRT from \fermiT+ 63 to \fermiT+ 100 sec. We consider a simple power-law model along with galactic and host X-ray absorption components for the joint BAT and XRT spectral analysis. We calculated a photon index of $\Gamma = -1.49 \pm 0.05$ (see Table \ref{SED_table}). Therefore, the soft and hard X-ray radiation during this temporal window should be originated from the same component. 
We calculated the $p$ value using the observed value of $\alpha_{\rm opt}-\beta_{\rm opt}$, $\alpha_{\rm x-ray}-\beta_{\rm x-ray}$ during this temporal window, and find $p$ = 2.10 $\pm$ 0.21, this is consistent with that calculated from SED 1.  

The SED 3 is produced from the data observed with BOOTES optical telescope, \swift UVOT, and \swift XRT from \fermiT+ 700 to \fermiT+ 1000 sec. In this case, the X-ray and optical could be described with a simple power law. The observed value of $\alpha_{\rm opt}-\beta_{\rm opt}$, $\alpha_{\rm x-ray}-\beta_{\rm x-ray}$  indicates that $\nu_{\rm c}$ was still beyond the $\nu_{\rm opt}$  and  $\nu_{\rm x-ray}$ spectral regime. We calculated the redshift and host extinction of the burst using this SED due to the availability of multi-band optical observations during this temporal window. Furthermore, we created two more SEDs (SED 4 and SED 5) at late epochs (23-31 ks and 51-65 ks) using optical and X-ray data. In these cases also, the X-ray and optical could be described with a simple power law, and it appears that the X-ray and the optical are on the same spectral segment ($\nu_{\rm opt}$ $<$ $\nu_{\rm x-ray}$ $<$ $\nu_{\rm c}$), though their light curve decay slopes are different (1.50 $\pm $0.02 and 1.11 $\pm$ 0.15 for X-ray and optical, respectively). However, they are still consistent within 3 Sigma.

\onecolumn
 
\begin{landscape}
\begin{tiny}
\begin{center}
\begin{longtable}{|c|c|c|c|c|c|c|c|c|c|c|c|} 
\caption{Afterglow modelling results from our theoretical fits for \thisgrbE using the combination of external reverse (RS) and forward shock (FS) model. We also present a sample of confirmed optical RS cases consistent with the thin shell in the ISM ambient medium to compare with the obtained parameters for \thisgrbE. For the present analysis, we have calculated the values of $T_{\rm dec}$ and $R_{\rm dec}$ for all such events. In the case of GRB 161023A, we assume $n$ = 1 cm$^{-3}$.}
\label{sample_Modelling} \\ \hline
\bf GRB/ &  {\boldmath$z$} & {\boldmath $T_{90}$} & \boldmath $p$ & \boldmath $\it \Gamma_{0}$& \boldmath $\epsilon_{\rm {e,f}}$  & \boldmath $\epsilon_{\rm {B,f}}$ & \boldmath $R_{\rm B}$  & \boldmath{$n_{0}$}  & \boldmath $E_{\rm K}$  & \boldmath$\eta_{\bf \gamma}$ & \boldmath $T_{\rm dec}/ R_{\rm dec}$  \\ 
 {\bf  (References)} &  & \bf  [sec]  &  & & $\bf [10^{-3}]$ & $\bf [10^{-5}]$ & \boldmath [$\epsilon_{\rm {B,r}}$ $/$ $\epsilon_{\rm {B,f}}$] & {$\bf [cm^{-3}]$} & $\bf [10^{52}  erg]$ &   & {\bf [sec] [$\bf \times ~10^{16}$ cm]}    \\ \hline
990123 & 1.60 & 63.3 &2.49 & 420 & 79.0 &  5 &  1156 & 0.3 & 108.0 & 0.2 $<$ $\eta_{\gamma}$ $<$ 0.9 &  36.36/ 14.80   \\ 
(1, 2, 3) &  &  & &  &  &   &   & &  &   &    \\ 
021211 & 1.006 & 2.41 &2.20 & 154 & 130.0  & 3  & 128 & 9.9 & 3.0 & $<$ 0.6&  38.45/ 2.73   \\
( 3, 4, 5 ) &  &  & &  &  &   &   & &  &   &   \\ 
060908 & 1.884 & 19.3 &2.24 & 107 & 14.0 &117 & 72 &  $190.0$ & 2.7 & 0.5 $<$ $\eta_{\gamma}$ $<$ 0.9 &   52.64/ 1.25  \\
(3, 6)&  &  & &  &  &   &   & &  &   &    \\ 
061126 &1.1588 & 191  &2.02 & 255 & 420.0 & 8 & 69 & 3.7  & 12.0 & 0.4 $<$ $\eta_{\gamma}$ $<$ 0.9 &  23.76/ 4.29  \\
(3,7)&  &  & &  &  &   &   & &  &   &    \\ 
080319B &0.937& $>$ 50 &2.57 & 286 & 68.0 &  4 & 16540 &  0.6  & 67.6 & $>$ 0.6 & 51.24/ 12.98 \\
(3, 8)&  &  & &  &  &   &   & &  &   &    \\ 
081007 & 0.5295 & 8.0 & 2.72 & 100 & - & -  & -  & 1 & 79  & 0.2 &  592.37/ 23.24  \\
(3, 9, 10, 11)&  &  & &  &  &   &   & &  &   &    \\ 
090102 & 1.547 & 27 &2.31 & 228 & 0.4 &  2 & 6666 &  359.0 &  816.0 & $<$ 0.4&  33.57/ 4.11 \\
(3, 12, 13)&  &  & &  &  &   &   & &  &   &    \\ 
090424 & 0.544 & 48 & 2.06 & 235  & 2.7 &  19 & 25 &  4.0 & 258 & $<$ 0.6 &  57.26/ 12.29 \\
(3, 10)&  &  & &  &  &   &   & &  &   &   \\ 
130427A & 0.34 & 162.83 &2.08 & 157& 3.3 & $22$ & 4 & $1.5$  & 521.0 & $<$ 0.8 & 255.34/ 28.18 5 \\
(3, 14, 15)&  &  & &  &  &   &   & &  &   &   \\ 
140512A & 0.725 & 154.8 &  2.25 & 112.3 & 290 &  0.00182 & 8187 & 9.7  & 765 & - &  490.05/ 21.50   \\ 
(16, 17, 18)&  &  & &  &  &   &   & &  &   &    \\ 
161023A &2.708& 80 & 2.32 & 140 & 4 & 1000 & 3.2 & - & 48 & - &  495.88/ 15.73  \\ 
(11, 19)&  &  & &  &  &   &   & &  &   &    \\ 
180418A & $<$ 1.31 & 1.5 &2.35 & 160 & 100 & 100 & 14 & 0.15 & 0.077 & -&  30.96/ 3.17 \\ 
(20, 21)&  &  & &  &  &   &   & &  &   &    \\ \hline
\bf 140102A & \bf 2.02$^{+0.05}_{-0.05}$ & \bf 3.58$^{+0.01}_{-0.01}$ & \bf 2.00$^{+0.01}_{-0.01}$ & \bf 218.98$^{+3.50}_{-3.67}$  & \bf 77.0$^{+6.7}_{-6.4}$ & \bf 420.0$^{+50.0}_{-40.0}$ & \bf 17.75 & \bf 0.70$^{+0.06}_{-0.05}$ & \bf 0.12 & \bf 0.99 & \bf 18.79/1.78  \\
(\bf Present work)&  &  & &  &  &   &   & &  &   &    \\ 
\hline
\end{longtable}
{(1) \cite{1999ApJ...518L...1B}; (2) \cite{1999GCN...224....1K}; (3) \cite{2014ApJ...785...84J}; (4) \cite{2006A&A...447..145V}; (5) \cite{2005ASPC..342..326P}; (6) \cite{2006GCN..5551....1P}; (7) \cite{2008ApJ...687..443G}; (8) \cite{2008GCN..7444....1V}; (9) \cite{2008GCN..8335....1B}; (10) \cite{2013ApJ...774..114J}; (11) \cite{2020ApJ...895...94Y}; (12) \cite{2009GCN..8766....1D}; (13) \cite{2010MNRAS.405.2372G}; (14) \cite{2013GCN.14455....1L}; (15) \cite{2013GCN.14470....1B}; (16) \cite{2014GCN.16310....1D}; (17) \cite{2014GCN.16258....1S}; (18) \cite{2016ApJ...833..100H}; (19) \cite{2018A&A...620A.119D}; (20) \cite{2019ApJ...881...12B}; (21) \cite{2020ApJ...905..112F}.}
\end{center}
\end{tiny}
\end{landscape}

\section{Summary and Conclusion}
\label{conclusions}

In this chapter, we have reported detailed early afterglow properties of \thisgrbE. Our afterglow modelling results suggest that the early ($\sim 30-100$ sec), bright optical emission of \thisgrbE can be well described with the RS model, and the late emission can be explained with the FS model. The RS model parameters for the early optical emission are useful to understand the magnetic energy available in the jet and the source environment \citep{2003ApJ...597..455K, 2000ApJ...542..819K}. We find that the value of $\epsilon_{\rm \bf{B,r}}$ is larger than $\epsilon_{\rm \bf{B,f}}$, which provides the value of magnetization parameter $R_{\rm B} \approx 18$. It demands a moderately magnetized baryonic jet-dominated outflow for \thisgrbE, similar to other cases of RS-dominated bursts \citep{2014ApJ...785...84J, 2015ApJ...810..160G}. We find a lower value of electron equipartition parameter for the reverse shock ($\epsilon_{\rm \bf{E,r}}$) than the commonly assumed value of $\epsilon_{\rm \bf{E,r}}$ =0.1.  \cite{2015ApJ...810..160G} suggest that a lower electron equipartition parameter in the external shock would lighten the `low-efficiency problem' of the internal shock model. We calculated the radiative efficiency ($\eta$= $E_{\rm \gamma, iso}$/($E_{\rm \gamma, iso}$+ $E_{\rm k}$)) value equal to 0.99 for \thisgrbE. Our model predicts a slightly lower X-ray flux during the early phase, similar to other cases of RS-dominated bursts. It might be some intrinsic property of the source, either the central engine activity or wavelength-dependent origin \citep{xray_access}. It will be beneficial to investigate the possible origin of excess X-ray emission for a larger sample in the near future. The closure relations indicate that the optical and X-ray emissions are consistent with $\nu_{\rm opt}$ $<$ $\nu_{\rm x-ray}$ $<$ $\nu_{\rm c}$ spectral regime for a slow cooling and ISM ambient medium. However, the observed LAT flux during the first SED lies below the extrapolated power-law decay slope, indicating the presence of a possible spectral break between BAT and LAT frequencies. We compare the physical parameters of \thisgrbE with other well-known cases of RS thin shell in the ISM-like medium and find that GRB 990123 and GRB 061126 have $T_{\rm dec}$ $<$ \tninty even after a larger value of the magnetization parameter ($R_{\rm B}$ $>>$ 1, i.e., thin shell case) and observed $T_{\rm dec}$ values being spread over more than two magnitudes in time. We also notice that the deceleration radius of these events spread over more than a magnitude (Table \ref{thin_sample}), suggesting a diverse behaviour of ejecta surrounding possible progenitors.

Overall, \thisgrbE provides a detailed insight into prompt spectral evolution and early optical afterglow along with GeV emission. In the future, many more observations of such early optical afterglows and their multiwavelength modelling of RS-dominated GRBs using different RS parameters might help to resolve open questions like low-efficiency problem, degree of magnetization, ejecta behaviour, environment, etc.               
\newcommand{\thisgrbG}{GRB~150309A\xspace}
\newcommand{\af}{\sw{afterglowpy}}

\chapter{\sc Dark GRBs and Orphan Afterglows}
\label{ch:61} 
\blfootnote{This chapter is based on the results published/accepted in A. J. Castro-Tirado, \& \textbf{Gupta, Rahul}, et al. 2023, \textit{AA} and \textbf{{Gupta}, Rahul} et al., 2022, {\textit{Journal of Astrophysics and Astronomy}, {\textbf{43}}, 11}.}

\ifpdf
    \graphicspath{{Chapter5/Chapter5Figs/PNG/}{Chapter5/Chapter5Figs/PDF/}{Chapter5/Chapter5Figs/}}
\else
    \graphicspath{{Chapter5/Chapter5Figs/EPS/}{Chapter5/Chapter5Figs/}}
\fi

\ifpdf
    \graphicspath{{Chapter5/Chapter5Figs/JPG/}{Chapter5/Chapter5Figs/PDF/}{Chapter5/Chapter5Figs/}}
\else
    \graphicspath{{Chapter5/Chapter5Figs/EPS/}{Chapter5/Chapter5Figs/}}
\fi

\normalsize

GRBs are also categorized based on the absence of emission in a particular wavelength; for example, the lack of optical emission leads to ``Dark bursts," and the absence of gamma-ray emission leads to ``orphan GRBs."

{\bf Dark GRBs:} The first GRB afterglow (in X-ray) associated with GRB 970228 was discovered in 1997 by BeppoSAX mission \citep{1997Natur.387..783C}. Later on, an optical afterglow was also detected from ground-based follow-up observations for the same burst at redshift $z$ = 0.695 \citep{1997Natur.386..686V}. However, soon after the first discovery of the optical afterglow, no optical counterpart associated with GRB 970828 was detected despite deep searches \citep{1998ApJ...493L..27G}, and the number of such GRBs (with an X-ray counterpart, but no optical counterpart) is increasing. These bursts are defined as ``dark burst’ or ``optically dim burst’ \citep{2004ApJ...617L..21J, 2003A&A...408L..21P}. In the first instance, the non-detection of optical counterparts was explained due to the delayed follow-up observations (the counterpart had faded significantly below the telescope's sensitivity limit) due to the unavailability of precise localization. However, after the launch of the \swift mission in 2004, rapid follow-up observations of afterglows helped reduce the fraction of dark GRBs, but still, a significant fraction of dark bursts exist. 

In the present era of \swift mission, dark GRBs have been defined in the framework of the most accepted fireball model of GRBs. \cite{2003ApJ...592.1018D} propose to define the dark burst using the ratio of optical to X-ray flux. \cite{2004ApJ...617L..21J} propose to use the optical to X-ray spectral index ($\beta_{\rm OX}$) to define the dark GRBs ($\beta_{\rm OX} < 0.5$). \cite{2009ApJ...699.1087V} propose to use the optical ($\beta_{\rm O}$) and X-ray ($\beta_{\rm X}$) spectral indices depending on the spectral regime. For example, $\beta_{\rm X}$ = $\beta_{\rm O}$ + 0.5, if the cooling frequency ($\nu_{c}$) of synchrotron spectrum is located in between the X-ray and optical frequencies and $\beta_{\rm X}$ = $\beta_{\rm O}$, for all other possible spectral regimes. Therefore, the possible range of $\beta_{\rm OX}$ in all the possible spectral regime is $\beta_{\rm X}$ -0.5 $\leq$ $\beta_{\rm OX}$ $\leq$ $\beta_{\rm X}$. In this context, they classified the dark GRBs by $\beta_{\rm OX} < \beta_{\rm X}-0.5$.  

There are various possible different factors responsible for the optical darkness of afterglows \citep{2011A&A...526A..30G}. i) optical afterglows could be intrinsically faint; ii) GRBs could be detected at a high redshift, because of which the Lyman-$\alpha$ forest emission will affect the optical emission; iii) Obscuration scenario, this could be due to dust in the GRBs host galaxies at larger distances or along the line of sight so that this could cause for a very reddened optical afterglow.

{\bf Orphan Afterglows:} In the present era of GRBs, many space-based missions such as \swift, \fermi, \kw, \AstroSat, {\it INTEGRAL}, etc., are continuously searching the whole sky for new GRBs candidates (prompt emission) with a large field of view (FOV). However, suppose a burst has a viewing angle ($\theta_{obs}$) greater than the jet opening angle ($\theta_{core}$). In that case, i.e., the case of off-axis observations, no gamma-ray emission will be detected as the prompt emission is beamed within an angle $1/\Gamma_{0} <$ $\theta_{core}$, where $\Gamma_{0}$ is the bulk Lorentz factor, \citep{2002ApJ...576..120T, 2014PASA...31...22G}. Therefore, space-based missions can only discover those GRBs whose jet is directed towards the Earth. But if the beaming angle intercepts the line of sight, multiwavelength afterglow can be detected. Such afterglows without any prompt emission detection are known as ``orphan afterglows." In the current era of survey telescopes having large FOV such as the Zwicky Transient Facility (ZTF), and coming facilities like the Large Synoptic Survey Telescope (LSST) help to discover more number of orphan afterglows.

{\bf Role of 3.6\,m DOT for dark and orphan GRBs observations:} Considering India's longitudinal advantage for the follow-up observations of GRBs, deep follow-up observations of possible afterglows of GRBs was occasionally performed \citep{2020GCN.27653....1K, 2020GCN.29148....1P, 2021GCN.29364....1G, 2021GCN.29490....1G, 2021RMxAC..53..113G} using India's largest 3.6-meter Devasthal Optical Telescope and other facilities located at Devasthal observatory of ARIES Nainital. The optical and near-infrared (NIR) back-end instruments of 3.6\,m DOT \citep{2020JApA...41...33S, 2018BSRSL..87...42P} offer spectral and imaging capabilities from optical to NIR wavelength and are very important for deep observations of afterglows and other fast fading transients. ARIES has a long history of more than two decays for the afterglow follow-up observations, including dark GRBs \citep{2003A&A...408L..21P}. Deep photometric observations of afterglows are essential for identifying the associated supernovae bumps observed in nearby long bursts and the dark nature of afterglows, revealing the cause of orphan GRBs, jet break, total energy, and their environment. On the other hand, spectroscopic observations are helpful for the redshift measurements of GRBs.

In the present chapter, we have studied a detailed analysis of two dark bursts (\thisgrbG and GRB 210205A) and an orphan afterglow (ZTF21aaeyldq). \thisgrbG is an intense (3.98 $\times$ 10$^{-5}$ erg cm$^{-2}$ using \fermi-GBM) two-episodic burst observed early on to $\sim$ 114 days post-burst. For this burst, an X-ray counterpart of the burst was discovered by \swift X-ray telescope \citep[XRT;][]{2005SSRv..120..165B} but no optical afterglow was discovered in spite of deep search using BOOTES and GTC telescopes. However, observations in the near-infrared (NIR) bands using CIRCE instrument mounted on GTC telescope reveals the detection of a very red afterglow. These observations suggest that \thisgrbG belongs to the ``dark bursts' subclass. For GRB 210205A, an X-ray afterglow was detected by \swift XRT, but we do not notice any optical counterpart despite deep follow-up observations using the 3.6\,m DOT telescope. In the case of ZTF21aaeyldq, no prompt emission was reported by any space-based satellites, and the ZTF survey discovered this source. The decay behaviour of the source confirms its orphan afterglow characteristic.

The outline of the present chapter is as follows: In \S~\ref{multiwavlength observation and data analysis dark}, we present the multiwavelength observations and data reduction of dark GRBs, \thisgrbG, and GRB 210205A, respectively. In \S~\ref{results for both the bursts dark}, we have given the results and discussion of dark GRBs. In \S~\ref{multiwavlength observation and data analysis Orphan}, we present the multiwavelength observations and data reduction of orphan afterglow (ZTF21aaeyldq). In \S~\ref{results for both the bursts orphan}, we have given the results and discussion of orphan afterglow. Finally, a brief summary and conclusion of the chapter are given in \S~\ref{conclusiones}. 

\section{Dark GRBs:}
\label{multiwavlength observation and data analysis dark}

\subsection{\thisgrbG}

\subsubsection{X-ray and Ultra-Violet observations}

The XRT instrument of \swift mission started follow-up observations of the BAT localization region at 23:05:18.2 UT (131.5 s after the BAT trigger) to search for X-ray afterglow. An uncatalogued counterpart (X-ray) candidate was discovered at RA, DEC = 18h 28m 24.81, +86d 25' 43.6" (J2000) within the \swift BAT error circle with a 90\% uncertainty radius of 1.4 arcsec \footnote{{https://www.swift.ac.uk/xrt\_positions/00634200/}} in the initial window timing mode (WT) exposure \citep{Cummings15}. The position of this fading afterglow was observed up to $\sim$ 4 $\times$ 10$^{5}$ s post-BAT detection. This work utilised X-ray afterglow data products, including both light curve and spectrum, obtained from the \swift online repository (available at \url{https://www.swift.ac.uk/}) hosted and managed by the University of Leicester \citep{2007A&A...469..379E, 2009MNRAS.397.1177E}. We conducted the analysis of the X-ray afterglow spectra acquired from the \swift XRT using the \sw{XSPEC} package \citep{1996ASPC..101...17A}. The XRT spectra were analysed within the energy range of 0.3-10 \keV. For the XRT spectral analysis, we used an absorbed power-law model to explain the spectral properties of the X-ray afterglow of GRB 150309A. We considered both the photoelectric absorption from the Milky Way galaxy (\sw{phabs} model in \sw{XSPEC}) and the host galaxy of the GRB (\sw{zphabs} model in \sw{XSPEC})) along with the power-law model for the afterglow. The absorption due to our Galaxy was set as a fixed spectral parameter with a hydrogen column density of $\rm NH_{\rm Gal} = 9.05 \times 10^{20} \, \rm cm^{-2}$ \citep{2013MNRAS.431..394W}. However, the intrinsic absorption  ({$\rm NH_{\rm z}$}) at the redshift value equal to two (the mean $z$ value for a typical long-duration GRB) was permitted to vary freely. In addition, the photon index value of the power-law component was left to vary. We have used \sw{C-Stat} statistics for the spectral fitting of XRT data.

The Ultra-Violet and Optical Telescope \citep[UVOT;][]{2005SSRv..120...95R} onboard \swift began observing the XRT localization region 140 s after the BAT trigger to search for UV/Optical afterglow. However, no credible UV or optical counterpart candidate was discovered within the \swift XRT error circle \citep{Cummings15, 2015GCN.17559....1O}. We extracted the UV afterglow data obtained using \swift mission following the method described in \cite{2021MNRAS.505.4086G}.

\subsubsection{Optical and near-IR afterglow observations}

\begin{figure}[ht!] 
\begin{center}
\includegraphics[angle=0,scale=0.25]{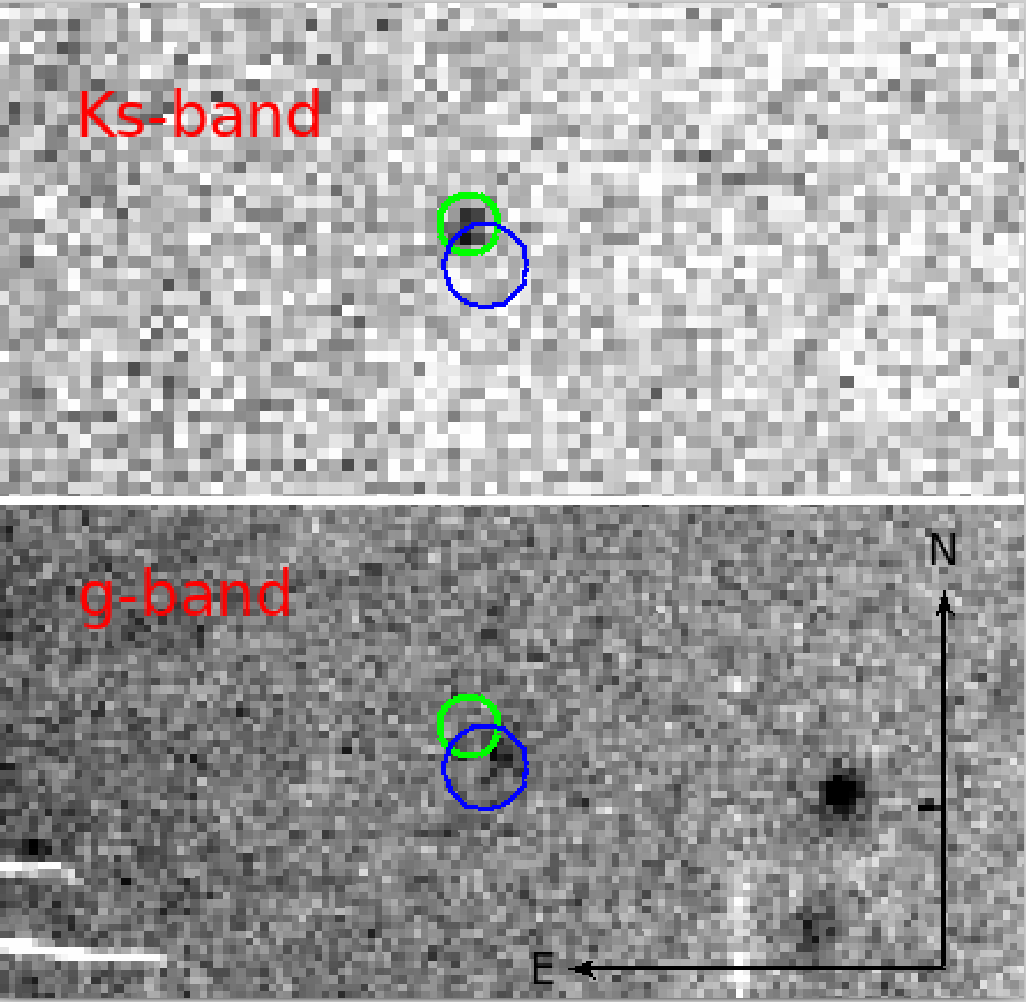} 
\caption{{The field of \thisgrbG.} {Top panel:} The $K_{\rm S}$-band discovery image (in green circle) of the afterglow of \thisgrbG taken at the GTC (+CIRCE) in 2015. The near-IR afterglow is close to the centre of the image. {Bottom panel:} The late-time $g$-filter image of the field observed using the GTC in July 2016. The potential host galaxy is close to the centre of the image (within the XRT error circle of radius 1.4 arcsec, shown with a blue circle).  North is shown in the upward direction, and East is shown towards the left direction.}
\label{carta JHK}
\end{center} 
\end{figure}

For \thisgrbG, soon after the detection of the X-ray counterpart by \swift XRT, many ground-based optical observatories (including BOOTES robotic telescope) started follow-up observations to search for the optical counterpart of \thisgrbG, although no optical afterglow (from early to late phases) candidate consistent with the BAT position was detected. Our later optical non-detections are consistent with the optical limits given by \citep{2015GCN.17565....1R} and \citep{2015GCN.17570....1M}.

We triggered the Target of opportunity (ToO) follow-up observations in the near-IR (JHK) starting 5.0~hr after the prompt discovery with the 10.4~m Gran Telescopio Canarias (GTC) equipped with the CIRCE instrument at the Spanish Observatory of La Palma. We detected a potential NIR counterpart candidate using GTC, although we could not find any afterglow emission at optical wavelengths. Additional near-IR observations were made at GTC on July 3 2015 (i.e., 114 days post-burst). Table \ref{optical/NIR observations} displays the optical and NIR observing log of \thisgrbG, the magnitude values listed in the table are given in the AB system. The $K_{\rm S}$-filter image of the \thisgrb field taken with the GTC (+CIRCE) in 2015 is shown in Figure \ref{carta JHK}.

To determine the magnitudes from the optical and nIR frames, we utilized the DAOPHOT routine under Image Reduction and Analysis Facility (IRAF)\footnote{IRAF is distributed by the NOAO, National  Optical Astronomy Observatories, which are operated by USRA, the Association of Universities for Research in  Astronomy, Inc., under cooperative agreement with the US National Science  Foundation. NSF.}.

\subsubsection{Search for potential host galaxy of \thisgrbG}
\label{Host Galaxy}

We performed deep optical imaging in $griz$ filters using the GTC telescope (see Table \ref{optical/NIR observations}). The deeper late-time optical photometric observations were gathered on July 19, August 24, August 25 (2015), July 07, July 30 (2016), and March 11 (2021). We carried out image pre-processing, such as dark subtraction and flat-fielding, using IRAF routines. Further, we performed the photometry of cleaned GTC images using standard IRAF software. To calibrate the instrumental magnitude, we completed the field calibration using the SDSS DR12 catalogue \cite{ala15}. AT the position RA= 18:28:24.67, and DEC= +86:25:44.16 (J2000), we detected a faint galaxy with the following magnitude values $g$ = 25.56 $\pm$ 0.27, $r$ = 25.26 $\pm$ 0.27, $i$ = 24.80 $\pm$ 0.22, and $z$ $\geq$ 24.4. The position of the candidate host galaxy is consistent with the XRT position (as shown in Figure \ref{carta JHK}). We also performed spectroscopy observations (covering the 3700 -- 7750 \AA wave range) with an exposure time of 3x1200 s on July 07 2016, using GTC. We used standard procedures for analyzing the OSIRIS spectra. Table \ref{optical/NIR observations} shows the optical and nIR photometry of the potential host. The reported magnitude values are not corrected for Galactic reddening. The calculated Galactic reddening value is E(B-V) = 0.1206 \citep{2011ApJ...737..103S}. The potential host galaxy of \thisgrbG is shown in Figure \ref{carta JHK}. A detailed method of host galaxy data analysis of GTC data is described in \cite{2021MNRAS.505.4086G}.

\begin{landscape}
\begin{scriptsize}
\begin{center}
\begin{longtable}{@{}lccccc@{}} 
\caption{The log of optical and near-infrared (nIR) images obtained on the \thisgrb field (top panel for the afterglow and bottom panel for the potential host galaxy). The magnitude values in different filters are uncorrected for the reddening due to our galaxy. The non-detection upper limits values are listed at 3 $\sigma$. The magnitudes listed are given in the AB system. $^{*}$ marker denotes the magnitudes in the Vega system.
} 
\label{optical/NIR observations} \\ \hline
Date   & T-\fermiT& Telescope/ & Filter/ & Exposure Time &  
Magnitude/ Upper limit \\ 
(UT, mid)  &  (s) & Instrument & Grism   &    (s)  & \\  
\hline 
09 Mar 2015, 23:05:53& 362  & 0.6m BOOTES-2/TELMA  & $--$ &       60   & $\geq$18.0   \\
09 Mar 2015, 23:05:53& 362  & 0.3m BOOTES-1        & $--$ &       60   & $\geq$16.5   \\
09 Mar 2015, 23:11:08& 677  & 0.6m BOOTES-2/TELMA  & $--$ &       120   & $\geq$20.2   \\
10 Mar 2015, 00:03:21& 3810  & 0.6m BOOTES-2/TELMA  & $i$ &       5400 &  $\geq$21.0   \\
10 Mar 2015, 01:22:34& 8563  & 0.6m BOOTES-2/TELMA  & $z$ &       3600 &  $\geq$19.2   \\
10 Mar 2015, 03:21:26& 15695  & 0.6m BOOTES-2/TELMA  & $Y$ &       9300 &  $\geq$18.0   \\
10 Mar 2015, 04:15&  18909 & GTC (CIRCE)    & $K_{\rm S}$ &      300   & 19.28 $\pm$ 0.11$^{*}$     \\ 
10 Mar 2015, 04:25& 19509  & GTC (CIRCE)    & $H$  &      300   & $\geq$ 21.4$^{*}$   \\
10 Mar 2015, 04:35& 20109  & GTC (CIRCE)    & $J$ &      300   &  $\geq$21.3$^{*}$   \\
10 Mar 2015, 05:09& 22149  & GTC (CIRCE)    & $K_{\rm S}$ &      300   &  19.50 $\pm$ 0.20$^{*}$   \\ 
\hline
03 Jul 2015, 00:30& 115.0625 (days)  & GTC (CIRCE)    & $K_{\rm S}$ &   1800   & $\geq$21.5$^{*}$   \\
03 Jul 2015, 00:40&  115.0696 (days) & GTC (CIRCE)    & $H$  &   1800   & $\geq$22.0$^{*}$   \\
\hline
Date   & T-\fermiT& Telescope/ & Filter/ & Exposure Time &  
Magnitude/ Upper limit \\ 
(UT, mid)  &  (days) & Instrument & Grism   &    (s)  & \\  
\hline 
19 Jul 2015, 01:31:50 & 131.1054  & GTC (OSIRIS)   & $r$  &   1800 (15x120 s)  & 25.26 $\pm$ 0.27 \\
24 Aug 2015, 00:34:18 & 167.0654  & GTC (OSIRIS)   & $i$  &   2160 (24x90 s)   &  24.89 $\pm$ 0.16 \\
25 Aug 2015, 23:33:37 & 169.0233  & GTC (OSIRIS)   & $i$  &   2160 (24x90 s)  &  25.09 $\pm$ 0.39 \\
07 Jul 2016, 01:32:06 & 485.1058 & GTC (OSIRIS)   &R1000B &  3600 (3x1200 s)  &   --    \\
30 Jul 2016, 00:26:21 &  508.0600 & GTC (OSIRIS)   & $g$  &   1200 (8x150 s)  &  25.56 $\pm$ 0.27 \\
30 Jul 2016, 01:20:48 & 508.0979  & GTC (OSIRIS)   & $i$  &   900 (10x90 s)  &  24.80 $\pm$ 0.22  \\
11 Mar 2021, 05:54:10 & 2193.2879  & GTC (OSIRIS)   & $z$  &   1890 (42x45 s) &   $\geq$24.4   \\
\end{longtable}
\end{center}
\end{scriptsize}
\end{landscape}

\subsection{GRB 210205A}

\subsubsection{X-ray afterglow observations and analysis}

The spacecraft slewed immediately to the burst location to search for the X-ray and optical/UV afterglows of GRB 210205A. The X-ray telescope (XRT) of \swift detected a new uncatalogued X-ray source (RA, Dec = 347.2214, 56.2943 (J2000)) $\sim$ 134.7 seconds since BAT detection \citep{2021GCN.29397....1D}. For the present work, we retrieved the X-ray data (both light curve and spectrum) products from the \swift XRT online repository \footnote{https://www.swift.ac.uk/} and followed the analysis methodology discussed in \cite{2021MNRAS.505.4086G}. The X-ray afterglow light curve has been presented in Figure \ref{xrtafterglow_210205A}. The evolution of X-ray photon indies in 0.3-10 \keV energy range has also been presented in Figure \ref{xrtafterglow_210205A}. The X-ray afterglow light curve could be best described with a power-law function with a temporal index of 1.06$^{+0.14}_{-0.12}$ (it is in good agreement with the result of the light curve fitting available at the XRT page\footnote{https://www.swift.ac.uk/xrt\_live\_cat/01030629/}). 

\begin{figure}[ht!]
\centering
\includegraphics[scale=0.34]{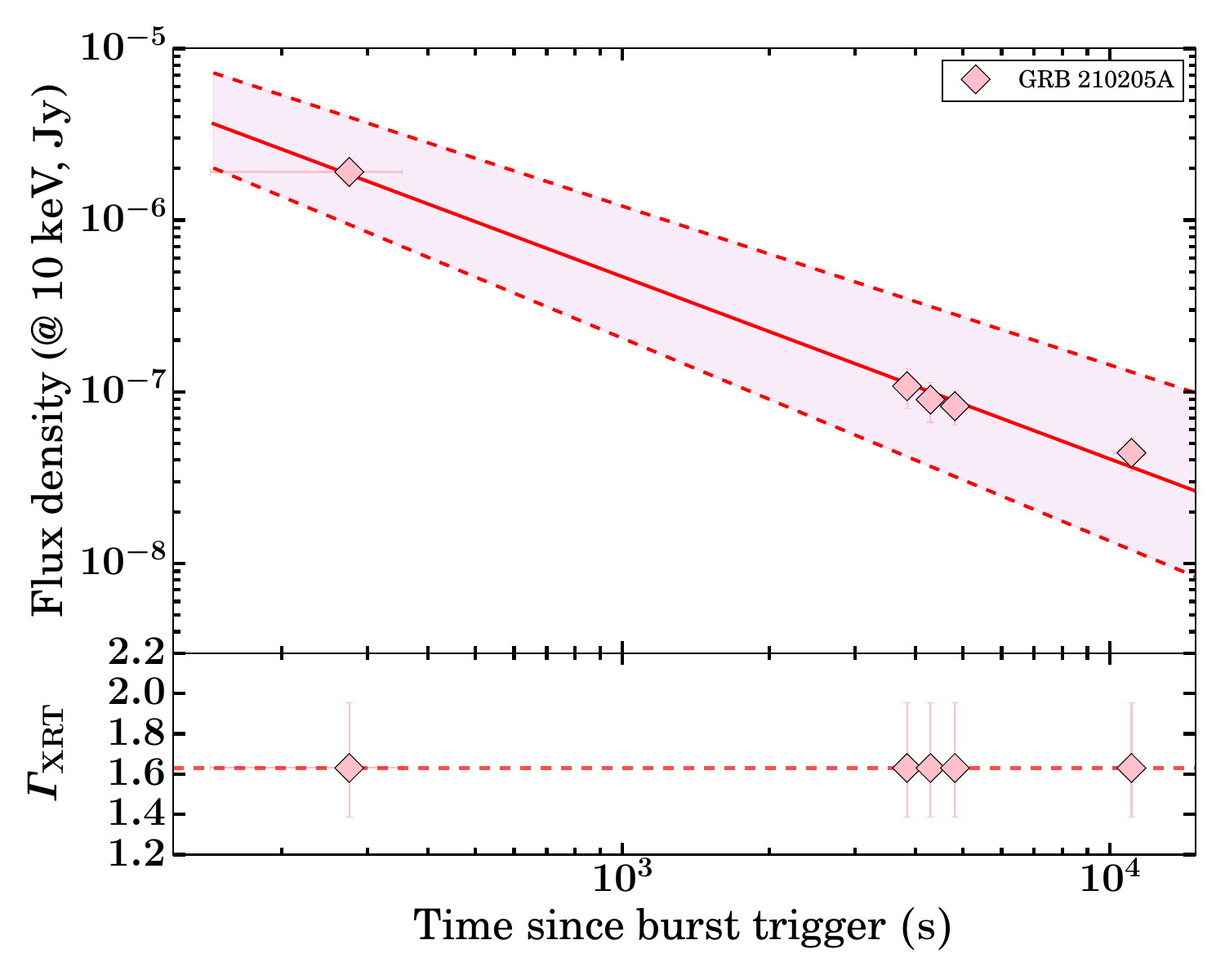}
\caption{{\it Top panel:} Temporal evolution of the X-ray afterglow of GRB 210205A along with a simple power-law model fit. {\it Bottom panel:} Evolution of photon indices in 0.3-10 \keV energy range.}
\label{xrtafterglow_210205A}
\end{figure}

\subsubsection{Optical follow-up observations and analysis}

\begin{table*}[t]
\large
\caption{The photometric observations log of GRB 210205A taken with 3.6\,m DOT. The tabulated magnitudes are in the AB magnitude system and have not been corrected for foreground extinction.}
\begin{center}
\begin{tabular}{c c c c c c}
\hline
\bf $\rm \bf T_{mid}$ (days) & \bf Exposure (s)  & \bf Magnitude  &\bf Filter & \bf Telescope & \bf References\\
\hline
1.0921 & 2 x 300 & $> 22.8$ & R & 3.6\,m DOT & Present work \\
1.1033 & 2 x 300 & $> 22.6$ & I & 3.6\,m DOT & Present work \\
\hline
\vspace{-2em}
\end{tabular}
\end{center}
\label{tab:observationslog:210205A}
\end{table*}

\swift Ultra-violet and Optical telescope \citep{2021GCN.29397....1D}, and Xinglong GWAC-F60A telescope \citep{2021GCN.29398....1X} started searching for the early optical emission, but no optical afterglow associated with GRB 210205A was detected. We performed the search for any new optical source using 0.6m Burst Observer and Optical Transient Exploring System (BOOTES) robotic telescope $\sim$ 1.18 hours post burst. We did not detect any new source within the \swift XRT enhanced position \citep{2021GCN.29399....1O, 2021GCN.29400....1H}. Furthermore, many other ground-based telescopes also performed deeper observations, but no optical afterglow candidate was reported \citep{2021GCN.29401....1F, 2021GCN.29402....1L, 2021GCN.29406....1H}.

\begin{figure}[ht!]
\centering
\includegraphics[angle=0,scale=0.35]{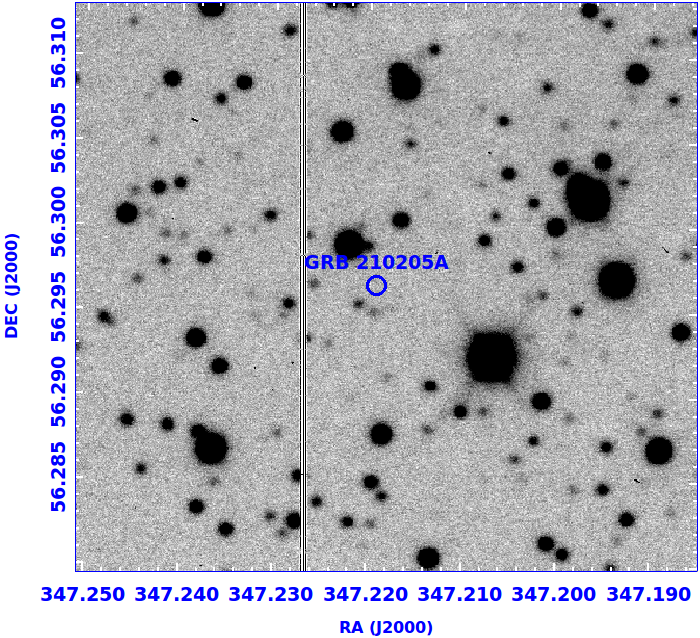}
\caption{The R-band finding chart of GRB 210205A obtained $\sim$ 1.10 days post burst using the 3.6\,m DOT. The field of view is $\sim$1$'$ $\times$ 1$'$, and the blue circle indicates the five arcsec uncertainty region at XRT ground localization.}
\label{fig:210205A}
\end{figure}

Further, we started the search for the optical afterglow of this XRT localized burst using the 15 $\mu$m pixel size 4K$\times$4K Charge-coupled device (CCD) Imager placed at the axial port of the newly installed 3.6\,m DOT of ARIES Nainital. The 4K$\times$4K CCD Imager is capable of deep optical imaging within a field of view of 6.5$'$ $\times$ 6.5$'$ \citep{2018BSRSL..87...42P, 2022JApA...43...27K}. Multiple frames with exposure times of 300 s each were taken in R, and I filters $\sim$ 1.10 days post burst \citep{2021GCN.29526....1P}. We do not find evidence of an optical afterglow source inside the XRT error circle, in agreement with other optical non-detections (see Table \ref{tab:observationslog:210205A}). We constrain the 3$\sigma$ upper limits ($>$ 22.8 mag in R and $>$ 22.6 mag in I filters, respectively). A finding chart obtained using 4K$\times$4K CCD Imager is shown in Figure \ref{fig:210205A}.

\section{Results and Discussion} 
\label{results for both the bursts dark} 

\subsection{\thisgrbG}

In the following subsections, we present the results of the afterglow analysis of \thisgrbG using X-ray to NIR band observations.  

\subsubsection{The nature of X-ray afterglow of \thisgrbG} 
\label{x-ray}

The X-ray afterglow observations using \swift/XRT shows a very steep decline $\rm \alpha_{\rm x1}$ = $-2.13^{+0.09}_{-0.09}$ from \swiftT+ 123 s to \swiftT+397 s followed by a shallower decline rate of $-1.05^{+0.03}_{-0.03}$ after \swiftT+397 s (see Figure \ref{fig:xrtlc} of the appendix). 
Further, we carried out \swift XRT spectral analysis for the temporal bins selected before and after the temporal break following \citep{2021MNRAS.505.4086G}. We calculated the spectral parameters for all the PC mode observations (interval after the temporal break) as the photon index value of $\Gamma_{\rm XRT} = 1.78^{+0.10}_{-0.10}$, and the intrinsic hydrogen density column $N_{\rm H, z} = 2.43^{+0.62}_{-0.57} \times 10^{22}$~cm$^{-2}$, considering $z$ = 2. Our intrinsic hydrogen density column measurement shows clear evidence of excess absorption over Galactic hydrogen column density. For the interval before the break, we have frozen the  $N_{\rm H, z}$ from the PC mode data and calculated $\Gamma_{\rm XRT} = 1.90^{+0.05}_{-0.05}$. Further, we also checked the evolution of photon indices from the \swift burst analyser page \footnote{{https://www.swift.ac.uk/burst\_analyser/00634200/}} and noted that photon indices do not change significantly during the entire emission phase. The unavailability of obvious spectral evolution between the data before and after the break plus the large change in the temporal index rules out the possibility of relating the temporal break at \swiftT+397 $\pm$ 48.60 s to the crossing of the cooling break frequency across the X-ray wavelength.

\begin{figure}[ht!]
\centering
\includegraphics[scale=0.33]{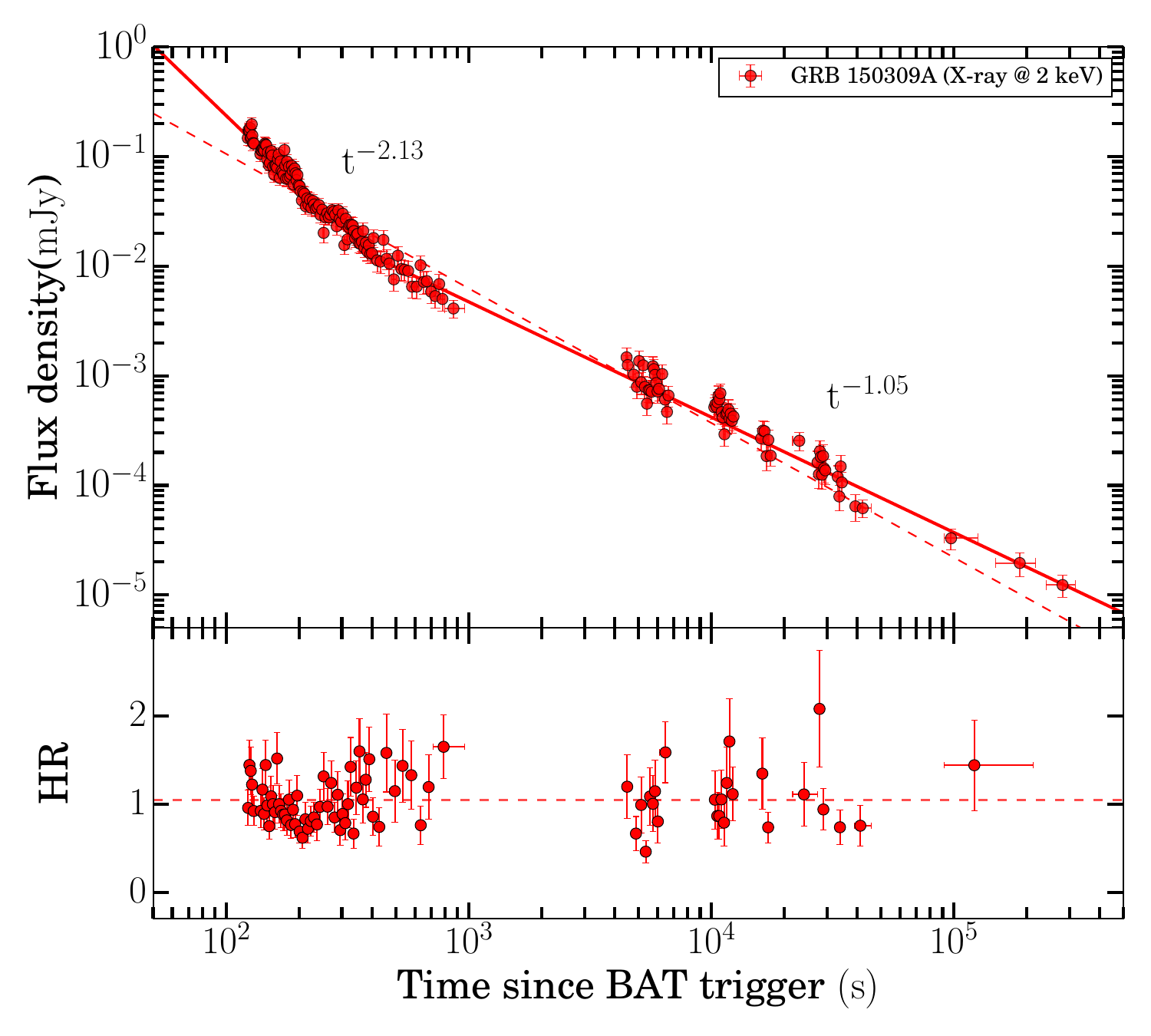}
\caption{{The X-ray flux light curve of \thisgrbG.} Top panel shows the XRT flux light curves at 2 \keV energy range. The light curve has been fitted with power-law and broken power-law (best fit) models. Bottom panel: Evolution of Hardness ratio in the XRT energy channel. The vertical red dashed line indicates the mean value of hardness ratio for \thisgrbG.}
\label{fig:xrtlc}
\end{figure}

The early steep decay emission observed before the break time ($<$ 400 s) could be interpreted as the tail (high-latitude emission) of the prompt emission. In this case, if the emission is purely non-thermal and consists of a single power-law, the temporal index can be expressed as $\alpha $\,=2\,+\,$\beta$ \citep{2000ApJ...541L..51K}. For the present burst, $\beta_{1}$ (X-ray spectral index during step decay phase)\,=0.90$\pm 0.05$, which results in an expected temporal index of $\sim$\,2.90, inconsistent with the observed decay (2.13\,$\pm 0.09$). Although, $\alpha $-$\beta$ relation during the early steep decay phase can be explained considering that the peak energy (\Ep) and flux during this phase rely on the viewing angle \cite{2011A&A...529A.142R}. This multi-peaked behaviour observed during the prompt emission can also cause the temporal indices inconsistent with the simple correlation between $\alpha $-$\beta$ during the high-latitude emission \citep{2009MNRAS.399.1328G}.

At later times ($>$400 s), the observed temporal and spectral indices ($\alpha_{2}$\,=\,1.05\,$\pm$\,0.03, $\beta_{2}$\,=0.78$\pm$\,0.10) are dominated by non-thermal emission because of the interaction of ejecta with the ambient medium (see section \ref{sed}). To investigate the nature of the circumburst medium (ISM or wind), we apply the closure relations between the temporal and spectral index during the shallow decay phase.

1.- Case when $\nu_{\rm X}$\,$<$\,$\nu_{\rm c}$:
\begin{itemize}
\setlength{\itemindent}{.5in}
        \item $\alpha_{\rm ISM}$\,=\,3$\beta_{2}$/2\,= 1.17$\pm0.15$
        \item $\alpha_{\rm wind}$\,=\,(3$\beta_{2}$\,+\,1)/2\,=\,1.67$\pm0.15$
\end{itemize}

2.- Case when $\nu_{\rm X}$\,$>$\,$\nu_{\rm c}$ (both mediums are indistinguishable):
\begin{itemize}
\setlength{\itemindent}{.5in}
        \item $\alpha$\,=\,(3$\beta_{2}$\,-\,1)/2\,=\,0.67$\pm0.15$
\end{itemize}

The case for an ISM environment with $\nu_{\rm X}$\,$<$\,$\nu_{\rm c}$ seems to be the best fit for our data. All these relations assume an electron spectral index, $p$\,$>$\,2. The electron spectral index obtained through the closure relations is $p$\,= 2.56 $\pm0.20$. Based on closure relations, all the cases with wind ($\rho$ $\propto$ $k^{-2}$) environment seems to be completely ruled out. Further, using the calculated values of temporal ($\alpha_{2}$\,=\,1.05\,$\pm$\,0.03) and spectral ($\beta_{2}$\,=0.78$\pm$\,0.10) indices during $\nu_{\rm X}$\,$<$\,$\nu_{\rm c}$ spectral regime and equation 1 of \cite{2014A&A...567A..84M}, we computed the density profile ($k$ = 0 for ISM and $k$ = 2 for wind) of the environment of \thisgrbG. We obtained $k$ $\sim$ -1.3, suggesting an intermediate density between ISM and wind ambient medium.

Additionally, we computed the value of jet kinetic energy ($E_{\rm K, iso}$) for an ISM environment with $\nu_{\rm X}$\,$<$\,$\nu_{\rm c}$ spectral regime utilizing the equation 11 of \cite{2018ApJS..236...26L}. We assumed typical micro-physical parameters (jet is due to Wiebel shocks and $\epsilon_{e}$ $\approx$ $\sqrt {\epsilon_{B}}$) to those found in the case of other well-studied GRBs, such as $\epsilon_{B}$ (energy fraction in the magnetic field) = 0.01, $\epsilon_{e}$ (energy fraction in electrons) = 0.1, and number density $n_{0}$ = 1 \citep{2002ApJ...571..779P}. We obtained $E_{\rm K, iso}$ = 5.36 $\times ~10^{53}$ erg for \thisgrbG.

\subsubsection{A potential near-IR afterglow of \thisgrbG and its confirmation} 
\label{sin emision} 

The fact that the XRT discovered the X-ray counterpart of \thisgrbG \citep{Cummings15}, enabled us to quickly identify a potential near-IR afterglow within the XRT X-ray error box, at coordinates RA (J2000) = 18 28 25.00, Dec(J2000) = +86 25 45.10 ($\pm$ 0$^{\prime\prime}$.4). The source remained stable (within errors) in brightness during the CIRCE observation window, with $K_{\rm S}$ = 19.28 $\pm$ 0.11 (Vega) and $K_{\rm S}$ = 19.50 $\pm$ 0.20 (Vega) at the beginning and end of the observation. The object was very red (Figure \ref{carta JHK}), and only upper limits were derived at the bluest nIR bands: $J$ $\geq$ 21.3 (Vega) and $H$ $\geq$ 21.4 (Vega). We calibrated the nIR observations utilizing the 2MASS Catalogue.

In order to confirm whether the highly reddened object (H-$K_{\rm S}$) $\geq$ 2 is the afterglow or an extremely reddened host galaxy, a second epoch observation at GTC (+CIRCE) was conducted on July 3. No source was detected with the following upper limits: $H$ $\geq$ 22.0 (Vega) and $K_{\rm S}$ $\geq$ 21.5 (Vega), implying that the afterglow faded by more than 2 mag in the $K_{\rm S}$-band after 114 days. This fading behaviour confirms that the highly reddened object is the NIR afterglow of \thisgrbG. 

\subsubsection{Spectral Energy Distribution of \thisgrbG} 
\label{sed}    

\begin{figure}[ht!]
\begin{center} 
\includegraphics[angle=0,scale=0.4]{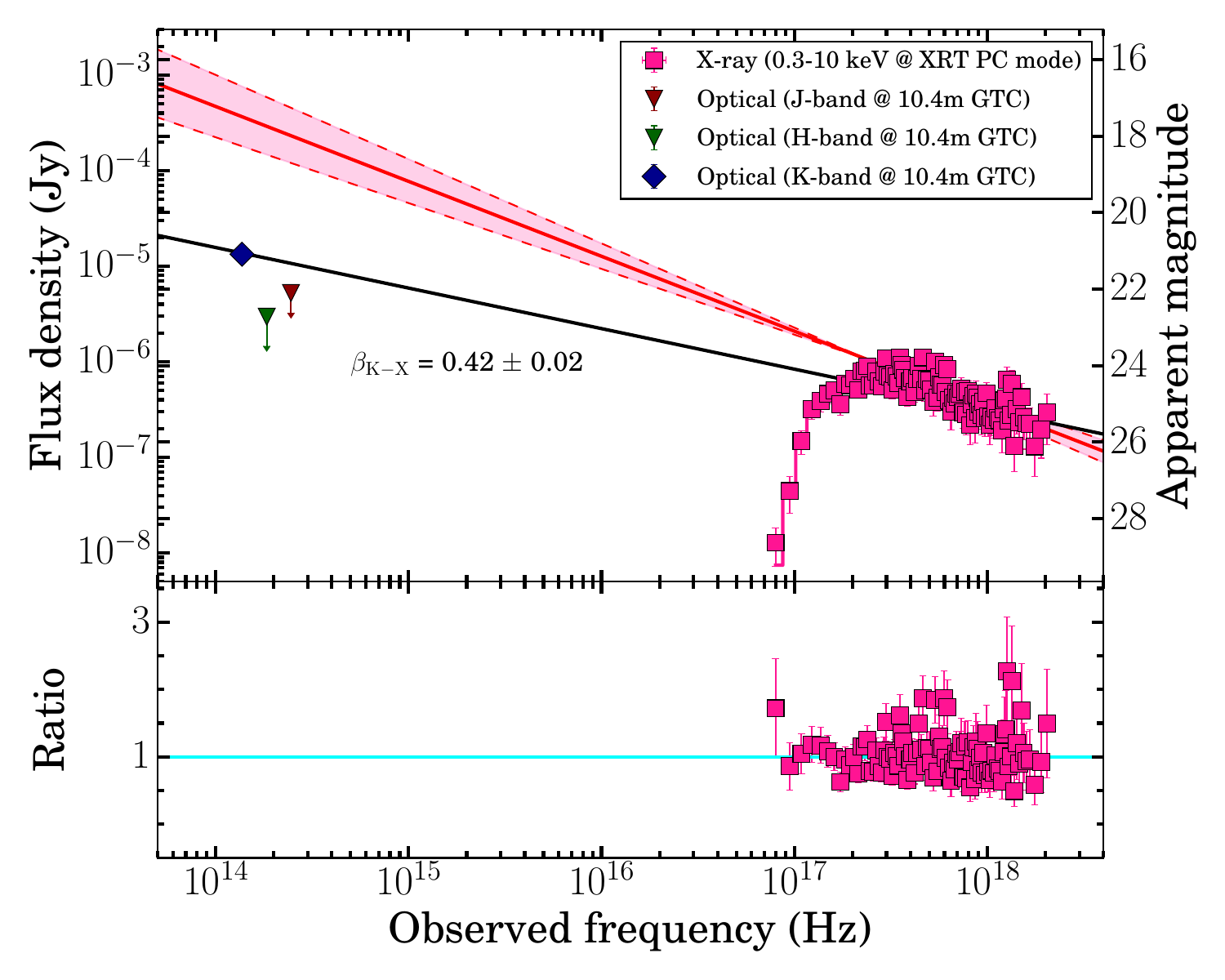} 
\caption{{The  The SED of \thisgrbG afterglow.} {Top panel:} The SED at \fermiT + 5.2 hours using the simultaneous X-ray data as well as the near-IR upper limits (and the K-band detection) reported in this paper. The solid red line indicates the X-ray spectral slop obtained using the best fit time-averaged PC mode spectrum, and the pink shaded region within the red dashed lines indicates the associated uncertainty with the X-ray spectral slop. The solid black line indicates the joint NIR-X-ray spectral index ($\beta_{\rm K-X}$). The left and right sides of the Y-axis, representing the flux density and magnitude values, are in the AB system. {\it Bottom panel:} Ratio of X-ray observed data and best-fit model. The horizontal cyan line indicates the ratio equal to one.}  
\label{OA sed}
\end{center} 
\end{figure} 

SED analysis using multi-wavelength data is very useful in constraining the location of synchrotron break frequencies and host extinction. We created the SED at \fermiT + 5.2 hours at the time of the GTC (+CIRCE) observations (to maximize the near-simultaneous broadband coverage). We have shown the SED for \thisgrbG in Figure \ref{OA sed}. The closure relations for the X-ray afterglow support that the cooling break frequency is beyond the X-ray spectral coverage ($>$ 10 \keV) for an ISM-like ($\rho$ = constant) ambient medium and slow-cooling regime at the epoch of the SED, supporting that no break in SED between X-ray and optical/NIR energies. Therefore, we extrapolated the unabsorbed spectral index of X-ray afterglow towards NIR wavelengths to calculate the extinction in the host galaxy. We noted that $K_{\rm S}$ band Galactic extinction corrected detection and upper limits in J and H filters using GTC telescope are situated much below the extrapolated X-ray spectral slope (see Figure \ref{OA sed}). Our SED analysis suggests that the host galaxy of the burst is highly extinguished, and \thisgrbG is a dark burst. We determined the $K_{\rm S}$ band host extinction (A$_{K_{\rm S}}$) = 3.60$^{+0.80}_{-0.76}$ mag) utilizing the X-ray to NIR SED. The calculated host extinction in the $K_{\rm S}$ filter corresponds to A$_{\rm V}$  = 34.67$^{+7.70}_{-7.32}$ mag, suggesting that \thisgrbG is one of the most extinguished bursts to date. We constrain the host galaxy extinction considering the Milky Way extinction law. We noted that such a high value of optical extinction (A$_{\rm V}$ =  34.67$^{+7.70}_{-7.32}$ mag) had been previously observed in the case of a few dark bursts, for example, GRB 051022 \citep{2007A&A...475..101C}, GRB 070306 \citep{2008ApJ...681..453J}, GRB 080325 \citep{2010ApJ...719..378H}, GRB 100614A, GRB 100615A \citep{2011A&A...532A..48D}.

\subsubsection{Possible origin of the optical darkness}

There are many possible explanations proposed for GRBs to be dark. In the case of \thisgrbG, the optical non-detections up to deeper limits indicate dark behaviour following the definition proposed at the early stages of their discovery \citep{1998ApJ...493L..27G}. In addition, we calculated the joint NIR-X-ray spectral index ($\beta_{\rm K-X}$) for \thisgrbG (the closure relation support for no spectral break between NIR to X-ray and cooling frequency lies beyond the X-ray energies) using the spectral energy distribution at \swiftT+5.2 hour and noted $\beta_{\rm K-X}$ = 0.42 $\pm$ 0.02 (see Figure \ref{sed}). Although, we could not estimate the optical-to-X-ray spectral index ($\beta_{\rm OX}$) at 11 hours post-trigger due to the unavailability of the optical observations near this epoch for \thisgrbG. The calculated value of $\beta_{\rm K-X}$ also supports that \thisgrbG is a dark burst such as GRB 051022 \citep{2007A&A...475..101C} and follows the definition described in \citet{2004ApJ...617L..21J}. Further, we plotted the evolution of NIR-X-ray spectral index vs X-ray spectral index for \thisgrbG (see Figure \ref{betaox}). We have also shown the $\beta_{\rm OX}$-$\beta_{\rm X}$ data points for other well-studied samples of \swift GRBs \cite{2012MNRAS.421.1265M, 2015MNRAS.449.2919L} for the comparison. Furthermore, \cite{2009ApJ...699.1087V} suggested the possible range of joint optical-X-ray spectral slope in different possible regimes: $\beta_{\rm X}$ -0.5 $\leq$ $\beta_{\rm OX}$ $\leq$ $\beta_{\rm X}$ following the external forward shock synchrotron model. We found \thisgrbG well satisfy the definition of darkness suggested by \cite{2009ApJ...699.1087V} in  $\nu < \nu_{\rm c}$ spectral regime.

\begin{figure}[ht!]
\centering
\includegraphics[angle=0,scale=0.35]{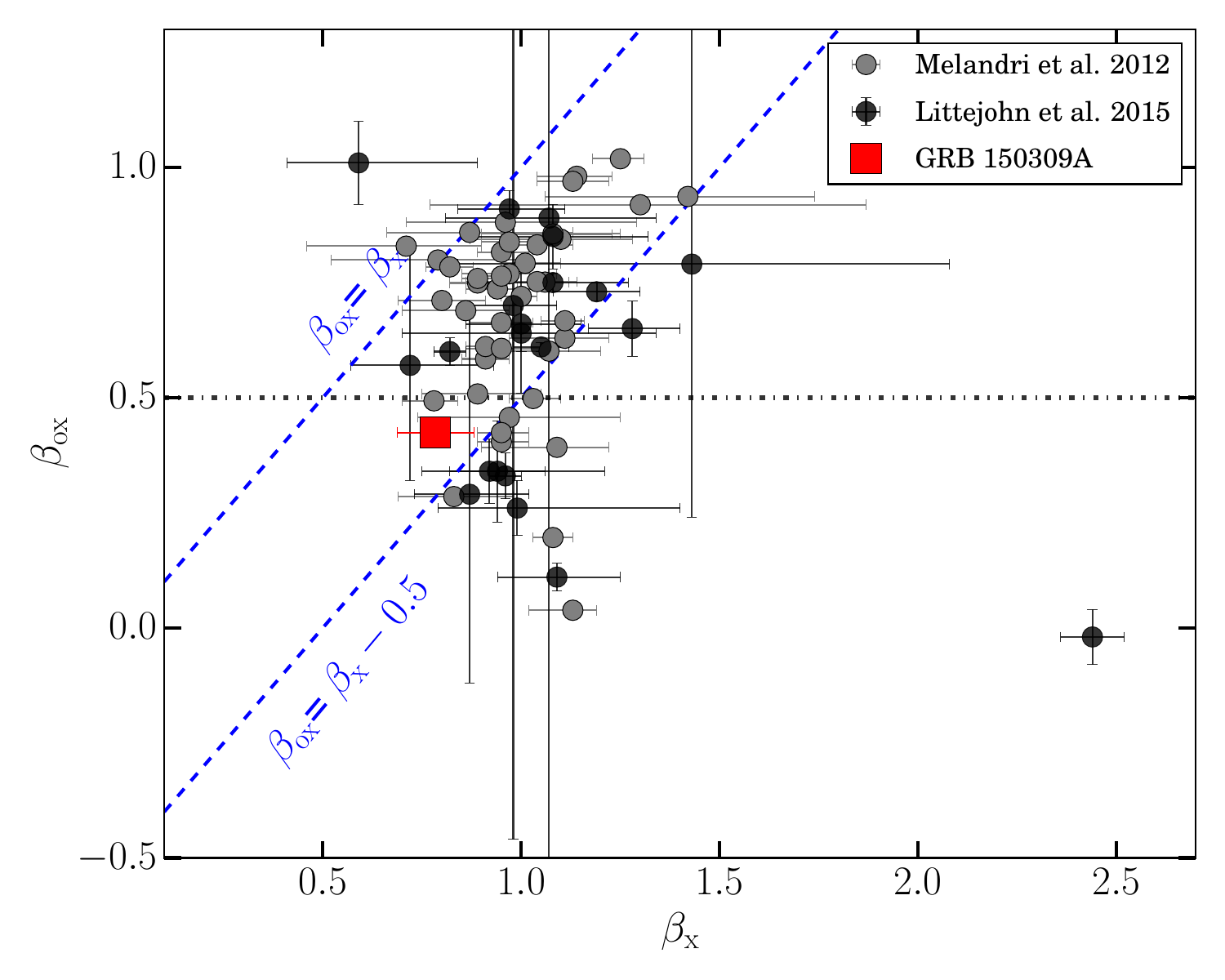}
\caption{The evolution of NIR-X-ray spectral index as a function of X-ray spectral index for \thisgrbG (presented with a red square). We have also shown the $\beta_{\rm OX}$-$\beta_{\rm X}$ data points for other well-studied samples of \swift GRBs \cite{2012MNRAS.421.1265M, 2015MNRAS.449.2919L} for the comparison. The horizontal black dashed line shows the $\beta_{\rm OX}$ $=$ 0.5.}
\label{betaox}
\end{figure}

We detect a source at the edge of the \swift XRT error circle (1.4 arcsec) with $K_{\rm S}$ = 19.28 $\pm$ 0.11 mag (Vega) in the first set of $K_{\rm S}$-band observations. We do not detect the afterglow either in the $H$ or the $J$ band data. We detected it again at $K_{\rm S}$ = 19.50 $\pm$ 0.20 mag (Vega) in the second set of $K_{\rm S}$ data (the magnitude difference is $\leq$1-sigma, so no apparent fading on this timescale of 20 minutes). The position offset of the $K_{\rm S}$-band afterglow is 1.1-arcsec $\pm$ 1.5-arcsec (including 1.4-arcsec error circle from \swift/XRT, and about 0.25-arcsec from CIRCE/2MASS astrometry), so this is definitely in the positional uncertainty region. 
The clear evidence of fading nature of the NIR source and its position consistent with the X-rays afterglow suggest that it is certainly the counterpart of the burst. Perhaps most interesting is the non-detection in H (bracketed in time by detection in the $K_{\rm S}$-band). The 2-sigma upper limit in H band observations is H $\geq$ 21.4 mag (Vega). This means that the counterpart is very red, with H-K $\geq$ 2.1 mag (95 \% confidence). Only two scenarios are possible: i) either a very dusty host galaxy with deeply embedded GRB (A$_V$ =  34.67$^{+7.70}_{-7.32}$ mag) or a very high-redshift Ly-alpha dropout ($z$ $\geq$10).

{\bf Dust extinguished scenario:} 
\label{obscuration} 

The observed characteristic of \thisgrbG indicates that it is positioned undoubtedly in the dark burst region of the $\beta_{\rm OX}$ $-$ $\beta_{\rm X}$ diagram (see Figure \ref{betaox}). It is clear that the optical afterglow of \thisgrbG could not be detected because of the obstruction of sight.
We carried out SED analysis (at \swiftT + 5.2 hours) to determine the host extinction at NIR wavelengths. Our SED analysis suggests $K_{\rm S}$ band brightness equal to $\sim$ 17.49 mag (the intrinsic afterglow brightness without any extinction obtained from the extrapolation of the unabsorbed spectral index of X-ray afterglow towards $K_{\rm S}$ band), from which we found a large host extinction at NIR wavelengths, with A$_{K_{\rm S}}$ = 3.60$^{+0.80}_{-0.76}$ mag (see section \ref{sed}). 

{\bf A high redshift scenario:} 

From our deep afterglow/host observations using GTC, no optical counterpart or confirmed associated host galaxy (see section \ref{host SED}) is detected. Therefore, we could not calculate the exact redshift of the burst. To determine the redshift and to examine the high redshift possibility for optical darkness, we utilized prompt emission Amati correlation (the correlation between rest frame \Ep and isotropic gamma-ray energy). We have used time-averaged spectral parameters and bolometric fluence in the source frame for different values of redshift ranging from $z$= 0.1 to $z$= 10. The Amati correlation for \thisgrbG at different redshift values is shown in Figure \ref{amati}. For the comparison, we have also shown the other long bursts studied by \cite{2012MNRAS.421.1256N}. This analysis indicates that the isotropic gamma-ray energy of \thisgrbG is such that it would not be excessive even for $z \geq$ 10. The very faint/non-detection of the afterglow in NIR/UV/optical filters also supports a high-$z$ origin. In addition, the prompt emission duration of \thisgrbG would be $\sim$ 5 s in the source frame at $z \geq$ 10, consistent with a long GRB. 

However, the high redshift origin of \thisgrbG is challenged based on the following argument: considering a high redshift origin of \thisgrbG, the soft X-rays photons in the source frame will be shifted out of the XRT energy coverage (0.3-10 \keV). Therefore, even a large column density in the source will result in little attenuation. Furthermore, we utilized the X-ray afterglow spectrum to constrain the redshift of \thisgrbG. We fitted the X-ray PC mode spectral data and derived (observer frame) considering $z$ = 0. The measured column density is higher than that of Galactic column density, and this excess column density is useful to estimate the limit on $z$. We have used the following equation to constrain the limit on $z$ \cite{2007AJ....133.2216G}:

\begin{equation}
\log{(1+z)} < 1.3 - 0.5 \,\log_{10}{(1 + \Delta N_H)},
\end{equation}

where $\Delta N_H$ represents the difference between column density measured considering $z$ = 0 ($N_H \sim 23.06 \times 10^{20} \mathrm{cm^{-2}}$) and Galactic ($N_H \sim 9.05 \times 10^{20} \mathrm{cm^{-2}}$) taken from \cite{2013MNRAS.431..394W}. The above equation indicates that \thisgrbG has a redshift value $z<4.15$.

\begin{figure}[ht!]
\centering
\includegraphics[angle=0,scale=0.4]{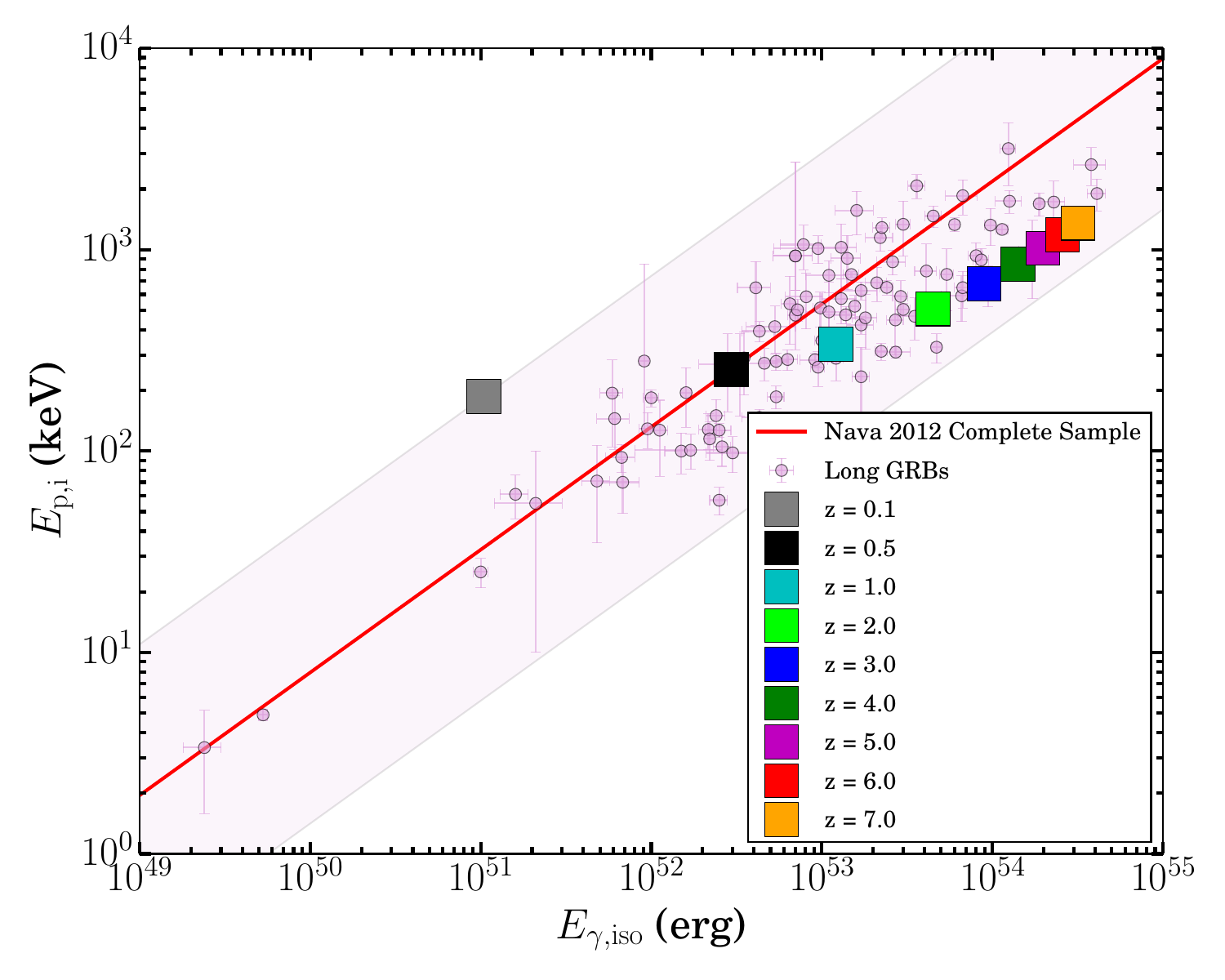}
\caption{Amati correlation for \thisgrbG. As there is no redshift measurement is available, we have vary the redshift from 0.1 to 10. For the comparison, we have also shown the other long bursts studied by \cite{2012MNRAS.421.1256N}. The solid red line shows the best-fit line, and the pink shaded band shows the associated 2-$\sigma$ uncertainty using the study of a sample of long bursts by \cite{2012MNRAS.421.1256N}.}
\label{amati}
\end{figure}

{\bf Nature of the potential host galaxy of \thisgrbG:} 
\label{host SED} 

\begin{figure*}
\centering
\includegraphics[scale=0.3]{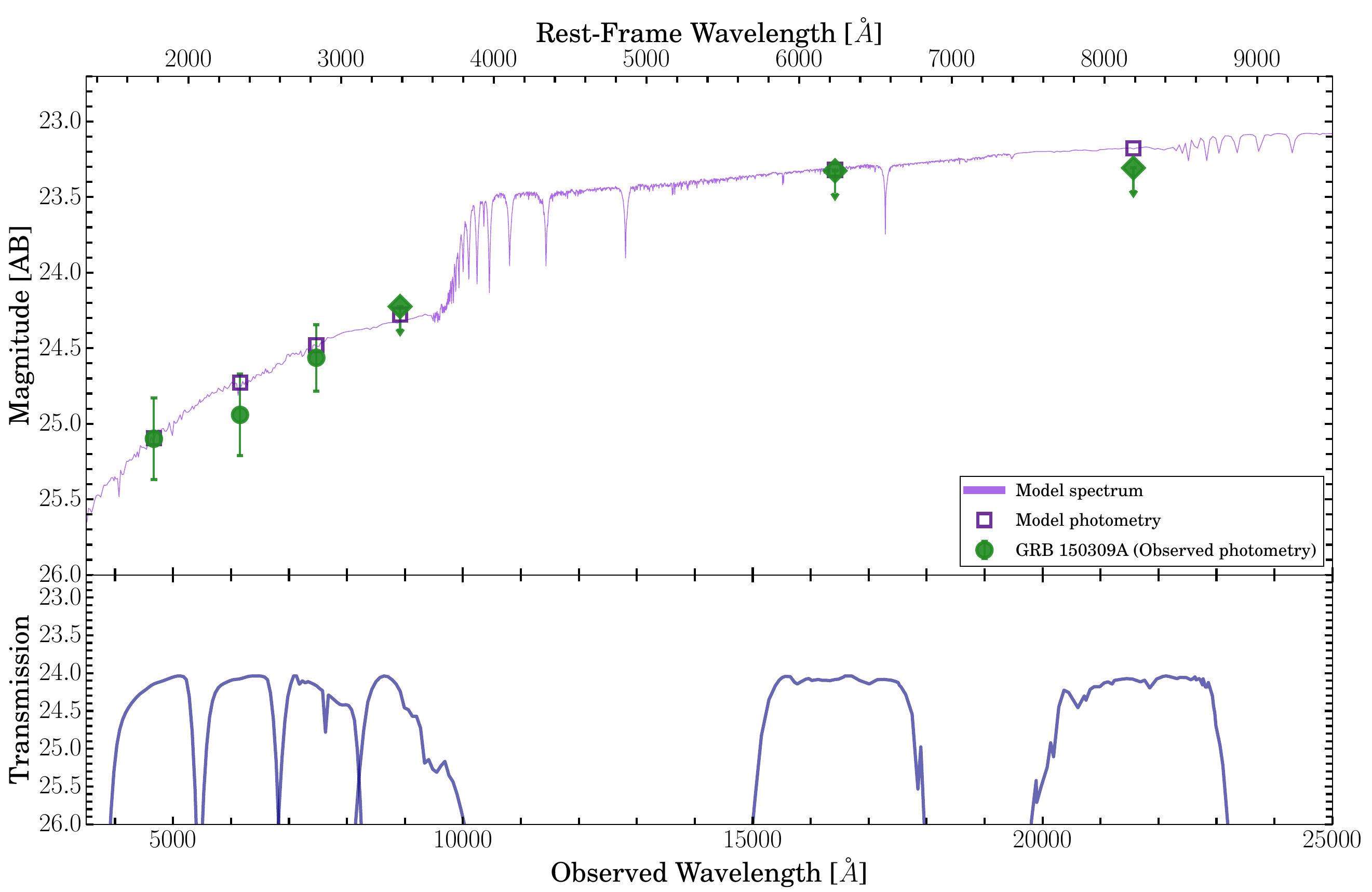}
\caption{Spectral energy distribution modelling of the potential host galaxy (from the g-band to the $K_{\rm S}$ band) of \thisgrbG using \sw{Prospector} software. The SED fitting shows no evidence of internal reddening. The bottom plot shows the transmission curves of the corresponding filters.}
\label{PotentialHG-SED}
\end{figure*}	  

Deeper observations searching the host of \thisgrbG found a potential host with an angular separation of $\sim$ 1.1$''$ from the NIR afterglow position (see section \ref{Host Galaxy}). The potential host is a faint ($r\sim 25.26 \pm 0.27\,\rm{mag}$) one with typical $g-i$ colour measured for galaxies at moderate redshift values detected in the optical bands (see Tab. \ref{optical/NIR observations}). The observed spectrum of the potential host galaxy taken using GTC was very noisy, and there is no emission-like feature up to $z$=1.08, discarding the low-redshift possibility (the non-detection of UV emission also discards the low redshift possibility). So, we executed the SED modelling of the potential galaxy of \thisgrbG utilizing photometric data and \sw{Prospector} software. \sw{Prospector} is a python-based stellar population code developed for the host galaxy SED modelling using both photometric and spectroscopic observations \citep{2017ApJ...837..170L, 2021ApJS..254...22J}. In order to model the observed data, \sw{Prospector} applies a library of FSPS (Flexible Stellar Population Synthesis) models \citep{2009ApJ...699..486C}. We used the dynamic nested sampling fitting routine \sw{dynesty} on the observed photometry of the potential host galaxy and calculated the posterior distributions of host galaxy parameters. To determine the stellar population properties of the potential host galaxy, we used the \sw{parametric$\_$sfh} model. This model enabled us to calculate several key host properties, including the stellar mass formed ($M_\star$) in solar mass units, host galaxy age ($t$), stellar metallicity ($\log Z/Z_\odot$), dust attenuation (A$_{\rm V}$), and the star formation timescale ($\tau$). In our SED analysis, we used these host galaxy model parameters as free variables. A detailed method of host galaxy SED modelling is described in \cite{2022JApA...43...82G}. Due to the limited number of data points, we also included the limiting mag values of $z$, $H$, and $K_{\rm S}$ filters for the SED analysis. We assumed redshift as a free parameter to constrain the photometric redshift of the potential host galaxy. Figure \ref{PotentialHG-SED} and Figure \ref{PotentialHG-SED_corner} (in the appendix) show the SED fitting and corresponding corner plot of the potential host galaxy of \thisgrbG, respectively. We determined the following parameters using SED modelling: stellar mass formed ($M_\star$) = 10.66$^{+0.44}_{-0.54}$, stellar metallicity ($\log Z/Z_\odot$) = -2.27$^{+1.12}_{-0.83}$, age of the galaxy ($t$) = 5.80$^{+4.77}_{-3.66}$, dust attenuation in rest-frame (A$_{\rm V}$) = 0.97$^{+0.34}_{-0.37}$ mag, and star formation timescale ($\tau$) = 1.68$^{+0.82}_{-0.83}$. Considering the redshift 1.88 obtained from SED modelling, we calculated the physical offset $\sim$ 9.5 kpc between the centre of the potential host galaxy and the NIR afterglow position. The measured physical distance is very large compared to those typically observed for long GRBs, suggesting that the galaxy might not be related with \thisgrbG. To further explore the nature of the potential galaxy, we calculated the chance coincidence probability of the candidate galaxy following the method described in \cite{2002AJ....123.1111B}. Using the measured brightness of the galaxy in the r-band and the observed offset values, we derived a chance alignment (P$_{cc}$) of about 5 \%. The derived P$_{c}$ value is small but still significantly high, indicating the candidate galaxy to be the host. In literature, authors have used a diverse range of P$_{cc}$ values, i.e., $\sim$ 1-10\%, to establish the association of faint galaxies with such transients. For example, \cite{2002AJ....123.1111B, 2010ApJ...722.1946B, 2022MNRAS.515.4890O}, use Pcc$<$10\%, \cite{2013ApJ...769...56F} use Pcc$<$5\%, and \cite{2014MNRAS.437.1495T} use 1-2\%, to identify if a galaxy is related to afterglow or not. In light of the above analysis, it is hard to decipher whether the observed candidate galaxy is the host of \thisgrbG.

\begin{figure*}
\centering
\includegraphics[scale=0.27]{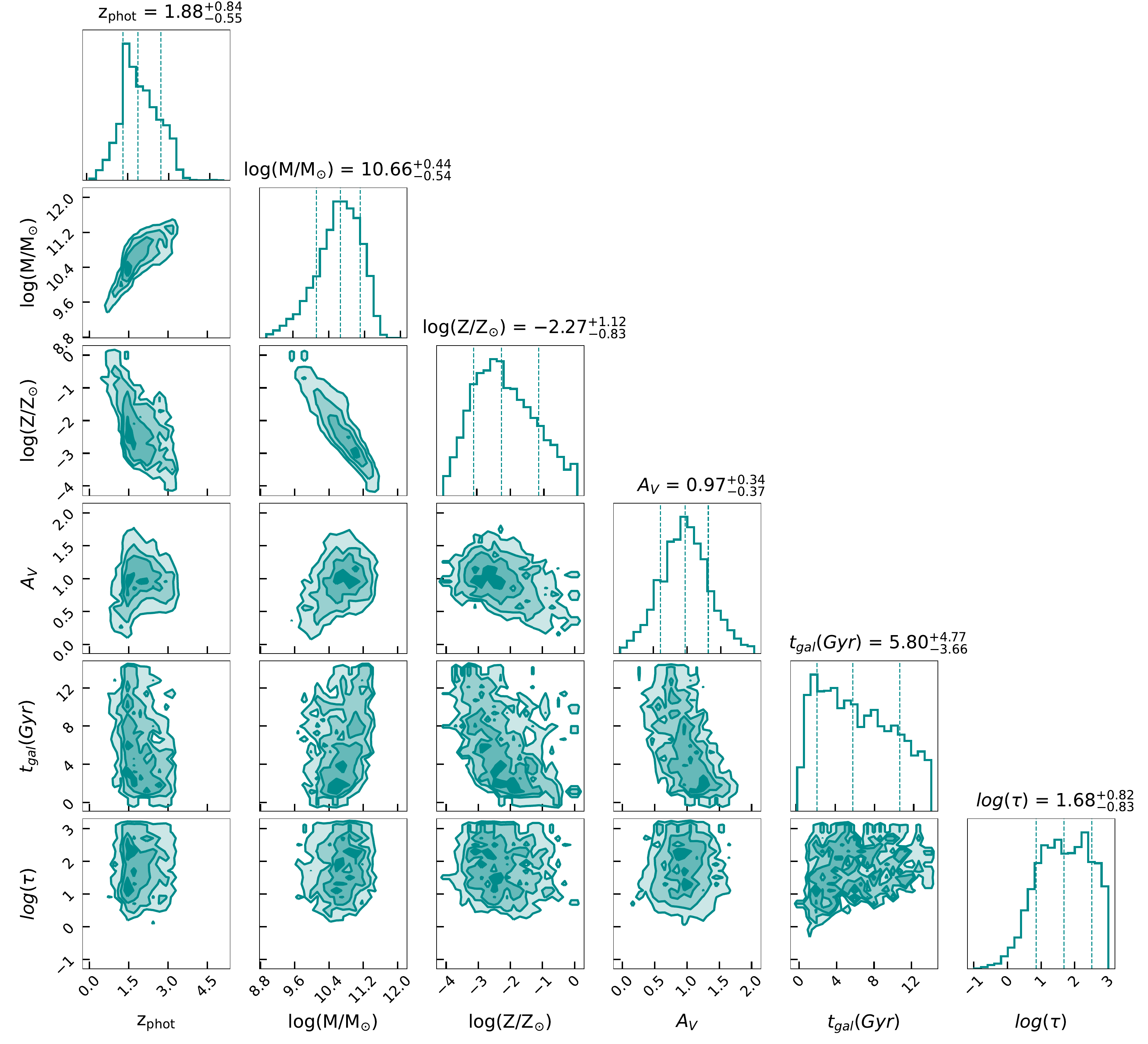}
\caption{The posterior distributions for the SED model parameters of the potential host galaxy of \thisgrbG obtained using nested sampling via \sw{dynesty} using \sw{Prospector} software.}
\label{PotentialHG-SED_corner}
\end{figure*}

\subsection{GRB 210205A}

\subsubsection{The nature of X-ray afterglow of GRB 210205A} 

As no redshift has been reported for this GRB, we modelled the time-averaged X-ray afterglow spectrum (\swiftT + 143 to \swiftT + 39716 s) considering redshift equal to 2, roughly mean redshift value for long bursts. The spectrum could be modelled using an absorption power-law with the following spectral parameters: hydrogen column density for the host galaxy ($\rm NH_{\rm host})= 5.77^{+6.26}_{-4.68} \times 10^{22}{\rm cm}^{-2}$ and $\beta_{\rm X}$ = 1.17$^{+ 0.41}_{-0.37}$.
To constrain the spectral regime, we implemented the closure relations for ISM and wind-like medium. We find that X-ray emission is explained with an adiabatic deceleration without an energy injection case. The closure relations also indicate that the X-ray afterglow could be best described with $\nu_{X-ray} > \nu_{c}$ spectral regime for a constant as well as wind ambient medium with the electron energy index p $\sim$ 2.34.

Further, we compared the XRT flux of GRB 210205A at 11 hours and 24 hours post burst in 0.3-10 \keV energy range with a complete sample of X-ray afterglows of long GRBs detected by \swift XRT till August 2021. We noticed that the X-ray afterglow of GRB 210205A is faint in comparison to the typical X-ray afterglows at both epochs (see Figure \ref{xrtflux}). 

\begin{figure}[ht!]
\centering
\includegraphics[scale=0.34]{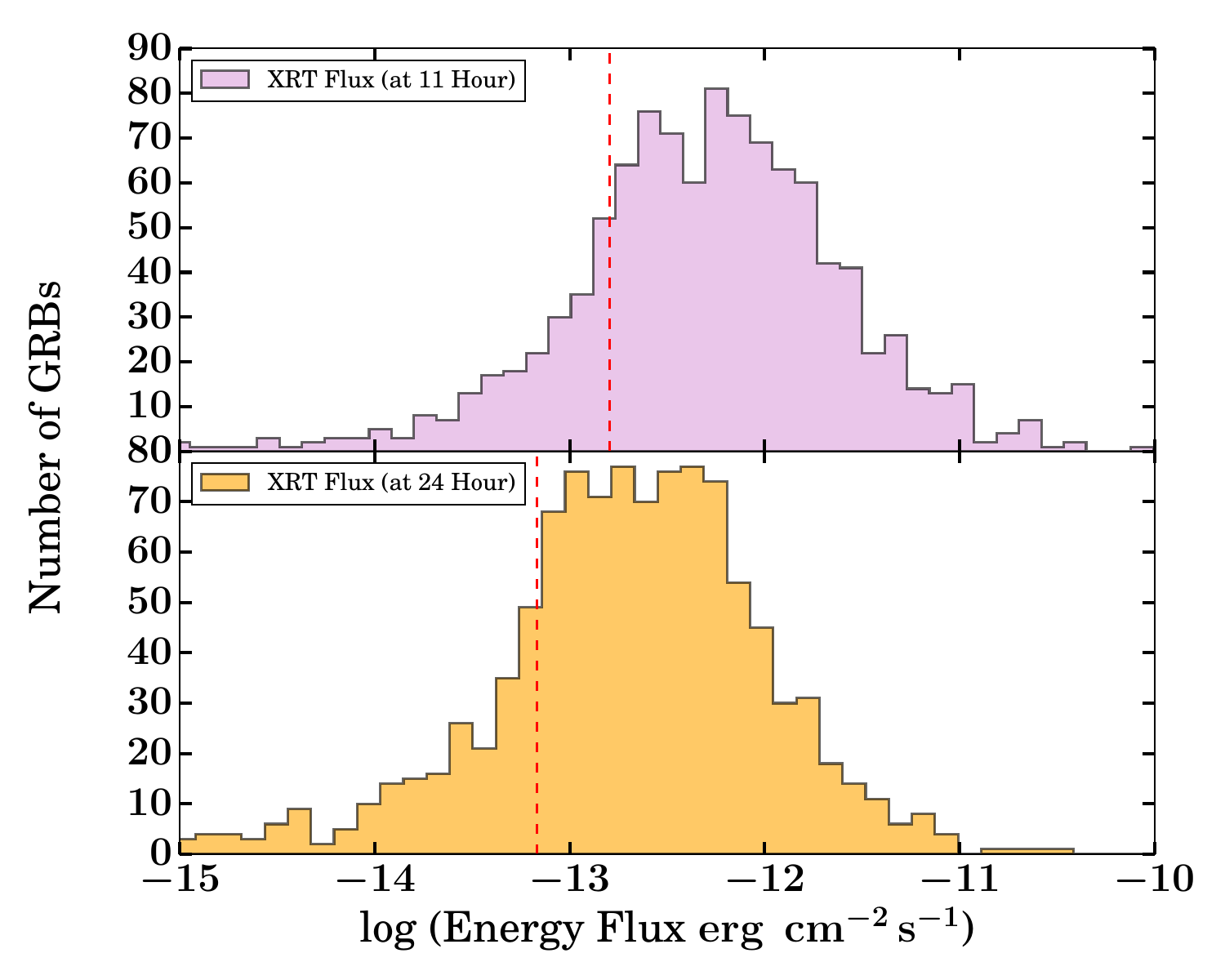}
\caption{{\it Top panel:} The energy flux distribution for a complete sample of X-ray afterglows, detected by \swift XRT at 11 hours after the burst detection in 0.3-10 \keV energy range. {\it Bottom panel:} similar as a top panel but flux calculated at 24-hour post burst. The vertical red dashed lines represent the position of GRB 210205A in respective panels.}
\label{xrtflux}
\end{figure}

\subsubsection{Spectral energy distribution of GRB 210205A}

\begin{figure}[ht!]
\centering
\includegraphics[angle=0,scale=0.35]{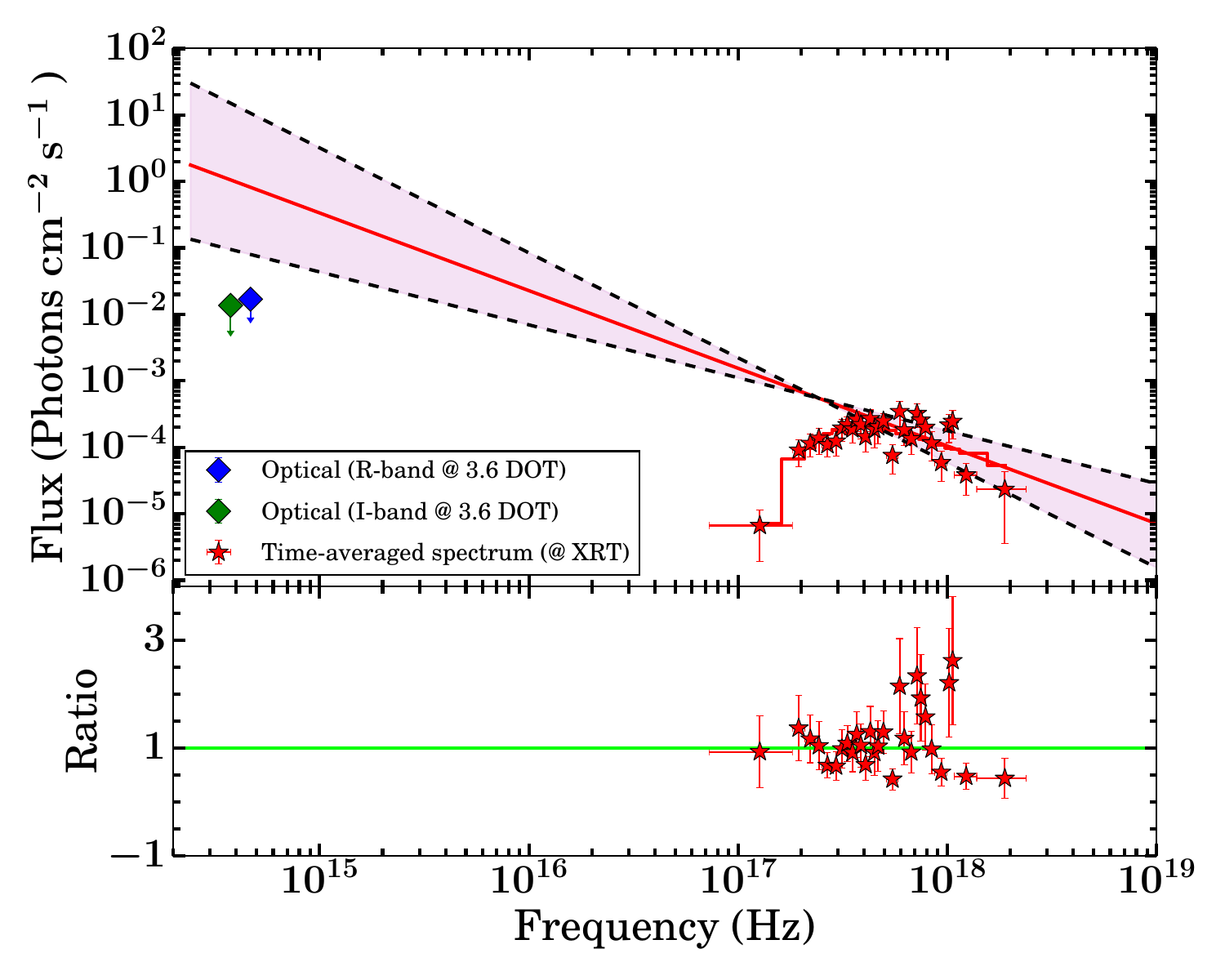}
\caption{{\it Top panel:} The spectral energy distribution for the afterglow of GRB 210205A. The solid red line shows the best fit for the time-averaged XRT spectrum, and the plum-shaded colour region shows its uncertainty region. Our R and I filters foreground corrected optical data points are shown in blue and green diamonds, respectively. {\it Bottom panel:} Ratio of data and model obtained after the spectral fitting of X-ray data.}
\label{SED:210205A}
\end{figure}

Spectral energy distribution (SED) is helpful to constrain the afterglow behaviour. Considering no spectral break between X-ray and optical frequencies, we extrapolated the X-ray spectral index towards optical frequencies to constrain the upper limit of the intrinsic flux of optical afterglow. We found that our deep limiting magnitude values (Galactic extinction corrected) obtained using 3.6\,m DOT telescope in R and I filters lie below the extrapolated X-ray power-law slope (see Figure \ref{SED:210205A}). This suggests that it requires absorption, and GRB 210205A could be a potential dark GRB candidate. We calculated the lower limit of extinction (the host extinction in I filter (A$_{\rm I}$) $>$ 0.25 mag) using the SED.

\begin{figure}[ht!]
\centering
\includegraphics[angle=0,scale=0.35]{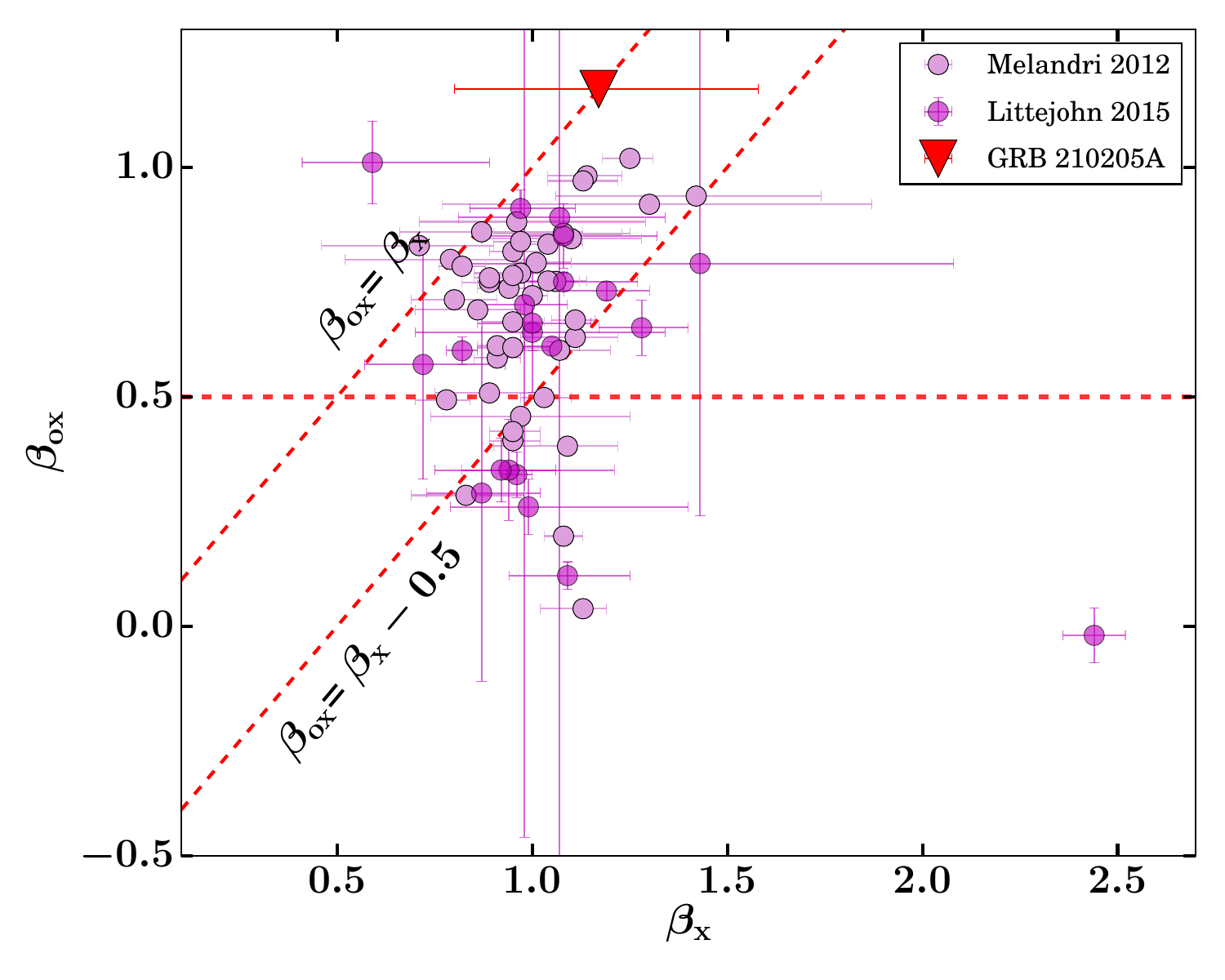}
\caption{Distribution of X-ray spectral indices as a function of optical-X-ray spectral indices for GRB 210205A (shown with a red triangle). For the comparison, data points for bright \swift long GRBs taken from \cite{2012MNRAS.421.1265M, 2015MNRAS.449.2919L} are also shown.}
\label{betaox:210205A}
\end{figure}

Furthermore, we constrain the upper limit on X-ray-to-optical spectral slope ($\beta_{\rm OX}$ $<$ 1.17), as the closure relation suggest for $\nu > \nu_{c}$ spectral regime. We compared the value of $\beta_{\rm OX}$ as a function of $\beta_{\rm X}$ of GRB 210205A along with a large sample of dark population studied by \cite{2012MNRAS.421.1265M, 2015MNRAS.449.2919L}. We find a hint that GRB 210205A satisfies the definition of dark GRBs given by \cite{2009ApJ...699.1087V}. The distribution of $\beta_{\rm OX}$ as a function of $\beta_{\rm X}$ for GRB 210205A is shown in Figure \ref{betaox:210205A}.

\subsubsection{Possible origin of the optical darkness}

It is clear that neither optical afterglow is detected nor any host galaxy associated with the burst was reported to measure the redshift of GRB 210205A. So, we have used the Amati correlation to explore the possibility of a high redshift origin of the burst. We used BAT fluence value to constrain the lower limit on isotropic gamma-ray energies for a range of redshift values $z$= 0.1 to $z$= 5. Figure \ref{amati:210205A} shows the position of GRB 210205A for different values of redshift in the Amati correlation plane of long GRBs along with other data points taken from \cite{2012MNRAS.421.1256N}. This analysis indicates that GRB 210205A might be a high redshift burst (with large associated uncertainties). Our analysis suggests that the source was not highly extinguished; therefore, both possibilities, i.e., intrinsically faint or a high redshift origin, could be possible reasons for the optical darkness of this burst.

\begin{figure}[ht!]
\centering
\centering
\includegraphics[angle=0,scale=0.4]{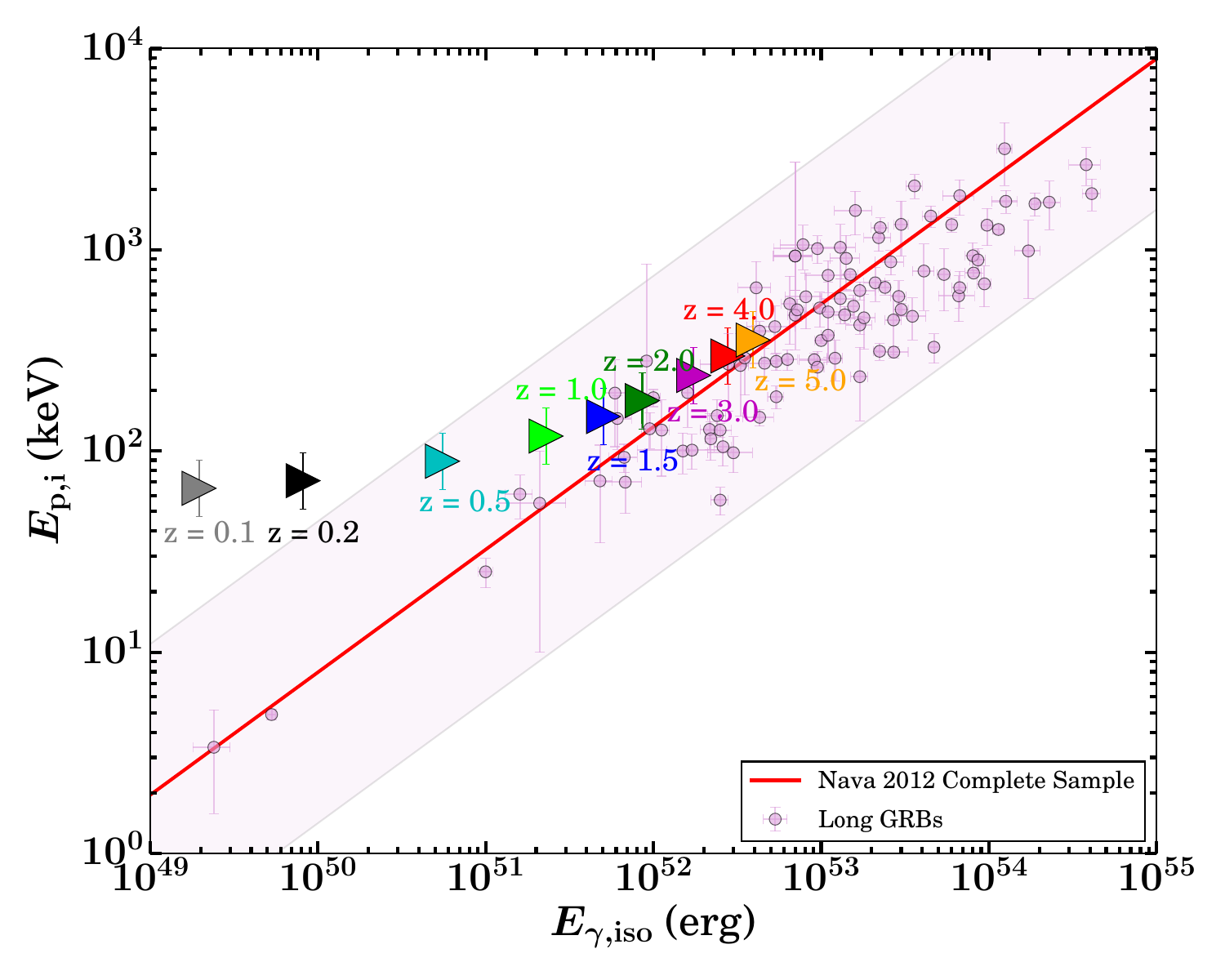}
\caption{GRB 210205A in Amati correlation plane of the long GRBs. The Redshift of the source has been varied from 0.1 to 5 as no spectroscopic or photometric redshift is available. For the comparison, data points for long GRBs taken from \cite{2012MNRAS.421.1256N} are also shown. The solid red line and shaded regions indicate the best-fit line and its associated 2-$\sigma$ uncertainty region for the complete sample of long GRBs studied by \cite{2012MNRAS.421.1256N}.}
\label{amati:210205A}
\end{figure}

\section{Orphan afterglows: ZTF21aaeyldq (AT2021any)}
\label{multiwavlength observation and data analysis Orphan}

\subsection{Optical observations and analysis}

The Zwicky Transient Facility (ZTF) announced the discovery of ZTF21aaeyldq (AT2021any) at the location RA= 08:15:15.34, DEC= -05:52:01.2 (J2000) with $r$ filter magnitude of 17.90 (AB). The object was discovered at 06:59:45.6 UT on 16$^{th}$ January 2021, only 22 minutes after the last non-detection ($r >$  20.28). ZTF carried out two additional observations in the same filter over the next 3.3 hours, and it confirmed that the transient has rapidly faded by two magnitudes \citep{2021GCN.29305....1H}. \cite{2021GCN.29305....1H} suggested that the colour of the source is moderately red ($g-r$ $\sim$ 0.3 mag). They also searched for the counterpart/host galaxy, but no such object is visible in deep Legacy Imaging Survey pre-imaging down to a limit of 24 magnitudes \citep{2019AJ....157..168D}. These characteristics suggest that ZTF21aaeyldq is a fast-fading, hostless, and Young optical transient. 

\cite{2021GCN.29307....1D} confirm the afterglow behaviour of ZTF21aaeyldq using photometric observations and measured the redshift using spectroscopic observations taken with OSIRIS instrument mounted on 10.4\,m GTC at $\sim$ 16.60 hours post the first detection. They identified many strong absorption lines such as Ly-alpha, SII, OI, SiII, SiIV, CII, CIV, FeII, AlII, and AlIII at a common redshift of $z$ = 2.514 (redshift of ZTF21aaeyldq).

We explored the \swift GRBs archive web page\footnote{https://swift.gsfc.nasa.gov/archive/grb\_table/} hosted by NASA's Goddard Space Flight Center, \fermi GBM Sub-threshold archive page \footnote{https://gcn.gsfc.nasa.gov/fermi\_gbm\_subthresh\_archive.html}, \fermi GBM GRBs catalog\footnote{https://heasarc.gsfc.nasa.gov/W3Browse/fermi/fermigbrst.html}, and GCN Circulars Archive\footnote{https://gcn.gsfc.nasa.gov/gcn3\_archive.html} for searching the associated gamma-ray counterpart of ZTF21aaeyldq between the last non-detection by ZTF (at 06:39:27 UT on 16 Jan 2021) and the first ZTF detection (at 06:59:46 UT on 16 Jan 2021). No GRB associated with ZTF21aaeyldq is detected using any space-based $\gamma$-ray telescopes during this temporal window. However, on the same day ($\sim$ 46 minutes before the last non-detection of ZTF) \AstroSat Cadmium Zinc Telluride Imager (CZTI) and Large Area X-ray Proportional Counter (LAXPC) detected a burst (GRB 210116A) with a \tninty duration of 9.5$^{+4.1}_{-1.8}$ s \citep{2021GCN.29342....1N} but due to unavailability of precise localization of GRB 210116A, it could not be confirmed or rule out the association between both the events. Considering the typical GRBs energy fluence threshold value of $\leq$ $\rm 10^{-6} erg ~{\rm cm}^{-2}$ \citep{2013ApJ...769..130C}, we constrain the isotropic equivalent gamma-ray energy $E_{\rm \gamma, iso}$ $\leq 1.52 \times 10^{52}$ erg.

\begin{figure}[ht!]
\centering
\includegraphics[angle=0,scale=0.35]{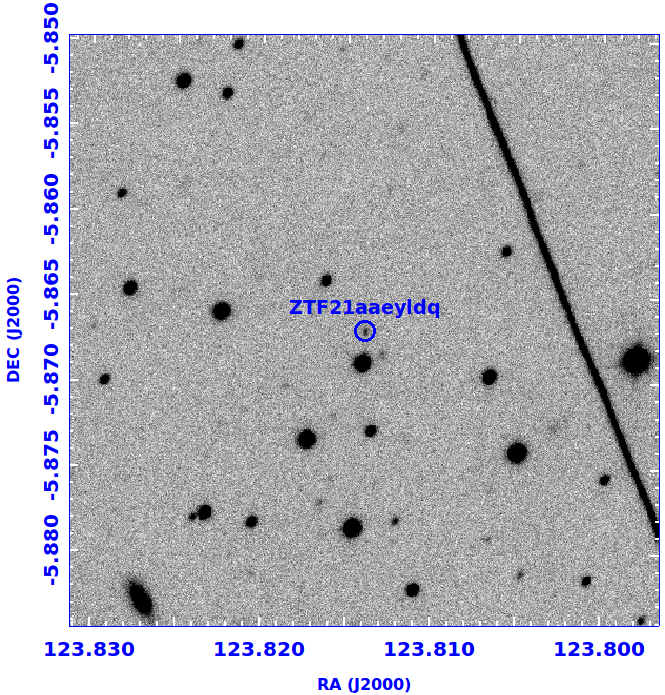}
\caption{The R-band finding chart of ZTF21aaeyldq obtained using the 3.6\,m DOT (taken $\sim$ 0.65 days post ZTF last non-detection) is shown. The field of view is $\sim$1$'$ $\times$ 1$'$, and the blue circle indicates the five arcsec uncertainty region at ZTF ground localization.}
\label{fig:ZTF}
\end{figure}

We observed ZTF21aaeyldq using the 4K$\times$4K CCD Imager installed at the axial port of the 3.6\,m DOT, started $\sim$ 15.258 hours after the ZTF first detection \citep{2021GCN.29308....1K}. We obtained two images with an exposure time of 300 seconds each in Bessel R and I filters. We reduced the data following the method discussed in \cite{2021MNRAS.505.4086G}. We also detected an uncatalogued source in both R, and I filters at the location reported by \cite{2021GCN.29305....1H}. A finding chart (R-filter) showing the detection of ZTF21aaeyldq, obtained using the 4K$\times$4K CCD Imager mounted at the 3.6\,m DOT, is presented in Figure \ref{fig:ZTF}. Further, we again observed the source in the r filter at $\sim$ 9.40 days after the first discovery using the same instrument and telescope. We acquired a consecutive set of 12 images with an exposure time of 300 seconds each. We do not detect any optical counterpart up to a magnitude limit of 23.98 in the stacked image \citep{2021GCN.29364....1G}.

In addition to follow-up of ZTF21aaeyldq using 3.6\,m DOT, we also performed the observations of this source using the 2.2\,m CAHA telescope located in Almeria (Spain) equipped with the CAFOS instrument. We obtained multiple images in BVRI filters with an exposure time of 240 s for each frame, starting at 00:02:24 UT on 17$^{th}$ January 2021. The optical counterpart is clearly detected in R and I filters and marginally visible in the V filter at the ZTF location. Using this telescope, we again monitored the object at the second epoch in the same filter system BVRI, starting at 04:08 UT. However, the source could not be detected during our second epoch of observations. We reduced the data using the same methodology as we did for 3.6\,m DOT. We calibrated the instrumental magnitudes with the nearby stars present in the USNO-B1.0 catalogue (same stars used for 3.6\,m DOT data calibration). We have listed the optical photometry log of our observations of ZTF21aaeyldq along with those obtained using different gamma-ray coordination networks (GCNs) in Table \ref{tab:observationslog_ZTF21aaeyldq}.

\begin{table*}[t]
\scriptsize
\caption{The optical photometric observations log of ZTF21aaeyldq taken with 3.6\,m DOT and 2.2\,m CAHA, including data from GCNs. The tabulated magnitudes are in the AB magnitude system and have not been corrected for foreground extinction. All the upper limits are given with three sigma.}
\begin{center}
\begin{tabular}{c c c c c c}
\hline
\bf $\rm \bf T_{mid}$ (days) & \bf Exposure (s)  & \bf Magnitude  &\bf Filter & \bf Telescope & \bf References\\
\hline
 0.0141 & -- & 17.90 $\pm$ 0.06 & r & ZTF & \cite{2021GCN.29305....1H} \\
 0.6998 & 1 x 60 & 21.64 $\pm$ 0.03  & r & 10.4\,m GTC  & \cite{2021GCN.29307....1D} \\
0.8451 & 5 x 300 &  21.86 $\pm$ 0.04  & r & NOT &  \cite{2021GCN.29310....1Z} \\
0.8261 & 5 x 300  &  22.28 $\pm$ 0.05  & g  & NOT  & \cite{2021GCN.29310....1Z} \\
1.8230 & --  & 23.39 $\pm$ 0.11 & g & 2.2\,m MPG &  \cite{2021GCN.29330....1G} \\
 2.0199 & --  &  23.47 $\pm$  0.06 & g & 2.2\,m MPG & \cite{2021GCN.29330....1G} \\
 2.8700 & --  &  23.84 $\pm$ 0.09  & g & 2.2\,m MPG & \cite{2021GCN.29330....1G} \\
0.8641 & 5 x 300 & 21.65 $\pm$ 0.03 & i & NOT &  \cite{2021GCN.29310....1Z}  \\
1.0556 & 3 x 180  & 22.10 $\pm$ 0.20  & i & Lowell  & \cite{2021GCN.29309....1A}\\
0.9969 & 1 x 900 & 21.62 $\pm$ 0.20  & J & LBT  &  \cite{2021GCN.29327....1R}\\
0.9969 & 1 x 900 & 21.36 $\pm$  0.24 & H & LBT  &  \cite{2021GCN.29327....1R}\\
1.7584 & 10 x 360 & 22.89 $\pm$ 0.12 & Rc & 2.2\,m CAHA  & \cite{2021GCN.29321....1K} \\
\hline 
0.7259 & 1 x 240 & $>$ 22.97 & B & 2.2\,m CAHA  & Present work \\
0.8964 & 1 x 240 & $>$ 22.13 & B & 2.2\,m CAHA  & Present work \\
0.7297 & 1 x 240 & 22.17 $\pm$ 0.39 & V & 2.2\,m CAHA  & Present work \\
0.8998 & 1 x 240 & $>$ 21.40 & V & 2.2\,m CAHA  & Present work \\
0.7332 & 1 x 240 & 21.22 $\pm$ 0.12 & R & 2.2\,m CAHA  & Present work \\
0.9032 & 1 x 240 & $>$ 20.96 & R & 2.2\,m CAHA  & Present work \\
0.7367 & 1 x 240 & 21.23 $\pm$ 0.19 & I & 2.2\,m CAHA  & Present work \\
0.9066 & 1 x 240 & $>$ 20.01 & I & 2.2\,m CAHA  & Present work \\

0.6498 & 1 x 300 & 21.25 $\pm$ 0.06 & R & 3.6\,m DOT & Present work\\

0.6546 & 1 x 300 & 21.36 $\pm$ 0.08 & I & 3.6\,m DOT & Present work\\
9.3963 & 12 x 300 & $> 23.98 $ & r & 3.6\,m DOT & Present work\\
\hline
\vspace{-2em}
\end{tabular}
\end{center}
\label{tab:observationslog_ZTF21aaeyldq}
\end{table*}

\subsection{X-ray afterglow observations and analysis}

After the independent discovery of the afterglow candidate by ZTF, several ground-based telescopes detected the source in different filters\footnote{https://gcn.gsfc.nasa.gov/other/ZTF21aaeyldq.gcn3}. In addition to multi-band optical/NIR observations, \cite{2021GCN.29313....1H} reported the detection of X-ray afterglow based on a target-of-opportunity (ToO) observations obtained using \swift XRT, $\sim$ 19.82 hours after the last ZTF non-detection. We obtained and reduced the XRT data using the online tool known as Build \swift-XRT products\footnote{https://www.swift.ac.uk/user\_objects/} provided by the \swift team. \swift-XRT observed the source at three different epochs (Obs Ids: 00013991001, 00013991002, and 00013991003) with a total exposure time of 8.2 ks (see Table \ref{tab:ZTFXRTobservationslog}). However, the source is only detected at the first epoch ($\rm \Delta T_{mid}$ $\sim$ 24.50 hours post-ZTF last non-detection) at the location of ZTF21aaeyldq with a count rate of 6.66$^{+1.84}_{-1.84} ~\times 10^{-3}$  s$^{-1}$ in 0.3-10 \keV. 
Considering the Galactic hydrogen column density value $\rm NH_{\rm Gal}= 7.75 \times 10^{20}{\rm cm}^{-2}$ \citep{2013MNRAS.431..394W} and X-ray photon index value of 2, we calculated the unabsorbed flux equal to 2.99 $\rm \times 10^{-13} erg ~{\rm cm}^{-2} ~{\rm s}^{-1}$ and X-ray luminosity equal to 1.60 $\rm \times 10^{46} erg ~{\rm s}^{-1}$ at $z$ equal to 2.514. The measured X-ray luminosity is typical of GRB X-ray afterglows at this epoch. We have used the Portable, Interactive Multi-Mission Simulator (\sw{PIMMS}) tool of NASA's HEASARC page\footnote{https://heasarc.gsfc.nasa.gov/cgi-bin/Tools/w3pimms/w3pimms.pl}to calculate the unabsorbed flux. During the last two epochs of observations, no X-ray emission has been detected at the ZTF afterglow positions. We obtained 3-$\sigma$ upper limits on count rate equal to 0.0043513 ($\rm \Delta T_{mid}$ $\sim$ 4.24 days post-ZTF last non-detection) and 0.0040133 ($\rm \Delta T_{mid}$ $\sim$ 9.35 days post-ZTF last non-detection) at the second and third epoch of observations, respectively. Considering the same values of $\rm NH_{\rm Gal}$ and X-ray photon index, we converted these count rates into an upper limit on the flux density of $<$ 1.95 $\rm \times 10^{-13} erg ~{\rm cm}^{-2} ~{\rm s}^{-1}$ and $<$ 1.80 $\rm \times 10^{-13} erg ~{\rm cm}^{-2} ~{\rm s}^{-1}$, respectively.

\begin{table*}[ht!]
\scriptsize
\caption{The observations log for the X-ray afterglow of ZTF21aaeyldq taken with \swift XRT.}
\begin{center}
\begin{tabular}{c c c c c c}
\hline
\bf Obs Ids & \bf $\rm \bf T_{mid}$ (days) &  \bf Count Rate  &\bf Energy range  & \bf Flux & \bf Telescope\\
\hline
00013991001 & 1.02 & 6.66$^{+1.84}_{-1.84}$  & 0.3-10 \keV & 2.99 $\rm \times 10^{-13} erg ~{\rm cm}^{-2} ~{\rm s}^{-1}$ & \swift XRT \\
00013991002 & 4.24 & $<$ 0.0043513 & 0.3-10 \keV & $<$ 1.95 $\rm \times 10^{-13} erg ~{\rm cm}^{-2} ~{\rm s}^{-1}$ & \swift XRT \\
00013991003 & 9.35 & $<$ 0.0040133 & 0.3-10 \keV & $<$ 1.80 $\rm \times 10^{-13} erg ~{\rm cm}^{-2} ~{\rm s}^{-1}$ & \swift XRT \\
\hline
\vspace{-2em}
\end{tabular}
\end{center}
\label{tab:ZTFXRTobservationslog}
\end{table*}

\section{Results and Discussion} 
\label{results for both the bursts orphan}

\subsection{Afterglow behaviour of ZTF21aaeyldq}

The afterglow light curve of ZTF21aaeyldq has been shown in Figure \ref{lightcurvecomparision}. The photometric magnitudes have been corrected for foreground extinction before converting them into flux units. The optical afterglow light curve of ZTF21aaeyldq has been continuously fading since the first detection by ZTF, typical characteristic of afterglows. We fitted the early $r$ band data taken from \cite{2021GCN.29305....1H, 2021GCN.29307....1D, 2021GCN.29310....1Z} and found that the $r$ band light curve is best described with a single power-law model with a temporal decay slope $\alpha =  0.89 \pm 0.03$, consistent with the slope reported by \cite{2021GCN.29321....1K}. Later on, \cite{2021GCN.29344....1K} observed the afterglow candidate ZTF21aaeyldq with the 2.2\,m telescope at Calar Alto, Spain, in the Rc filter at 2.8 days after the first detection, and with Gamma-ray Burst Optical/Near-infrared Detector (GROND) mounted at the 2.2\,m MPG telescope at 3.9 days after the first ZTF detection, respectively. They detected the source clearly in each stacked frame. Further, they fitted the optical data taken from various GCNs \citep{2021GCN.29305....1H, 2021GCN.29307....1D, 2021GCN.29310....1Z, 2021GCN.29330....1G} along with their observations. They found that a broken power-law is better fitting the data ($\chi^{2}$/dof = 0.12) with temporal index before the break ($\alpha_{1}$) equal to 0.95 $\pm$ 0.03, temporal index after the break ($\alpha_{2}$) equal to 2.30 $\pm$ 0.76, and break time T$_{b}$ equal to 0.82 $\pm$ 0.08 days. Our late-time observations using 3.6\,m DOT ($\sim$ 9.40 days post first detection) are also consistent with the temporal index after the break suggested by \citep{2021GCN.29344....1K}. However, we could not confirm if the break is achromatic/chromatic due to the unavailability of simultaneous multi-wavelength data. In any case, the jet break is a typical characteristic of GRB afterglows, and it indicates that the nature of ZTF21aaeyldq is a GRB afterglow. Since this event was established as an orphan afterglow based on no detection of any associated GRBs between the last non-detection and the first detection of optical emission by ZTF, we extended our analysis by comparing the properties with other orphan events with a redshift measurement. The light curve evolution of ZTF21aaeyldq with other known orphan afterglows (ZTF20aajnksq/AT2020blt and ZTF19abvizsw/AT2019pim) are shown in Figure \ref{lightcurvecomparision}. In the case of ZTF20aajnksq, we collected data from \cite{2020ApJ...905...98H} and for ZTF19abvizsw ($\rm T_{0}$: at 07:35:02 UT on 1$^{st}$ September 2019, the last non-detection\footnote{{https://www.wis-tns.org/object/2019pim}}), we obtained optical observations from \cite{2020ApJ...905..145K}. Based on the above, we noticed that ZTF21aaeyldq has many similar features to classical GRBs afterglows. The measured redshift of ZTF21aaeyldq is typical of long GRBs, and the strong absorption lines identified in the optical spectrum are usually present in GRBs afterglows at measured redshift. The detection of an X-ray counterpart with typical X-ray luminosity also indicates the afterglow nature of ZTF21aaeyldq. In addition to these, the presence of a break in the optical light curve with typically expected temporal indices due to jet break confirms the afterglow behaviour of the source.  

\begin{figure}[ht!]
\centering
\includegraphics[angle=0,scale=0.35]{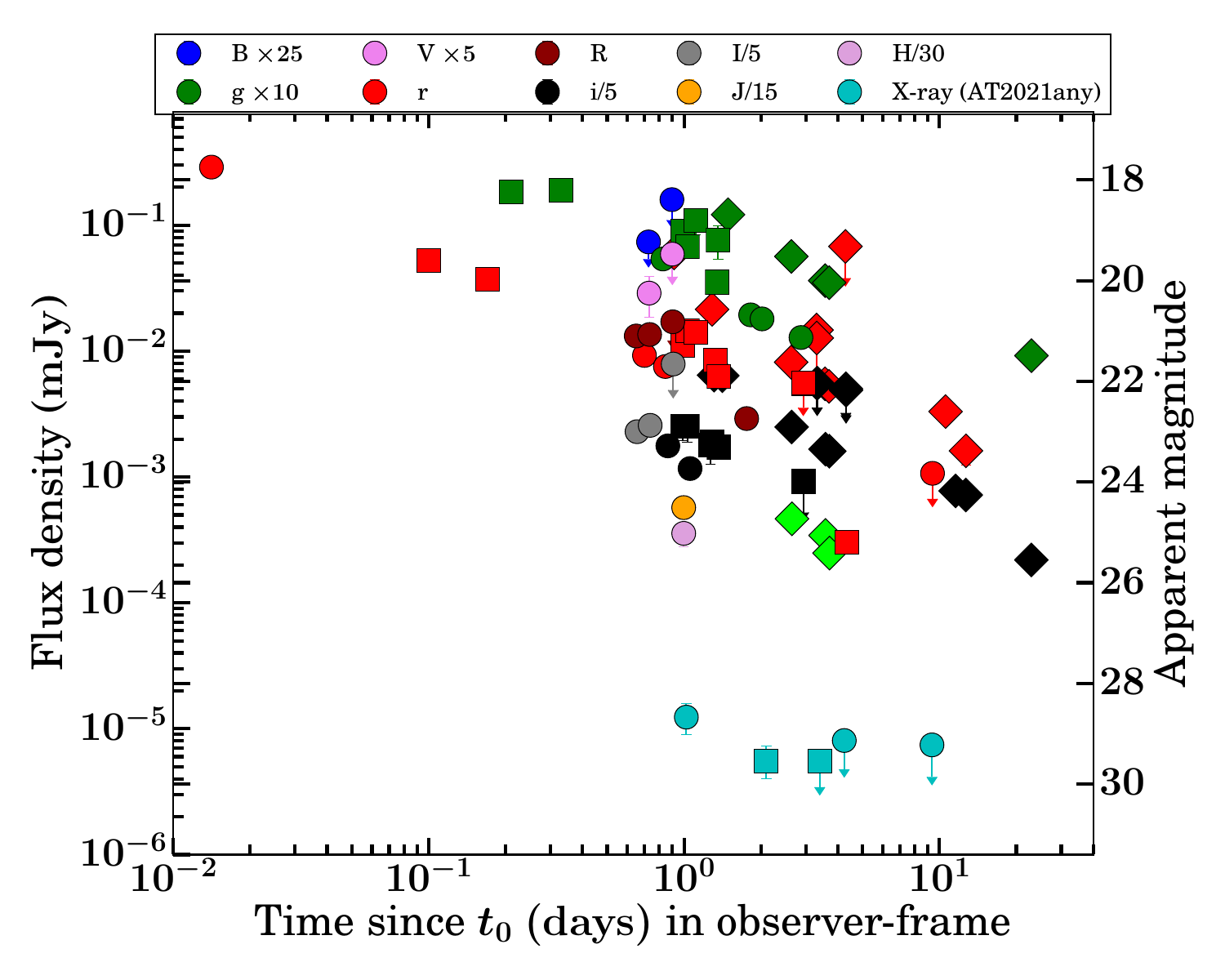}
\caption{{Comparison between afterglow light curves of orphan afterglows with a measured redshift known so far:} The X-ray and optical/NIR afterglow light curve of ZTF21aaeyldq (shown with circles of different colours). For the comparison, we have also demonstrated X-ray and optical afterglow light curves of ZTF20aajnksq ($z \sim$ 2.90), shown with squares and ZTF19abvizsw ($z \sim$ 1.26), pictured with diamonds. We have used the same colours and offsets for ZTF20aajnksq and ZTF19abvizsw also. The lime diamonds denote the z-filter observations (with an offset factor of 35) of ZTF19abvizsw.}
\label{lightcurvecomparision}
\end{figure}

\subsection{Afterglow modelling of ZTF21aaeyldq}

According to the standard external shock fireball model for the afterglows, the X-ray and optical emission from afterglows can be described with synchrotron emission for constant or WIND-like external medium \citep{1998ApJ...497L..17S}. The broadband synchrotron spectral energy distribution consists of three break frequencies: the synchrotron cooling frequency $\nu_{\rm c}$, the synchrotron peak frequency $\nu_{\rm m}$, and the synchrotron self-absorption frequency $\nu_{\rm a}$. The self-absorption frequency does not affect the X-ray and optical data at early epochs, and it mainly affects the low-frequency observations of afterglows. Depending on the ordering of these break frequencies, we can constrain the spectral regimes of afterglows at a particular epoch \citep{2003BASI...31...19P}. 

Detailed multiwavelength modelling is helpful to constrain physical parameters associated with the afterglow. We performed detailed multiwavelength modelling of the light curve of the ZTF21aaeyldq afterglow using the publicly available $\af$ Python package. It is an open-source numerical and analytic modelling tool to calculate the multiwavelength light curve and spectrum using synchrotron radiation from an external shock for the afterglows of GRBs \citep{2020ApJ...896..166R}. $\af$ package has capabilities to produce the light curves and spectrum of afterglows considering both structured jets and off-axis observers. We have used the Markov chain Monte Carlo (MCMC) Ensemble sampler using \sw{emcee} python package for fitting the multiwavelength light curve to get the model parameters and associated errors.

\begin{figure}[ht!]
\centering
\includegraphics[angle=0,scale=0.35]{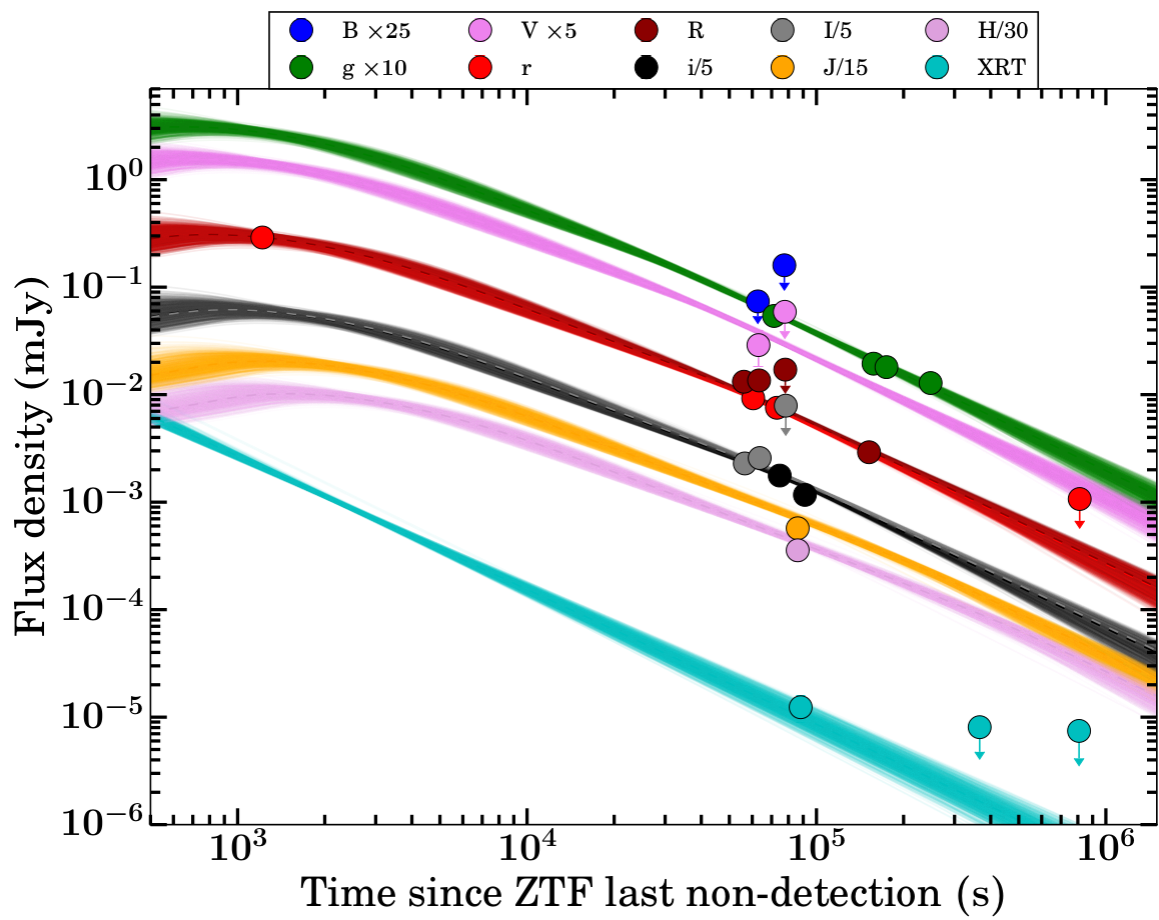}
\caption{Multiwavelength observations data of ZTF21aaeyldq along with the best fit afterglow modelling results obtained using $\af$. The shaded regions indicate the uncertainty region around the median light curve.}
\label{modelling}
\end{figure}

\begin{figure*}[ht!]
\centering
\includegraphics[angle=0,scale=0.4]{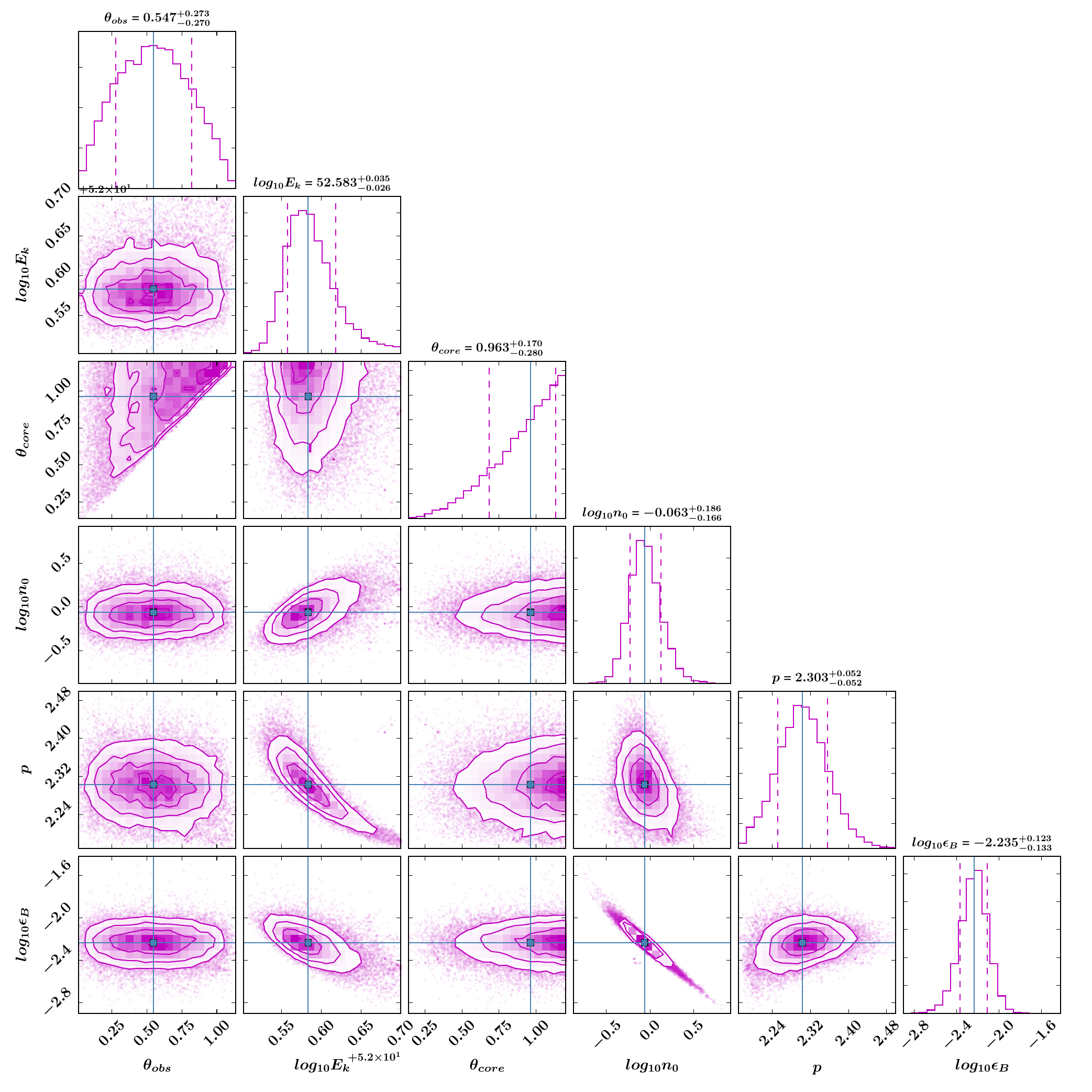}
\caption{Posterior distribution and parameter constraints for the 15000 simulations obtained using broadband afterglow modelling of ZTF21aaeyldq using \linebreak $\af$.}
\label{modelling_pd}
\end{figure*}

\begin{table*}
\caption{The input parameters, range, best-fit value, and associated errors of multi-wavelength afterglow modelling of ZTF21aaeyldq were performed using the $\af$ python package.}
\begin{center} 
\begin{tabular}{c c c c c}
\hline
Model Parameter & Unit & Prior Type  & Range & Best fit Value \\
\hline \hline
$\theta_{obs}$ & rad & $\sin(\theta_{obs})$ & [0.01, 1.20]  & $0.55_{-0.27}^{+0.27}$ \\
$log_{10}(E_k)$ & erg &  uniform & [51, 52.7] & $52.58_{-0.03}^{+0.03}$\\ 
$\theta_{core}$ & rad & uniform &  [0.01, 1.20] & $0.96_{-0.28}^{+0.17}$  \\ 
$log_{10}(n_0)$ & cm$^{-3}$ &  uniform  &  [-6, 1.6] & $-0.06_{-0.17}^{+0.19}$  \\ 
$p$ & - &  uniform & [2.0001, 2.5] & $2.30_{-0.05}^{+0.05}$   \\ 
$log_{10}(\epsilon_{B})$ & - &  uniform & [-3.6, -1.2] & $-2.23_{-0.13}^{+0.12}$ \\  
$\xi$ & -& -& 1 & - \\
\hline
\hline
\end{tabular}
\end{center}
\label{tab:model}
\end{table*}

There are various jet structures possible for GRBs afterglows. $\af$ has capabilities to produce the light curves for some commonly used jet structures such as top-hat, Gaussian, Power-law, etc. We have used the top-hat jet type structure to model the afterglow data of ZTF21aaeyldq using the $\af$ package. For top hat jet-like afterglows, we consider six free parameters ($\theta_{obs}$: viewing angle, $E_{k}$: on-axis isotropic equivalent energy, $\theta_{core}$: half-width of the jet core, $n_{0}$: number density of ISM medium, $p$: electron distribution power-law index, and $\epsilon_{B}$: thermal energy fraction in the magnetic field). We have fixed the $\epsilon_{e}$ (thermal energy fraction in electrons) value = 0.10 \citep{2002ApJ...571..779P}. We have considered sine prior for viewing angle, and for all the remaining parameters, we have used uniform priors. In addition, we assume that the fraction of electrons that get accelerated is equal to one, and the redshift is equal to 2.514 for the model fitting. Furthermore, due to the large scales, we have set $E_{k}$, $n_{0}$ and $\epsilon_{B}$ parameters on the log scale. We have listed the priors types and the bounds of the parameters for each model parameter in Table \ref{tab:model}.  

In Figure \ref{modelling}, we have shown the optical and X-ray afterglow light curves of ZTF21aaeyldq along with the best-fit model obtained using $\af$.  The observed X-ray and optical data points are shown with circles in the figure. The dark-colour dashed lines show the light curve generated using the median value from parameter distribution from the MCMC routine. The shaded coloured bands show the uncertainty bands around the median light curve. Figure \ref{modelling_pd} shows the posterior distribution for the best-fit results of microphysical parameters obtained using MCMC simulation. The distribution of median posterior and 16\% and 84\% quantiles for each modelled parameter are also presented in the same plot. We modelled the optical and X-ray afterglow and obtained the following best fit micro-physical parameters: the model constrained the jet on-axis isotropic equivalent energy $log_{10}E_{k}$ = 52.58$^{+0.04}_{-0.03}$ erg, the jet structure parameter $\theta_{core}$ = 0.96$^{+0.17}_{-0.28}$ radian, viewing angle $(\theta_{obs})$ = 0.55$^{+0.27}_{-0.27}$ radian, the ambient number density $log_{10}n_{0}$ = -0.06$^{+0.19}_{-0.17}$ cm$^{-3}$, the parameters electron energy index $p$ = 2.30$^{+0.05}_{-0.05}$, and $log_{10}\epsilon_{B}$ = -2.24$^{+0.12}_{-0.13}$. These parameters are consistent with the typical afterglow parameters of other well-studied GRBs \citep{2002ApJ...571..779P}.

Furthermore, we created the SEDs at three epochs (see Figure \ref{modelling_SED}). The first is at 0.65 days (the epoch of 3.6\,m DOT observations), the second is at 0.73 days (the epoch of CAHA observations), and the third is at 1.02 days (the epoch of detection of the X-ray emission). During the first and second SEDs, the cooling frequency ($\nu_{c}$) lies close to our optical observations taken with 3.6\,m DOT and 2.2\,m CAHA telescope, and later on, it seems that $\nu_{c}$ passes through the optical observations.

\begin{figure}[ht!]
\centering
\includegraphics[angle=0,scale=0.35]{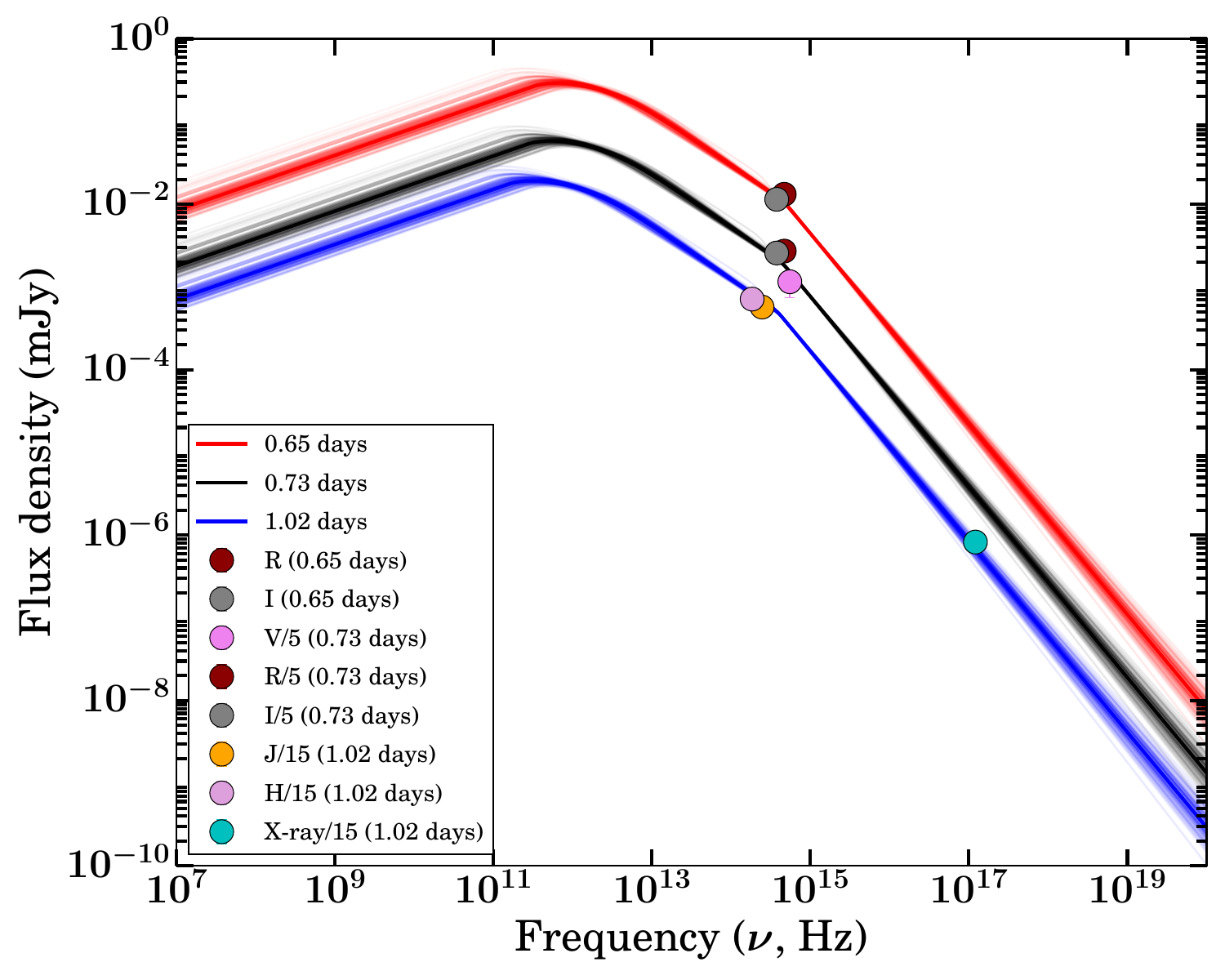}
\caption{The spectral energy distributions of ZTF21aaeyldq at three different epochs obtained for the best-fit model parameters using $\af$. The SEDs show the evolution of break frequencies of synchrotron emission.}
\label{modelling_SED}
\end{figure}

\subsection{Possible explanations of the nature of the orphan afterglow}

We noticed that the temporal evolution of optical data of ZTF21aaeyldq is consistent with that of typical GRB afterglows. The jet break signature in the optical data indicates that the source observing angle was within the jet opening angle. There could be three possible reasons for the origin of ZTF21aaeyldq. The simplest explanation for any orphan afterglow is that the source is observed on-axis ($\theta_{obs}$ $<$ $\theta_{core}$), however, the gamma-ray counterpart was unambiguously missed by space-based gamma-ray GRBs missions, either due to their sensitivity or source was not in their field of view. Our afterglow modelling suggests that this could be the case for the ZTF21aaeyldq. A schematic diagram showing a top-hat jet for the on-axis view is presented in Figure \ref{jet}. Other natural explanations for orphan afterglows could be off-axis observations ($\theta_{obs}$ $>$ $\theta_{core}$) and a dirty fireball.

Some authors use the off-axis jet scenarios to explain the orphan afterglows \citep{2002ApJ...570L..61G}. The off-axis models are also useful to explain the short GRBs from structured jets \citep{2018MNRAS.473L.121K}, low-luminosity bursts \citep{2007ApJ...662.1111L} and the X-ray plateaus present in the X-ray afterglow light curves \citep{2020MNRAS.492.2847B}. However, in the case of ZTF21aaeyldq, our afterglow modelling suggests that ZTF21aaeyldq was a classical burst viewed on-axis \citep{2022MNRAS.512.1391S}, so we discard the off-axis possibility. In addition, \cite{2020ApJ...896..166R} suggests a few other signatures of off-axis observations, such as an early shallow decay and a large value of change in temporal indices before and after the break ($\Delta \alpha$). However, in the case of ZTF21aaeyldq, we did not notice an early shallow decay, and also $\Delta \alpha$ value is not very large. Therefore, we discard the possibility of off-axis observations.

The third possibility to explain the orphan afterglow is a dirty fireball, i.e., fireball ejecta with a low value of bulk Lorentz factor ($\Gamma_0 \sim$ 5) and with large numbers of baryons. For a dirty fireball, gamma-ray photons of prompt emission are absorbed due to the optical thickness. In such a case, we expect a large value of deceleration time due to a low value of $\Gamma_0$. The deceleration time for a constant density medium is defined as $T_{\rm dec}$ = 59.6 $\times ~ n_{0}^{-1/3}$ $\Gamma_{0, 2.5}^{-8/3}$  $E_{k, 55}^{1/3}$ s \citep{2021MNRAS.505.1718J}. Considering $\Gamma_0$ equal to 100, we calculated $T_{\rm dec}$ = 210.68 s using the parameters obtained from afterglow modelling, but for $\Gamma_0$ equal to 10, we calculated $T_{\rm dec}$ = 1.13 days. We looked for the GCNs circular to find the bursts within two days before the ZTF discovery of ZTF21aaeyldq, but we did not find any coincident GRBs. In addition, we constrain the limit on bulk Lorentz factor ($\Gamma_{0}$ $<$ 202) using the prompt emission correlation between $\Gamma_{0}$-$E_{\gamma, \rm iso}$\footnote{$\Gamma_{0}$ $\approx$ 182 $\times$ $E_{\gamma, \rm iso, 52}^{0.25}$} \citep{2010ApJ...725.2209L}. In light of the above discussion, we discard the possibility of a dirty fireball.

\begin{figure}[ht!]
\centering
\includegraphics[angle=0,scale=0.4]{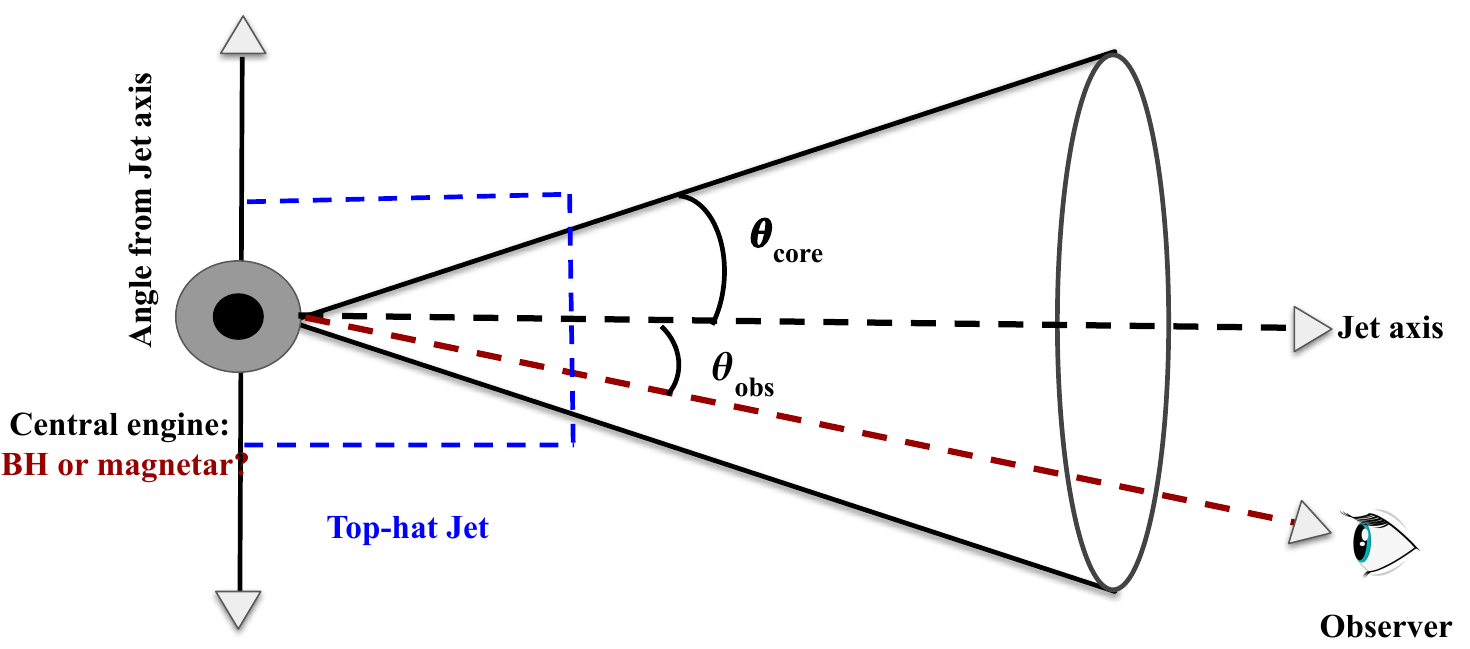}
\caption{A schematic diagram of a top-hat jet of a GRB for the on-axis view ($\theta_{obs}$ $<$ $\theta_{core}$) is presented, one of the possible scenarios for ZTF21aaeyldq.}
\label{jet}
\end{figure}

\section{Summary and Conclusion} 
\label{conclusiones} 


Optical follow-up observations are helpful to constrain the jet geometry, environment, total energy, and other microphysical parameters of GRBs afterglows. India has a longitudinal advantage and a long history of these ToO observations. The recent installation of 3.6\,m DOT at Devasthal Observatory of ARIES help us to move one step forward to these scientific goals. The back-end instruments of DOT are crucial for deep imaging and spectroscopy of afterglows. In this work, we present the analysis of three exciting afterglow sources (two dark and one orphan GRB) observed using 3.6\,m DOT and 10.4\,m GTC along with other telescopes.

Detection of the very red afterglow is crucial to explore the early Universe as they are probes to look for very high redshifts or test beds studying dusty environments surrounding GRBs. In this work, we apprise the discovery of a very red counterpart ($K_{\rm S}$-band) of \thisgrbG $\sim$ 5.2 hours post burst with the CIRCE instrument mounted at the GTC, but it was not detected in any bluer filters of \swift UVOT/BOOTES, suggesting for a very high redshift origin. However, our present analysis discarded this possibility based on a few arguments, including spectral analysis of X-ray afterglow, constrain $z <$ 4.15; SED modelling of the potential nearby galaxy constraining moderate redshift values and offset analysis between the centroid of the potential host to the location of the afterglow etc. Furthermore, we also performed the SED modelling of the potential host galaxy. Our results demand a relatively lower value of extinction and redshift along with typical physical parameters with a rather large physical offset from the galaxy's centre. The considerable difference between A$_{\rm V}$ values obtained from the host galaxy SED modelling and the one estimated from afterglow SED analysis of \thisgrbG is an indicative of local dust surrounding the progenitor or anomalous extinction within the host galaxy as described in \cite{2009AJ....138.1690P, 2013ApJ...778..128P}. Based on our analysis of the potential host, it is hard to decipher whether the observed candidate galaxy is the host of \thisgrbG.

Our analysis of the afterglow SED shows that significant reddening is required to describe the observed $K_{\rm S}$ band (NIR) afterglow of \thisgrbG, assuming the cooling frequency of external synchrotron model is beyond the optical/NIR and X-rays frequencies. Following this method, we calculated the $K_{\rm S}$ band host reddening (A$_{K_{\rm S}}$) = 3.60$^{+0.80}_{-0.76}$ mag, equivalent to A$_{\rm V}$ =  34.67$^{+7.70}_{-7.32}$ mag. Our analysis also indicates that \thisgrbG is one of the most intense dark GRBs detected so far. Our results suggest that the environment of \thisgrbG demands a high extinction towards the line of sight. Hence, dust obscuration is the most probable reason for the optical darkness of \thisgrbG. 

For GRB 210205A, an X-ray afterglow was detected by \swift XRT, but we do not find any optical counterpart despite deep follow-up observations using the 3.6\,m DOT telescope.
We investigate the possible reason for the optical darkness of the source. We noticed that GRB 210205A has a fainter X-ray emission compared to well-known X-ray afterglows detected by \swift XRT telescope. We obtained the optical-to-X-ray spectral index and compared it with a large sample of GRBs. We found a hint that the source satisfied the recent definition of the darkness of the afterglow. We also measure the host galaxy extinction using afterglow SED, suggesting either intrinsic faintness of the source or high redshift origin as a possible reason for the optical darkness of the afterglow.

The synergy between state-of-the-art nIR camera such as CIRCE and the largest diameter optical telescope available so far (the GTC), along with the recently launched JWST, makes an ideal combination for studying the population of dark GRBs to a great extent, and determining if extinction in the host galaxy or very high redshift is the reason for a significant fraction of the afterglows being beyond the reach of optical telescopes in about 20\% of events \citep{2011A&A...526A..30G, 2008MNRAS.388.1743T, 2011A&A...532A..48D}. Furthermore, the community is developing several larger optical telescopes, for example, the Extremely Large Telescope (ELT) and Thirty Meter Telescope (TMT). Our study provides insights for future observations of similar fainter afterglows of dark GRBs using upcoming larger telescopes \citep{2013JApA...34..157P}.

In the case of ZTF21aaeyldq, no gamma-ray prompt emission counterpart was reported by any space-based satellites, and the ZTF survey discovered this source. Based on optical identification of the afterglow, XRT observations and other follow-up observations found a fading source characterized with typical afterglow nature with redshift measured using the 10.4\,m GTC. The following key observational facts confirm the afterglow nature of ZTF21aaeyldq. (i) Fading nature of the source consistent with typical temporal index seen in case of afterglows. (ii) jet break in the optical light curve (iii) The detection of X-ray afterglow with typical luminosity. We performed detailed multiwavelength modelling of the afterglow, including data taken with 3.6\,m DOT. Our results suggest that afterglow could be described with a slow cooling case for an ISM-like ambient medium. The micro-physical afterglow parameters of ZTF21aaeyldq are consistent with the typical afterglow parameters of other well-studied GRBs. Our modelling also suggests that ZTF21aaeyldq was viewed on-axis ($\theta_{obs}$ $<$ $\theta_{core}$), however, the gamma-ray counterpart was unambiguously missed by space-based gamma-ray GRBs missions, either due to their sensitivity (weak prompt emission) or source was not in their field of view (Earth occultation). It helps us rule out other hypotheses for the possible origin of orphan afterglows, such as off-axis view and a low-Lorentz factor jet. We also compared the light curve evolution of ZTF21aaeyldq with other known orphan afterglows with a measured redshift. We found that the nature of the afterglow of ZTF21aaeyldq is very similar to other well-known orphan afterglows ZTF20aajnksq and ZTF19abvizsw. The comparison also indicates that during the late epochs, ZTF21aaeyldq was fainter than other known cases of orphan afterglows. Further, we measure the energetics of prompt emission (using the limit on $E_{\rm \gamma, iso}$ $\leq 1.52 \times 10^{52}$ erg) and afterglow phases of ZTF21aaeyldq and estimated a radiative efficiency ($\eta$= $E_{\rm \gamma, iso}$/($E_{\rm \gamma, iso}$+ $E_{k}$)) $\leq$ 28.6 \%. This suggests that most energy is used as kinetic energy and supports the internal shocks as the likely origin of prompt emission. We expect to detect more similar sources in the current era of survey telescopes with large fields of view like ZTF and others. Our results also highlight that the 3.6\,m DOT has a unique capability for deep follow-up observations of similar and new transients to deeper limits as a part of time-domain astronomy.

\chapter{\sc Host galaxies, and Environments of GRBs}
\label{ch:6} 
\blfootnote{This chapter is based on the results published in: \textbf{{Gupta}, Rahul} et al., 2022, {\textit{Journal of Astrophysics and Astronomy}, {\textbf{43}}, 82}.}

\ifpdf
    \graphicspath{{Chapter6/Chapter6Figs/PNG/}{Chapter6/Chapter6Figs/PDF/}{Chapter6/Chapter6Figs/}}
\else
    \graphicspath{{Chapter6/Chapter6Figs/EPS/}{Chapter6/Chapter6Figs/}}
\fi

\ifpdf
    \graphicspath{{Chapter6/Chapter6Figs/JPG/}{Chapter6/Chapter6Figs/PDF/}{Chapter6/Chapter6Figs/}}
\else
    \graphicspath{{Chapter6/Chapter6Figs/EPS/}{Chapter6/Chapter6Figs/}}
\fi

\normalsize

Followed by the prompt emission and longer-lasting multi-wavelength afterglow phases \citep{2004RvMP...76.1143P, 2019Natur.575..455M}, late-time observations of the host galaxies of GRBs are of crucial importance to examine the burst environment and, in turn, about the possible progenitors.   

GRBs can be used to study the galaxies both at high (the most distant, GRB 090423 with spectroscopic $z \sim$ 8.2) and low (nearest, GRB 170817A $z \sim$ 0.0097) redshifts due to their intrinsic brightness (much higher signal to noise ratio; \citealt{2012A&A...542A.103B}). The GRB host galaxy characteristics of long and short GRBs are largely different such as morphology, stellar population, offsets, etc. but share common properties too for a fraction of observed populations. These observed characteristics are likely associated with the physical conditions surrounding possible progenitors producing GRBs. Long GRBs are generally localized in active star-forming and young stellar population dwarf galaxies. Since long GRBs are likely to be related to the death of massive stars, they are widely cited as robust and potentially unbiased tracers of the star formation and metallicity history of the Universe up to $z$ $\sim$ 8 \citep{2009ApJ...691..182S}. Host galaxy observations of long GRBs suggest that they preferentially occur in low-metallicity galaxies \citep{2011MNRAS.414.1263M}. On the other hand, short GRBs are expected in any type of galaxy associated with an old stellar population \citep{2009ApJ...690..231B}. Their locations are relative to their host centres and have a median physical offset of about five kpc, which is about four times larger than the median offset for long bursts \citep{2013ApJ...776...18F}. Therefore, host parameters can constrain the nature of GRBs' possible progenitors and environments.

In the pre-\swift era (before 2004), there were few bursts with measured redshifts. In this era, the host galaxies were intensely studied once the redshift was known to be low ($z$ $\le$ 0.3). In \swift era, the number of GRBs with measured redshift values increases but still $\sim$ 25 \% of the localized ones and could still be biased against dusty events\footnote{https://www.mpe.mpg.de/~jcg/grbgen.html}. \cite{2009ApJ...691..182S} studied the host galaxy properties for a large sample of GRB hosts and suggested that GRB hosts are similar to normal star-forming galaxies in both the nearby and the distant universe. \cite{2016ApJ...817....7P, 2016ApJ...817....8P} examined an unbiased sample \footnote{introduced as the \swift Gamma-Ray Burst Host Galaxy Legacy Survey} of the host galaxies of long GRBs (mainly photometric) and proposed that the dusty bursts are generally found in massive host galaxies. It gives a clue that the massive galaxies (star-forming) are typical and homogeneously dusty at higher redshift. On the other hand, low-mass galaxies (star-forming) have a small amount of dust in their interstellar medium (to some level). Also, \cite{2010MNRAS.405...57S, 2018A&A...617A.105J} presented comparative studies of the host galaxies of GRBs and compared their properties with those of core-collapse supernovae (CCSNe). More recently, \cite{2021MNRAS.503.3931T} presented a comprehensive study of a large sample of core-collapse supernova (CCSN) host galaxies and compared it with the host galaxies of the nearest long GRBs and superluminous supernova (SLSN) and found a hint that host-galaxy mass or specific star-formation rate is more fundamental in driving the preference for SLSNe and long GRBs in unusual galaxy environments.

Deeper optical photometric follow-up observations of energetic transients such as afterglows of long and short GRBs are frequently carried out \citep{2021RMxAC..53..127K, 2020GCN.29148....1P, 2022JApA...43...11G, 2021GCN.29490....1G, 2021GCN.31299....1G, 2022MNRAS.511.1694G, 2022MNRAS.513.2777K} using the recently commissioned largest Indian optical telescope, i.e., 3.6\,m DOT situated at Devasthal observatory of ARIES Nainital and the back-end instruments \citep{2018BSRSL..87...42P, 2018BSRSL..87...58O, 2022JApA...43...27K}. Observations of galaxies and other objects of low surface brightness are also carried out using the 3.6\,m DOT \citep{2022JApA...43...27K, 2022JApA...43....7P}. In this work, we performed the spectral energy distribution modelling of a sample of five host galaxies of GRBs observed by the 3.6\,m DOT/back-ends and compared the results with other well-studied samples of host galaxies. We observed the host galaxies of these five bursts subject to the availability of the observing time of the CCD IMAGER and clear sky conditions (see section \ref{sample}). This work demonstrates the capabilities of deep follow-up observations of such faint and distant hosts of explosive transients using the 3.6\,m DOT. We have arranged this chapter in the following sections. In section \ref{sample}, we present our host galaxies sample (with brief details about each burst) and their multi-band photometric observations taken with 3.6\,m DOT. In section \ref{sedmodelling}, we present the host galaxy spectral energy distribution modelling of our photometric data along with those obtained from literature using \sw{Prospector} software (version 1.1.0). In section \ref{results_host}, we have given the SED modelling results and comparison with other known host galaxies. Finally, in section \ref{Summary and Conclusion_host}, we have presented the summary and conclusion of the present chapter.

\section{Sample of Host galaxies, observations with the 3.6m DOT}
\label{sample}

In this section, we provide details of multi-band photometric observations of the host galaxies of the sample (five bursts with peculiar features between redshifts 0.0758 to 2.02, see details below), using India's largest 3.6\,m DOT telescope and back-end instruments like 4Kx4K CCD Imager \citep{2018BSRSL..87...42P} and the TANSPEC \citep{2018BSRSL..87...58O}. Deeper optical observations of the host galaxies of GRB 030329, GRB 130603B, GRB 140102A, and GRB 190829A in several optical filters ($B, V, R, I$) were obtained using the 4Kx4K CCD Imager. Details about observations of each of these four host galaxies are described below in respective sub-sections. We performed the optical photometric data analysis for the host galaxies observations using IRAF/DAOPHOT using methods described in \citep{2022JApA...43...27K, 2021MNRAS.505.4086G, 2019MNRAS.485.5294P}. In the case of GRB 200826A, photometric optical-NIR observations in $I, J$, and $K$ filters were performed using the TANSPEC \citep{2018BSRSL..87...58O}, and the detail about the data reduction are described in the respective subsection below. A photometric observation log for each burst of our sample is given in Table \ref{tab:observationslog_host} of the appendix. The redshift distribution of all the GRBs with a measured redshift up to January 2021 (data obtained from GRBweb catalogue page\footnote{https://user-web.icecube.wisc.edu/~grbweb\_public/Summary\_table.html} provided by P. Coppin), along with those discussed presently are shown in Figure \ref{redshift}.

\begin{figure}[ht!]
\centering
\includegraphics[scale=0.35]{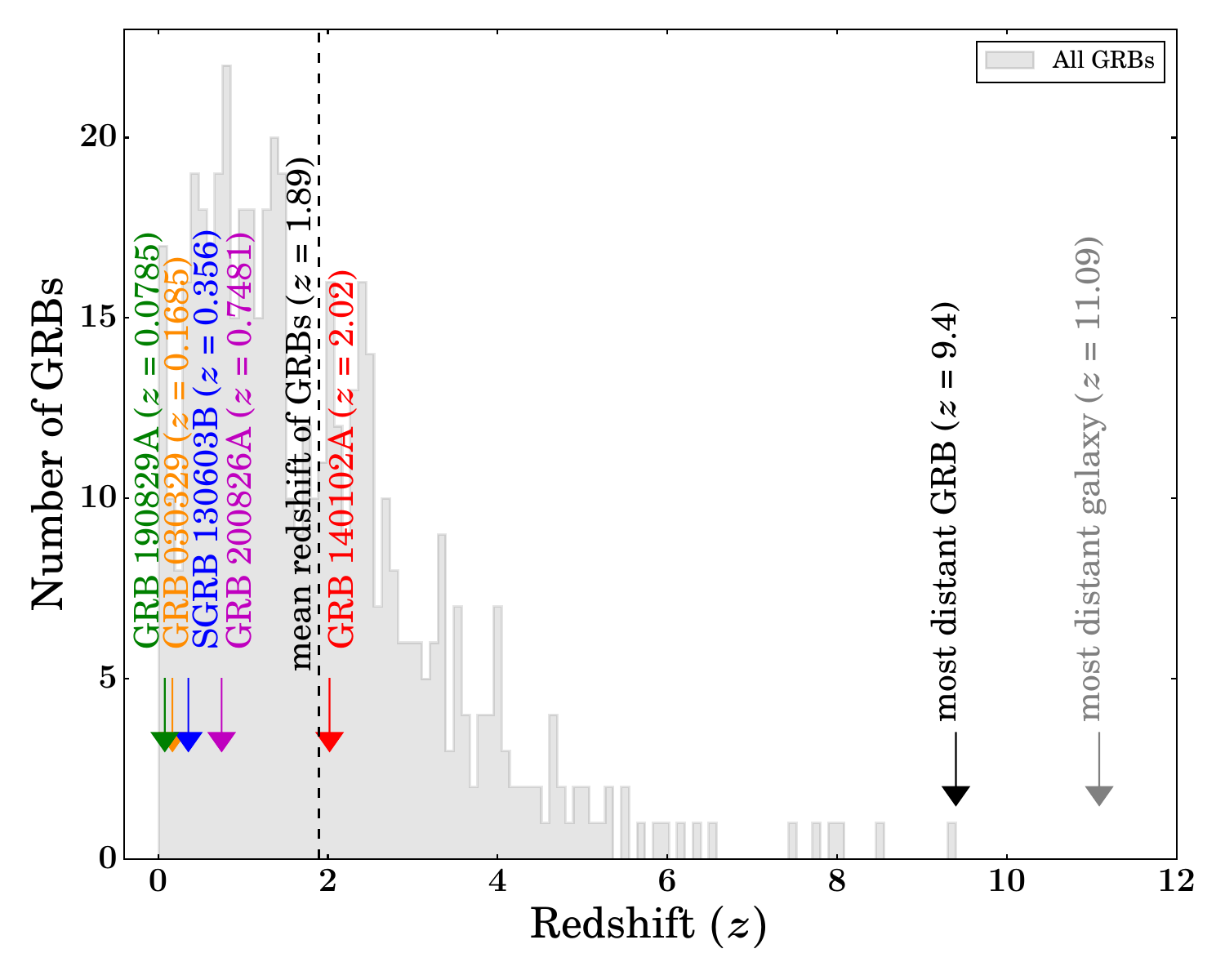}
\caption{The redshift distribution for all the GRBs with a measured redshift value till January 2021 (shown with grey colour). The position of each burst of our sample is also shown. The vertical black dashed line represents the mean value of the redshift for all the GRBs with a measured redshift.} 
\label{redshift}
\end{figure}

\begin{table*}
\scriptsize
\caption{Log of the host galaxy panchromatic observations of our sample taken with 3.6\,m DOT and those reported in the literature. The magnitude shown with star markers is in the Vega system. NOT, CAHA, GTC, and LDT denote the Nordic Optical Telescope, Centro Astronómico Hispano-Alemán, Gran Telescopio Canarias, and Lowell Discovery Telescope, respectively.}
\begin{center}
\begin{tabular}{c c c c c c}
\hline
\bf Date & \bf Exposure (s) & \bf Magnitude (AB) & \bf Filter & \bf Telescope & \bf References \\ \hline

\multicolumn{6}{c}{\textbf{GRB 030329}} \\ \hline 
24.03.2004 & 5x900 & $23.45 \pm 0.10$ & U & 2.56m NOT & \cite{2005AA...444..711G} \\
24.03.2004 & 3x600 & $23.26 \pm 0.07$ & B & 2.56m NOT & \cite{2005AA...444..711G} \\
05.01.2004& 113x60 & $22.42 \pm 0.16$ & J & 3.5m CAHA & \cite{2005AA...444..711G} \\
06.01.2004& 109x60 & $22.55 \pm 0.24$ & H & 3.5m CAHA & \cite{2005AA...444..711G} \\
07.01.2004& 99x60 & $>21.57 $ & K & 3.5m CAHA & \cite{2005AA...444..711G} \\ 
\bf 23.03.2017&\bf 3x600 & $\bf 22.81 \pm 0.13$ &\bf V &\bf 3.6\,m DOT &\bf Present work \\
\bf 23.03.2017&\bf 4x600 & $\bf 22.83 \pm 0.07$ &\bf R &\bf 3.6\,m DOT &\bf Present work \\ \hline

\multicolumn{6}{c}{\textbf{GRB 130603B}} \\ \hline 
05.06.2013 & 5x300 & $20.69 \pm 0.15$ & I & 1.5m OSN & \cite{2019MNRAS.485.5294P} \\
22.06.2013 & 15x60 & $19.69 \pm 0.13$ & K & 3.5m CAHA & \cite{2019MNRAS.485.5294P} \\
22.06.2013 & 15x60 & $20.06 \pm 0.09$ & J & 3.5m CAHA & \cite{2019MNRAS.485.5294P} \\
22.06.2013 & 15x60 & $19.68 \pm 0.13$ & H & 3.5m CAHA & \cite{2019MNRAS.485.5294P} \\
22.06.2013 & 15x60 & $20.11 \pm 0.07$ & Z & 3.5m CAHA & \cite{2019MNRAS.485.5294P} \\
05.07.2013 & 4x50 & $22.01 \pm 0.03$ & g & 10.4m GTC & \cite{2019MNRAS.485.5294P} \\
05.07.2013 & 4x50 & $20.97 \pm 0.01$ & r & 10.4m GTC & \cite{2019MNRAS.485.5294P} \\
05.07.2013 & 4x50 & $20.65 \pm 0.02$ & i & 10.4m GTC & \cite{2019MNRAS.485.5294P} \\
\bf 23.03.2017 &\bf 2x300 &\bf $22.13 \pm 0.05$ &\bf B &\bf 3.6\,m DOT & \cite{2019MNRAS.485.5294P} \\
\bf 23.03.2017 &\bf 2x300 &\bf $20.72 \pm 0.02$ &\bf R  &\bf 3.6\,m DOT  & \cite{2019MNRAS.485.5294P}  \\ \hline

\multicolumn{6}{c}{\textbf{GRB 140102A}} \\ \hline

13.05.2014   &   59x65.0   & $   21.18 \pm   0.26 $  &  H  & CAHA & \cite{2021MNRAS.505.4086G}\\
18.07.2017   &   7x120.0   & $   25.13 \pm   0.16 $  &  g  & 10.4m GTC & \cite{2021MNRAS.505.4086G}\\
18.07.2017   &   7x90.0    & $   24.47 \pm   0.13 $  &  r  & 10.4m GTC & \cite{2021MNRAS.505.4086G}\\
 18.07.2017   &   7x90.0    & $   24.17 \pm   0.13 $  &  i  & 10.4m GTC & \cite{2021MNRAS.505.4086G}\\
 18.07.2017   &   6x60.0    & $   23.88 \pm   0.18 $  &  z  & 10.4m GTC & \cite{2021MNRAS.505.4086G}\\
\bf 16.01.2021   & \bf 3x300, 2x900   & \bf  $\geq$ 24.10  &\bf  R  &\bf  3.6\,m DOT& \cite{2021MNRAS.505.4086G} \\ \hline

\multicolumn{6}{c}{\textbf{GRB 190829A}} \\ \hline 

26.09.2000 & --&$18.64 \pm 0.05$ &u & SDSS & \cite{2005AJ....129.1755A} \\
26.09.2000 & --& $16.686 \pm 0.005$ & g& SDSS & \cite{2005AJ....129.1755A}\\
26.09.2000 & --&$15.729 \pm 0.004$ & r& SDSS & \cite{2005AJ....129.1755A}\\
26.09.2000 &-- &$15.229 \pm 0.004$ & i& SDSS & \cite{2005AJ....129.1755A}\\
26.09.2000 & --&$14.872 \pm 0.007$ & z& SDSS & \cite{2005AJ....129.1755A}\\
12.12.2000&-- & $13.798 \pm 0.096^{*}$& J & 2MASS &\cite{2006AJ....131.1163S} \\
12.12.2000&-- & $12.877 \pm 0.092^{*}$& H & 2MASS &\cite{2006AJ....131.1163S} \\
12.12.2000&-- & $12.320 \pm 0.104^{*}$& K & 2MASS &\cite{2006AJ....131.1163S} \\
\bf 03.10.2020 &\bf 2x300 & $\bf 17.05 \pm 0.05^{*}$ &\bf B &\bf 3.6\,m DOT &\bf Present work \\
\bf 03.10.2020 &\bf 2x200 & $\bf 15.75 \pm 0.02^{*}$ &\bf R &\bf 3.6\,m DOT &\bf Present work \\
\bf 03.10.2020 &\bf 2x200 & $\bf 15.26 \pm 0.03^{*}$ &\bf I &\bf 3.6\,m DOT &\bf Present work \\ \hline 

\multicolumn{6}{c}{\textbf{GRB 200826A}} \\ \hline 

28.08.2020 & 3600 &  $21.11 \pm  0.16^{*}$  & J & Palomar Hale 200-in (P200) & \cite{2021NatAs...5..917A} \\
13.09.2020 & 5x180 & $23.45 \pm 0.24$ & u & 4.3m LDT & \cite{2021NatAs...5..917A} \\
13.09.2020 & 4x180 & $23.36 \pm 0.05$ & g & 4.3m LDT & \cite{2021NatAs...5..917A} \\
13.09.2020 & 10x150 & $22.86 \pm 0.18$ & r & 4.3m LDT & \cite{2021NatAs...5..917A} \\
13.09.2020 & 6x180 & $22.13 \pm 0.05$ & z & 4.3m LDT & \cite{2021NatAs...5..917A} \\
\bf 04.11.2020 &\bf 12x300 & $\bf 22.71 \pm 0.10$ &\bf i &\bf 3.6\,m DOT &\bf Present work \\
\bf 04.11.2020 &\bf 2100 &\bf $>$ 20.56$^{*}$ &\bf J &\bf 3.6\,m DOT &\bf Present work \\
\bf 11.11.2020 &\bf 2100 &\bf $>$ 19.55$^{*}$ &\bf K &\bf 3.6\,m DOT & {\bf Present work} \\ 
\hline
\vspace{-1em}
\end{tabular}
\end{center}
\label{tab:observationslog_host}
\end{table*}

\subsection{GRB 030329 (associated SN 2003dh):}

GRB 030329 was triggered by many detectors on-board the High Energy Transient Explorer (HETE-2) mission at 11:37:14.67 UT on 2003 March 29. The prompt emission light curve of this GRB consists of two merging emission pulses with a total duration of $\sim$ 25 s in 30-400 \keV energy band. Later on, a multiwavelength follow-up observations campaign of GRB 030329 revealed the discovery of optical \citep{2003GCN..1985....1P}, X-ray \citep{2003GCN..1996....1M}, and radio counterparts \citep{2003GCN..2014....1B}. \cite{2003GCN..2020....1G} measured the redshift of the burst ($z$) = 0.1685 using the early spectroscopy observations taken with the Very Large Telescope (VLT). Furthermore, a late-time bump in the optical light curve along with spectroscopic observations confirms the detection of associated broad-line type Ic supernova \citep{2003Natur.423..847H}, establishing the relationship between the afterglow of long GRBs and supernovae. \cite{2008MNRAS.387.1227O} utilizes the spectral evolutionary models to constrain the progenitor and suggested a collapsar scenario for the progenitor of GRB 030329/SN 2003dh.

We performed the host galaxy observations of GRB 030329 using 4K $\times$ 4K Imager mounted on the axial port 3.6\,m DOT in March 2017. Multiple frames with an exposure time of 600 s were taken in R and V filers. The host galaxy of GRB 030329/SN 2003dh is clearly detected in both filters. A finding chart taken in the R filer is shown in the upper-left panel of Figure \ref{host_findingchart}.

\subsection{GRB 130603B (associated kilonova emission):}

This burst was detected by Burst Alert Telescope (BAT) on-board \swift mission at 15:49:14 UT on 03$^{rd}$ June 2013 at the position RA = 172.209, DEC = +17.045 degree (2000) with an uncertainty of three arcmin. The prompt emission BAT light curve consists of the fast-rising and exponential decay (FRED) like single structure with \tninty duration of 0.18 $\pm$ 0.02 s (in 15-350 \keV), classifying this burst as a short-duration GRB \citep{2013GCN.14741....1B}. \cite{2013GCN.14744....1T} obtained the afterglow spectrum using 10.4m GTC and reported the redshift of the burst $z$ = 0.356. Later on, the late time near-infrared (NIR) observations reveal the detection of the kilonova emission (the first known case) accompanied by short GRB 130603B, supporting the merger origin of short GRBs \citep{2013Natur.500..547T}. \cite{2014A&A...563A..62D} studied the environment and proposed that the explosion site of this burst is not similar to those seen in the case of long GRBs.

We observed the host galaxy of short GRB 130603B using 3.6\,m DOT in B and R filters with an exposure time of 2x300 s in each on $23^{rd}$ March 2017. We detected a bright galaxy (see the upper-right panel of Figure \ref{host_findingchart}) in both filters. 

\begin{figure*}[ht!]
\centering
\includegraphics[angle=0,scale=0.25]{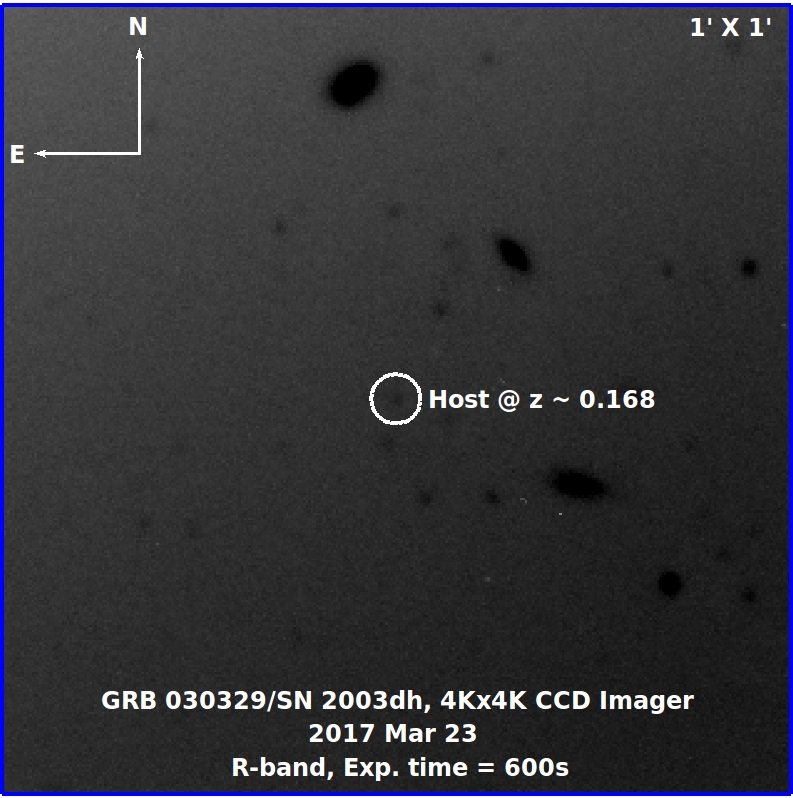}
\includegraphics[angle=0,scale=0.25]{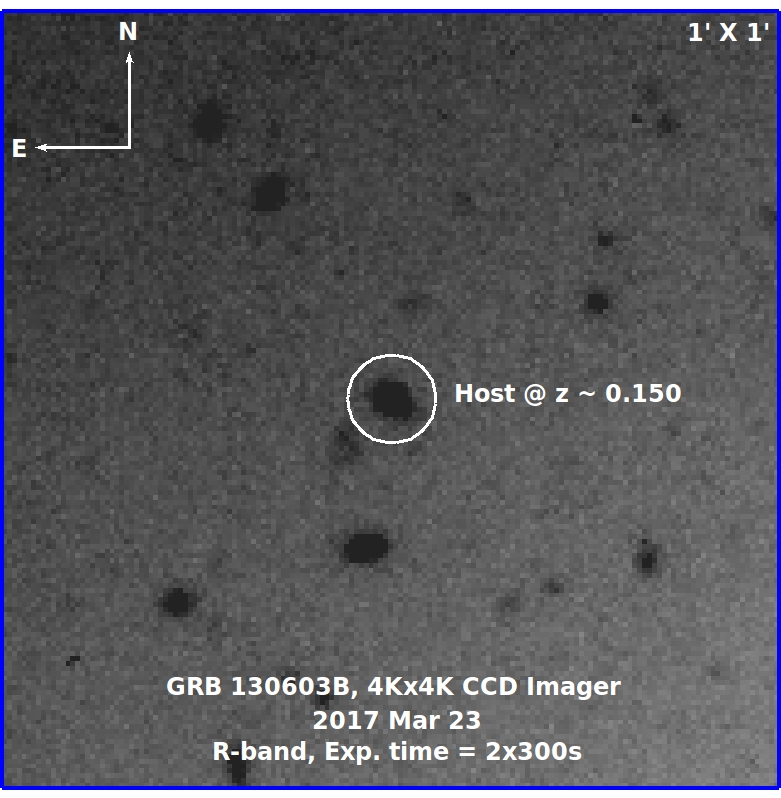}
\includegraphics[angle=0,scale=0.25]{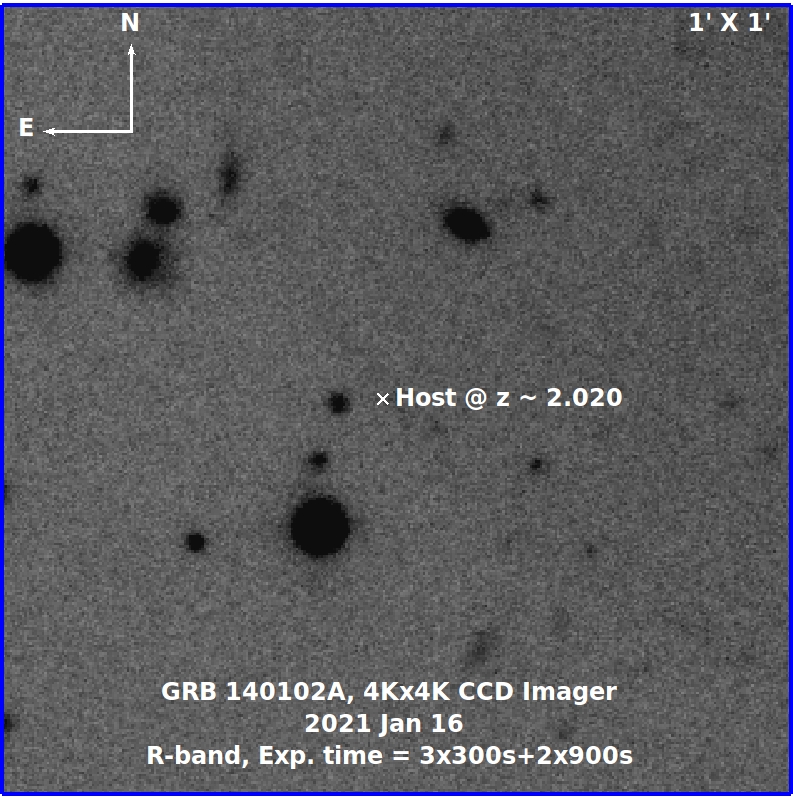}
\includegraphics[angle=0,scale=0.25]{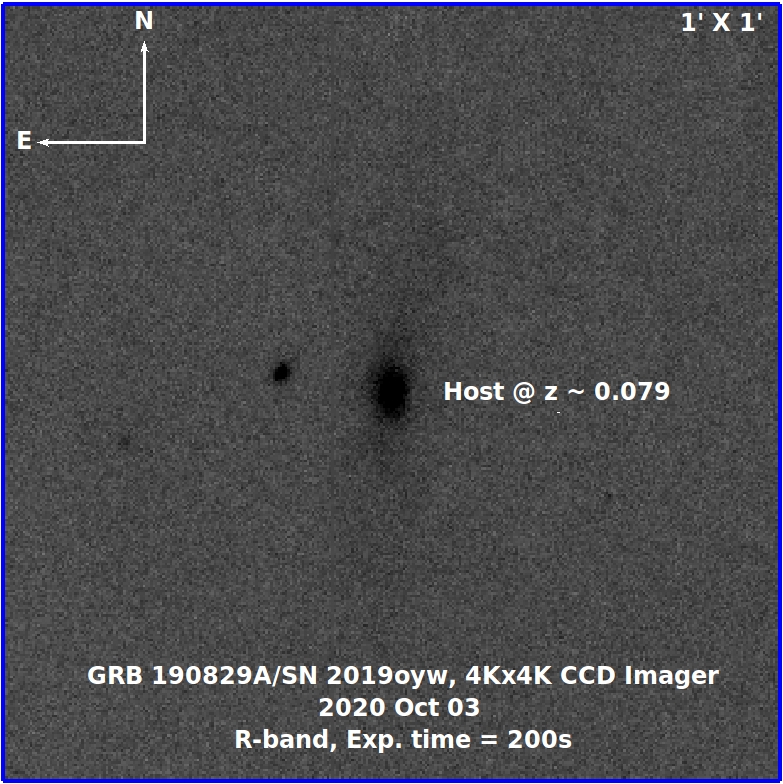}
\caption{The R-band finding charts of the host galaxies of GRB 030329 (top left), GRB 130603B (top right), GRB 140102A (bottom left), and GRB 190829A (bottom right) were obtained using 4K$\times$4K CCD Imager mounted on the 3.6\,m DOT \citep{2018BSRSL..87...42P, 2022JApA...43...27K}. The position of the host galaxies in the frames is marked with circles.}
\label{host_findingchart}
\end{figure*}

\subsection{GRB 140102A (early reverse shock emission):}

GRB 140102A was jointly detected by \fermi (by both Gamma-ray Burst Monitor (GBM) and Large Area Telescope (LAT)) and \swift BAT detectors. We carried out an early follow-up of the optical afterglow of GRB 140102A reveals the rarely observed reverse shock signature in the light curve. We also constrain the redshift of the burst $z$ = 2.02 using joint X-ray and optical SED. We calculated the magnetization parameter using the afterglow modelling of GRB 140102A and suggested that the jet composition could be dominant with a moderately magnetized outflow, in this case, \citep{2021MNRAS.505.4086G}. 
Furthermore, we carried out the host galaxy observations of GRB 140102A in the R filter (with a total exposure time of 45 minutes) using 3.6\,m DOT on 16 January 2021. However, we could not detect the host galaxy (see the lower-left panel of Figure \ref{host_findingchart}), but we could constrain the deep limiting magnitude. 

\subsection{GRB 190829A (nearest Very high energy (VHE) burst):}

GRB 190829A was detected by \fermi GBM (at 19:55:53 UT), and \swift BAT (at 19:56:44.60 UT) on $29^{th}$ August 2019. \cite{2019ATel13052....1D} reported the detection of very high energy (V.H.E.) emission from the source using H.E.S.S. observations. We studied the prompt emission characteristics of the two-episodic low-luminous GRB 190829A and found that the first episode is an Amati outlier, showing the peculiar nature of the prompt emission \citep{2020ApJ...898...42C, 2021RMxAC..53..113G}. Furthermore, we calculated the redshift of the burst $z$ = 0.0785 using 10.4m GTC spectroscopic observations, making the event the nearest V.H.E. detected GRB along with the detection of associated supernova \citep{2021A&A...646A..50H}. However, it is still an unsolved question that V.H.E. emission is originated due to their environment or its emission mechanism \citep{2020A&A...633A..68D}.

The associated host galaxy of GRB 190829A is a significantly bright SDSS galaxy (SDSS J025810.28-085719.2). We observed this galaxy using 3.6\,m DOT in B, R, and I filters on 03$^{rd}$ October 2020. The host galaxy is clearly detected in each filter of our observations. A finding chart for the host galaxy of GRB 190829A is shown in the lower-right panel of Figure \ref{host_findingchart}.

\subsection{GRB 200826A (shortest long burst):}
GRB 200826A was detected by \fermi GBM at 04:29:52.57 UT on 26 August 2020 with a \tninty duration of 1.14 s in the GBM 50-300 \keV energy range \citep{2020GCN.28287....1M}. Late-time optical follow-up observations revealed the bump in the light curve, consistent with the supernova emission. Although the prompt properties of the burst are typical to those of short GRB, late-time follow-up observations confirm a collapsar origin \citep{2021NatAs...5..917A}. \cite{2021NatAs...5..917A} suggested that the burst is the shortest long burst with SN bump and lies on the brink between a successful and a failed collapsar.

We have obtained the optical and NIR photometric data of the host galaxy of GRB 200826A using the TIFR-ARIES Near-Infrared Spectrometer \citep[TANSPEC;][]{2018BSRSL..87...58O} mounted on the 3.6\,m DOT, Nainital, India during the nights of 2020 November. TANSPEC is a unique instrument that provides simultaneous wavelength coverage from 0.5-2.5 $\mu$m in imaging and spectroscopic modes. We have given exposures of one hour, 35 minutes, 35 minutes in I (4$^{th}$ November 2020), J (4$^{th}$ November 2020), and K (11$^{th}$ November 2020) bands, respectively. In the I band, 12 frames of 5 min exposure, whereas, in $J$ and $K$ bands, three sets of $20\times5$ sec exposure at seven dithered positions (total of 35 minutes in $J$ and $K$ bands) were taken with TANSPEC. We have used standard data reduction procedures for image cleaning, photometry, and astrometry \citep[for details, see ][]{2020MNRAS.498.2309S}. The host galaxy is detected in the I band as $22.71 \pm 0.10$ mag, and there was an upper limit of J $>$ 20.56 mag and K $>$ 19.55 mag. A finding chart of the host galaxy of GRB 200826A taken with TANSPEC is shown in Figure \ref{host_findingchart_tanspec}. 

\begin{figure}[ht!]
\centering
\includegraphics[angle=0,scale=0.3]{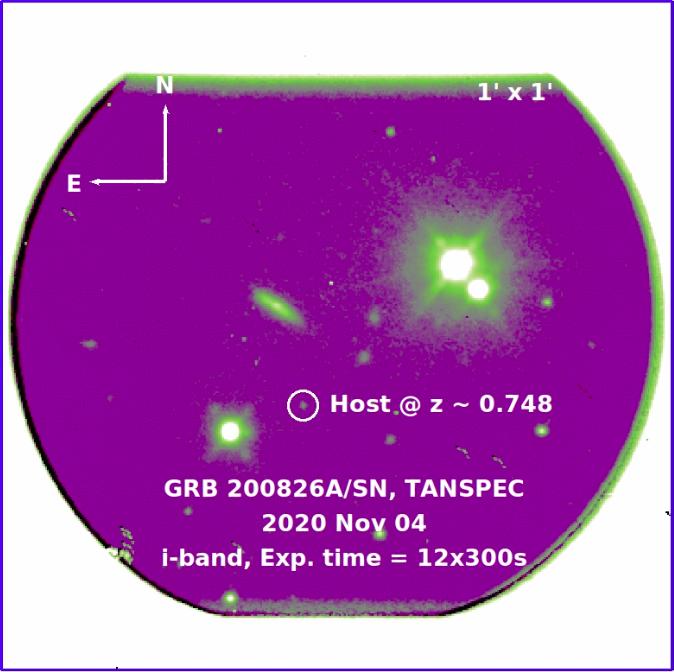}
\caption{The i-band finding chart of the host galaxy of GRB 200826A/SN was obtained using the TANSPEC \citep{2018BSRSL..87...58O}. The position of the host galaxy in the frame is marked with a circle.}
\label{host_findingchart_tanspec}
\end{figure}

\section{Results and Discussion}
\label{results_host}

\subsection{Panchromatic SED modelling}
\label{sedmodelling}

In our previous studies \citep{2019MNRAS.485.5294P, 2021MNRAS.505.4086G}, \sw{LePHARE} software is used for the modelling of the host galaxies, and it suffers from a major limitation, i.e., using only chi-square statistics to the best-fit solution. The results of \sw{LePHARE} software are primarily affected by degeneracy among parameters as it could not provide the posterior distributions. Therefore, in this work, we utilized an advanced software called \sw{Prospector} (version 1.1.0) for SED fitting to constrain the host galaxies properties of our sample of five hosts. \sw{Prospector} \citep{2017ApJ...837..170L, 2021ApJS..254...22J} software (stellar population modelling code) for modelling the SEDs using the measured photometric magnitudes values of the host galaxies. \sw{Prospector} utilizes a library of Flexible Stellar Population Synthesis models \citep{2009ApJ...699..486C}. It is advanced software that determines the best-fit solution to the SED model fitting using \sw{Dynesty} (implements dynamic nested sampling algorithm) and produces the posterior distributions for the model parameters\footnote{ https://github.com/bd-j/prospector}. The posterior distributions are useful to verify the degeneracy between the model parameters. We performed the SED fitting to photometric data for each of the host galaxies of our sample at their respective fixed redshift values. We have used the \sw{parametric$\_$sfh} model to calculate the stellar population properties such as stellar mass formed ( $M_\star$, in units of solar mass), stellar metallicity ($\log Z/Z_\odot$), age of the galaxy ($t$), rest-frame dust attenuation for a foreground screen in mags (A$_{\rm V}$), and star formation timescale ($\tau$) for an exponentially declining star formation history (SFH). We have set these host galaxy model parameters free to determine the posterior distribution and consider uniform priors across the allowed parameter space within Flexible Stellar Population Synthesis models. We have fixed the maximum values of the prior of age of the galaxies corresponding to the age of the Universe at their respective measured redshift values. For the host SED modelling using stellar population models, we consider Milky Way extinction law \citep{1989ApJ...345..245C} and Chabrier initial mass function \citep{2003PASP..115..763C}. We calculated the star formation rate using the equation given in Section 4.1 of \cite{2020ApJ...904...52N}:


This section presents the results of the host galaxies modelling of our sample. We corrected the values of the observed magnitudes (AB system) for the foreground galactic extinction for each galaxy following \cite{2011ApJ...737..103S} and used them as input to \sw{Prospector} for the SED modelling of galaxies. The best-fit SEDs are shown in Figure \ref{sed_host} (corner plots are shown in Figure \ref{corner_Host} of the appendix), and their results thus obtained are tabulated in Table \ref{tab:sedresults}.

\begin{figure}
\centering
\includegraphics[scale=0.29]{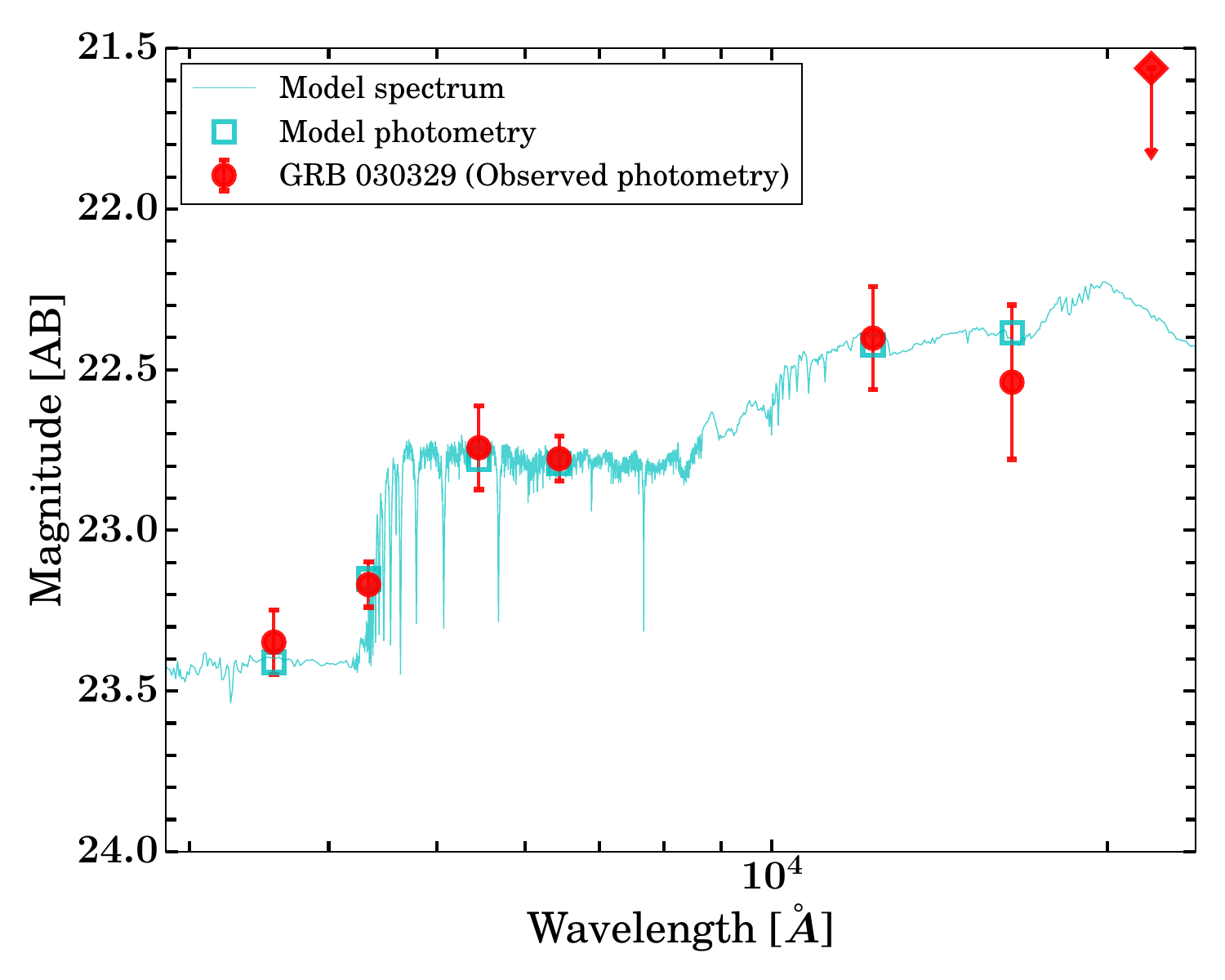}
\includegraphics[scale=0.29]{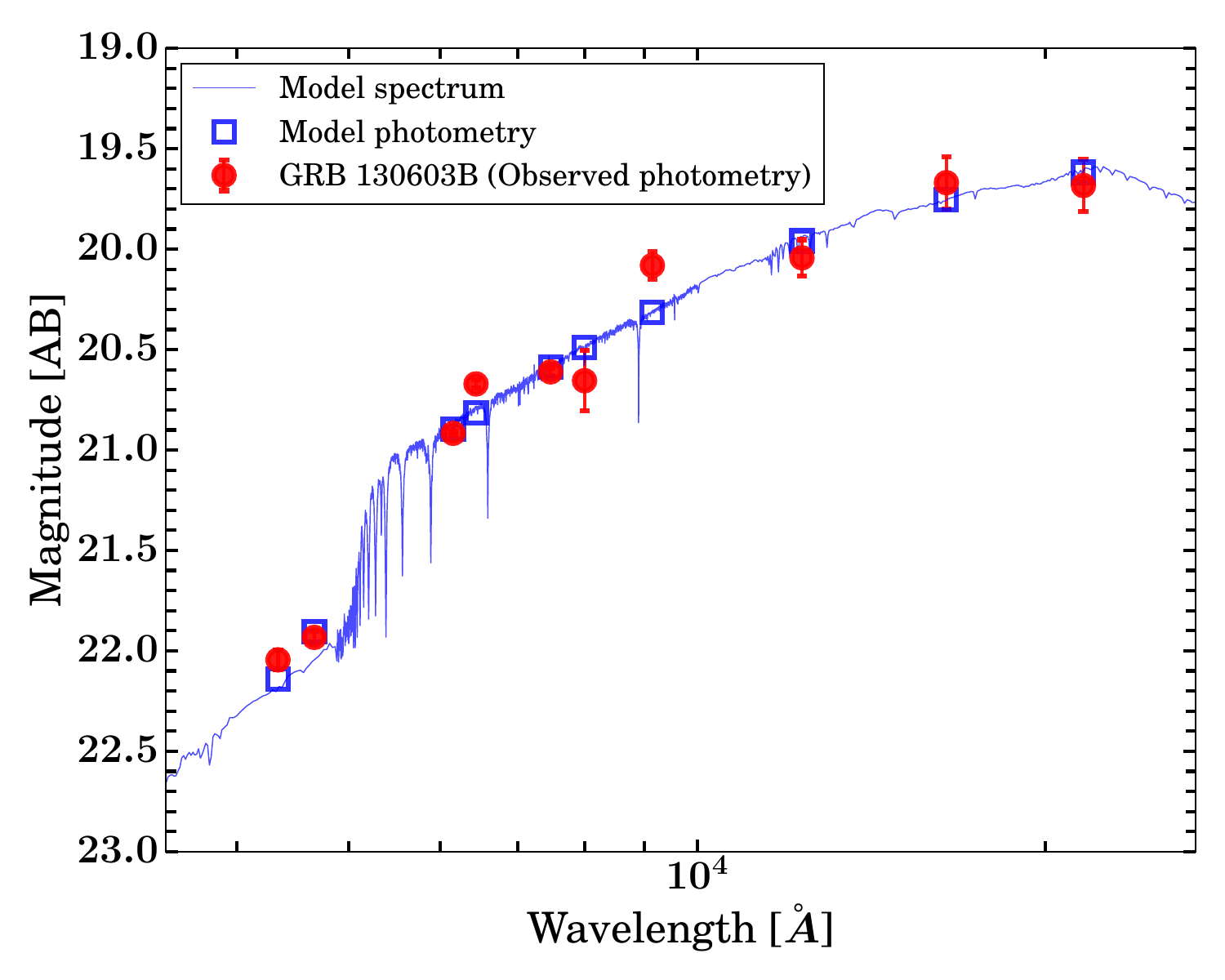}
\includegraphics[scale=0.29]{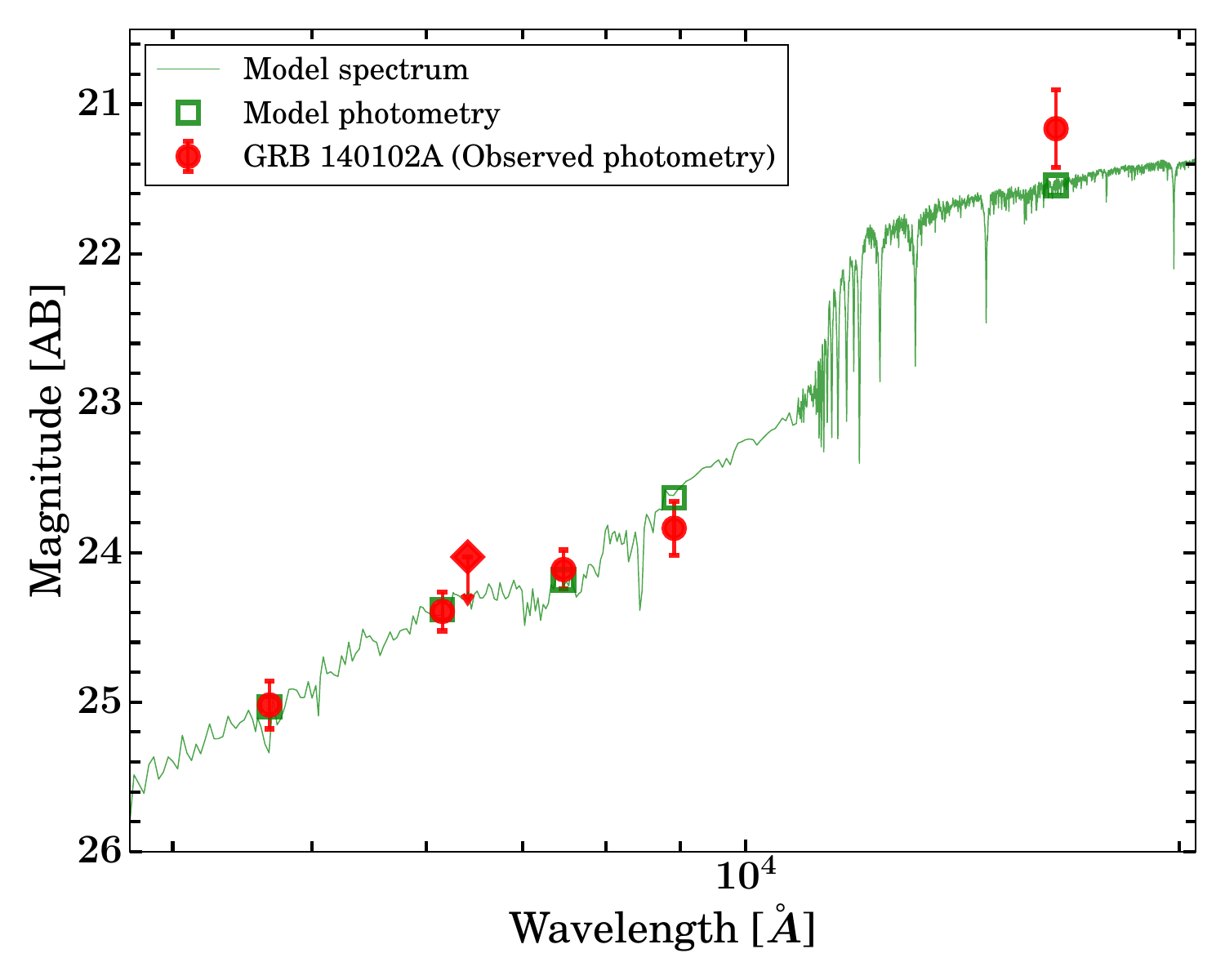}
\includegraphics[scale=0.29]{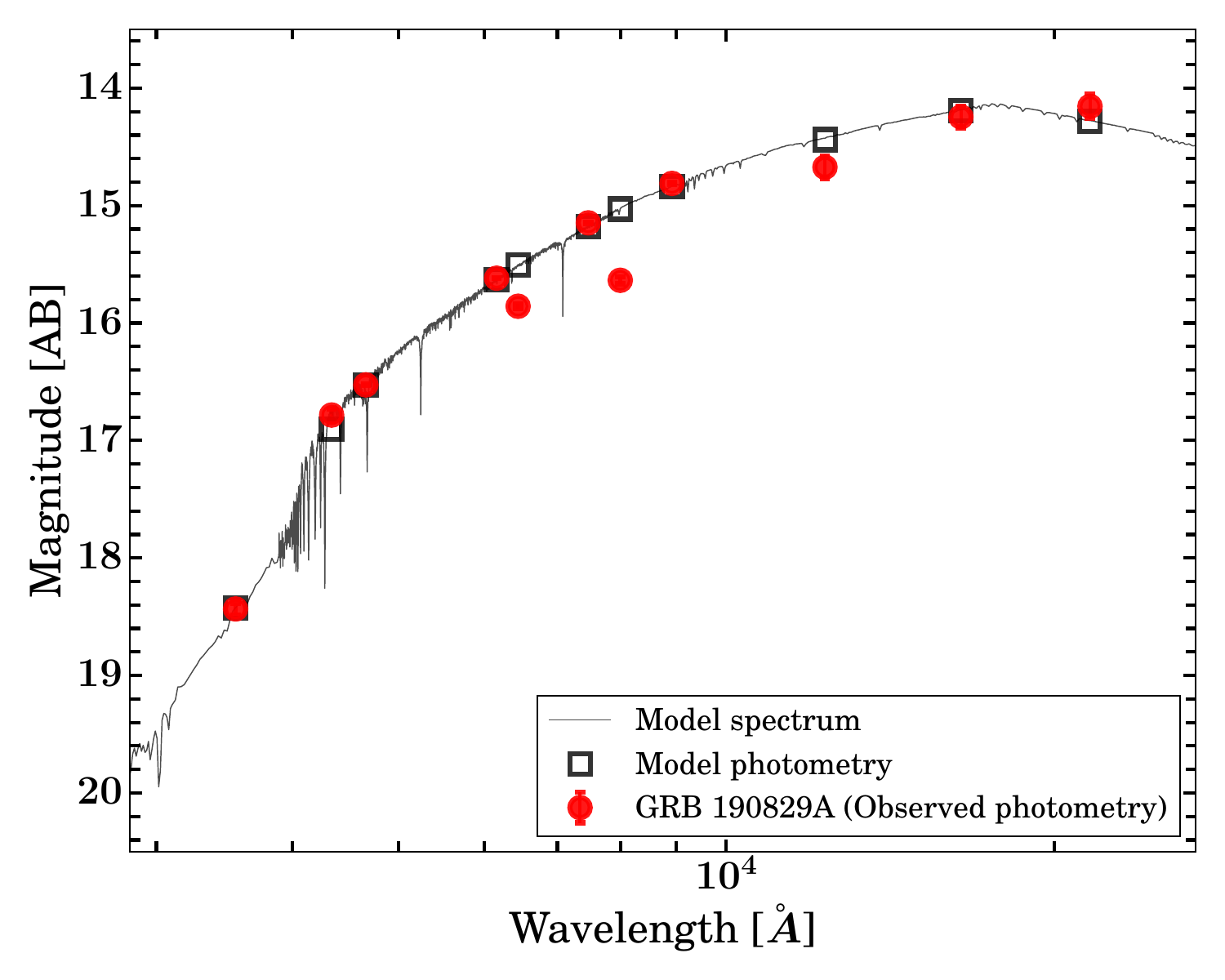}
\includegraphics[scale=0.29]{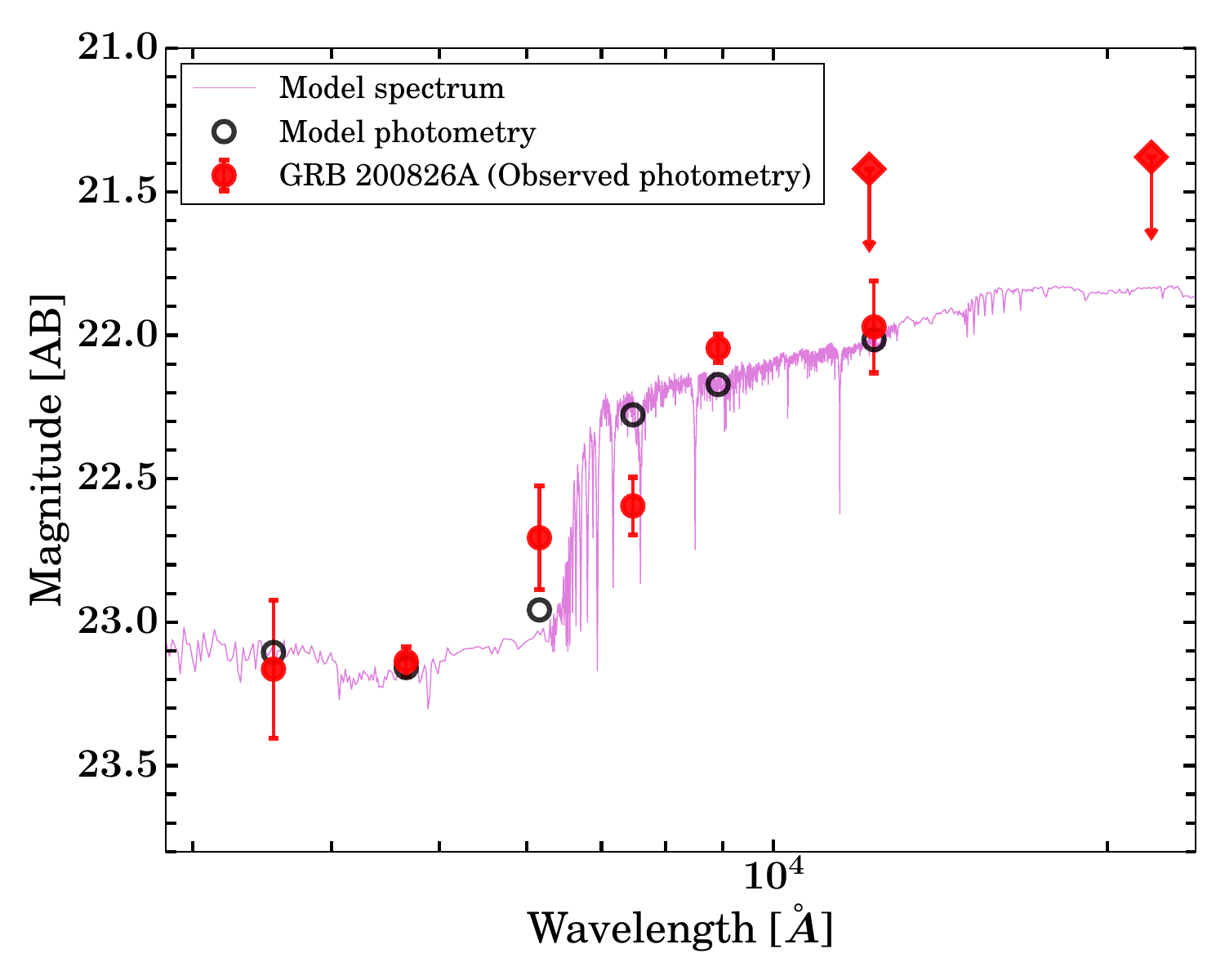}
\caption{Spectral energy distribution modelling of the host galaxies of our sample: The best fit modelled spectrum and photometry obtained using \sw{Prospector} are presented in cyan, blue, green, black, magenta for GRB 030329, GRB 130603B, GRB 140102A, GRB 190829A, and GRB 200826A, respectively. The photometry data of the host galaxies (corrected for galactic extinction) in the AB system is shown with red circles.}
\label{sed_host}
\end{figure}

\begin{table}
\scriptsize
\caption{The stellar population properties of the host galaxies of our sample were obtained using spectral energy density modelling using \sw{Prospector}. NH$_{\rm host}$ denotes the intrinsic column density.}
\begin{center}
\begin{tabular}{c c c c c c c c c}
\hline
\bf GRB &\bf RA & \bf DEC &$\bf  z$ & \bf log(M/M$_{\odot}$) & \bf log(Z/Z$_{\odot}$) & $\bf A_{V}$ & $\bf t_{gal}$  & \bf  NH$_{\rm host}$ \\ 
 &\bf  (J2000) & \bf  (J2000) & &  &  &\bf  (Mag) & \bf  (Gyr) &  ($\rm \bf cm^{-2}$)\\ \hline
030329 & 161.21 & 21.52 & 0.1685 &7.98$^{+0.12}_{-0.14}$ & -0.29$^{+0.26}_{-0.18}$& 0.12$^{+0.15}_{-0.08}$ & 1.21$^{+0.59}_{-0.44}$&  2 $\rm \times 10^{20}$\\
130603B &172.20 &17.07 & 0.356&10.63$^{+0.09}_{-0.10}$ & -1.50$^{+0.40}_{-0.36}$& 1.65$^{+0.29}_{-0.23}$ &7.09$^{+1.76}_{-2.14}$ & 4.5 $\rm \times 10^{21}$\\
140102A &211.92 &1.33 & 2.02 &11.51$^{+0.29}_{-0.27}$ &-0.27$^{+0.95}_{-1.16}$ &1.23$^{+0.34}_{-0.33}$ & 2.23$^{+0.69}_{-0.81}$ &6.1 $\rm \times 10^{21}$ \\
190829A & 44.54& -8.96& 0.0785 &12.04$^{+0.09}_{-0.10}$ & -2.39$^{+0.24}_{-0.21}$ &2.37$^{+0.22}_{-0.20}$ &9.91$^{+1.85}_{-2.21}$ & 1.12 $\rm \times 10^{22}$ \\
200826A &6.78 &34.03 & 0.748&9.92$^{+0.08}_{-0.10}$ &-0.37$^{+0.19}_{-0.21}$ &  0.19$^{+0.17}_{-0.11}$&4.74$^{+1.53}_{-1.90}$ &6 $\rm \times 10^{20}$ \\
\hline
\vspace{-2em}
\end{tabular}
\end{center}
\label{tab:sedresults}
\end{table}

\begin{figure}
\centering
\includegraphics[scale=0.13]{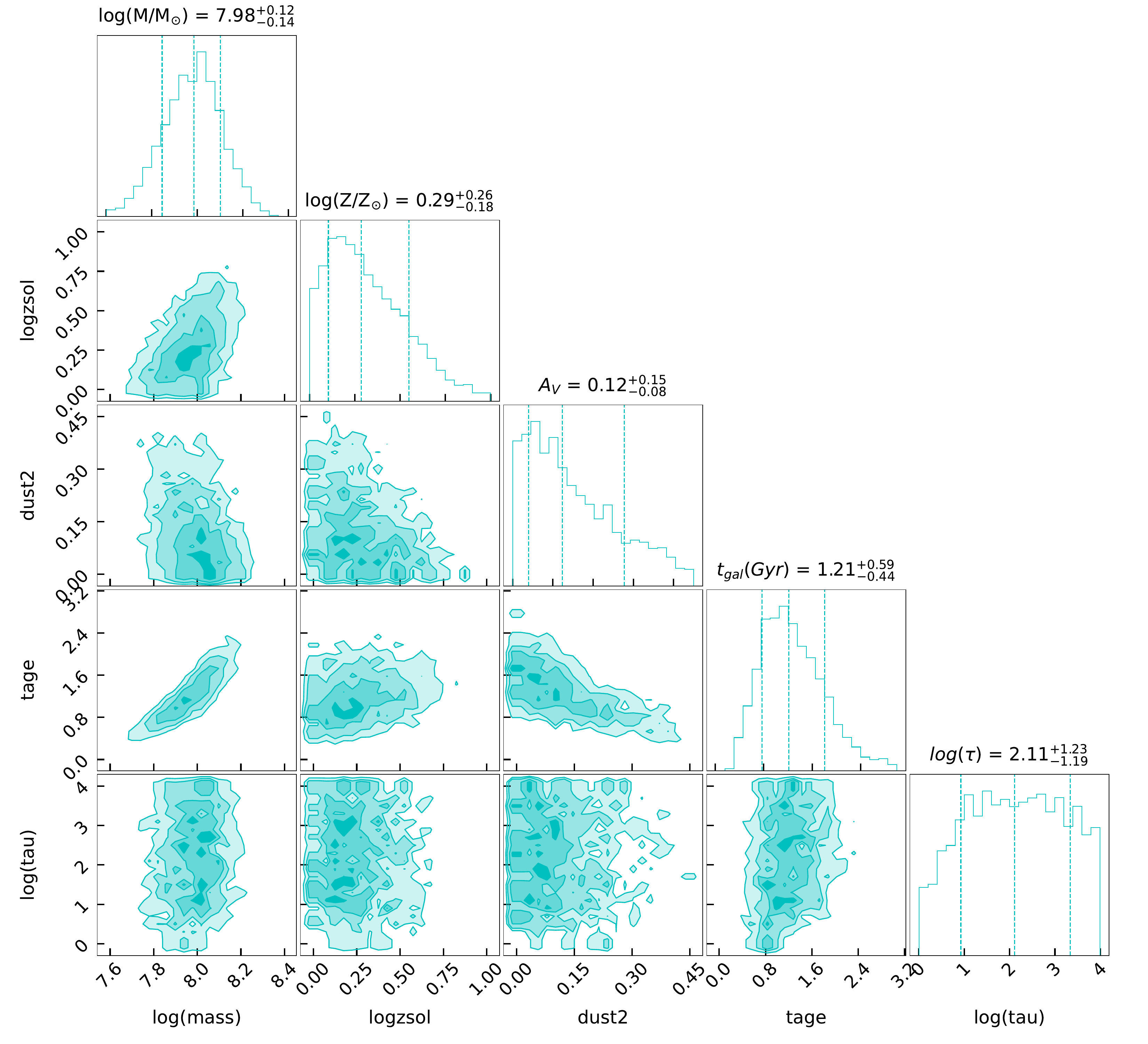}
\includegraphics[scale=0.13]{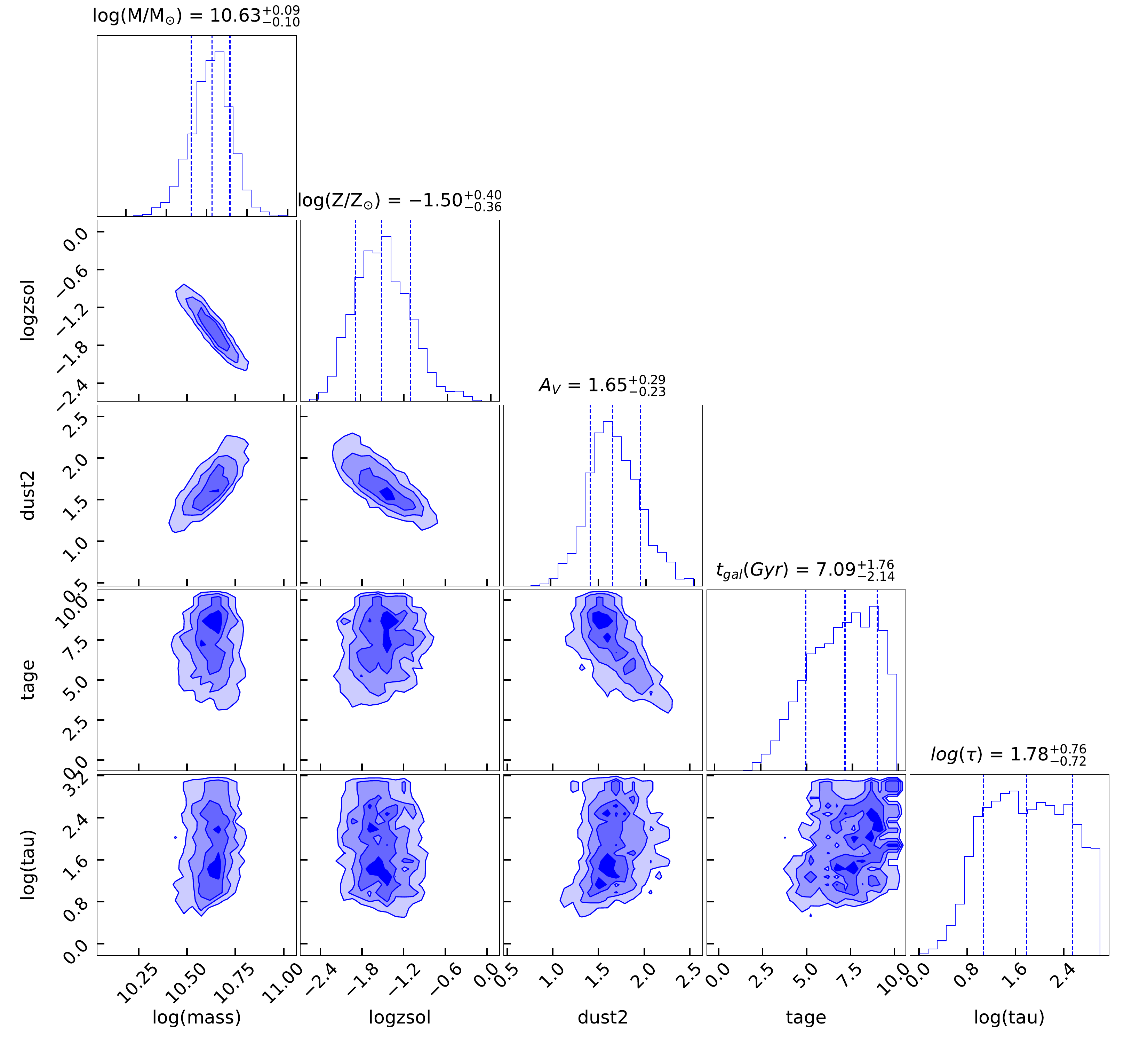}
\includegraphics[scale=0.13]{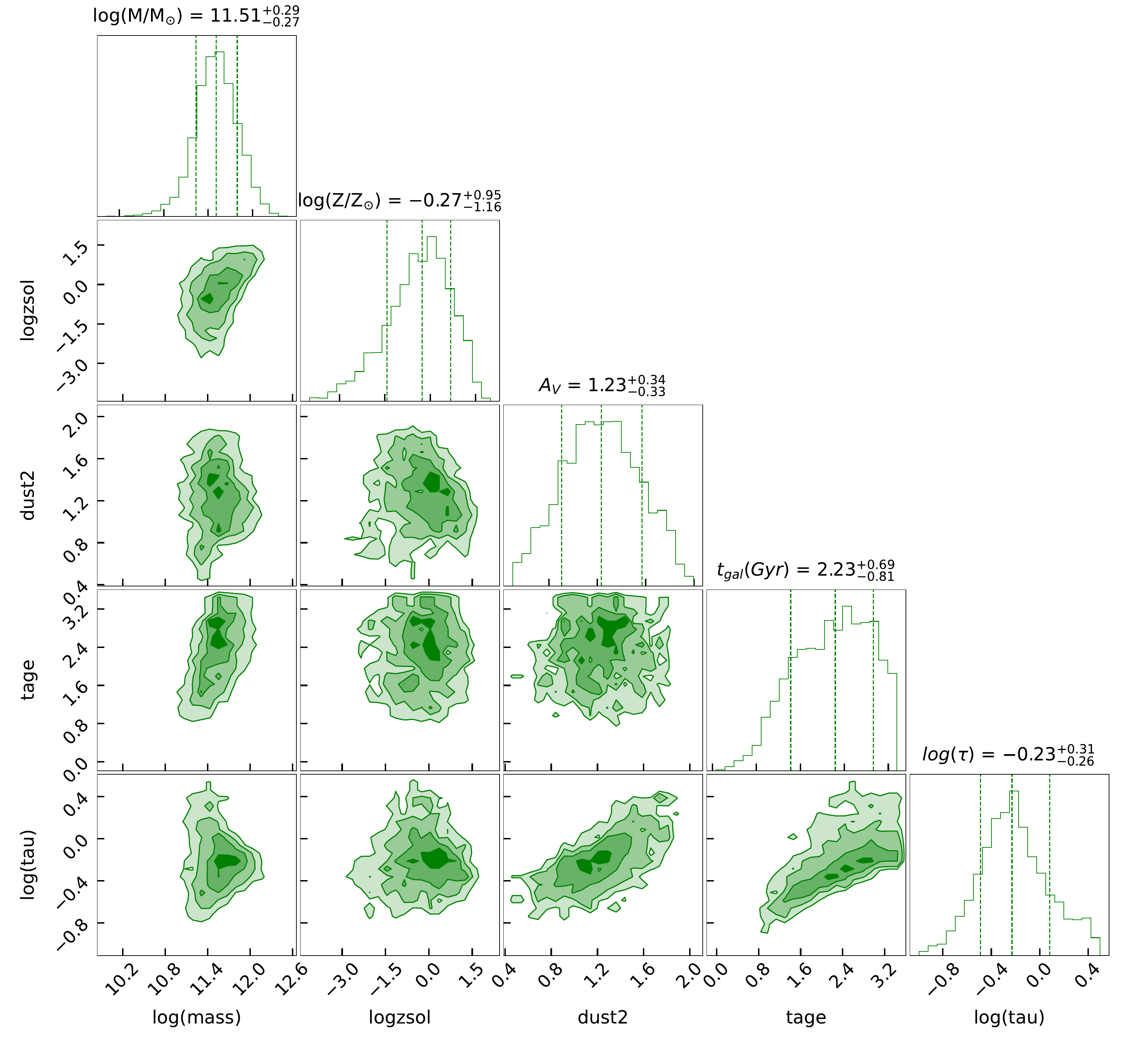}
\includegraphics[scale=0.13]{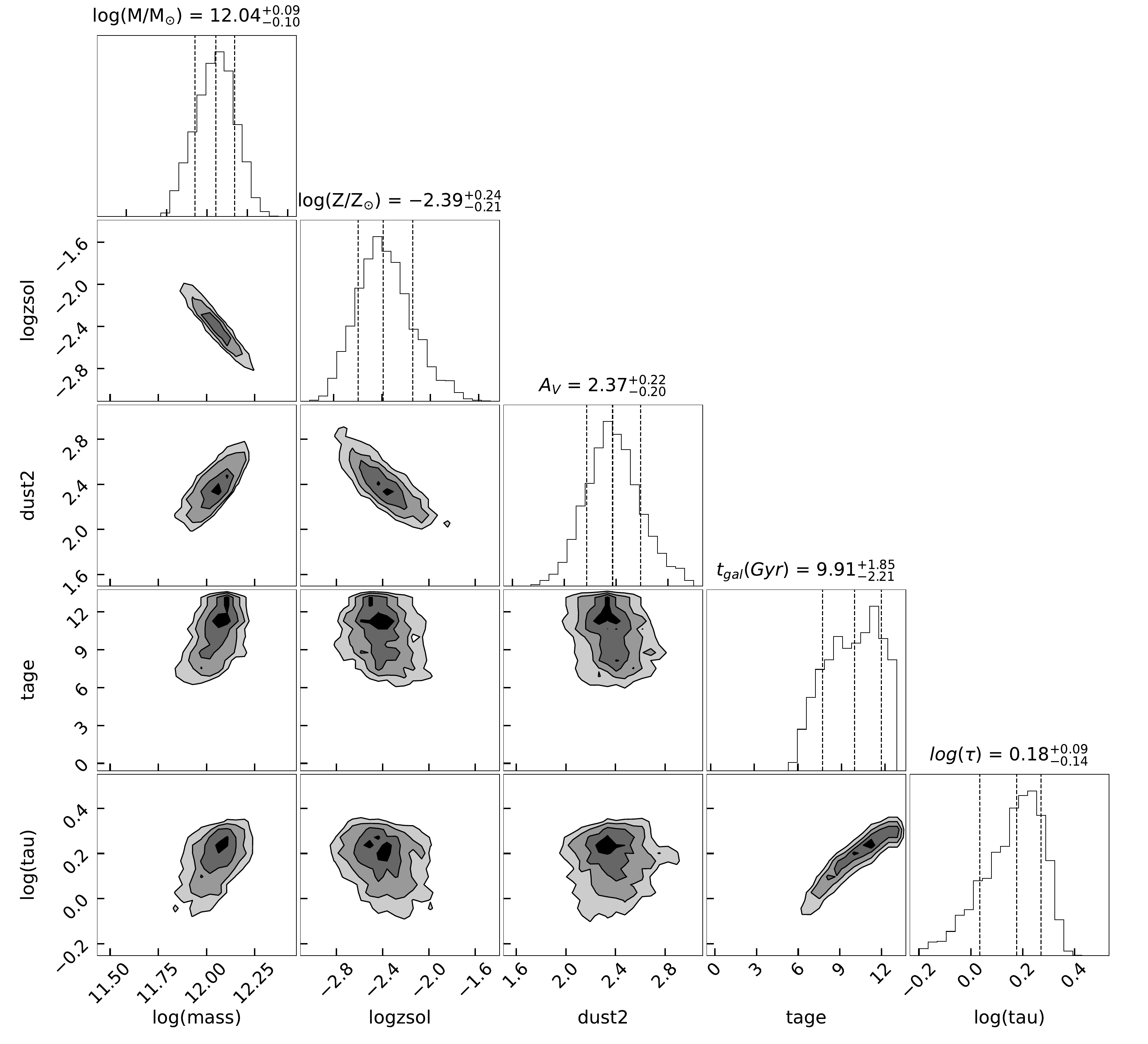}
\includegraphics[scale=0.13]{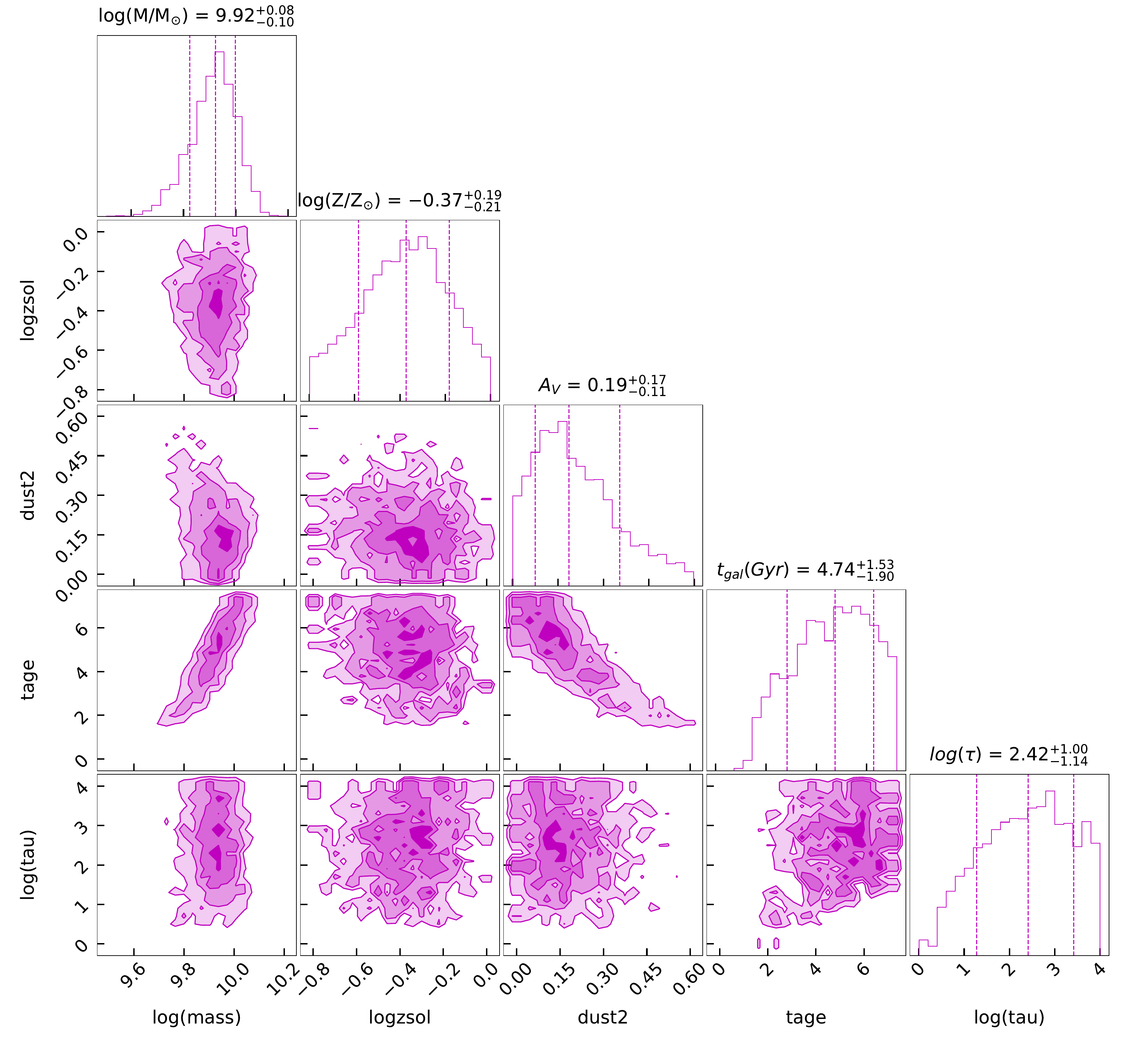}
\caption{The posterior distributions for the SED model parameters of GRB 030329 (cyan), GRB 130603B (blue), GRB 140102A (green), GRB 190829A (black), and GRB 200826A (magenta) obtained using nested sampling via \sw{dynesty} using \sw{Prospector} software.}
\label{corner_Host}
\end{figure}

{\bf GRB 030329:}

We modelled the SED (constructed using the data taken using 3.6\,m DOT along with those published in \cite{2005AA...444..711G}) of the host galaxy of GRB 030329 (a low redshift galaxy) using \sw{Prospector}. The best-fit SED using nested sampling via \sw{dynesty} method provides the following physical parameters: stellar mass formed (log(M/M$_{\odot}$)) = 7.98$^{+0.12}_{-0.14}$, stellar metallicity ($log (Z/Z_\odot$) = -0.29$^{+0.26}_{-0.18}$, age of the galaxy ($t_{gal}$) = 1.21$^{+0.59}_{-0.44}$ Gyr, rest-frame dust attenuation ($A_{V}$) = 0.12$^{+0.15}_{-0.08}$ mag, and log($\tau$) = 2.11$^{+1.23}_{-1.19}$. Furthermore, we derive the star formation rate of the galaxy = 1.57 $\times 10^{-1}$ M$_{\odot} \rm ~yr^{-1}$ and it is consistent with the value report by \cite{2003Natur.423..847H} based on the host galaxy emission line properties. Our analysis suggests a low mass and low star formation galaxy for GRB 030329 \citep{2005AA...444..711G}. 

{\bf GRB 130603B:}

We performed the modelling of the photometric data of the host galaxy obtained using 3.6\,m DOT along with those taken with other facilities (OSN, CAHA, and GTC) using the \sw{LePHARE} software and presented the results in \citep{2019MNRAS.485.5294P}. We find that the burst's environment is undergoing moderate star formation activity \citep{2019MNRAS.485.5294P}. However, we noted that \sw{LePHARE} software has some major limitations for such analysis, which we discussed in section \ref{sedmodelling}. Therefore, in this work, we performed the SED modelling of GRB 130603B using an advanced software called \sw{Prospector}, and we derive the following host parameters: log(M/M$_{\odot}$) = 10.63$^{+0.09}_{-0.10}$, $log(Z/Z_\odot)$ = -1.50$^{+0.40}_{-0.36}$, age of the galaxy = 7.09$^{+1.76}_{-2.14}$ Gyr, $A_{V}$ = 1.65$^{+0.29}_{-0.23}$ mag, and log($\tau$) = 1.78$^{+0.76}_{-0.72}$. We calculated the star formation rate of the galaxy = 11.57 M$_{\odot} \rm ~yr^{-1}$. Our results suggest that the host galaxy of GRB 130603B has a high-mass galaxy with moderate star formation activity, consistent with those reported in \cite{2019MNRAS.485.5294P}.

{\bf GRB 140102A:}

In our recent work \citep{2021MNRAS.505.4086G}, we performed the SED modelling of the host galaxy of GRB 140102A using \sw{LePHARE} software (uses $\chi^{2}$ statistics) with PEGASE2 stellar synthesis population models library. We obtained the best-fit solution with the following host galaxy parameters: age of the stellar population in the host galaxy = $9.1 \pm 0.1$ Gyr, mass = ($1.9 \pm 0.2) \times 10^{11}$ M$_{\odot}$, and SFR = $20 \pm 10$ M$_{\odot} \rm ~yr^{-1}$ with a relatively poor chi-square value ($\chi^{2}$ = 0.1). This indicates that the error bars are overestimated or the model is too flexible which may cause a large degeneracy in model parameters. For the present work, we collected the photometric observations for the host galaxy of GRB 140102A from our recent work \cite{2021MNRAS.505.4086G}, and performed the modelling using \sw{Prospector} due to the limitation of \sw{LePHARE} software as mentioned above. We have frozen the redshift $z$ = 2.02 obtained from the afterglow SED of GRB 140102A, to model the host SED using \sw{Prospector}. We find the stellar mass of log(M/M$_{\odot}$) = 11.88$^{+0.34}_{-0.32}$, the stellar metallicity of $log (Z/Z_\odot)$ = -0.19$^{+0.78}_{-1.03}$, age of the galaxy ($t_{gal}$) = 8.51$^{+3.30}_{-3.58}$ Gyr, dust extinction of $A_{V}$ = 1.35$^{+0.25}_{-0.25}$ mag, and with a moderate star formation rate of 52.90 M$_{\odot} \rm ~yr^{-1}$. The results indicate that the host was a high-mass galaxy with a high star-formation rate, consistent with those obtained from \sw{LePHARE}.

{\bf GRB 190829A:}

GRB 190829A has a very bright and nearby SDSS host galaxy. We modelled the SED using the data observed using 3.6\,m DOT along with those reported in the literature (see Table 2 in the appendix). We find stellar mass of log(M/M$_{\odot}$) = 12.04$^{+0.09}_{-0.10}$, stellar metallicity of $log (Z/Z_\odot)$ = -2.39$^{+0.24}_{-0.21}$, age of the galaxy ($t_{gal}$) = 9.91$^{+1.85}_{-2.21}$ Gyr, dust extinction of $A_{V}$ = 2.37$^{+0.22}_{-0.20}$ mag, and with a moderate star formation rate of 6.87 M$_{\odot} \rm ~yr^{-1}$. The SED modelling indicates that the host of GRB 190829A is a massive galaxy with a very high star-formation rate.

{\bf GRB 200826A:}

We modelled the host galaxy of GRB 200826A \citep{2022ApJ...932....1R} using the observed magnitude values using TANSPEC mounted on the main axis of 3.6\,m DOT along with data published in \cite{2021NatAs...5..917A}, and find stellar mass of log(M/M$_{\odot}$) = 9.92$^{+0.08}_{-0.10}$, stellar metallicity of $log(Z/Z_\odot)$ = -0.37$^{+0.19}_{-0.21}$, age of the galaxy ($t_{gal}$) = 4.74$^{+1.53}_{-1.90}$ Gyr, dust extinction of $A_{V}$ = 0.19$^{+0.17}_{-0.11}$ mag, and with a moderate star formation rate of 3.49 M$_{\odot} \rm ~yr^{-1}$. These parameters are typical to those observed for long GRBs host galaxies and consistent with \cite{2021NatAs...5..917A}.

\subsection{Comparison with a known sample of host galaxies}

\begin{figure}
\includegraphics[scale=0.28]{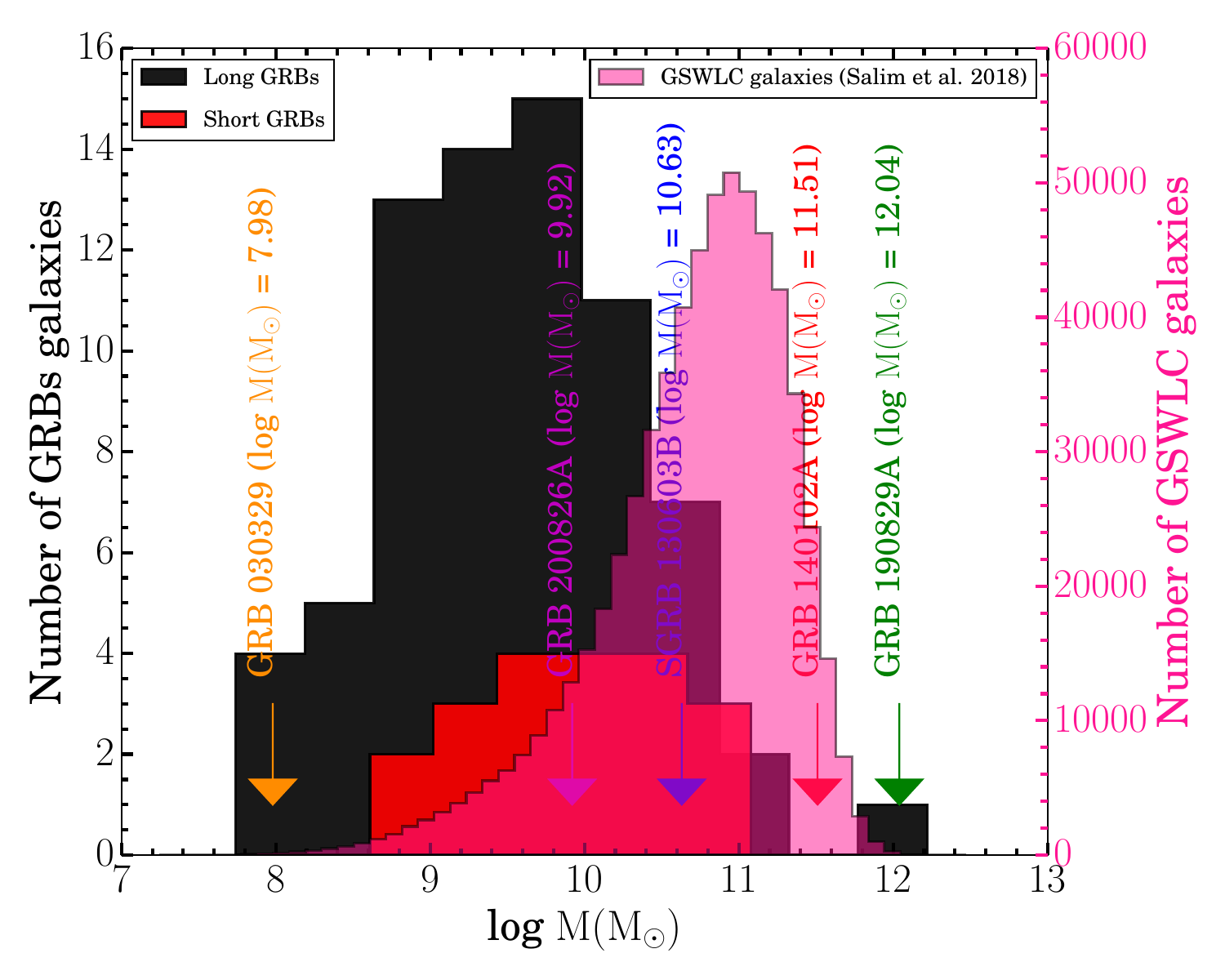}
\includegraphics[scale=0.28]{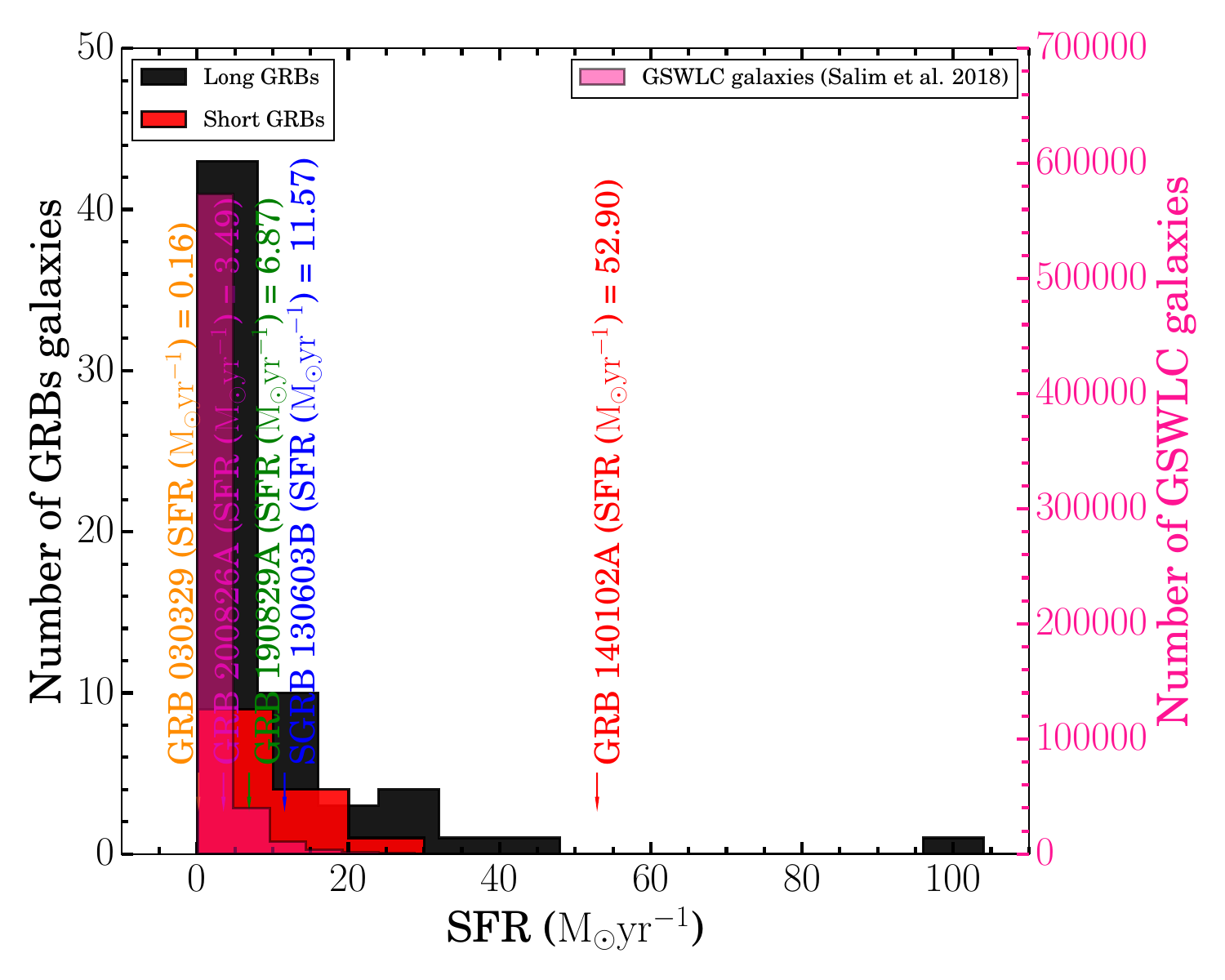}
\includegraphics[scale=0.22]{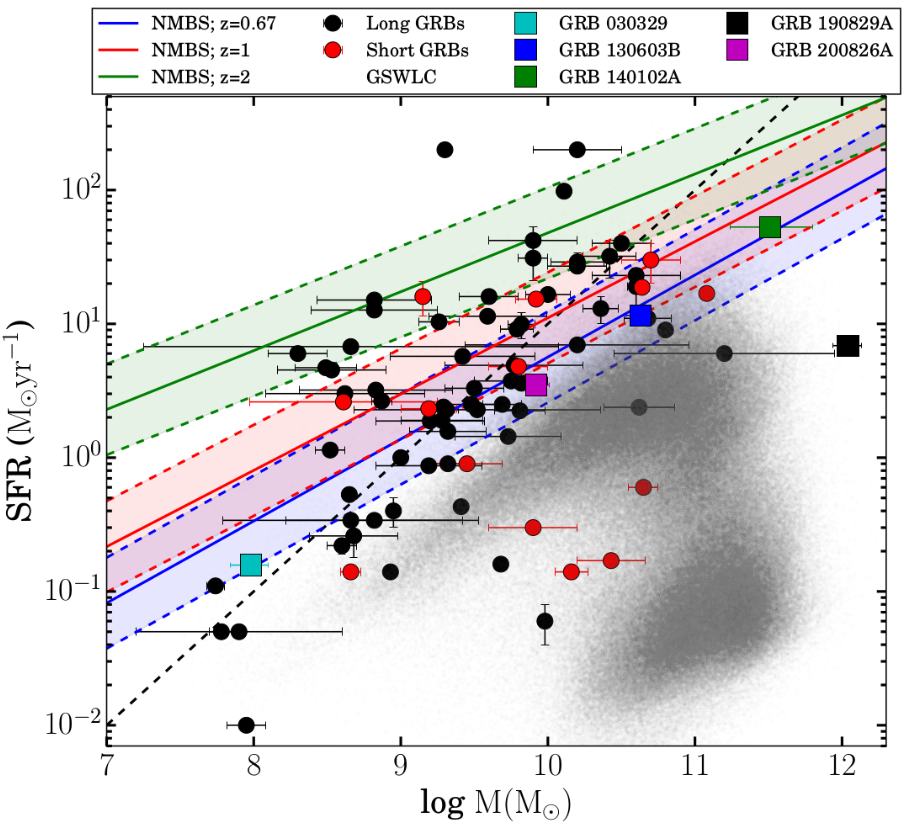}
\includegraphics[scale=0.22]{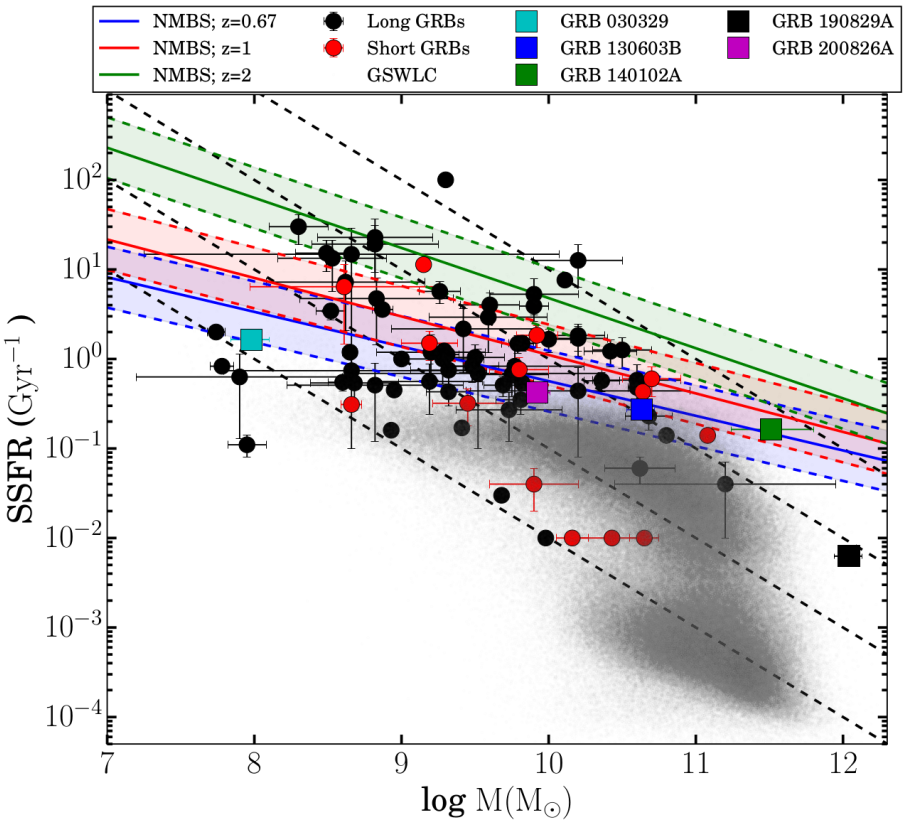}
\caption{{\it Top left panel:} The distribution of stellar mass for the host galaxies of GRBs and galaxies studied in GALEX-SDSS-WISE Legacy Catalog (right side Y-scale). {\it Top right panel:} The distribution of SFR for the host galaxies of GRBs and galaxies studied in the GALEX-SDSS-WISE Legacy Catalog (right side Y-scale). The positions of the host galaxies presented in our sample are shown with downward arrows. {\it Bottom left panel:} The stellar mass of the host galaxies versus star formation rate for our sample, obtained from the host SED modelling. Square markers indicate the host galaxies' position for our sample. The dashed line indicates a constant specific star formation rate of 1 Gyr$^{-1}$. {\it Bottom right panel:} The stellar mass of the host galaxies versus specific star formation rate for our sample. Square markers indicate the host position for our sample. The dashed lines indicate the constant star formation rates of 0.1, 1, 10, and 100 M yr$^{-1}$ from left to right. Black and red circles show the long GRBs and short GRBs host galaxies with star formation rates (bottom left) and specific star formation rates (bottom right) calculated from GHostS from 1997 to 2014 \citep{2006AIPC..836..540S, 2009ApJ...691..182S}. The grey dots show the location of GSWLC galaxies in both the correlations \citealt{2016ApJS..227....2S, 2018ApJ...859...11S}. The coloured solid lines and shaded regions in both the correlations indicate the best-fit power-law functions and their dispersion calculated for normal star-forming galaxies from NEWFIRM Medium-Band Survey (NMBS) sample at redshift values of 0.67 (the mean redshift of the sample), 1 and 2 (the mean redshift of long GRBs), respectively \citep{2012ApJ...754L..29W}.}
\label{host_sfr}
\end{figure}

We compared the host galaxy's properties of our sample (see Table \ref{tab:sedresults}) with other well-studied samples of GRB host galaxies from the literature \citep{2006AIPC..836..540S, 2009ApJ...691..182S} and GALEX-SDSS-WISE Legacy Catalog\footnote{This catalogue comprises the properties of $\sim$ 700,000 galaxies with measured redshifts values below 0.3 using SDSS.} (GSWLC; \citealt{2016ApJS..227....2S, 2018ApJ...859...11S}). The physical parameters (stellar mass and star formation rate) distribution for the host galaxies of GRBs and GSWLC are shown in Figure \ref{host_sfr} (upper left and right panels). The positions of the host galaxies presented in our sample are shown with downwards arrows. We noticed that the average value of the stellar mass of the galaxies in the GSWLC is higher than the average values of the stellar mass of the host galaxies of long and short GRBs. On the other hand, the average value of the star formation rate (SFR) of the galaxies in the GSWLC is lower than the average values of the stellar mass of the host galaxies of long and short GRBs. Furthermore, we studied possible correlations between the stellar mass of the host galaxies versus SFR and the stellar mass of the host galaxies versus specific star formation rates (SSFRs). For normal star-forming galaxies, the correlation between the stellar mass and SFR is defined as the "main sequence." This correlation reveals the possible procedures of the star formation histories of galaxies. If the correlation is tighter, it suggests that the star formation history traces stellar mass growth more smoothly. On the other hand, if the correlation is weaker (high scatter), it suggests a random star formation history \citep{2007ApJ...670..156D, 2007ApJ...660L..43N, 2011MNRAS.410.1703F}. Hence, GRB host galaxy properties can be characterized by comparing them with the main sequence. The relation between the stellar mass of the host galaxies as a function of SFRs for GRBs and GSWLC is shown in Figure \ref{host_sfr} (bottom left panel). The position of the host galaxies presented in our sample is shown with coloured squares. The coloured solid lines and shaded regions in both the correlations indicate the best-fit power-law functions and their dispersion calculated for normal star-forming galaxies from NEWFIRM Medium-Band Survey (NMBS) sample at redshift values of 0.67 (the mean redshift of the sample), 1 and 2 (the mean redshift of long GRBs), respectively \citep{2012ApJ...754L..29W}. We noticed that the host galaxies' physical properties of GRBs are more common to normal star-forming galaxies at the high-redshift Universe in comparison to the low-redshift Universe \citep{2013ApJ...778..128P, 2019MNRAS.488.5029H}.

In addition, we found that GRBs in our sample follow Mass-SFR correlation (see Figure \ref{host_sfr}) as described previously by \cite{2006AIPC..836..540S, 2009ApJ...691..182S}. Furthermore, we notice that the star formation rate of GRB 130603B, GRB 140102A, GRB 190829A, and GRB 200826A in our sample are higher in comparison to the median value of 2.5 M$_{\odot} \rm ~yr^{-1}$ \citep{2009ApJ...691..182S}. On the other hand, GRB 030329 has a low star formation rate. The host galaxies of GRB 130603B, GRB 140102A, GRB 190829A, and GRB 200826A have higher mass than the galaxies with semi-star formation rates. We also compared the SSFRs of the host galaxies of our sample with the GRB's host galaxies sample studied by \cite{2006AIPC..836..540S, 2009ApJ...691..182S} and normal star-forming galaxies (GSWLC; \citealt{2016ApJS..227....2S, 2018ApJ...859...11S}). The SSFR indicates the intensity of star formation in particular galaxies. The correlation between the stellar mass of the galaxies and SSFRs suggests how the galaxies compose their stellar populations \citep{2015A&A...577A.112L}. We noticed that other than GRB 030329, all other four host galaxies have lower SSFR in comparison to the average value of 0.8 G$\rm yr^{-1}$ (see Figure \ref{host_sfr}), suggesting a lower intensity of star-formation for these host galaxies. On the other hand, the observed higher value specific star formation rate for the host galaxy of GRB 030329 indicates a young, starbursting galaxy \citep{2006ApJ...653L..85C}. The relation between SSFR-Mass also indicates that the physical properties of the host galaxies of GRBs are more common to normal star-forming galaxies at the high-redshift \citep{2013ApJ...778..128P, 2019MNRAS.488.5029H}.

{\bf Host galaxies of GRBs, and Supernovae:}

Long GRBs usually occur at high redshift; however, some of the nearby long bursts are associated with broad-line type Ic supernovae (stripped-envelope). However, it is still not understood that all long GRBs are connected with broad-line type Ic supernovae, and we could only detect the near ones due to the observational constraints \citep{2017AdAst2017E...5C}. Therefore, the examination of the host galaxies properties of long GRBs and supernovae will be helpful to explore their environment and progenitors. Recently, \cite{2021MNRAS.503.3931T} compared the host galaxy properties of long bursts with core-collapse supernovae and superluminous supernovae and suggested that cumulative properties of the host galaxies of long GRBs without supernovae and with supernovae are not much different. Out of four long GRBs in our sample, three GRBs (GRB 030329/SN 2003dh, GRB 190829A/ SN 2019oyw, and GRB 200826A) were associated with broad-line type Ic supernovae. We compared (mass as a function of SFR) the results of these GRBs/SNe with those published in \cite{2021MNRAS.503.3931T}. We find that GRB 030329 and GRB 200826A follow the correlation plane of long GRBs and CCSN; however, the host galaxy of GRB 190829A lies on the right side of the distribution (see Figure \ref{host_sfr_2021}). We searched the host galaxy (SDSS) of GRB 190829A in GSWLC sample studied by \cite{2016ApJS..227....2S, 2018ApJ...859...11S} and found that the SFR and Stellar-mass values of the host galaxy (ObjID: 1237652899156721762) of GRB 190829A are log (SFR)= 0.395 $\pm$ 0.103 M$_{\odot} \rm ~yr^{-1}$, log(M/M$_{\odot}$ = 11.256 $\pm$ 0.012, respectively {\footnote{The model fit used by \cite{2016ApJS..227....2S, 2018ApJ...859...11S} is underfitting the observed data (reduced chi-square = 2.15)}}. These values also indicate that the host galaxy of GRB 190829A is a massive and high star-forming galaxy, consistent with our results.

\begin{figure}[ht!]
\centering
\includegraphics[scale=0.35]{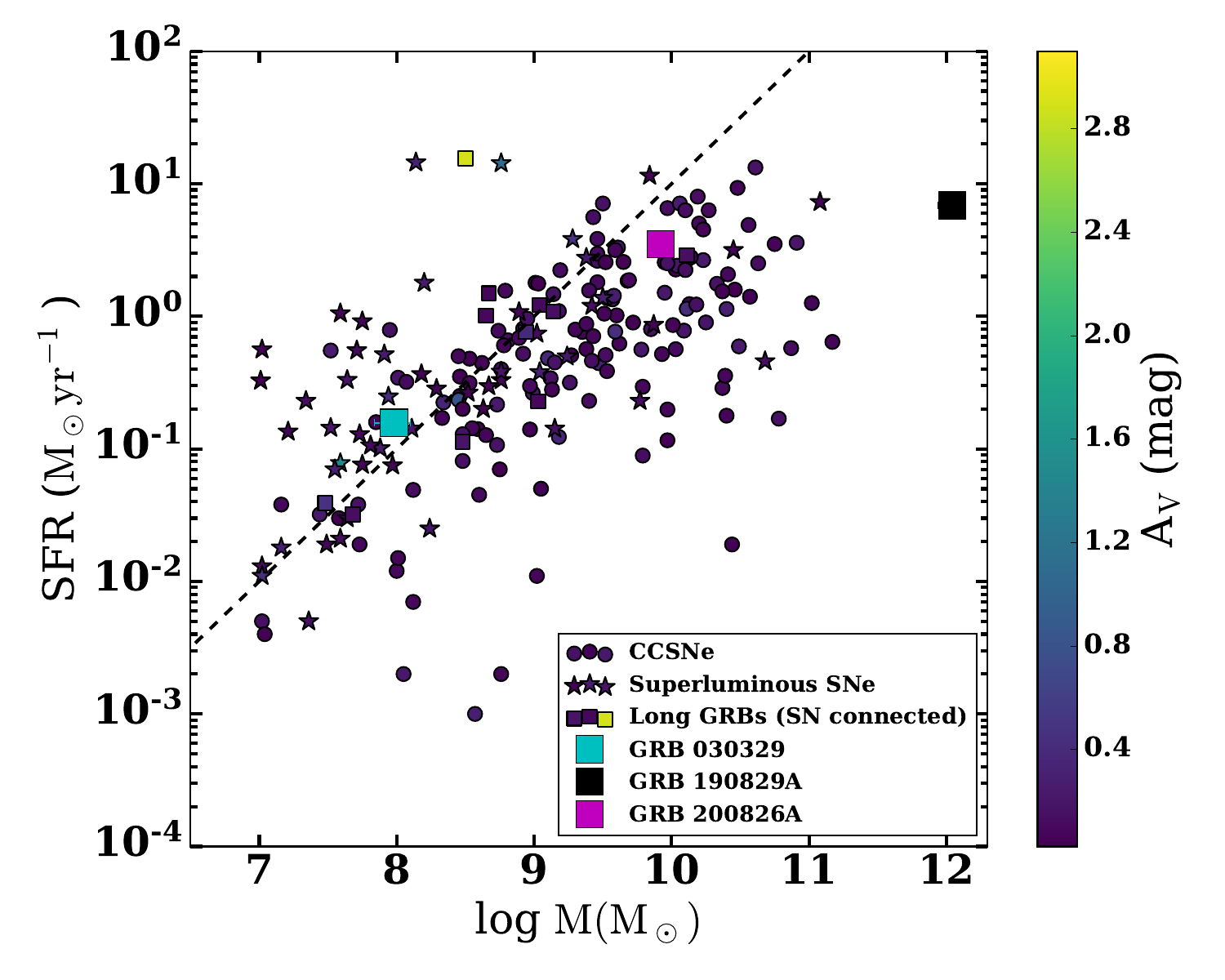}
\caption{The stellar mass of the host galaxies versus star formation rate for GRBs connected with supernovae in our sample, obtained from the host SED modelling. Circles, stars, and squares markers indicate the position of host galaxies of CCSN, superluminous supernovae, and long GRBs connected with supernovae, respectively, taken from \cite{2021MNRAS.503.3931T}. The right side Y-scale shows the corresponding A$_{V}$ values.  
}\label{host_sfr_2021}
\end{figure}

{\bf Dust and gas in the host galaxies:}

\begin{figure}[ht!]
\centering
\includegraphics[scale=0.29]{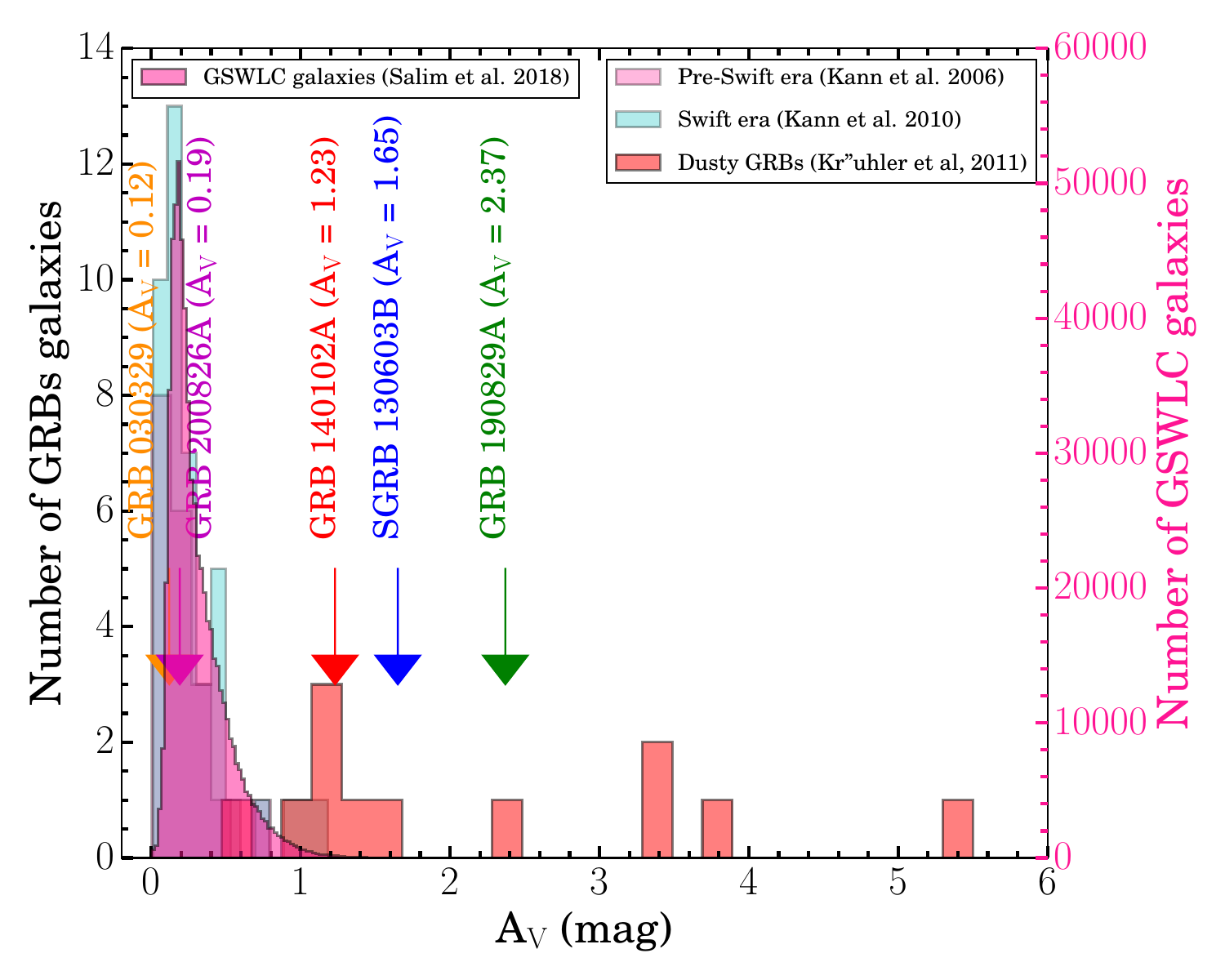}
\includegraphics[scale=0.29]{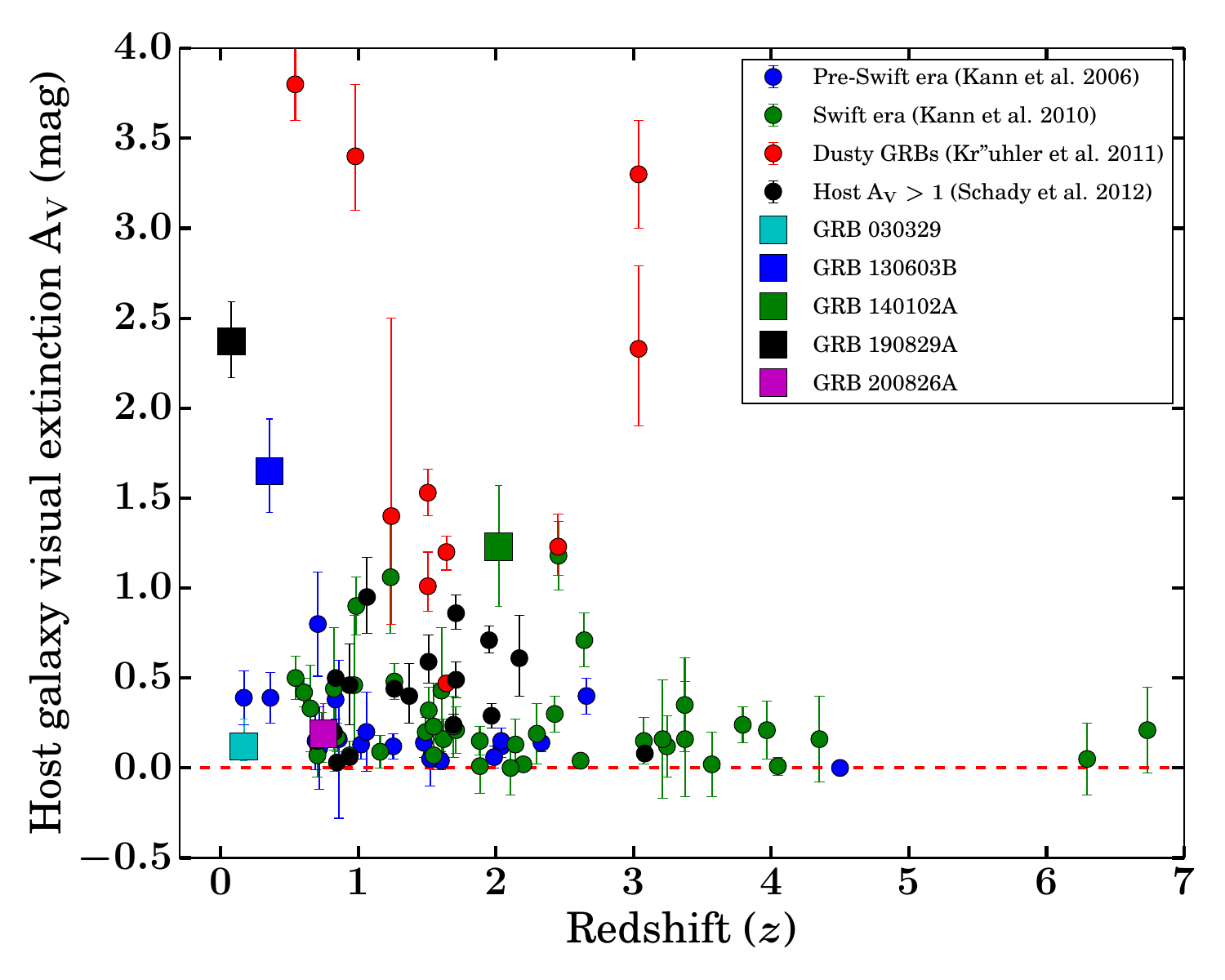}
\includegraphics[scale=0.29]{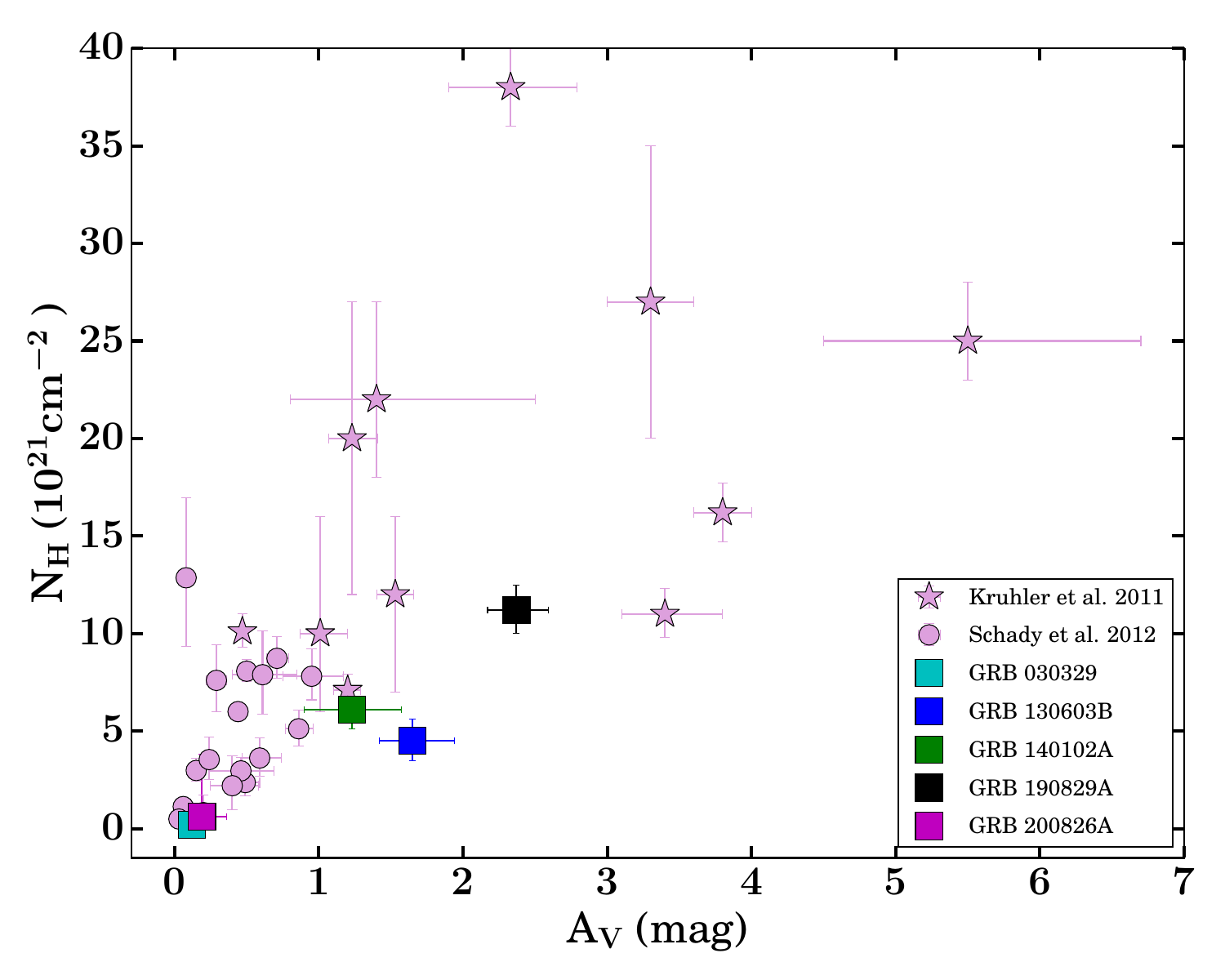}
\caption{{Top panel:} The distribution of visual extinction (in the source frame) of the host galaxies in our sample. For the comparison, the data points for GRBs in pre-\swift \citep{2006ApJ...641..993K}, post-\swift \citep{2010ApJ...720.1513K} era, dusty GRBs \citep{2011A&A...534A.108K} and GSWLC \citep{2016ApJS..227....2S, 2018ApJ...859...11S} are also shown. {Middle panel:} Redshift evolution of visual extinction. The horizontal red dashed line shows A$_{V}$ = 0. {Bottom panel:} The dust extinction as a function of X-ray column density in the local environment of the host galaxies of our sample, along with other data points taken from \cite{2011A&A...534A.108K, 2012A&A...537A..15S}.}
\label{host_av_nh}
\end{figure}

We calculated the dust extinction in the local environment of the host galaxies of our sample and compared them with a larger sample of the host galaxy of GRBs \citep{2006ApJ...641..993K, 2010ApJ...720.1513K, 2011A&A...534A.108K}/ GSWLC \citep{2016ApJS..227....2S, 2018ApJ...859...11S} galaxies taken from the literature. We determine the visual dust extinction in rest-frame (A$_{v}$) using the host galaxy SED modelling of each burst in our sample (see section \ref{sedmodelling} for more details). Figure \ref{host_av_nh} (top panel) shows the distribution of visual extinction (in the source frame) of the host galaxies in our sample. For the comparison, we have also shown the distribution of visual extinction for GRBs in pre-\swift \citep{2006ApJ...641..993K}, post-\swift \citep{2010ApJ...720.1513K} era, dusty GRBs \citep{2011A&A...534A.108K} and GSWLC \citep{2016ApJS..227....2S, 2018ApJ...859...11S}. We find that A$_{v}$ values are distributed over a wide range for our sample, and part of host galaxies (GRB 130603B, GRB 140102A, and GRB 190829A) are extinguished by dust. Moreover, these galaxies (dusty) typically have higher stellar mass, consistent with previous studies \citep{2009AJ....138.1690P, 2011A&A...534A.108K}. In addition, we also noted that the A$_{v}$ values $\leq$ 1 for other than the dusty sample, suggesting that the dusty galaxies ($\sim$ 20 \%) are highly extinguished and might cause optical darkness \citep{2004ApJ...617L..21J, 2011A&A...526A..30G, 2022JApA...43...11G}.

Figure \ref{host_av_nh} (middle panel) shows the evolution of host visual extinction as a function of redshift for our sample along with those data points published by \citep{2006ApJ...641..993K, 2010ApJ...720.1513K, 2011A&A...534A.108K, 2012A&A...537A..15S}. We noticed that the visual extinction is decreasing with redshift, although it might be due to the selection effect as there are only a few GRBs with redshift $\geq$ 5, and the observation of dusty galaxies at such high redshift is very difficult \citep{2010ApJ...720.1513K}.   

Furthermore, we used the spectral analysis results of the X-ray afterglow data from the literature to constrain the intrinsic hydrogen column density (N$_{\rm H}$ ($z$)) of each burst in our sample. The X-ray afterglow spectra of GRBs could be typically described using a simple absorption power-law model consisting of three components: a Galactic absorption (N$_{\rm H}$), host absorption (N$_{\rm H}$ ($z$)), and a power-law component due to synchrotron emission. We obtained N$_{\rm H}$ ($z$) values of GRB 030329 from \cite{2004A&A...423..861T}, GRB 130603B from \swift X-ray telescope web-page maintained by Phil Evans, GRB 140102A from \cite{2021MNRAS.505.4086G}, GRB 190829A from \cite{2020ApJ...898...42C}, and for GRB 200826A from \cite{2021NatAs...5..917A}, respectively. The distribution of host dust extinction and gas column densities in the local environment for our sample, along with other data points taken from the literature, are shown in Figure \ref{host_av_nh} (bottom panel). We noticed that GRB 190829A has a considerable amount of dust and gas in the local environment of its host galaxy. The observed considerable amount of dust and gas might be related to the associated VHE emission from GRB 190829A, a similar dusty environment is also seen in the case of other VHE detected bursts such as GRB 190114C \citep{2020A&A...633A..68D} and GRB 201216C \citep{2022MNRAS.513.1895R}. However, due to the limited number of VHE-detected GRBs, it is still an unsolved problem whether VHE-detected GRBs require unique environments to emit the VHE emission or whether VHE emission is only due to the burst emission mechanisms such as Synchrotron Self Compton \citep{2019Natur.575..455M, 2019Natur.575..464A, 2020A&A...633A..68D, 2021RMxAC..53..113G}.

\section{Summary and Conclusion}
\label{Summary and Conclusion_host}

The observed gamma-ray prompt emission properties of GRBs do not always depict about nature of their progenitors and environments and, in turn, about unambiguous classification. Recently, the origin of a few short bursts (e.g., GRB 090426 and GRB 200826A) from the collapse of massive stars \citep{2009A&A...507L..45A, 2021NatAs...5..917A} and long GRBs (e.g., GRB 211211A, and GRB 060614) from the merger of two compact objects \citep{2022Natur.612..228T, 2022Natur.612..223R, 2015NatCo...6.7323Y} are confirmed. These examples suggest that at least some of the short GRBs might be originated from collapsars, and some of the long GRBs might be originated from compact mergers. Therefore, the late-time observations of the host galaxies are crucial in examining the burst environment and, in turn, the possible progenitors, especially for the hybrid cases. In this paper, we present the photometric observations of the five interesting GRBs' host galaxies observed using India's largest optical telescope (3.6\,m DOT) to constrain the environment of these bursts, the nature of possible progenitors and explore the deep observations capabilities of 3.6\,m DOT. Our optical-NIR multi-band data of these five hosts, along with those published ones, were used to perform multi-band modelling of the host galaxies using \sw{Prospector} software (version 1.1.0). We noted that the host galaxies in our sample have a wide range of physical parameters (see Table \ref{tab:sedresults}). The host galaxies of GRB 130603B, GRB 140102A, GRB 190829A, and GRB 200826A have a massive stellar population galaxy with a high star formation rate. However, GRB 030329 has a low-mass galaxy with a low star formation rate; such host galaxies having a low mass with a low star formation rate are rare \citep{2010ApJ...721.1919C}. We compared the stellar population properties (such as SFR, SSFR, Mass, etc.) of the host galaxies of our sample with a large sample of long and short bursts along with those taken from literature specifically with GSWLC. We find that all the bursts in our sample satisfy the typical known correlation between host galaxy parameters. We noted that the GRBs generally occur in host galaxies that have less massive and high star-forming galaxies than GSWLC galaxies. Further, the host galaxies' physical properties of GRBs are more common to normal star-forming galaxies at higher redshifts.

In addition, we obtained the X-ray hydrogen column densities from the X-ray afterglow observations of these bursts and studied their distribution with optical dust extinction. We find that GRB 190829A has a considerable amount of dust and gas in the local environment of its host galaxy. A dusty environment is also seen in the case of other VHE-detected bursts such as GRB 190114C \citep{2020A&A...633A..68D} and GRB 201216C \citep{2022MNRAS.513.1895R}. It suggests that VHE-detected GRBs might require a unique local environment for VHE emission to occur.  Unfortunately, due to the small size of the present sample, it is difficult to quantify the selection effects, which further limits a robust statistical analysis. Our results demonstrate that the back-end instruments (such as IMAGER and TANSPEC) of 3.6\,m DOT have a unique capability for optical-NIR deep observations of faint objects such as host galaxies of GRBs and other interesting transients in the near future. Also, in the near future, systematic studies (with a larger sample) of the host galaxies along with prompt emission and afterglow properties of hybrid GRBs may play a crucial role in understanding their progenitors.


\chapter{\sc Summary and future prospects}
\label{ch:7} 

\ifpdf
    \graphicspath{{Chapter7/Chapter7Figs/PNG/}{Chapter7/Chapter7Figs/PDF/}{Chapter7/Chapter7Figs/}}
\else
    \graphicspath{{Chapter7/Chapter7Figs/EPS/}{Chapter7/Chapter7Figs/}}
\fi

\ifpdf
    \graphicspath{{Chapter7/Chapter7Figs/JPG/}{Chapter7/Chapter7Figs/PDF/}{Chapter7/Chapter7Figs/}}
\else
    \graphicspath{{Chapter7/Chapter7Figs/EPS/}{Chapter7/Chapter7Figs/}}
\fi

\normalsize

\section{Summary and Conclusion}

In this thesis (titled ``Multiwavelength Observations of Gamma-ray Bursts" ), we explore the comprehensive study of GRBs using observations (obtained from different space-based and ground-based telescopes) across different wavelengths of the electromagnetic spectrum. We studied several open enigmas of GRBs, such as What is the jet composition? Is it a baryon-dominated or Poynting-flux-dominated outflow? What is the underlying emission process that gives rise to observed radiation? Where and how does the energy dissipation occur in the outflow? Is it via internal shocks or magnetic reconnections? How to solve the radiative efficiency problem? What are the possible causes of Dark GRBs and orphan afterglows? How to investigate the local environment of GRBs?  We performed a detailed multi-scale study of GRBs' spectral, temporal, and polarimetric characteristics (both prompt emission and afterglow) targeted to resolve some of these enigmas. The research work in this thesis utilizes novel methods and observational techniques, including light curve analysis/modelling, spectral analysis, and polarization measurements, to extract valuable information from the observed data. The results obtained from these analyses are compared with theoretical models to improve our knowledge of the physical processes accountable for GRBs and their associated phenomena. 

The main aim of our study on the prompt is to explore the radiation physics and jet compositions of GRBs. This is a long debatable problem, and prompt emission spectroscopy alone could not provide the solutions independently. To examine the jet composition and radiation physics of prompt emission, we have used a unique time-resolved spectro-polarimetric technique using the prompt observations of some of the bright GRBs discovered using {\it Fermi} and {\it AstroSat} CZTI instruments.

In chapter \ref{ch:3}, we conducted a time-resolved spectral analysis of a very luminous burst (GRB 210619B) detected using {\it Fermi} GBM and ASIM instruments. This GRB has a very bright and hard pulse followed by a fainter and softer pulse. Our spectral analysis shows that the first pulse has very hard values of low-energy photon indices followed by softer values during the weaker episodes. This indicates a possible thermal to non-thermal transition of jet composition within the burst. For another one of the most energetic bursts (GRB 190530A), we carried out a detailed time-resolved spectro-polarimetric analysis using simultaneous  {\it Fermi} and {\it AstroSat} CZTI observations. During the first two pulses, the values of low-energy photon indices were in agreement with poynting flux-dominated synchrotron scenarios. Our polarization analysis was also consistent with the same scenarios, i.e., synchrotron emission in an ordered orientation of magnetic field vectors. Our study suggests that spectro-polarimetry of the prompt emission of GRBs can solve the emission mechanisms and jet composition of GRBs.

The community is working towards the next gamma-ray missions (e.g., COSI, eAstroGAM, AMEGO, AMEGO-X). Our work provides information to those future missions, especially the time-resolved polarization measurements under this work giving a good insight into the upcoming GRB polarimeters such as COSI, POLAR 2, and Daksha missions. COSI will measure GRBs polarisation in soft gamma rays for around 40 GRBs in two years of operation. Detail information about Daksha and POLAR 2 mission is given in section \ref{Daksha and POLAR 2}.

The internal shock model of GRBs is inefficient in converting the jet's kinetic energy into gamma-ray radiation, known as a low-efficiency problem. This problem can be resolved using very early optical afterglow follow-up observations and modelling of rarely observed reverse shock emissions in earlier phases. 

In chapter \ref{ch:5}, we present our detailed analysis of very early optical afterglow data (even before the XRT and UVOT) of GRB 140102A. We have used a robotic telescope named BOOTES (Burst Observer and Optical Transient Exploring System) for the very early follow-up observations of this GRB. Our broadband afterglow modelling shows that early optical emission is dominated by a reverse shock moving toward ejecta. Late-time emission is dominated by forward shock emission driving in the opposite direction of ejecta. Our modelling constrains a lower electron equipartition parameter of the reverse shock component, resulting in a very high value of radiative efficiency. Therefore, early observations and modelling of such cases may help to lighten the low-efficiency problem. Furthermore, we compared the physical properties of GRB 140102A with a complete sample of thin shell reverse shock-dominated GRBs and noted that such GRBs might have a wide range of magnetization.

In chapter \ref{ch:61}, we studied the dark GRBs and orphan afterglow properties using late-time broadband afterglows data. We examine the characteristics of two dark bursts (GRB 150309A and GRB 210205A). We have used a meter to the 10-meter class telescope for the follow-up observations of these bursts. The detailed analysis of these GRBs reveals that GRB 150309A is one of the most dust-extinguished GRBs to date, and local dust in the host galaxy might be the potential reason for its optical darkness. In the case of GRB 210205A, we estimated a lower value of dust extinction in the local host using joint optical to X-ray SED analysis; therefore, either intrinsic faintness or high redshift might be the possible origin of its optical darkness. We examined an orphan afterglow's characteristics (AT20221any) and compared its brightness with other known cases of orphan afterglows. Our detailed multiwavelength modelling of AT20221any suggests that it was observed on-axis. Still, no gamma-ray detecting satellite could detect the prompt emission either because the source was not in their field of view or due to limited sensitivity for fainter GRB detection.  

The community is developing several larger optical telescopes, such as the Extremely Large Telescope (ELT) and Thirty Meter Telescope (TMT). Such telescopes will be very useful for fainter sources observations like afterglows of GRBs. Our study on dark GRBs is very important for future observations of similar sources using upcoming larger telescopes. On the other hand, the community is also developing many survey telescopes, including LSST and others. One of the key scientific goals of such survey telescopes is the discovery of orphan afterglows. Our study on orphan afterglows provides insight into such scientific goals. 

In chapter \ref{ch:6}, we studied the properties of the host galaxy of a sample of five peculiar GRBs observed using 3.6\,m DOT and 10.4\,m GTC telescopes to understand the environments of GRBs. We carried out our detailed modelling of photometric data of these host galaxies. We compared the physical parameters of these galaxies with well-studied host galaxies of long and short GRBs. We noted that most of the bursts in our sample have a massive galaxy with a high star formation rate (SFR), and only one burst (GRB 030329) belongs to a rare low-mass host galaxy with a low SFR. Our study demonstrated the capabilities of 3.6\,m DOT for faint sources observations such as host galaxies of GRBs.

In conclusion, the thesis titled ``Multiwavelength Observations of Gamma-ray Bursts" highlights the importance of multiwavelength observations in comprehensively studying GRBs. The combination of data obtained from different wavelengths provides a holistic view of these cataclysmic events, allowing us to examine the physical mechanisms, progenitor systems, and environments associated with GRBs. By utilising sophisticated data analysis techniques, the thesis provides beneficial insights into the properties and behaviour of GRBs. The comparison of observational data with theoretical models allows for the refinement and validation of existing theories, bringing us closer to a comprehensive understanding of the physics underlying these extraordinary cosmic events.

\section{Future prospects}

The research completed in this thesis opens avenues for future investigations, including the utilization of advanced observational facilities, improved instrumentation, and refined theoretical models. Each new GRB mission revealed new discoveries and surprises to astronomers. Since the first discovery of GRB using the Vela mission, the GRB field is significantly evolved and entered into the multi-messenger era. The discovery of kilonova and supernova emissions from both short as well as long GRBs puzzled the classical classification scheme of GRBs. In this multi-messenger era, simultaneous follow-up observations of gravitational waves, electromagnetic signals (from VHE to radio wavelength) from both long and short GRBs, and neutrinos will play a crucial role in unveiling the origin of GRBs. In the near future, the following major GRB-dedicated mission/telescopes to study temporal/spectral/polarimetric properties of GRBs are planned:

\subsection{SVOM mission}

Space-based multi-band astronomical Variable Objects Monitor (SVOM) is a dedicated GRB mission developed by French Space Agency, the Chinese National Space Administration along with other international groups \citep{2022IJMPD..3130008A}. SVOM will be capable of detecting and proving fast localization of GRBs for rapid follow-up observations of the multi-wavelength afterglows. SVOM is expected to be launched in December 2023 with four payloads (two wide field and other two narrow field instruments): the wide field gamma-ray/X-ray camera ECLAIRs, the wide field gamma-ray spectrometer Gamma Ray Monitor (GRM), the narrow field soft X-ray detector Microchannel X-ray Telescope (MXT), and the narrow field optical detector Visible Telescope (VT).   

ECLAIRs is a coded masked hard X-ray detector (working in 4-150 keV energy range). ECLAIRs will be used for GRB trigger and provide arc-minute localization of GRB position. It will also send alerts to the spacecraft (to slew in the direction of GRBs) and ground-based telescopes to perform follow-up observations in other wavelengths. GRM will mainly be used for the detailed wide range (15 – 5000 keV) spectral analysis of GRBs. Considering the good sensitivity and wide field of view of GRM, it will also be used to trigger GRBs. ECLAIRs and GRM together provide 4 keV to 5 MeV spectral coverage for prompt emission. MXT will be used for quick observations of X-ray afterglows in soft X-rays (0.2-10 keV), similar to the XRT instrument. VT will be used for quick (in less than 300 sec post ECLAIRs trigger) observations of SVOM GRBs and provide sub-arc-second localization or faint limit on the optical counterpart. In addition to four on-board instruments on the SVOM mission, there are many wide-field ground-based telescopes such as Ground-based Wide Angle Camera (GWAC) and Ground Follow-up Telescopes (GFT). GWAC is used to search the prompt optical emission utilizing its wide field of view. GFT are the network of two robotic meter-class telescopes for follow-up observations. SOVM mission will be very crucial for many unexplored aspects of GRBs such as broadband spectral analysis of prompt emission, quick identification of dark GRBs which will help to probe the high redshift universe, measuring temporal and spectral behaviour of afterglows of GRBs from early to late epochs, etc.   

\subsection{THESEUS}

Transient High-Energy Sky and Early Universe Surveyor (THESEUS) is also a dedicated GRB mission developed by European Space Agency. The main objective of THESEUS mission is to study and characterize GRBs and explore the early universe. This mission also aims to advance multi-messenger science in particular and time-domain astronomy in general. THESEUS is expected to launch by 2032 with three payloads (two wide fields and one narrow field instrument): wide field (1.5 sr) X-Gamma ray Imaging Spectrometer (XGIS, working in 2 keV to 20 MeV energy range), wide field (1 sr) Soft X-ray Imager (SXI, working in 0.3-6 keV energy range), and narrow field (15x15 arcmin) InfraRed Telescope (IRT, working in 0.7-1.8 $\mu$m). XGIS is a coded aperture detector and provides localization $<$ 15 arc minutes. The main goal of XGIS is the detection of GRBs and other high energy transients and provide broadband energy spectrum \citep{2021arXiv210208701L}. SXI is a set of four lobster-eye telescopes and provides localization $<$ 1-2 arc minutes. SXI will also be triggering GRBs with increased sensitivity in soft X-ray which will help to probe GRBs at high redshift. The THESEUS spacecraft will be capable of rapid slewing (several degrees in a few minutes). The fast slewing capability will be utilized by IRT for the early search of the infrared afterglow of GRBs. IRT will rapidly localize the afterglows up to a few arcsec which will help to determine their on-board redshift. Furthermore, it will also help to probe the spectroscopic properties of afterglow and associated host galaxies.

\subsection{Daksha and POLAR 2}
\label{Daksha and POLAR 2}

Daksha is a high-energy Indian mission dedicated to studying and characterising GRBs and electromagnetic counterparts of gravitational waves. It will consist of two identical satellites to cover the whole sky in $\sim$ 1 keV to $\sim$ 1 MeV energy range. Daksha is expected to launch in around 2028. The high sensitivity (higher than \swift mission) and broad spectral coverage (including soft X-ray) of this mission will help to probe the GRB prompt emission in detail. Additionally, it is expected that Daksha will be capable of measuring the prompt emission polarization for 5-8 bursts per year with observed fluence value greater than $10^{-4}$ erg $cm^{-2}$, which will help to discriminate between the different radiation models of prompt emission \citep{2022arXiv221112052B}.  

POLAR 2 is a follow-up of the Chinese POLAR mission. POLAR 2 is a next-generation polarimeter and is expected to launch in 2024. The previous polarization measurement using recently dedicated X-ray polarimeters such as \AstroSat, POLAR, and GAP indicate the polarization properties of GRBs may change within the burst's intervals. Therefore, a more sensitive polarimeter capable of time-resolved polarization measurement is required to discriminate between the different radiation models of prompt emission. POLAR 2 will have around an order of effective area higher than POLAR, making it more sensitive for polarization measurement. POLAR 2 will be capable of measuring the prompt emission polarization for around 50 bursts per year \citep{2020SPIE11444E..2VH}.   

\subsection{CTA}

Recently, the afterglows of seven GRBs have been discovered in the VHE domain by ground-based MAGIC and HESS Cerenkov telescopes. We studied the characteristics of these VHE bursts and compared their properties with other well-studied GRBs \citep{2021RMxAC..53..113G, 2023ApJ...942...34R}. However, the origin of such sources is not well understood. The most accepted radiation model working at the VHE domain of GRBs afterglows is expected to be Inverse Compton. Is VHE emission common in all GRBs? If yes, why is it not detected by MAGIC and HESS or if not, what kind of local environment causing such extreme energy? Can VHE emission during prompt emission be detected? How do we detect such emissions at high redshift? Such questions can be addressed using more sensitive Cerenkov telescopes.

Cherenkov Telescope Array (CTA) will be the next generation Cerenkov telescope with a very large area and observe the very high energy emission from astronomical sources in the 20 GeV to 300 TeV energy range. CTA will consist of three kinds of telescopes (a total of around 100 telescopes installed in the northern and southern hemispheres) for wider energy coverage:  Small-Sized, Medium-Sized, and Large-Sized telescopes. The wide energy converges of CTA indicate that GRBs can be detected by CTA. \citep{2011arXiv1109.5680B} calculated the possible detection rate of GRBs using CTA based on observed temporal and spectral properties of \fermi LAT detected GRBs. They noted that CTA will be detecting between 0.35 and 1.6 GRBs per year, although the detection rate depends on the intrinsic VHE GRB population and sensitivity that is achieved by the CTA. CTA will be capable of detecting VHE emission from GRBs even larger distances and for longer duration post trigger than HESS and MAGIC. This new VHE window with increased sensitivity and detection rate of CTA will help to investigate the GRB science in more detail and provide information about temporal and spectral evolution during the VHE component. It is expected that CTA will be detecting VHE emission from short GRBs also, which will further help to connect the VHE emission and gravitational waves and provide unique information on binary compact merger sources in this multi-messenger era.

\subsection{SKA}

The radio afterglows of GRBs can be observed even after months or years after the first detection by gamma-ray satellite. However, there is a large fraction of GRBs ($\sim$ 70 \%) for which no radio afterglows have been discovered. The radio afterglows could not discover or monitored for a very long time due to the limitation of the sensitivity current radio telescope. Radio follow-up observations play a crucial role in understanding GRB jet physics. The early radio afterglow observations help to probe the signature of reverse shock and the late-time follow-up of radio afterglows helps to study the jet expansion, estimating the jet's true energy, shock efficiency, and other shock properties.  

Square Kilometre Array (SKA) is the next-generation radio telescope developed by the international team. SKA will be more than sensitive than any existing radio telescope and is expected to have its first light in 2027. SKA will consist of three kinds of antennas elements to cover a wide range of frequencies from 50 MHz to 14 GHz: SKA-low array (50 to 350 MHz), SKA-mid array (350 MHz to 14 GHz), and SKA-survey array (350 MHz – 4 GHz). SKA will provide well-sample radio afterglows light curves in radio frequencies (from 50 MHz to 14 GHz) and help to probe the true energy of GRBs using calorimetry and transition from the relativistic phase to the non-relativistic phase of GRBs. Additionally, the SKA-survey array will help to discover radio orphan afterglows.

The synchronous observations of GRBs and related transients utilizing these upcoming major ground and space-based facilities will play a pivotal role in unravelling the intricate details of these energetic and cosmic sources. Our study on the prompt emission of GRBs, as well as the comprehensive analysis of their early-to-late time afterglow observations, modelling, and host galaxy investigations, provides valuable insights for future observations of similar sources using upcoming larger telescopes. Overall, the research presented in this thesis contributes significantly to the field of GRB and paves the way for further progress in our understanding of this extraordinary cosmic phenomenon under a larger perspective of time-domain astronomy. Looking ahead, the combined observations of GRBs across various messengers hold immense potential for unravelling the mysteries of GRBs. These multi-messenger observations of GRBs and related transients will shed light on the fundamental processes occurring in the Universe, opening new avenues of exploration and discovery.  

\appendix
\chapter{\sc Prompt emission: physical mechanisms}

\ifpdf
    \graphicspath{{Appendix/appendix_AFigs/PNG/}{Appendix/appendix_AFigs/PDF/}{Appendix/appendix_AFigs/}}
\else
    \graphicspath{{Appendix/appendix_AFigs/EPS/}{Appendix/appendix_AFigs/}}
\fi

\normalsize

\section{\thisgrb: FIGURES and TABLES}

\renewcommand{\thefigure}{A\arabic{figure}}
\setcounter{figure}{0}

\begin{figure}[ht!]
\centering
\includegraphics[scale=0.29]{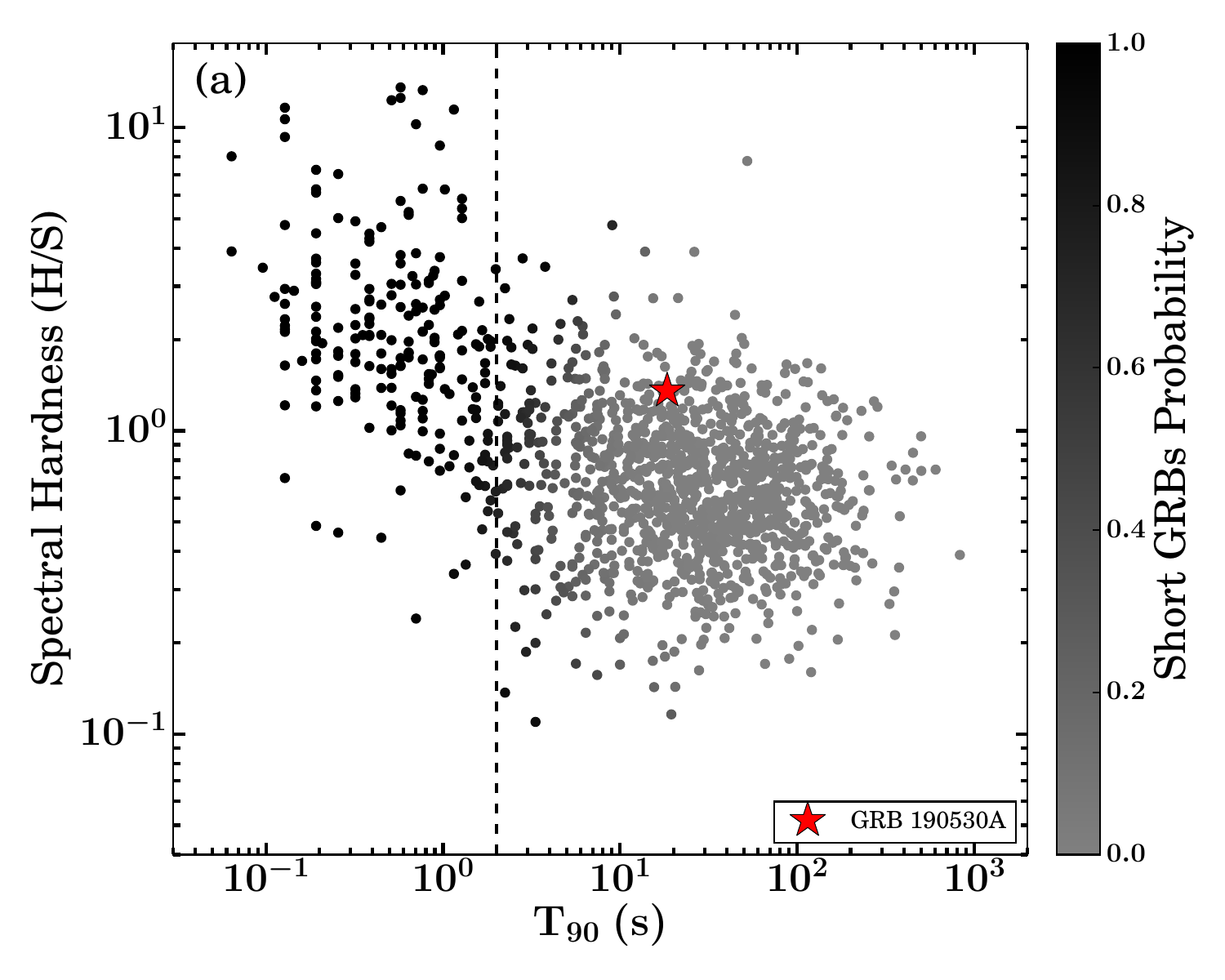}
\includegraphics[scale=0.29]{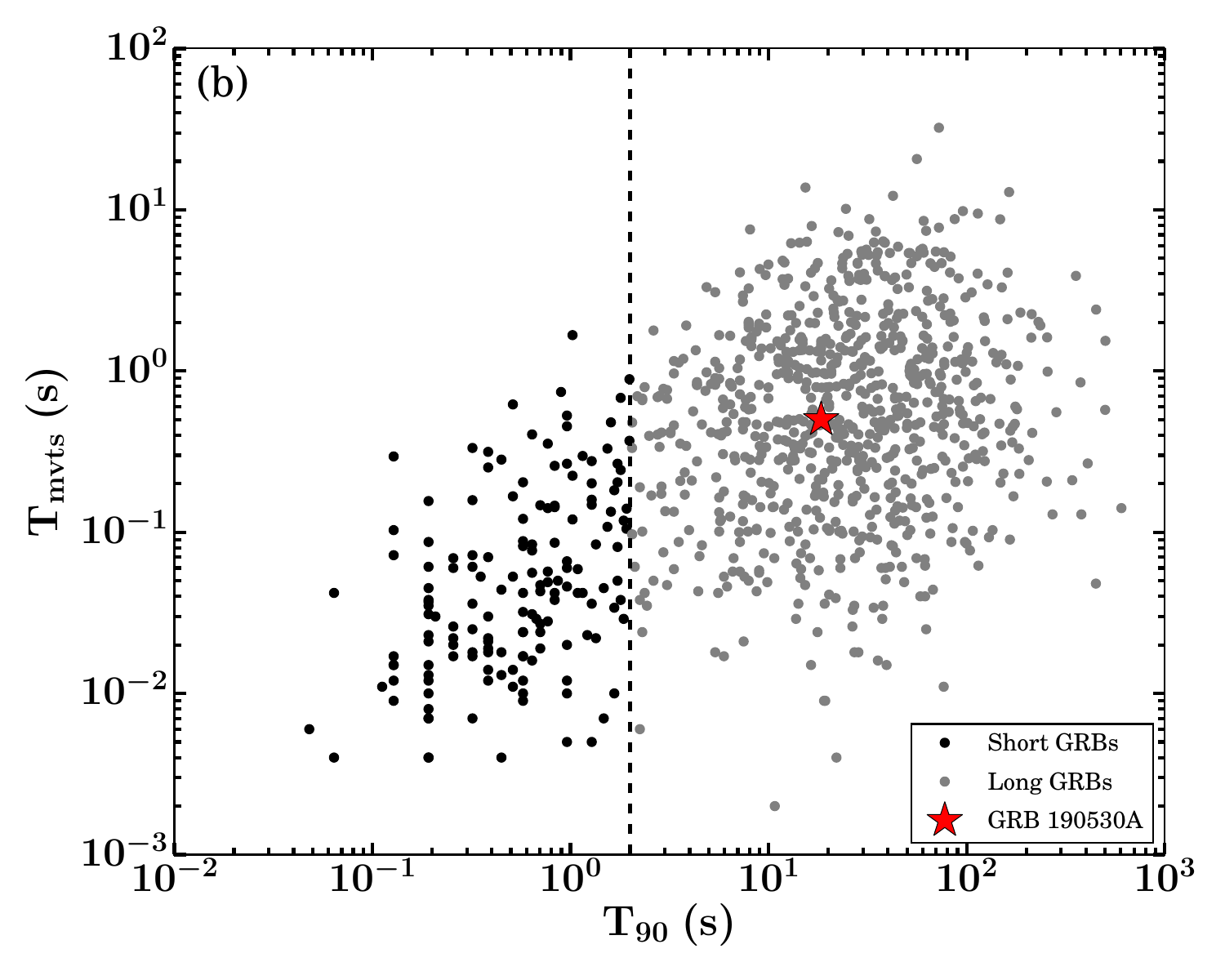}
\caption{{Prompt emission characteristics of \thisgrb (shown with a red star):} (a) The spectral hardness as a function of \tninty duration for \thisgrb along with the data points for short (black circles) and long bursts (grey circles) used in Goldstein et al. (2017). The right side colour scale shows the probability of a GRB belonging to the short bursts class. The vertical dashed lines show the boundary between short and long GRBs. (b) Minimum variability time scale (\mvts) as a function of \tninty duration for \thisgrb along with the short and long GRBs sample studied by \cite{2015ApJ...811...93G}.}
\label{fig:prompt_properties_GRB190530A}
\end{figure}

\begin{figure}[H]
\centering
\includegraphics[scale=0.07]{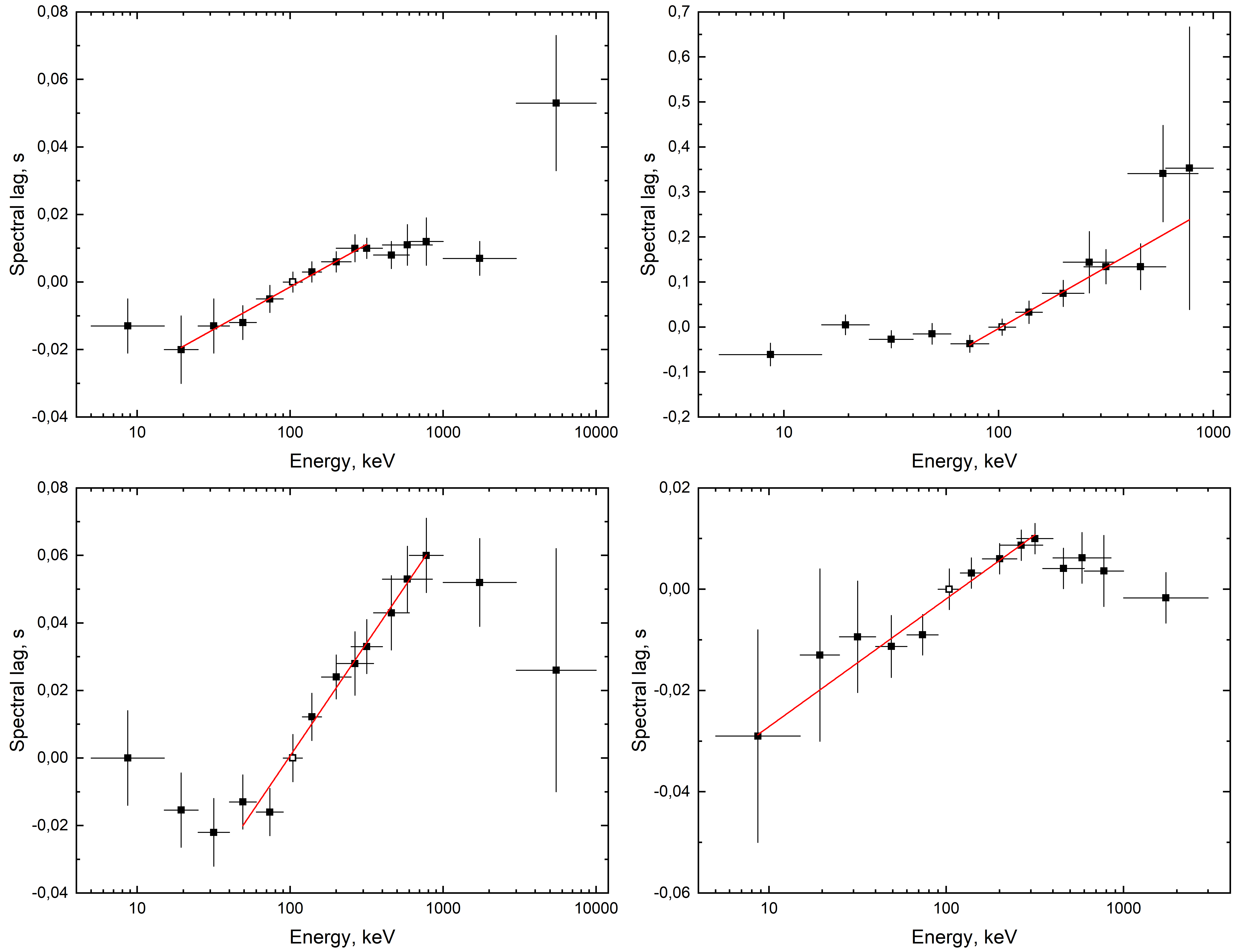}
\caption{Spectral evolution of the total emission episode of \thisgrb (top left), it first (top right), second (bottom left), and third (bottom right) episodes, based on \fermi GBM data. The horizontal axis -- the energy in units of \keV, the vertical axis -- the spectral lag in units of seconds relative to the (90, 120) \keV channel, shown by the unfilled symbol. Red lines represent logarithmic function fits.}
\label{fig:lags_GRB190530A}
\end{figure}

\begin{figure}[H]
\centering
\includegraphics[scale=0.29]{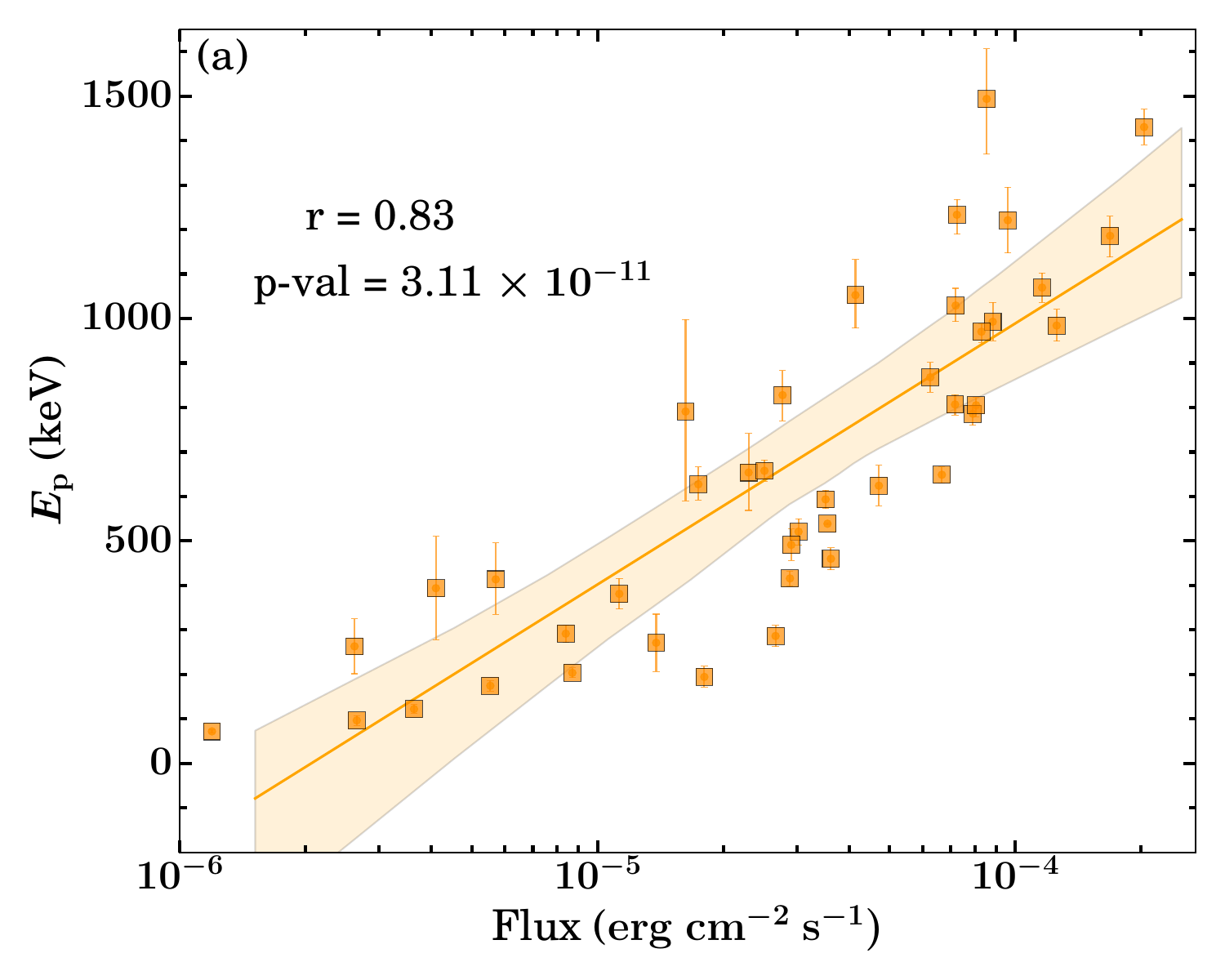}
\includegraphics[scale=0.29]{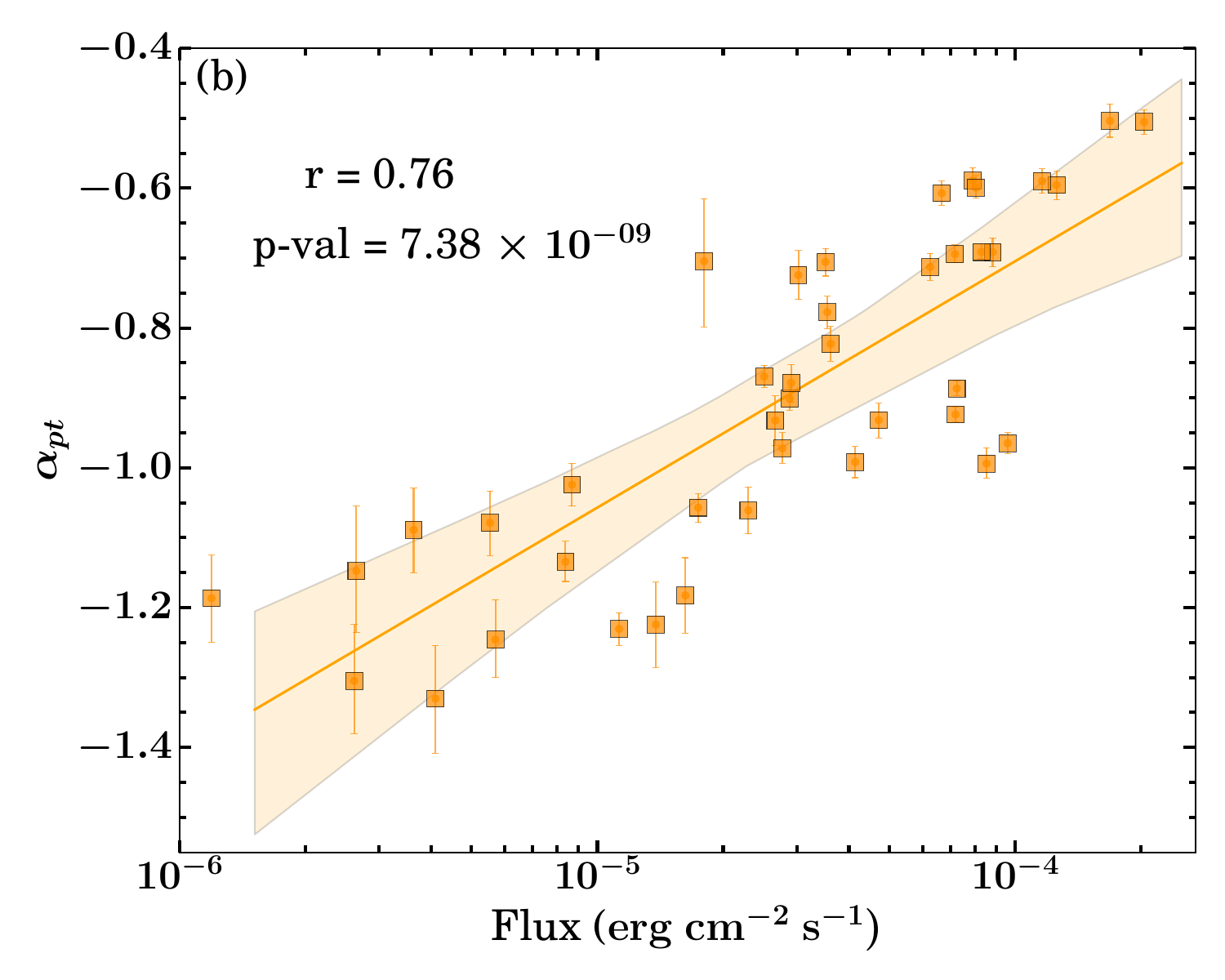}
\includegraphics[scale=0.29]{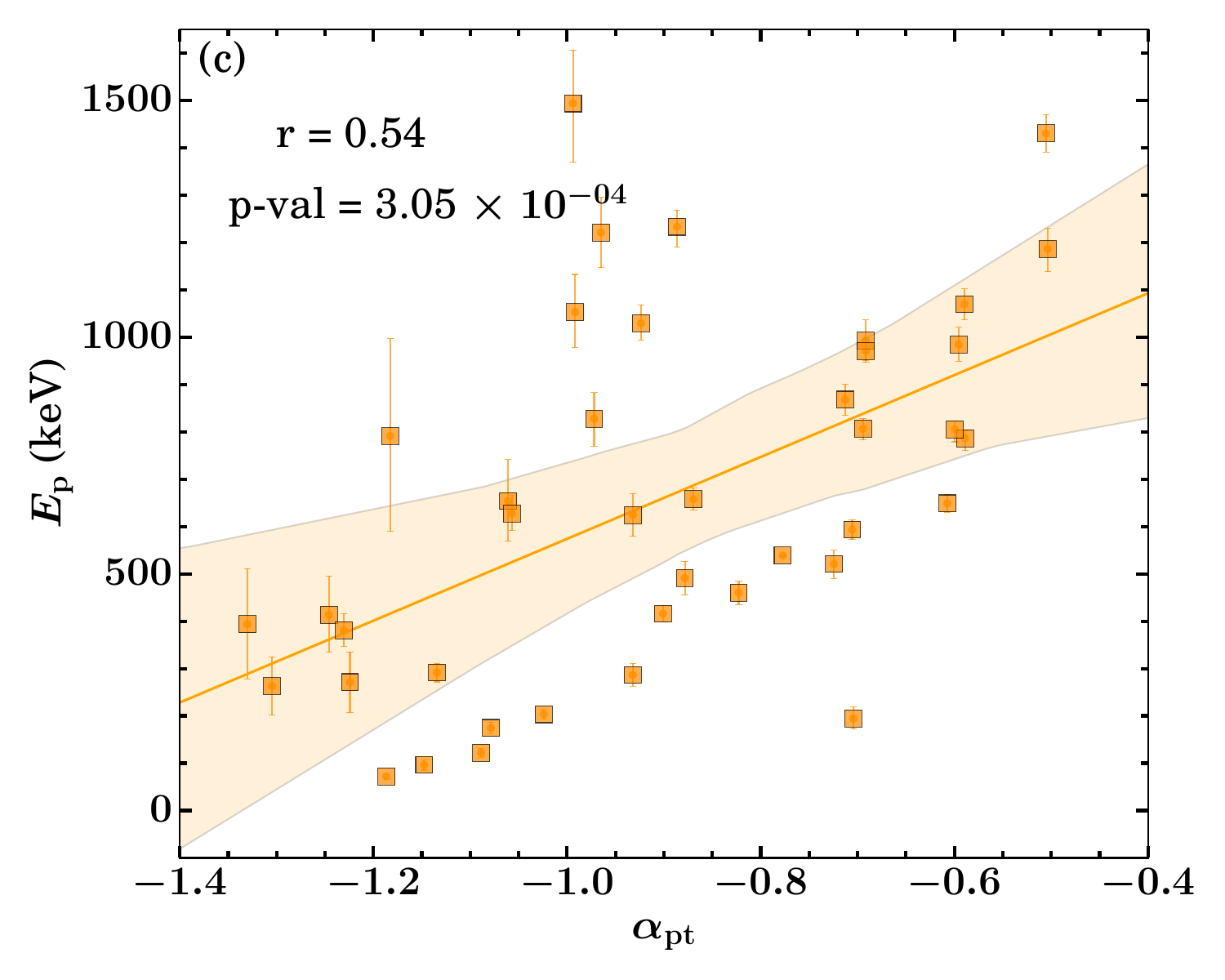}
\caption{{Correlation between spectral parameters obtained from time-revolved spectral analysis using the \fermi GBM data.} (a) Peak energy (\Ep) of \sw{Band} function versus and flux, (b) low-energy spectral index ($\alpha_{\rm pt}$) of \sw{Band} function versus flux, (c) Peak energy (\Ep) as a function of low-energy spectral index ($\alpha_{\rm pt}$). Correlation shown in (a), (b), and (c) are obtained using \fermi GBM observations and modelling with \sw{Band} function. The best fit lines are shown with orange solid lines and shaded grey region show the 2 $\sigma$ confidence interval of the correlations.}
\label{spc_GRB190530A}
\end{figure}

\begin{landscape}
\begin{table}
\begin{scriptsize}
\begin{center}
\begin{longtable}{|c|c|c|c|c|c|c|c|}
\caption{The joint GBM - LAT best fit (shown with boldface) spectral model parameters for the time-integrated spectrum (0 - 25 s) of \thisgrb.}
\label{tab:TAS_GRB190530A}
\\ \hline
\textbf{Model} & \multicolumn{4}{c|}{\textbf{Parameters}} & \textbf{-Log(Likelihood)} & \textbf{AIC} & \textbf{BIC} \\ \hline 
Band & $\it \alpha_{\rm pt}$=  -0.99$^{+0.01}_{-0.01}$ & \multicolumn{2}{c|}{$\it \beta_{\rm pt}$= -3.15$^{+0.04}_{-0.03}$} & \Ep = 822.00$^{+7.67}_{-10.36}$ & 4368.60 & 8757.68 & 8798.82\\ \hline 
\sw{SBPL} & $\alpha_{1}$=  -1.05$^{+0.01}_{-0.01}$ & \multicolumn{2}{c|}{$\alpha_{2}$= -3.23$^{+0.03}_{-0.03}$} & $E_{0}$= 854.47$^{+16.73}_{-18.20}$ & 4462.02 & 8944.52 & 8985.66\\ \hline 
CPL & $\it \alpha_{\rm pt}$=  -1.00$^{+0.01}_{-0.01}$ & \multicolumn{3}{c|}{$E_{0}$ =849.58$^{+11.40}_{-10.01}$} & 4569.47 & 9157.33 & 9194.39\\ \hline 
\sw{bknpow} & $\it \alpha_{1,2}$= 1.12$^{+0.01}_{-0.01}$, 2.70$^{+0.02}_{-0.02}$ & \multicolumn{3}{c|}{$E_{\rm b1}$= 596.18$^{+5.24}_{-5.34}$} & 5115.49 & 10251.45 & 10292.59\\ \hline 
\textbf{\sw{bkn2pow}} & \bf $\it \bf \alpha_{1,2,3}$= 1.03$\bf^{+0.01}_{-0.01}$, 1.42$\bf^{+0.01}_{-0.01}$, 3.04$\bf^{+0.02}_{-0.02}$ & \multicolumn{3}{c|}{$\bf E_{\rm \bf b1,b2}$= \bf 136.65$^{\bf+2.90}_{-2.88}$, 888.36$\bf^{+12.71}_{-11.94}$} &\bf 4331.84 &\bf 8688.35 &\bf 8737.61\\ \hline 
{Band+BB} & $\alpha_{\rm pt}$= -1.01$^{+0.01}_{-0.01}$ & $\beta_{\rm pt}$ = -3.19$^{+0.04}_{-0.04}$ & \Ep= 871.78$^{+11.10}_{-11.94}$ & $\rm k{\it T}_{\rm BB}$= 35.38$^{+2.28}_{-2.10}$ & 4334.66 & 8694.00 & 8743.26\\ \hline 
\sw{SBPL}+BB & $\alpha_{1}$= -1.07$^{+0.01}_{-0.01}$ & $\alpha_{2}$= -3.38$^{+0.04}_{-0.04}$ & $E_{0}$= 1017.30$^{+25.61}_{-26.37}$ & $\rm k{\it T}_{\rm BB}$= 32.66$^{+1.42}_{-1.50}$ & 4378.32 & 8781.31 & 8830.57\\ \hline 
\sw{bknpow}+BB & $\it \alpha_{1,2}$= 1.17$^{+0.01}_{-0.01}$, 2.90$^{+0.03}_{-0.05}$ & $E_{\it b1}$= 757.92$^{+8.84}_{-11.17}$ & \multicolumn{2}{c|}{$\rm k{\it T}_{\rm BB}$ = 42.38$^{+0.82}_{-0.76}$} & 4445.97 & 8916.61 & 8965.87\\ 
\hline
\end{longtable}

\bigskip

\begin{longtable}{|c|c|c|c|c|c|c|c|c|c|c|}
\caption{Results of time-resolved spectral fitting of \thisgrb for \sw{Band} and \sw{Band}+ \sw{BB} functions using \fermi GBM data. Temporal binning are performed based on constant binning method of 1 s. The best fit model is shown in bold for each bin. Flux values (in erg $\rm cm^{-2}$ $\rm s^{-1}$) are calculated in 8 \keV-30 MeV energy range.}
\label{TRS_Table_coarser} \\
\hline
\bf Sr. no. & $\rm \bf t_1$,$\rm \bf t_2$ \bf (s) & \boldmath $\it \alpha_{\rm pt}$ & \boldmath $\it \beta_{\rm pt}$ & \boldmath \Ep (\keV) &  \bf (Flux $\times 10^{-06}$)  &\bf \sw{-Log(likelihood)/BIC} &\textbf{GoF}& \boldmath ${\rm k}{\it T} ~(keV)$ & \bf \sw{-Log(likelihood)/BIC}&\textbf{GoF} \\
\hline
1& 0,1&$-1.20_{-0.03}^{+0.03}$&$-2.42_{-0.18}^{+0.19}$&$308.32_{-25.30}^{+24.23}$& 7.43 &$1041.25/2107.14$& 0.76 & 30.76$_{-1.34}^{+1.45}$& $1041.25/2119.46$& 0.68\\
2& 1,2&$-1.03_{-0.03}^{+0.03}$&$-2.86_{-0.24}^{+0.24}$&$211.55_{-9.47}^{+9.43}$&  8.14&$\bf 1114.92/2254.47$ & 0.02 & $36.52_{-5.60}^{+4.12}$& $1112.82/2262.60$& 0.02 \\
3& 2,3&$-1.12_{-0.04}^{+0.04}$&$-2.77_{-0.28}^{+0.29}$&$147.51_{-9.69}^{+9.22}$&4.29  &$\bf 1000.59/2025.82$ & 0.64 &$6.60_{-2.35}^{+0.76}$ & $999.21/2035.37$& 0.61\\
4& 3,4&$-1.11_{-0.10}^{+0.10}$&$-2.44_{-0.18}^{+0.18}$&$90.78_{-10.22}^{+10.23}$& 2.27 &$\bf 910.28/1845.20$ & 0.97 & $4.91_{-1.64}^{+0.48}$& $909.70/1856.35$& 0.97\\
5& 4,5&$-1.22_{-0.09}^{+0.08}$&$-4.38_{-0.16}^{+2.09}$&$75.42_{-4.83}^{+5.32}$&  1.04&$\bf 895.46/1815.56$ &0.50 & $25.94_{-1.40}^{+1.42}$ & $895.43/1827.82$& 0.49 \\
6& 5,6&$-1.28_{-0.08}^{+0.08}$&unconstrained&$79.70_{-6.01}^{+6.06}$& 0.93 &$\bf 890.51/1805.65$& 0.41 &$4.23_{-0.18}^{+0.21}$ & $890.51/1817.97$& 0.38\\
7& 6,7&$-1.05_{-0.09}^{+0.09}$&unconstrained&$67.87_{-3.61}^{+3.63}$&0.91  &$\bf 888.94/1802.51$ &0.54 & $10.15_{-5.24}^{+19.83}$& $888.94/1814.83$& 0.49\\
8& 7,8&$-1.33_{-0.07}^{+0.08}$&unconstrained&$73.01_{-1.41}^{+1.27}$& 0.94 &$\bf 913.64/1851.92$ & 0.02 & $9.29_{-3.09}^{+0.96}$& $912.21/1861.37$& 0.04 \\
9& 8,9&$-1.02_{-0.01}^{+0.01}$&$-2.74_{-0.13}^{+0.13}$&$1127.57_{-51.46}^{+51.26}$&52.85 &$\bf 1323.85/2672.33$& 0.03 &$0.61_{-1.28}^{+0.31}$ & $1323.85/2684.65$& 0.03 \\
10& 9,10&$-0.94_{-0.01}^{+0.01}$&$-2.88_{-0.12}^{+0.12}$&$923.78_{-33.56}^{+32.53}$& 62.25 &$1368.76/2762.15$ & 0.06 &$0.82_{-0.41}^{+1.40}$& $1368.76/2774.47$& 0.04\\
11& 10,11&$-0.88_{-0.02}^{+0.02}$&$-2.46_{-0.09}^{+0.09}$&$419.46_{-17.23}^{+17.69}$&28.97  &$\bf 1249.27/2523.18$& 0.35 & $19.69_{-4.22}^{+3.18}$& $1240.81/2518.58$& 0.47 \\
12& 11,12&$-1.23_{-0.02}^{+0.02}$&$-2.43_{-0.21}^{+0.20}$&$349.15_{-30.28}^{+29.87}$& 11.14 &$\bf 1114.14/2252.92$& 0.20 & $5.86_{-1.19}^{+0.71}$& $1112.30/2261.55$& 0.20 \\
13& 12,13&$-1.01_{-0.01}^{+0.01}$&unconstrained&$637.91_{-25.47}^{+27.10}$& 17.90 &$\bf 1163.90/2352.43$& 0.65 & $57.64_{-1.34}^{+1.37}$& $1163.90/2364.75$& 0.57\\
14& 13,14&$-0.84_{-0.01}^{+0.01}$&unconstrained&$677.40_{-19.82}^{+20.03}$&  28.52&$\bf 1229.01/2482.66$& 0.53 &$34.06_{-23.64}^{+19.49}$ & $1228.96/2494.88$& 0.48\\
15& 14,15&$-0.92_{-0.01}^{+0.01}$&unconstrained&$709.16_{-21.16}^{+20.59}$& 33.00 &$1349.78/2724.20$& 0.01 &$1.25_{-0.91}^{+0.48}$ & $1349.79/2736.54$& 0.01\\
16& 15,16&$-0.75_{-0.01}^{+0.01}$&$-4.20_{-0.35}^{+0.38}$&$1066.21_{-19.44}^{+19.53}$&85.46  &$\bf 1393.07/2810.78$& 0.07 & $17.67_{-3.99}^{+2.60}$& $1389.33/2815.61$& 0.11\\
17& 16,17&$-0.70_{-0.01}^{+0.01}$&unconstrained&$1029.19_{-17.56}^{+16.12}$& 85.83 &$\bf 1427.67/2879.98$ & 0.01 &$4.30_{-2.75}^{+0.51}$ & $1423.82/2884.59$& 0.01 \\
18& 17,18&$-0.69_{-0.01}^{+0.01}$&unconstrained&$894.11_{-15.05}^{+16.03}$&  73.49&$\bf 1393.67/2811.97$& 0.01 & $6.20_{-4.42}^{+1.05}$& $1392.84/2822.64$& 0.01\\
19& 18,19&$-0.71_{-0.01}^{+0.01}$&unconstrained&$714.18_{-12.50}^{+12.58}$&  60.07&$\bf 1399.09/2822.81$& 0.01 & unconstrained& $1399.09/2835.13$& 0.01\\
20& 19,20&$-1.10_{-0.02}^{+0.02}$&$-3.45_{-0.69}^{+0.69}$&$549.74_{-27.86}^{+28.03}$&15.46  &$\bf 1158.58/2341.79$& 0.29 & $16.11_{-5.23}^{+2.29}$& $1151.61/2340.18$& 0.45 \\
\hline
\end{longtable}

\end{center}
\end{scriptsize}
\end{table}
\end{landscape}

\begin{landscape}
\begin{table}[p]
\begin{scriptsize}
\begin{center}
\begin{longtable}{|c|c|c|c|c|c|c|c|c|}
\caption{Results of time-resolved spectral fitting of \thisgrb for \sw{bkn2power} function using \fermi GBM data. Temporal binning is performed based on the constant binning method of 1 s. The best fit model is shown in bold for each bin.}
\label{TRS_Table_coarser_bkn2pow}\\
\hline
\bf Sr. no. & $\rm \bf t_1$,$\rm \bf t_2$ \bf (s) & \boldmath $\it \alpha_{\rm 1}$ & \boldmath $\it \alpha_{\rm 2}$ & \boldmath $\it \alpha_{\rm 3}$ & \boldmath $E_{\rm break, 1}$ (\keV) & \boldmath \Ep or $E_{\rm break, 2}$(\keV)  &\bf \sw{-Log(likelihood)/BIC}&\textbf{GoF} \\
\hline
1& 0, 1&unconstrained&$1.39_{-0.02}^{+0.02}$&$2.34_{-0.12}^{+0.11}$& $14.99_{-0.79}^{+1.23}$ &$223.22_{-20.67}^{+19.58}$ &$\bf 1028.51/2093.97$& 0.89\\
2& 1, 2&$0.58_{-0.21}^{+0.28}$&$1.27_{-0.02}^{+0.02}$&$2.38_{-0.06}^{+0.06}$& $15.75_{-1.62}^{+2.05}$ &$143.91_{-6.83}^{+6.73}$ &$ 1112.09/2261.14$& 0.04\\
3& 2,3&$1.17_{-0.06}^{+0.06}$&$1.61_{-0.05}^{+0.05}$&$2.68_{-0.16}^{+0.16}$& $38.47_{-5.16}^{+5.39}$ &$164.67_{-15.90}^{+16.42}$ &$ 1000.42/2037.80$& 0.61 \\
4& 3,4&$1.10_{-0.17}^{+0.19}$&$1.56_{-0.07}^{+0.07}$&$2.38_{-0.11}^{+0.11}$& $21.47_{-4.33}^{+4.86}$ &$85.06_{-9.91}^{+10.27}$ &$ 909.36/1855.68$& 0.96\\
5& 4,5&$0.84_{-0.13}^{+1.11}$&$1.79_{-0.04}^{+0.09}$&$3.00_{-0.36}^{+0.37}$& $19.27_{-1.17}^{+8.86}$ &$110.85_{-15.33}^{+18.06}$ &$ 893.04/1823.03$& 0.53 \\
6& 5,6&unconstrained&$1.67_{-0.04}^{+0.04}$&$3.33_{-0.30}^{+0.30}$& $11.38_{-0.21}^{+0.19}$ &$113.85_{-0.98}^{+1.06}$ &$ 890.09/1817.14$& 0.33\\
7& 6,7&$0.17_{-0.18}^{+0.84}$&$1.62_{-0.05}^{+0.05}$&$3.61_{-0.38}^{+0.39}$& $13.53_{-0.92}^{+2.03}$ &$98.90_{-7.70}^{+8.33}$ &$ 891.95/1820.86$& 0.37\\
8& 7,8&unconstrained&$1.72_{-0.04}^{+0.04}$&$3.20_{-0.54}^{+0.52}$& $10.37_{-0.79}^{+0.81}$ &$115.27_{-16.75}^{+16.55}$ &$ 916.47/1869.89$& 0.02 \\
9& 8,9&$1.01_{-0.01}^{+0.01}$&$1.34_{-0.02}^{+0.02}$&$2.62_{-0.08}^{+0.07}$& $111.75_{-10.65}^{+10.17}$ &$1038.47_{-62.60}^{+62.84}$ &$1319.88/2676.71$& 0.04\\
10& 9,10&$0.92_{-0.01}^{+0.01}$&$1.34_{-0.02}^{+0.02}$&$2.75_{-0.07}^{+0.07}$& $107.54_{-6.94}^{+7.05}$ &$933.16_{-42.24}^{+43.84}$ &$\bf 1354.17/2745.29$& 0.31\\
11& 10,11&$0.90_{-0.02}^{+0.03}$&$1.33_{-0.03}^{+0.02}$&$2.39_{-0.05}^{+0.05}$& $66.84_{-5.32}^{+5.17}$ &$357.54_{-19.22}^{+18.32}$ &$ 1247.16/2531.28$& 0.44\\
12& 11,12&$0.31_{-0.20}^{+0.23}$&$1.38_{-0.02}^{+0.02}$&$2.16_{-0.06}^{+0.06}$& $14.81_{-0.45}^{+0.50}$ &$200.63_{-14.20}^{+14.73}$ &$ 1113.07/2263.10$& 0.24\\
13& 12,13&$0.89_{-0.13}^{+0.13}$&$1.18_{-0.01}^{+0.01}$&$2.83_{-0.11}^{+0.11}$& $20.61_{-3.89}^{+4.10}$ &$478.51_{-20.66}^{+20.83}$ &$ 1185.90/2408.75$& 0.20\\
14& 13,14&$0.93_{-0.01}^{+0.01}$&$1.41_{-0.04}^{+0.04}$&$3.34_{-0.17}^{+0.17}$& $162.73_{-13.59}^{+13.44}$ &$752.28_{-37.88}^{+37.04}$ &$ 1235.15/2507.26$& 0.46\\
15& 14,15&$0.87_{-0.02}^{+0.02}$&$1.46_{-0.02}^{+0.02}$&$3.77_{-0.33}^{+0.32}$& $98.70_{-4.95}^{+4.95}$ &$1138.55_{-104.06}^{+98.93}$ &$\bf 1302.00/2640.95$& 0.01 \\
16& 15,16&$0.78_{-0.01}^{+0.01}$&$1.13_{-0.02}^{+0.02}$&$3.02_{-0.06}^{+0.06}$& $123.96_{-9.19}^{+9.10}$ &$920.13_{-24.13}^{+24.74}$ &$ 1415.49/2867.93$& 0.02\\
17& 16,17&$0.81_{-0.01}^{+0.01}$&$1.16_{-0.03}^{+0.03}$&$3.41_{-0.08}^{+0.08}$& $201.73_{-15.79}^{+15.57}$ &$970.73_{-25.05}^{+24.58}$ &$ 1443.64/2924.23$& 0.01 \\
18& 17,18&$0.76_{-0.01}^{+0.01}$&$1.17_{-0.02}^{+0.02}$&$3.42_{-0.09}^{+0.09}$& $144.39_{-9.66}^{+9.79}$ &$890.79_{-22.25}^{+22.15}$ &$ 1404.30/2845.56$& 0.01 \\
19& 18,19&$0.74_{-0.01}^{+0.01}$&$1.39_{-0.02}^{+0.02}$&$3.61_{-0.12}^{+0.12}$& $137.24_{-5.39}^{+5.23}$ &$941.15_{-28.94}^{+27.47}$ &$ 1400.04/2837.04$& 0.01\\
20& 19,20&$1.10_{-0.02}^{+0.02}$&$1.50_{-0.03}^{+0.03}$&$3.40_{-0.36}^{+0.37}$& $81.36_{-8.00}^{+8.15}$ &$735.67_{-59.17}^{+61.97}$ &$1150.73/2338.42$ & 0.43\\
\hline
\end{longtable}
\end{center}
\end{scriptsize}
\end{table}
\end{landscape}

\begin{landscape}
\begin{table}[ht!]
\begin{tiny}
\begin{center}
\begin{longtable}{|c|c|c|c|c|c|c|c|c|c|c|}
\caption{Same as Table \ref{TRS_Table_coarser} but temporal binning are performed based on Bayesian Block algorithm.}  
\label{TRS_Table_band}\\
\hline
\bf Sr. no. & $\rm \bf t_1$,$\rm \bf t_2$ \bf (s) & \boldmath $\it \alpha_{\rm pt}$ & \boldmath $\it \beta_{\rm pt}$ & \boldmath \Ep (\keV) &  \bf (Flux $\times 10^{-06}$)  &\bf \sw{-Log(likelihood)/BIC} & \textbf{GoF} &\boldmath ${\rm k}{\it T} ~(keV)$ & \bf \sw{-Log(likelihood)/BIC}& \textbf{GoF} \\
\hline
1& 0.000106, 0.282062&$-1.30_{-0.07}^{+0.08}$&$-2.83_{-0.99}^{+0.68}$&$263.00_{-61.64}^{+61.92}$& 2.62 &$\bf 81.66/187.95$& 0.35 & $8.33_{-1.69}^{+1.10}$ & 78.77/194.49 &  0.36 \\
2&0.282062, 0.509457  &$-1.25_{-0.05}^{+0.06}$&unconstrained&$413.58_{-78.62}^{+82.23}$& 5.70 &$\bf -12.61/-0.58$& 0.89 & $6.69_{-0.80}^{+0.72}$ & -17.11/2.74& 0.88\\
3&0.509457, 1.148330 &$-1.13_{-0.03}^{+0.03}$&$-2.85_{-0.41}^{+0.40}$&$291.50_{-19.82}^{+19.42}$& 8.39 &$804.74/1634.12$& 0.23  & $69.08_{-7.85}^{+6.26}$ & 799.99/1636.93& 0.27\\
4& 1.148330, 1.947446 &$-1.02_{-0.03}^{+0.03}$&$-2.66_{-0.19}^{+0.18}$&$203.43_{-10.91}^{+10.90}$& 8.70 &$\bf  993.30/2011.23$& 0.03  & $32.70_{-5.85}^{+3.70}$ & 992.12/2021.19& 0.02 \\
5& 1.947446, 2.336272 &$-1.08_{-0.05}^{+0.05}$&$-3.21_{-0.52}^{+0.51}$&$174.37_{-11.61}^{+12.26}$& 5.54 &$\bf 374.64/773.92$& 0.70 & $6.87_{-2.86}^{+0.89}$ & 374.08/785.12& 0.74 \\
6& 2.336272, 3.005074 &$-1.09_{-0.06}^{+0.06}$&$-2.61_{-0.22}^{+0.23}$&$122.03_{-9.57}^{+9.76}$& 3.64 &$\bf 724.79/1474.21$& 0.57 & $5.21_{-1.97}^{+0.47}$ & 723.80/1484.55& 0.55 \\
7& 3.005074, 3.657240 &$-1.15_{-0.09}^{+0.09}$&$-2.37_{-0.17}^{+0.16}$&$96.51_{-11.12}^{+10.73}$& 2.65 &$\bf 669.09/1362.82$& 0.70 & $29.66_{-18.38}^{+7.65}$ & 668.95/1374.86& 0.69 \\
8& 3.657240, 8.002803 &$-1.19_{-0.06}^{+0.06}$&$-2.67_{-0.19}^{+0.18}$&$71.82_{-4.42}^{+4.33}$& 1.19 &$\bf 1982.52/3989.68$& 0.85 & $4.47_{-0.98}^{+0.41}$ & 1980.80/3998.55& 0.84\\
9&8.002803, 8.122972 &$-1.33_{-0.08}^{+0.08}$&unconstrained&$393.70_{-115.97}^{+116.96}$& 4.10 &$\bf -494.69/-964.75$& 0.47 & $9.17_{-2.64}^{+1.13}$ & -496.56/-956.16& 0.46\\
10& 8.122972, 8.225416  &$-1.18_{-0.05}^{+0.05}$&$-2.50_{-0.52}^{+0.44}$&$790.88_{-201.06}^{+206.51}$& 16.24  &$\bf -482.81/-940.98$& 0.47 & $21.75_{-14.98}^{+6.32}$ & -482.99/-929.02& 0.45\\
11& 8.225416, 8.436160&$-1.06_{-0.03}^{+0.03}$&$-2.58_{-0.37}^{+0.36}$&$653.78_{-85.17}^{+87.65}$& 22.99 &$\bf 105.11/234.87$& 0.36  &$21.57_{-7.45}^{+3.53}$ & 103.22/243.39& 0.40 \\
12&8.436160, 8.609198  &$-0.99_{-0.02}^{+0.02}$&unconstrained&$1052.68_{-74.37}^{+79.83}$& 41.42 &$\bf 49.78/124.20$&  0.53 & $0.50_{-0.16}^{+1.31}$ & 49.78/136.52& 0.49 \\
13&8.609198, 8.704119  &$-0.99_{-0.02}^{+0.02}$&$-4.01_{-0.54}^{+1.27}$&$1493.73_{-124.56}^{+113.07}$& 85.37 &$\bf -273.71/-522.78$& 0.05  &$1.10_{-0.58}^{+1.46}$ & -273.70/-510.45& 0.03\\
14&8.704119, 8.947630  &$-0.96_{-0.01}^{+0.01}$&$-3.24_{-0.37}^{+0.36}$&$1220.97_{-73.90}^{+73.97}$& 95.91 &$\bf 436.00/896.63$& 0.16  &$41.08_{-11.41}^{+7.67}$ & 427.60/892.15& 0.31\\
15&8.947630, 9.661208  &$-0.92_{-0.01}^{+0.01}$&$-3.00_{-0.16}^{+0.14}$&$1029.21_{-36.11}^{+38.79}$& 71.96 & $1151.30/2327.23$& 0.62  & $1.01_{-0.39}^{+6.23}$ & 1151.30/2339.55& 0.59  \\
16&9.661208, 9.912693  &$-0.93_{-0.03}^{+0.02}$&$-2.54_{-0.16}^{+0.16}$&$624.17_{-45.10}^{+45.71}$& 47.13 &$\bf 362.97/750.59$& 0.02  & $23.45_{-6.42}^{+4.00}$ & 359.19/755.34& 0.02\\
17&9.912693, 11.082932  &$-0.90_{-0.02}^{+0.02}$&$-2.43_{-0.08}^{+0.08}$&$415.68_{-17.25}^{+16.54}$& 28.87 &$\bf 1393.56/2811.75$& 0.25  & $102.00_{-1.37}^{+1.38}$ & 1393.56/2824.07& 0.18 \\
18&11.082932, 12.220607 &$-1.23_{-0.02}^{+0.02}$&$-2.40_{-0.20}^{+0.19}$&$381.01_{-34.18}^{+34.91}$& 11.27 &$\bf 1227.44/2479.51$& 0.48  & $24.67_{-9.10}^{+4.71}$ & 1226.42/2489.79& 0.50\\
19&12.220607, 12.748669 &$-1.06_{-0.02}^{+0.02}$&$-4.46_{-0.08}^{+2.12}$&$627.33_{-35.72}^{+38.90}$& 17.42 &$\bf 758.81/1542.25$& 0.02  & $10.02_{-6.52}^{+1.85}$ & 758.51/1553.98& 0.02 \\
20&12.748669, 13.396796 &$-0.87_{-0.02}^{+0.02}$&unconstrained&$657.71_{-23.09}^{+24.19}$& 25.08 &$\bf 960.57/1945.77$& 0.15  & $53.05_{-1.46}^{+1.32}$ & 960.57/1958.09& 0.15  \\
21&13.396796, 13.542326 &$-0.72_{-0.03}^{+0.03}$&$-3.83_{-0.65}^{+1.28}$&$520.89_{-30.95}^{+29.20}$& 30.26 &$\bf -101.70/-178.76$& 0.09 & $84.16_{-10.83}^{+9.60}$ & -106.21/-175.46& 0.07 \\
22&13.542326, 13.836101  &$-0.97_{-0.02}^{+0.02}$&unconstrained&$827.80_{-57.81}^{+55.25}$& 27.72 &$\bf 381.11/786.85$& 0.46 & $21.90_{-6.74}^{+3.59}$ & 379.73/796.42& 0.52\\
23&13.836101, 14.219772  &$-0.71_{-0.02}^{+0.02}$&unconstrained&$593.45_{-20.75.64}^{+20.34}$& 35.13 &$\bf 623.29/1271.22$& 0.30 & $39.07_{-10.68}^{+7.37}$  & 619.54/1276.03& 0.38 \\
24&14.219772, 14.658330 &$-0.88_{-0.03}^{+0.03}$&$-2.75_{-0.29}^{+0.28}$&$491.14_{-35.24}^{+35.56}$& 29.07 &$713.56/1451.76$& 0.03 & $27.52_{-2.55}^{+2.53}$ & {\bf 688.39/1413.74}& 0.24\\
25&14.658330, 15.192231  &$-0.89_{-0.01}^{+0.01}$&$-4.60_{-0.07}^{+1.53}$&$1233.27_{-43.23}^{+34.78}$& 72.45 &$954.15/1932.93$& 0.01 & $141.81_{-1.42}^{+1.36}$ & 954.15/1945.25& 0.01 \\
26& 15.192231, 15.338963 &$-0.51_{-0.02}^{+0.02}$&unconstrained&$1430.37_{-39.35}^{+40.02}$& 203.24 &$186.79/398.22$& 0.01 & $8.02_{-1.04}^{+1.13}$ & {\bf 162.21/361.37}& 0.01 \\
27& 15.338963, 15.687352 &$-0.59_{-0.01}^{+0.02}$&$-4.47_{-0.21}^{+1.27}$&$785.94_{-25.83}^{+17.10}$& 79.11 &$\bf 660.06/1344.76$ & 0.40 & $1.19_{-0.65}^{+1.22}$ & 660.06/1357.07& 0.38 \\
28& 15.687352, 16.251414 &$-0.93_{-0.03}^{+0.03}$&$-2.15_{-0.07}^{+0.07}$&$286.27_{-24.07}^{+24.34}$& 26.66 &$847.89/1720.42$& 0.11 & $20.19_{-1.75}^{+1.75}$ & 825.21/1687.38& 0.27 \\
29&16.251414, 16.407026 &$-0.69_{-0.02}^{+0.02}$&$-3.81_{-0.50}^{+0.54}$&$992.80_{-42.60}^{+43.88}$& 88.26 &$\bf 67.38/159.39$& 0.67 & $9.87_{-1.17}^{+1.12}$ & 60.67/158.29& 0.79 \\
30& 16.407026, 16.524165 &$-0.59_{-0.02}^{+0.02}$&unconstrained&$984.19_{-34.64}^{+36.28}$& 125.52 &$\bf -70.94/-117.25$& 0.83 & $32.06_{-13.17}^{+5.73}$ & -72.26/-107.56& 0.83 \\
31& 16.524165, 16.705623 &$-0.59_{-0.02}^{+0.02}$&unconstrained&$1069.46_{-32.93}^{+32.73}$& 115.88 &$\bf 229.87/484.38$& 0.38 & $10.15_{-2.78}^{+2.08}$ & 228.01/492.97& 0.36\\
32&16.705623, 16.787528   &$-0.50_{-0.02}^{+0.02}$&unconstrained&$1185.83_{-47.53}^{+43.98}$& 168.56 &$\bf -250.30/-475.96$& 0.01 & $4.81_{-3.38}^{+0.53}$ & -250.74/-464.53& 0.01 \\
33&16.787528, 17.114033  &$-0.60_{-0.01}^{+0.02}$&$-4.47_{-0.25}^{+1.13}$&$805.07_{-26.23}^{+17.72}$& 80.52 &$\bf 632.33/1289.29$& 0.07  & $0.87_{-0.48}^{+1.25}$ & 632.33/1301.61& 0.09\\
34& 17.114033, 17.332842  &$-0.71_{-0.02}^{+0.02}$&unconstrained&$867.86_{-33.23}^{+33.24}$& 62.57 &$\bf 300.86/626.35$&  0.02 & $9.84_{-1.60}^{+1.41}$&  296.81/630.57& 0.02\\
35& 17.332842, 17.924038 &$-0.69_{-0.01}^{+0.01}$&$-4.86_{-0.03}^{+1.09}$&$970.42_{-23.81}^{+17.70}$& 82.93 &$\bf 1050.02/2124.69$& 0.10  & $3.90_{-2.84}^{+0.62}$ & 1050.18/2137.31& 0.08\\
36&17.924038, 18.441228  &$-0.69_{-0.01}^{+0.01}$&$-3.85_{-0.48}^{+0.51}$&$806.72_{-23.57}^{+22.02}$& 71.72 &$\bf 931.43/1887.50$& 0.15  & unconstrained & 931.43/1899.82& 0.12\\
37&18.441228, 18.542279  &$-0.78_{-0.02}^{+0.02}$&unconstrained&$538.64_{-1.40}^{+1.32}$& 35.51 &$-299.09/-573.54$& 0.04  & $25.50_{-3.02}^{+2.62}$ & {\bf -319.09/-601.22}& 0.37\\
38& 18.542279, 18.880490&$-0.61_{-0.02}^{+0.02}$&$-3.97_{-0.51}^{+0.55}$&$648.72_{-18.48}^{+18.02}$& 66.69 &$625.26/1275.16$& 0.53  & $60.01_{-5.89}^{+5.98}$ & {\bf 615.12/1267.20}& 0.74\\
39&18.880490, 19.209439  &$-0.82_{-0.02}^{+0.02}$&$-2.89_{-0.24}^{+0.26}$&$459.88_{-25.13}^{+24.29}$& 36.18 &$523.32/1071.28$& 0.29  & $25.45_{-1.92}^{+1.88}$ & {\bf 490.26/1017.47}& 0.95\\
40& 19.209439, 19.316852 &$-0.70_{-0.09}^{+0.09}$&$-2.26_{-0.14}^{+0.14}$&$194.21_{-22.45}^{+24.49}$& 18.01 &$\bf -419.57/-814.50$& 0.56   & $0.88_{-0.70}^{+0.28}$ & -419.58/-802.20& 0.47 \\
41& 19.316852, 19.790851&$-1.22_{-0.06}^{+0.06}$&$-1.99_{-0.11}^{+0.10}$&$271.00_{-64.36}^{+64.39}$& 13.81 &$586.79/1198.21$& 0.35  & $11.74_{-1.98}^{+1.52}$ & 578.98/1194.92& 0.38 \\ \hline
\end{longtable}
\end{center}
\end{tiny}
\end{table}
\end{landscape}

\begin{landscape}
\begin{table}[ht!]
\begin{tiny}
\begin{center}
\begin{longtable}{|c|c|c|c|c|c|c|c|c|c|}
\caption{Same as Table \ref{TRS_Table_coarser_bkn2pow} but temporal binning are performed based on Bayesian Block algorithm.} 
\label{TRS_Table_bkn2pow}\\
\hline
\bf Sr. no. & $\rm \bf t_1$,$\rm \bf t_2$ \bf (s) & \boldmath $\it \alpha_{\rm 1}$ & \boldmath $\it \alpha_{\rm 2}$ & \boldmath $\it \alpha_{\rm 3}$ & \boldmath $E_{\rm break, 1}$ (\keV) & \boldmath \Ep or $E_{\rm break, 2}$(\keV) &  \bf (Flux $\times 10^{-06}$)  &\bf \sw{-Log(likelihood)/BIC}& \textbf{GoF} \\
\hline 
1& 0.000106, 0.282062&$0.57_{-0.20}^{+0.47}$&$1.58_{-0.06}^{+0.07}$&$2.57_{-0.44}^{+0.43}$& $20.78_{-2.58}^{+4.44}$ &$267.78_{-2.32}^{+2.29}$ & 2.26&$77.87/192.70$& 0.40\\
2&0.282062, 0.509457  &$0.47_{-0.19}^{+0.48}$&$1.44_{-0.04}^{+0.05}$&$2.71_{-0.36}^{+0.37}$& $19.43_{-2.18}^{+3.75}$ &$323.15_{-63.26}^{+64.76}$ & 4.93&$-17.23/2.50 $& 0.88\\
3&0.509457, 1.148330 &unconstrained&$1.33_{-0.02}^{+0.02}$&$2.52_{-0.12}^{+0.13}$& $13.87_{-0.13}^{+0.14}$ &$215.70_{-14.95}^{+15.66}$ &7.88 &$\bf 794.98 /1626.91 $& 0.42\\
4& 1.148330, 1.947446 &$0.46_{-0.18}^{+0.21}$&$1.27_{-0.02}^{+0.02}$&$2.30_{-0.05}^{+0.06}$& $15.30_{-0.31}^{+0.31}$ &$131.78_{-7.07}^{+6.49}$ & 9.53 &$ 989.30/2015.56 $& 0.09\\
5& 1.947446, 2.336272 &$0.35_{-0.21}^{+0.56}$&$1.35_{-0.03}^{+0.03}$&$2.67_{-0.14}^{+0.15}$& $13.58_{-1.39}^{+2.24}$ &$148.30_{-11.90}^{+11.91}$ & 4.85&$ 376.09/789.13 $& 0.73\\
6& 2.336272, 3.005074 &$1.22_{-0.08}^{+0.07}$&$1.63_{-0.09}^{+0.10}$&$2.51_{-0.14}^{+0.14}$& $40.00_{-8.46}^{+8.53}$ &$125.82_{-17.05}^{+17.46}$ & 3.70&$ 725.85/1488.66 $& 0.60\\
7& 3.005074, 3.657240 &$1.24_{-0.16}^{+0.17}$&$1.59_{-0.07}^{+0.07}$&$2.37_{-0.13}^{+0.12}$& $25.92_{-5.48}^{+6.05}$ &$93.44_{-12.14}^{+11.48}$ & 2.46&$668.18 /1373.31 $& 0.67\\
8& 3.657240, 8.002803 &$0.91_{-0.20}^{+0.26}$&$1.57_{-0.04}^{+0.04}$&$2.43_{-0.06}^{+0.06}$& $15.20_{-1.60}^{+2.01}$ &$64.82_{-4.58}^{+4.38}$ & 1.22&$1984.51 /4005.98 $& 0.75\\
9&8.002803, 8.122972 &$0.46_{-0.19}^{+0.65}$&$1.51_{-0.06}^{+0.06}$&$2.93_{-0.61}^{+0.59}$& $18.39_{-2.65}^{+4.69}$ &$330.09_{-63.33}^{+61.71}$ &3.10 &$ -496.51/-956.07 $& 0.28\\
10&8.122972, 8.225416  &$1.08_{-0.15}^{+0.21}$&$1.34_{-0.06}^{+0.06}$&$2.38_{-0.29}^{+0.29}$& $34.64_{-16.86}^{+21.45}$ &$538.39_{-105.60}^{+108.55}$ & 13.99&$ -482.93/-928.91 $& 0.40\\
11& 8.225416, 8.436160&$1.05_{-0.05}^{+0.05}$&$1.43_{-0.05}^{+0.05}$&$2.64_{-0.23}^{+0.23}$& $84.96_{-19.03}^{+19.23}$ &$687.39_{-46.84}^{+45.03}$ &22.30 &$ 102.61/242.19 $& 0.40 \\
12&8.436160, 8.609198  &unconstrained&$1.12_{-0.01}^{+0.02}$&$3.18_{-0.23}^{+0.23}$& $10.75_{-1.53}^{+4.53}$ &$797.24_{-54.77}^{+57.77}$ & 13.75&$ 58.22/153.39 $& 0.41\\
13&8.609198, 8.704119  &$0.83_{-0.13}^{+0.14}$&$1.17_{-0.03}^{+0.03}$&$2.85_{-0.20}^{+0.20}$& $41.93_{-12.84}^{+13.76}$ &$1132.80_{-123.51}^{+120.41}$ & 79.60&$ -274.05/-511.15 $& 0.03\\
14&8.704119, 8.947630  &$0.96_{-0.02}^{+0.02}$&$1.42_{-0.04}^{+0.04}$&$3.05_{-0.16}^{+0.17}$& $167.56_{-17.24}^{+18.23}$ &$1479.67_{-124.99}^{+131.35}$ &98.18 &$ 431.44/899.83 $& 0.36 \\
15&8.947630, 9.661208  &$0.90_{-0.02}^{+0.02}$&$1.30_{-0.02}^{+0.02}$&$2.76_{-0.07}^{+0.08}$& $109.80_{-8.23}^{+8.46}$ &$988.01_{-46.76}^{+49.80}$ & 74.45&$\bf 1142.16/2321.27 $& 0.81\\
16&9.661208, 9.912693  &$0.92_{-0.03}^{+0.03}$&$1.36_{-0.04}^{+0.04}$&$2.55_{-0.12}^{+0.11}$& $83.94_{-10.83}^{+10.50}$ &$620.87_{-58.94}^{+57.32}$ & 47.42&$ 361.55/760.06 $& 0.06 \\
17&9.912693, 11.082932  &$0.92_{-0.02}^{+0.02}$&$1.36_{-0.03}^{+0.02}$&$2.38_{-0.05}^{+0.05}$& $68.22_{-4.90}^{+4.75}$ &$365.79_{-18.82}^{+19.24}$ & 29.69&$ 1389.78/2816.52 $& 0.31\\
18&11.082932, 12.220607 &$0.48_{-0.22}^{+0.33}$&$1.37_{-0.02}^{+0.02}$&$2.14_{-0.06}^{+0.05}$& $15.13_{-1.13}^{+1.57}$ &$209.76_{-15.81}^{+15.44}$ &11.61 &$1229.89 /2496.73 $& 0.49\\
19&12.220607, 12.748669 &$0.74_{-0.19}^{+0.22}$&$1.23_{-0.02}^{+0.02}$&$2.67_{-0.16}^{+0.15}$& $19.54_{-3.14}^{+3.46}$ &$465.04_{-33.77}^{+31.40}$ & 17.03&$ 766.07/1569.10 $& 0.02 \\
20&12.748669, 13.396796 &$0.83_{-0.17}^{+0.18}$&$1.05_{-0.01}^{+0.01}$&$2.70_{-0.09}^{+0.09}$& $19.90_{-6.33}^{+6.79}$ &$442.12_{-17.33}^{+18.50}$ &25.81 &$977.60/1992.16 $& 0.03\\
21&13.396796, 13.542326 &unconstrained&$0.92_{-0.02}^{+0.02}$&$2.56_{-0.12}^{+0.12}$& $6.55_{-1.28}^{+1.19}$ &$317.46_{-18.10}^{+18.37}$ &32.84 &$ -101.63/-166.30$& 0.06\\
22&13.542326, 13.836101  &$0.97_{-0.04}^{+0.04}$&$1.30_{-0.04}^{+0.04}$&$3.19_{-0.29}^{+0.30}$& $88.41_{-16.93}^{+17.60}$ &$821.11_{-75.92}^{+78.51}$ &28.17 &$ 381.38/799.71 $& 0.50\\
23&13.836101, 14.219772  &$0.77_{-0.02}^{+0.02}$&$1.38_{-0.04}^{+0.04}$&$3.16_{-0.17}^{+0.18}$& $129.86_{-7.05}^{+6.81}$ &$662.85_{-39.75}^{+39.29}$ &37.68 &$ 626.86/1290.67 $& 0.34 \\
24&14.219772, 14.658330 &$0.88_{-0.02}^{+0.02}$&$1.56_{-0.03}^{+0.03}$&$3.34_{-0.35}^{+0.35}$& $99.75_{-6.54}^{+6.87}$ &$878.35_{-94.82}^{+89.30}$ &28.31 &$688.86/1414.67 $& 0.29\\
25&14.658330, 15.192231  &$0.83_{-0.02}^{+0.02}$&$1.20_{-0.02}^{+0.02}$&$3.29_{-0.14}^{+0.14}$& $85.71_{-8.28}^{+8.65}$ &$1242.21_{-55.87}^{+52.35}$ & 73.55 &$\bf 944.50/1925.95 $& 0.03\\
26& 15.192231, 15.338963 &$0.19_{-0.06}^{+0.06}$&$0.71_{-0.01}^{+0.01}$&$3.17_{-0.10}^{+0.10}$& $7.17_{-1.74}^{+1.75}$ &$1021.15_{-29.69}^{+30.46}$ & 203.54&$183.75 /404.45 $& 0.01 \\
27& 15.338963, 15.687352 &$0.70_{-0.02}^{+0.02}$&$1.34_{-0.05}^{+0.05}$&$3.26_{-0.13}^{+0.13}$& $204.82_{-14.66}^{+14.01}$ &$838.77_{-39.94}^{+40.87}$ &80.39 &$671.73/1380.42 $& 0.28\\
28& 15.687352, 16.251414 &$0.99_{-0.03}^{+0.03}$&$1.65_{-0.03}^{+0.03}$&$4.32_{-0.73}^{+0.76}$& $74.92_{-4.73}^{+4.59}$ &$952.13_{-5.48}^{+5.40}$ &19.15 &$\bf 822.72/1682.39 $& 0.34\\
29&16.251414, 16.407026 &$0.22_{-0.20}^{+0.32}$&$0.89_{-0.02}^{+0.02}$&$2.95_{-0.12}^{+0.12}$& $18.36_{-2.64}^{+3.46}$ &$702.41_{-30.15}^{+31.25}$ & 78.26&$ 65.37/167.70 $& 0.84 \\
30& 16.407026, 16.524165 &$0.66_{-0.03}^{+0.03}$&$1.27_{-0.05}^{+0.05}$&$3.96_{-0.27}^{+0.28}$& $189.76_{-20.48}^{+19.34}$ &$1235.85_{-73.37}^{+71.17}$ & 126.94&$ -63.64/-90.33 $&  0.63 \\
31& 16.524165, 16.705623 &$0.69_{-0.03}^{+0.03}$&$0.97_{-0.04}^{+0.04}$&$3.26_{-0.13}^{+0.13}$& $148.00_{-31.18}^{+31.97}$ &$892.21_{-38.97}^{+37.92}$ & 116.25&$ 228.55/494.06 $& 0.66 \\
32&16.705623, 16.787528   &$0.60_{-0.03}^{+0.03}$&$1.04_{-0.07}^{+0.07}$&$3.42_{-0.19}^{+0.18}$& $201.81_{-37.64}^{+37.06}$ &$1105.90_{-65.19}^{+61.52}$ & 168.73&$-243.39 /-449.83 $& 0.01 \\
33&16.787528, 17.114033  &$0.64_{-0.02}^{+0.02}$&$1.20_{-0.04}^{+0.04}$&$3.16_{-0.11}^{+0.12}$& $137.79_{-10.87}^{+11.02}$ &$791.75_{-33.91}^{+34.40}$ & 82.75 &$633.13 /1303.21 $& 0.19\\
34& 17.114033, 17.332842  &$0.69_{-0.11}^{+0.11}$&$0.98_{-0.02}^{+0.02}$&$3.74_{-0.24}^{+0.23}$& $42.63_{-12.47}^{+12.16}$ &$794.57_{-32.43}^{+31.65}$ & 58.16&$296.58 /630.11 $& 0.01\\
35& 17.332842, 17.924038 &$0.77_{-0.01}^{+0.01}$&$1.14_{-0.03}^{+0.03}$&$3.43_{-0.11}^{+0.10}$& $157.94_{-15.52}^{+15.05}$ &$943.25_{-29.66}^{+28.16}$ & 82.62&$ 1056.42/2149.80 $& 0.09\\
36&17.924038, 18.441228  &$0.75_{-0.01}^{+0.02}$&$1.30_{-0.03}^{+0.03}$&$3.27_{-0.13}^{+0.12}$& $155.01_{-10.38}^{+10.62}$ &$885.85_{-37.61}^{+36.76}$ &73.37 &$ 934.56/1906.07 $& 0.23 \\
37&18.441228, 18.542279  &$0.63_{-0.06}^{+0.06}$&$1.53_{-0.04}^{+0.04}$&$4.50_{-1.32}^{+1.29}$& $85.41_{-7.61}^{+7.68}$ &$1160.60_{-173.62}^{+167.14}$ & 39.28&$-318.51/-600.06 $& 0.33\\
38&  18.542279, 18.880490&$0.71_{-0.02}^{+0.01}$&$1.62_{-0.05}^{+0.05}$&$3.85_{-0.28}^{+0.28}$& $199.46_{-9.29}^{+9.43}$ &$1039.63_{-65.84}^{+65.72}$ &68.04 &$622.30 / 1281.55$& 0.67 \\
39&18.880490, 19.209439  &$0.79_{-0.03}^{+0.03}$&$1.50_{-0.03}^{+0.03}$&$3.14_{-0.17}^{+0.17}$& $85.44_{-5.37}^{+5.04}$ &$687.54_{-18.57}^{+18.94}$ & 36.01&$492.69/1022.33 $& 0.91\\
40& 19.209439, 19.316852 &$0.70_{-0.15}^{+0.16}$&$1.26_{-0.10}^{+0.10}$&$2.25_{-0.11}^{+0.11}$& $36.06_{-7.83}^{+8.49}$ &$158.09_{-21.90}^{+21.12}$ &16.70 &$ -419.90/-802.85 $& 0.53 \\
41& 19.316852, 19.790851&$1.04_{-0.08}^{+0.07}$&$1.61_{-0.03}^{+0.03}$&$3.01_{-1.14}^{+1.10}$& $34.35_{-3.75}^{+3.95}$ &$803.15_{-253.48}^{+253.73}$ & 9.72&$\bf  577.01/1190.97 $& 0.44
\\ \hline
\end{longtable}
\end{center}
\end{tiny}
\end{table}
\end{landscape}

\section{\thisgrbB: FIGURES and TABLES}

\begin{figure}[ht!]
\centering
\includegraphics[scale=0.65]{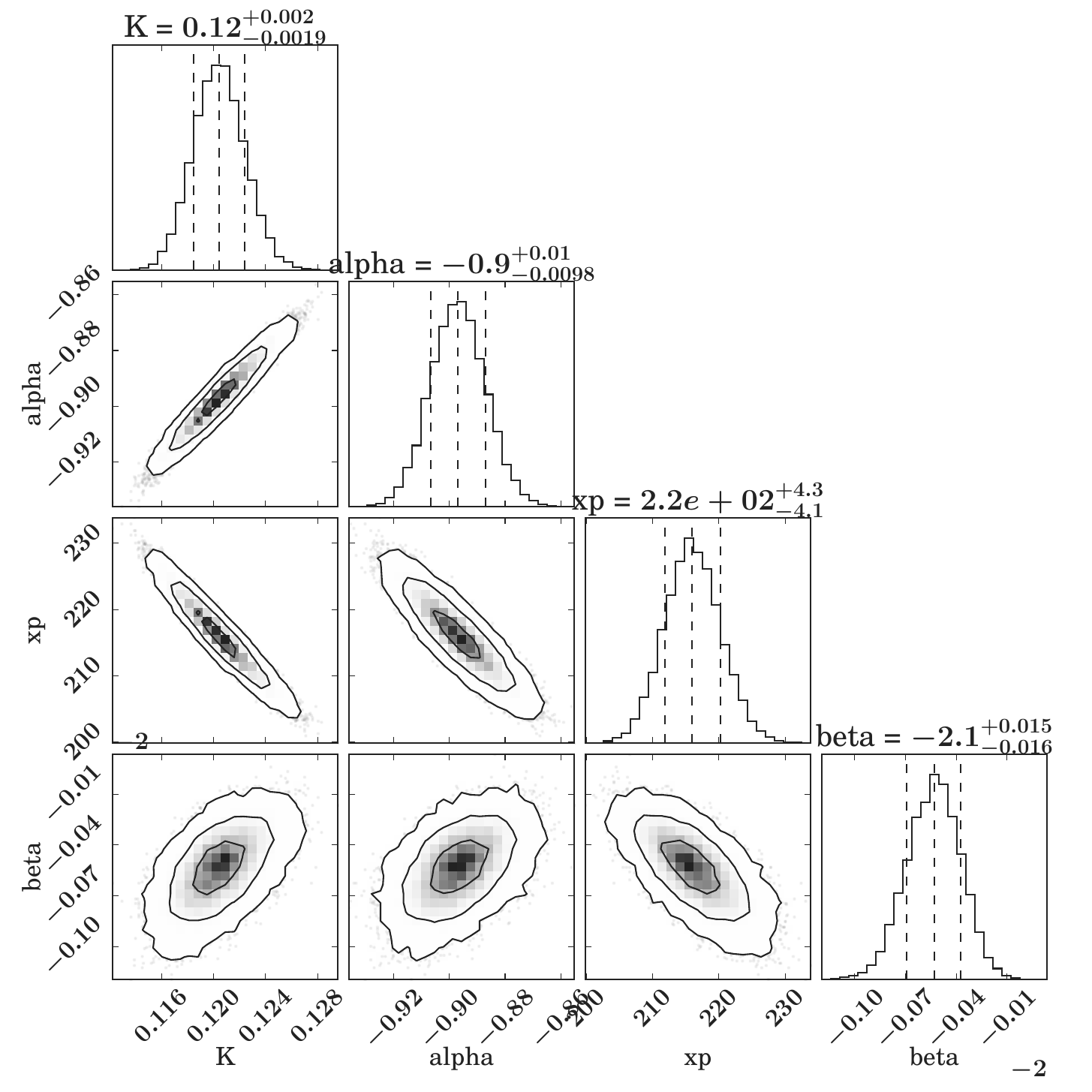}
\caption{The corner plot for the best fit time-averaged (\fermiT to \fermiT+ 67.38\,s) spectrum with 5000 number of simulations.}
\label{TAS_corner_GRB210619B}
\end{figure}



\begin{landscape}
\begin{tiny}
\begin{center}
\begin{longtable}{|c|c|c|ccccc|cccc|c|} 
\caption{Results of the time-resolved prompt emission spectral analysis of \thisgrbB using \sw{Band} and \sw{CPL} functions and \fermi GBM data. The energy flux values (in erg $\rm cm^{-2}$ $\rm s^{-1}$) are calculated in 8 \keV-40\,MeV energy range. The notation S denotes the significance of the spectrum of each time bin.}
\label{TRS_Table_Bayesian_GRB210619B} \\ \hline
T$_{\rm start}$ (s) & T$_{\rm stop}$ (s) & S & \boldmath $\it \alpha_{\rm pt}$ & \boldmath $\it \beta_{\rm pt}$ & \boldmath \Ep (\keV) &  \bf (Flux $\times 10^{-06}$)  & \bf DIC$_{\rm Band}$ & \boldmath $\it \Gamma_{\rm CPL}$ & \boldmath $E_{\rm c}$ (\keV) &  \bf (Flux $\times 10^{-06}$)  & \bf DIC$_{\rm CPL}$ & \rm \bf $\Delta$ DIC\\ \hline
0.068 & 0.213 & 39.61 & $-0.17_{-0.10}^{0.09}$ & $-1.86_{-0.06}^{+0.06}$ & $467.71_{-49.46}^{+50.36}$ & 77.32 & 100.11 & $-0.45_{-0.05}^{+0.05}$ & $530.00_{-49.55}^{+49.80}$ & 32.6 & 151.29 & -51.18 \\
0.213 & 0.302 & 41.06 & $-0.24_{-0.09}^{+0.09}$ & $-1.83_{-0.06}^{+0.06}$ & $446.77_{-50.90}^{+51.20}$ & 107.36 & -453.79 & $-0.55_{-0.04}^{+0.04}$ & $603.21_{-56.25}^{+56.72}$ & 46.82 & -403.6 & -50.19 \\
0.302 & 0.397 & 55.82 & $-0.36_{-0.07}^{+0.06}$ & $-2.01_{-0.07}^{+0.07}$ & $496.58_{-43.73}^{+44.12}$ & 114.85 & -290.18 & $-0.57_{-0.03}^{+0.03}$ & $554.48_{-42.29}^{+42.52}$ & 60.7 & -230.91 & -59.27 \\
0.397 & 0.502 & 73.81 & $-0.35_{-0.06}^{+0.06}$ & $-2.20_{-0.09}^{+0.09}$ & $493.20_{-39.94}^{+40.49}$ & 115.28 & -67.78 & $-0.58_{-0.03}^{+0.03}$ & $544.08_{-38.24}^{+37.99}$ & 78.1 & -33.02 & -34.76 \\
0.502 & 1.008 & 213.07 & $-0.35_{-0.02}^{+0.02}$ & $-2.10_{-0.03}^{+0.03}$ & $434.48_{-14.15}^{+14.07}$ & 178.35 & 2468.1 & $-0.61_{-0.01}^{+0.01}$ & $524.01_{-16.24}^{+16.30}$ & 110.24 & 3082.47 & -614.37 \\
1.008 & 1.172 & 110.46 & $-0.40_{-0.05}^{+0.05}$ & $-2.19_{-0.06}^{+0.06}$ & $367.26_{-22.66}^{+23.03}$ & 115.13 & 649.43 & $-0.61_{-0.03}^{+0.03}$ & $387.09_{-20.13}^{+20.27}$ & 72.5 & 729.02 & -79.59 \\
1.172 & 1.921 & 250.46 & $-0.47_{-0.02}^{+0.02}$ & $-2.30_{-0.03}^{+0.03}$ & $364.12_{-9.40}^{+9.48}$ & 106.9 & 2915.46 & $-0.64_{-0.01}^{+0.01}$ & $367.53_{-8.83}^{+8.93}$ & 73.2 & 3248.94 & -333.48 \\
1.921 & 2.078 & 104.55 & $-0.50_{-0.05}^{+0.05}$ & $-2.32_{-0.08}^{+0.08}$ & $286.95_{-16.52}^{+16.71}$ & 69.7 & 419.03 & $-0.68_{-0.03}^{+0.03}$ & $297.70_{-17.37}^{+17.22}$ & 47.5 & 476.94 & -57.91 \\
2.078 & 2.463 & 143.64 & $-0.36_{-0.04}^{+0.04}$ & $-2.15_{-0.04}^{+0.04}$ & $221.87_{-10.09}^{+10.13}$ & 64.49 & 1722.38 & $-0.66_{-0.02}^{+0.02}$ & $260.37_{-10.09}^{+10.33}$ & 37.8 & 1875.05 & -152.67 \\
2.463 & 2.849 & 157.93 & $-0.51_{-0.03}^{+0.03}$ & $-2.25_{-0.05}^{+0.05}$ & $295.65_{-12.54}^{+12.67}$ & 73.82 & 1813.37 & $-0.69_{-0.02}^{+0.02}$ & $314.59_{-12.51}^{+12.76}$ & 48.2 & 1949.14 & -135.77 \\
2.849 & 3.159 & 119.25 & $-0.57_{-0.05}^{+0.05}$ & $-2.26_{-0.07}^{+0.07}$ & $259.66_{-15.65}^{+16.17}$ & 49.37 & 1300.49 & $-0.75_{-0.02}^{+0.03}$ & $293.17_{-15.22}^{+15.12}$ & 32.5 & 1355.46 & -54.97 \\
3.159 & 3.556 & 114.24 & $-0.55_{-0.05}^{+0.05}$ & $-2.19_{-0.06}^{+0.06}$ & $190.54_{-11.22}^{+11.27}$ & 33.53 & 1573.62 & $-0.80_{-0.03}^{+0.03}$ & $238.38_{-13.52}^{+13.62}$ & 20.0 & 1651.65 & -78.03 \\
3.556 & 3.905 & 93.9 & $-0.60_{-0.06}^{+0.06}$ & $-2.28_{-0.08}^{+0.08}$ & $182.74_{-12.13}^{+11.97}$ & 23.57 & 1246.25 & $-0.81_{-0.03}^{+0.03}$ & $217.83_{-14.53}^{+14.30}$ & 15.3 & 1285.31 & -39.06 \\
3.905 & 4.386 & 89.92 & $-0.55_{-0.08}^{+0.08}$ & $-2.12_{-0.06}^{+0.06}$ & $156.45_{-13.59}^{+13.47}$ & 21.25 & 1685.38 & $-0.86_{-0.03}^{+0.04}$ & $224.93_{-16.10}^{+16.59}$ & 11.8 & 1736.74 & -51.36 \\
4.386 & 4.878 & 72.66 & $-0.70_{-0.08}^{+0.08}$ & $-2.15_{-0.08}^{+0.08}$ & $157.93_{-15.92}^{+15.93}$ & 14.04 & 1650.0 & $-0.97_{-0.04}^{+0.04}$ & $244.94_{-23.72}^{+23.45}$ & 8.05 & 1679.4 & -29.4 \\
4.878 & 5.855 & 78.09 & $-0.75_{-0.07}^{+0.07}$ & $-2.41_{-0.14}^{+0.14}$ & $164.70_{-13.68}^{+13.73}$ & 7.24 & 2533.13 & $-0.88_{-0.04}^{+0.04}$ & $183.12_{-14.66}^{+14.69}$ & 5.06 & 2547.54 & -14.41 \\
5.855 & 6.519 & 53.16 & $-0.60_{-0.10}^{+0.10}$ & $-2.28_{-0.12}^{+0.13}$ & $135.87_{-13.49}^{+13.28}$ & 6.06 & 1836.63 & $-0.85_{-0.06}^{+0.06}$ & $163.89_{-18.66}^{+18.28}$ & 3.75 & 1852.18 & -15.55 \\
6.519 & 8.47 & 73.61 & $-0.84_{-0.05}^{+0.05}$ & $-2.47_{-0.14}^{+0.15}$ & $180.81_{-11.83}^{+11.74}$ & 4.39 & 3434.79 & $-0.92_{-0.04}^{+0.04}$ & $198.40_{-17.00}^{+16.87}$ & 3.19 & 3448.67 & -13.88 \\
8.47 & 9.594 & 67.88 & $-0.61_{-0.06}^{+0.06}$ & $-2.13_{-0.08}^{+0.08}$ & $207.45_{-15.92}^{+15.68}$ & 10.09 & 2693.48 & $-0.80_{-0.04}^{+0.04}$ & $247.71_{-21.32}^{+21.19}$ & 5.42 & 2725.47 & -31.99 \\
9.594 & 10.083 & 35.24 & $-0.81_{-0.14}^{+0.14}$ & $-2.07_{-0.21}^{+0.20}$ & $208.97_{-49.91}^{+51.72}$ & 7.61 & 1424.88 & $-0.96_{-0.06}^{+0.06}$ & $291.76_{-46.09}^{+45.79}$ & 3.61 & 1437.42 & -12.54 \\
10.083 & 11.346 & 37.26 & $-1.00_{-0.09}^{+0.09}$ & $-2.25_{-0.26}^{+0.25}$ & $206.89_{-37.22}^{+37.08}$ & 3.36 & 2684.86 & $-1.08_{-0.06}^{+0.06}$ & $277.93_{-44.76}^{+44.64}$ & 1.99 & 2691.42 & -6.56 \\
13.652 & 14.807 & 47.61 & $-0.83_{-0.07}^{+0.07}$ & $-2.18_{-0.14}^{+0.14}$ & $225.54_{-26.34}^{+25.82}$ & 5.68 & 2628.19 & $-0.96_{-0.05}^{+0.05}$ & $292.08_{-38.63}^{+38.30}$ & 3.3 & 2641.13 & -12.94 \\
14.807 & 15.716 & 66.97 & $-0.86_{-0.05}^{+0.05}$ & $-2.31_{-0.14}^{+0.15}$ & $238.77_{-21.99}^{+22.26}$ & 8.62 & 2384.08 & $-0.95_{-0.04}^{+0.04}$ & $287.80_{-27.58}^{+27.73}$ & 5.71 & 2395.35 & -11.27 \\
15.716 & 16.624 & 50.21 & $-0.94_{-0.07}^{+0.07}$ & $-2.55_{-0.23}^{+0.23}$ & $158.62_{-15.74}^{+15.61}$ & 3.78 & 2314.14 & $-1.02_{-0.06}^{+0.06}$ & $187.58_{-24.39}^{+24.90}$ & 2.87 & 2319.26 & -5.12 \\
16.624 & 19.933 & 62.17 & $-0.93_{-0.09}^{+0.09}$ & $-2.26_{-0.13}^{+0.13}$ & $115.98_{-13.21}^{+13.11}$ & 2.54 & 4089.06 & $-1.09_{-0.05}^{+0.05}$ & $169.93_{-18.91}^{+19.06}$ & 1.54 & 4107.09 & -18.03 \\
19.933 & 20.414 & 30.45 & $-0.81_{-0.16}^{+0.16}$ & $-2.54_{-0.28}^{+0.27}$ & $120.73_{-18.61}^{+18.56}$ & 2.65 & 1302.6 & $-0.95_{-0.11}^{+0.11}$ & $139.61_{-27.72}^{+27.80}$ & 1.97 & 1306.19 & -3.59 \\
20.414 & 22.098 & 83.25 & $-0.78_{-0.06}^{+0.06}$ & $-2.40_{-0.10}^{+0.10}$ & $130.48_{-7.73}^{+7.68}$ & 4.97 & 3285.78 & $-0.93_{-0.04}^{+0.04}$ & $151.82_{-11.91}^{+11.72}$ & 3.46 & 3315.72 & -29.94 \\
22.098 & 22.991 & 44.62 & $-0.88_{-0.10}^{+0.10}$ & $-2.73_{-0.21}^{+0.22}$ & $98.14_{-7.82}^{+7.86}$ & 2.24 & 2266.88 & $-0.99_{-0.08}^{+0.08}$ & $109.19_{-15.68}^{+15.81}$ & 1.8 & 2270.88 & -4.0 \\
31.821 & 34.636 & 31.63 & $-1.05_{-0.12}^{+0.12}$ & $-2.74_{-0.29}^{+0.29}$ & $102.95_{-12.24}^{+12.10}$ & 0.83 & 3794.2 & $-1.13_{-0.09}^{+0.09}$ & $133.40_{-25.91}^{+25.80}$ & 0.68 & 3793.86 & 0.34 \\
35.169 & 35.633 & 38.92 & $-0.94_{-0.06}^{+0.06}$ & $-2.54_{-0.30}^{+0.29}$ & $289.46_{-32.92}^{+32.93}$ & 6.06 & 1492.65 & $-0.95_{-0.06}^{+0.06}$ & $298.59_{-42.07}^{+41.31}$ & 4.49 & 1494.31 & -1.66 \\
35.633 & 36.354 & 38.32 & $-0.86_{-0.08}^{+0.08}$ & $-2.86_{-0.31}^{+0.31}$ & $180.31_{-18.21}^{+17.92}$ & 2.88 & 1972.32 & $-0.90_{-0.07}^{+0.07}$ & $175.68_{-24.39}^{+24.20}$ & 2.48 & 1970.36 & 1.96 \\
36.354 & 36.638 & 31.9 & $-0.68_{-0.14}^{+0.14}$ & $-2.43_{-0.23}^{+0.24}$ & $142.74_{-19.41}^{+19.11}$ & 4.62 & 683.44 & $-0.83_{-0.10}^{+0.10}$ & $150.54_{-27.45}^{+27.34}$ & 3.06 & 687.04 & -3.6 \\
36.638 & 36.961 & 52.31 & $-0.56_{-0.10}^{+0.10}$ & $-2.25_{-0.13}^{+0.13}$ & $178.78_{-19.76}^{+19.23}$ & 11.33 & 1081.59 & $-0.79_{-0.06}^{+0.06}$ & $210.28_{-25.39}^{+25.90}$ & 6.96 & 1097.6 & -16.01 \\
36.961 & 37.44 & 47.35 & $-0.92_{-0.10}^{+0.10}$ & $-2.51_{-0.24}^{+0.24}$ & $169.29_{-24.24}^{+23.96}$ & 5.38 & 1552.48 & $-1.03_{-0.06}^{+0.06}$ & $212.09_{-30.73}^{+30.22}$ & 4.03 & 1559.1 & -6.62 \\
37.44 & 38.25 & 75.05 & $-0.74_{-0.06}^{+0.06}$ & $-2.70_{-0.18}^{+0.18}$ & $141.42_{-8.28}^{+8.41}$ & 5.62 & 2240.16 & $-0.82_{-0.05}^{+0.05}$ & $134.20_{-10.28}^{+10.14}$ & 4.52 & 2247.91 & -7.75 \\
38.25 & 39.556 & 66.59 & $-0.76_{-0.09}^{+0.09}$ & $-2.46_{-0.14}^{+0.14}$ & $100.32_{-7.94}^{+8.08}$ & 3.53 & 2905.42 & $-0.96_{-0.05}^{+0.05}$ & $121.76_{-11.60}^{+11.72}$ & 2.55 & 2920.88 & -15.46 \\
39.556 & 41.849 & 64.94 & $-1.00_{-0.07}^{+0.07}$ & $-2.31_{-0.12}^{+0.12}$ & $123.41_{-11.14}^{+11.25}$ & 3.05 & 3676.57 & $-1.14_{-0.05}^{+0.05}$ & $187.30_{-21.69}^{+21.62}$ & 1.98 & 3693.33 & -16.76 \\
41.849 & 44.578 & 54.62 & $-1.04_{-0.17}^{+0.18}$ & $-2.41_{-0.27}^{+0.26}$ & $92.25_{-20.53}^{+19.47}$ & 1.85 & 3815.0 & $-1.25_{-0.05}^{+0.05}$ & $168.82_{-23.01}^{+23.58}$ & 1.28 & 3873.71 & -58.71 \\
44.578 & 45.205 & 34.78 & $-1.06_{-0.11}^{+0.11}$ & $-2.84_{-0.28}^{+0.28}$ & $98.25_{-9.77}^{+9.71}$ & 1.96 & 1733.51 & $-1.11_{-0.10}^{+0.10}$ & $120.29_{-22.03}^{+21.62}$ & 1.66 & 1732.15 & 1.36 \\
45.869 & 46.996 & 56.31 & $-0.91_{-0.13}^{+0.13}$ & $-2.57_{-0.22}^{+0.21}$ & $84.94_{-10.13}^{+10.06}$ & 2.58 & 2607.69 & $-1.10_{-0.07}^{+0.07}$ & $115.68_{-14.22}^{+14.13}$ & 1.94 & 2619.28 & -11.59 \\
46.996 & 47.344 & 51.11 & $-0.78_{-0.08}^{+0.08}$ & $-3.19_{-0.28}^{+0.28}$ & $137.16_{-8.78}^{+8.71}$ & 4.83 & 1027.46 & $-0.81_{-0.07}^{+0.07}$ & $121.35_{-13.39}^{+13.26}$ & 4.4 & 1022.74 & 4.72 \\
47.6 & 47.974 & 76.9 & $-0.59_{-0.05}^{+0.06}$ & $-2.82_{-0.21}^{+0.22}$ & $193.50_{-10.68}^{+10.54}$ & 11.73 & 1421.85 & $-0.65_{-0.04}^{+0.04}$ & $158.08_{-10.98}^{+10.83}$ & 9.77 & 1424.93 & -3.08 \\
47.974 & 48.695 & 72.5 & $-0.66_{-0.08}^{+0.08}$ & $-2.44_{-0.11}^{+0.12}$ & $115.31_{-8.21}^{+8.12}$ & 6.4 & 2157.85 & $-0.88_{-0.05}^{+0.05}$ & $132.80_{-11.48}^{+11.45}$ & 4.58 & 2177.07 & -19.22 \\
48.695 & 49.243 & 41.06 & $-0.92_{-0.14}^{+0.14}$ & $-2.43_{-0.19}^{+0.19}$ & $90.52_{-10.98}^{+10.76}$ & 3.04 & 1640.68 & $-1.13_{-0.10}^{+0.10}$ & $134.34_{-26.72}^{+26.46}$ & 2.15 & 1651.54 & -10.86 \\
49.243 & 50.071 & 31.76 & $-1.17_{-0.14}^{+0.13}$ & $-2.53_{-0.30}^{+0.29}$ & $111.41_{-20.52}^{+20.29}$ & 1.88 & 2164.14 & $-1.26_{-0.09}^{+0.09}$ & $178.14_{-40.38}^{+40.32}$ & 1.37 & 2169.89 & -5.75 \\
50.431 & 50.517 & 33.99 & $-0.37_{-0.14}^{+0.14}$ & $-2.21_{-0.19}^{+0.19}$ & $252.46_{-40.36}^{+41.02}$ & 24.98 & -670.88 & $-0.62_{-0.08}^{+0.08}$ & $269.88_{-38.82}^{+39.72}$ & 14.6 & -663.99 & -6.89 \\
50.517 & 51.143 & 111.14 & $-0.53_{-0.04}^{+0.04}$ & $-2.65_{-0.15}^{+0.15}$ & $256.88_{-11.44}^{+11.30}$ & 19.45 & 2254.79 & $-0.59_{-0.03}^{+0.03}$ & $203.12_{-9.28}^{+9.53}$ & 15.1 & 2272.57 & -17.78 \\
51.143 & 51.339 & 48.03 & $-0.59_{-0.13}^{+0.13}$ & $-2.12_{-0.11}^{+0.11}$ & $134.42_{-17.31}^{+17.36}$ & 14.33 & 440.09 & $-0.87_{-0.07}^{+0.07}$ & $180.85_{-26.03}^{+25.98}$ & 7.31 & 465.91 & -25.82 \\
51.339 & 51.804 & 49.62 & $-0.75_{-0.10}^{+0.10}$ & $-2.55_{-0.21}^{+0.21}$ & $132.36_{-12.83}^{+12.90}$ & 5.35 & 1493.78 & $-0.88_{-0.07}^{+0.07}$ & $140.52_{-16.10}^{+16.14}$ & 4.02 & 1500.38 & -6.6 \\
51.804 & 52.376 & 68.71 & $-0.66_{-0.08}^{+0.08}$ & $-2.63_{-0.15}^{+0.16}$ & $116.88_{-7.24}^{+7.27}$ & 5.75 & 1853.42 & $-0.79_{-0.06}^{+0.06}$ & $112.55_{-9.99}^{+9.95}$ & 4.48 & 1867.32 & -13.9 \\
52.376 & 52.651 & 31.12 & $-0.99_{-0.17}^{+0.16}$ & $-2.66_{-0.28}^{+0.27}$ & $87.95_{-12.01}^{+11.95}$ & 2.8 & 630.89 & $-1.14_{-0.12}^{+0.12}$ & $122.50_{-27.16}^{+26.87}$ & 2.23 & 633.62 & -2.73 \\
52.651 & 53.803 & 42.75 & $-0.62_{-0.20}^{+0.21}$ & $-2.38_{-0.13}^{+0.13}$ & $62.61_{-7.19}^{+7.18}$ & 2.01 & 2655.99 & $-1.04_{-0.09}^{+0.09}$ & $92.16_{-13.91}^{+14.01}$ & 1.36 & 2694.36 & -38.37 \\
53.803 & 54.719 & 56.65 & $-0.64_{-0.09}^{+0.09}$ & $-2.83_{-0.18}^{+0.18}$ & $93.91_{-4.88}^{+4.75}$ & 2.77 & 2387.05 & $-0.74_{-0.07}^{+0.07}$ & $82.11_{-7.98}^{+7.84}$ & 2.34 & 2394.05 & -7.0 \\
54.719 & 56.207 & 55.52 & $-0.83_{-0.09}^{+0.09}$ & $-2.94_{-0.23}^{+0.23}$ & $77.76_{-4.39}^{+4.40}$ & 1.74 & 3054.06 & $-0.91_{-0.07}^{+0.07}$ & $77.23_{-7.70}^{+7.80}$ & 1.5 & 3055.92 & -1.86 \\
58.382 & 59.675 & 64.29 & $-0.60_{-0.14}^{+0.14}$ & $-2.64_{-0.14}^{+0.15}$ & $63.29_{-4.54}^{+4.54}$ & 2.34 & 2832.41 & $-0.90_{-0.07}^{+0.07}$ & $69.32_{-6.64}^{+6.58}$ & 1.82 & 2859.66 & -27.25 \\
59.675 & 60.966 & 45.17 & $-0.64_{-0.23}^{+0.23}$ & $-2.56_{-0.14}^{+0.14}$ & $45.87_{-4.27}^{+4.25}$ & 1.41 & 2723.52 & $-1.11_{-0.11}^{+0.11}$ & $66.40_{-10.29}^{+10.45}$ & 1.08 & 2789.89 & -66.37 \\
60.966 & 61.509 & 38.74 & $-0.54_{-0.16}^{+0.17}$ & $-3.10_{-0.24}^{+0.23}$ & $55.32_{-3.21}^{+3.23}$ & 1.66 & 1541.7 & $-0.72_{-0.13}^{+0.13}$ & $47.45_{-6.29}^{+6.22}$ & 1.47 & 1538.24 & 3.46 \\
61.509 & 62.273 & 35.2 & $-0.75_{-0.26}^{+0.27}$ & $-2.62_{-0.16}^{+0.16}$ & $41.18_{-3.87}^{+3.98}$ & 1.31 & 1948.82 & $-1.21_{-0.14}^{+0.14}$ & $66.40_{-13.28}^{+13.36}$ & 1.01 & 2008.03 & -59.21 \\
62.273 & 63.528 & 30.68 & $-0.46_{-0.28}^{+0.28}$ & $-2.54_{-0.13}^{+0.13}$ & $37.16_{-3.34}^{+3.30}$ & 0.88 & 2601.14 & $-1.24_{-0.14}^{+0.14}$ & $69.83_{-15.51}^{+15.71}$ & 0.68 & 2667.76 & -66.62 \\ \hline
\end{longtable}

\bigskip

\begin{longtable}{|c|c|ccccc|ccc|}
\caption{Joint \swift BAT, \fermi GBM, and ASIM spectral analysis results of the bright emission pulse of \thisgrbB using \sw{Band}, \sw{CPL} and \sw{Band+pow}, \sw{CPL+pow} functions. Flux values (in erg $\rm cm^{-2}$ $\rm s^{-1}$) are calculated in 8 \keV-10\,MeV energy range. T$_{\rm start}$ and T$_{\rm stop}$ have been referred with respect to the ASIM reference time (T$_{\rm 0, ASIM}$). The errors given are $1{\sigma}$.}
\label{Joint_spectralanalysis_Table}\\
\hline
T$_{\rm start}$ (s) & T$_{\rm stop}$ (s)  & \boldmath $\it \alpha_{\rm pt}$ & \boldmath $\it \beta_{\rm pt}$ & \boldmath \Ep (\keV) &  \bf (Flux $\times 10^{-06}$)  & \bf $\chi^2_{\nu}$  (d.o.f.) & \bf {\tt Band+pow} & \bf {\tt cutoffpl}  & \bf {\tt cutoffpl+pow}  \\ \hline
0.00& 1.00 & $-0.41{\pm}0.08$      & $-1.79_{-0.06}^{+0.04}$ & $590_{-70}^{+100}$ & $24.1_{-1.6}^{+1.7}$    &  1.09\,(352)  &  1.10\,(350) & 1.45\,(353)  & 1.56\,(351)    \\ \hline
1.00& 1.70 & $-0.36{\pm}0.02$      & $-2.12{\pm}0.02$      & $422{\pm}14$         & $121.0_{-2.6}^{+2.7}$   &  1.30\, (417) &  1.31 (415)  & 3.25 (418)   & 2.17 (416)             \\ \hline
1.70& 2.18 & $-0.52{\pm}0.02$      & $-2.37_{-0.04}^{+0.03}$ & $380_{-14}^{+15}$  & $87.6{\pm}2.3$          &  1.20\, (381) &  1.21 (379)  & 1.86 (382)   & 1.62 (380)              \\ \hline
2.18& 2.79 & $-0.51{\pm}0.02$      & $-2.40{\pm}0.04$      & $317{\pm}10$         & $69.0_{-1.8}^{+1.9}$    &  1.30\, (382) &  1.31 (380)  & 1.93 (383)   &  1.73 (381)           \\ \hline
2.79& 3.60 & $-0.49{\pm}0.02$      & $-2.28{\pm}0.03$      & $250_{-7}^{+8}$      & $51.2{\pm}1.5$    &    1.48 (389) &   1.49 (387) & 2.29 (390)   &  2.02 (388)           \\ \hline
3.60& 4.50 & $-0.61{\pm}0.03$      & $-2.37_{-0.06}^{+0.05}$ & $202{\pm}7$        & $23.9_{-0.9}^{+1.0}$    &    1.24 (362) &   1.25 (360) & 1.57 (363)   &  1.53 (361)           \\ \hline
\end{longtable}

\end{center}
\end{tiny}
\end{landscape}

\begin{table*}
        \centering
    \caption{{\it ASIM} time-resolved spectral analysis results for \thisgrbB . Time intervals have been referred with respect to the ASIM reference time (T$_{\rm 0, ASIM}$).}
    \label{tab:asim}
    \begin{scriptsize}
    \begin{tabular}{ccccc}
    \hline
    Interval      & Time interval & Photon Index &  ${\chi}^2$\,(d.o.f.) &  Flux ($0.5-10$\,MeV)  \\
                  & (s)           &              &                       &  ($10^{-5}\,{\rm erg}\,{\rm cm}^{-2}{\rm s}^{-1}$)                 \\
    \hline
1                 & $0.00-1.00$      & $2.11^{+0.12}_{-0.11}$       &  11.69\,(16)      &  $1.31^{+0.23}_{-0.19}$  \\
2                 & $1.00-1.70$      & $2.25^{+0.03}_{-0.03}$    &  15.92\,(12)       &  $6.89^{+0.27}_{-0.26}$    \\
3                 & $1.70-2.18$      & $2.49^{+0.06}_{-0.06}$    &  6.99\,(12)       &  $4.02^{+0.30}_{-0.28}$   \\
4                 & $2.18-2.79$      & $2.60^{+0.09}_{-0.08}$    &  7.43\,(12)       &  $2.56^{+0.56}_{-0.46}$  \\
5                 & $2.79-3.60$      & $2.56^{+0.11}_{-0.10}$       &  13.31\,(12)      &  $1.81^{+0.51}_{-0.40}$   \\
6                 & $3.60-4.50$      & $2.97^{+0.37}_{-0.30}$       &  9.31\,(12)       &  $0.69^{+0.12}_{-0.11}$    \\
    \hline
    \end{tabular}
    \end{scriptsize}
\end{table*}

\begin{table}[ht!]
\centering
\caption{The results of energy-resolved spectral lag analysis for \thisgrbB. We considered 15-25 \keV and 8-30 \keV energy channels as a reference light curve for the \swift BAT and \fermi GBM observations, respectively. The each lag values is calculated for 50000 numbers of simulations to fit the cross-correlation function.}
\begin{scriptsize}
\begin{tabular}{cc|cc}
\hline
\bf  &  \bf {\swift BAT} &  &  \bf {\fermi GBM} \\
\hline
\bf {Energy range \ (\keV)} &  \bf {Spectral lag (ms)} & \bf {Energy range \ (\keV)} &  \bf {Spectral lag (ms)} \\
\hline 
25-50  &  $-38.28_{-73.51}^{+73.46}$ & 30--50  & $-37.16_{-68.66}^{+68.89}$  \\ 
50-100 &  $-160.25_{-66.42}^{+66.53}$&  50--100  &  $-95.46_{-60.67}^{+61.31}$ \\ 
100-200 & $-239.73_{-46.06}^{+46.26}$ &   100--150 &  $-207.11_{-60.08}^{+60.83}$ \\ 
200-350   &$-386.78_{-36.90}^{+37.23}$ &  150--200 &  $-308.86_{-49.87}^{+50.02}$ \\ 
  &  &  200-250 & $-378.55_{-44.99}^{+44.84}$  \\
 &&  250--300  & $-405.58_{-43.91}^{+44.32}$  \\
 &&  300--400  & $-473.87_{-32.91}^{+32.98}$  \\
 &&  400--500  & $-545.70_{-31.87}^{+32.09}$  \\
 &&  500--600  & $-555.08_{-27.89}^{+27.65}$  \\
 &&  600--700  & $-579.31_{-27.19}^{+27.29}$  \\
 &&  700--800  &  $-511.07_{-26.68}^{+26.40}$ \\
 &&  800--900  & $-786.49_{-21.40}^{+21.37}$  \\
\hline
\end{tabular}
\label{tab:spectral lag_GRB210619B}
\end{scriptsize}
\end{table}

\chapter{\sc Early afterglow: physical mechanisms}

\ifpdf
    \graphicspath{{Appendix/appendix_BFigs/JPG/}{Appendix/appendix_BFigs/PDF/}{Appendix/appendix_BFigs/}}
\else
    \graphicspath{{Appendix/appendix_BFigs/EPS/}{Appendix/appendix_BFigs/}}
\fi

\normalsize

\section{\thisgrbE: FIGURES and TABLES}

\renewcommand{\thefigure}{B\arabic{figure}}
\setcounter{figure}{0}

\begin{figure}[ht!]
\centering
\includegraphics[height=19cm,width=15.5cm]{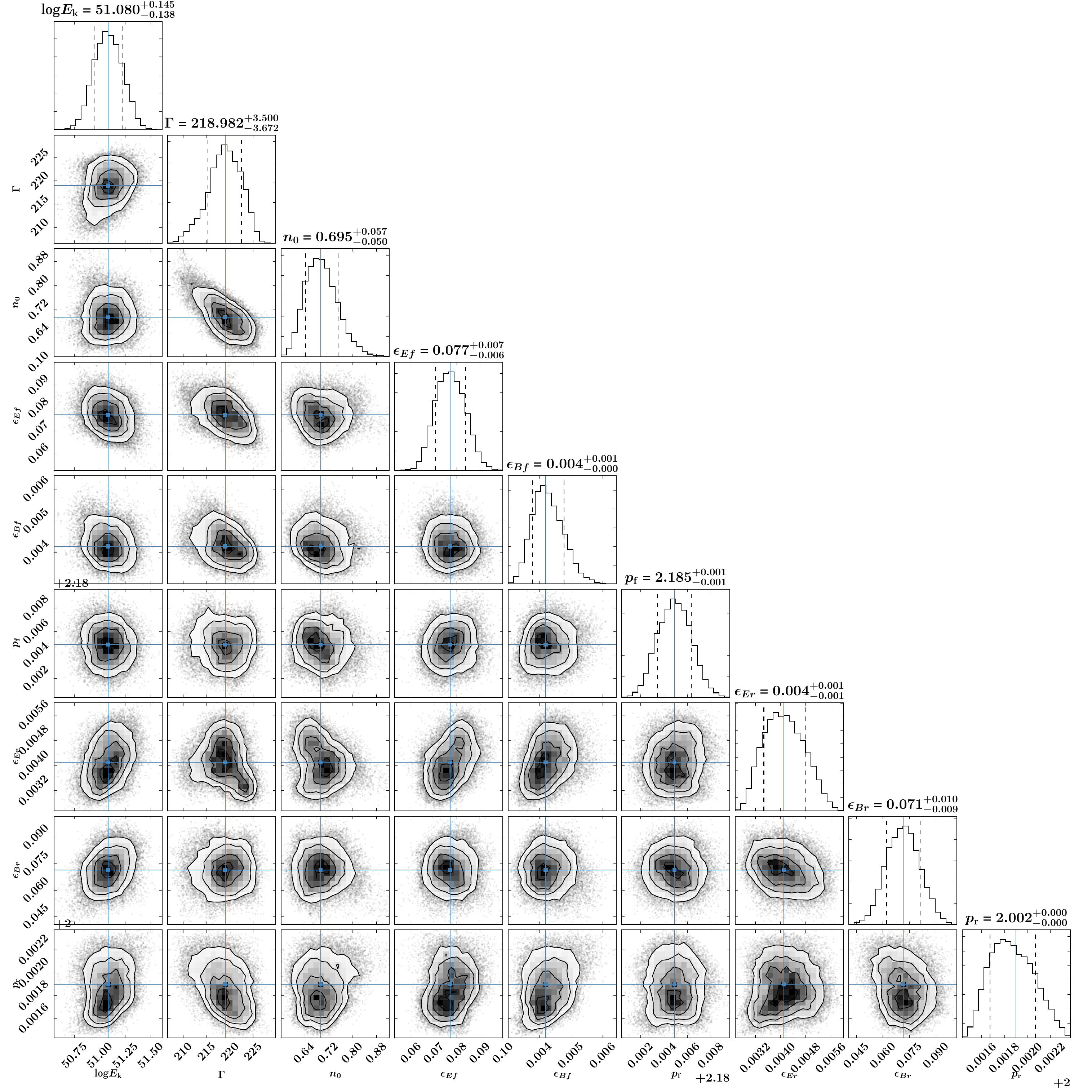}
\caption{The corner plot has been obtained from the \sw{PyMultiNest} simulation for each parameter using the thin shell RS and FS afterglow model. The best-fit parameter values in this figure are shown in blue.} 
\label{corner}
\end{figure}


\begin{landscape}
\begin{tiny}
\begin{center}
\begin{longtable}{|c|c|c|c|c|c|c|c|}
\caption{\fermi LAT high energy emission ($>100$ MeV) in different  temporal bins for a fit with a power-law model for \thisgrbE. All the photon and energy fluxes has been calculated in the 100.0 - 10000.0 MeV energy range.}
\label{tab:lat_sed_GRB140102A}\\
\hline
\bf Sr. no. & \bf Time   & \bf Index           & \bf Energy flux                        & \bf  Photon flux              & \bf Test Statistic & \bf Log(likelihood) \\
& \bf (sec)    &                &( $\rm  \bf 10^{-8} ~ergs$  $\rm \bf cm^{-2}$ $\rm \bf sec^{-1}$)   & ($\rm \bf \times 10^{-5} ~photons$   $\rm \bf cm^{-2}$ $\rm \bf sec^{-1}$) & \bf (TS)&\\ \hline
(0)& 0 - 5   & $-1.95 \pm 0.49$ & $ 13.4 \pm 9.85$& $  16.8 \pm 9.45$& 27& 23.98\\
(1)& 5 - 10 & $-2.47 \pm 0.61 $ & $ 9.99 \pm 6.39$ &    $22.4 \pm 10.2$  &33 & 17.96\\ 
(2)& 10 - 100 & $-2.14 \pm 0.50$ &$0.75 \pm 0.50 $&$1.20  \pm 0.60$ &23& 40.87\\
(3) &100 - 1000  &$-1.33 \pm 0.32 $ & $0.13 \pm 0.08$& $0.06 \pm 0.05$ &32&  208.10\\
(4) &2905 - 7460 &$-2$ (fixed)& $ <0.02$& $ < 0.03$& 1 & 847.28 \\
\hline 
\end{longtable}

\bigskip

\begin{longtable}{|c|c|c|c|c|c|c|c|c|}
\caption{The best fit model (shown with boldface) between different models used for the time-averaged joint spectral analysis of \fermi LAT standard, LAT- LLE, \fermi GBM (NaI +  BGO), and \swift BAT data of GRB 140102A. BB correspond to the \sw{Black Body} model.}
\label{tab:TAS}\\
\hline
\textbf{Time (sec)/ Detectors} & \textbf{Model} & \multicolumn{4}{c|}{\textbf{Parameters}} & \textbf{-Log(Likelihood)} & \textbf{AIC} & \textbf{BIC} \\ \hline
\multirow{6}{*}{\begin{tabular}[c]{@{}c@{}}0-5\\ \\ \\ (BAT+ GBM +LAT)\end{tabular}} & \sw{bkn2pow} & $\it \Gamma_{1,2,3}$= 0.69$^{+0.06}_{-0.04}$, 1.26$^{+0.01}_{-0.01}$, 2.47$^{+0.03}_{-0.03}$ & \multicolumn{3}{c|}{$E_{\rm b1,b2}$= 26.92$^{+1.18}_{-1.12}$, 146.31$^{+3.24}_{-3.94}$} & 2407.72 & 4844.22 & 4903.75 \\ \cline{2-9} 
 & \sw{SBPL} & $\alpha_{1}$=  -0.92$^{+0.02}_{-0.01}$ & \multicolumn{2}{c|}{$\alpha_{2}$= -2.85$^{+0.06}_{-0.03}$} & $E_{0}$= 163.28$^{+5.94}_{-8.15}$ & 2435.64 & 4895.86 & 4946.97 \\ \cline{2-9} 
 & Band & $\it \alpha_{\rm pt}$= -0.85$^{+0.02}_{-0.01}$ & \multicolumn{2}{c|}{$\it \beta_{\rm pt}$= -2.72$^{+0.03}_{-0.05}$} & \Ep = 190.21$^{+2.91}_{-5.55}$ & 2426.26 & 4877.11 & 4928.22 \\ \cline{2-9} 
 & \sw{bkn2pow}+BB & $\it \Gamma_{1,2,3}$= 0.73$^{+0.01}_{-0.01}$, 1.49$^{+0.03}_{-0.03}$, 2.60$^{+0.02}_{-0.06}$ & $E_{\it b1,b2}$= 26.39$^{+0.16}_{-0.15}$, 254.72$^{+15.42}_{-21.30}$ & \multicolumn{2}{c|}{$\rm k{\it T}_{\rm BB}$ = 32.91$^{+1.19}_{-1.18}$} & 2403.39 & 4839.80 & 4907.71 \\ \cline{2-9} 
 & \sw{SBPL}+BB & $\alpha_{1}$= -0.58$^{+0.10}_{-0.09}$ & $\alpha_{2}$= -2.93$^{+0.07}_{-0.04}$ & $E_{0}$= 143.99$^{+10.33}_{-12.83}$ & $\rm k{\it T}_{\rm BB}$= 7.48$^{+0.25}_{-0.38}$ & 2402.90 & 4834.59 & 4894.11 \\ \cline{2-9} 
 & \textbf{Band+BB} & $\alpha_{\rm pt}$= -0.58$^{+0.04}_{-0.04}$ & $\rm \beta_{\rm pt}$ = -2.72$^{+0.02}_{-0.07}$ & \Ep= 186.57$^{+4.51}_{-4.16}$ & $\rm k{\it T}_{\rm BB}$= 7.90$^{+0.25}_{-0.28}$ & 2398.90 & 4826.58 & 4886.10 \\ \cline{2-9}
 & Band + PL & $\alpha_{1}$= -0.85$^{+0.01}_{-0.01}$ & $\alpha_{2}$= -2.72$^{+0.13}_{-0.03}$ & \Ep= 190.14$^{+1.87}_{-4.27}$ & $\it \Gamma_{\rm PL}$= -1.59$^{+0.75}_{-0.38}$ & 2425.75 & 4880.28 & 4939.80  \\ \hline
\end{longtable}

\end{center}
\end{tiny}
\end{landscape}

\begin{table*}
\caption{Log of \swift UVOT observations $\&$ photometry of \thisgrbE afterglow. No correction for Galactic extinction is applied.}
\label{table:UVOT}
\begin{center}
    \begin{scriptsize}
\begin{tabular}{|c|c|c|c||c|c|c|c||}
\hline
\bf $\rm \bf  T_{\rm \bf  mid}$  &\bf Exp.& {\bf  Magnitude}  &\bf Filter  & \bf $\rm \bf T_{\rm \bf mid}$ &\bf Exp. & {\bf  Magnitude} & \bf Filter   \\
\bf  (sec) &   \bf  (sec)&  &  &\bf  (sec) &  \bf  (sec)&  &      \\
\hline \hline
67.82& 5.0 & $15.67^{+0.07}_{-0.07}$& $White$ & 49.90& 5.0& $13.97^{+0.11}_{-0.10}$ & $V$    \\
70.32& 10.0 & $15.70^{+0.05}_{-0.05}$& $White$ & 54.47& 4.1& $13.98^{+0.12}_{-0.11}$ & $V$   \\
72.82&5.0 & $15.74^{+0.07}_{-0.07}$ & $White$ & 617.37& 19.8& $17.21^{+0.36}_{-0.27}$ &$V$    \\
77.82& 5.0& $15.83^{+0.07}_{-0.07}$ & $White$ & 791.14& 19.8& $17.61^{+0.52}_{-0.35}$&$V$   \\
82.82&5.0 & $15.99^{+0.08}_{-0.07}$ & $White$ & 1132.96& 192.42&$ 17.80^{+0.41}_{-0.29}$& $V$   \\
82.82&15.0 & $15.95^{+0.04}_{-0.04}$& $White$ & 1479.24& 192.11& $>$18.58 & $V$   \\
87.82&5.0 & $16.02^{+0.08}_{-0.07}$ & $White$ & 1910.55& 364.92& $>$ 18.89& $V$  \\
92.82&5.0 &$16.12^{+0.08}_{-0.07}$ & $White$ &2255.54 &19.8&$>$18.36 &$V$  \\
97.82& 5.0& $16.42^{+0.09}_{-0.08}$ & $White$ & 6733.22& 199.8& $19.63^{+0.64}_{-0.40}$& $V$    \\
100.32& 20.0& $16.35^{+0.04}_{-0.04}$& $White$ &11470.66 &412.95 & $19.66^{+0.41}_{-0.30}$ & $V$   \\
102.82& 5.0&$16.29^{+0.08}_{-0.08}$ & $White$ & 53979.81& 199.8& $19.60^{+0.60}_{-0.39}$& $V$    \\
107.82& 5.0& $16.50^{+0.09}_{-0.08}$ & $White$ & 11470.66& 412.9 & $19.67^{+0.41}_{-0.30}$& $V$   \\
112.82 &5.0 & $16.42^{+0.09}_{-0.08}$ & $White$ & 53979.81&339.7& $>$ 20.61& $V$   \\
117.82& 5.0& $16.43^{+0.09}_{-0.08}$  & $White$ & 543.32&19.8 &$17.93^{+0.31}_{-0.24}$ & $B$  \\
120.32& 20.0& $16.56^{+0.04}_{-0.04}$& $White$ & 716.08 & 19.8 & $18.34^{+0.43}_{-0.31}$ & $B$   \\
122.82& 5.0& $16.58^{+0.09}_{-0.08}$ & $White$ &  1231.49 & 192.2& $18.17^{+0.25}_{-0.21}$&$B$  \\
127.82& 5.0 &$16.75^{+0.10}_{-0.09}$ & $White$ & 1577.50 & 191.8& $18.38^{+0.35}_{-0.26}$& $B$   \\
132.82& 5.0& $16.74^{+0.10}_{-0.09}$ & $White$ &  2008.91 &364.92 & $19.12^{+0.75}_{-0.44}$& $B$  \\
137.82& 5.0& $16.76^{+0.10}_{-0.09}$& $White$ & 6117.32  &199.8 &$19.89^{+0.34}_{-0.26}$ &$B$\\
142.82& 5.0&$16.81^{+0.10}_{-0.09}$ & $White$ & 7553.81 &199.8 & $20.26^{+0.73}_{-0.43}$& $B$  \\
142.82&25.0 & $16.83^{+0.04}_{-0.04}$& $White$& 36270.45 & 907.0& $>$21.50& $B$  \\
147.82& 5.0 &$16.86^{+0.10}_{-0.09}$ & $White$ & 61293.68& 7811.2& $>$ 22.19& $B$  \\
152.82& 5.0& $16.98^{+0.11}_{-0.10}$& $White$ & 288.13 &  20.0& $17.63^{+0.15}_{-0.13}$ & $U$ \\
157.82& 5.0& $16.84^{+0.10}_{-0.09}$& $White$ & 308.13  & 20.0 & $17.66^{+0.15}_{-0.13}$& $U$ \\
162.82& 5.0& $16.93^{+0.10}_{-0.09}$& $White$ &  328.13 & 20.0 & $17.61^{+0.15}_{-0.13}$& $U$  \\
167.82& 5.0 &$17.17^{+0.12}_{-0.11}$& $White$ & 348.13 & 20.0 & $17.67^{+0.15}_{-0.13}$& $U$  \\
170.32& 30.0& $17.06^{+0.04}_{-0.04}$& $White$ &368.13 & 20.0 & $17.78^{+0.0.16}_{-0.14}$ & $U$  \\
172.82&5.0& $17.12^{+0.11}_{-0.10}$& $White$ & 388.13 &20.0  &$18.10^{+0.20}_{-0.17}$ & $U$  \\
177.82&5.0 &$17.17^{+0.12}_{-0.10}$ & $White$ & 408.13 & 20.0 & $18.22^{+0.21}_{-0.18}$ &$U$ \\
182.82&5.0& $17.07^{+0.11}_{-0.10}$& $White$ & 428.13 & 20.0  & $17.71^{+0.15}_{-0.13}$& $U$ \\
187.82& 5.0&$17.29^{+0.12}_{-0.11}$ & $White$ & 448.13 & 20.0 & $18.10^{+0.20}_{-0.17}$& $U$    \\
192.82& 5.0&$17.20^{+0.12}_{-0.11}$ & $White$ & 468.13 & 20.0& $18.12^{+0.21}_{-0.17}$&$U$  \\
197.82&5.0 &$17.18^{+0.12}_{-0.11}$ & $White$ & 488.13 & 20.0 &$18.24^{+0.22}_{-0.18}$ &$U$   \\
200.20&29.76 & $17.31^{+0.05}_{-0.05}$& $White$ & 508.13& 20.0& $18.11^{+0.20}_{-0.17}$ & $U$    \\
202.82& 5.0& $17.24^{+0.12}_{-0.11}$& $White$ & 523.02 & 9.8 &$18.43^{+0.37}_{-0.28}$ &$U$  \\
207.82& 5.0& $17.44^{+0.13}_{-0.12}$& $White$ & 308.13 & 60.0  & $17.63^{+0.08}_{-0.08}$&$U$  \\
212.70& 4.8&$17.47^{+0.14}_{-0.12}$ & $White$ & 368.13& 60.0 &$17.86^{+0.09}_{-0.09}$ &$U$  \\
567.83&  19.75& $18.48^{+0.12}_{-0.11}$& $White$ & 438.13&80.0 & $18.04^{+0.09}_{-0.08}$ & $U$    \\
740.69& 19.75& $18.82^{+0.16}_{-0.14}$& $White$ &503.02 & 49.8& $18.23^{+0.13}_{-0.12}$ & $U$    \\
931.12& 149.76& $19.03^{+0.07}_{-0.06}$& $White$ &  691.33& 19.8& $18.34^{+0.23}_{-0.19}$ & $U$    \\
1255.78&192.09 & $19.23^{+0.15}_{-0.14}$& $White$ & 1206.62& 192.0& $19.20^{+0.31}_{-0.24}$ & $U$   \\
1602.07&192.42 & $19.46^{+0.21}_{-0.18}$& $White$ & 1552.79&191.9 & $18.90^{+0.27}_{-0.22}$ & $U$    \\
2033.22& 364.84& $19.53^{+0.21}_{-0.18}$& $White$ &1984.04 &364.7 & $19.19^{+0.35}_{-0.27}$ & $U$   \\
6321.93& 199.73& $20.68^{+0.23}_{-0.19}$& $White$ & 5911.80& 199.8& $20.16^{+0.22}_{-0.18}$ & $U$    \\
7758.75& 199.77& $21.15^{+0.57}_{-0.37}$& $White$ &7348.47 & 99.89200& $20.85^{+0.47}_{-0.32}$ & $U$    \\
58406.17& 211.27& $>$ 23.89 & $White$  &24964.78&725.6 &$21.39^{+0.48}_{-0.33}$ &$U$ \\
178945.23& 12160.06 & $>$ 23.94 & $White$  & 64529.61&906.8&$>$ 24.15&$U$ \\
&  &  &  & 89761.87&7192.6 &$>$ 23.59&$U$\\
\hline
\end{tabular}
\end{scriptsize}
\end{center}
\end{table*}


\begin{landscape}
\begin{scriptsize}
\begin{center}
\begin{longtable}{|c|c|c|c|c||c|c|c|c|c|}
\caption{Log of optical photometry observations of \thisgrbE afterglow. The magnitudes have not been corrected for Galactic Extinction E(B-V)= 0.0295. Magnitudes are in AB based system. Clear filter was calibrated to r filter.}
\label{optical_data}  \\ 
\hline 
 \bf $\rm \bf T_{start}$ (sec) & $\rm \bf T_{stop}$\bf (sec) & \bf Exposure (sec)  & \bf Magnitude  &\bf Filter &\bf $\rm \bf T_{start}$ (sec) & $\rm \bf T_{stop}$\bf (sec) & \bf Exposure (sec)  & \bf Magnitude  &\bf Filter\\
\hline 
\endfirsthead
\caption{continued.}\\
\hline
 \bf $\rm \bf T_{start}$ (sec) & $\rm \bf T_{stop}$\bf (sec) & \bf Exposure (sec)  & \bf Magnitude  &\bf Filter &\bf $\rm \bf T_{start}$ (sec) & $\rm \bf T_{stop}$\bf (sec) & \bf Exposure (sec)  & \bf Magnitude  &\bf Filter\\
\hline 
\endhead
\hline
\endfoot
29.2     &   29.7     &   0.5     & $   12.68 \pm   0.23 $  &    C& 144.3    &   144.8    &   0.5     & $   15.24 \pm   0.06 $  &    C   \\ 
 34.3     &   34.8     &   0.5     & $   12.97 \pm   0.08 $  &    C&  146.0 &   146.5    &   0.5     & $   15.43 \pm   0.07 $  &    C   \\ 
38.9     &   39.4     &   0.5     & $   13.20 \pm   0.12 $  &    C&  147.8  &   148.3    &   0.5     & $   15.57 \pm   0.09 $  &    C   \\ 

     40.7     &   41.2     &   0.5     & $   13.35 \pm   0.27 $  &    C&  149.5    &   150.0    &   0.5     & $   15.59 \pm   0.25 $  &    C   \\ 

     42.5     &   43.0     &   0.5     & $   13.37 \pm   0.19 $  &    C&  151.3    &   151.8    &   0.5     & $   15.47 \pm   0.15 $  &    C   \\ 

     44.2     &   44.7     &   0.5     & $   13.45 \pm   0.14 $  &    C&  153.0    &   153.5    &   0.5     & $   15.63 \pm   0.20 $  &    C   \\ 

     45.9     &   46.4     &   0.5     & $   13.45 \pm   0.34 $  &    C&  154.8    &   155.3    &   0.5     & $   15.66 \pm   0.08 $  &    C   \\ 

     47.7     &   48.2     &   0.5     & $   13.63 \pm   0.19 $  &    C&  156.5    &   157.0    &   0.5     & $   15.75 \pm   0.15 $  &    C   \\ 

     49.4     &   49.9     &   0.5     & $   13.54 \pm   0.10 $  &    C&  158.2    &   158.7    &   0.5     & $   15.57 \pm   0.20 $  &    C   \\ 

     51.2     &   51.7     &   0.5     & $   13.73 \pm   0.14 $  &    C&  160.0    &   160.5    &   0.5     & $   15.59 \pm   0.19 $  &    C   \\ 

     52.9     &   53.4     &   0.5     & $   13.78 \pm   0.17 $  &    C&  161.7    &   162.2    &   0.5     & $   15.57 \pm   0.25 $  &    C   \\ 

  54.7     &   55.2     &   0.5     & $   13.87 \pm   0.14 $  &    C&  188.1    &   193.9    &   4x0.5   & $   16.04 \pm   0.08 $  &    C   \\ 

  56.4     &   56.9     &   0.5     & $   13.83 \pm   0.17 $  &    C&  195.2    &   200.9    &   4x0.5   & $   16.07 \pm   0.08 $  &    C   \\ 

  58.1     &   58.6     &   0.5     & $   13.97 \pm   0.09 $  &    C&  202.2    &   212.3    &   4x0.5   & $   16.09 \pm   0.12 $  &    C   \\

  59.9     &   60.4     &   0.5     & $   14.03 \pm   0.12 $  &    C&  213.6    &   219.3    &   4x0.5   & $   16.15 \pm   0.25 $  &    C   \\ 

  61.6     &   62.1     &   0.5     & $   13.99 \pm   0.11 $  &    C&  220.5    &   228.1    &   5x0.5   & $   16.01 \pm   0.13 $  &    C   \\

  63.4     &   63.9     &   0.5     & $   14.21 \pm   0.15 $  &    C&   240.6    &   246.4    &   4x0.5   & $   16.26 \pm   0.08 $  &    C   \\ 

  65.1     &   65.6     &   0.5     & $   14.16 \pm   0.09 $  &    C&  247.7    &   253.3    &   4x0.5   & $   16.36 \pm   0.11 $  &    C   \\ 

  66.8     &   67.3     &   0.5     & $   14.28 \pm   0.23 $  &    C&  255.5    &   261.3    &   4x0.5   & $   16.61 \pm   0.06 $  &    C   \\ 
  
  68.6     &   69.1     &   0.5     & $   14.17 \pm   0.05 $  &    C&  262.6    &   268.4    &   4x0.5   & $   16.57 \pm   0.06 $  &    C   \\
  
  70.4     &   70.9     &   0.5     & $   14.26 \pm   0.12 $  &    C&  269.6    &   277.4    &   4x0.5   & $   16.69 \pm   0.11 $  &    C   \\ 

  72.1     &   72.6     &   0.5     & $   14.40 \pm   0.16 $  &    C&  278.7    &   284.4    &   4x0.5   & $   16.45 \pm   0.10 $  &    C   \\ 

  73.8     &   74.3     &   0.5     & $   14.42 \pm   0.09 $  &    C&  285.6    &   291.4    &   4x0.5   & $   16.40 \pm   0.09 $  &    C   \\ 

  75.6     &   76.1     &   0.5     & $   14.49 \pm   0.19 $  &    C&  292.6    &   298.3    &   4x0.5   & $   16.75 \pm   0.06 $  &    C   \\ 

  77.3     &   77.8     &   0.5     & $   14.46 \pm   0.13 $  &    C&  299.6    &   307.1    &   4x0.5   & $   16.56 \pm   0.09 $  &    C   \\ 

  79.1     &   79.6     &   0.5     & $   14.56 \pm   0.12 $  &    C&  308.3    &   314.1    &   4x0.5   & $   16.45 \pm   0.10 $  &    C   \\ 

  80.8     &   81.3     &   0.5     & $   14.51 \pm   0.05 $  &    C&  315.3    &   321.1    &   4x0.5   & $   16.72 \pm   0.07 $  &    C   \\ 

  82.6     &   83.1     &   0.5     & $   14.49 \pm   0.13 $  &    C&  322.3    &   326.3    &   3x0.5   & $   16.49 \pm   0.14 $  &    C   \\ 

  84.3     &   84.8     &   0.5     & $   14.68 \pm   0.15 $  &    C&  327.8    &   339.1    &   2x5.0   & $   16.52 \pm   0.08 $  &    C   \\ 

  86.1     &   86.6     &   0.5     & $   14.73 \pm   0.12 $  &    C&  340.3    &   351.6    &   2x5.0   & $   17.03 \pm   0.06 $  &    C   \\ 

  87.8     &   88.3     &   0.5     & $   14.54 \pm   0.12 $  &    C&  352.8    &   364.1    &   2x5.0   & $   16.97 \pm   0.04 $  &    C   \\ 

  89.5     &   90.0     &   0.5     & $   14.63 \pm   0.11 $  &    C&  365.3    &   376.5    &   2x5.0   & $   16.76 \pm   0.04 $  &    C   \\ 

  96.9     &   97.4     &   0.5     & $   14.84 \pm   0.09 $  &    C&  377.8    &   389.0    &   2x5.0   & $   16.89 \pm   0.05 $  &    C   \\ 

  98.6     &   99.1     &   0.5     & $   15.02 \pm   0.12 $  &    C&  390.3    &   401.5    &   2x5.0   & $   16.95 \pm   0.05 $  &    C   \\

  100.3    &   100.8    &   0.5     & $   14.89 \pm   0.12 $  &    C&  485.3    &   496.6    &   2x5.0   & $   17.35 \pm   0.10 $  &    C   \\

  102.1    &   102.6    &   0.5     & $   15.10 \pm   0.15 $  &    C&  497.9    &   509.1    &   2x5.0   & $   17.34 \pm   0.06 $  &    C   \\

  111.1    &   111.6    &   0.5     & $   15.18 \pm   0.18 $  &    C&  510.4    &   527.2    &   2x5.0   & $   17.43 \pm   0.07 $  &    C   \\

  112.9    &   113.4    &   0.5     & $   15.10 \pm   0.12 $  &    C&  528.4    &   539.7    &   2x5.0   & $   17.36 \pm   0.09 $  &    C   \\

  114.7    &   115.2    &   0.5     & $   15.23 \pm   0.16 $  &    C&  540.9    &   552.2    &   2x5.0   & $   17.39 \pm   0.04 $  &    C   \\

  116.4    &   116.9    &   0.5     & $   15.34 \pm   0.08 $  &    C&  553.4    &   572.2    &   2x5.0   & $   17.35 \pm   0.06 $  &    C   \\ 

  118.2    &   118.7    &   0.5     & $   15.26 \pm   0.16 $  &    C&  727.6    &   757.6    &   30.0    & $   17.75 \pm   0.05 $  &    C   \\ 
  119.9    &   120.4    &   0.5     & $   15.28 \pm   0.21 $  &    C&  758.9    &   788.9    &   30.0    & $   17.78 \pm   0.03 $  &    C   \\ 

  121.6    &   122.1    &   0.5     & $   15.28 \pm   0.15 $  &    C&  1082.6   &   1112.6   &   30.0    & $   18.17 \pm   0.04 $  &    C   \\ 

  123.4    &   123.9    &   0.5     & $   15.32 \pm   0.19 $  &    C&  1113.9   &   1143.9   &   30.0    & $   18.11 \pm   0.05 $  &    C   \\ 

  125.1    &   125.6    &   0.5     & $   15.16 \pm   0.17 $  &    C&  1729.3   &   1849.3   &   120.0   & $   18.43 \pm   0.06 $  &    C   \\ 

  126.9    &   127.4    &   0.5     & $   15.28 \pm   0.10 $  &    C&  2037.1   &   2157.1   &   120.0   & $   18.62 \pm   0.06 $  &    C   \\ 

  128.6    &   129.1    &   0.5     & $   15.37 \pm   0.11 $  &    C&  2341.6   &   2461.6   &   120.0   & $   18.82 \pm   0.04 $  &    C   \\ 

  130.4    &   130.9    &   0.5     & $   15.22 \pm   0.24 $  &    C&  3163.0   &   3283.0   &   120.0   & $   18.79 \pm   0.05 $  &    C   \\ 

  132.1    &   132.6    &   0.5     & $   15.58 \pm   0.10 $  &    C&  3470.7   &   3590.7   &   120.0   & $   18.96 \pm   0.05 $  &    C   \\ 

  133.9    &   134.4    &   0.5     & $   15.50 \pm   0.08 $  &    C&  3775.2   &   3895.2   &   120.0   & $   19.02 \pm   0.05 $  &    C   \\ 

  135.6    &   136.1    &   0.5     & $   15.37 \pm   0.27 $  &    C&  4619.7   &   4739.7   &   120.0   & $   18.93 \pm   0.04 $  &    C   \\ 

  137.3    &   137.8    &   0.5     & $   15.54 \pm   0.18 $  &    C&  4978.0   &   5098.0   &   120.0   & $   19.14 \pm   0.04 $  &    C   \\ 

  139.1    &   139.6    &   0.5     & $   15.70 \pm   0.09 $  &    C&  5282.5   &   5402.5   &   120.0   & $   19.01 \pm   0.05 $  &    C   \\ 

  140.8    &   141.3    &   0.5     & $   15.56 \pm   0.20 $  &    C&  448.7    &   468.7    &   20.0    & $   17.66 \pm   0.08 $  &    g   \\ 

  142.6    &   143.1    &   0.5     & $   15.60 \pm   0.08 $  &    C&   622.3    &   642.3    &   20.0    & $   17.94 \pm   0.04 $  &    g  \\

  643.6    &   663.6    &   20.0    & $   18.04 \pm   0.08 $  &    g &   1853.2   &   2033.2   &   180.0   & $   18.56 \pm   0.07 $  &    r   \\ 

  854.9    &   914.9    &   60.0    & $   18.12 \pm   0.05 $  &    g   &   3286.8   &   3466.8   &   180.0   & $   19.29 \pm   0.08 $  &    r   \\ 

  1209.9   &   1269.9   &   60.0    & $   18.27 \pm   0.06 $  &    g   &  4768.9   &   4948.9   &   180.0   & $   19.32 \pm   0.07 $  &    r   \\ 
  2159.4   &   2339.4   &   180.0   & $   19.08 \pm   0.04 $  &    g&  918.6    &   978.6    &   60.0    & $   18.05 \pm   0.13 $  &    i   \\ 

  3592.9   &   3772.9   &   180.0   & $   19.47 \pm   0.05 $  &    g&  1273.6   &   1333.6   &   60.0    & $   18.20 \pm   0.11 $  &    i   \\ 

  5100.3   &   5280.3   &   180.0   & $   19.65 \pm   0.05 $  &    g&  2467.1   &   2647.1   &   180.0   & $   18.67 \pm   0.07 $  &    i   \\ 

  405.3    &   425.3    &   20.0    & $   17.35 \pm   0.10 $  &    r&  3900.6   &   4080.6   &   180.0   & $   18.60 \pm   0.05 $  &    i   \\ 

  426.6    &   446.6    &   20.0    & $   17.30 \pm   0.07 $  &    r&  5408.0   &   5588.0   &   180.0   & $   19.13 \pm   0.12 $  &    i   \\ 

  578.9    &   598.9    &   20.0    & $   17.75 \pm   0.16 $  &    r&  1335.9   &   1515.9   &   180.0   & $   17.96 \pm   0.15 $  &    Z   \\ 

  600.2    &   620.2    &   20.0    & $   17.71 \pm   0.10 $  &    r&  2649.4   &   2889.4   &   240.0   & $   18.81 \pm   0.23 $  &    Z   \\ 

  792.7    &   852.7    &   60.0    & $   18.17 \pm   0.08 $  &    r&  4083.0   &   4323.0   &   240.0   & $   19.27 \pm   0.21 $  &    Z   \\ 

  1147.7   &   1207.7   &   60.0    & $   18.57 \pm   0.08 $  &    r&
  
  5590.4   &   5830.4   &   240.0   & $   18.98 \pm   0.14 $  &    Z   \\
  
    5668   &   5768   &   100.0    & $   18.99 \pm   0.01 $  &    R&
  
  5794  &   5894    &   100.0   & $   19.05 \pm   0.05 $  &    R   \\
  
    5913  &  6013    &   100.0    & $   19.06 \pm   0.05 $  &    R&
  
   96259   &   98238   &   300.0 $\times$ 6  & $   22.21 \pm   0.27 $  &    R  \\
  \hline
\end{longtable}
\end{center}
\end{scriptsize}
\end{landscape}


\addtocounter{page}{1}
\appendix
\bibliographystyle{Classes/mnras} 
\renewcommand{\bibname}{References} 
\bibliography{bib/Thesis_Rahul}
\addcontentsline{toc}{chapter}{\sc References} 
\end{document}